\RequirePackage{luatex85}

\documentclass[
    a4paper,
    11pt, 
    twoside, 
    openright 
]{memoir}
\chapterstyle{veelo}
\usepackage{iftex}

\usepackage{graphicx}
\usepackage{float}
\usepackage[shortlabels]{enumitem}

\ifPDFTeX
\usepackage[utf8]{inputenc}
\usepackage[T1]{fontenc}
\fi
\usepackage{microtype}

\setitemize{itemsep=0em}

\newcommand\numberthis{\addtocounter{equation}{1}\tag{\theequation}}

\usepackage{calrsfs}
\DeclareMathAlphabet{\pazocal}{OMS}{zplm}{m}{n}

\usepackage{amsmath}
\usepackage{amssymb}
\usepackage{amsthm}
\usepackage{thmtools}
\usepackage{mathtools}

\ifPDFTeX

\usepackage{calrsfs}
\DeclareMathAlphabet{\pazocal}{OMS}{zplm}{m}{n}

\usepackage{libertine}
\usepackage[scaled=0.96]{zi4} 

\fi

\usepackage{microtype}

\usepackage{abstract}
\usepackage[backend=bibtex,style=numeric-comp,doi=false,url=false,maxnames=10]{biblatex}

\usepackage[ruled,vlined,linesnumbered,algo2e,resetcount,algochapter]{algorithm2e}
\usepackage{algorithm}

\usepackage{tikz}

\usepackage{xspace}
\usepackage{hyphenat} 

\usepackage{accents}

\newcommand\LL{\mathbf{L}}
\newcommand\Lpseudo{\mathbf{L}^\dagger}
\newcommand\DD{\mathbf{D}}
\newcommand\AAA{\mathbf{A}}
\newcommand\bb{\mathbf{b}}
\newcommand\xx{\mathbf{x}}
\newcommand\ww{\mathbf{w}}
\newcommand\ones{\mathbf{1}}
\newcommand\Reff{R_{\textrm{eff}}}

\newcommand{\vecnorm}[1]{\left\lVert#1\right\rVert}

\newcommand\distI{\textrm{dist}}
\newcommand\stretchI{\textrm{stretch}}
\newcommand\avStretchI{\textrm{avg-stretch}}

\DeclarePairedDelimiterX{\set}[2]{\{}{\}}{#1\,\delimsize|\,\mathopen{}#2}
\DeclarePairedDelimiter{\abs}{\lvert}{\rvert}
\DeclarePairedDelimiterX{\floor}[1]{\lfloor}{\rfloor}{#1}
\newcommand{\fraction}{\operatorname{frac}}
\newcommand{\nil}{\textsc{nil}}
\newcommand{\expdis}{\operatorname{Exp}}
\newcommand{\str}{\operatorname{stretch}}
\newcommand{\avestr}{\operatorname{avg-stretch}}
\newcommand{\capacity}{\operatorname{cap}}

\newcommand{\poly}{\operatorname{poly}}
\newcommand{\polylog}{\operatorname{polylog}}
\newcommand{\dist}{\textrm{dist}}

\newcommand{\lev}{\ell}

\newcommand{\degree}{\operatorname{deg}}
\newcommand{\fract}{\beta}
\newcommand{\diam}{\Delta}
\renewcommand{\Pr}{\mathbb{P}}
\DeclareMathOperator{\mincut}{\textrm{min-cut}}

\newcommand{\R}{\mathbb{R}}
\newcommand{\vect}[1]{\ensuremath{\mathbf{#1}}}
\newcommand{\mat}[1]{\ensuremath{\mathbf{#1}}}
\newcommand{\1}{\textbf{1}}
\newcommand{\child}{\textrm{c}}
\newcommand{\EE}{\mathcal{E}}
\newcommand{\buffer}{\textrm{X}}

\DeclareMathOperator{\expec}{\mathbb{E}}
\DeclareMathOperator{\maxflow}{max-flow}
\DeclareMathOperator{\congestion}{cong}
\global\long\def\cutsp{\textsc{VertexSparsify}}
\newcommand\dd{\mathbf{d}}

\newcommand{\expect}[1]{\mathbb{E}\left[#1\right]}
\newcommand{\Gs}{G^*}

\newcommand{\kk}{k}
\newcommand{\cc}{\mathbf{c}}

\newcommand{\la}{\lambda}

\newcommand{\calO}{O}
\newcommand{\rr}{\mathbb{R}}
\newcommand{\ra}{\rightarrow}
\newcommand{\calT}{\mathcal{K}}
\newcommand{\nn}{\mathbb{N}}

\def \eps {{\epsilon}}
\newcommand{\Property}{\mathcal{P}}

\newcommand\bbtil{\tilde{\bb}}
\newcommand\xxtil{\tilde{\xx}}
\newcommand{\LP}{\LL^\dag}
\newcommand\Otil{\tilde{O}}
\newcommand\SC{\mathbf{SC}}
\newcommand\boldone{\boldsymbol{1}}
\newcommand\boldzero{\boldsymbol{0}}
\def\prob#1#2{\mathbb{P}_{#1}\left[ #2 \right]}
\newcommand\cchi{\boldsymbol{\chi}}
\newcommand\er{R_{\textrm{eff}}}
\newcommand{\proj}[2]{\mathbf{P}({#2})}
\newcommand\II{\mathbf{I}}
\renewcommand\AA{\mathbf{A}}
\newcommand\BB{\mathbf{B}}
\newcommand\Htil{\tilde{H}}
\newcommand\mhat{{\hat{{m}}}}
\newcommand{\pnew}{\mathit{p}^{\mathrm{new}}}
\newcommand\pp{\mathbf{p}}
\newcommand\ff{\mathbf{f}}
\newcommand\WW{\mathbf{W}}
\newcommand\qq{\mathbf{q}}
\newcommand\rrvec{\mathbf{r}}
\newcommand{\escape}{X(u,U)}
\newcommand\pmf[3]{f_{s(w),#1}^{#2,#3}}
\newcommand\pmfg[3]{g_{#1}^{#2,#3}}
\newcommand\LLtil{\tilde{\LL}}
\def\norm#1{\left\| #1 \right\|}
\newcommand{\newS}{\tilde{S}}
\newcommand\vv{\mathbf{v}}

\newcommand\XX{\mathbf{X}}
\newcommand\PPi{\boldsymbol{\Pi}}

\def\expec#1#2{{\mathbb{E}}_{#1}\left[ #2 \right]}
\newcommand\HH{\mathbf{H}}

\DontPrintSemicolon
\SetKw{KwAnd}{and}
\SetProcNameSty{textsc}
\SetFuncSty{textsc}

\SetKwInOut{Input}{Input}
\SetKwInOut{Output}{Output}

\let\oldnl\nl
\newcommand{\nonl}{\renewcommand{\nl}{\let\nl\oldnl}}

\SetKwProg{Procedure}{Procedure}{}{}
\SetKwFunction{Insert}{insert}
\SetKwFunction{Delete}{delete}
\SetKwFunction{Relax}{relax}
\SetKwFunction{Query}{query}
\SetKwFunction{Distance}{distance}
\SetKwFunction{Partition}{partition}
\SetKwFunction{Initialize}{initialize}
\SetKwFunction{Increase}{increase}
\SetKwFunction{UpdateLevels}{updateLevels}

\usepackage{cleveref} 

\title{Dynamic Graph Algorithms and Graph Sparsification: New
Techniques and Connections}
\author{Gramoz Goranci~M.\,Sc.}

\hypersetup{
    pdftitle = {Dynamic Graph Algorithms and Graph Sparsification: New
Techniques and Connections},
    pdfauthor = {Gramoz Goranci},
    pdfsubject = {Dissertation},
    pdfkeywords = {theoretical computer science, efficient algorithms}
}

\setsecnumdepth{subsection}
\settocdepth{subsection}

\usetikzlibrary{decorations.pathreplacing,positioning,shapes.misc,calc}

\DontPrintSemicolon
\SetProcNameSty{textsc}
\SetFuncSty{textsc}
\SetAlFnt{\small}

\allowdisplaybreaks[1]

\declaretheorem[numberwithin=section]{theorem}
\declaretheorem[numberlike=theorem]{lemma}
\declaretheorem[numberlike=theorem]{corollary}
\declaretheorem[numberlike=theorem]{proposition}
\declaretheorem[numberlike=theorem]{claim}
\declaretheorem[numberlike=theorem]{invariant}
\declaretheorem[numberlike=theorem]{conjecture}
\declaretheorem[numberlike=theorem]{definition}
\declaretheorem[style=remark,numberlike=theorem]{remark}
\declaretheorem[numberlike=theorem]{fact}
\declaretheorem[numberlike=theorem]{observation}

\begin{document}

\frontmatter

\makepagestyle{titlepage}
\makeoddhead{titlepage}{}{}{~~~~~~~~~~~~~~~~~~~~~~~~~~~~~~~~~~~~~~~~~~~~~~~~~~~~~~~~~~~~~~~~~~~~~~~~~~~~~~~~~~~~~~~~~~~~~~~~~~~~~~~~~~~~~~~~~~~~~~~~~~~\includegraphics{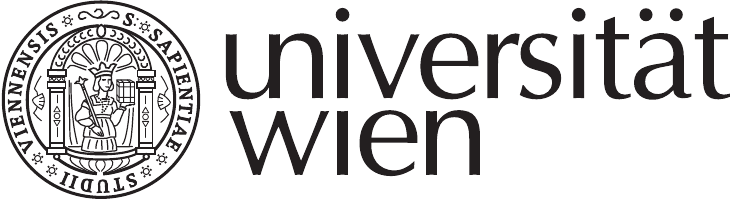}}
\calccentering{\unitlength}                         
\begin{adjustwidth*}{\unitlength}{-\unitlength}     
\begin{adjustwidth}{-1cm}{-1cm}

\thispagestyle{titlepage}
{\centering

\sffamily

~
\vfill

\vfill

\HUGE \textbf{\textsc{Dissertation / Doctoral Thesis}}\\

\vfill

\normalsize Titel der Dissertation / Title of the Doctoral Thesis \\

\huge \textbf{\thetitle}

\vfill

\normalsize verfasst von / submitted by \\
\Large \theauthor

\vfill

\normalsize angestrebter akademischer Grad / in partial fulfillment of the requirements for the degree of \\
\Large Doktor der Technischen Wissenschaften (Dr.\,techn.)

\vfill

\normalsize

\begin{tabbing}
field of study as it appears on the student record sheet:  \hspace{0.5em} \= Informatik \kill
Wien, 2019 / Vienna, 2019 \\
\\
Studienkennzahl lt. Studienblatt / \>  A 786 880 \\
degree programme code as it appears on the student \>  \\
record sheet: \>  \\
Dissertationsgebiet lt. Studienblatt /  \> Informatik \\
field of study as it appears on the student record sheet:  \> \\
Betreuerin / Supervisor: \> Univ.-Prof.\ Dr.\ Monika Henzinger \\
\end{tabbing}

\vspace{-7ex}

~
} 

\end{adjustwidth}
\end{adjustwidth*}

\normalfont


\cleardoublepage
\begin{abstract}
Graphs naturally appear in several real-world contexts including social networks, the web network, and telecommunication networks. While the analysis and the understanding of graph structures have been a central area of study in algorithm design, the rapid increase of data sets over the last decades has posed new challenges for designing efficient algorithms that process large-scale graphs. These challenges arise from two usual assumptions in classical algorithm design, namely that graphs are static and that they fit into a single machine. However, in many application domains, graphs are subject to frequent changes over time, and their massive size makes them infeasible to be stored in the memory of a single machine. 

Driven by the need to devise new tools for overcoming such challenges, this thesis focuses in two areas of modern algorithm design that directly deal with processing massive graphs, namely dynamic graph algorithms and graph sparsification. We develop new algorithmic techniques from both dynamic and sparsification perspective for a multitude of graph-based optimization problems which lie at the core of Spectral Graph Theory, Graph Partitioning and Metric Embeddings. Our algorithms are faster than any previous one and design smaller sparsifiers with better (approximation) quality. More importantly, this work introduces novel reduction techniques that show unexpected connections between seemingly different areas such as dynamic graph algorithms and graph sparsification. In particular we obtain the following results:
\begin{itemize}
\item The first dynamic algorithm for maintaining approximate solutions to Laplacian systems in sub-linear update and query time and an extension of the technique to dynamically maintaining variants of Vertex Spectral Sparsifiers and All-Pair Effective Resistances in undirected, weighted graphs. We also prove conditional lower bounds that certify that there are no efficient dynamic algorithms for maintaining Effective Resistances exactly. 
\item The first dynamic algorithm for maintaining low-stretch spanning trees with sub-polynomial stretch and sub-linear update time in undirected, unweighted graphs and an extension of the technique to dynamically maintaining low-diameter clustering. 
\item The current best-known algorithms for incrementally maintaining global Minimum Cut, approximate All-Pair Maximum Flow and Sparsest Cut in undirected, unweighted graphs. A key primitive behind our algorithms is a new notion of Local Sparsifiers, a stronger variant of the well-studied notion of Vertex Sparsifiers. 
\item The current best-known algorithm for constructing vertex sparsifiers that are minors of the input graph and preserve shortest path distances approximately and reachability information exactly. We derive upper-bounds on the quality and size of such sparsifiers and also prove lower-bounds that better explain the trade-off between these two quantities.
\end{itemize}
\end{abstract}

\cleardoublepage
\begin{abstract}
Graphen sind passende Modelle in mehreren realen Kontexten, unter anderem in sozialen Netzwerken, dem Web-Netzwerk und in Telekommunikationsnetzen. Die Analyse und das Verst\"andnis von Graphstrukturen sind ein zentraler Gesichtspunkt im Design von Algorithmen. Jedoch stellt das rasante Wachstum an Datenmengen neue Herausforderungen an das Design von effizienten Algorithmen f\"ur riesige Graphen. Diese Herausforderungen entspringen aus zwei Annahmen des klassischen Algorithmendesigns, und zwar dass Graphen statisch sind und in den Speicher einer einzelnen Machine passen. Jedoch sind Graphen in vielen Anwendungen in konstanter Ver\"anderung und oftmals zu gro\ss, um im Speicher einer einzelnen Maschine gespeichert werden zu k\"onnen. 

Getrieben durch den Bedarf, neue L\"osungen f\"ur diese Herausforderungen zu finden, fokussiert sich diese Dissertation auf zwei Bereiche des modernen Algorithmendesigns, um L\"osungen f\"ur diese Probleme zu finden; n\"amlich dynamische Graphalgorithmen und Graphsparsifikation. Wir entwickeln neue algorithmische Techniken f\"ur beide Bereiche, um graphbasierte Optimierungsprobleme unter anderem in Spectral Graph Theory, Graph Partitioning und Metric Embeddings effizienter l\"osen zu k\"onnen. Unsere Algorithmen sind schneller als jegliche vorherige und wir entwickeln kleinere Sparsifier mit besserer~Approximationsqualit\"at. Au\ss erdem entwickelt diese Arbeit neuartige Reduktionstechniken, welche unerwartete Zusammenh\"ange zwischen scheinbar verschiedenen Bereichen, wie zum Beispiel dynamische Graphalgorithmen und Graphsparsifikation aufzeigen, insbesondere erreichen wir die folgende Resultate: 
\begin{itemize}

\item
Den ersten dynamischen Algorithmus f\"ur die Aufrechterhaltung von approximativen L\"osungen f\"ur Laplacian Systeme mit sub-linearer Update und Query Time und eine Erweiterung der Technik, um dynamisch Vertex Spectral Sparsifiers und All-Pair Effective Resistances in ungerichteten, gewichteten Graphen aufrechtzuerhalten. Wir beweisen au\ss erdem Conditional Lower Bounds, welche beweisen, dass es keinen effizienten dynamischen Algorithmus geben kann, der Effective Resistances exakt berechnet. 
\item
Den ersten dynamischen Algorithmus, der Low-stretch aufspannende B\"aume mit sub-polynomialem Stretch und sub-lineare Update Time in ungerichteten, ungewichteten Graphen aufrecht erh\"alt. Au\ss erdem eine Erweiterung der Technik, um dynamische Cluster mit niedrigem Diameter aufrechtzuerhalten. 

\item
Den besten bekannten Algorithmus f\"ur inkrementellen global Minimum Cut, approximativen All-Pair Maximum Flow und Sparsest Cut in ungerichteten, ungewichteten Graphen. Ein wichtiger Baustein f\"ur diese Algorithmen ist ein neuentwickeltes Konzept namens Local Sparsifier, eine st\"arkere Variante der bekannten Vertex Sparsifier.

\item
Den besten bekannten Algorithmus f\"ur die Konstruktion von Vertex Sparsifiers, die Minors des Inputgraphen sind und eine Approximation der k\"urzesten Wege und die exakte Erreichbarkeit aufrecht erhalten. Wir entwickeln obere Schranken f\"ur die Qualit\"at und Gr\"o\ss e solcher Sparsifier und beweisen untere Schranken, welche den Kompromiss der beiden Zielfunktionen besser erkl\"aren.
\end{itemize} 
\end{abstract}

\cleardoublepage
\renewcommand{\abstractname}{Acknowledgments}
\begin{abstract}
\fontsize{10.8pt}{12.96pt}

{\selectfont
First and foremost, I would like to express my sincerest gratitude to my advisor Monika Henzinger, for her guidance and patience over these last four years of my Ph.D. She has always been there to listen to my ideas, generously shared her research insights with me and encouraged me to work on topics that I was most excited about. She gave never-ending support, especially when things did not go smoothly. She has been, and continues to be, a great source of inspiration for me, both in terms of academic and personal development.

I am deeply thankful to Harald R\"acke, for being a great mentor and collaborator, and for hosting me several times at TU Munich. His approach and intuition to problem solving continue to amaze me, and I feel fortunate to have learned from his insights.

I am particularly grateful to Richard Peng, who hosted me during my research stay at Georgia Tech in Atlanta. I have benefited immensely from his generosity, research insights in numerical and graph algorithms, his academic advice, guidance and our joint collaborations. 
 
Special thanks to Robert Krauthgamer and Danupon Nanongkai who have agreed to review this thesis and be part of my thesis committee. I have enjoyed and gained a lot from research conversations with both of them. I would also like to thank Thatchaphol Saranurak and Mikkel Thorup for the remarkable research collaborations.

It has been a great experience to spend these years as part of the TAA group. I want to thank Sebastian for being a great office mate, for the collaboration and the academic advise; Pan for all the collaboration and exceptional support; Darek for working with me, being an inspiring friend and for playing music together; Marco for being the co-author of my first paper; Veronika for being a caring colleague and convincing me to come to Vienna for my Ph.D.; Stefan for being there throughout our joint path and for all the conversations on life; Wolfgang for his support, and for organizing several entertaining non-academic events; Alex for being a kind and cheerful office mate; Valon and Labinot for the numerous conservations about life and politics during lunches at Mensa; Ulli, Birgit and Christina for keeping the administrative workload low and allowing me to practice my German with them, former and current colleagues, and visitors for the pleasant atmosphere they have created. I have had a very rewarding experience during my visit in Atlanta, and for this I thank David, Saurabh, Yu, Matthew and Di for being great friends and colleagues.

I am indebted to the endless support and love my family has given me during these years of graduate school. I thank my parents, Suzana and Kamuran, my siblings Gladiola and Shkamb and their spouses Labi and Ida, and my in-laws Shpresa, Nail and Pellumb, thank you all for being there for me! A special thank you goes to my grandparents, especially to my late grandfather Rifat, who was a great source of inspiration for me and infused within me the passion for science.

Finally, I would like to wholeheartedly thank my fianc\`e Edona for her love, support and for cheering me up (especially after conference notifications), and reminding me often that life is a strict superset of research. Thank you for also proof-reading many of my manuscripts, grant proposals and parts of this thesis. 

}

\end{abstract}

\cleardoublepage
\renewcommand{\abstractname}{Funding Acknowledgments}
\begin{abstract}
The research leading to these results has received funding from the European Research Council under the European Union's Seventh Framework Programme (FP/2007-2013) / ERC Grant Agreement no. 340506, and the Doctoral Programme ``Vienna Graduate School on Computational Optimization'' funded by the Austrian Science Fund (FWF), project no. W1260-N35.
\end{abstract}

\cleardoublepage
\renewcommand{\abstractname}{Bibliographic Note}
\begin{abstract}
Most of the results of this thesis were already published in conference proceedings and journal articles. Therefore, the chapters of this thesis are based on the following publications and manuscripts:
\begin{itemize}
\itemsep0.5em
\item \textbf{\Cref{cha:ESA2018_ER}:} \fullcite{GoranciHP18}. 

\item \textbf{\Cref{cha:STOC2019_DER}:} \fullcite{DurfeeGGP19}.

\item \textbf{\Cref{cha:STOC2019_LSST}:} \fullcite{GoranciK18:arxiv}.

\item \textbf{\Cref{cha:TALG2018_IMC}:} \fullcite{GoranciHT18}.

\item \textbf{\Cref{cha:Man2019_LS}:} \fullcite{GoranciHS19}.

\item \textbf{\Cref{cha:ICALP2016_DM}:} \fullcite{cheung2016}.

\item \textbf{\Cref{cha:ESA2017_RM}:} \fullcite{GoranciHP17}.

\end{itemize}

\end{abstract}

\cleardoublepage
\tableofcontents*

\mainmatter

\chapter{Introduction}

Recent technological developments, in particular the increased popularity of the Internet, have lead to an enormous increase in data volumes. While the access to this large amount of data has allowed us to gain complex insights and identify various patterns about data, performing basic computational tasks and storing the data have posed major challenges that require new treatments beyond the usage of traditional applications and tools. This leaves computer scientists the mandate to address such challenges by developing models that better suit the modern technological advances.

A considerable fraction of the generated data can be modeled using \emph{Graphs}. A graph is a collection of nodes and a collection of edges, where each edge connects a pair of nodes. Graphs are ubiquitous structures in mathematics and computer science, and naturally appear in several real-world contexts including social networks (e.g., the Facebook graph), the web network, and telecommunication networks. In comparison to other data representations, graphs are particularly desirable since (suitably) visualizing them often offers ways to identify interesting patterns in the data, e.g., detecting communities in social networks. Another reason why graph representations are found appealing is because algorithms for manipulating and storing them have been thoroughly studied since the early days of algorithm design. Nevertheless, a large number of these graph algorithms work under the assumptions that graphs are static, i.e., that they do not undergo changes and that they can be stored into the memory of a single machine. Unfortunately, these assumptions fail to capture graphs that appear in many important real-world scenarios.

As a motivating example consider a map graph, where each node corresponds to a city and an edge between any two nodes represents the route connecting them. This map graph also includes the length of each route by labeling the edges with the corresponding distance. A fundamental question in algorithm design is to understand the metric structure of the map graph; more concretely, we want to compute the shortest path distance between any two given nodes in the map graph. This task has been addressed by many classical algorithms and the running time complexity of this problem is shown to be cubic on the number of nodes in the graph. However, it does not take too much effort to realize that real-world map graphs undergo changes. For example, due to construction work, it may occur that some road connecting two cities is blocked, which in turn implies that the edge connecting these cities is deleted from the map graph. The deletion of such an edge might affect the shortest path between cities, thus implying that the old solution is incorrect for the new map graph. One obvious way to correct the solution is to recompute shortest paths from scratch in the new graph. However, this trivial solution comes at the expense of high computational burden, which is not feasible for small devices with limited resources, e.g., navigation systems. A set of natural questions that arise are the following: Can we design methods that perform better than re-computation from scratch? If yes, what is the best possible speed-up that we can achieve? Do we have to pay in the quality of the solution to get better performance?

Another common challenge that we face when dealing with huge graphs are computational and storage resources. This is due to the fact that the size of a graph can be as large as quadratic in the number of nodes. A traditional approach to address this issue has been to \emph{compress} large graphs into smaller ones while preserving properties or features of interest. These compressed versions of graphs are particularly desirable since any computational task on the original graph can be now performed on the compressed graph, thus leading to significant savings in computational and storage resources. Graph compression is commonly studied from two perspectives: (1) reducing the number of edges of a graph, (2) and reducing the number of nodes. While the first approach has been successfully employed for improving the running time of many basic graph problems, its practical applicability is somewhat limited due to the fact that most large networks are already sparse. As a result, compression tools that reduce the number of nodes have received increasing attention over the last decade. To illustrate, let us go back to the graph map example. Suppose that among all cities in the graph, we are interested only in a subset of nodes that are ``important'' to us. This is relevant in many practical scenarios, e.g., one desires to preserve distance information only among big cities while ignoring the small ones. The following questions naturally arise: Can we compress the map graph into a graph only on the big cities while preserving distances? What is the incurred loss when transferring from the large map graph to the smaller one? What is the trade-off between the loss and the size of the compressed graph?

All of the above questions and their variants will be addressed in this thesis. In particular, we will provide provable algorithmic tools from both dynamic and compression perspective for a multitude of graph-based optimization problems that arise in Spectral Graph Theory, Graph Partitioning and Metric Embeddings. More importantly, this thesis establishes novel reduction techniques that reveal unexpected connections between time-evolving graphs and graph compression. In what follows, we will first review results in dynamic graph algorithms and then discuss our contributions in the area of graph sparsification. 

\section{Dynamic Graph Algorithms}

Suppose we are given a graph $G=(V,E)$ and a property $\mathcal{P}$ with respect to $G$. Furthermore, assume that the structure of $G$ is slightly perturbed, that is, an edge is either inserted or deleted from $G$. Can we efficiently update the property $\mathcal{P}$ in the perturbed graph rather than recomputing it from scratch? This basic question has been asked for many important graph properties for decades and the area that exclusively studies these questions is called \emph{Dynamic Graph Algorithms}. 

More concretely, a \emph{dynamic graph algorithm} is a data structure that supports the following operations on a given input graph $G$:

\begin{itemize} 
\item \textsc{Preprocess}$(G)$: preprocess the graph $G$
\item \textsc{Insert}$(u,v)$: insert the edge $(u,v)$ to $G$
\item \textsc{Delete}$(u,v)$: delete the edge $(u,v)$ from $G$
\item \textsc{Query}$(\mathcal{P})$: query the property $\mathcal{P}$ 
\end{itemize} 

In some variants of dynamic graph algorithms, the query operation might not be supported and the goal there is simply to maintain a correct property $\mathcal{P}$ with respect to the current graph at any point in time. A dynamic algorithm is characterized with three different time measures: (1) \emph{processing time}, which denotes the time to support operation \textsc{Preprocess}$(G)$; (2) \emph{update time}, which denotes the time to support operations \textsc{Insert}$(u,v)$ and \textsc{Delete}$(u,v)$, and (3) \emph{query time}, which denotes the time to support operation \textsc{Query}$(\mathcal{P})$. Update and query times can either be \emph{worst-case}, that is, the time spent to process each update or query individually, or \emph{amortized}, that is, the running time amortized over a sequence of operations. 

Depending on the types of update operations we support, dynamic algorithms are classified into three main categories: (i) \emph{fully dynamic}, if
update operations consist of both edge insertions and deletions; (ii) \emph{incremental}, if update operations consist of edge insertions only; and (iii) \emph{decremental}, if update operations consist of edge
deletions only. When studying the update times in algorithms of type (ii) and (iii), it is common to consider \emph{total update time}, which is the time spent over a sequence of $\Theta(m)$ insertions or deletions, where $m$ denotes the number of final or initial edges in the graph, respectively. Dynamic algorithms can either be \emph{deterministic} or \emph{randomized}, and usually algorithms with better running times are obtained if randomization is allowed. A common assumption in randomized dynamic algorithm is that the adversary is \emph{oblivious}, that is, the sequence of updates and queries is fixed in advance by the adversary, and the choices are are revealed to the algorithm one by one. 

There has been outstanding progress on devising efficient dynamic graph algorithms, especially during the last two decades, and the graph properties that have been considered include connectivity~\cite{HenzingerK99,HenzingerT97,HenzingerKNS15,KapronKM13,Thorup00}, reachability~\cite{DemetrescuI05,HenzingerK95,King99,Roditty08,RodittyZ16,
RodittyZ02,Sankowski04,ChechikHILP16}, shortest paths~\cite{DemetrescuI04, Thorup04,Bernstein13,BernsteinR11,HenzingerKN16,HenzingerKNSODA14,
HenzingerKN14}, matching~\cite{BaswanaGS15,BhattacharyaHI18,GuptaP13,NeimanS16,OnakR10,
BhattacharyaHN16}, (global) minimum cut~\cite{Thorup07,ThorupSWAT00,Henzinger97,KargerSODA94}, minimum spanning tree~\cite{Frederickson85,HenzingerK01,HolmLT01}, spanner~\cite{AusielloFI06,Elkin11,BernsteinFH19,BaswanaKS12,BodwinK16}, cut and spectral sparsifier~\cite{AbrahamDKKP16}, etc. However, despite this volume of work, there is a large number of questions that remain poorly understood. The situation is even worse if we consider several ``non-basic'' graph problems e.g., variants of graph partitioning, where no non-trivial solutions are known. Driven by this, in this thesis we study dynamic algorithms for new graph properties that appear to be important in different application domains but have not been considered so far. We also make progress on fundamental basic graph problems by improving their long-standing running time guarantees. 

\subsection{Dynamic Algorithms for Spectral Primitives}

In this thesis we study algorithms for dynamically maintaining solutions to Laplacian systems and Effective Reistances (see Chapters~\ref{cha:ESA2018_ER} and \ref{cha:STOC2019_DER}). Laplacian systems are an important subclass of linear systems
which arise in many natural contexts and have found applications in machine learning, computer graphics and image processing. Solving Laplacian systems has received considerable attention after the seminal work of Spielman and Teng~\cite{SpielmanT14} who devised the first near-linear time solver. A formal definition of such systems is given below. 

Given a graph $G=(V,E)$ with $n$ nodes and $m$ edges, let $\LL := \DD - \AAA$ denote the \emph{Laplacian matrix} of $G$, where $\DD$ and $\AAA$ are the associated degree and adjacency matrix of $G$, respectively. Matrix $\LL$ together with a vector $\bb \in \mathbb{R}^{n}$ form a system of linear equations $\LL \xx = \bb$, which is referred to as \emph{Laplacian system}. Let $\Lpseudo$ denote the pseudo-inverse of $\LL$. The \emph{solution vector} $\xx \in \mathbb{R}^{n}$ satisfies $\xx = \Lpseudo \bb$, and it exists if and only if $\bb^{\top} \ones = 0$, where $\ones$ is the all-ones $n$-dimensional vector. Let $\ones_u \in \mathbb{R}^{n}$ denote the indicator vector of a vertex $u$ such that $\ones_u(v) = 1$ if $v = u$ and $0$ otherwise.  

We introduce a dynamic model for solving Laplacian systems that supports insertions and deletions of edges in the underlying graph~(which correspond to modifying entries in $\LL$), modifications to vector $\bb$, and query access to one or few coordinates of an approximate solution vector, all in sublinear time. Concretely, we obtain the following result.

\begin{theorem} \label{intro:thm: dynamicLaplacian}
Given any error parameter $\epsilon \in (1/m, 1)$, there is a fully-dynamic algorithm for solving Laplacian systems on undirected, unweighted bounded-degree graphs while supporting insertions and deletions of edges, modifications to vector $\bb$, as well as query access to one or few entries of a vector $\tilde{\xx}$ such that $\vecnorm{\tilde{\xx}-\Lpseudo \bb}_{\LL} \leq \epsilon \vecnorm{\Lpseudo \bb}_{\LL}$, all in $\tilde{O}(n^{11/12} \epsilon^{-5})$\footnote{Throughout this thesis, we use $\tilde{O}(\cdot)$ to hide polylogarithmic factors, i.e., $\tilde{O}(f(n)) = O(f(n) \cdot \textrm{poly}(\log f(n))).$} expected amortized time. These guarantees hold against an oblivious adversary.
\end{theorem}

A spectral primitive closely related to Laplacian systems is effective resistance, a graph property that has received  increasing attention recently due to its application in speeding up algorithms for several cornerstone graph problems~\cite{ChristianoKMST11,Madry13,Madry16,Schild18}. Given a graph $G=(V,E,\ww)$ with $\ww$ assigning non-negative weights to edges in $E$, and any pair of vertices $u,v \in V$, we let \[ \Reff^G(u,v) := (\ones_u - \ones_v)^{\top} \Lpseudo (\ones_u - \ones_v) \] denote the \emph{effective resistance} between $u$ and $v$ in $G$. When $G$ is viewed as a resistor network, where resistances are the inverse of the edge weights, effective resistance between $u$ and $v$ can be thought of as the energy of the flow when routing one unit of current from $u$ to $v$. 

We study fully-dynamic graph algorithms for maintaining All-Pair Effective Resistances. Surprisingly enough, we show that this graph property admits sub-linear update and query times while achieving very high approximation accuracy to effective resistance. This is in stark contrast to related graph measures like shortest path, for which (conditional) hardness results are known in the fully-dynamic setting~\cite{HenzingerKNS15}, and maximum flow, which remains poorly understand from the dynamic perspective.

\begin{theorem} \label{intro:thm: dynamicER}
For any given error parameter $\epsilon \in (0,1)$, there is a fully-dynamic algorithm for maintaining $(1 \pm \epsilon)$-approximation to effective resistances in undirected, unweighted graphs while supporting insertions and deletions of edges as well as pair-wise effective resistance queries, all in $\tilde{O}\left(\min\{m^{3/4},n^{5/6}\} \epsilon^{-4} \right)$ expected amortized time. Our guarantees hold against an oblivious adversary.
\end{theorem}

We extend the above result in two directions. First, the above algorithm can be extended to also handle weighted graphs, albeit with a bound of $\tilde{O}(n^{5/6} \epsilon^{-4})$ on the expected amortized update and query time. Second, if we restrict to weighted graphs that admit \emph{small} separators, e.g., planar graphs, our \emph{worst-case} running time guarantees improve to $\tilde{O}(\sqrt{n} \epsilon^{-2})$. The key idea behind Theorem~\ref{intro:thm: dynamicER} and its corresponding extensions is dynamically maintaining an approximation to \emph{Schur complements} (also known as \emph{vertex spectral sparsifiers}). Roughly speaking, given a graph $G$ and a subset of vertices $K$, a Schur complement is a graph with vertex set $K$ that preserves effective resistances among any pair of vertices from $K$ in $G$. 

Despite the fact that our results share the same idea at a high level, there are subtle differences between their implementations. For general graphs, our techniques crucially rely on the fact that Schur complements can be viewed as a sum of random walks. This allows us to subsample vertices from the original graph and then construct a Schur complement with respect to this subsampled vertex set. The subsampling makes sure that the random walks are short and thus they can be maintained dynamically using elementary data-structures. On the other hand, for planar graphs we exploit the fact that they admit sub-linear separators (hence the $\sqrt{n}$ dependency on the running time), as well as the fact that approximate Schur complements can be computed in nearly-linear time~\cite{DurfeeKPRS17}. Inspired by the seminal work of Lipton, Rose and Tarjan~\cite{LRT79} on \emph{nested dissection}, these two ingredients are then brought together to dynamically maintain Schur complements for this family of graphs.  

All results we presented above guarantee only approximate answers to effective resistance queries. An obvious question is whether there are dynamic algorithms that can \emph{exactly} report effective resistances while still achieving sub-linear update and query time. We show that this is likely not the case. In particular, assuming a certain believable conjecture, we prove that there are no algorithms that simultaneously achieve sub-linear update and query time. 

\begin{theorem} \label{intro:thm: hardnessER}
No incremental or decremental algorithm can maintain the (exact) $(s,t)$ effective resistance in general graphs on $n$ vertices with both $O(n^{1-\delta})$ worst-case update time and $\tilde{O}(n^{2-\delta})$ worst-case query time for any $\delta > 0$, unless the \emph{OMv} conjecture~\cite{HenzingerKNS15} is false.
\end{theorem}

The preceding result can be extended to graphs that admit small separators, albeit with guarantees of $O(n^{1/2 - \delta})$ and $O(n^{1-\delta})$ on the update and query time, respectively. At the heart of our reductions, that prove these results, is a relation between effective resistance and the problem of detecting cycles of certain length in a graph. We defer the reader to Chapter~\ref{cha:ESA2018_ER} for more details. 

\subsection{Dynamic Low-Stretch Trees}
In this thesis we study algorithms for dynamically maintaining Low-Stretch Spanning Trees and Spanners~(see Chapter~\ref{cha:STOC2019_LSST}). Trees are the simplest class of graphs. From the algorithmic point of view, they are very appealing since many graph-based problems admit somewhat easier solutions when restricted to tree instances. In order to be able to exploit such a desirable behavior of trees, the problem of approximating general graphs by trees while preserving relevant graph properties, has been extensively studied in algorithm design. One notable example is Low-Stretch Spanning Tree, which at a high level is a spanning tree of a given input graph that preserves distances on average with a small stretch. Such trees are a central concept in Metric Embeddings and have found numerous applications in fast solvers for symmetric diagonally dominant (SDD) linear systems~\cite{KelnerOSZ13}, in the construction of competitive oblivious routing schemes~\cite{Racke08} and in approximation algorithms~\cite{PelegR98}. A formal definition of low-stretch trees is given below.

Given a graph $G=(V,E,\ww)$ and any $u,v \in V$, let $\distI_G(u,v)$ denote the length of a shortest path between $u$ and $v$ in $G$. Let $T$ be a spanning tree of $G$. We define the \emph{stretch} of an edge $(u,v) \in E$ with respect to $T$ to be $\stretchI_T(u,v) := \frac{\distI_T(u,v)}{\ww(u,v)}$. The \emph{average stretch} over all edges of $G$ with respect to $T$ is given by
\[
	\avStretchI_T(G) := \frac{1}{|E|} \sum_{(u,v) \in E} \stretchI_T(u,v).
\] 
We say that $T$ is a \emph{low-stretch} spanning tree whenever the average stretch is sub-polynomial or poly-logarithmic in the number of nodes $n = |V|$.

Motivated by the fundamental importance of low-stretch spanning trees as well as their powerful applications, we considered the maintenance of this object from a dynamic point of view. Indeed, designing dynamic algorithm for such trees was posed as an open problem by Baswana et al.~\cite{BaswanaKS12}. However, despite the extensive research in dynamic algorithms in recent years, no progress was made in this direction. In this thesis, we show the first non-trivial guarantees for this problem.

\begin{theorem} \label{intro:thm: dynamicLSST}
Given any unweighted, undirected graph with $n$ nodes undergoing edge insertions and deletions, there is a fully dynamic algorithm for maintaining a spanning tree of expected average stretch $n^{o(1)}$ that has expected amortized update time $n^{1/2+o(1)}$. These guarantees hold against an oblivious adversary.
\end{theorem}

The above algorithm can be slightly modified to give average stretch $O(t)$ and update time $n^{1+o(1)}/t$ for $t \geq \sqrt{n}$. This shows that the $\sqrt{n}$ barrier in the running time is not inherent, at least if a very large stretch is tolerable. One of the major building blocks of our algorithm is to dynamically maintain a clustering of a graph into small-diameter clusters (also known as \emph{low-diameter decomposition}). This is implemented using the random-shift clustering due to Miller, Peng and Xu~\cite{MillerPX13} together with many adaptations to make it work in the dynamic setting. We then employ a dynamic version of the hierarchy of low-diameter clusters due to Alon, Karp, Peleg, and West~\cite{AlonKPW95}, which in turn requires a sophisticated amortization approach to control propagation of updates within the hierarchy. Additionally our algorithm uses dynamic cut sparsifiers to reduce the problem to sparse graphs. While it is known that cuts and distances are dual to each other in similar settings~\cite{AndersenF09}, our argument requires a slight deviation from common approaches.    

The dynamic random-shift clustering could be of independent interest. Indeed, a direct consequence of this technique improves the previously best-known guarantees for dynamically maintaining graph spanners. Roughly speaking, a \emph{graph spanner} is a (sparse) subgraph of a given graph $G$ that preserves all pair shortest path distances of $G$ up to a multiplicative error.

\begin{theorem} \label{intro:thm: dynamicSpanner}
Let $t \geq 1$ be a parameter. Given any unweighted, undirected graph with $n$ nodes undergoing edge insertions and deletions, there is a fully dynamic algorithm for maintaining a spanner of stretch $(2t-1)$ and expected size $O(n^{1+1/t} \log n)$ that has expected amortized update time $O(t \log^2 n)$. These guarantees hold against an oblivious adversary.
\end{theorem}

Compared to the state-of-the art result of Baswana et al.~\cite{BaswanaKS12}, the above theorem improves upon the size of the spanner and the update time by a factor of $t$. Independently of our work, Saranurak and Wang~\cite{SaranurakW19} obtained similar guarantees for dynamically maintaining low-diameter clusters using different techniques. Concretely, they employ expander decomposition as a subroutine and use pruning to maintain a valid decomposition under edge deletions. We believe that our solution is arguably simpler than their expander pruning approach. 

\subsection{Dynamic Graph Partitioning}

In this thesis we study incremental algorithms for maintaining Global Minimum Cut and Sparsest Cut, both being core concepts in Graph Partitioning (see Chapters~\ref{cha:TALG2018_IMC} and~\ref{cha:Man2019_LS}). Graph partitioning problems typically involve partitioning the input graph into smaller components while minimizing the number of connections between these components. These problems have historically occupied a central place in understanding network flows~\cite{FF56}, packet routing~\cite{PelegU89} and VLSI layout. They have also been employed in many divide-and-conquer approaches for solving clustering problems. In what follows we start by defining the global minimum cut problem and then later discuss the results related to the sparsest cut problem. 

Given an unweighted, undirected graph $G=(V,E)$, and a subset of vertices $S \subseteq V$, the \emph{edge cut} $E(S, V \setminus S)$ is a set of edges that have one endpoint in $S$ and the other in $V \setminus S$. Let $\lambda(S) = |E(S, V \setminus S)|$ denote the size of the edge cut. A \emph{global minimum cut} is a subset $S$ whose edge cut size is the smallest among all subsets of vertices in $G$. Let $\lambda(G)$ denote the edge cut size of the global minimum cut in $G$. There has been extensive work on designing algorithms for computing global minimum cuts in the static setting and it is known that the problem can be solved in nearly linear-time~\cite{Karger00,KawarabayashiT19,HenzingerRW17}.

The first work on dynamic Global Minimum Cut is due to Karger~\cite{KargerSODA94}, who gave the first non-trivial running time guarantees for the problem. When both insertions and deletions of edges are supported, Thorup~\cite{Thorup07} achieves a $(1+o(1))$ approximation to the value of global minimum cut in $\tilde{O}(\sqrt{n})$ update and query time, and these bounds are the best-known to date. However, none of these works applies to maintain the exact value of global minimum cut and Thorup~\cite{Thorup07} even poses this question as an open problem. One exception here is the work by Henzinger~\cite{Henzinger97}, who obtains an exact incremental algorithm with $\tilde{O}(\lambda(G))$ amortized update time, where $\lambda(G)$ is the value of the global minimum cut in the graph after all insertions are processed. Note that $\lambda(G)$ can be as large as $O(n)$ and the main question is whether a truly sub-linear running time can be achieved. In the following result, we show that this is indeed the case by providing an exponential speed up on the update time of Henzinger~\cite{Henzinger97}.

\begin{theorem} \label{intro:thm: dynamicMinCut}
Given any unweighted, undirected graph with $n$ nodes undergoing edge insertions, there is a \emph{deterministic} incremental algorithm for \emph{exactly} maintaining the value of a global minimum cut $\lambda(G)$ in $O(\log^{3} n \log \log n)$ amortized time and $O(1)$ query time.
\end{theorem}

The above result stays in sharp contrast to a \emph{polynomial} conditional lower-bound for the fully dynamic \emph{weighted} global minimum cut problem due to Nanongkai and Saranurak~\cite{NanongkaiS16}. The high-level idea behind our result is to combine a sparsification routine of Kawarabayshi and Thorup~\cite{KawarabayashiT19} or its recent improvement by Henzinger, Rao and Wang~\cite{HenzingerRW17}, and an exact incremental algorithm of Henzinger~\cite{Henzinger97}. We remark that the combination itself is not immediate and it entails opening the black-boxes used in these works and skillfully extending them to obtain our desirable guarantees.  

Motivated by the recent work on \emph{space-efficient} dynamic algorithms~\cite{BhattacharyaHNT15}, we also consider efficient maintenance of global minimum cut using only $\tilde{O}(n)$ space. The results we obtain achieve $O(n \log n)$ space while still being able to support insertions and (approximate) queries in poly-logarithmic and constant time, respectively. Note that this setting differs from the standard \emph{graph stream model}, which typically allows $\tilde{O}(n)$ space while ignoring relevant measures like update and query time. 

We next discuss our contribution related to the Sparsest Cut problem, which is a well-studied NP-hard problem, that often serves as a prime example when discussing applications of metric embeddings in combinatorial optimization. Given an unweighted, undirected graph $G=(V,E)$, and a subset of vertices $S \subseteq V$, we define the \emph{uniform sparsity} of the \emph{cut} $(S,V \setminus S)$ as $\Phi_G(S):= \frac{E_G(S, V \setminus S)}{|S| \cdot |V \setminus S|}$. The \emph{uniform sparsest cut of} $G$ is the cut $(S, V \setminus S)$ with smallest possible sparsity. Let $\Phi(G)$ denote the value of the sparsest cut in $G$. In the literature, there are several efficient algorithms for approximating $\Phi(G)$ with a multiplicative factor of $O(\log^{c} n)$, where $c \in [1/2,1]$~\cite{AroraRV09,Sherman09,KhandekarRV09}. However, prior to our work, nothing was known about the complexity of this problem in the dynamic setting.

We make the first positive progress towards understanding the Uniform Sparsest Cut problem from the dynamic point of view. In the insertions-only model, we show that we can maintain a poly-logarithmic approximation to the sparsest cut in sub-linear update time. As a by-product of our techniques, our algorithm provides a trade-off between the approximation error and the update time.

\begin{theorem} \label{intro:thm: dynamicSparsestCut}
Let $t \geq 1$ be a parameter. Given any unweighted, undirected graph with $n$ nodes undergoing edge insertions, there is a randomized incremental algorithm for maintaining an $O(\log^{8t} n)$ approximation to the value of uniform sparsest cut $\Phi(G)$ in $\tilde{O}(n^{2/(t+1)})$ worst-case update time time and $O(1)$ query time. Our algorithm extends to weighted graphs with polynomially bounded weights. 
\end{theorem}

The key idea behind the proof of Theorem~\ref{intro:thm: dynamicSparsestCut} is a new notion of sparsifiers, called \emph{local sparsifiers}. These sparsifiers are a stronger version of the well-studied notion of \emph{vertex sparsifiers}. Concretely, in Vertex Sparsification, given a graph $G=(V,E)$ and a subset of vertices $K$, referred to as \emph{terminals}, the goal is to construct a graph $H=(V',E')$ with $V' \supseteq K$ and $|V(H)|$ is ``small'' such that $H$ preservers some  graph property $\mathcal{P}$ that involves the terminals $K$ in $G$ (see the next section for an in-depth treatment on vertex sparsifiers). A \emph{local sparsifier} is a data-structure generalization of vertex sparsifiers; formally, given a graph $G=(V,E)$, the goal is to build a data-structure that supports the following operations: 
\begin{enumerate} 
\itemsep0em
\item \textsc{Preprocess}$(G)$: preprocess the graph $G$
\item \textsc{QuerySparsifier}$(G,K)$: compute and output a vertex sparsifier $H$ of $G$ that preserves some property $\mathcal{P}$ among vertices in $K$. 
\end{enumerate}
In other words, this definition suggests that local sparsifiers allow us to extract vertex sparsifiers for any set of terminals in $K$. Note that operation (2) is a very strong requirement, as there are $\Theta(2^{n})$ different terminal sets. 

We show that a variant of \emph{tree cut sparsifiers} due to Peng, R\"acke, Shah and T\"aubig~\cite{Peng16,RackeST14} can be used to construct local sparsifier that preserve cut-structure of the graph up to poly-logarithmic factors while achieving $\tilde{O}(m)$ preprocessing time and $\tilde{O}(|K|)$ query time. In particular, this implies that the uniform sparsest cuts are also preserved within the same approximation. We then design a reduction that converts such an efficient local sparsifier into an incremental algorithm that maintains a poly-logarithmic approximation to the uniform sparsest cut. The same technique allows us to obtain very fast incremental algorithms for the approximate All-Pair Maximum Flow problem with similar guarantees to those in Theorem~\ref{intro:thm: dynamicSparsestCut}.  This is quite intriguing since nothing was known about the dynamic Max-Flow problem in general graphs, even when allowing poly-logarithmic approximation.

In fact, our reduction relating local sparsifiers and incremental graph algorithms applies to a larger family of graph properties. For example, using variants of the \emph{distance oracle} due to Thorup and Zwick~\cite{ThorupZ05} we construct efficient local sparsifiers, which in turn imply a \emph{deterministic} incremental algorithm that approximates All-Pair Shortest Paths up to a constant factor in sub-linear update and query time.

Another important problem in dynamic algorithms is to understand the complexity of $(1+\epsilon)$-approximate maximum flow problem in the dynamic setting. Even when restricted to the weaker \emph{offline dynamic} model, where edge updates and queries are given in advance, the problem remains poorly understood. On the other hand, over the last years there has been increasing interest in proving conditional polynomial lower-bounds for dynamic problems. A property that most of these lower-bounds share is that they apply to the offline dynamic model. Driven by this, we develop a framework that connects offline dynamic problems and vertex sparsification. Specifically, we show that if there are efficient $(1+\epsilon)$ vertex sparsifiers of size $\tilde{O}(\textrm{poly}(|K|, 1/\epsilon))$ that preserve cuts, then the approximate offline maximum flow problem admits sub-linear update and query times. This would imply that no $\Omega(n^{1-o(1)})$ lower bound can be shown for the approximate offline max flow problem. For other connections we refer the reader to Chapter~\ref{cha:Man2019_LS}.

\section{Graph Sparsification}
A \emph{graph sparsifier} is a ``compressed'' version of a large input graph that preserves properties like distance or reachability information, cut value or graph spectrum. Traditionally, graph sparsifiers have been studied from two perspectives: (1) those that reduce the number of edges of a graph, referred to as \emph{edge sparsifiers}, and (2) those that reduce the number of nodes, referred to as \emph{vertex sparsifier}. Edge sparsifiers have been successfully applied for improving the running time of many basic graph-based optimization problems, and the most notable examples include transitive reductions~\cite{AhoGU72}, spanners~\cite{AlthoferDDJS93}, cut sparsifiers~\cite{BenczurK96} and spectral sparsifier~\cite{SpielmanT11}. In this thesis, we focus on vertex sparsifiers. Concretely, given a graph $G=(V,E)$ and a subset of vertices $K$, referred to as \emph{terminals}, the goal is to construct a graph $H=(V',E')$ satisfying the following properties:
\begin{itemize}
\item $V' \supseteq K$ and $|V'|$ is ``small'', ideally $|V'| = O( \textrm{poly} (|K|))$,
\item $H$ (approximately) preserves properties like reachability, distance, cuts or multi-commodity flows defined among terminals in $K$; often it is desirable that $H$ is structurally similar to $G$, e.g., when $G$ is planar, so is $H$
\end{itemize}

When $H$ preserves some property approximately, the approximation ratio is referred to as the \emph{quality} of the sparsifier. The usefulness of such a sparsification tool is apparent from an algorithm point of view; once $H$ is computed, we can perform algorithmic tasks only in $H$ instead of $G$, which in turn leads to savings in computational and storage resources. Besides their practical relevance, vertex sparsifiers have also found applications within other sub-areas of Theoretical Computer Science, namely approximation algorithms~\cite{Moitra09,EnglertGKRTT14}, dynamic graph algorithms~\cite{GoranciHP18,GoranciHP17a}, and network routing~\cite{Chuzhoy12a}. 

In what follows, which constitutes Chapters~\ref{cha:ICALP2016_DM} and~\ref{cha:ESA2017_RM} of this thesis, we will discuss our contributions on vertex sparisifers that are structurally similar to the input graph and at the same time preserve distances in undirected graphs or reachability information in directed graphs.

\subsection{Distance Approximating Minors}

We study vertex sparsifiers that are obtained using minor operations while preserving distances among terminal pairs approximately. Minors are particularly desirable since they preserve structural properties of the input graph, e.g., a minor of a planar graph is another planar graph. Formally, given a weighted graph $G=(V,E,\ww)$ and a designated subset of terminals $K$, an $\alpha$-\emph{distance approximating minor} of $G$ is a weighted graph $H = (V',E',\ww')$ such that 
\begin{itemize} 

\item  $V' \subseteq K$ and $V'$ is small, ideally $|V'|= O(\textrm{poly}(|K|))$, 

\item $H$ is a minor of $G$, i.e., $H$ is obtained from $G$ by deleting edges and vertices any by contracting edges. No terminal can be deleted, and no two terminals can be contracted together.

\item Terminal distances are preserved up to an $\alpha$ factor, i.e., for any pair of vertices $u,v \in K$, we have
\[
	\distI_G(u,v) \leq \distI_{H}(u,v) \leq \alpha \cdot \distI_G(u,v).
\]
\end{itemize}

Vertices in $V' \setminus K$ are usually referred to as \emph{non-terminals} or \emph{Steiner vertices}. Gupta~\cite{Gupta01} introduced the strongest version of the problem which requires that $V' = K$, also known as the \emph{Steiner Removal Problem}. In this setting, he showed that trees admit sparsifiers with quality $8$. Kamma, Krauthgamer and Nguyen~\cite{KammaKN15} showed that general graphs admit sparsifiers with quality $O(\log^{5}(|K|))$. This bound has been subsequently improved to $O(\log^{2}(|K|))$ by Cheung~\cite{Cheung18} and finally to $O(\log(|K|))$ by Filtser~\cite{Filtser18}.

At the other extreme, Krauthgamer, Nguyen and Zondiner~\cite{KrauthgamerNZ14} considered the setting where distances are preserved \emph{exactly}, i.e., $\alpha=1$ and Steiner vertices are allowed, also known as \emph{distance preserving minors}. They showed that general graphs admit distance preserving minors with $O(|K|^{4})$ extra non-terminals. A natural question to ask is what is the trade-off between the quality and the number of non-terminals? We make progress on this question from both lower and upper bound perspectives. Specifically, we start by presenting the following lower bound result.

\begin{theorem} \label{intro:thm: damLB}
Let $c > 0$ be a constant. For infinitely many $k \in \mathbb{N}$, there exists a graph with $k$ terminals which does not admit an $(\alpha-\epsilon)$-distance approximating minor with $k^{\gamma}$ non-terminals, for all $\epsilon > 0$, where $\alpha, \gamma$ are given in the table below.
\begin{center}
\begin{tabular}{c|c|c|c|c|c|c|c}
$\alpha$ & $2$ & $2.5$ & $3$ & $10/3$ &  $11/3$ &  $4$ & $4.2$  \\ \hline
$\gamma$ & $2$ & $5/4$ & $6/5$ & $10/9$ & $11/10$ & $12/11$ & $21/20$  \\
\end{tabular}
\end{center}
\end{theorem}

To obtain the above result we introduce a novel black-box reduction technique that converts lower bounds for the SPR problem~\cite{ChanXKR06} into super-linear lower-bounds on the number of non-terminals for distance approximating minors with the same quality. At the heart of our graph constructions are variants of \emph{Steiner Systems}~\cite{Wilson75}, which are useful concepts studied in combinatorial design. We believe that this connection might be of independent interest.

From the upper bound perspective, we ask the question of whether one can construct $(1+\epsilon)$-distance approximating minors with less than $O(|K|^{4})$ non-terminals. For planar graphs, we show this can be actually achieved.

\begin{theorem} \label{intro:thm: damUB}
Given a weighted planar graph $G=(V,E,\ww)$, and a set of terminals $K$, there exists an algorithm that computes an $(1+\epsilon)$-distance approximating minor $H=(V',E',\ww')$ of $G$ with $|V'| = O(|K|^{2} \epsilon^{-2} \log^{2} |K|)$ non-terminals. 
\end{theorem}

Key to the above result is the notion of \emph{terminal path cover}. At a high level such a cover is a set of shortest paths in the graph whose union (1) contains the terminal set and (2) approximately preserves shortest path distances among terminals. We show that \emph{distance oracles} for planar graphs due to Thorup~\cite{ThorupJACM04} can be extended to construct terminal path covers for planar graphs. This, combined with the counting argument for branching events in shortest paths of Coppersmith and Elkin~\cite{CoppersmithE06} proves the claimed guarantees. We remark that our result has been subsequently extended to minor-free graphs by Gupta and DiRenzo~\cite{GuptaR16}.

It is an important question whether one can improve the bound on the number of non-terminals in Theorem~\ref{intro:thm: damUB} while keeping the same quality. In fact, any sub-quadratic bound on the number of non-terminals would imply non-trivial bounds for dynamic planar all-pairs shortest path problem in the offline setting. We refer the reader to Chapter~\ref{cha:Man2019_LS} for a detailed treatment on this connection. 

\subsection{Reachability Preserving Minors}

Sparsification in directed graphs is usually a much harder task when compared to the undirected counterpart, with many basic graph properties admitting no non-trivial results. In this thesis, we focus on one of most basic graph properties, namely \emph{reachability}, and study it from the vertex sparsification point of view. 

Formally, given a directed graph $G=(V,E)$ and a designed subset of terminals $K$, a \emph{reachability preserving minor} of $G$ is a directed graph $H=(V',E')$ such that (1) $V' \supseteq K$ and $V'$ is small, ideally $|V'| = O(\textrm{poly}(|K|))$, (2) $H$ is a minor of $G$ and (3) for any pair of vertices $u,v \in K$, there is a directed path from $u$ to $v$ in $H$ if and only if there is a directed path from $u$ to $v$ in $G$.

We initiate the study of constructing such sparsifiers and provide the first non-trivial guarantees on the problem. Our lower bound shows that, in general, it is not possible to construct reachability preserving minors with a sub-quadratic number of non-terminals. 

\begin{theorem} \label{intro:thm: rpmLB}
For infinitely many $k \in \mathbb{N}$ there exists a directed planar graph $G$ with $k$ terminals such that any reachability preserving minor of $G$ must use $\Omega(k^{2})$ non-terminals.
\end{theorem}  

In fact, the graph instance for proving the above lower bound is a \emph{directed acylic grid} with terminals distributed on the boundary of the grid. Our argument essentially proves that all internal, non-boundary vertices of the grid must be retained, if we want to preserve reachability information among terminals. Similar ideas for proving lower-bounds on distance preserving minors for undirected graphs were employed by Krauthgamer, Nguyen, and Zondiner~\cite{KrauthgamerNZ14}. 

We complement the lower bound by showing that planar graphs admit reachability sparsifiers with at most $O(|K|^2 \log |K|)$ terminals. For general graphs our bounds are worse only by another $|K|$ factor. Surprisingly, the gaps between the best upper and lower bounds are tighter when compared to distance preserving minors in the undirected setting.

\begin{theorem} \label{intro:thm: rpmUB}
Given a directed graph $G=(V,E)$, and a set of terminals $K$, there exists an algorithm that computes a rechability preserving minor $H=(V',E')$ of $G$ with $|V'| = O(|K|^{3})$ non-terminals. When $G$ is a planar directed graph, the number of non-terminals improves to $|V'| = O(|K|^{2} \log |K|)$.
\end{theorem} 

\subsection{Structure of Thesis}
We start with the fully-dynamic all-pairs effective resistances problem in Chapter~\ref{cha:ESA2018_ER}. We obtain a $(1+\epsilon)$-approximation with $\tilde{O}(\sqrt{n} \epsilon^{-2})$ update and query time on graphs that admit small separators. In the setting where exact effective resistances are required, we show two conditional lower-bounds, one applying to general graphs and the other to graphs that admit small separators, which justify our upper bound that only supports approximate queries. In Chapter~\ref{cha:STOC2019_DER}, we study dynamic algorithms for maintaining vertex spectral sparsifiers with respect to a carefully chosen set of terminals. We show the applicability of this technique to (1) dynamic Laplacian solvers with $\tilde{O}(n^{11/12} \epsilon^{-5})$ update and query time on unweighted, bounded degree graphs and (2) dynamic $(1 + \epsilon)$-approximate all-pairs effective resistances with $\tilde{O}(\min(m^{3/4},n^{5/6} \epsilon^{-2}) \epsilon^{-4})$ update and query time on undirected, unweighted graphs, and $\tilde{O}(n^{5/6} \epsilon^{-6})$ on undirected, weighted graphs. 

We then shift our focus to studying tree-based graph approximations in the dynamic setting. In Chapter~\ref{cha:STOC2019_LSST}, we develop an algorithm that dynamically maintains a spanning tree with $n^{o(1)}$ average stretch and $O(n^{1/2 + o(1)})$ update time on undirected, unweighted graphs. As a by-product of our techniques, we give the best-known running time and size guarantees for the dynamic spanner problem. 

In Chapters~\ref{cha:TALG2018_IMC} and~\ref{cha:Man2019_LS} we study incremental algorithms for graph partitioning problems. Concretely, in Chapter~\ref{cha:TALG2018_IMC}, we show an incremental algorithm for exactly maintaining the value of a global minimum cut in $O(\log^{3} n \log \log n)$ update time and $O(1)$ query time. We also design incremental algorithms with small approximation errors that are both time and space efficient. In Chapter~\ref{cha:Man2019_LS}, we introduce the notion of local sparsifiers and design efficient variants of such sparsifiers for graph properties like distances and cuts. We then show a technique that converts these sparsifiers into incremental algorithms for efficiently maintaining approximate solutions to a range of graph problems including all-pairs minimum cuts and uniform sparsest cut.

The last part of this thesis is devoted to graph sparsification. In Chapter~\ref{cha:ICALP2016_DM}, we study distance approximating minors from both a lower and upper bound perspective. For example, we show that for distortion $3-\epsilon$ there are $k$-terminal graphs for which any distance approximating minor needs to retain at least $\Omega(k^{6/5})$ non-terminals. For planar graphs, we show that there are $(1+\epsilon)$-distance approximating minors with $\tilde{O}(k^{2} \epsilon^{-2})$ non-terminals. In Chapter~\ref{cha:ESA2017_RM}, we consider reachability preserving minors. We prove that, in general, it is not possible to construct such sparsifiers with a sub-quadratic number of non-terminals and show a matching upper bound on planar graphs, up to a logarithmic factor. We also prove new guarantees for vertex sparsifiers that preserve distance and cuts on planar graphs with terminals lying on the same face.

\chapter[Dynamic Effective Resistances and Approximate Schur Complement on Separable Graphs][Dynamic Schur Complement on Seperable Graphs]{Dynamic Effective Resistances and Approximate Schur Complement on Separable Graphs}\label{cha:ESA2018_ER}

We consider the problem of dynamically maintaining (approximate) all-pairs effective resistances in separable graphs, which are those that admit an $n^{c}$-separator theorem for some $c<1$. We give a fully dynamic algorithm that maintains  $(1+\varepsilon)$-approximations of the all-pairs effective resistances of an $n$-vertex graph $G$ undergoing edge insertions and deletions with $\tilde{O}(\sqrt{n}/\varepsilon^2)$ worst-case update time and $\tilde{O}(\sqrt{n}/\varepsilon^2)$ worst-case query time, if $G$ is guaranteed to be $\sqrt{n}$-separable (i.e., it is taken from a class satisfying a $\sqrt{n}$-separator theorem) and its separator can be computed in $\tilde{O}(n)$ time. Our algorithm is built upon a dynamic algorithm for maintaining \emph{approximate Schur complement} that approximately preserves pairwise effective resistances among a set of terminals for separable graphs, which might be of independent interest.

We complement our result by proving that for any two fixed vertices $s$ and $t$, no incremental or decremental algorithm can maintain the $s-t$ effective resistance for $\sqrt{n}$-separable graphs with worst-case update time $O(n^{1/2-\delta})$ and query time $O(n^{1-\delta})$ for any $\delta>0$, unless the Online Matrix Vector Multiplication (OMv) conjecture is false. 

We further show that for \emph{general} graphs, no incremental or decremental algorithm can maintain the $s-t$ effective resistance problem with worst-case update time $O(n^{1-\delta})$ and query-time $O(n^{2-\delta})$ for any $\delta >0$, unless the OMv conjecture is false.

\section{Introduction}\label{sec:intro_DERP}
Effective resistances and the closely related electrical flows  are basic concepts for resistor networks~\cite{doyle84} and were found to be very useful in the design of graph algorithms, e.g., for computing and approximating maximum flow~\cite{ChristianoKMST11,Madry13,Madry16}, random spanning tree generation~\cite{MST15fast,Schild18}, multicommodity flow~\cite{KMP12faster}, oblivious routing~\cite{HHNRR08}, and graph sparsification~\cite{SpielmanS11,DinitzKW15}. They also have found applications in social network analysis, e.g., for measuring the similarity of vertices in social networks~\cite{LZ18:kirchhoff}, in machine learning, e.g., for Gaussian sampling~\cite{cheng2015efficient} and in chemistry, e.g., for measuring chemical distances~\cite{klein1993resistance}. Previous research has studied the problem of how to quickly compute and approximate the effective resistances (or equivalently, \emph{energies} of electrical flows), as such algorithms can be used as a crucial subroutine for other graph algorithms. For example, one can $(1+\varepsilon)$-approximate the $s-t$ effective resistance in  $\tilde{O}(m+n\varepsilon^{-2})$~\cite{DurfeeKPRS17} and $\tilde{O}(m \log (1/\varepsilon))$~\cite{CohenKMPPRX14} time, respectively, in any $n$-vertex $m$-edge weighted graph, for any two vertices $s,t$. There are also algorithms that find $(1+\varepsilon)$-approximations to the effective resistance between every pair of vertices in $\tilde{O}(n^2/\varepsilon)$ time~\cite{JS18:sketch}. In order to exactly compute the $s-t$ (or single-pair) and all-pairs effective resistance(s), the current fastest algorithms run in times $O(n^\omega)$ (by using the fastest matrix inversion algorithm~\cite{bunch1974,ibarra1982}) and $O(n^{2+\omega})$, respectively, where $\omega < 2.373$ is the matrix multiplication exponent~\cite{williams2012}. In planar graphs, the algorithms for exactly computing $s-t$ and all-pairs effective resistance(s) run in times $O(n^{\omega/2})$~(by the nested dissection method for solving linear system in planar graphs \cite{LRT79}) and $O(n^{2+\omega/2})$, respectively.

A natural algorithmic question is how to efficiently maintain the effective resistances \emph{dynamically}, i.e., if the graph undergoes edge insertions and/or deletions, and the goal is to support the update operations and query for the effective resistances as quickly as possible, rather than having to recompute it from scratch each time. Besides the potential applications in the design of other (dynamic) algorithms, it is also of practical interest, e.g., to quickly report the (dis)similarity between any two nodes in a social network in which its members and their relationship are constantly changing. So far our understanding towards this question is very limited: for exact maintenance, the only approach (for single-pair effective resistance) we are aware of is to invoke the dynamic matrix inversion algorithm which gives $O(n^{1.575})$ update time and $O(n^{0.575})$ query time or $O(n^{1.495})$ update time and $O(n^{1.495})$ query time~\cite{Sankowski04}; for $(1+\varepsilon)$-approximate maintenance, we can maintain the spectral sparsifier of size $n\poly(\log n, \varepsilon^{-1})$ with $\poly(\log n, \varepsilon^{-1})$ update time~\cite{AbrahamDKKP16}, while answering each query will cost $\Theta(n\poly(\log n, \varepsilon^{-1}))$ time.

In this chapter, we study the problem of dynamically maintaining the (approximate) effective resistances in \emph{separable graphs}, which are those that satisfies an $n^c$-separator theorem for some $c<1$. Interesting classes of separable graphs include planar graphs, minor free graphs, bounded-genus graphs, almost planar graphs (e.g., road networks)~\cite{lipton79}, most $3$-dimensional meshes~\cite{MTTV1997separators} and many real-world networks (e.g., phone-call graphs, Web graphs, Internet router graphs)~\cite{BBK03compact}. In the static setting, effective resistances (or electrical flows) in planar/separable graphs have been utilized by Miller and Peng~\cite{MillerP13} to obtain the first $\tilde{O}(\frac{m^{6/5}}{\varepsilon^{\Theta(1)}})$ time algorithm for approximate maximum flow in such graphs, and have also been studied by Anari and Oveis Gharan~\cite{NO15:TSP} in the analysis of an approximation algorithm for Asymmetric TSP. We now give the necessary definitions to state our results.

\paragraph{Effective Resistances.} Let $G=(V,E,\vect{w})$ be a undirected weighted graph with $\vect{w}(e)>0$ for any $e\in E$. Let $\mat{A}$ denote its weighted adjacency matrix and $\mat{D}$ denote the weighted degree diagonal matrix. Let $\mat{L}=\mat{D}-\mat{A}$ denote the \emph{Laplacian} matrix of $G$. Let $\mat{L}^\dagger$ denote the Moore-Penrose pseudo-inverse of the Laplacian of $G$. Let $\1_u \in \R^V$ denote the indicator vector of vertex $u$ such that $\1_u(v)=1$ if $v=u$ and $0$ otherwise. Let $\boldsymbol{\chi}_{s,t}=\1_s-\1_t$. Given any two vertices $u,v\in V$, the \emph{$s-t$ effective resistance} is defined as $R^G_{\textrm{eff}}(s,t):=\boldsymbol{\chi}_{s,t}^\top\mat{L}^\dagger\boldsymbol{\chi}_{s,t}.$

\paragraph{Separable graphs.} Let $\mathcal{C}$ be a class of graphs that is closed under taking subgraphs. We say that $\mathcal{C}$ satisfies a \emph{$f(n)$-separator theorem} if there are constants $\alpha <1$ and $\beta>0$ such that every graph in $S$ with $n$ vertices has a cut set with at most $\beta f(n)$ vertices that separates the graph into components with at most $\alpha n$ vertices each~\cite{lipton79}. 
In this chapter we are particularly interested in the class of graphs that satisfies an $n^{1/2}$-separator theorem, which include the class of planar graphs, $K_t$-minor free graphs and bounded-genus graphs, etc., though our approach can also be generalized to other class of graphs that satisfies a $n^{c}$-separator theorem, for some $c<1$. In the following, we call a graph \emph{$f(n)$-separable} if it is a member of a class that satisfies an $f(n)$-separator theorem.

We would like to quickly maintain the exact or a good approximation of the $s-t$ effective resistances in a $\sqrt{n}$-separable graph that undergoes edge insertions and deletions, for all pairs $s,t\in V$. We call this the \emph{dynamic all-pairs effective resistances problem.} Our goal is to solve this problem with both small update and query times. More precisely, our data structure supports the following operations. 
\begin{itemize}
	\item \textsc{Insert}$(u,v,w)$: Insert the edge $(u,v)$ of weight $w$ to $G$, provided that the updated graph remains $\sqrt{n}$-separable.
	\item \textsc{Delete}$(u,v)$: Delete the edge $(u,v)$ from $G$.   
	\item \textsc{EffectiveResistance}$(s,t)$: Return the exact or approximate value of the effective resistance between $s$ and $t$ in the current graph $G$.
\end{itemize}  

We remark that our algorithm can be extended to handle \sloppy operations  \textsc{Increase}($u,v,\Delta$) and \textsc{Decrease}($u,v,\Delta$) that increases and decreases the weight of any existing edge $(u,v)$ by $\Delta$, respectively, as one can simply delete the edge first and then insert it again with the corresponding new weight. For our lower bound, we will consider the \emph{incremental} (or \emph{decremental}) $s-t$ effective resistance problem, that is, $s,t$ are two vertices fixed at the beginning, and only operations \textsc{Insert} \& \text{Decrease} (or \textsc{Delete} \& \textsc{Increase}) and \textsc{EffectiveResistance} are allowed. The basic idea is that in the incremental (or decremental) setting, the effective resistances are monotonically decreasing (or increasing)~(see e.g., \cite{ChristianoKMST11}). 
For any $\varepsilon \in (0,1)$, we say that an algorithm is a $(1+\varepsilon)$-approximation to $R^G_{\textrm{eff}}(s,t)$ if \textsc{EffectiveResistance}$(s,t)$ returns a positive number $k$ such that $(1-\varepsilon) \cdot R^G_{\textrm{eff}}(s,t) \leq k \leq (1+\varepsilon) R^G_{\textrm{eff}}(s,t)$. 

\subsection{Our Results}
We give a fully dynamic algorithm for maintaining $(1+\varepsilon)$-approximations of all-pairs and single-pair effective resistance(s) with small update and query times for any $\sqrt{n}$-separable graph, if its separator can be computed fast. Throughout, all the running times of our algorithms are measured in \emph{worst-case} performance. All our algorithms are randomized, and the performance guarantees hold with probability at least $1-n^{-c}$ for some $c\geq 1$. Specifically, we show the following theorem. 
\begin{theorem}\label{thm:upperbound}
	Let $G$ denote a dynamic $n$-vertex graph under edge insertions and deletions. Assume that $G$ is $\sqrt{n}$-separable and its separator can be computed in $s(n)$ time, throughout the updates. There exist fully dynamic algorithms that maintain  $(1+\varepsilon)$-approximations of 
	\begin{itemize}
		\item the all-pairs effective resistances with $\tilde{O}(\frac{\sqrt{n}}{\varepsilon^2}+\frac{s(n)}{\sqrt{n}})$ update time and $\tilde{O}(\frac{\sqrt{n}}{\varepsilon^2})$ query time;
		\item the $s-t$ effective resistance with $\tilde{O}(\frac{\sqrt{n}}{\varepsilon^2}+\frac{s(n)}{\sqrt{n}})$ update time and $O(1)$ query time.
	\end{itemize} 
	In particular, if $s(n)=\tilde{O}(n)$, then our update times are $\tilde{O}(\frac{\sqrt{n}}{\varepsilon^2})$.
\end{theorem}

By using the well known facts that a balanced separator of size $O(\sqrt{n})$ for planar graphs (and bounded-genus graphs) can be computed in $O(n)$ time~\cite{lipton79}, and for $K_t$-minor-free graphs (for any fixed integer $t>0$) in $O(n^{1+\delta})$ time, for any constant $\delta > 0$~\cite{kawar2010}, we obtain dynamic algorithms for the effective resistances for planar and minor-free graphs with $\tilde{O}(\sqrt{n}/\varepsilon^2)$ and $\tilde{O}(\sqrt{n}/\varepsilon^2+n^{1/2+\delta})$ update time, respectively. 

The performance of our dynamic algorithm in planar graphs almost matches the best-known dynamic algorithm for $(1+\varepsilon)$-approximate all-pairs shortest path in planar graphs with $\tilde{O}(\sqrt{n})$ update and query time~\cite{AbrahamCG12}, though our approaches are different. This is interesting as the shortest path corresponds to flows with controlled $\ell_1$ norm while the energy of electrical flows (i.e., effective resistance) corresponds to those with minimum $\ell_2$ norm.     

In order to design a dynamic algorithm for effective resistances of separable graphs (i.e., to prove Theorem~\ref{thm:upperbound}), we give a fully dynamic algorithm that efficiently maintains an \emph{approximate Schur complement}~\cite{KyngLPSS16,KyngS16,DurfeeKPRS17} of such graphs (see Section~\ref{sec:dynamic_ASC}), which might be of independent interest.
Approximate Schur complement can be treated as a \emph{vertex sparsifier} that preserves pairwise effective resistances among a set of terminals (see Section~\ref{sec:asc_vs}). Therefore, our algorithm is a dynamic algorithm for \emph{vertex effective resistance sparsifiers} with sublinear (in $n$) update time for separable graphs. The problem of dynamically maintaining graph \emph{edge sparsifiers} has received attention very recently. For example, Abraham et al. presented fully dynamic algorithms that maintain cut and spectral sparsifiers with poly-logarithmic update times~\cite{AbrahamDKKP16}. Formally, we prove the following theorem. 

\begin{theorem}\label{thm:dynamic_asc}
	For an $n$-vertex $\sqrt{n}$-separable graph $G$ whose separator can be computed in $s(n)$ time, and a terminal set $K \subseteq V$ with $|K|\leq O(\sqrt{n})$, there exists a fully dynamic algorithm that maintains a $(1+\delta)$-approximate Schur complement with respect to $K'$ such that $K \subseteq K'$ and $|K'|=O(\sqrt{n})$, while achieving $\tilde{O}(\sqrt{n}/\delta^{2}+\frac{s(n)}{\sqrt{n}})$ update time. 
	Furthermore, our algorithm supports terminal additions as long as $|K|\leq O(\sqrt{n})$.
\end{theorem}

We complement our algorithm by giving a conditional lower bound for any \emph{incremental} or \emph{decremental} algorithm that maintains \emph{single-pair} effective resistance of a $\sqrt{n}$-separable graph. Our lower bound is established from the \emph{Online Matrix Vector Multiplication} (\emph{OMv}) \emph{conjecture} (see Section \ref{sec: prelim_DERP}).
\begin{theorem}\label{thm:lowerbound_separable}
	No incremental or decremental algorithm can maintain the (exact) $s-t$ effective resistance in $\sqrt{n}$-separable graphs on $n$ vertices with both $O(n^{\frac12-\delta})$ worst-case update time and $O(n^{1-\delta})$ worst-case query time for any $\delta >0$, unless the OMv conjecture is false.
\end{theorem}

We note that there are very few conditional lower bounds for dynamic \emph{planar/separable} graphs, as most known reductions are highly non-planar. The only recent result that we are aware of is by Abboud and Dahlgaard~\cite{AbboudD16}, who showed that under some popular conjecture, no algorithm for dynamic shortest paths or maximum weight bipartite matching in planar graphs has both updates and queries in amortized $O(n^{1/2-
	\delta})$ time, for any $\delta>0$.

We also give a stronger conditional lower bound for the same problem in \emph{general} graphs, which shows that it is hard to maintain effective resistances with both sublinear (in $n$) update and query times for general graphs, even for the incremental or decremental setting. 
\begin{theorem}\label{thm:lowerbound_general}
	No incremental or decremental algorithm can maintain the (exact) $s-t$ effective resistance in general graphs on $n$ vertices with both $O(n^{1-\delta})$ worst-case update time and $O(n^{2-\delta})$ worst-case query time for any $\delta >0$, unless the OMv conjecture is false.
\end{theorem}

We remark that both lower bounds for separable and general graphs hold for any algorithm with sufficiently high accurate approximation ratio, and both lower bounds for incremental algorithms hold even if only edge insertions are allowed (see Section~\ref{sec: lowerBound}). 

\paragraph*{Comparison to~\cite{GoranciHP17a}}
In our previous work~\cite{GoranciHP17a}, we gave a fully dynamic algorithm for $(1+\varepsilon)$-approximating all-pairs effective resistances for planar graphs with $\tilde{O}(r/\varepsilon^2)$ update time and $\tilde{O}((r+n/\sqrt{r})/\varepsilon^2)$ query time, for any $r$ larger than some constant. The algorithm can also be generalized to $\sqrt{n}$-separable graphs, and we also provided a conditioned lower bound for any approximation algorithm of the $s-t$ effective resistance in general graphs in the \emph{vertex-update} model. However, besides the apparent improvement of the performance of the dynamic algorithm (i.e., we reduce the best trade off between update time and query time from $\tilde{O}(n^{2/3})$ and $\tilde{O}(n^{2/3})$ to $\tilde{O}(n^{1/2})$ and $\tilde{O}(n^{1/2})$), our current work also improves over and differs from~\cite{GoranciHP17a} in the following perspectives.
\begin{itemize}
\item Our algorithm dynamically maintains the approximate Schur complement of a separable graphs by maintaining a separator tree of such graphs, rather than their \emph{$r$-divisions} as used in~\cite{GoranciHP17a}. In fact, we do not believe purely $r$-divisions based algorithms will achieve the performance as guaranteed by our new algorithm. This is evidenced by previous dynamic algorithms for maintaining reachability in directed planar graphs by Subramanian \cite{Subramanian93}, $(1+\varepsilon)$-approximating to all-pairs shortest paths by~Klein and Subramanian~\cite{KleinS98}, exactly maintaining $s-t$ max-flow in planar graphs by Italiano et al.~\cite{ItalianoNSW11}, all of which are based on $r$-divisions and have running times of order $n^{2/3}$ (and some of which have been improved by using other approaches). 
\item Our current lower bound is much stronger than the previous one: the previous lower bound only holds for general graphs and the \emph{vertex-update} model, where nodes, not edges, are turned on or off, and its proof was based on a simple relation between $s-t$ connectivity and $s-t$ effective resistance $R^G_{\textrm{eff}}(s,t)$ (i.e., if $s,t$ is connected iff $R^G_{\textrm{eff}}(s,t)$ is not infinity). In contrast, our new lower bounds hold for separable graphs (and also general graphs) and the edge-update model. The corresponding proofs exploit new reductions from the OMv problem to the $5$-length cycle detection and triangle detection problems in separable graphs and general graphs, respectively, which might be of independent interest, and the latter problems are related to the effective resistances (see Section~\ref{sec:proof_lower_separable}).    
\end{itemize}


\subsection{Our Techniques}
Our dynamic algorithm for maintaining an Approximate Schur complement (ASC) w.r.t. a set of terminals for separable graphs is built upon maintaining a \emph{separator tree} of such graphs and two properties (called \emph{transitivity} and \emph{composability}) of ASCs. Such a tree can be constructed very efficiently by recursively partitioning the subgraphs using separators. Slightly more formally, each node in the tree corresponds to a subgraph of the original graph and contains a subset of vertices as its boundary vertices which in turn are treated as terminals. For each node $H$, we will maintain an ASC $H'$ of $H$ w.r.t its terminals. We will guarantee throughout all the updates that the ASC of any node can be computed efficiently in a bottom-up fashion, by the above two properties of ASCs. This stems from the fact that we only need to recompute the ASCs of nodes that lie on a path from a \emph{constant} number leaves to the node of interest. Since each such path has length $O(\log n)$ and the recomputation of ASC of one node takes time $\tilde{O}(\sqrt{n})$, the update time will be guaranteed to be $\tilde{O}(\sqrt{n})$. For the detailed implementation, we need to overcome the difficulty that the error in the approximation ratio might accumulate through this recursive computation and an update might require to change the set of boundary vertices of many nodes, thus resulting in a prohibitive running time. We remark that though the idea of using separator tree of planar/separable graphs is standard~(e.g.,~\cite{EppsteinGIS96}), the main novelty of our algorithm is to use such a tree as the backbone to dynamically maintain the approximate Schur complement.

To obtain our dynamic algorithms for all-pairs effective resistance, we appropriately declare and add new terminals whenever we get a new query, and then run the above dynamic algorithm for ASC with respect to the corresponding terminal set.

To obtain our lower bound, we provide new reductions from the Online Boolean Matrix-Vector Multiplication (OMv) problem to the incremental or decremental single-source effective resistance problem. More specifically, given an OMv instance with vectors $\vect{u,v}$ and a matrix $\mat{M}$, we construct a $\sqrt{n}$-separable graph $G$ such that $\vect{u}\mat{M}\vect{v}=1$ if and only if there exists a cycle of length $5$ incident to some vertex $t$ in $G$. This $5$-length cycle detection problem in turn can be solved by inspecting the diagonal entry corresponding to $t$ of the inverse of a matrix that is defined from $G$. Furthermore, the diagonal entry of this matrix is inherently related to the effective resistance~\cite{MNSUW17:spectrum}. By appropriately dynamizing the graph $G$ and using the time bounds for the OMv problem from the conjecture, we get the conditional lower bound for separable graphs. 

For general graphs, the lower bound is proved in a similar way, except that the constructed graph is different and we instead use a relation between effective resistance and triangle detection problem. That is, we first reduce the OMv problem to the $t$-triangle detection problem such that the OMv instance satisfies $\vect{u}\mat{M}\vect{v}=1$ if and only if there exists a triangle incident to some vertex $t$ in the constructed $G$. The latter problem can again be solved by checking the diagonal entry corresponding to $t$ of some matrix, which in turn encodes the effective resistance of between $t$ and a properly specified vertex $s$.

\paragraph*{Other Related Work.}
Previous work on dynamic algorithms for planar or plane graphs include: shortest paths~\cite{KleinS98,AbrahamCG12,ItalianoNSW11}, $s-t$ min-cuts/max-flows~\cite{ItalianoNSW11}, reachability in directed graphs~\cite{Subramanian93,IKLS17decremental,DiksS07}, ($k$-edge) connected components~\cite{EppsteinGIS96,HIKLRS2017contracting}, the best swap and the minimum spanning forest~\cite{EppsteinGIS96}. There also exist work on dynamic algorithms for $\sqrt{n}$-separable graphs, e.g., on transitive closure and $(1+\varepsilon)$-approximation of all-pairs shortest paths~\cite{karczmarz18}.

It is interesting to note that for the (simpler) offline dynamic effective resistance problems, i.e., the sequence of updates and queries are given as an input, Li et al.~\cite{LPYZ18} recently gave an incremental algorithm with $O(\frac{\poly\log n}{\varepsilon^2})$ amortized update and query time for general graphs.

\section{Preliminaries} \label{sec: prelim_DERP}


Let $G=(V,E,\vect{w})$ be an undirected weighted graph such that $\vect{w}(e)>0$ for any $e\in E$. We fix an arbitrary orientation of edges and treat $G$ as a \emph{resistor network} such that each edge $e\in E$ represents a resistor with \emph{resistance} $\vect{r}(e):=1/\vect{w}(e)$. For any vertex pair $s,t$, the $s-t$ flow is a function $\vect{f}: E \rightarrow \R^+$ satisfying the \emph{conservation condition}, i.e., for any vertex $v\in V\setminus\{s,t\}$, $\sum_{u:(u,v)\in E}\vect{f}(u,v)=\sum_{u:(v,u)\in E} \vect{f}(v,u)$. The \emph{energy of an $s-t$ flow} is defined as $\EE_{G}(\vect{f},s,t):=\sum_{e\in E}\vect{r}(e)\vect{f}(e)^2$. The \emph{$s-t$ electrical flow} $\vect{f}^*$ is defined as the $s-t$ flow that minimizes the energy $\EE_{G}(\vect{f},s,t)$ among all $s-t$ flows $\vect{f}$ with unit flow value, i.e., $\sum_{v\in V}\vect{f}(s,v)=1$. Let $\EE_G(s,t)$ denote the energy of the $s-t$ electrical flow, that is, $\EE_{G}(s,t):=\EE_{G}(\vect{f}^*,s,t)$. An electrical flow $\vect{f}$ naturally corresponds to a \emph{potential} $\phi$ in the sense that we can assign each vertex $u$ a potential $\phi(u)$ such that for any $e = (u,v)$, $\vect{f}(e)=\frac{\phi(u)-\phi(v)}{\vect{r}(e)}$. 

It is well known that the $s-t$ effective resistance $R^G_{\textrm{eff}}(s,t)$ as defined in Section~\ref{sec:intro_DERP} satisfies that $R^G_{\textrm{eff}}(s,t)=\phi(s)-\phi(t)$, which is the potential difference between $s,t$ when we send one unit of the (unique) $s-t$ electrical flow from $s$ to $t$. Furthermore, it holds that for any $s,t$, the energy of the $s-t$ electrical flow is equivalent to the $s-t$ effective resistance, that is, $\EE_{G}(s,t)=R^G_{\textrm{eff}}(s,t)$~(see e.g., \cite{doyle84}). In the following, we will mainly focus on how to dynamically maintain (approximation of) effective resistance $R^G_{\textrm{eff}}(s,t)$.

\paragraph*{Properties of Separable Graphs.} 
Let $G=(V,E)$ be a sparse, $O(\sqrt{n})$-separable graph. For an edge-induced subgraph $H$ of $G$, any vertex that is incident to vertices not in $H$ is called a \emph{boundary vertex}. We let $\partial(H)$ denote the set of \emph{boundary vertices} belonging to $H$. All other vertices incident to edges from only $H$ will be called \emph{interior vertices} of $H$.

A hierarchical decomposition of $G$ is obtained by recursively partitioning the graph using separators into edge-disjoint subgraphs (called regions), where the removal of each separator partitions the subgraph into two two edge-disjoint subgraph. This decomposition is represented by a binary (decomposition) tree $\mathcal{T}(G)$, which we refer to as a \emph{separator tree} of $G$. For any subgraph $H$ of $G$, we use $H \in \mathcal{T}(G)$ to denote that $H$ is a node of $\mathcal{T}(G)$ (to avoid confusion with the vertices of $G$, we refer to the vertices of $\mathcal{T}(G)$ as nodes). The \emph{height} $\eta(H)$ of a node is the number of edges in the longest path between that node and a leaf. In addition, let $S(H)$ denote a balanced separator of the subgraph $H$. Formally, $\mathcal{T}(G)$ satisfies the following properties:
\begin{enumerate}
\itemsep0em
\item The root node of $\mathcal{T}(G)$ is the graph $G$.
\item A non-leaf node $H \in \mathcal{T}(G)$ has exactly two children $\child_1(H)$, $\child_2(H)$ and a balanced separator $S(H)$ such that $\child_1(H) \cup \child_2(H) = H$, $V(\child_1(H)) \cap V(\child_2(H)) = S(H)$ and $E(\child_1(H)) \cap E(\child_2(H)) = \emptyset$.
\item For a node $H \in \mathcal{T}(G)$, the set of boundary vertices $\partial(H) \subseteq V(H)$ is defined recursively as follows:
\begin{itemize}[noitemsep]
\item If $H$ is the root of $\mathcal{T}(G)$, i.e., $H=G$, then $\partial(G) = S(G)$. 
\item Otherwise, $\partial(H) = S(H) \cup (\partial(P) \cap V(H))$, where $P$ is the parent of $H$ in $\mathcal{T}(G)$.
\end{itemize}
\item For each node $H \in \mathcal{T}(G)$ and its children $\child_1(H)$, $\child_2(H)$, we have $\partial(\child_1(H)) \cup \partial(\child_2(H)) \supseteq \partial(H)$.
\item The number of boundary vertices per node $H \in \mathcal{T}(G)$, i.e., $|\partial(H)|$, is bounded by $O(\sqrt{n})$.
\item There are $O(\sqrt{n})$ leaf subgraphs in $\mathcal{T}(G)$, each having at most $O(\sqrt{n})$ edges.
\item The height of the tree $\mathcal{T}(G)$ is $O(\log n)$, i.e., $\eta(G) = O(\log n)$.
\item Each edge $e \in E$ is contained in a unique leaf subgraph of $\mathcal{T}(G)$.
\end{enumerate}

The lemma below shows that a separator tree can be constructed with an additional $\log n$ factor overhead in the running time for computing a separator. We include its proof here for the sake of completeness.

\begin{lemma} \label{lem: sepTreeTime}
Let $G=(V,E)$ be a $O(\sqrt{n})$-separable graph whose balanced separator can be computed in $s(n)$ time. There is an algorithm that computes a separator tree $\mathcal{T}(G)$ in $O(s(n) \log n)$ time. 
\end{lemma}
\begin{proof}
For some constant $c \geq 1$, let $S(G)$ be a $\alpha$-balanced separator of size $c\sqrt{n}$, where $\alpha = 2/3$. First, we let $G$ be the root node of $\mathcal{T}(G)$. Let $G_1$ and $G_2$ be the two disjoint components of $G$ obtained after the removal of the vertices in $S$.  We define the children $\child_1(G), \child_2(G)$ of $G$ as follows: $V(\child_i(G)) = V(G_i) \cup S(G)$, $E(\child_i(G)) = E(G_i)$, for $i =1,2$, and whenever an edge connects two vertices in $S(G)$, we arbitrarily append it to either $E(\child_1(G))$ or $E(\child_2(G))$. By construction, property (2) in the definition of $\mathcal{T}(G)$ holds. We continue by repeatedly splitting each child $\child_i(G)$ in the same way as we did for $G$, until there are $O(\sqrt{n})$ components, each of size $O(\sqrt{n})$. The components at this level form the \emph{leaf nodes} of $\mathcal{T}(G)$. Note that the height of $\mathcal{T}(G)$ is bounded by $O(\log n)$ as the size of any child of a node $H$ is at most $2/3$ fraction of the size of $H$.

We define the boundary vertices for each node in $\mathcal{T}(G)$ according to property (3) in the definition of separator trees. To get the bound on the number of boundary vertices per node $H \in \mathcal{T}(G)$, note that the size of $\partial(H)$ is bounded by 
\[
	\left(c \cdot \sum_{i=0}^{O(\log n)}\sqrt{(2/3)^i}\right) \sqrt{n} = O(\sqrt{n}).
\]   


Finally, let $t(n)$ be the maximum time required to construct the separator tree of a $O(\sqrt{n})$-separable graph with $n$ vertices. Then, for some suitably chosen $n_0$, 

\[ t(n) \leq
  \begin{cases}
    s(n) + \max\{t(n_1) + t(n_2)\}  & \quad \text{if } n >  n_0,\\
    0 & \quad \text{if } n \leq n_0,
  \end{cases}
\]
where the maximum is over $n_1, n_2$ such that
\begin{align*}
	& n   \leq n_1 + n_2 \leq n + 2c\sqrt{n}, \quad \text{ and } \quad  \frac{1}{3}n \leq n_i \leq \frac23 n + c\sqrt{n} \quad \text{ for } i = 1,2.
\end{align*}

By a similar analysis as the proof of Theorem 1 of~\cite{EppsteinGIS96}, we can guarantee that $t(n) \leq O(s(n) \log n)$. \qedhere
\end{proof}

\paragraph*{The OMv Conjecture.} 
Our lower bound will be built upon the following OMv problem and conjecture.
\begin{definition}
	In the \emph{Online Boolean Matrix-Vector Multiplication (OMv)} problem, we are given an integer $n$ and an $n\times n$ Boolean matrix $\mat{M}$. Then at each step $i$ for $1\leq i\leq n$, we are given an $n$-dimensional column vector $\vect{v}_i$, and we should compute $\mat{M}\vect{v}_i$ and output the resulting vector before we proceed to the next round.
\end{definition}
\begin{conjecture}[OMv conjecture~\cite{HenzingerKNS15}]~\label{conjecture}
	For any constant $\varepsilon>0$, there is no $O(n^{3-\varepsilon})$-time algorithm that solves OMv with error probability at most $1/3$. 
\end{conjecture}
We will work on a related problem which is called the $\vect{u}\mat{M}\vect{v}$ problem.
\begin{definition}\label{def:umv_problem}
	In the $\vect{u}\mat{M}\vect{v}$ problem with parameters $n_1,n_2$, we are given a matrix $M$ of size $n_1\times n_2$ which can be preprocessed. After preprocessing, a vector pair $\vect{u},\vect{v}$ is presented, and our goal is to compute $\vect{u}^\top\mat{M}\vect{v}$.
\end{definition}
\begin{theorem}[\cite{HenzingerKNS15}]\label{thm:umv_hard}
	Unless the OMv conjecture~\ref{conjecture} is false, there is no algorithm for the $\vect{u}\mat{M}\vect{v}$ problem with parameters $n_1,n_2$ using polynomial preprocessing time and computation time $O(n_1^{1-\delta}n_2+n_1n_2^{1-\delta})$ that has an error probability at most $1/3$, for some constant $\delta$.
\end{theorem}

\paragraph*{Spectral and Resistance Sparsifiers.}
Below we present two notion of edge sparsifiers. The first requires that the quadratic form of the original and sparsified graph are close. The second requires that all-pairs effective resistances of the corresponding graphs are close. 
\begin{definition}[Spectral Sparsifier] \label{def: specSpar_Planar} Given a graph $G=(V,E,\vect{w})$ and $\varepsilon \in (0,1)$, we say that a subgraph $H=(V,E_H,\vect{w}_H)$ is an $(1 \pm \varepsilon)$-\emph{spectral sparsifier} of $G$ if 
	\[ \forall \vect{x} \in \mathbb{R}^{n},~(1-\varepsilon)\vect{x}^\top\mat{L}_G\vect{x} \leq \vect{x}^\top{\mat{L}_H}\vect{x} \leq (1+\varepsilon)\vect{x}^\top\mat{L}_G\vect{x}. \]
\end{definition}
\begin{definition}[Resistance Sparsifier] Given a graph $G=(V,E,\vect{w})$ and $\varepsilon \in (0,1)$, we say that a subgraph $H=(V,E_H,\vect{w}_H)$ is an  $(1 \pm \varepsilon)$-\emph{resistance sparsifier} of $G$ if
	\[ \normalfont
	\forall u,v \in V,~(1-\varepsilon)R^G_{\textrm{eff}}(u,v) \leq R_{\textrm{eff}}^H(u,v) \leq (1+\varepsilon) R^G_{\textrm{eff}}(u,v).
	\]
\end{definition}
The following lemma shows that Definition~\ref{def: specSpar_Planar} is equivalent to approximating the pseudoinverse Laplacians. We include its proof here for the sake of completeness.
\begin{lemma} \label{lem: equivPseudo}
Assume $G$ is connected. Then the following statements are equivalent:
\begin{enumerate}
\itemsep0em
\item  $\forall \vect{x} \in \mathbb{R}^{n},~(1-\varepsilon)\vect{x}^\top\mat{L}_G\vect{x} \leq \vect{x}^\top{\mat{L}_H}\vect{x} \leq (1+\varepsilon)\vect{x}^\top\mat{L}_G\vect{x}.$
\item $\displaystyle \forall \vect{x} \in \mathbb{R}^{n},~\frac{1}{(1+\varepsilon)}\vect{x}^\top\mat{L}_G^\dagger\vect{x} \leq \vect{x}^\top\mat{L}_H^\dagger\vect{x} \leq \frac{1}{(1-\varepsilon)}\vect{x}^\top\mat{L}_G^\dagger\vect{x}.$
\end{enumerate}
\end{lemma}
\begin{proof}[Proof of Lemma~\ref{lem: equivPseudo}]
Since $\mat{L}_G$ is symmetric we can diagonalize it and write \[ \mat{L}_G = \sum_{i=1}^{n-1} \lambda^{G}_i \vect{u}_i \vect{u}_i^\top, \]
where $\lambda^{G}_1 \geq \ldots \geq \lambda^{G}_{n-1}$ are the non-zero sorted eigenvalues of $\mat{L}_G$ and $\vect{u}_1,\ldots,\vect{u}_{n-1}$ are a corresponding set of orthonormal eigenvectors. The \emph{Moore-Penrose Pseudoinverse} of $\mat{L}_G$ is then defined as 
\[
   \mat{L}_G^\dagger = \sum_{i=1}^{n-1} \frac{1}{\lambda^{G}_i} \vect{u}_i \vect{u}_i^\top.
\]

We next show that for every $\vect{x} \in \mathbb{R}^{n}$, $(1-\varepsilon)\vect{x}^\top\mat{L}_G\vect{x} \leq \vect{x}^\top{\mat{L}_H}\vect{x}$ is equivalent to $\vect{x}^\top\mat{L}_H^\dagger\vect{x} \leq \frac{1}{(1-\varepsilon)}\vect{x}^\top\mat{L}_G^\dagger\vect{x}$. The other equivalence can be shown in a symmetric way. 

For every $\vect{x} \in \mathbb{R}^{n}$, by definition of $\mat{L}_G$ and $\mat{L}_H$ we have \[(1-\varepsilon)\vect{x}^\top\mat{L}_G\vect{x}  \leq \vect{x}^\top{\mat{L}_H}\vect{x} \Longleftrightarrow (1-\varepsilon)\sum_{i=1}^{n-1} \lambda^{G}_i(\vect{u}_i^\top \vect{x})^{2} \leq \sum_{i=1}^{n-1} \lambda^{H}_i(\vect{u}_i^\top \vect{x})^{2}.\]

We next show that 
\begin{equation}
\label{eq: equivalPseudo}
\begin{split}
	\forall \vect{x} \in \mathbb{R}^{n},~(1-\varepsilon)\sum_{i=1}^{n-1} \lambda^{G}_i(\vect{u}_i^\top \vect{x})^{2} & \leq \sum_{i=1}^{n-1} \lambda^{H}_i(\vect{u}_i^\top \vect{x})^{2}  \\
	 & \Longleftrightarrow (1-\varepsilon)\lambda^{G}_i \leq \lambda^{H}_i,~ \forall i=1,\ldots,n-1.
\end{split}
\end{equation}

Since for every $\mat{x} \in \mathbb{R}^{n}$, $(\vect{u}_i^\top\vect{x})^{2} \geq 0$, $i=1,\ldots,n-1$, the if-direction of the equivalence in (\ref{eq: equivalPseudo}) follows immediately. For the only-if direction, we proceed by contraposition. To this end, assume that there exists some $i \in \{1,\ldots,n-1\}$ such that $(1-\varepsilon) \lambda_i^{G} > \lambda_i^{H}$. Then there exists a vector $\vect{x} = \vect{u_i} \in \mathbb{R}^{n}$ such that 
\[
   	(1-\varepsilon)\sum_{i=1}^{n-1} \lambda^{G}_i(\vect{u}_i^\top \vect{x})^{2} = (1-\varepsilon) \lambda_i^{G} > \lambda_i^{H} = \sum_{i=1}^{n-1} \lambda^{H}_i(\vect{u}_i^\top \vect{x})^{2},
\]
where the first and last inequality follow from the fact that $\vect{u}_i$'s are orthonormal eigenvectors, i.e., $\vect{u}^\top_i \vect{u}_i = 1$ and $\vect{u}^\top_i \vect{u}_j = 0$, $\forall i \neq j$. This gives a contradiction and thus proves the only-if direction. 
Now, for every $\vect{x} \in \mathbb{R}^{n}$ we have

\begin{align*}
(1-\varepsilon)\vect{x}^\top\mat{L}_G\vect{x}  \leq \vect{x}^\top{\mat{L}_H}\vect{x} & \Longleftrightarrow (1-\varepsilon)\lambda^{G}_i \leq \lambda^{H}_i,~ \forall i=1,\ldots,n-1 \\
& \Longleftrightarrow \frac{1}{\lambda^{H}_i} \leq \frac{1}{(1-\varepsilon)}\cdot \frac{1}{\lambda^{G}_i},~ \forall i=1,\ldots,n-1 \\
& \Longleftrightarrow \sum_{i=1}^{n-1} \frac{1}{\lambda^{H}_i}(\vect{u}_i^\top \vect{x})^{2} \leq \frac{1}{(1-\varepsilon)}\sum_{i=1}^{n-1} \frac{1}{\lambda^{G}_i}(\vect{u}_i^\top \vect{x})^{2} \\
& \Longleftrightarrow \vect{x}^\top\mat{L}_H^\dagger\vect{x} \leq \frac{1}{(1-\varepsilon)}\vect{x}^\top\mat{L}_G^\dagger\vect{x},
\end{align*}
where the penultimate equivalence can be proven in a similar way to equivalence in (\ref{eq: equivalPseudo}).
\end{proof}

In our algorithm we use the following observations: (1) Since, by definition, the effective resistance between any two nodes $u$ and $v$ is the quadratic form defined by the pseudo-inverse of the Laplacian computed at the vector $\1_s-\1_t$, i.e., $R^G_{\textrm{eff}}(u,v) = (\1_s-\1_t)^\top\mat{L}^\dagger(\1_s-\1_t)$, it follows that the effective resistances between any two nodes in $G$ and $H$ are the same up to a $1/(1 \pm \varepsilon)$ factor. By definitions for resistance and spectral sparsifiers, and Lemma~\ref{lem: equivPseudo} we have the following fact.

\begin{fact}
	Let $\varepsilon\in (0,1)$ and let $G$ be a graph. Then every $(1\pm\varepsilon)$-spectral sparsifier of $G$ is an $1/(1\pm \varepsilon)$-resistance sparsifier of $G$. 
\end{fact}

(2) The lemma below suggests that given a graph, by decomposing the graph into several  pieces and computing a good sparsifier for each piece, one can obtain a good sparsifier for the original graph which is the union of the sparsifiers for all pieces. 

\begin{lemma}[\cite{AbrahamDKKP16}, Lemma~4.18] \label{lemm: decomposability} Let $G=(V,E,\vect{w})$ be a weighted graph whose set of edges is partitioned into $E_1,\ldots,E_\ell$. Let $H_i$ be a $(1\pm \varepsilon)$-spectral sparsifier of $G_i=(V,E_i)$, where $i=1,\ldots,\ell$. Then $H=\bigcup_{i=1}^{\ell} H_i$ is a $(1\pm \varepsilon)$-spectral sparsifier of $G$. 
\end{lemma}

\paragraph*{Schur Complement and Approximate Schur Complement.}
For a given connected graph $G=(V,E)$ and a set $K\subset V$ of terminals with $1\leq |K|\leq |V|-1$, let $F = V \setminus K$ be the set of non-terminal vertices in $G$. The partition of $V$ into $F$ and $K$ naturally induces the following partition of the Laplacian $\mat{L}_G$ into blocks:
\[ \mat{L}_G = 
\begin{bmatrix}
\mat{L}_{[F,F]} & \mat{L}_{[F,K]} \\
\mat{L}_{[K,F]} & \mat{L}_{[K,K]} \\
\end{bmatrix}
\]
We remark that since $G$ is connected and $F$ and $K$ are non-empty, one can show that $\mat{L}_{[F,F]}$ is invertible. We have the following definition of Schur complement.
\begin{definition}[Schur Complement] \label{def: Schur} The (unique) \emph{Schur complement} of a graph Laplacian $\mat{L}_G$ with respect to a terminal set $K$ is
	\[
	\SC(G,K) := \mat{L}_{[K,K]} - \mat{L}_{[K,F]}\mat{L}^{-1}_{[F,F]}\mat{L}_{[F,K]}.
	\]
\end{definition} 
It is well known that the matrix $\SC(G,K)$ is a Laplacian matrix for some graph $G'$. 


\begin{definition}[Approximate Schur Complement (ASC)] \label{def: approxSchurComplement} Given a graph $G=(V,E,\mat{w})$, $K \subset V$ and its Schur complement $\SC(G,K)$, we say that a graph $H=(K,E_H,\mat{w}_H)$ is a $(1\pm \varepsilon)$\emph{-approximate Schur complement (abbr. $(1\pm \varepsilon)$-ASC)} of $G$ with respect to $K$ if 
	\[
	\forall \vect{x} \in \mathbb{R}^{k},~(1-\varepsilon)\vect{x}^\top\SC(G,K)\vect{x} \leq \vect{x}^\top{\mat{L}_H}\vect{x} \leq (1+\varepsilon)\vect{x}^\top\SC(G,K)\vect{x}. 
	\]
	Moreover, we say that $H$ is an $1$\emph{-ASC} of $G$ with respect to $K$ if $\mat{L}_H = \SC(G,K)$.
\end{definition}

Note that $(1\pm \varepsilon)$-ASC is a spectral sparsifier of Schur complement. 
Furthermore, approximate Schur complement can be computed efficiently as guaranteed in the following lemma~\cite{DurfeeKPRS17}. 
\begin{lemma} \label{lem: ApproxSchur}
	Fix $\varepsilon \in (0,1/2)$ and $\gamma \in (0,1)$, and let $G=(V,E,\vect{w})$ be a graph with $K \subset V$ and  $|K|=k$. There is an algorithm \textsc{ApproxSchur}$(G,K, \varepsilon, \delta)$ that computes a $(1 \pm \varepsilon)$-\emph{ASC} $H$ of $G$ with respect to $K$ such that the following statements hold  probability at least $1-\gamma$: 
	\begin{enumerate}
	\itemsep0em
		\item The graph $H$ has $O(k\varepsilon^{-2}\log(n/\gamma))$ edges.
		\item The total running time for computing $H$ is $\tilde{O}( m \log^{3}(n / \gamma) + n \varepsilon^{-2} \log^{4} (n/\gamma))$.
	\end{enumerate}	
\end{lemma}

\section{Useful Properties of Approximate Schur Complement}
\label{sec:asc_vs}
In this section we show that Approximate Schur complement can be treated as a vertex effective resistance sparsifier, which is a small graph that (approximately) preserves the pairwise effective resistances among terminal vertices of the original graph. Then we show two important properties called \emph{transitivity} and \emph{composability} properties of ASCs, which will be exploited in our dynamic algorithms for ASCs and effective resistances.


\paragraph*{ASC as Vertex Resistance Sparsifier.} 
To maintain all-pairs effective resistances efficiently, it will be useful to consider the following notion of \emph{vertex sparsifier} that preserves pairwise effective resistances among a set of terminals. 
\begin{definition}[Vertex Resistance Sparsifier (VRS)] Given a graph $G=(V,E,\vect{w})$ with $K \subset V$, we say that a graph $H=(K,E_H,\vect{w}_H)$ is an $(1 \pm \varepsilon)$-\emph{vertex resistance sparsifier}~\emph{(abbr. $(1\pm \varepsilon)$-VRS)} of $G$ with respect to $K$ if
	\[ \normalfont
	\forall s,t \in K,~(1-\varepsilon) R^G_{\textrm{eff}}(s,t) \leq R^H_{\textrm{eff}}(s,t) \leq (1+\varepsilon) R^G_{\textrm{eff}}(s,t).
	\]
\end{definition} 

We show that ASC can be treated as a vertex resitance sparsifier. For this, we recall the following lemma which shows that the quadratic form of the pseudo-inverse of the Laplacian $\mat{L}$ is preserved by taking the quadratic form of the pseudo-inverse of its Schur complement, for demand vectors supported on the terminals. 

\begin{lemma}[\cite{MillerP13}, Lemma~5.1] \label{lemm: SchurComp} Let $\bb$ be a demand vector of a graph $G$ whose vertices are partitioned into terminals $K$, and non-terminals $F$ such that only terminals have non-zero entries in $\bb$. Let $\bb_K$ be the restriction of $\bb$ on the terminals and let $\SC(G,K)$ be the Schur complement of $\mat{L}_G$ with respect to $K$. Then 
	\[
	\bb^\top \mat{L}_G^{\dagger} \bb =  \bb_K^\top \SC(G,K)^{\dagger} \bb_K.
	\]
\end{lemma}

Using interchangeability between graphs and their Laplacians, we can interpret the above result in terms of graphs as well. The lemma below relates ASCs and vertex resistance sparsifiers. We include its proof here for the sake of completeness. 
\begin{lemma}~\label{lemma:exact_schur} Let $G=(V,E,\mat{w})$ be a graph with $K \subset V$. If $H$ is an $(1\pm \varepsilon)$-\emph{ASC} of $G$ with respect to $K$, then $H$ is an \emph{$1/(1 \pm \varepsilon)$-VRS} of $G$ with respect to $K$.
\end{lemma}
\begin{proof}
	Let $k = |K|$. First, note that by Definition~\ref{def: approxSchurComplement} and Lemma~\ref{lem: equivPseudo} we have
	\[
	\forall \vect{x} \in \mathbb{R}^{k},~\frac{1}{(1+\varepsilon)}\vect{x}^\top\SC(G,K)^\dagger \vect{x} \leq \vect{x}^\top{\mat{L}_H}^\dagger \vect{x} \leq \frac{1}{(1-\varepsilon)}\vect{x}^\top\SC(G,K)^\dagger \vect{x}. 
	\]
	
	Next, let $(s,t) \in K$ be any terminal pair. Consider the demand vector $\vect{\chi}_{s,t} \in \mathbb{R}^{k}$ and extend this vector to $\vect{\chi}_{s,t}' = \begin{bmatrix} \vect{0} &  \vect{\chi}_{s,t} \end{bmatrix}^\top \in \mathbb{R}^{n}$. By definition of effective resistance and Lemma \ref{lemm: SchurComp}  we get that 
	\begin{align*}
	R_{\textrm{eff}}^H(s,t)  & = \vect{\chi}_{s,t}^\top{\mat{L}_H}^\dagger \vect{\chi}_{s,t}  \leq \frac{1}{(1-\varepsilon)} \vect{\chi}_{s,t}^\top \SC(G,K)^\dagger \vect{\chi}_{s,t} \\ 
	& = \frac{1}{(1-\varepsilon)}\vect{\chi}_{s,t}'^\top \mat{L}_G^\dagger \vect{\chi}_{s,t}'  = \frac{1}{(1-\varepsilon)} R^G_{\textrm{eff}}(s,t).
	\end{align*}
	For the lower-bound on $R_{\textrm{eff}}^H(s,t)$, using the same reasoning, we get that
	\begin{align*}
	R_{\textrm{eff}}^H(s,t)  & = \vect{\chi}_{s,t}^\top{\mat{L}_H}^\dagger \vect{\chi}_{s,t}  \geq \frac{1}{(1+\varepsilon)} \vect{\chi}_{s,t}^\top \SC(G,K)^\dagger \vect{\chi}_{s,t} \\ 
	& = \frac{1}{(1+\varepsilon)}\vect{\chi}_{s,t}'^\top \mat{L}_G^\dagger \vect{\chi}_{s,t}' = \frac{1}{(1+\varepsilon)} R^G_{\textrm{eff}}(s,t). \qedhere
	\end{align*} 
\end{proof}


\paragraph*{Transitivity and Composability of ASCs.}
In the following, we show a \emph{transitivity} property of ASCs and then show how the ASCs of two neighboring nodes of the separator tree $\mathcal{T}(G)$ can be combined to give the ASC of their parent (called \emph{composability}), which will enable us to compute the ASCs of all nodes of $\mathcal{T}(G)$ in a bottom-up fashion. 

\paragraph{Transitivity of ASCs.} To show the transitivity property the ASCs, we will use the following lemma which establishes the connection between the Schur complement and the Laplacian of the original graph. 

\begin{lemma}[\cite{MillerP13}, Lemma B.2] \label{lem: Schur}
Let $\mat{L}_G$ be the Laplacian of $G$ and let $\SC(G,K)$ be its Schur complement. For any $x \in \mathbb{R}^{k}$ the following holds
\[
	\mat{x}^\top \SC(G,K) \mat{x} = \min_{\mat{y}}
	 \begin{bmatrix*}
           \mat{y} \\
           \mat{x} \\
         \end{bmatrix*}^\top \mat{L}_G \begin{bmatrix*}
           \mat{y} \\
           \mat{x} \\
         \end{bmatrix*}.
\]
\end{lemma}


We are now ready to show the following transitive property of ASCs.

\begin{lemma}[Transitivity of ASCs] \label{lem: multiComb}

If $H'$ is an $(1 \pm \varepsilon)$-\emph{ASC} of $G$ with respect to $K'$, and $H$ is an $(1 \pm \varepsilon)$-\emph{ASC} of $H'$ with respect to $K$, where $K' \supseteq K$, then $H$ is an $(1 \pm \varepsilon)^2$-\emph{ASC} of $G$ with respect to $K$.
\end{lemma}
\begin{proof}
Let $k=|K|$ and $k'=|K'|$. By the assumption of the lemma, the following inequalities hold:
\[
 \forall \vect{x} \in \mathbb{R}^{k'},~(1-\varepsilon)\vect{x}^\top\SC(G,K')\vect{x} \leq \vect{x}^\top{\mat{L}_{H'}}\vect{x} \leq (1+\varepsilon)\vect{x}^\top\SC(G,K')\vect{x},
 \]
and
\[
 \forall \vect{x} \in \mathbb{R}^{k},~(1-\varepsilon)\vect{x}^\top\SC(H',K)\vect{x} \leq \vect{x}^\top{\mat{L}_H}\vect{x} \leq (1+\varepsilon)\vect{x}^\top\SC(H',K)\vect{x}.
 \]
We need to show that
\[
	\forall \vect{x} \in \mathbb{R}^{k},~ (1-\varepsilon)^2 \vect{x}^\top{\SC(G,K)}\vect{x} \leq \vect{x}^\top{\mat{L}_H}\vect{x} \leq (1+\varepsilon)^2 \vect{x}^\top{\SC(G,K)}\vect{x}.
\]

We first show the upper bound on $\vect{x}^\top{\mat{L}_H}\vect{x}$. Note that since $K' \supseteq K$, using Gaussian elimination, $\SC(G,K)$ can be constructed by first constructing $\SC(G,K')$ from $G$ and then constructing $\SC(G,K)$ from $\SC(G,K')$ using Gaussian elimination. Thus $\SC(G,K)$ is the Schur complement of $\SC(G,K')$ with respect to $K$. For any $\vect{x} \in \mathbb{R}^{k}$, let $\vect{y}$ be the vector that attains the minimum value in Lemma~\ref{lem: Schur} for $\SC(G,K')$. If we define $\vect{x'} = \begin{bmatrix} \vect{y} &  \vect{x} \end{bmatrix}^\top \in \mathbb{R}^{k'}$, then we get
\begin{align*}
 \vect{x}^\top{\mat{L}_H}\vect{x} & \leq (1+\varepsilon)\vect{x}^\top\SC (H',K)\vect{x} 
  \leq (1+\varepsilon) \vect{x'}^\top \mat{L}_{H'} \vect{x'} \\ 
 & \leq (1+\varepsilon)^2 \vect{x'}^\top\SC(G,K')\vect{x'} 
  = (1+\varepsilon)^2 \vect{x}^\top{\SC(G,K)}\vect{x}.
\end{align*}
We now give the lower bound on $\vect{x}^\top{\mat{L}_H}\vect{x}$. Recall that $\SC(H',K)$ is the Schur complement of $\mat{L}_{H'}$ with respect to $K$. For any vertex $\vect{x} \in \mathbb{R}^{k}$, let $\vect{y}$ be the vector given by Lemma~\ref{lem: Schur} for $\mat{L}_{H'}$. If we define $\vect{x''} = \begin{bmatrix} \vect{y} &  \vect{x} \end{bmatrix}^\top \in \mathbb{R}^{k'}$, then we get
\begin{align*}
 \vect{x}^\top{\mat{L}_H}\vect{x} & \geq (1-\varepsilon)\vect{x}^\top\SC (H',K)\vect{x} 
  = (1-\varepsilon) \vect{x''}^\top \mat{L}_{H'} \vect{x''} \\ 
 & \geq (1-\varepsilon)^2 \vect{x''}^\top\SC(G,K')\vect{x''} 
  \geq (1-\varepsilon)^2 \vect{x}^\top{\SC(G,K)}\vect{x}. \qedhere
\end{align*}\end{proof}

\paragraph{Composability of ASCs.} To show the composability of ASCs, we first review an equivalent way of defining Schur complements. The main idea is to view $\SC(G,K)$ as a multi-graph where each multi-edge corresponds to a walk in $G$ that starts and ends at $K$, but has all intermediate vertices in $V \setminus K$. We call such a walk a \emph{terminal-free} walk that starts and ends in $K$. Formally, a terminal-free walk \[u_0, \ldots, u_\ell\] of length $\ell$, with $u_0,u_\ell \in K$ and $u_i \in V \setminus K$, for $i=1,\ldots,\ell$ corresponds to a multi-edge between $u_0$ and $u_\ell$ in $\SC(G,K)$ with weight given by
\begin{equation} \label{eq: weightSchur}
	\ww^{\SC(G,K)}_{u_0,\ldots,u_{\ell}} = \frac{\prod_{0 \leq i \leq \ell} \ww({u_i, u_i+1})}{\prod_{0 < i < \ell} \dd(u_i)},
\end{equation}
where $\dd(u) = \sum_{v : (u,v) \in E} \ww(u,v)$ denotes the weighted degree of a vertex $u$.

This connection is formally proven in the lemma below.

\begin{lemma}[\cite{DurfeePPR17:arxiv}, Lemma 5.4] \label{lem: EquivSchur} 
Given a graph $G$ and a partition of its vertices into $K$ and $V \setminus K$, the graph $G^{K}$ obtained by forming an union over all multi-edges corresponding to terminal-free walks that start and end in $K$, with weights given by Equation (\ref{eq: weightSchur}) is exactly $\SC(G,K)$.
\end{lemma}

We next show that if a graph can be viewed as a combination of two graphs along some subset of shared terminals, combining the respective sparsifiers of these two graphs in the same way gives a sparsifer for the original graph.

Formally, Let $G_1=(V_1,E_1)$ and $G_2=(V_2,E_2)$ be edge-disjoint graphs with terminals $K_1$ and $K_2$, respectively. Furthermore, assume that all vertices in the intersection of $V_1$ and $V_2$, if exist, are terminals in both graphs. That is, $(V_1 \cap V_2) \subset K_i$, for $i=\{1,2\}$. The \emph{merge} of $G_1$ and $G_2$ is the graph $G =(V_1 \cup V_2, E_1 \cup E_2)$ with terminals $K_1 \cup K_2$ formed by identifying the terminals in $S$. We denote this operation by $G := G_1 \oplus G_2$.

\begin{lemma}[Composability of Schur complement]\label{lemm: exactMerge}

Let $G := G_1 \oplus G_2$. If $H_1$ is an $1$\emph{-ASC} of $G_1$ with respect to $K_1$, and $H_2$ is an $1$\emph{-ASC} of $G_2$ with respect to $K_2$, then $H := H_1 \oplus H_2$ is an $1$\emph{-ASC} of $G$ with respect to $K$.
\end{lemma}
\begin{proof}
Note that $H_i = \SC(G_i, K_i)$, for $i=\{1,2\}$, and recall that the $G_1$ and $G_2$ share the terminals in some non-empty subset $S$, i.e., $S \subset K_i$, for $i=\{1,2\}$. To prove the lemma, we need to show that 

\[ \SC(G_1, K_1) \oplus \SC(G_2, K_2) = \SC(G,K). \] 

We do so by making use of Lemma~\ref{lem: EquivSchur}. More specifically, we argue that every multi-edge (along with its corresponding weight) in $\SC(G,K)$ is contained either in $\SC(G_1, K_1)$ or $\SC(G_2, K_2)$. We distinguish the following cases.

(1) For any two terminals $t$ and $t'$ in $K_1 \setminus S$, we have that $\SC(G_1, K_1)$ contains all the multi-edges between $t$ and $t'$ in $\SC(G,K)$. This is because $G_1$ and $G_2$ are edge-disjoint, and there is no terminal-free walk between $t$ and $t'$ in $G$ that does not use a terminal in $S$. The same reasoning can be applied to terminal pairs in $K_2 \setminus S$. 

(2) For any two terminals $s$ and $t$ in $S \times K$, we have that the corresponding multi-edges in $\SC(G,K)$, are either contained in $\SC(G_1, K_1)$ or $\SC(G_2, K_2)$. If $t \in K_1 \setminus S$ or $t \in K_2 \setminus S$, then the same reasoning as in case (1) applies. However, if $t \in S$, then $S(G_1,K_1)$ contains all the multi-edges that correspond to terminal-free walks between $s$ and $t$ that use the edges in $G_1$, and $S(G_2,K_2)$ contains all the multi-edges that correspond to terminal-free walks between $s$ and $t$ that use the edges in $G_2$.

(3) For any two terminals $t$ and $t'$ in $(K_1 \setminus S) \times (K_2 \setminus S)$, there is no terminal-free walk between $t$ and $t'$ in $G$ that does not use a terminal in $S$, since $S$ is a separator of $G$. Thus there are no multi-edges between $t$ and $t'$ in $\SC(G,K)$, so the merge $\SC(G_1, K_1) \oplus \SC(G_2, K_2)$ correctly does not add such edges.   
\end{proof}

\begin{lemma}[Composition of ASCs] \label{lem: approxMerge}

Let $G := G_1 \oplus G_2$. If $H_1'$ is an $(1\pm \varepsilon)$\emph{-ASC} of $G_1$ with respect to $K_1$, and $H_2'$ is an $(1 \pm \varepsilon)$\emph{-ASC} of $G_2$ with respect to $K_2$, then $H' := H_1' \oplus H_2'$ is an $(1 \pm \varepsilon)$\emph{-ASC} of $G$ with respect to $K$.
\end{lemma}
\begin{proof}
First, let $H_1$ be an $1$-ASC of $G_1$ with respect to $K_1$, and $H_2$ be an $1$-ASC of $G_2$ with respect to $K_2$. By Lemma~\ref{lemm: exactMerge}, $H := H_1 \oplus H_2$ is an $1$-ASC of $G$ with respect to $K$, i.e., $\mat{L}_H = \SC(G,K)$. Now note that we can treat $H_i$ and $H_i'$, for $i=\{1,2\}$ as graphs defined on the same vertex set $V(H)$, by adding appropriate isolated vertices. By assumption, each $H_i'$ is an $(1\pm \varepsilon)$-spectral sparsifier of $H_i$ and thus, applying the Decomposition Lemma~\ref{lemm: decomposability} gives that $H':= H_1' \oplus H_2'$ is an $(1\pm \varepsilon)$-spectral sparsifier of $H$, or equivalently, $H'$ is an $(1 \pm \varepsilon)$-ASC of $G$.
\end{proof}

\section{Dynamic Effective Resistances on Separable Graphs}
In this section, we first present our fully dynamic algorithm for maintaining a $(1\pm \delta)$-approximate Schur complement (i.e., prove Theorem~\ref{thm:dynamic_asc}) and then use it give a dynamic algorithm for $(1+\varepsilon)$-approximating all-pairs effective resistances in separable graphs and prove Theorem~\ref{thm:upperbound}. For simplicity, we assume that the separator of $G$ can be computed in $\tilde{O}(n)$ time. 
\subsection{Dynamic Approximate Schur Complement}\label{sec:dynamic_ASC}

Let $\delta \in (0,1)$. Let $K\subset V$ be a set of terminals with $|K|\leq O(\sqrt{n})$. We give a data-structure for maintaining a $(1 \pm \delta)$-ASC of a $O(\sqrt{n})$-separable graph $G$ with respect to a set $K'$ of $O(\sqrt{n})$ vertices (which contains the terminal set $K$) that supports \textsc{Insert} and \textsc{Delete} operations as defined before. In addition, it supports the following operation:
\begin{itemize}

\item \textsc{AddTerminal}$(u)$: Add the vertex $u$ to the terminal set $K$, as long as $\abs{K} \leq O(\sqrt{n})$. 
\end{itemize}
%


\paragraph*{Data Structure.} Throughout we compute and maintain a balanced separator $S(G)$ of $G$ that contains $K$ and satisfies that $|S(G)|\leq O(\sqrt{n})$. We let $K'=S(G)$ and we will maintain a $(1\pm \delta)$-ASC of $G$ w.r.t. $K'$. By definition of boundary vertices,  $K'=\partial(G)$. Let $\delta'=\frac{\delta}{c\log n+1}$ for some constant $c$. In our dynamic algorithm, we will maintain a separator tree $\mathcal{T}(G)$ (see {Section~\ref{sec: prelim_DERP}) such that for each node $H \in \mathcal{T}(G)$, we maintain its separator $S(H)$ and a set $\buffer(H)$ of edges of $H$, which is initially empty, and an ASC $H'$ of $H$ w.r.t. $\partial(H)$. 
Throughout the updates, the set $\buffer(H)$ will denote the subset of edges which are only contained in $H$ while contained in neither of its children. Let $\mathcal{D}(G,\delta)$ denote such a data-structure. We recompute $\mathcal{D}(G,\delta)$ every $\Theta(\sqrt{n})$ operations using the initialization below.


\paragraph*{Initialization.} We show how to efficiently compute the ASC $H'$ for each node $H$ from $\mathcal{T}(G)$. We do this in a bottom-up fashion by first calling 
Algorithm~\ref{alg: approxSchurNode} on each leaf node and then on the non-leaf nodes, where $\textsc{ApproxSchur}$ is the procedure from Lemma~\ref{lem: ApproxSchur}.

In what follows, whenever we compute an approximate Schur complement, we assume that procedure \textsc{ApproxSchur} from Lemma~\ref{lem: ApproxSchur} is invoked on the corresponding subgraph and its boundary vertices, with error $\delta'$ and error probability $\gamma = 1/n^3$. In the following, we will assume that all the calls to the \textsc{ApproxSchur} are correct.




\begin{algorithm2e}[t]
\label{alg: approxSchurNode}
\caption{\textsc{ApproxSchurNode}$(H, \partial(H),\delta')$}
Set $\gamma \gets 1/n^{3}$ \\
\If{$H$ is a leaf} 
{
	Set $H' \gets \textsc{ApproxSchur}(H,\partial(H),\delta', \gamma)$
} 

\If{$H$ is a non-leaf}
{
    Let $\child_1(H),\child_2(H)$ be the children of $H$ \\
    Let $\child_i(H)'$ be the ASC of $\child_i(H)$, for $i=1,2$  \\ 
    Set $R \gets \child_1(H)' \oplus_{\phi} \child_2(H)'$ and $E(R) \gets E(R) \cup \buffer(H)$  \\
    Set $H' \gets \textsc{ApproxSchur}(R,\partial(H),\delta', \gamma)$
}

\Return $H'$
\end{algorithm2e}

%
%

The following lemma shows that after invoking Algorithm~\ref{alg: approxSchurNode} in a bottom-up fashion, we have computed the ASC for every node in $\mathcal{T}(G)$.
\begin{lemma} \label{lem: dsCorrectness}

Let $H \in \mathcal{T}(G)$ be a node of height $\eta(H) \geq 0$ and $\emph{X}(H) = \emptyset$. Then $H' = \textsc{ApproxSchur}$ $\textsc{Node}(H,\partial(H),\varepsilon)$ is an $(1\pm \delta')^{\eta(H)+1}$\emph{-ASC} of $H$ with respect to $\partial(H)$. 
\end{lemma}
\begin{proof}
We proceed by induction on $\eta(H)$. For the base case, i.e., $\eta(H) = 0$, $H$ is a leaf node. By Lemma~\ref{lem: ApproxSchur} and Algorithm~\ref{alg: approxSchurNode}, $H'$ is indeed a $(1 \pm \delta')$-ASC of $H$ with respect to $\partial(H)$. 
	
	Let $H$ be a non-leaf node, i.e. $\eta(H) > 0$. Let $\child_1(H), \child_2(H)$ and $\child_1'(H), \child_2'(H)$ be defined as in Algorithm~\ref{alg: approxSchurNode}. By properties (2), (3) and (4) of $\mathcal{T}(G)$ and the fact that $X(H)=\emptyset$, we have $H = \child_1(H) \oplus \child_2(H)$. 
	By induction hypothesis, it follows that $\child_i(H)'$ is an $(1\pm \delta')^{\eta(\child_i(H)) + 1}$-ASC of $\child_i(H)$, for $i=1,2$. Using Lemma~\ref{lem: approxMerge} and since $\eta(\child_i(H)) + 1 = \eta(H)$, for $i=1,2$, we get that $R := \child_1(H)' \oplus \child_2(H)'$ is an $(1\pm \delta')^{\eta(H)}$-ASC of $H$ with respect to $V(R) := \partial(\child_1(H)) \cup \partial(\child_2(H))$. Now, since $V(R) \supseteq \partial(H)$ by property (4) of $\mathcal{T}(G)$ and by Lemma~\ref{lem: ApproxSchur}, it follows that $H'$ is an $(1 \pm \delta')$-ASC of $R$ with respect to $\partial(H)$. Finally, applying Lemma~\ref{lem: multiComb} on $R$ and $H'$ we get that $H'$ is an $(1 \pm \delta')^{\eta(H) + 1}$-ASC of $H$.
\end{proof}

Next we analyze the running time of the initialization and recomputation procedure. The lemma below shows that the ASC of any node in $\mathcal{T}(G)$ can be computed in $\tilde{O}(\sqrt{n}/\delta^{2})$.

\begin{lemma} \label{lemm: computeAscNode}
	Let $H \in \mathcal{T}(G)$ and assume that $\abs{\emph{\buffer}(H)} \leq O(\sqrt{n})$. Then we can compute an \emph{ASC} $H'= \textsc{ApproxSchurNode}(H,\partial(H),\varepsilon)$ of $H$ in $\tilde{O}(\sqrt{n}/\delta^{2})$ time.
\end{lemma}
\begin{proof}
	We distinguish two cases. First, if $H$ is a leaf node, then by property (5) of $\mathcal{T}(G)$, we have that $\abs{E(H)} \leq O(\sqrt{n})$. The latter along with Lemma~\ref{lem: ApproxSchur} (2) imply the time to compute $H'$ is $\tilde{O}(\sqrt{n}/\delta'^{2})$. Second, if $H$ is a non-leaf node, then by Lemma~\ref{lem: ApproxSchur} (1) we know that $\abs{E(\child_i(H)')} \leq \tilde{O}(\sqrt{n}/\delta'^{2})$, for $i = 1,2$. Since by assumption $\abs{\buffer(H)} \leq O(\sqrt{n})$, we get that $\abs{R \cup \buffer(H)} \leq \tilde{O}(\sqrt{n}/\delta'^{2})$. Thus, the time to compute $H'$ on top of $R \cup \buffer(H)$ is bounded by $\tilde{O}(\sqrt{n}/\delta'^{2}) = \tilde{O}(\sqrt{n}/\delta^{2})$ (again by Lemma~\ref{lem: ApproxSchur} (2) and the choice of $\delta'$).
\end{proof}

We now analyze the running time for initializing our data-structure. Let $T_{\mathcal{D}(G)}$ denote the time required to compute $\mathcal{D}(G)$. 

\begin{lemma} \label{lem: DSrunningTime}
	
	The time $T_{\mathcal{D}(G)}$ required to compute $\mathcal{D}(G)$ is $\tilde{O}(n / \delta^{2})$.
\end{lemma}
\begin{proof}
	By Lemma~\ref{lem: sepTreeTime} recall that we can construct $\mathcal{T}(G)$ in $\tilde{O}(n)$ time. Note that by construction of the separator tree, the number of non-leaf nodes is bounded by the number of leaf nodes. Since there there are $O(\sqrt{n})$ leaf nodes, the total number of nodes in $\mathcal{T}(G)$ is $O(\sqrt{n})$. By Lemma~\ref{lemm: computeAscNode} we get that the time needed to compute an ASC $H'$ for every node $H \in \mathcal{T}(G)$ is $\tilde{O}(\sqrt{n}/\delta^{2})$. Combining the above bounds gives that $T_{\mathcal{D}(G)}$ is $\tilde{O}(n / \delta^{2})$.
\end{proof}
Since $\delta'=\frac{\delta}{c\log n+1}$ and $\eta(G)=O(\log n)$, the graph $G'$ is a $(1\pm\delta)$-ASC of $G$ w.r.t. $\partial(G)$.

\paragraph*{Handling Edge Insertions.} We now describe the \textsc{Insert} operation. Let us consider the insertion of an edge $e=(u,v)$ of weight $w$. We maintain a stack $Q$, which is initially set to empty. We then update the root node by adding $(u,v)$ with weight $w$ to $G$, and push $G$ onto $Q$. During the traversal of $\mathcal{T}(G)$, our procedure maintains two pointers that point to the current node $H$ (initially set to $G$) and a node $N$ (if any exists) that represents the node for which $u$ and $v$ belong to different children of $N$, respectively. As long as we have not found such a node $N$, and the current node $H$ is not a leaf, we proceed as follows. 

We examine the child of $H$ that contains both $u$ and $v$ (if there is more than one, then we just pick one of them). If $u$ and $v$ belong to the same child, say $\child(H)$, then we add this edge to $\child(H)$ and update the current node $H$ to $\child(H)$. We then push $H$ onto $Q$. If, however, $u$ and $v$ belong to different children, then we set $N$ to be the current node $H$ and add the edge $(u,v)$ to $\buffer(N)$, since $u$ and $v$ cannot appear together in the nodes of the lower levels. At this point, this forces $u$ and $v$ to become boundary vertices in $N$ and all other nodes descending from $N$ that contain either $u$ or $v$. We handle this by making use of the $\textsc{AddBoundary}()$ procedure, depicted in Algorithm~\ref{alg: addBoundary}. Finally, we recompute the ASCs of the affected nodes in a bottom-up fashion using the stack $Q$ (as shown in Algorithm~\ref{alg: updateApproxSchur}). This procedure is summarized in Algorithm~\ref{alg: insert}. We remark that for simplicity, we let $Q.\textsc{Push}(H)$ denote the event of pushing the pointer to $H$ to the stack $Q$, for any node $H$.

\begin{algorithm2e}[t]
\caption{\textsc{UpdateApproxSchur}$(\textsc{Stack} Q)$}
\label{alg: updateApproxSchur}

\While{$Q \neq \emptyset$} 
{
  Set $H \gets Q.\textsc{Pull}()$ \\
  Set $H' \gets \textsc{ApproxSchurNode}(H,\partial(H),\varepsilon)$ \\
}

\end{algorithm2e}

\begin{algorithm2e}[t]
\caption{\textsc{Insert}$(u,v,w)$}
\label{alg: insert}
Let $Q$ be an initially empty stack. \\
Set $E(G) \gets E(G) \cup \{(u,v)\}$,  $Q.\textsc{Push}(G)$, $H \gets G$ and  $N \gets \nil$ \\
\While {$N = \nil$ and $H$ is a non-leaf}
{
	\eIf{there exists a child of $H$ that contains both $u$ and $v$}
	{
		Let $\child(H)$ denote any such a child \\
		Set $E\left(\child(H)\right) \gets E\left(\child(H)\right) \cup \{(u,v)\}$ \\
		Set $H \gets \child(H)$ \\
		$Q.\textsc{Push}(H)$ \\
	}
	{
	    Set $N \gets H$ \\
        Set $\buffer(N) \gets \buffer(N) \cup \{(u,v)\}$ \\
        \textsc{AddBoundary}$(u,N)$, \textsc{AddBoundary}$(v,N)$ \\
	}
}

\tcp{Update the ASCs of the nodes in $Q$} 
\textsc{UpdateApproxSchur}$(Q)$
\end{algorithm2e}

After the pre-processing step and after each insertion/deletion of an edge, our augmented separator tree $\mathcal{T}(G)$ satisfies the following invariant.

\begin{invariant} For every edge $e$ in the current graph $G$, exactly one of the following two holds:
\begin{itemize}
\item there is a leaf node $H \in \mathcal{T}(G)$ such that $e \in E(H)$,
\item there is an internal node $H \in \mathcal{T}(G)$ such that $e \in \emph{X}(H)$.
\end{itemize}
\end{invariant}


The following lemma guarantees that the updated graph $G'$ (i.e., the sparsifier of the root node $G$) is good approximation to the Schur complement of $G$ with respect to the boundary, after the execution of $\textsc{Insert}(u,v)$ in Algorithm~\ref{alg: insert}. 

\begin{lemma}\label{lemma:insertion} Let $G'$ be the updated sparsifier of the root node $G$, after the insertion of edge $(u,v)$. Then $G'$ is an $(1 \pm \delta)$-\emph{ASC} of $G$ with respected to $\partial(G)$.
\end{lemma}
\begin{proof}
	We proceed inductively as in the proof of Lemma~\ref{lem: dsCorrectness} and show that for any node $H$, the corresponding sparsifier $H'$ is an $(1\pm \delta')^{\eta(H)+1}$-ASC of $H$ with respect to $\partial(H)$. Since the base case remains the same, let us consider a non-leaf node $H$. If $\buffer(H) = \emptyset$, then the correctness follows from the inductive step of Lemma~\ref{lem: dsCorrectness}. However, $\buffer(H) \neq \emptyset$ implies that $H \neq \child_1(H) \oplus \child_2(H)$. This is because $H$ is the last node for the edges of $\buffer(H)$ whose endpoints were contained in the same node in $\mathcal{T}(G)$. Recall that the endpoints of all the edges in $\buffer(H)$ were declared boundary vertices for $H$ and all descendants containing them. Thus we have that
	\[
	H = \left(\child_1(H) \oplus \child_2(H)\right) \cup \buffer(H).
	\]
	By induction hypothesis, it follows that $c_i(H)'$ is an $(1\pm \delta')^{\eta(\child_i(H)) + 1}$-ASC of $c_i(H)$, for $i=1,2$. Using Lemma~\ref{lem: approxMerge} and since $\eta(\child_i(H)) + 1 = \eta(H)$, for $i=1,2$, we get that $R := \child_1(H)' \oplus_{\phi} \child_2(H)'$ is an $(1\pm \delta')^{\eta(H)}$-ASC of $H \setminus \buffer(H)$ with respect to $V(R):= \partial(\child_1(H)) \cup \partial(\child_2(H))$. First, since $V(R) \supseteq V(\buffer(H))$ by construction, Lemma~\ref{lem: approxMerge} implies that $R' := R \cup \buffer(H)$ is an $(1\pm \delta')^{\eta(H)}$-ASC of $(H \setminus \buffer(H)) \cup \buffer(H) = H$ with respect to $V(R)$. Second, since $V(R) \supseteq \partial(H)$ by property (4) of $\mathcal{T}(G)$ and by Lemma~\ref{lem: ApproxSchur}, it follows that $H'$ is an $(1 \pm \delta')$-ASC of $R'$ with respect to $\partial(H)$. Finally, applying Lemma~\ref{lem: multiComb} on $R'$ and $H'$ we get that $H'$ is an $(1 \pm \delta')^{\eta(H) + 1}$-ASC of $H$. The statement of the lemma then follows from the facts that $\delta'=\frac{\delta}{c\log n+1}$ and $\eta(G)=O(\log n)$.
\end{proof}

For the running time of $\textsc{Insert}(u,v,w)$, we distinguished two cases. 

First, suppose that the insertion of the edge $(u,v)$ does not trigger a re-computation of the data-structure. Note that the stack $Q$ (in Algorithm~\ref{alg: insert}) contains all nodes in the path starting from the root node $G$, and then repeatedly choosing \emph{exactly one} child of the current node that contains both $u$ and $v$, until the node $N$ is reached. Since the height of $\mathcal{T}(G)$ is $O(\log n)$, it follows that $\abs{Q} \leq O(\log n)$. Additionally, by Lemma~\ref{lemm: computeAscNode}, the time to re-compute an ASC of any node is bounded by $\tilde{O}(\sqrt{n}/\delta^{2})$. Thus we get that the time needed to update the ASCs of the nodes in $Q$ is $\tilde{O}(\sqrt{n}/\delta^{2})$. As we will shortly argue, the running time of $\textsc{AddBoundary}()$ is also bounded by $\tilde{O}(\sqrt{n}/\delta^{2})$. Combining the above, we get that the running time of $\textsc{Insert}(u,v)$ is $\tilde{O}(\sqrt{n}/\delta^{2})$.

Second, suppose that the edge $(u,v)$ triggers a re-computation of the data-structure. Then by Lemma~\ref{lem: DSrunningTime}, we recompute $\mathcal{D}(G, \delta)$ in $\tilde{O}(n/\delta^{2})$ time. Since we recompute that data-structure every $\Theta(\sqrt{n})$ insertions, the amortized update time per insertion is $\tilde{O}(\sqrt{n}/\delta^{2})$. The above bounds combined give that the amortized time per edge insertion is bounded by $\tilde{O}(\sqrt{n}/\delta^{2})$. This bound can be made worst-case by keeping two copies of the data structure and performing periodical rebuilds.

\paragraph*{Handling Terminal Additions to the Boundary.} We now describe the \textsc{AddTerminal}$(u)$ operation. 
It is implemented by simply invoking \textsc{AddBoundary}$(u,G)$, where $G$ is the root of $\mathcal{T}(G)$. For the procedure $\textsc{AddBoundary}(u,H)$, we maintain a stack $Q$, which is initially set to empty. As long as the current $H$ is a node in $\mathcal{T}(G)$, we first check whether $u \in \partial(H)$. If this is the case, then we simply do nothing as the ASC $H'$ of $H$ with respect to $\partial(H)$ contains $u$. Otherwise, we add $u$ to $\partial(H)$, and push the node $H$ to $Q$. Next, if $H$ is not a leaf-node, let $\child(H)$ be the \emph{unique} child that contains $u$. We then set $\child(H)$ to be our current node $H$ and perform the same steps as above, until we reach some leaf-node, in which case we set $H$ to $\nil$. Finally, we recompute the ASCs of the affected nodes in a bottom-up fashion using the stack $Q$. This procedure is summarized in Algorithm~\ref{alg: addBoundary}. 

The correctness of this procedure can be shown similarly to the correctness of $\textsc{Insert}()$. For the running time, the crucial observation is that if $u \not \in \partial(H)$, for some non-leaf node $H$, then by property (2) of $\mathcal{T}(G)$, it follows that $u$ is assigned to an \emph{unique} child of $H$. Thus, in the worst-case, the stack $Q$ contains all the nodes in the path between $H$ and some leaf-node. Note that $\abs{Q} = O(\log n)$ and by Lemma~\ref{lemm: computeAscNode}, time to re-compute an ASC of any node is $\tilde{O}(\sqrt{n}/\delta^{2})$. Combining the above, we get that the running time of $\textsc{AddBoundary}(u,H)$ is $\tilde{O}(\sqrt{n}/\delta^{2})$.

\begin{algorithm2e}[t]
\caption{\textsc{AddBoundary}$(u,v,w)$}
\label{alg: addBoundary}
Let $Q$ be an initially empty stack. 
\While {$N = \nil$}
{
	\If{$u \not \in \partial(H)$}
	{
		Set $\partial(H) \gets \partial(H) \cup \{u\}$ \\
	    $Q.\textsc{Push}(H)$ \\
		\If {$H$ is a non-leaf} 
		{
			Let $c(H)$ be the \emph{unique} child that contains $u$ \\
			Set $H \gets \child(H)$ \\
		}
	}
	
	\If{$H$ is a leaf}
	{
		Set $H \gets \nil$ \\
	}
}

\tcp{ Update the ASCs of the nodes in $Q$}
\textsc{UpdateApproxSchur}$(Q)$.
\end{algorithm2e}

\paragraph*{Handling Edge Deletions.} We now describe the \textsc{Delete} operation. Let us consider the deletion of an edge $e = (x,y)$. Our procedure is symmetric to the \textsc{Insert}() operation. We maintain a stack $Q$, which is initially set to empty. We then update the root node by deleting $(u,v)$ from $G$, and pushing $G$ onto $Q$. During the traversal of $\mathcal{T}(G)$, our procedure maintains the current node $H$ (initially set to $G$) and determines the node $N$ that represent the lowest-level node in $\mathcal{T}(G)$ that contains the edge $(u,v)$. Note that $N$ is not necessarily a leaf-node. As long as we have not found such a node we proceed as follows. 

We examine the \emph{unique} child of $H$ that contains the edge $(u,v)$ (by property (2) of $\mathcal{T}(G)$).  If there exists such a child $c(H)$, then we delete $(u,v)$ from $c(H)$ and update the current node $H$ to $c(H)$. We then push $H$ to $Q$. If, however, such a child does not exist, then we set $N$ to be the current node $H$. Next, if $N$ is a non-leaf node, we remove the edge $(u,v)$ from $\buffer(N)$. Finally, we recompute the ASCs of the affected nodes in a bottom-up fashion using the stack $Q$. This procedure is summarized in Algorithm~\ref{alg: delete}.

Similarly to the \textsc{Insert}$()$ operation, we can show  that the worst-case running time of \textsc{Delete}$(u,v)$ operation is $\tilde{O}(\sqrt{n}/\delta^{2})$.

Finally, recall that we set $\gamma=1/n^3$ as the error probability of \textsc{ApproxSchur} from Lemma~\ref{lem: ApproxSchur}. This will guarantee that throughout all updates, our algorithm succeeds with probability at least $1-O(n)\cdot \frac{1}{n^3}\geq 1- O(\frac{1}{n^2})$ as the total number of nodes in $\mathcal{T}(G)$ is $O(\sqrt{n})$, each update involves recomputation of the ASCs of $O(\log n)$ nodes and our algorithm recomputes the data structure every $\Theta(\sqrt{n})$ operations.

\begin{remark} We can easily generalize the above framework to $O(\sqrt{n})$-separable graphs for which the separator can be computed in $s(n)$ time, since the only place we need such computation is to initialize or re-compute the data structure $\mathcal{D}(G,\delta)$ (after every $\Theta(\sqrt{n})$ operations). This implies that the update time will become $\tilde{O}((s(n)+n/\delta^{2})/\sqrt{n})$ and the query time remains the same as before.  
\end{remark}

\begin{algorithm2e}[t]
\caption{\textsc{Delete}$(u,v)$}
\label{alg: delete}
Let $Q$ be an initially empty stack. \\
Set $E(G) \gets E(G) \setminus \{(u,v)\}$,  $Q.\textsc{Push}(G)$, $H \gets G$ and  $N \gets \nil$. \\
\While {$N = \nil$}
{
	\eIf{If there exists a (\emph{unique}) child $\child(H)$ of $H$ that contains $(u,v)$}
	{
		$E\left(\child(H)\right) \gets E\left(\child(H)\right) \setminus \{(u,v)\}$. \\
		Set $H \gets \child(H)$. \\
		$Q.\textsc{Push}(H)$. \\
	}
	{
	    Set $N \gets H$. \\
	}
	
	\If{$N$ is a non-leaf} 
	{ 
	  $\buffer(N) \gets \buffer(N) \setminus \{(u,v)\}$. \\
	}	
}

\nonl\texttt{// Update the ASCs of the nodes in $Q$} \\
\textsc{UpdateApproxSchur}$(Q)$.

\end{algorithm2e}

\subsection{Extension to Dynamic All-Pairs Effective Resistance}

We next explain how to use a dynamic ASC algorithm to obtain a fully-dynamic algorithm for maintaining an $(1+\varepsilon)$-approximation to all-pairs (resp., single-pair) effective resistance(s) in a $O(\sqrt{n})$-separable graph $G$ and prove Theorem~\ref{thm:upperbound}. The data-structure support the operations $\textsc{Insert}(u,v,r)$, $\textsc{Delete}(u,v)$, and  \textsc{EffectiveResistance}$(s,t)$ as defined in the beginning of this chapter. 

Our dynamic effective resistance algorithm uses the above dynamic algorithm for maintaining a $(1\pm \delta)$-ASC as a subroutine. Formally, to maintain $(1+\varepsilon)$-approximations of effective resistances, we will invoke the dynamic ASC algorithm with parameters $\delta = \varepsilon/4$.
To answer the queries of the effective resistance of any two given vertices, we use the following result due to Durfee et al.~\cite{DurfeeKPRS17}.

\begin{theorem} \label{thm: spielmanTeng}
	Fix $\delta \in (0,1/2)$ and let $G=(V,E , \vect{w})$ be a weighted graph with two distinguished vertices $s,t \in V$. There is an algorithm \textsc{EstimateEffRes}$(G,s,t)$ that computes a value $\psi$ such that
	\[ \normalfont
	(1-\delta) R^G_{\textrm{eff}}(s,t) \leq \psi \leq (1+\delta)R^G_{\textrm{eff}}(s,t), 
	\]
	in time $\tilde{O}(m + n/\delta^{2})$ with probability at least $1-n^c$ for some constant $c\geq 1$.
\end{theorem}

For simplicity, we focus on the case that the separator of the separable graph can be computed in $\tilde{O}(n)$ time. The algorithm and analysis can be easily generalized to handle the case when the computation time for separator is $s(n,m)$, by the same argument as before.

We now describe the query operation. We first consider how to maintain all-pairs effective resistances. Given $s$ and $t$, we start by calling  $\textsc{AddTerminal}(s)$ and $\textsc{AddTerminal}(t)$ from the dynamic ASC data-structure. This ensures that both $s$ and $t$ are boundary nodes at the root node $G$ (if they were not previously). Thus we obtain a $(1\pm \delta)$-ASC, denoted as $G'$, of the root node $G$ with respect to $\partial(G)$ and run on $G'$ a nearly linear time algorithm for estimating the $s-t$ effective resistance (see Theorem~\ref{thm: spielmanTeng}). Let $\psi$ denote such an estimate. This procedure is summarized in Algorithm~\ref{alg: effectiveResistance}.

\begin{algorithm2e}[t]
\caption{\textsc{EffectiveResistance}$(s,t)$}
\label{alg: effectiveResistance}
\textsc{AddTerminal}$(s)$, \textsc{AddTerminal}$(t)$   \\
Let $G'$ be the ASC of the root node $G$ with respect to $\partial(G)$ \\
Set $\psi \gets$ \textsc{EstimateEffRes}$(G',s,t)$ \\
Return $\psi$ \\
\end{algorithm2e}

For the correctness, by Lemma~\ref{lemma:exact_schur}, we have that $G'$ preserves all-pairs effective resistances among vertices in $\partial(G)$ of $G$ up to an $1/(1 \pm \delta) \approx (1 \pm 2\delta)$ factor. Since we ensured that $s$ and $t$ are included in $\partial(G)$, the $s-t$ effective resistance is approximated within the same factor. By Theorem~\ref{thm: spielmanTeng}, it follows that the estimate $\psi$ approximates the effective resistance between $s$ and $t$ in $G'$, up to a $(1\pm \delta)$ factor. Combining the above guarantees, we get $\psi$ gives an $(1 \pm 2\delta)(1\pm \delta) \leq (1 \pm \varepsilon)$-approximation to $R^G_{\textrm{eff}}(s,t)$, by the choice of $\delta$.

Once the query is answered, we then undo all the changes that we have performed in $\mathcal{T}(G)$ i.e., we bring the data-structure to its state before the query operation. This ensures that the number of terminals at the root node $G$ does not accumulate over a large sequence of query operations.

For the running time, first recall that each $\textsc{AddTerminal}()$ operation can be implemented in $\tilde{O}(\sqrt{n}/\delta^{2})$. Now, as $\abs{V(G')} \leq O(\sqrt{n})$ and $\abs{E(G')} \leq \tilde{O}(\sqrt{n}/\delta^{2})$, by Theorem~\ref{thm: spielmanTeng} it follows that estimate $\psi$ can be computed in $\tilde{O}(\sqrt{n}/\delta^{2})$ time. Combining the time bounds we get that that the worst-case time to answer an $\textsc{EffectiveResistance}(s,t)$ query is $\tilde{O}(\sqrt{n}/ \delta^{2})$. Finally, note that in the same time bound, we can also undo all the changes we have made.

For the single-pair $s-t$ effective resistance, the two vertices $s,t$ are fixed throughout all the operations. For each edge insertion or deletion, we first update the data structure in the same way as for the all-pairs version, and then we compute the $s-t$ effective resistance $R^G_{\textrm{eff}}(s,t)$ and store the answer. For the query for $R^G_{\textrm{eff}}(s,t)$, we simply report the stored answer. The update time is still $\tilde{O}(\sqrt{n}/\delta^2)$, while the query time is only $O(1)$.

\section{Lower Bounds for Dynamic Effective Resistances} \label{sec: lowerBound}

\subsection{A Lower Bound for $O(\sqrt{n})$-Separable Graphs}\label{sec:proof_lower_separable}
In this section, we prove a conditional lower bound for incrementally or decrementally maintaining the $s-t$ effective resistance in $O(\sqrt{n})$-separable graphs and give the proof of Theorem~\ref{thm:lowerbound_separable}. Our proof actually holds for any algorithm that maintains a $(1+O(\frac{1}{n^{36}}))$-approximation of $s-t$ effective resistance. 

We first consider the incremental case, in which only edge insertions are allowed. 

\paragraph*{The reduction.} We reduce the $\vect{u}\mat{M}\vect{v}$ problem~(see Definition \ref{def:umv_problem}) with parameters $n_1=n_2:=n_0$ to the $s-t$ effective resistance problem as follows. Let $\mat{M}$ be the $n_0\times n_0$ Boolean matrix of the $\vect{u}\mat{M}\vect{v}$ problem. Let $n= n_0^2+2n_0+2$. Let $\kappa = 3(n-1)^6$.

Given the matrix $\mat{M}$, we construct a graph $G_\mat{M}=(V_\mat{M}, E)$ as follows. 
\begin{itemize}
	\item  For each pair $1\leq i,j\leq n_0$, we create two vertices $a_{ij}$ and $b_{ij}$, and add an edge $(a_{ij},b_{ij})$ if and only if $M_{ij}=1$. 
	\item For each row $i$, we create a vertex $u_i$ and add edge $(u_i,a_{ik})$ for each $1\leq k\leq n_0$. For each column $j$, we create a vertex $v_j$ and add edge $(v_j,b_{kj})$ for each $1\leq k\leq n_0$. 
\end{itemize}		
This finishes the definition of $G_\mat{M}$. Note that $V_\mat{M}=\{a_{ij},b_{ij}, 1\leq i,j\leq n_0 \}\cup\{u_i,1\leq i\leq n_0 \} \cup\{v_j, 1\leq j\leq n_0\}$. For any vertex $x\in V_\mat{M}$, let $\deg_{G_\mat{M}}(x)$ denote the degree of $x$ in $G_\mat{M}$. 


Now we add two new vertices $t$ and $s$ to $G_\mat{M}$. For any $x\in \{a_{ij},b_{ij}, 1\leq i,j\leq n_0 \}$, add an edge $(s,x)$ with weight $\kappa-\deg_{G_\mat{M}}(x)$. Denote the resulting graph by $G$ and note that $G$ contains $|V_\mat{M}\cup\{s,t\}|=n_0^2+2n_0+2 = n$ vertices.

Assume that $G$ is started in a dynamic effective resistance data structure. We also maintain a number of counters in the data structure. More specifically, we initialize a global counter $Y:=0$. For each vertex $x\in \{u_i, 1\leq i\leq n_0 \}\cup \{v_j, 1\leq j\leq n_0\}$, we maintain a counter $c(x)$ which is initialized to be $0$. 
We now explain how we use this data structure to determine $\vect{u}\mat{M}\vect{v}$.

\begin{itemize}
	\item Once $\vect{u}$ arrives, for any $i$ such that $\vect{u}_i=1$, we insert an edge $(t, u_i)$ with weight $1$, increase $Y$ and $c(u_i)$ by $1$. 
	\item Once $\vect{v}$ arrives, for any $j$ such that $\vect{v}_j=1$, we insert an edge $(t, v_j)$ with weight $1$, increase $Y$ and $c(v_j)$ by $1$. 
	\item Insert an edge $(s,t)$ with weight $\kappa - Y$. For each vertex $x\in \{u_i, 1\leq i\le n_0 \}\cup\{v_j, 1\leq j\leq n_0 \}$, insert an edge $(s,x)$ with weight $\kappa-c(x)-\deg_{G_\mat{M}}(x)$.
	\item We perform one effective resistance query $\textsc{EffectiveResistance}(s,t)$ to obtain the (approximate) $s-t$ effective resistance in the final graph. Let $\lambda=\textsc{EffectiveResistance}(s,t)$. If $\lambda\leq \frac{1}{\kappa}+\frac{Y}{\kappa^3}+\frac{Y(n_0+1)}{\kappa^5} -\frac{1}{\kappa^6}$, then return $1$; otherwise, return $0$. 
\end{itemize}

\paragraph*{Analysis.} Note that throughout the whole sequence of updates (which are only edge insertions) and queries, the dynamic graph $G$ is always $O(\sqrt{n})$-separable, since the set $S:=\{u_1,\cdots, u_{n_0}\}\cup\{v_1,\cdots,v_{n_0} \}\cup\{s,t\}$ is a balanced separator of size $O(\sqrt{n})$.

We have the following lemma that shows an important property of our reduction. The proof of the lemma is deferred to the end of this section.
\begin{lemma}\label{lemma:separable_LB}
	For $\kappa = 3(n-1)^6$, assume that $\textsc{EffectiveResistance}(s,t)$ returns an $(1+\frac{1}{\kappa^{6}})$-approximation of the $s-t$ effective resistance in the final graph $G$. Then the following holds:
	\begin{itemize}
		\item  If $\vect{u}\mat{M}\vect{v}=1$, then $\lambda \leq \frac{1}{\kappa}+\frac{Y}{\kappa^3}+\frac{Y(n_0+1)}{\kappa^5} -\frac{1}{\kappa^6}$;
		\item  If $\vect{u}\mat{M}\vect{v}=0$, then $\lambda >\frac{1}{\kappa}+\frac{Y}{\kappa^3}+\frac{Y(n_0+1)}{\kappa^5} -\frac{1}{\kappa^6}$.
	\end{itemize}
\end{lemma}

Note that by the above lemma, the $\vect{u}\mat{M}\vect{v}$ problem can be solved according to our estimator $\lambda$. Thus, the lower bound for the incremental setting in Theorem~\ref{thm:lowerbound_separable} follows by Theorem~\ref{thm:umv_hard} and by noting that the total number of updates is $O(n_0)=O(\sqrt{n})$ and the total number of queries is $1$.

In the following we prove Lemma~\ref{lemma:separable_LB}. The proof is based on a connection between the $5$-length cycle detection problem and the effective resistance problem.
\begin{proof}[Proof of Lemma~\ref{lemma:separable_LB}]
	Let $G$ denote the final graph of our reduction. Let $H:=G[V_\mat{M}\cup\{t\}]$ denote the subgraph induced by vertex set $V_\mat{M}\cup\{t\}$. We observe that in the graph $H$, there is a cycle of length $5$ containing vertex $t$ if and only if $\vect{u}\mat{M}\vect{v}=1$. 
	
	On the other hand, we can use our estimator $\lambda$ to distinguish if $H$ contains a $5$-length cycle incident to $t$ or not. 
	We let $\mat{A}\in \mathbb{R}^{(n-1)\times (n-1)}$ denote the adjacency matrix of the graph $H$. Note that all entries in $A$ are either $1$ or $0$.
	
	The first claim relates the $5$-length cycle detection to the trace of a matrix related to $\mat{A}$. Recall that we let $X_{uv}$ denote the entry of matrix $X$ with row index corresponding to vertex $u$ and column index corresponding to vertex $v$.
	\begin{claim}\label{claim:cycle_trace}
		Let $\mat{B}=\kappa\cdot \mat{I} - \mat{A}$. If $H$ contains a $5$-length cycle incident to $t$, then $(\mat{B}^{-1})_{tt} \leq \frac{1}{\kappa}+\frac{Y}{\kappa^3}+\frac{Y(n_0+1)}{\kappa^5}-\frac{1.1}{\kappa^6}$. If $H$ does not contain a $5$-length cycle incident to $t$, then $(\mat{B}^{-1})_{tt} \geq \frac{1}{\kappa}+\frac{Y}{\kappa^3}+\frac{Y(n_0+1)}{\kappa^5} -\frac{0.9}{\kappa^6}$.
	\end{claim}
	\begin{proof}
		First we note that $B$ is invertible, as it is strictly symmetric
		diagonally dominant. Furthermore, it holds that 
		$\kappa\cdot\mat{B}^{-1}=(I-\frac{1}{\kappa}\cdot \mat{A})^{-1}$ and thus by the Neumann series expansion, we have \[\kappa\cdot \mat{B}^{-1} =(I-\frac{1}{\kappa}\cdot \mat{A})^{-1}= \sum_{i=0}^\infty (-\frac{1}{\kappa})^i \cdot \mat{A}^i.\]
		This further implies that 
		\begin{equation}
		\label{eqn:inverse_B}
		\begin{split}
		(\kappa\cdot \mat{B}^{-1})_{tt} 
		& = \1_t^\top(\sum_{i=0}^\infty (-\frac{1}{\kappa})^i \cdot \mat{A}^i)\1_t=\sum_{i=0}^\infty (-\frac{1}{\kappa})^i \cdot\1_t^\top(\mat{A}^i)\1_t \\
		& = \sum_{i=0}^\infty (-\frac{1}{\kappa})^i \cdot (\mat{A}^i)_{tt}.	
		\end{split}
		\end{equation}

		Now observe that since $\kappa=3 (n-1)^6$, the first six terms of the above power series dominate. More precisely, note that $(\mat{A}^i)_{tt}$ is the number of $i$-length paths from $t$ to $t$, which is at most $(n-1)^i$. Thus 
		\begin{eqnarray*}
			\sum_{i=6}^\infty \abs{(-\frac{1}{\kappa})^i \cdot (\mat{A}^i)_{tt}}
			\leq 
			\sum_{i=6}^\infty \frac{1}{\kappa^i} (\mat{A}^i)_{tt} 
			\leq 
			\sum_{i=6}^\infty \frac{1}{\kappa^i} (n-1)^i
			\leq
			\frac{0.9}{\kappa^5}.
		\end{eqnarray*}
		
		
		
		Now observe that $(\mat{A}^0)_{tt}=\mat{I}_{tt}=1$; that $\mat{A}_{tt}=0$ since $H$ is a simple graph; that $(\mat{A}^2)_{tt}=\deg_H(t)=Y$, where the last equation follows from the definition of $Y$; that $(\mat{A}^3)_{tt}=0$ since there is no triangle containing $t$; and that $(\mat{A}^4)_{tt}=\sum_{w: (w,t)\in E}\sum_{x:(x,w)\in E}1=\sum_{w: (w,t)\in E} \deg_{G_\mat{M}}(w) = \det_H(t)\cdot (n_0+1) = Y(n_0+1)$. Therefore, 
		\begin{itemize}
			\item If $H$ contains a $5$-length cycle incident to $t$, then $(\mat{A}^5)_{tt}\geq 2$, and thus
			\begin{align*}
				(\kappa\cdot \mat{B}^{-1})_{tt} & \leq 1+\frac{Y}{\kappa^2} + \frac{Y(n_0+1)}{\kappa^4} - \frac{2}{\kappa^5} + \frac{0.9}{\kappa^5} \\ 
				&= 1+\frac{Y}{\kappa^2}+ \frac{Y(n_0+1)}{\kappa^4} -\frac{1.1}{\kappa^5} 
			\end{align*}
			\item If $H$ has no $5$-length cycle incident to $t$, then $(\mat{A}^5)_{tt}=0$, and thus
			\begin{eqnarray*}
				(\kappa\cdot \mat{B}^{-1})_{tt} \geq 1+\frac{Y}{\kappa^2} + \frac{Y(n_0+1)}{\kappa^4} - \frac{0.9}{\kappa^5}
			\end{eqnarray*}
		\end{itemize}
		This completes the proof of the claim.
	\end{proof}
	
	The following claim relates $s-t$ effective resistance to $\mat{B}^{-1}$. The proof almost follows from Lemma 23 in~\cite{MNSUW17:spectrum}, while we include a proof here for the sake of completeness.
	\begin{claim}\label{claim:trace}
		Let $\Lambda=\mathcal{E}_G(s,t)$ and $\mat{B}=\kappa\cdot \mat{I} - \mat{A}$. Then it holds that 
		$\Lambda=(\mat{B}^{-1})_{tt}.$
	\end{claim}
\begin{proof}
	Let $\mat{L}$ denote the Laplacian matrix of $G$ and let $\vect{v}\in \mathbb{R}^{V_\mat{M}\cup\{t\}}$ denote the vector with entries corresponding to weights between $s$ and $u$ for each $u\in V_\mat{M}\cup\{t\}$, i.e., $\vect{v}_u = \kappa-\deg_H(u)$. 
	
	Now the key observation is that 
	\begin{equation*}
	\mat{L}=\left(\begin{array}{cc}
	\mat{B} & -\vect{v}\\
	-\vect{v}^\top & \deg_G(s)
	\end{array}	\right)
	\end{equation*}
	
	For any $\vect{x}\in\mathbb{R}^{V_\mat{M}\cup\{t\}\cup\{s\}}$, let $\widehat{\vect{x}}\in\mathbb{R}^{V_\mat{M}\cup\{t\}}$ be the vector containing the first entries corresponding to vertices in $V_\mat{M}\cup\{t\}$ of $\vect{x}$. Let $\vect{y}$ be the solution of the Laplacian system $\mat{L}\vect{y}=\1_s-\1_t$. Thus, $\vect{y}=\mat{L}^\dag(\1_s-\1_t)$. It also holds that 
	\begin{eqnarray*}
		\mat{B}\cdot \widehat{\vect{y}} -\vect{v}\cdot y_{s} = -\widehat{\vect{1}}_t
	\end{eqnarray*}
	In addition, we know that $\mat{L}\vect{1}=\vect{0}$, and thus $\mat{B}\cdot \widehat{\1}=\vect{v}$. This further implies that, $\widehat{\vect{y}}=\mat{B}^{-1}\cdot \vect{v}\cdot y_s-\mat{B}^{-1}\widehat{\1}_t=y_s\cdot \widehat{\1}-\mat{B}^{-1}\widehat{\1}_t$. Thus, 
	\begin{align*}
		(\1_s-\1_t)^\top \mat{L}^{\dag} (\1_s-\1_t) & =(\1_s-\1_t)^\top\vect{y}=y_s-\widehat{\1}_t^\top\cdot \widehat{\vect{y}} \\
		& =y_s-\widehat{\1}_t^\top\cdot (y_s\cdot \widehat{\1}-\mat{B}^{-1}\widehat{\1}_t) =\widehat{\1}_t^\top\mat{B}^{-1}\widehat{\1}_t
	\end{align*}
	
	Therefore, 
	\begin{eqnarray*}
		\Lambda=\mathcal{E}_G(s,t) = (\1_s-\1_t)^\top \mat{L}^{\dag} (\1_s-\1_t) =  \widehat{\1}_t^\top\mat{B}^{-1}\widehat{\1}_t =(\mat{B}^{-1})_{tt}
	\end{eqnarray*}
\end{proof}
	
	Finally, by the above two claims, if $\vect{u}\mat{M}\vect{v}=1$, then $H$ contains a $5$-length cycle incident to $t$, and thus $\Lambda=(\mat{B}^{-1})_{tt} \leq \frac{1}{\kappa}+\frac{Y}{\kappa^3}+\frac{Y(n_0+1)}{\kappa^5}-\frac{1.1}{\kappa^6}$; if $\vect{u}\mat{M}\vect{v}=0$, then $H$ does not contain any $5$-length cycle incident to $t$, and thus $\Lambda=(\mat{B}^{-1})_{tt} \geq \frac{1}{\kappa}+\frac{Y}{\kappa^3}+\frac{Y(n_0+1)}{\kappa^5} -\frac{0.9}{\kappa^6}$. The statement of the lemma then follows by the fact that $\lambda$ is a $(1+\frac{1}{\kappa^6})$-approximation of $\Lambda$, and that $\frac{1}{\kappa^6}(\frac{1}{\kappa}+\frac{Y}{\kappa^3}+\frac{Y(n_0+1)}{\kappa^5}-\frac{0.9}{\kappa^6})<\frac{0.1}{\kappa^6}$.
\end{proof}

For the lower bound for the decremental setting, 
we start with a graph where $t$ is initially connected to $s$ with weight $\kappa-2n_0$ and to all vertices $x\in\{u_i,1\leq i\leq n_0 \} \cup\{v_j,1\leq j\leq n_0 \}$ with weights $\kappa-1-\deg_{G_\mat{M}}(x)$. When the vectors $\vect{u},\vect{v}$ arrive, we need to increase the weights of some edges $(s,x)$ and $(s,t)$ depending if the corresponding entry of $\vect{u},\vect{v}$ is $1$ or $0$, so that every vertex in $G$ has the same weighted degree $\kappa$. We omit further details here.

\subsection{A Lower Bound for General Graphs}
In this section, we prove Theorem~\ref{thm:lowerbound_general}, which gives a lower bound for incremental and decremental $s-t$ effective resistance problem in general graphs. 

\begin{proof}[Proof of Theorem~\ref{thm:lowerbound_general}]
	We only consider here the incremental setting, where only edge insertions are allowed. For the decremental setting, the correctness follows from a similar construction and similar arguments for decremental lower bound in the proof of Theorem~\ref{thm:lowerbound_separable}. 
	
	We reduce the $\vect{u}\mat{M}\vect{v}$ problem with parameters $n_1=n_2:=n_0$ to the $s-t$ effective resistance problem as follows. Let $\mat{M}$ be the $n_0\times n_0$ Boolean matrix of the $\vect{u}\mat{M}\vect{v}$ problem. Let $n=2n_0+2$ and let $\kappa = 3(n-1)^5$.
	
	We first create a bipartite graph $G_\mat{M}=((R,C),E)$ where $R=(r_1,\cdots,r_{n_0})$ and $C=(c_1,\cdots, c_{n_0})$ corresponding to the rows and columns of $\mat{M}$, respectively. We add an edge $(r_i, c_j)$ in $E$ iff $\mat{M}_{ij}=1$. This finishes the definition of $G_\mat{M}$. For each vertex $x\in R\cup C$, let $\deg_{G_\mat{M}}(x)$ denote the degree of vertex $x$ in $G_\mat{M}$.
	
	Now we add tow new vertices $s,t$ to $G_\mat{M}$. Denote the resulting graph by $G$ and note that $G$ contains $|R\cup C\cup\{s,t\}|=2n_0+2$ vertices. 
	
	Assume that $G$ is started in a dynamic effective resistance data structure. We also initialize a global counter $Y$ to be $0$ and for each vertex $x\in R\cup C$, we initialize a counter $c(x)$ to be $0$. We now explain how we use this data structure to determine $\vect{u}\mat{M}\vect{v}$.
	
	\begin{itemize}
		\item Once $\vect{u}$ arrives, for any $i$ such that $\vect{u}_i=1$, we insert an edge $(t, r_i)$ with weight $1$, and increase $Y$ and $c(r_i)$ by $1$. 
		\item Once $\vect{v}$ arrives, for any $j$ such that $\vect{v}_j=1$, we insert an edge $(t, c_j)$ with weight $1$, and increase $Y$ and $c(c_j)$ by $1$. 
		\item Insert an edge $(s,t)$ with weight $\kappa-Y$. For each $x\in V_\mat{M}$, insert an edge $(s,x)$ with weight $\kappa-c(x)-\deg_{G_\mat{M}}(x)$. 
		\item We perform one effective resistance query $\textsc{EffectiveResistance}(s,t)$ to obtain the (approximate) $s-t$ effective resistance in the final graph. Let $\lambda=\textsc{EffectiveResistance}(s,t)$. If $\lambda\leq \frac{1}{\kappa}+\frac{Y}{\kappa^3}-\frac{1}{\kappa^4}$, then return $1$; otherwise, return $0$.
	\end{itemize}
	
	We have the following lemma similar to Lemma~\ref{lemma:separable_LB}. 
	\begin{lemma}\label{lemma:correctness_LB}
		For $\kappa=3(n-1)^5$, assume that $\textsc{EffectiveResistance}(s,t)$ returns a $(1+\frac{1}{\kappa^{4}})$-approximation of the $s-t$ effective resistance in the final graph $G$. Then the following holds:
		\begin{itemize}
			\item If $\vect{u}\mat{M}\vect{v}=1$, then $\lambda \leq \frac{1}{\kappa}+\frac{Y}{\kappa^3}-\frac{1}{\kappa^4}$;
			\item If $\vect{u}\mat{M}\vect{v}=0$, then $\lambda >\frac{1}{\kappa}+\frac{Y}{\kappa^3}-\frac{1}{\kappa^4}$.
		\end{itemize}
	\end{lemma}
	Given the above Lemma, we can then solve the $\vect{u}\mat{M}\vect{v}$ problem according to the value of our estimator $\lambda$. Thus, the statement of the theorem follows by noting that the total number of updates is $O(n_0)=O(n)$ and the total number of queries is $1$, and by Theorem~\ref{thm:umv_hard}. Now we give a sketch of the proof of the above lemma.
	\begin{proof}[Proof Sketch of Lemma~\ref{lemma:correctness_LB}]
		The proof is almost the same as the proof of Lemma~\ref{lemma:separable_LB}. Here we point out the main difference. Let $G$ denote the final graph of our reduction. Let $H:=G[R\cup C\cup\{t\}]$ denote the subgraph induced by vertex set $R\cup C\cup\{t\}$. We observe that in the graph $H$, there is a triangle incident to vertex $t$ iff $\vect{u}\mat{M}\vect{v}=1$. Now we use our estimator $\lambda$ to distinguish if $H$ contains a triangle incident to $t$ or not.
		
		We let $\mat{A}\in \mathbb{R}^{(n-1)\times (n-1)}$ denote the adjacency matrix of the graph $H$. Note that all entries in $A$ are either $1$ or $0$. Let $\mat{B}=\kappa\cdot \mat{I} - \mat{A}$. Again, by the Neumann series expansion of $\mat{B}^{-1}$, we could derive the same expression of $(\kappa\cdot \mat{B}^{-1})_{tt}$ as Equation~\ref{eqn:inverse_B}, that is 
		\begin{eqnarray*}
			(\kappa\cdot \mat{B}^{-1})_{tt} = \sum_{i=0}^\infty (-\frac{1}{\kappa})^i \cdot (\mat{A}^i)_{tt}.	
		\end{eqnarray*}
		
		Now observe that since $\kappa=3(n-1)^5$, the first four terms of the above power series dominate. More precisely, by the fact that $(\mat{A}^i)_{tt}\leq (n-1)^i$ for any $i\geq 4$, we have that
		\begin{eqnarray*}
			\sum_{i=4}^\infty \abs{(-\frac{1}{\kappa})^i \cdot (\mat{A}^i)_{tt}}
			\leq 
			\sum_{i=4}^\infty \frac{1}{\kappa^i} (\mat{A}^i)_{tt} 
			\leq 
			\sum_{i=4}^\infty \frac{1}{\kappa^i} (n-1)^i
			\leq
			\frac{0.9}{\kappa^3}.
		\end{eqnarray*}
		
		Furthermore, it holds that $(\mat{A}^0)_{tt}=\mat{I}_{tt}=1$; that $\mat{A}_{tt}=0$ since $H$ is a simple graph; and that $(\mat{A}^2)_{tt}=\deg_H(t)=Y$, where the last equation follows from the definition of $Y$. Therefore, 
		\begin{itemize}
			\item If $H$ contains a triangle incident to $t$, then $(\mat{A}^3)_{tt}\geq 2$, and thus
			\begin{eqnarray*}
				(\kappa\cdot \mat{B}^{-1})_{tt} \leq 1+\frac{Y}{\kappa^2}-\frac{2}{\kappa^3} + \frac{0.9}{\kappa^3} = 1+\frac{Y}{\kappa^2}-\frac{1.1}{\kappa^3}
			\end{eqnarray*}
			\item If $H$ has no triangle incident to $t$, then $(\mat{A}^3)_{tt}=0$, and thus
			\begin{eqnarray*}
				(\kappa\cdot \mat{B}^{-1})_{tt} \geq 1+\frac{Y}{\kappa^2} - \frac{0.9}{\kappa^3}
			\end{eqnarray*}
		\end{itemize}
		
		That is, if $H$ contains a triangle incident to $t$, then $(\mat{B}^{-1})_{tt} \leq \frac{1}{\kappa}+\frac{Y}{\kappa^3}-\frac{1.1}{\kappa^4}$. If $H$ does not contain a triangle incident to $t$, then $(\mat{B}^{-1})_{tt} \geq \frac{1}{\kappa}+\frac{Y}{\kappa^3}-\frac{0.9}{\kappa^4}$.
		
		Now let $\Lambda=\mathcal{E}_G(s,t)$. Then by the same argument for proving Claim~\ref{claim:trace}, we have that $\Lambda=(\mat{B}^{-1})_{tt}.$ 
		
		Finally, by the above two claims, if $\vect{u}\mat{M}\vect{v}=1$, then $H$ contains a triangle incident to $t$, and thus $\Lambda=(\mat{B}^{-1})_{tt} \leq \frac{1}{\kappa}+\frac{Y}{\kappa^3}-\frac{1.1}{\kappa^4}$; if $\vect{u}\mat{M}\vect{v}=0$, then $H$ does not contain any triangle incident to $t$, and thus $\Lambda=(\mat{B}^{-1})_{tt} \geq \frac{1}{\kappa}+\frac{Y}{\kappa^3}-\frac{0.9}{\kappa^4}$. The statement of the lemma then follows by the fact that $\lambda$ is a $(1+\frac{1}{\kappa^4})$-approximation of $\Lambda$ and that $\frac{1}{\kappa^4}(\frac{1}{\kappa}+\frac{Y}{\kappa^3}-\frac{0.9}{\kappa^4})\leq \frac{0.1}{\kappa^4}$.
	\end{proof} \qedhere
\end{proof}

\section{Conclusion}
In this chapter, we studied the problem of dynamically maintaining All-Pairs Effective Resistances in graphs that admit small separators, e.g., planar graphs. We show a fully-dynamic algorithm that reports a $(1+\epsilon)$ approximation to any effective resistance query on a graph undergoing edge insertions and deletions with $\tilde{O}(\sqrt{n}\epsilon^{-2})$ update and query time. We also prove two conditional lower bounds, one applying to graphs with small separators and the other to general graphs, which show the hardness of the problem in the exact setting and justify our upper bounds that only support approximate queries. 

Our work leaves several interesting open problems for future work. For example, it is interesting to improve upon the update query and time of our dynamic All-Pair Effective Resistances problem in planar graphs while keeping the same approximation guarantee. While we do believe that poly-logarithmic running times should be achievable for this problem, this may require developing some new ideas and techniques that go beyond the standard $\sqrt{n}$ barrier, which also appears for the dynamic planar APSP problem~\cite{AbrahamCG12}. Another interesting direction is to extend our lower-bound for separable graphs to the more restricted family of planar graphs. Finally, the most important problem is whether there is a non-trivial fully-dynamic algorithm for maintaining All-Pairs Effective Resistances in general graphs. In Chapter~\ref{cha:STOC2019_DER} we make substantial progress on this question and present the first algorithm that achieves sub-linear update and query time.

\chapter[Fully Dynamic Spectral Vertex Sparsifiers and Applications][Dynamic Spectral Vertex Sparsifiers]{Fully Dynamic Spectral Vertex Sparsifiers and Applications}\label{cha:STOC2019_DER}

We study \emph{dynamic} algorithms for maintaining spectral vertex sparsifiers
of graphs with respect to a set of terminals $K$ of our choice.
Such objects preserve pairwise resistances, solutions to systems of linear
equations, and energy of electrical flows between the terminals in $K$.
We give a data structure that supports insertions and deletions of edges,
and terminal additions, all in sublinear time.
We then show the applicability of our result to the following problems.

(1) A data structure for dynamically maintaining the solutions to Laplacian systems $\LL \xx = \bb$, where $\LL$ is the graph Laplacian matrix and $\bb$ is a demand vector.
For a bounded degree, unweighted graph, we support modifications to
both $\LL$ and $\bb$ while providing access to $\epsilon$-approximations
to the energy of routing an electrical flow with demand $\bb$, as well
as query access to entries of a vector $\tilde{\xx}$ such that
$\vecnorm{\xxtil-\LP \bb}_{\LL}\le \epsilon \vecnorm{\LP \bb}_{\LL}$
in $\tilde{O}(n^{11/12}\epsilon^{-5})$ expected amortized update and query time. 

(2) A data structure for maintaining fully dynamic All-Pairs Effective Resistance. For an intermixed sequence of edge insertions, deletions, and resistance queries, our data structures returns $(1 \pm \epsilon)$-approximation to all the resistance queries against an oblivious adversary with high probability. Its expected amortized update and query times are  $\Otil(\min(m^{3/4},n^{5/6} \epsilon^{-2}) \epsilon^{-4})$ on an unweighted graph, and  $\Otil(n^{5/6}\epsilon^{-6})$ on weighted graphs.

The key ingredients in these results are
(1) the intepretation of Schur complement as a sum of random walks, and
(2) a suitable choice of terminals based on the behavior of these random
walks to make sure that the majority of walks are local, even when the graph itself
is highly connected and
(3) maintenance of these local walks and numerical solutions
using data structures.

These results together represent the first data structures for maintain
key primitives from the Laplacian paradigm for graph algorithms in
sublinear time without assumptions on the underlying graph topologies.
The importance of routines such as effective resistance, electrical flows,
and Laplacian solvers in the static setting make us optimistic that some of our
components can provide new building blocks for dynamic graph algorithms.

\section{Introduction}

Problems arising from analyzing and understanding graph structures have
motivated the development of many powerful tools for storing and compressing
graphs and networks.
One such tool that has received a considerable amount of attention over
the past two decades is graph sparsification~\cite{BenczurK96,BatsonSST13}.
Roughly speaking, a graph sparsifier is a ``compressed'' version of a large input graph that preserves important properties like distance information~\cite{PelegS89}, cut value~\cite{BenczurK96} or graph spectrum~\cite{SpielmanT11}.
Graph Sparsifiers fall into two main categories:
\emph{edge sparsifiers}, which are graphs that reduce the number of edges, and
\emph{vertex sparsifiers}, which are graphs that reduce the number of vertices.
Both categories have many applications in approximation algorithms~\cite{FakcharoenpholRT04,Racke08},
machine learning~\cite{LoukasV18,WagnerGKM18},
and most recently efficient graph algorithms~\cite{SpielmanT14,Madry10,Sherman13,KelnerLOS14}.
While edge sparsifiers have played an instrumental role in obtaining nearly linear time algorithms~\cite{BatsonSST13}, their practical applicability is somewhat limited due to the fact most of the large networks are already sparse.
On the other hand, vertex sparsifiers address the ``real'' compression of
large networks by reducing the number of vertices.

While vertex sparisifers in general are significantly more difficult to generate~\cite{Moitra09,charikar,mm10},
a notable exception is vertex sparsifiers for quadratic minimization problems,
otherwise known as Schur complements.
Concretely, given an undirected, weighted graph $G$, a subset of terminal vertices
$K$ and its corresponding Laplacian matrix, a graph $H$ with $V(H) = K$ is a vertex resistance sparsifier of $G$ with respect to $K$ if the Laplacian matrix of $H$ is obtained by the Schur complement of the Laplacian of $G$ with respect to $K$.
Schur complement is a central concept in physics and linear algebra with a
wide range of applications including multi-grid solvers,
Markov chains and finite-element analysis~\cite{DorflerB13},
and have also recently found extensive applications in
graph algorithms~\cite{KyngLPSS16,KyngS16,DurfeeKPRS17,DurfeePPR17:arxiv,SchildRS18,Schild18}.

Most of the massive graphs in the real world, such as
social networks, the web graph, are subject to frequent changes over time.
This dynamic behavior of graph has been studied for several important graph
problems, where the basic idea is to maintain problem solutions as graphs undergo edge insertions and deletions in time faster than recomputing the solution from scratch.
Dynamic graph algorithms have also been formulated for many problems that
involve edge sparsifiers~\cite{HolmLT01,Thorup07,KapronKM13,GoranciHT18},
as well important variants of edge sparsifiers themselves,
including minimum spanning trees~\cite{HolmRW15,NanongkaiS17,Wulff-Nilsen17,NanongkaiSW17},
spanners~\cite{BaswanaKS12},
spectral sparsifiers~\cite{AbrahamDKKP16},
and low-stretch spanning trees~\cite{GoranciK18:arxiv}.
However, despite the increasing importance of high quality vertex sparsifiers
in graph algorithms, to the best of our knowledge very little is known about
their maintenance in the dynamic setting.

In this chapter we give the first non-trival \emph{dynamic} algorithms for maintaining Schur complements of general graphs with respect to a set of terminal of our choice. Our data-structure maintains at any point of time a $(1 \pm \epsilon)$ approximation to the Schur complement while supporting insertions and deletions of edges, and arbitrary vertex additions to the terminal set. To the best of our knowledge, prior dynamic Schur complement algorithms were only known for minor-free graphs~\cite{GoranciHP17a,GoranciHP18}.

\begin{restatable}{lemma}{Dynamic}
\label{lem:Dynamic}
Given an error threshold $\epsilon>0$, an unweighted undirected multi-graph $G=(V,E)$ with $n$ vertices, $m$ edges, a subset of terminal vertices $K'$  and a parameter $\beta \in (0,1)$ such that $|K'|=O(\beta m)$, there is a data-structure \textsc{DynamicSC}$(G,K', \beta)$ for maintaining a graph $\tilde{H}$ with $\LL_{\tilde{H}} \approx_\epsilon \SC(G, K)$ for some $K$ with $K'\subseteq K$, $|K|=O(\beta m)$, while supporting $O(\beta m)$ operations in the following running times: 
\begin{itemize}
\item \textsc{Initialize}$(G, K', \beta)$: Initialize the data-structure, in $\Otil(m \beta^{-2} \epsilon^{-4})$ expected amortized time.
\label{case:initialize}
\item \textsc{Insert$(u,v)$}: Insert the edge $(u,v)$ to $G$ in $\tilde{O}(1)$ amortized time.
\label{case:Insert}
\item \textsc{Delete$(u,v)$}: Delete the existing edge $(u,v)$ from $G$ in $\tilde{O}(1)$ amortized time.
\label{case:Delete}
\item \textsc{AddTerminal$(u)$}: Add $u$ to $K'$ in $\tilde{O}(1)$ amortized time.
\label{case:AddTerminal}
\end{itemize}
\end{restatable}

Our algorithm extends to weighted graphs, albeit with slightly larger running time guarantees. Concretely we give an algorithm that maintains an approximate Schur Complement with $\tilde{O}(m\beta^{-4}\epsilon^{-4})$ expected amortized time for initializing the data-structure, and $O(1)$ amortized time for the remaining operations. We discuss such extensions in Section~\ref{sec:DynamicSCWeighted}.

The key algorithmic components behind the result in unweighted graphs are (1) the interpretation of Schur complement as a sum of random walks and (2) randomly picking a terminal vertex subset onto which the vertex resistance sparsifiers is constructed. Specifically, in a novel way we combine random walk based methods for generating resistance vertex sparsifiers~\cite{DurfeePPR17:arxiv} with results in combinatorics that bound the speed at which such walks spread among vertices~\cite{BarnesF96}. Our result in the weighted case essentially follows the same idea except that the speed at which random walks visit different vertices in weighted networks could be very slow. To control this, we instead exploit an event driven simulation of random walks that interacts well with other parts of our data structure and leads to comparable running time guarantees.

We show the applicability of our dynamic Schur complement to two cornerstone problems in graph Laplacian literature,
namely \emph{dynamic} Laplacian solver~\cite{SpielmanT14}
and \emph{dynamic} All-Pair Effective Resistances~\cite{SpielmanS11}.

Solving linear systems lies at the heart of many problems arising in scientific computing, numerical linear algebra, optimization and computer science. An important subclass of linear systems are Laplacian systems, which arise in many natural contexts, including computation of voltages and currents in electrical network. Solving Laplacian system has received increasing attention over the past years after the breakthrough work of Spielman and Teng~\cite{SpielmanT14} who gave the first near-linear time algorithm. Motivated by fast Laplacian solvers in different model of computations~\cite{AndoniKP18:arxiv,PengS14}, we initiate the study of algorithms for dynamically solving Laplacian systems. Concretely, given a graph Laplacian $\LL\in \mathbb{R}^{n\times n}$ and a vector $\bb\in\mathbb{R}^n$  in the range of $\LL$, the goal is to maintain an $\xx$ such that $\LL \xx=\bb$, while off-diagonals of $\LL$ and the entries of $\bb$ change over time. To allow for sub-linear query times, here we focus on querying one (or a few) coordinates of $\xx$. Formally, given any index $u \in \{1,\ldots,n\}$, the goal is to output $\xxtil_u$ for some approximation $\xxtil$ of $\LP \bb$. Our contribution is the first sub-linear dynamic Laplacian solver in bounded degree graphs.


\begin{theorem}
\label{thm:Solver}
For any given error threshold $m^{-1} < \epsilon < 1$,
there is a data-structure for maintaining an unweighted, undirected bounded degree $G=(V,E)$ with $n$ vertices, $m$ edges and a vector $\bb\in \mathbb{R}^n$ that supports the following operations
in $\Otil(n^{11/12} \epsilon^{-5})$ expected amortized time:
\begin{itemize}
	\item \textsc{Insert}$(u,v)$: Insert the edge $(u,v)$ with resistance $1$ in $G$.
	\item \textsc{Delete}$(u,v)$: Delete the edge $(u,v)$ from $G$.
	\item \textsc{Change}$(u,\bb'_u,v,\bb'_v)$: Change $\bb_u$ to $\bb'_u$ and $\bb_v$ to $\bb'_v$ while keeping $\bb$ in the range of $\LL$.
	\item \textsc{Solve}$(u)$: Return $\tilde{\xx}_u$ with $\tilde{\xx}$ such that $\vecnorm{\xxtil-\LL^{\dag} \bb}_{\LL}\le \epsilon\vecnorm{\LL^{\dag} \bb}_{\LL}$. 
\end{itemize}
\end{theorem}

Note that the $\tilde{\xx}$ in the theorem above is not guaranteed to be inside the range of $\LL$ and it only preserves the differences between vertices in the same connected component. 

We observe that conditioning on the vector $\bb$ having small support, i.e., a small number of non-zero elements, leads to a dynamic solver by just including the corresponding vertices into the Schur complement, and maintaining a dynamic Schur complement onto these vertices augmented with some carefully chosen additional terminals. Upon receipt of a query index, we add the corresponding vertex to the current Schur complement and simply solve a linear system there. However, note that the demand vector may have a large number of non-zero entries, thus preventing us from obtaining a sub-linear time algorithm with this approach. We alleviate this by projecting this vector onto the set of current terminals and showing that such projection can be maintained dynamically while introducing controllable error in the approximation guarantee. 



Another application of our technique is dynamic maintainance of effective resistance, a well studied quantity that has direct applications in random walks, spanning trees~\cite{MST15fast} and graph sparsification~\cite{SpielmanS11}. We maintain (approximate) All-Pair Effective Resistances of a graph $G$ among any pair of query vertices while supporting an intermixed sequence of edge insertions and deletions in $G$. Our study is also motivated in part by the wide usage of \emph{commute distances}, a random walk-based similarity measure that has been successfully employed in important practical applications such as link predictions~\cite{Liben-NowellK07}. Since commute distance is a scaled version of effective resistance, our dynamic algorithm readily extends to this graph measure while achieving the same approximation and running time guarantees.



\begin{theorem}
\label{thm:UnweightedER}
For any given error threshold $\epsilon > 0$,
there is a data-structure for maintaining an unweighted, undirected multi-graph $G=(V,E)$ with up to $m$ edges that supports the following operations
in $\tilde{O}(m^{3/4}\epsilon^{-4})$ expected amortized time:
\begin{itemize}
	\item \textsc{Insert}$(u,v)$: Insert the edge $(u,v)$ with resistance $1$ in $G$.
	\item \textsc{Delete}$(u,v)$: Delete the edge $(u,v)$ from $G$.
	\item \textsc{EffectiveResistance}$(s,t)$: Return a $(1 \pm \epsilon)$-approximation to the effective resistance between $s$ and $K$ in the current graph $G$. 
\end{itemize}
\end{theorem}

Our algorithm can also handle weighted graphs, \sloppy albeit with a bound of $\tilde{O}(m^{5/6}\epsilon^{-4})$ on the expected amortized update and query time. By running this algorithm on the output of a dynamic spectral sparsifier~\cite{AbrahamDKKP16}, we obtain a bound of $\tilde{O}(n^{5/6} \epsilon^{-6})$ per operation, which is truly sub-linear irrespective of graph density.

We are optimistic that our algorithmic ideas could be useful for dynamically
maintaining a wider range of graph properties.
Both the results that we give dynamic algorithms for,
vertex sparsifiers and Schur complements, have wide ranges of applications
in static settings, with the latter being at the core of the `Laplacian paradigm' of graph
algorithms~\cite{Spielman10,Teng10:survey}.
While it's less clear that solutions across multiple Laplacian solves can be
propagated to each other as the input dynamically changes, repeated sparsification
on the other hand represents a routine that composes and interacts well
with a much wider range of primitives.
As a result, we are optimistic that it can be used as a building block in dynamic
versions of many existing applications of Laplacian solvers.

\subsection{Related Works}
\label{subsec:related}

The recent data structures for maintaining effective resistances in planar
graphs~\cite{GoranciHP17a,GoranciHP18} drew direct connections between
Schur complements and data structures for maintaining them
in dynamic graphs.
This connection is due to the  preservation of effective resistances
under vertex eliminations (Fact~\ref{fact:SchurER}).
From this perspective, the Schur complement can be viewed as a
vertex sparsifier for preserving resistances among a set of terminal vertices.

The power of vertex or edge graph sparsifiers,
which preserve certain properties while reducing problem sizes,
has long been studied in data structures~\cite{Eppstein91,EppsteinGIN97}.
Ideas from these results are central to recent works on offline
maintenance for $3$-connectivity~\cite{PengSS17}, generating
random spanning trees~\cite{DurfeeKPRS17}, and new notions of centrality for networks~\cite{LZ18:kirchhoff}.
Our result is the first to maintain such vertex sparsifiers,
specifically Schur complements, for \emph{general} graphs in online settings.

While the ultimate goal is to dynamically maintain (approximate)
minimum cuts and maximum flows, 
effective resistances represent a natural `first candidate' for this
direction of work due to them having perfect vertex sparsifiers.
That is, for any subset of terminals, there is a sparse graph on them
that approximately preserves the effective resistances among all
pairs of terminals.
This is in contrast to distances, where it's not known whether
such a graph can be made sparse, or in contrast to cuts, where the existence of
such a dense graph is not known~(assuming that we are not content with large constant or poly-logarithmic approximations).


\paragraph*{Dynamic Graph Algorithms.}

The maintenance of graph properties in dynamic algorithms has been a major area of ongoing research in data structures. The problems being maintained include $2-$ or $3-$connectivity~\cite{EppsteinGIN97,HolmLT01,HolmRW15},
shortest paths~\cite{HenzingerKN14,HenzingerKN16,BernsteinC16,AbrahamCK17},
global minimum cut~\cite{Henzinger97,Thorup07,LackiS11,GoranciHT18},
maximum matching~\cite{OnakR10,GuptaP13,BhattacharyaHN16},
and maximal matching~\cite{BaswanaGS15,NeimanS16,Solomon16}. Perhaps most closely related to our work are dynamic algorithms that maintain properties related to paths~\cite{Frederickson85,
EppsteinGIN97,HolmLT01,KapronKM13,Wulff-Nilsen17,NanongkaiSW17,NanongkaiS17}. In particular, the work of Wulff-Nilsen~\cite{Wulff-Nilsen17} also utilizes
the behavior of random walks under edge deletions to keep track of
low-conductance cuts.


Dynamic algorithms for evaluating algebraic functions such as matrix determinant and matrix inverse has also been considered~\cite{Sankowski04}. One application of such algorithms is that they can be used to dynamically maintain single-pair effective resistance. Specifically, using the dynamic matrix inversion algorithm, one can dynamically maintain \emph{exact} $(s,t)$-effective resistance in $O(n^{1.575})$ update time and $O(n^{0.575})$ query time.


\paragraph*{Vertex Sparsifiers.}

Vertex sparsifiers have been studied in more general settings for
preserving cuts and flows among terminal
vertices~\cite{Moitra09,charikar,KrauthgamerR13}.
Efficient versions of such routines have direct applications in
data structures, even when they only work in restricted
settings: terminal sparsifiers on quasi-bipartite graphs~\cite{andoni}
were core routines in the data structure for
maintaining flows in bipartite undirected graphs~\cite{AbrahamDKKP16}.

Our data structure utilizes vertex sparsifiers, but in even more
limited settings as we get to control the set of vertices to sparsify onto.
Specifically, the local maintenance of this sparsifier under insertions
and deletions hinges upon the choice of a random subset of terminals,
while vertex sparsifiers usually need to work for any subset of terminals.
Evidence from numerical algorithms~\cite{KyngLPSS16,DurfeePPR17:arxiv} suggest
this choice can significantly simplify interactions between algorithmic
components.
We hope this flexibility can motivate further studies of vertex sparsifiers
in more restrictive, but still algorithmically useful settings.

\paragraph*{Organization. }
The chapter is organized as follows. We discuss preliminaries
in Section~\ref{sec:Preliminaries_DSC} and give an overview of the key techniques in Section~\ref{sec:Overview_DSC}. After that we give a data-structure for dynamic Schur complement on unweighted graphs in Section~\ref{sec:DynamicSchurComplement}, which can be applied to the dynamic All-Pairs Effective Resistance problem. In Section~\ref{sec:DynamicSCWeighted}, we extend our data-structure to weighted graphs. In Section~\ref{sec:DynamicSolver}, we give a data-structure for dynamic projection of a vector onto a subset of vertices of an unweighted bounded degree graph, which we combine with dynamic Schur complement to give a dynamic Laplacian solver.
In Section~\ref{sec:SchurComplement}, we provide details on the graph approximation guarantees and properties of projections that our random walk sampling and other routines rely on. Finally, in Section~\ref{sec:approx_sample}, we provide an algorithm for approximately sampling the sum of reciprocals of the edge weights of a random walk which allows us to generate long random walks without going through each step.

\section{Preliminaries}
\label{sec:Preliminaries_DSC}

In our dynamic setting, an undirected, weighted multi-graph undergoes both insertions and deletions of edges. We let $G = (V, E, \ww)$ always refer to the \emph{current} version of the graph.  
We will use $n$ and $m$  to denote bounds on the number
of vertices and edges at any point, respectively.

For an unweighted, undirected multi-graph $G$, let $\AA_G$ denote its adjacency matrix and let $\DD_G$ its degree diagonal matrix~(counting edge multiplicities for both matrices). The graph \emph{Laplacian} $\LL_{G}$ of $G$ is then defined as $\LL_G = \DD_G-\AA_G$. Let $\LL_G^{\dag}$ denote the Moore-Penrose pseudo-inverse of $\LL_{G}$. We often omit the subscript when the underlying graph is clear from the context. We also need to define the indicator vector $\boldone_{u} \in \mathbb{R}^{V}$ of a vertex $u$ such that $\boldone_u(v) = 1$ if $v = u$, and $\boldone_u(v) = 0$ otherwise. Let $\dd(u) = \sum_{v : (u,v) \in E} \ww(u,v)$ be the weighted degree of a vertex $u$. We refer the reader to Chapter~\ref{cha:ESA2018_ER} for definitions concerning electrical flows. 

A \emph{walk} in $G$ is a sequence of vertices such that
consecutive vertices are connected by edges. A \emph{random walk} in $G$ is a walk that starts at a starting vertex $w_0$, and at step $i \geq 1$, the vertex $w_i$ is chosen randomly among the neighbors of $w_{i-1}$. If graph $G$ is unweighted, then each of its neighbors becomes $w_i$ with equal probability. If $G$ is weighted, the probability $\prob{w}{w_i=u \mid w_0,\ldots,w_{i-1}}$ is proportional to the edge weight $\ww(w_{i-1},u)$. 




\paragraph*{Effective Resistance. }

For our algorithm, it will be useful to define effective resistance using linear algebraic structures. Specifically, given any two vertices $u$ and $v$ in $G$, if $\cchi(u,v) := \boldone_u - \boldone_v$, then the \emph{effective resistance} between $u$ and $v$ is given by
\[
\er^{G}\left(u, v \right)
:=
\cchi_{u, v}^\top
\LL_{G}^{\dag}
\cchi_{u, v}.
\]


Linear systems in graph Laplacian matrices can be solved in
nearly-linear time~\cite{KoutisMP11}.
One prominent application of these solvers is the approximation of
effective resistances.

\begin{lemma} \label{lemm:efficientEffectiveResistance}
Fix $\epsilon \in (0,1)$  and let $G=(V,E)$ be any graph with two arbitrary distinguished vertices $u$ and $v$. There is an algorithm that computes a value $\phi$ such that
\[
	\normalfont (1-\epsilon)\er^G(u,v) \leq \phi \leq (1+\epsilon)\er^G(u,v),
\]
in $\tilde{O}(m + n/\epsilon^{2})$ time with high probability.
\end{lemma}

%




\paragraph*{Schur complement.}

Given a graph $G=(V,E)$, we can think of the \emph{Schur complement} as the partially eliminated state of $G$. This relies on some partitioning of $V$ into two disjoint subset of vertices $K$, called \emph{terminals} and $F := V \setminus K$, called \emph{non-terminals}, which in turn partition the Laplacian $\LL$ into $4$ blocks:
\begin{equation}
\label{eq: laplacianPartition}
\LL
:=
\left[
\begin{array}{cc}
\LL_{\left[F, F\right]}
&
\LL_{\left[F, K\right]}\\
\LL_{\left[K, F\right]}
&
\LL_{\left[K, K\right]}
\end{array}
\right].
\end{equation}

The \emph{Schur complement} onto $K$, denoted by
$\SC(G, K)$ is the matrix after eliminating
the variables in $F$.
Its closed form is given by
\[
\SC\left(G, K \right)
=
\LL_{\left[K, K\right]}
-
\LL_{\left[K, F\right]}
\LL_{\left[F, F\right]}^{-1}
\LL_{\left[F, K\right]}.
\]

It is well known that $\SC(G,K)$ is a Laplacian matrix of a graph on vertices in $K$. To simplify our exposition, we let $\SC(G,K)$ denote both the Laplacian and its corresponding graph. An important property of Schur complement which we exploit in this work is to view the Schur complement as a collection of random walks. This particular feature will be discussed in more detail in Section~\ref{sec:Overview_DSC}. The key role of Schur complements in our algorithms stems from the fact that they can be viewed as vertex sparsifiers that preserve pairwise effective resistances.
\begin{fact}[Vertex Resistance Sparsifier]
\label{fact:SchurER}
For any graph $G=(V,E)$, any subset of vertices $K$,
and any pair of vertices $u, v \in K$,
\[ \normalfont
\er^{G}\left(u, v \right)
=
\er^{\SC\left(G, K \right)}
\left(u, v\right).
\]
\end{fact}

\paragraph*{Spectral Approximation} 

\begin{definition}[Spectral Sparsifier] \label{def: specSpar} Given a graph $G=(V,E,\ww)$ and $\epsilon \in (0,1)$, we say that a graph $H=(V,E',\ww')$ is a $(1 \pm \epsilon)$-\emph{spectral sparsifier} of $G$ (abbr. $H \approx_{\epsilon} G$) if $E' \subseteq E$, and for all $\vect{x} \in \mathbb{R}^{n}$ 
	\[ (1-\varepsilon)\vect{x}^\top\LL_{G}\vect{x} \leq \vect{x}^\top{\LL_{H}}\vect{x} \leq (1+\varepsilon)\vect{x}^\top\LL_{G}\vect{x}. \]
\end{definition}

In the dynamic setting, Abraham et al.~\cite{AbrahamDKKP16} recently
showed that $(1\pm\epsilon)$-spectral sparsifiers of a dynamic graph $G$
can be maintained efficiently. This algorithm will be invoked in several occasions throughout this chapter.

\begin{lemma}[\cite{AbrahamDKKP16},~Theorem 4.1]
\label{lem:DynamicSpectralSparsifier}
Given a graph $G$ with polynomially bounded edge weights, with high probability, we can dynamically maintain a $(1 \pm \epsilon)$-spectral sparsifier of size $\Otil(n \epsilon^{-2})$ of $G$ in $O(\log^{9} n \epsilon^{-2})$ expected amortized time per edge insertion or deletion. The running time guarantees hold against an oblivious adversary.
\end{lemma}

The above result is useful because matrix approximations also
preserve approximations of their quadratic forms. As a consequence of this fact, we get the following lemma.

\begin{lemma} \label{lem:ApproxER} If $H$ is a $(1 \pm \epsilon)$-spectral sparsifier of $G$, then for any pair of vertices $u$ and $v$
	\[ \normalfont
	(1-\varepsilon)\er^G(u,v) \leq \er^H(u,v) \leq (1+\varepsilon) \er^G(u,v).
	\]
\end{lemma} 

\subsection{Projection matrix and its properties}
We next define a matrix that naturally appears when performing Gaussian elimination on the non-terminal vertices. Concretely, given a graph $G=(V,E)$ and terminals $K \subseteq V$, the \emph{matrix-projection} of the non-terminals $F= V \setminus K$ onto $K$ is given by
\[
\proj{G}{K}
:=
\left[
\begin{array}{cc}
-\LL_{\left[K, F\right]}\LL_{\left[F,F\right]}^{-1}
&
\II_{K}
\end{array}
\right].
\]
We next review some useful properties about the matrix projection $\proj{G}{K}$. Consider the laplacian system $\LL \xx = \bb$, where $\LL$ is partitioned into block-matrices as in Equation~(\ref{eq: laplacianPartition}). This in turn partitions the solution vector into non-terminals and terminals, i.e., $\xx = \begin{bmatrix} \xx_F \; \xx_K \end{bmatrix}^{\top}$.

\begin{restatable}{lemma}{SolveByScAndProj}
\label{fac:solve_by_sc_and_proj}
Let $\xx_K$ be a solution vector such that $\SC(G,K)\xx_K=\proj{G}{K}\bb$. Then there exists an extension $\xx$ of $\xx_K$ such that $\LL\xx=\bb$.
\end{restatable}
\begin{proof}We assume without loss of generality that the underlying graph $G$ is connected. Consider the following extended linear system
\[
\begin{bmatrix}
\LL_{[F,F]} & \LL_{[F,K]} \\
\mathbf{0} & \SC(G,K)
\end{bmatrix} 
\begin{bmatrix}
\xx_F \\
\xx_K
\end{bmatrix} =
\begin{bmatrix}
\II_F & \mathbf{0} \\
\multicolumn{2}{c}{\proj{G}{K}}
\end{bmatrix} 
\begin{bmatrix}
\bb_F \\
\bb_K
\end{bmatrix}
\]

Using the definitions of Schur complement and projection matrix, we can rewrite the above equation as follows:

\[
\begin{bmatrix}
\LL_{[F,F]} & \LL_{[F,K]} \\
\mathbf{0} & \LL_{[K,K]} - \LL_{[K,F]} \LL^{-1}_{[F,F]} \LL_{[F,K]}
\end{bmatrix} 
\begin{bmatrix}
\xx_F \\
\xx_K
\end{bmatrix} =
\begin{bmatrix}
\II_F & \mathbf{0} \\
-\LL_{[K,F]}\LL^{-1}_{[F,F]} & I_K 
\end{bmatrix} 
\begin{bmatrix}
\bb_F \\
\bb_K
\end{bmatrix}
\]
Multiplying both sides from the left with  
\[
\begin{bmatrix}
\II_F & \mathbf{0} \\
\LL_{[K,F]}\LL^{-1}_{[F,F]} & I_K 
\end{bmatrix},
\]
we get that 
\[
\begin{bmatrix}
\LL_{[F,F]} & \LL_{[F,K]} \\
\LL_{[K,F]} & \LL_{[K,K]}
\end{bmatrix} 
\begin{bmatrix}
\xx_F \\
\xx_K
\end{bmatrix} =
\begin{bmatrix}
\bb_F \\
\bb_K
\end{bmatrix} \text{ or }
\LL \xx = \bb,
\]
what we wanted to show.

\end{proof}

The following lemma draws a connection between the projection matrix and certain probabilities which will allow us to take a combinatorial view on several cases. 

\begin{restatable}{lemma}{StopVertexDistribution}
\label{fac:StopVertexDistribution} Consider a graph $G=(V,E)$. For any subset of vertices $K \subseteq V$, a vertex $v \in K$, and a vertex $u \in F = V \setminus K$, let $\prob{u}{t_v < t_{K \setminus v}}$ be the probability that the random walk that starts at $u$ hits $v$ before hitting any other vertex from $K \setminus v$. Then we have that
\[ 
\left[ \proj{G}{K}\boldone_u \right] (v)  = \prob{u}{t_v < t_{K \setminus v}}.
\] 

In fact, $\{ \proj{G}{K}\boldone_u \}_{v \in K}$ is a probability distribution for any fixed vertex $v \in F$.
\end{restatable}

\begin{proof}
First, note that if there is no path from vertices in $K$ to $F = V \setminus K$, then the lemma holds trivially. Thus assume $K$ and $F$ are connected by paths. Next, let \[\LL_{[F,F]}=\DD_F-\AA_F,
\] where $\DD_F$ is the diagonal of $\LL_{[F,F]}$ and $\AA_F$ is the negation of the off-diagonal entries, and then expand $\LL_{[F,F]}^{-1}$ using the Jacobi series:
\begin{align*}
\LL_{[F,F]}^{-1} & =(\DD_F-\AA_F)^{-1}=\DD^{-1/2} \left(\II-\DD_F^{-1/2}\AA_F \DD_F^{-1/2}\right)^{-1}\DD_F^{-1/2}\\
& =\DD_F^{-1/2}\left(\sum_{\ell=0}^{\infty}(\DD_F^{-1/2}\AA_F \DD_F^{-1/2})^\ell \right)\DD_F^{-1/2}
=\sum_{\ell=0}^{\infty}(\DD_F^{-1}\AA_F)^\ell \DD_F^{-1}.
\end{align*}
The above series converges due to the fact that $\LL_{[F,F]}$ is strictly diagonally dominant. Concretely, the latter implies $(\AA_F\DD_F^{-1})^\ell$ tends to zero as $\ell$ tends to infinity. Substituting this in the definition of $\proj{G}{K}$ and letting $\boldone_u = \begin{bmatrix}\boldone_u^{F} & \boldone_u^{K} \end{bmatrix}^{\top}$ we get that

\begin{align*}
\proj{G}{K} \boldone_u & = \begin{bmatrix} -\sum_{\ell = 0}^{\infty}\LL_{[K,F]} (\DD_F^{-1}\AA_F)^\ell \DD_F^{-1} & \II_K \end{bmatrix} \begin{bmatrix} \boldone_u^{F} \\ \boldone_u^{K} \end{bmatrix} \\
& = \sum_{\ell = 0}^{\infty}-\LL_{[K,F]} (\DD_F^{-1}\AA_F)^\ell \DD_F^{-1} \boldone_u^{F}.
\end{align*}
In particular, it follows that for any $v \in K$
\[ \left[ \sum_{\ell = 0}^{\infty}-\LL_{[K,F]} (\DD_F^{-1}\AA_F)^\ell \DD_F^{-1} \boldone_u^{F} \right] (v) =\sum_{\substack{u_0=u,\ldots,u_{\ell-1}\in F, \\ u_\ell=v}} \frac{\prod_{i=0}^{\ell-1} \ww(u_i, u_{i+1})}{\prod_{i=1}^{\ell-1} \dd(u_i)}. \qedhere
\]
\end{proof}

Given a demand vector $\bb \in \mathbb{R}^{n}$, we say that $\proj{G}{K} \cdot \bb$ is the projection of $\bb$ onto $K$. In general, the projection of $\bb$ is shorter than the original vector $\bb$. However, for the sake of exposition, often we consider $\proj{G}{K}\cdot \bb$ to be an $n$-dimensional vector by assuming that all coordinates in $F=V \setminus K$ are $0$.
\begin{restatable}{lemma}{minenergytoS}
\label{lem:min_energy_to_S}
Consider a graph $G=(V,E)$. Let $K \subseteq V$ be a subset of vertices, and let $u \in F = V \setminus K$. Consider the demand vector $\boldone_u - \proj{G}{K} \boldone_u$ that requests to send one unit of flow from $u$ to $K$ according to the probability distribution $\{\proj{G}{K}\boldone_u \}_{v \in K}$. Then the minimum energy needed to route this demand is given by
\[
	\vecnorm{\boldone_u - \proj{G}{K} \boldone_u}_{\LL^{\dagger}}^2 = (\boldone_u - \proj{G}{K} \boldone_u)^{\top} \LL^{\dagger} (\boldone_u - \proj{G}{K} \boldone_u). \qedhere
\]
\end{restatable}
\begin{proof}
Given a valid demand vector $\bb$ with $\bb^{\top} \boldone = 0$, Lemma~2.1 due to Miller and Peng~\cite{MillerP13} shows that the minimum energy for routing $\bb$ is given by $\bb^{\top}\LL^{\dagger} \bb$. Since by construction we have that $[\boldone_u - \proj{G}{K} \boldone_u]^{\top} \boldone = 0$, substituting this demand vector in place of $\bb$ gives the lemma.
\end{proof}

\section{Overview}
\label{sec:Overview_DSC}

The core building block of our algorithm is a fast 
routine that generates and maintains an  
approximate Schur complement onto a set of terminals $K$ of 
our choice under insertion and deletions of edges as well 
as terminal additions, with all of these operations being supported 
in sub-linear time. One of the key ideas is to view to the 
Schur complement as a sum of random walks, and then observe 
that sampling exactly one walk per edge in the 
original graph already yields the desired object. Concretely,
we build upon ideas introduced in sparsifying random walk 
polynomials~\cite{cheng2015efficient},
and Schur complements~\cite{KyngLPSS16,DurfeePPR17:arxiv} to
show that it suffices to keep a union of these walks. The 
following result is implicit in these works, and we review it
in Section~\ref{sec:SchurComplement} for the sake of completeness.

\begin{restatable}{theorem}{SparsifySchur}
\label{thm:SparsifySchur}
Let $G=(V,E,w)$ be an undirected, weighted multi-graph with a subset of vertices $K$. Furthermore, let $\epsilon \in (0,1)$, and let $\rho$ be some parameter related to the concentration of sampling given by
\[
\rho = O\left( \log{n}  \epsilon^{-2} \right).
\]
Let $H$ be an initially empty graph, and for every edge $e=(u,v)$ of 
repeat $\rho$ times the following procedure:
\begin{enumerate}
\itemsep0em
\item Simulate a random walk starting from $u$ until
it \emph{first} hits $K$ at vertex $t_1$,
\item Simulate a random walk starting from $v$ until
it \emph{first} hits $K$ at vertex $t_2$,
\item Combine these two walks (including $e$) to get a walk $u = (t_1=u_0,\ldots,u_\ell=t_2)$, where $\ell$ is the length of the combined walk.
\item Add the edge $(t_1, t_2)$ to $H$ with weight
\[
	1/\left( \rho \sum_{i=0}^{\ell-1} \left(1/\ww(u_i,u_{i+1}) \right )\right)
\]
\end{enumerate}
The resulting graph $H$ satisfies $\normalfont \LL_H \approx_{\epsilon} \SC(G,T)$ with high probability.
\end{restatable}

The output approximate Schur complement of $H$ onto $K$ has up to $\tilde{O}(m \epsilon^{-2})$ edges, and thus is very dense to be leveraged as a sparsifier for our applications. Fortunately, there already exist efficient dynamic spectral sparsifiers, and we can always afford to keep a sparsifier $\tilde{H}$ of $H$ whose size is only $\tilde{O}(|K| \epsilon^{-2})$. 

The performance of our data structure depends on how fast we can generate the random walks used to create $H$. Note that even on the length $n$ path with terminals $K$ concentrated on one end, the lengths of these walks may be as long as $\Omega(n^2)$. To overcome this we shorten the walks by augmenting $K$ with roughly $O(\beta m)$ random vertices from a carefully chosen distribution. This random augmentation of $K$ ensures that any vertex $v$ in $G$ is roughly $O(\beta^{-1})$ apart from a vertex in $K$, and then our problem reduces to understanding the rate at which a random walk spreads among \emph{distinct} edges. Concretely, our goal is to efficiently generate the first $k$ distinct edges visited by a walk in $G$. We distinguish the following cases.

\begin{enumerate}
\itemsep0em
\item For unweighted graphs, we utilize a
result by Barnes and Feige~\cite{BarnesF96} that shows that
with high probability a walk reaches $k$ distinct edges
in about $k^2$ steps.
\item For weighted graphs, we employ an event driven
simulation of walks. Specifically, by computing the exit probability on the current set of
edges visited so far, we sample the first $k$ new edges reached by the walk in $\poly(k)$ time. Then, because we know the order that each edges is first
reached, the first among them that belongs to $K$ gives the intersection
of the walk with $K$. 
\end{enumerate}

Following Point (1), our dynamic Schur complement \sloppy data-structure $H$ with respect to a randomly augmented $K$ is initialized by generating for each edge $e \in E$, $\rho$ random walk of length roughly $\beta^{-2}$. This operation costs roughly $O(m\beta^{-2})$. We then make the observation that the ability to add terminals into $K$ means we only need to consider insertions/deletions between vertices in $K$. Specifically, for each affected edge we append its endpoints to $K$. 
A further advantage of this approach is that additions to $K$ only shorten random walks in $H$, and the cost of shortening or truncating these random walks in $H$ can be charged to the cost of constructing them during the initialization. Thus, it follows that we can support terminal additions, and thus insert or delete edges in $O(1)$ amortized time. Maintaining a sparsifier $\tilde{H}$ of $H$ introduces only polylogarithmic overheads, so this step does not affect much our running times. We next discuss the applicability of this result.

The data-structure we presented readily gives a \emph{sub-linear} dynamic Laplacian solver for the case where $\bb$ has small support, namely
fewer than $\beta m$ vertices of $\bb$ are non-zero. This can be accomplished by simply appending the entries of $\bb$~(more precisely, their corresponding vertices) to the Schur complement $H$, and 
solving the system on $H$ upon receipt of an index query. The resulting solution vector can then be lifted back to the original Laplacian using Lemma~\ref{fac:solve_by_sc_and_proj}. However, note that our data-structure can only support up to $O(|K|) = O(\beta m)$ operations if we want to keep the the size of $H$ small. Thus, to limit the growth in $|K|$ we periodically rebuild the entire data structure~(i.e., we resample the set of new terminals completely) after $\beta m$ operations, which in turn gives an amortized update time of $O(m \beta^{-2}/ (\beta m)) = O(\beta^{-3})$. Combining this with the bound of $O(\beta m)$ on the query time we obtain the following trade-off
\[
	\tilde{O}(\beta^{-3} + \beta m),
\]
which is minimized when $\beta = m^{-1/4}$, thus giving an update and query time of $O(m^{3/4})$.

So it remains to address the case where $\bb$ has a large number of non-zero entries.
We overcome this difficulty by projecting this vector onto the current set of terminals $K$ using the matrix $\proj{G}{K}$ and analyzing the error incurred by this projection.
Our main observation is that the standard notion of error
in Laplacian solvers, namely the $\LL$-norm, corresponds to energies
of electrical flows. This allows us to incur error in some of the $\bb(u)$ values and then bound the energy of fixing them.
To find such flows, we once again consider our problem from a random walk
perspective, namely we view the projection of $\bb$ onto $K$ being equivalent to moving $\bb$ around via random walks~(Lemma~\ref{fac:StopVertexDistribution}). As such walks are short on unweighted graphs, we can relate their energies to the length of the walks times $\bb(u)^2$~(Lemma~\ref{lem:min_energy_to_S}). 

One final obstacle is that if we move some vertex $u$ from outside of $K$ into $K$, the walks affected may be from multiple $\bb(u)$s. To address this, we bound the `load' of a vertex, defined as
the number of walks that go through it, by the total
length of the walks. The latter follows from the uniform distribution of random walks being stationary. Thus, as long as we picked $K$ so that all the entries in $V \setminus K$ have small magnitudes, each move
of some $u$ into $K$ incurs some small error. Bounding the accumulation of such errors, and rebuilding appropriately gives the overall dynamic solver result.

One application of the dynamic Laplacian solver is that we can maintain the energy of electrical flow for routing $\bb$. This can also be viewed as an extension of our dynamic effective resistances data-structure, which can only maintain the energy of electrical flows for $\bb$ with two non-zeros.
Some further extensions in this direction that we believe would be useful
are providing implicit access to the dual electrical flows, as well as
finding the $k$ largest entries either in the flow edges or the solution
vector $\xx$.
However, such extensions will likely require a better
understanding of the graph sparsifier component~\cite{AbrahamDKKP16},
which is treated as a black box in this work.

For dynamically maintaining effective resistance in unweighted graphs, we essentially follow the same approach as with the dynamic solver for small support demand vectors, and thus obtain a running time of $O(m^{3/4})$ on both update and query time. For weighted graphs, we employ the weighted dynamic Schur complement algorithm~(following Point(2)) which gives slightly weaker guarantees, namely a bound of $\tilde{O}(m^{5/6})$ on the update and query time. Interestingly, this weighted version has another immediate advantage; by running the data-structure on the output of a dynamic spectral sparsifier~(Lemma~\ref{lem:DynamicSpectralSparsifier}), we obtained a bound of $\tilde{O}(n^{5/6})$ per operation, which is truly sub-linear irrespective of graph density. 

\section{Dynamic Schur Complement}
\label{sec:DynamicSchurComplement}

In this section we show how to dynamically maintain approximate Schur complements. We first restrict our attention to unweighted graphs~(i.e., prove Lemma~\ref{lem:Dynamic}), and then show how these result extend to the weighted case. We also present two applications of our data structures, namely dynamic maintenance of effective resistance on both unweighted~(Theorem~\ref{thm:UnweightedER}) and weighted graphs~(Theorem \ref{thm:WeightedER}). 

\subsection{Dynamic Schur Complement on Unweighted Graphs}
In this section we design a data-structure for maintaining approximte Schur complements under the assumption that the dynamic graph remains unweighted throughout the updates. Specifically, we have the following lemma.

\begin{lemma}[Restatement of Lemma~\ref{lem:Dynamic}]
Given an error threshold $\epsilon>0$, an unweighted undirected multi-graph $G=(V,E)$ with $n$ vertices, $m$ edges, a subset of terminal vertices $K'$  and a parameter $\beta \in (0,1)$ such that $|K'|=O(\beta m)$, there is a data-structure \textsc{DynamicSC}$(G,K', \beta)$ for maintaining a graph $\tilde{H}$ with $\LL_{\tilde{H}}\approx_\epsilon \SC(G, K)$ for some $K$ with $K'\subseteq K$, $|K|=O(\beta m)$, while supporting $O(\beta m)$ operations in the following running times: 
\begin{itemize}
\item \textsc{Initialize}$(G, K', \beta)$: Initialize the data-structure, in $\Otil(m \beta^{-2} \epsilon^{-4})$ expected amortized time.
\item \textsc{Insert$(u,v)$}: Insert the edge $(u,v)$ to $G$ in $\tilde{O}(1)$ amortized time.
\item \textsc{Delete$(u,v)$}: Delete the existing edge $(u,v)$ from $G$ in $\tilde{O}(1)$ amortized time.
\item \textsc{AddTerminal$(u)$}: Add $u$ to $K'$ in $\tilde{O}(1)$ amortized time.
\end{itemize}
\end{lemma}


To prove the lemma above, we first review the interpretation of Schur Complements using random walks, and then discuss how to generate and maintain these walks under edge updates and addition of terminal vertices. 

Given a graph $G=(V,E)$ and a subset of terminals $K$ recall that $\SC(G,K)$ was defined using an algebraic expression that involved the Laplcian of $G$. However, since it is still unclear how to exploit this expression in the dynamic setting we instead take a different, more `combinatorial', view on $\SC(G,K)$. Concretely, we will interpret $\SC(G,K)$ as a collection of random walks, each starting at an edge of $G$ and terminating in $K$, as described in Theorem~\ref{thm:SparsifySchur}.

Let $H$ be the output graph from the construction in Theorem~\ref{thm:SparsifySchur}. Recall that $H$ is an approximate Schur Complement onto $K$ that has up to
$\rho m = \tilde{O}(m\epsilon^{-2})$ edges~(that is, $\rho$ for each edge in $G$, where $\rho = O(\log n \epsilon^{-2})$ is the sampling parameter). As we will next show, $H$ does not change too much~(in amortized sense) upon inserting
or deleting an edge in $G$. We will be able to maintain $H$ such that the distribution of $H$ is the same as $H(G)$ of the current graph $G$.
Therefore, we can maintain these changes using a dynamic spectral 
sparsifier $\Htil$ of $H$~(Lemma~\ref{lem:DynamicSpectralSparsifier}), and whenever a query comes, we answer 
it on $\Htil$ in $\Otil(\abs{K} \epsilon^{-2}) = \Otil(\beta m \epsilon^{-2})$ time.

While it is widely known how to generate random walks efficiently, we note that the length of the walks generated in Theorem~\ref{thm:SparsifySchur} could be prohibitively large if $K$ is being picked arbitrarily. To see this, recall our example where we considered a path of length $n$ with terminals $K$ being places in one end. The length of such random walks may be as long as $\Omega(n^{2})$. To shorten these random walks, we augment $K'$ with a random subset of vertices, which results in a larger set $K$.
Coming back to the path example,
$\beta n$ uniformly random vertices will be roughly
$\beta^{-1}$ apart, and random walks will reach one of these $\beta n$ vertices in about $\beta^{-2}$ steps.
Because $G$ could be a multi-graph, and we want to support queries
involving any vertex, we pick $K$ as the end points of a uniform
subset of edges.
A case that illustrates the necessity of this choice is 
a path except one edge has $n$ parallel edges.
In this case it takes $\Theta(n)$ steps in expectation for
a random walk to move away from the end points of that edge.
This choice of $K$ completes the definition of our data structure,
which we summarize in Algorithm~\ref{alg:Initialize_Unweighted}, and will discuss throughout the rest of this section.

\begin{algorithm2e}[t]
\caption{$\textsc{InitializeUnweighted}(G, K', \beta)$}
\label{alg:Initialize_Unweighted}
\Input{Unweighted graph $G$, set of vertices $K'\subseteq V$ such that $|K'| \le O(m\beta)$, and $\beta\in (0, 1)$}
\Output{Approximate Schur Complement $H$ and union of $\beta$-shorted walks $W$} 
Set $K\leftarrow K'$, $H\leftarrow (V, \emptyset)$ and $W\leftarrow \emptyset$ \;
For each edge $e =(u,v)$ in $G$, let $K \leftarrow K \cup \{u,v\}$ with probability $\beta$ \; 
Let $\rho \gets O(\log n\epsilon^{-2})$ be the sampling overhead according to Theorem~\ref{thm:SparsifySchur} \;
\For{each edge $e=(u,v) \in E$ and each $i=1,\ldots,\rho$}
{ 
Generate a random walk $w_1(e,i)$ from $u$ until $\Theta(\beta^{-1}\log n)$ different edges have been hit, it reaches $K$, or it has hit every edge in its component \;
Generate a random walk $w_2(e,i)$ from $v$ until $\Theta(\beta^{-1}\log n)$ different edges have been hit, it reaches $K$, or it has hit every edge in its component \;
\If{both walks reach $K$ at $t_1$ and $t_2$ respectively}
{
Connect $w_1(e,i)$, $e$ and $w_2(e,i)$ to form a walk $w(e,i)$ between $t_1$ and $t_2$ \;
Let $\ell \leftarrow \ell(w_1(e,i))+\ell(w_2(e,i))+1$ be the length of $w(e,i)$ \;
Add an edge $(t_1, t_2)$ with weight $1/(\rho \ell)$ to $H$ \;
Add $w(e,i)$ to $W$ \;
}
}
\Return $H$ and $W$
\end{algorithm2e}

The performance of our data structures hinge upon the
properties of the random walks generated.
We start by formalizing such a structure involving the set of augmented
terminals described above while parameterizing it with a more general probability $\beta$ for including the endpoints of the edges.
\begin{definition}[$\beta$-shorted walks] \label{def:Walk}
Let $G$ be an weighted, undirected multi-graph and $\beta \in (0,1)$ a parameter.
A collection of $\beta$-\emph{shorted walks} $W$ on $G$ is a set of random
walks created as follows:
\begin{enumerate}
\itemsep0em
\item Choose a subset of terminal vertices $K$, obtained by including
the endpoints of each edge independently with probability at $\beta$.
\item For each edge $e \in E$, generate $\rho$ walks from its endpoints
either until $\Omega(\beta^{-1} \log{n})$ different edges have been hit, or they reach $K$, or they visited each edge that is in the same connected component as $e$.
\end{enumerate}
\end{definition}

As we will shortly seee, the main property of the collection $W$ is that its random walks are short. Moreover, we will also prove that all walks in $W$ will reach $K$ with high probability.
These guarantees are summarized in the following theorem.

\begin{restatable}{theorem}{RandomWalkProperties}
\label{thm:RandomWalkProperties}
Let $G=(V,E)$ be any undirected multi-graph,
and $\beta \in (0,1)$ a parameter. 
Any set of $\beta$-shorted walks $W$,
as described in Definition~\ref{def:Walk},
satisfies the following:
\begin{itemize}
\item With high probability, any random walk in $W$ starting in a connected
component containing a vertex from $K$ terminates at a vertex in $K$.
\end{itemize}
\end{restatable}

Note that Theorem \ref{thm:RandomWalkProperties} is conditioned upon the connected component having a vertex in $K$: this is necessary because walks stay inside a connected component.
However, this does not affect our queries:
our data-structure has an operation for making any vertex $u$ a
terminal, which we call during each query to ensure both $s$ and $K$
are terminal vertices.
Such an operation interacts well with Theorem~\ref{thm:RandomWalkProperties}
because it can only increase the probability of an edge's endpoints
being chosen.

Proving the theorem requires to determine the rate at which a random walk visits at least $\beta^{-1} \log n$ edges.  It turns out that a random walk of length $\tilde{O}(\beta^{-2})$ is highly likely to achieve this. For formally showing this, we need the following result by Barnes and Feige~\cite{BarnesF96}.


\begin{theorem}[\cite{BarnesF96}, Theorem 1.2]
\label{thm:ExpectedTime}
There is an absolute constant $c_{BF}$ such that for 
any undirected, unweighted, multi-graph $G$
with $n$ vertices and $m$ edges,
any vertex $u$ and any value $\mhat \leq m$,
the expected time for a random walk starting from $u$ to visit
at least $\mhat$ \emph{distinct} edges is at most $c_{BF} \mhat^2$.
\end{theorem}

The above theorem can be amplified into a with high probability
bound by repeating the walk $O(\log{n})$ times.

\begin{corollary} \label{cor:NumDistinctEdges}
In any undirected unweighted multi-graph $G$ with $m$ edges,
for any starting vertex $u$, any length $\ell$,
and a parameter $\delta \geq 1$,
a walk of length $c_{BF} \cdot \delta \cdot \ell \log n$ from $u$ visits
at least $\ell^{1/2}$ \emph{distinct} edges with probability at least $1 - n^{-\delta}$.
\end{corollary}

\begin{proof}
We can view each such walk as a concatenation of $\delta \log n$
sub-walks, each of length $c_{BF} \cdot \ell$.

We call a sub-walk \emph{good} if the number of distinct edges that
it visits is at least $\ell^{1/2}$.
Applying Markov's inequality to the result of Theorem~\ref{thm:ExpectedTime},
a walk takes more than $O(\ell)$ steps to visit $\ell^{1/2}$ distinct edges
with probability at most $1/2$.

This means that each subwalk fails to be good with probability at most $1/2$.
Thus, the probability that all subwalks fail to be good is at most
$2^{-\delta \log n} = n^{-\delta}$. The result then follows from an union bound over all starting vertices $u \in V$.
\end{proof}


We now have all the tools to prove Theorem~\ref{thm:RandomWalkProperties}.

\begin{proof}[Proof of Theorem~\ref{thm:RandomWalkProperties}]
For any walk $w$, we define $V(w)$~(respectively, $E(w)$) to be the set of distinct vertices~(respectively, edges) that a walk $w$ visits. Consider a random walk $w$ that starts at $u$ of length
\[
\ell = c_{BF} \cdot \delta^3 \cdot  \beta^{-2} \log^{3} n
\]
where $\delta$ is a constant related to the success probability.

If the connected component containing the walk has fewer than
\[
\delta \cdot \beta^{-1} \cdot \log{n}
\]
edges, then Corollary~\ref{cor:NumDistinctEdges}
gives that we have covered this entire component with high probability,
and the guarantee follows from the assumption that this component contains a vertex of $K$.

Otherwise, we will show that $w$ reached enough edges for one of their endpoints to be picked to be picked into $K$ with high probability.
The key observation is that because $w$ is generated independently from $K$, we can bound the probability of this walk not hitting $K$ by first generating $w$, and then $K$. Specifically, for any size threshold $z$, we have
\begin{align} \label{eqn: randomWalkProb}
 \prob{K, w}{V\left( w \right) \cap K  = \emptyset} & =
\prob{w, K}{V\left( w \right) \cap K = \emptyset}   \\
& \leq
\prob{w}{\left| E\left( w \right) \right| \leq z} + \prob{w: \left| E\left( w \right) \right| \geq z, K}
{V\left( w \right) \cap K = \emptyset} \nonumber.
\end{align}

By Corollary~\ref{cor:NumDistinctEdges} and the choice of $\ell$, if we set
\[
z = \delta \cdot \beta^{-1} \cdot \log{n},
\]
then the first term in Equation~(\ref{eqn: randomWalkProb})
is bounded by $n^{-\delta}$.
For bounding the second term, we can now focus on a particular
walk $\widehat{w}$
that visits at least $\delta \cdot \beta^{-1} \cdot \log{n}$
distinct edges, i.e.,
\[
\left|E\left(\widehat{w}\right)\right|
\ge
  \delta \cdot \beta^{-1} \log{n}.
\]

Recall that we independently added the end points of each
of these edges into $K$ with probability $\beta$.
If any of them is selected, we have a vertex that is both
in $V(\widehat{w})$ and $K$.
Thus the probability that $K$ contains
no vertices from $V(\widehat{w})$ is at most
\[
\left(
  1 - \beta
\right)^{|E(\widehat{w})|}
\leq
\left(
  1 - \beta
\right)^{\delta \cdot \beta^{-1} \log{n}}
\leq
e^{- \delta \log n} 
\leq
n^{-\delta},
\]
which completes the proof.
\end{proof}

Corollary~\ref{cor:NumDistinctEdges} together with Theorem \ref{thm:RandomWalkProperties} yield the following lemma. 
\begin{lemma} \label{lem: preprocessingTime}
Algorithm \ref{alg:Initialize_Unweighted} runs in $\Otil(m\beta^{-2}\epsilon^{-2})$ time and outputs a graph $H$ with $\LL_H\approx_\epsilon \SC(G,K)$, with high probability.
\end{lemma}
\begin{proof}
By Corollary \ref{cor:NumDistinctEdges}, the length of each walk generated in Algorithm~\ref{alg:Initialize_Unweighted} is bounded by $O(\beta^{-2}\log^3n)$. In addition, note that each step in a random walk can be simulated in $O(1)$ time. This is due to the fact that we can sample an integer in $[0,n-1]$ by drawing $x\in[0,1]$ uniformly and taking $\lfloor xn \rfloor$. Combining these with the fact that the algorithm generates $\rho m = \tilde{O}(m \epsilon^{-2})$ walks, it follows that the running time of the algorithm is dominated by $\Otil(m\beta^{-2}\epsilon^{-2})$. 

Note that the collection of generated walks form the set $W$ of $\beta$-shorted walks. By Theorem~\ref{thm:RandomWalkProperties}, with high probability, each of the walks that starts at a component containing a vertex in $K$ hits $K$. Conditioning on the latter, Theorem~\ref{thm:SparsifySchur} gives that with high probability, $\LL_H\approx_\epsilon \SC(G,K)$.
\end{proof}

\paragraph*{Handling edge updates and terminal additions.} We start by observing that there is always a one-to-one correspondence between the collection of $\beta$-shorted walks $W$ and our approximate Schur complement $H$.
Accordingly, our primary concern will be supporting the $\textsc{Insert}$, $\textsc{Delete}$, and $\textsc{AddTerminal}$ operations in the collection $W$.
However, as $W$ undergoes changes, we need to efficiently update the sparsifier $H$. To handle these updates, we would ideally have efficient access to which walks in $W$ are affected by the corresponding updates.

To achieve this, we index into walks that utilize a vertex
or an edge, and thus set up a reverse data structure pointing
from vertices and edges to the walks that contain them.
The following lemma says that we can modify this representation with minimal cost.
\begin{lemma}
\label{lem:ReversePointers}
For the collection of $\beta$-shorted walks $W$, let $W_v$ and $W_e$ be the specific walks of $W$ that contain vertex $v$ and edge $e$, respectively. 
We can maintain a data structure for $W$ such that for any vertex $v$ or edge $e$ it reports either
\begin{itemize}
\item All walks in $W_v$ or $W_e$ in $O(|W_v|)$ or $O(|W_e|)$ time, respectively, or
\end{itemize}
with an additional $O(\log{n})$ overhead for any changes made to $W$.
\end{lemma}

\begin{proof}
For every vertex~(respectively, edge), we can maintain a balanced binary search tree
consisting of all the walks that use it in time proportional
to the number of vertices~(respectively, edges) in the walks.
Supporting rank and select operations on such trees then gives the claimed bound.
\end{proof}

As a result, any update made to the collection of walks can be updated in the approximate Schur complement $H$ generated from these walks in $O(\log n )$ time. We now have all the necessary ingredients to prove Lemma~\ref{lem:Dynamic}.

\begin{algorithm2e}[h]
\caption{$\textsc{AddTerminal}(u)$}
\label{alg:add_Terminal}
\Input{Vertex $u$ such that $u \not \in K$}
Set $K \gets K \cup \{u\}$ \;
Shorten all random walks in $W$ to the first location they meet $u$ \;
Update the corresponding edges in $H$ and $\tilde{H}$ \;
\end{algorithm2e}


\begin{proof}[Proof of Lemma~\ref{lem:Dynamic}]

We give a two-level data-structure for dynamically maintaining Schur complements. Specifically, we keep the terminal set $K$ of size $\Theta(m\beta)$. This entails maintaining
\begin{enumerate}
\item an approximate Schur complement $H$ of $G$ with respect to $K$~(Theorem~\ref{thm:SparsifySchur}),
\item a dynamic spectral sparsifier $\tilde{H}$ of $H$~(Lemma~\ref{lem:DynamicSpectralSparsifier}).
\end{enumerate}
We implement the procedure $\textsc{Initialize}$ by running Algorithm~\ref{alg:Initialize_Unweighted}, which produces a graph $H$ and then computing a spectral sparsifier $\tilde{H}$ of $H$ using Lemma~\ref{lem:DynamicSpectralSparsifier}. Note that by construction of our data-structure, every update in $H$ will be handled by the black-box dynamic sparsifier $\tilde{H}$.

As we will shortly see, operations $\textsc{Insert}$ and $\textsc{Delete}$ will be reduced to adding terminals to the set $K$. Thus, the bulk of our effort is devoted to implementing the procedure $\textsc{AddTerminal}$. Let $u$ be a non-terminal vertex that we want to append to $K$. We set $K \gets K \cup \{u\}$, and then shorten all the walks at the first location they meet $u$. This shortening of walks induces in turn edge insertions and deletions to $H$, which are then processed by $\tilde{H}$. The pseudocode for this operation is summarized in Algorithm~\ref{alg:add_Terminal}. To quickly locate the first appearances of $u$ in the random walks from $W$, we make use of the data-structure from Lemma~\ref{lem:ReversePointers}. Let us first describe the construction of such data-structure during the preprocessing phase. Let $W_u$ be the balanced binary search tree consisting of all the walks that use the vertex $u$ in $W$. Fix $w \in W_u$. For any $t \geq 0$, if $w$ visits $u$ after $K$ steps, we check whether $W_u$ contains $w$ or not. If the latter holds, we know that $u$ has appeared before in $w$ and we do not need to add $w$ to $W_u$. Otherwise, we add $w$ to $W_u$ as this is the first time the walk $w$ visits $u$. After locating the first appearances of $u$, we cut the walks in these locations, delete the corresponding affected walks (together with their weight from $H$), and insert the new shorter walks to $H$. Note that we can simply use arrays to represent each random walk in $W$.


We next discuss the implementation of operations \textsc{Insert} and \textsc{Delete}. Specifically, upon insertion or deletion of an edge $e = (u,v)$ in $G$, we append both $u$ and $v$ to the terminal set $K$. Now, all the walks that pass through $u$ or $v$ in $W$ must be shorten at the first location they meet $u$ or $v$. For inserting an edge $(u,v)$ with weight $\ww(u,v)$ in $G$, we simply add $\rho$ trivial random walks~(i.e., the edge $(u,v)$) of weight $\frac{\ww(u,v)}{\rho}$ to $H$ (which sum up to the edge $(u,v)$ itself). For deleting the edge $(u,v)$ with weight $\ww(u,v)$ from $G$, simply delete these $\rho$ random walks between $u$ and $v$ in $H$ (which exist since we guaranteed that $u$ and $v$ are added as terminals to $H$). 




We next analyze the performance of our data-structure. Let us start with the pre-processing time. First, by Lemma~\ref{lem: preprocessingTime} we get that the cost for constructing $H$ on a graph with $m$ edges is bounded by $\Otil(m \beta^{-2} \epsilon^{-2})$. Next, since $H$ has $\Otil(m \epsilon^{-2})$ edges, constructing $\tilde{H}$ takes $\Otil(m \epsilon^{-4})$ time. Thus, the amortized time of \textsc{Initialize} operation is bounded by $\Otil(m\beta^{-2} \epsilon^{-4})$. 

We now analyze the update operations. By the above discussion, note that it suffices to bound the time for adding a vertex to $K$, which in turn (asymptotically) bounds the update time for edge insertions and deletions. The main observation we make is that adding a vertex to $K$ only shortens the existing walks, and Lemma~\ref{lem:ReversePointers} allows us to find such walks in time proportional to the amount of edges deleted from the walk. Since the walk needed to be generated in the \textsc{Initialize} operation, the deletion of these edges take equivalent time to generating them. Moreover, we note that (1) handling the updates in $\tilde{H}$ induced by $H$ introduces additional $O(\poly(\log n)\epsilon^{-2})$ overheads, and (2) adding or deleting $\rho$ edges until the next rebuild costs $\tilde{O}(\beta m \epsilon^{-2})$, since we process only up to $\beta m$ operations. These together imply that the amortized cost for adding a terminal can be charged against the pre-processing time, which is bounded by $\Otil(m\beta^{-2} \epsilon^{-4})$, up to poly-logarithmic factors. Thus it follows that the operations \textsc{AddTerminal}, \textsc{Insert} and \textsc{Delete} can be implemented in $\tilde{O}(1)$ amortized update time. 
\end{proof}

\subsection{Dynamic All-Pair Effective Resistance on Unweighted Graphs}
In this section we present the first application of our dynamic Schur complement data structure for unweighted graphs. Concretely, we design a dynamic algorithm that supports an intermixed sequence of edge insertions, deletions and pair-wise resistance queries, and returns a $(1 \pm \epsilon)$-approximation to all the resistance queries. 

We start by reviewing two natural attempts for solving this problem.
\begin{itemize}[noitemsep]
\item First, since spectral sparsifiers preserve effective resistances (Lemma~\ref{lem:ApproxER}), we could dynamically maintain a spectral sparsifier~(Lemma~\ref{lem:DynamicSpectralSparsifier}), and then compute the $(s,t)$ effective resistance on this sparsifier. This leads to a data structure with $\poly(\log n, \epsilon^{-1})$ update time
and $\Otil(n \epsilon^{-2})$ query time.
\item Second, by the preservation of effective resistances under
Schur complements (Fact~\ref{fact:SchurER}), we could also utilize Schur complements to obtain a faster query time among a set of $\beta m$
terminals, $K$, for some reduction factor $\beta \in (0,1)$,
at the expense of a slower update time.
Specifically, after each edge update, we recompute an approximate Schur complement of the sparsifier onto $K$ in
time~$\Otil(m \epsilon^{-2})$~\cite{DurfeeKPRS17},
after which each query takes $\Otil(\beta m \epsilon^{-2})$ time.
\end{itemize}

The first approach obtains sublinear update time, while the second
approach gives sublinear query time. Our algorithm stems from combining these two methods,
with the key additional observation being that adding more vertices
to $K$ makes the Schur complement algorithm more local. Specifically, using Lemma~\ref{lem:Dynamic} leads to a data-structure for dynamically maintaining all-pair effective resistances.

\begin{proof}[Proof of Theorem~\ref{thm:UnweightedER}]

Let $\mathcal{D}(\tilde{H})$ denote the data structure that maintains a dynamic (sparse) Schur complement $\tilde{H}$ of $G$~(Lemma~\ref{lem:Dynamic}). Since $\mathcal{D}(\tilde{H})$ supports only up to $\beta m$ operations, we rebuild $\mathcal{D}(\tilde{H})$ on the current graph $G$ after such many operations. Note that the operations \textsc{Insert} and \textsc{Delete} on $G$ are simply passed to $\mathcal{D}(\tilde{H})$. For processing the query operation $\textsc{EffectiveResistance}(s,t)$, we declare $s$ and $t$ terminals (using the operation \textsc{AddTerminal} of $\mathcal{D}(\tilde{H})$), which ensures that they are both now contained in $\tilde{H}$. Finally, we compute the (approximate) effective resistance between $s$ and $t$ in $\tilde{H}$ using Lemma~\ref{lemm:efficientEffectiveResistance}.

We now analyze the performance of our data-structure. Recall that the insertion or deletion of an edge in $G$ can be supported in $\tilde{O}(1)$ expected amortized time by $\mathcal{D}(\tilde{H})$. Since our data-structure is rebuilt every $\beta m$ operations, and rebuilding $\mathcal{D}(\tilde{H})$ can be implemented in $\tilde{O}(m\beta^{-2} \epsilon^{-4})$,  it follows that the amortized cost per edge insertion or deletion is 
\[
	\frac{\tilde{O}(m\beta^{-2} \epsilon^{-4})}{\beta m} = \tilde{O}(\beta^{-3} \epsilon^{-4}).
\]

The cost of any $(s,t)$ query is dominated by (1) the cost of declaring $s$ and $t$ terminals and (2) the cost of computing the $(s,t)$ effective resistance to $\epsilon$ accuracy on the graph $\tilde{H}$. Since (1) can be performed in $\tilde{O}(1)$ time, we only need to analyze (2). We do so by first giving a bound on the size of $K$. To this end, note that each of the $m$ edges in the current graph adds two vertices to $K$ with probability $\beta$ independently. By a Chernoff bound, the number of random augmentations added to $K$ is at most $2\beta m$ with high probability.
In addition, since $\mathcal{D}(\tilde{H})$ is rebuilt every $\beta m$ operations, the size of $K$ never exceeds $4\beta m$
with high probability. The latter also bounds the size of $\Htil$ by $\Otil(\beta m\epsilon^{-2})$
and gives that the query cost is $\tilde{O}(\beta m \epsilon^{-4})$.

Combining the above bounds on the update and query time, we obtain the following trade-off \[ \tilde{O}\left((\beta m + \beta^{-3})\epsilon^{-4}\right),\]
which is minimized when $\beta = m^{-1/4}$, thus giving an expected amortized update and query time of \[ \tilde{O}(m^{3/4}\epsilon^{-4}). \qedhere \]
\end{proof}

\subsection{Dynamic Schur Complement on Weighted Graphs}
\label{sec:DynamicSCWeighted}


In this section we present an extension of Lemma~\ref{lem:Dynamic} to weighted graphs while slightly increasing the running time guarantees. Concretely, we prove the following lemma. 

\begin{restatable}{lemma}{WeightedDynamic}
\label{lem:WeightedDynamic}
Given an error threshold $\epsilon>0$, a weighted, undirected multi-graph $G=(V,E,\ww)$ with $n$ vertices, $m$ edges, a subset of terminal vertices $K'$ and a parameter $\beta \in (0,1)$ such that $|K'|=O(\beta m)$, there is a data-structure \textsc{WeightedDynamicSC}$(G,K', \beta)$ for maintaining a graph $\tilde{H}$ with $\LL_{\tilde{H}} \approx_\epsilon \SC(G, K)$ for some $K$ with $K'\subseteq K$, $|K|=O(\beta m)$, while supporting $O(\beta m)$ operations in the following running times:
\begin{enumerate}
\itemsep0em
\item \textsc{Initialize}$(G, K', \beta)$: Initialize the data-structure in $\Otil(m \beta^{-4}\epsilon^{-4})$ expected amortized time.
\label{case:w_initialize}
\item \textsc{Insert$(u,v,w)$}: Insert the edge $(u,v)$ with weight $w$ to $G$ in $\tilde{O}(1)$ amortized time. 
\label{case:w_Insert}
\item \textsc{Delete$(u,v)$}: Delete the existing edge $(u,v)$ from $G$ in $\tilde{O}(1)$ amortized time.
\label{case:w_Delete}
\item \textsc{AddTerminal$(u)$}: Add $u$ to $K'$ in $\tilde{O}(1)$ amortized time.
\label{case:w_AddTerminal}
\end{enumerate}
\end{restatable}

While the extension of our data-structure to weighted graphs builds upon the ideas we used in the unweighted case, there are a few obstacles that force us to introduce new components in our algorithm in order to make such an extension feasible. To illustrate, consider path of constant length with edge weights alternating between $1$ and $n^{10}$. Recall that the running time our data-structure depends on the speed at which random walks visit distinct edges in a graph. Due to the structure of the edge weights, a random walk in this graph is expected to take $\Theta(n^{10})$ steps before hitting a constant number of different edges. This shows that the naive generation of random walks in weighted graphs may be computationally prohibitive for our purposes. 

To rectify the above issue, we make the important observation that it is not necessary to keep information for every single step of a random walk. Instead, it would suffice if we could efficiently determine the step at which the walk meets a new vertex along with the corresponding weight associated with the walk, which defines the edge weight that is added to the sparsifier. This high-level idea allows us to generate random walks much faster, and we next make this more precise.

Following the notation we used in the unweighted case, for an arbitrary vertex $v \in V$, a set of terminals $K \subseteq V$ and a parameter $\beta \in (0,1)$, a \emph{$\beta$-shorted walk} with respect to $v$ and $K$ is a random walk that starts at a given vertex $v \in V$ and halts whenever $\Omega(\beta^{-1} \log n)$ \emph{different} vertices have been hit, it reaches a vertex in $K$, or it has hit every edge in the connected component containing $v$. The main contribution of this section is summarized in the following lemma.

\begin{lemma}
\label{lem: generateSingleWalk}
Let $G=(V,E,\ww)$ be an undirected, weighted graph with polynomially bounded weights. Let $K \subseteq V$ be a set of terminals and $v \in V$ be an arbitrary vertex. Then there is an algorithm that generates a $\beta$-shorted random walk with respect to $v$ and $K$ and approximates its corresponding weight up to a $(1 + \epsilon)$ multiplicative error in $\tilde{O}(\beta^{-4} \epsilon^{-2})$ time. 
\end{lemma}

We first give an intuition behind the algorithm in the above lemma and briefly describe how this algorithm interacts with other parts of our dynamic data-structure. Let $w=(w_0,\ldots,w_t)$ be a random walk that starts at an endpoint of an edge, and define 
\begin{equation} \label{eq: weightWalk}
s(w) := \sum_{i=1}^t \frac{1}{\ww(w_{i-1},w_i)}, 
\end{equation}
to be its corresponding weight. Recall that before adding the walk $w$ to $H$, we must scale it proportionally to $1/s(w)$~(Theorem \ref{thm:SparsifySchur}). Observe that throughout our dynamic algorithm, the only modification we might do to $w$ is to truncate it at the first location it meets a new vertex $u$ that is being declared a terminal. Moreover, after this modification, note that the old value of $s(w)$ is no longer valid and we need to extract $s(w)$ that corresponds to the new walk. To allow efficient access to such information, we can view the walk $w$ as being split into sub-walk segments by the first locations $w$ meets new vertices and store the weights of each such sub-walks. As we will next see, this bookkeeping alone allows us to proceed with the same algorithm as in the unweighted case.


We next give the three main components for implementing the algorithm stated in Lemma~\ref{lem: generateSingleWalk}. 
\begin{enumerate}[(A)]
\itemsep0em 
\item Sample the number of steps needed for a random walk $w$ to visit a new vertex.\label{comp: 1}
\item Sample a new \emph{distinct} vertex that $w$ hits, and its corresponding edge. \label{comp: 2}
\item Sample the (approximate) weight of a random walk between two given vertices. \label{comp: 3}
\end{enumerate}
After describing each of them, we will see that their combination naturally leads to our desired result. 

Let us first discuss \ref{comp: 1}. For any $t \geq 0$, consider a $t$-step random walk $w$ and let $U=\{w_0,\ldots,w_t\}$ be the set of \emph{distinct} vertices that $w$ has visited up to step $t$. Define $u := w_t \in U$ to be the \emph{current} vertex of the walk $w$. Our goal is to efficiently sample the number of steps the walk $w$ needs to visit a vertex not in $U$. 
To this end, we start by introducing some useful notation. For any $i \geq 0$, let $\pnew(i)$ be the probability that $w$ meets a new vertex that is not in $U$ in $w_{t+1},\ldots,w_{t+i}$. For $v \in U$, let $\pp_i(v)$ be the probability that $w_{t+i}=v$, conditioned on $w$ not having met any new vertex in $w_{t+1},\ldots,w_{t+i-1}$. Then it can be easily verified that both $\pnew(i)$ and $\pp_{i}(v)$ are just linear combinations of $\pnew(i)$ and $\pp_{i-1}(v)$
\begin{align}
\label{eq: p} \pp_i(v) & =\sum_{u \in U} \left( \pp_{i-1}(u) \cdot \frac{\ww(u,v)}{\dd(u)} \right), \quad \forall v \in U.\\
\label{eq: pnew} \pnew(i) & =\pnew(i-1)+\sum_{u \in V \setminus U} \sum_{v \in U} \left( \pp_{i-1}(v)\cdot \frac{\ww(v,u)}{\dd(v)} \right). 
\end{align}

Next, using the linearity of the recurrences in~(\ref{eq: pnew}) and~(\ref{eq: p}) we can find a matrix $\WW$ of dimension $(k+1) \times (k+1)$, where $k = |U|$, satisfying the following equality

\begin{equation} \label{eq: poweringW}
\left[ \begin{array}{l} \pp_{i} \\ \pnew(i) \end{array} \right] = \WW \cdot \left[ \begin{array}{l} \pp_{i-1} \\ \pnew(i-1) \end{array} \right], \quad \forall i \geq 1.
\end{equation}

The main advantage introducing such a matrix is that it allows us to efficiently compute $\pnew(i)$ and $\pp_i$ using fast exponentiation via repeated squaring. Specifically, let $\pp_0$ be a unit vector of dimension $k$, where for the current vertex $u$ of the walk $w$ we have that $\pp_0(u) = 1$, and $0$ otherwise. Let $\hat{\pp}_0 = \left[ \pp_0 \quad \pnew(0)\right]^{\top}$ be the extended $k+1$ dimension vector, where $\pnew(0) = 0$. For any $i \geq 1$, repeatedly applying Equation~(\ref{eq: poweringW}) and letting $\hat{\pp}_i := \WW^{i} \hat{\pp}_0$ yields 

\begin{equation} \label{eq: powerEquiv} 
\hat{\pp}_i(v) = \pp_i(v),~\forall v \in U \quad \text{and} \quad \hat{\pp}_i(k+1) = \pnew(i). 
\end{equation}

Using the above relation, we can use fast exponentiation via repeated squaring to compute $\pnew(i)$ in $O(k^3 \log (i))$ time. This follows directly from the following well-known lemma, which we will exploit in a few other places throughout this work. 

\begin{lemma} \label{lem: binaryExponentation}
Let $\BB$ be a matrix of dimension $n \times n$, and $\BB^i$ denote the $i$-th power of $\BB$, for any $i \geq 1$. Then there is an algorithm that computes $\BB^{i}$ in $O(n^{3} \log (i))$ time. 
\end{lemma}

We now have all the tools to describe the sampling procedure for computing the number of steps that the walk needs to visit a vertex that is distinct from the vertices in $U$. We accomplish this using a ``binary search''-inspired subroutine, which works as follows. As an input, our algorithm is given a $(k+1) \times (k+1)$ matrix $\WW$~(as defined in Equation~(\ref{eq: poweringW})), the vector $\hat{\pp}_0$, and an integer $M$, which is an upper-bound on the cover time of $G$. The algorithm also maintains variables $\ell, r, \ell p, rp$ with the following initialization $\ell \gets 0$, $r \gets M$, $\ell p \gets 0$ and $rp \gets 1$. As long as $(\ell \neq r)$, it defines the average $\eta = \lfloor (\ell + r)/2 \rfloor$ and then proceeds to compute $\hat{\pp}_{\eta} = \WW^{\eta} \hat{\pp}_0$ using Lemma~\ref{lem: binaryExponentation}. Note that $\pnew(\eta) = \hat{\pp}_{\eta}(k+1)$ by Equation~(\ref{eq: powerEquiv}). Finally, the algorithm uses $\pnew(\eta)$ to randomly decide whether $w$ meets a new vertex in the next $\eta$ steps or not. In other words, it updates the maintained variables using the rule below:
\begin{enumerate}
\itemsep0em 
\item with probability $(\pnew(\eta)-\ell p)/(rp - \ell p)$, set $r \gets \eta $, and $rp = \pnew(\eta)$,
\item otherwise, with probability $(rp - \pnew(\eta)) / (rp - \ell p)$, set $\ell \gets \eta + 1$, $\ell p = \pnew(\eta)$.
\end{enumerate}
If $(\ell = r)$, then the algorithm returns $\ell$. This procedure is summarized in Algorithm~\ref{alg:binary_search}.
 
\begin{algorithm2e}[t]
\caption{\textsc{BinarySearch}$(\WW, \hat{\pp}_0, M)$}
\label{alg:binary_search}
\Input{A $(k+1) \times (k+1)$ matrix $\WW$, a $(k+1)$ dimensional vector $\hat{\pp}_0$, and an integer $M$}
\Output{An integer}
Set $\ell \gets 0$, $r \gets M$, $\ell p \gets 0$ and $rp \gets 1$ \;
\While{$(\ell \neq r)$}
{
\label{part:mideqlr} Set $\eta \leftarrow \lfloor (\ell+r)/2 \rfloor$ \;
Compute $\hat{\pp}_\eta =\WW^{\eta}\hat{\pp}_0$ using Lemma~\label{line: exactProb} \ref{lem: binaryExponentation} \;
Set $\pnew(\eta) = \hat{\pp}_\eta(k+1)$\;
\label{line: updateVariables} With probability $(\pnew(\eta)-\ell p)/(rp - \ell p)$, set $r \gets \eta $, and $rp = \pnew(\eta)$,
otherwise, with probability $(rp - \pnew(\eta)) / (rp - \ell p)$, set $\ell \gets \eta + 1$, $\ell p = \pnew(\eta)$ \;
}
\Return $\ell$
\end{algorithm2e}  

We next show the correctness of the above procedure. To do so, we first need the following notation. For a $t$-step random walk $w$ and a current vertex $u = w_t \in U$, let $\escape$ be the smallest number of steps of steps needed for $w$ to visit a vertex not in $U$, i.e., $X(u,U) = \min\{i \mid i \geq 1,~ w_{t+i} \not \in U \}$. Note that $\escape$ is a random variable, and $\escape \leq M$ by definition of $M$.

\begin{lemma} \label{lem: binarySearch}
Let $w$ be a $t$-step random walk, $U$ the set of distinct vertices $w$ visited, $u = w_{t} \in U$ the current vertex and $k = |U|$ the number of distinct vertices $w$ has visited so far. For $\WW$, $\hat{\pp}_0$, and $M$ defined as above, \textsc{BinarySearch$(\WW, \hat{\pp}_0, M)$} correctly samples $\escape$, i.e., the number of steps $w$ needs to visit a vertex not in $U$, in $O(k^{3}\log^2 M)$ time. 
\end{lemma}
\begin{proof}
By Equation~(\ref{eq: powerEquiv}) and Line~\ref{line: exactProb} in Algorithm~\ref{alg:binary_search}, note that $\pnew(\eta)$ is the probability that $w$ meets a new vertex in the fist $\eta$ steps. The correctness of \textsc{BinarySearch} can be proven using an inductive argument on the number of iterations of the while loop. Here, we just show the crucial parts for being able to apply such an argument. First, observe that right after Line~\ref{part:mideqlr} in the while loop, we have that \[ (rp - \ell p) = \prob{\escape}{\ell \leq \escape \leq r}. \] 
The latter holds because $\ell p = \prob{\escape}{\escape \leq \ell}$ and $rp = \prob{\escape}{\escape \leq r}$, which in turn can be verified for each assignment of $\ell$ and $r$. Next, we prove that conditioning on $\ell \leq \escape \leq r$ right after Line~\ref{part:mideqlr} in the while loop, Algorithm~\ref{alg:binary_search} samples $\escape$ from the correct distribution. This is true when $(\ell=r)$, since the condition of the while loop is no longer satisfied and the algorithm returns $\ell$. If, however $(\ell \neq r)$, then we need to compute the following probabilities: (1) $\prob{\escape}{\escape \le \eta \mid \ell \le \escape \le r}$ and (2) $\prob{\escape}{\escape > \eta \mid \ell \le \escape \le r}$. To determine (1), we get that
\begin{align*}
&\prob{\escape}{\escape\le \eta \mid \ell \le \escape \le r} \\
& =\frac{\prob{\escape}{(\escape\le \eta) \wedge (\ell \le \escape\le r)}}{\prob{\escape}{\ell \le \escape \le r}} \\[0.1cm]
& = \frac{\prob{\escape}{\ell \le \escape \le \eta}}{\prob{\escape}{\ell \le \escape \le r}} \\[0.1cm] 
& = \frac{(\pnew(\eta)- \ell p)}{(rp- \ell p)}.
\end{align*} 
The probability from case (2) can be shown similarly. Since Line~\ref{line: updateVariables} in Algorithm~\ref{alg:binary_search} updates the search boundaries $\ell$ and $r$ and their corresponding values $\ell p$ and $rp$ using probabilities (1) and (2), the correctness of the algorithm follows. 

For the running time, observe that the number of iterations until the condition of the while loop is no longer satisfied is bounded by $O(\log M)$. Moreover, the running time of one iteration is dominated by the time needed to compute $\WW^{\eta}\hat{\pp}_0$. Since $\WW$ is a $(k+1) \times (k+1)$ dimensional matrix and $\eta \leq M$, Lemma~\ref{lem: binaryExponentation} implies that the matrix powering step can be computed in $O(k^{3} \log M)$. Thus, it follows that Algorithm~\ref{alg:binary_search} can be implemented in $O(k^{3} \log^{2} M)$ time.
\end{proof}

We next explain how to sample a new distinct vertex, and its corresponding edge of a $t$-step random walk $w$, i.e., we discuss component~\ref{comp: 2}. Let $\escape$ be the index computed by \textsc{BinarySearch} routine. We first compute the probability dsitribution $\qq$ over vertices in $U$ after performing the next $(\escape-1)$ steps of the random walk $w$, conditioning on $w$ not leaving $U$. Afterwards we proceed to computing the probability distribution $\rrvec$ over the edges leaving $U$, i.e., edges in the cut $(U, V \setminus U)$, conditioning on $w_0,\ldots,w_t$ and $w_{t+\escape}$ being the first vertex not in $U$. Formally, for $v \in U$, $z \in V \setminus U$, we have
\begin{equation} \label{eq: distributionR}
	\rrvec(v,z) = \frac{\qq(v)\ww(v,z)}{R}, \text{ where } R := \sum_{v \in U,z \in V \setminus U} \qq(v) \ww(v,z).
\end{equation}
Finally, we sample $(w_{t+\escape-1},w_{t+\escape})$ according to $\rrvec$, where $w_{t+\escape}$ is the first vertex not in $U$. The lemma below shows that we can efficiently sample from $\rrvec$.

\begin{lemma} \label{lem: sampleNewVertex}
Let $w$ be a $t$-step random walk and let $U$ with $k = |U|$ be the set of distinct vertices $w$ has visited so far. Given the number of steps $\escape$ needed for $w$ to visit a vertex not in $U$, there exists an algorithm that samples an edge leaving $U$, and the first vertex not in $U$ in $O(k^{3} \log M)$ time. 
\end{lemma}
\begin{proof}
We start by showing how to compute the distribution $\qq$. To this end, recall that $\pp_i(v)$ is the probability that $w_{t+i} = v$, conditioned on $w$ not having met any vertex different from $U$ in $w_{t+1}, \ldots, w_{t+i-1}$. Thus, by Equation~(\ref{eq: poweringW}),  we can use the fast exponentiation routine~(Lemma~\ref{lem: binaryExponentation}) to compute the vector $\hat{\pp}_{\escape-1} = \WW^{\escape-1} \hat{\pp}_0$. Since by Equation~(\ref{eq: powerEquiv}) we have that $\hat{\pp}_{\escape-1}(v) = \pp_{\escape-1}(v)$ for each $v \in U$, it follows that $\qq(v)$  is exactly $\pp_{\escape-1}(v)$. Note that the running time for implementing this step is $O(k^{3} \log M)$ as $\escape \leq M$.

We next describe how to efficiently sample from the distribution $\rrvec$. First, it will be helpful to to sample a vertex $v \in U$ conditioning on the $w_{t+\escape}$ being the first vertex not in $U$. Specifically, we are interested in sampling a vertex $v \in U$ with probability 
\begin{equation} \label{eq: auxiliaryDistr}
\frac{\qq(v)\cdot \ww(v,V \setminus U)}{R}, \text{ where } \ww(v,V \setminus U) := \sum_{z \in V \setminus U} \ww(v,z).
\end{equation}

For being able to efficiently sample from this distribution, we need to compute $\ww(v,V \setminus U)$, which in turn may require examining up to $\Omega(n)$ edges incident to $v$. However, this is not sufficient for our purposes as our ultimate goal is to sample from $\rrvec$ in time only proportional to $k$. To alleviate this, observe that $\ww(v,V \setminus U) = (\dd(v) - \sum_{z \in U} \ww(v,z) )$. Thus, maintaining \emph{weighted} degree $\dd_v$ for each $v \in V$, allows us to compute $\ww(v,V \setminus U)$ in $O(k)$ time. Similarly, rearranging the sums in the definition of $R$ we get  
\[ R = \sum_{v \in U,~z \in V \setminus U} \qq(v) \ww(v,z) = \sum_{v \in U} \left( \qq(v) \cdot \ww(v, V \setminus U) \right),
\]
which in turn implies that $R$ can be computed in $O(k^2)$ time. The latter gives that the distribution defined in Euqation~(\ref{eq: auxiliaryDistr}) can be computed in $O(k^{2})$ time. For sampling a vertex $v \in U$ from this distribution we simply generate a uniformly-random value $x \in [0,1]$, and then perform binary search on the prefix sum array of the probability distribution. Since computing the prefix sum array and performing binary search can be done in $O(k)$ and $O(\log n)$ time, respectively, we get that sampling $v \in U$ according to distribution defined in Equation~(\ref{eq: auxiliaryDistr}) can be performed in $O(k^{2})$ time. 

We next explain how to sample an edge $(v,z)$, where $z \in V \setminus U$ and $v \in U$ is the vertex we sampled from above. The probability distribution from which $(v,z)$ is sampled is as follows
\begin{equation} \label{eq: lastProb}
 \frac{\ww(v,z)}{\ww(v, V \setminus U)}.
\end{equation}

To see the idea behind this choice, note that Equation~(\ref{eq: lastProb}) combined with Equation~(\ref{eq: auxiliaryDistr}) yields the distribution $\rrvec$ as defined in Equation~(\ref{eq: distributionR}), which ensures that the edge is sampled correctly. However, one complication we face with is that $v$ may be incident to $\Omega(n)$ edges. Remember that for sampling an edge one needs access to the prefix sum array, which is expensive for our purposes. A natural attempt is to compute such an array during preprocessing. Nevertheless, this alone does not suffice as the set $U$ will change over the course of our algorithm. Instead, for every vertex $v \in V$, we maintain an augmented Balanced Binary Tree~(BBT) on the edge weights incident to $v$. Augmented BBT is a data-structure that supports operations such as (1) computing prefix sums and (2) updating the edge weights incident to $v$, both in $O(\log n)$ time.

We employ the augmented BBT data-structure as follows. First, for each vertex $U$ and the sampled vertex $v \in U$, we update the weights of the edges from $v$ to $U$ to $0$ in the augmented BBT of $v$. We then sample a uniformly-random value $x \in [0,W]$, and use the prefix sums computation in the tree to determine the range in which $x$ lies together with the corresponding edge $(w_{t+\escape-1},w_{t+\escape})$, where $w_{t+\escape}$ is the first vertex not in $U$. After having sampled the edge, we undo all the changes we performed in the augmented BBT of $v$.  It follows that sampling an edge according to Equation~(\ref{eq: lastProb}) can be implemented in $O(k \log n)$. Putting together the above running times, we conclude that sampling an edge leaving $U$ as well as the first vertex not in $U$ can be implemented in $O(k^{3} \log M)$ time. 
\end{proof}

The last ingredient we need is an efficient way to sample the sum of weights in the random walk starting at $w_{t}$ and ending at $w_{t+\escape}$, where $\escape$ is the number of steps needed for the walk to leave the vertex set $U$~(Component~\ref{comp: 3}). In other words, we need to sample the following sum
\[
	\sum_{i=t+1}^{t+\escape} \frac{1}{\ww(w_{i-1},w_i)}.
\]
We accomplish this task by employing a doubling technique. To illustrate, for any pair of vertices $u,v \in V$ and $s(w)$ as defined in Equation~(\ref{eq: weightWalk}), let
\begin{equation} \label{eq: pmf}
\pmf{\ell}{u}{v}
\end{equation}
be the probability mass function of $s(w)$ conditioning on (1) $w$ being a random walk that starts at $u$ and ends at $v$, i.e., $w \sim w_{u,v}$ and (2) length of the walk $\ell(w)$ is $\ell$ in $G$. Then it can be shown that
\[
\pmf{\ell}{u}{v} = \sum_{y \in V} \left( \pmf{\ell/2}{u}{y} * \pmf{\ell/2}{y}{v} \right),
\]
where $*$ denotes the convolution between two probability mass functions. Equivalently, the convolution is the probability mass function of the sum of the two corresponding random variables. The above relation suggests that if (1) we have some \emph{approximate} representation of the probability mass functions $\pmf{\ell/2}{u}{v}$ for all $u,v\in V$, and (2) we are able to compute the convolution of the two mass functions under such representation, we can produce approximations for $\pmf{\ell}{u}{v}$, where $u,v \in V$. This idea is formalized in the following lemma.

\begin{restatable}{lemma}{approxsample}
\label{lem:approx_sample}
Let $G=(V,E,\ww)$ be a undirected, weighted graph with $\ww(e) = [1,n^c]$ for each $e \in E$, where $c$ is a positive constant. For any finite random walk $w$ of length $\ell$ with $\ell \leq n^{d}$, where $d$ is a positive constant, let $s(w)$ be the sum of the inverse of its edge weights, i.e.,
\[
    s(w) = \sum_{i=1}^{\ell} \frac{1}{\ww(w_{i-1},w_i)}.
\] 
Moreover, for any $u,v \in V$, let
\[
\pmf{\ell}{u}{v}
\] 
be the probability mass function of $s(w)$ conditioning on \emph{(1)} $w$ being a random walk that starts at $u$ and ends at $v$, and \emph{(2)} length of the walk $\ell(w)$ is $\ell$ in $G$.  Then, for any pair $u,v \in V$, there exists an algorithm that that samples from $\pmf{\ell}{u}{v}$ and outputs a sampled $s(w)$ up to a $(1 + \epsilon)$ multiplicative error in $\tilde{O}(n^{3} \epsilon^{-2})$ time.
\end{restatable}



We finally describe a procedure that generates a $\beta$-shorted walk with respect to some vertex $v$ and set of terminals $K$. Concretely, the algorithm  maintains (1) a set $U$, initialized to $\{v\}$, of the distinct vertices visited so far by a random walk $w$ starting at $v$, (2) the number of steps $t$ the walk $w$ has performed so far and (3) two lists $L_w$ and $L_s$, initially set to empty, containing the first occurrences of distinct vertices of $w$ and the weightes of the sub-walks induced by the distinct vertices, respectively. Next, as long as $w$ does not hit a vertex in $K$ or there are vertices in the component containing $v$ that are still not visited by $w$, for the next $\Theta(\beta^{-1} \log n)$ steps, the algorithm repeatedly generates a new vertex not in the current $U$ by using components~\ref{comp: 1},~\ref{comp: 2} and~\ref{comp: 3}. In each iteration, the maintained quantities $U$, $t$, $L_w$ and $L_s$ are updated accordingly. Note that this procedure indeed outputs all necessary information we need from a $\beta$-shorted walk. A detailed implementation of the algorithm is summarized in Figure~\ref{alg:GenSingleWalkWeighted}.

\begin{algorithm2e}[t]
\caption{\textsc{GenerateSingleWalk}$(G,K,v)$}
\label{alg:GenSingleWalkWeighted}
\Input{Weighted graph $G=(V,E,\ww)$ with $\ww(e) = [1,n^{c}]$ for each $e \in E$ and $c > 0$, a set of vertices $K \subseteq V$, a vertex $v\in V$ such that the component containing $v$ contains at least one vertex in $K$}
\Output{Two lists $L_w$ and $L_s$ containing the first occurrences of distinct vertices of a random walk $w$ starting at $v$ and the weights of the sub-walks induced by the distinct vertices, respectively}

Set $U \gets \{v\}$, $k \gets |U|$, and let $u \gets v$ be the current vertex \;
Let $t \gets 0$ be the index of current step of random walk $w$, i.e., $w_{t}=u$ \;
Let $L_w$ and $L_s$ be two lists, initially set to empty \;
\For{each $i = 1,\ldots, \Theta(\beta^{-1}\log n)$}
{
Let $\WW$ be a matrix of dimension $(k+1) \times (k+1)$ as defined in  Equation~(\ref{eq: poweringW}) \;
Set $\hat{\pp}_0 = \left[ \pp_0 \quad 0\right]^{\top}$, where $\pp_0(u) \gets 1$, and $\pp_0(\hat{u}) \gets 0$ for every $\hat{u} \in U \setminus u$ \;

Set $\escape \gets \textsc{BinarySearch}(\WW, \hat{\pp}_0, O(m^{3}))$ \; 

Compute the probability distribution $\qq$ over vertices in $U$ after $(\escape-1)$ steps of the random walk $w$, conditioning on $w$ not leaving $U$ \;

Compute the probability distribution $\rr$ over the the edges in $(U, V \setminus U)$ conditioning on $w_0,\ldots,w_t$ and $w_{t+\escape}$ being the first vertex not in $U$. Concretely, for $v \in U$, $z \in V \setminus U$,
\[
	\rrvec(v,z) = \frac{\qq(v)\ww(v,z)}{R}, \text{ where } R \gets  \left(\sum_{v \in U,~z \in V \setminus U} \qq(v) \ww(v,z) \right).
\]

Sample $(w_{t+\escape-1}, w_{t+\escape})$ according to $\rrvec(w_{t+\escape-1},w_{t+\escape})$ \; 
Set $e^{\textrm{new}} \gets (w_{t+\escape-1}, w_{t+\escape})$ \; 

Invoke Lemma~\ref{lem:approx_sample} in the inducted graph $G[U]$ to sample 

\[s = \sum_{j= t+1}^{t+\escape-1} \frac{1}{\ww(w_{j-1},w_{j})} . \]

Append $w_{t+\escape}$ to $L_w$ and $(s+1/\ww(e^{\textrm{new}}))$ to $L_s$ \;

\eIf{$w_{t+\escape} \in K$}{Go to Line~\ref{line: Return}}
{
Set $t \gets (t+\escape)$, $u \gets w_{t+\escape}$, $U \gets U \cup \{u\}$, and $k \gets (k + 1)$ \;

If $U$ covers the entire component, go to Line~\ref{line: Return}. Otherwise, $i\leftarrow (i+1)$ \; 
}
}
\Return lists $L_w$ and $L_s$. \label{line: Return}
\end{algorithm2e}

We now have all the necessary tools to prove Lemma~\ref{lem: generateSingleWalk}.

\begin{proof}[Proof of Lemma~\ref{lem: generateSingleWalk}]
We first show correctness. By Lemma~\ref{lem: binarySearch} it follows that~$\textsc{BinarySearch}$ correctly samples the number of steeps before a walk meets a new vertex. Next, Lemma~\ref{lem: sampleNewVertex} implies that the we can sample the new distinct vertex and its corresponding edge. Finally, by Lemma~\ref{lem:approx_sample} we know that the weight of each sub-walk of a $\beta$-shorted walk is approximated within a $(1 + \epsilon)$ multiplicative error. Bringing these approximation together we get that the weight of the $\beta$-shorted walk itself is approximated within the same multiplicative error.

We now analyse the running time of procedure \textsc{GenerateSingleWalk}. We start by bounding the cover time of $G$, which in turn bounds the number of steps for a random walk to meet a new vertex. To this end, note that it takes expected $O(m^2)$ time to meet a vertex in the same component~(\cite{ALLRK79}). Thus, if we perform a random walk of length $O(m^3)$ we are guaranteed that it covers ever vertex in the component containing the starting vertex, with high probability. 

Next, we analyze the running time for the steps executed within one iteration of the for loop. Observe that $k = |U| = O(\beta^{-1} \log n)$ at any point of time throughout our algorithm. The latter together with Lemma~\ref{lem: binarySearch} give that it takes $O(k^{3} \log ^{2} M) = \tilde{O}(\beta^{-3})$ time to sample the minimum number of steps for a random walk to visit a vertex not in $U$, where $M = O(m^{3})$ by the discussion above. Furthermore, by Lemma~\ref{lem: sampleNewVertex} we can sample the new vertex not in $U$, and its corresponding edge in $\tilde{O}(\beta^{-3})$ time. Finally, Lemma~\ref{lem:approx_sample} implies that the weight $s(w)$ of the random sub-walk between the current vertex and the new generated vertex can be approximately sampled in $\tilde{O}(\beta^{-3} \epsilon^{-2})$ time. The latter holds because Lemma~\ref{lem:approx_sample} is invoked on top of the graph $G[U]$ for which $|V(G[U])| = O(\beta^{-1} \log n)$. Combining the above running times, we get that one iteration can be implemented in $\tilde{O}(\beta^{-3} \epsilon^{-2})$ time. Since there are $O(\beta^{-1} \log n)$ iterations, we conclude that the overall running time of our procedure is $\tilde{O}(\beta^{-4} \epsilon^{-2})$.
\end{proof}

We now present the procedure for generating a Schur complement on weighted graphs. The idea behind this algorithm is the same as in the unweighted setting, except that now we use \textsc{GenerateSingleWalk} to extract the information needed to simulate $\beta$-shorted walks. For the sake of completeness we summarize the details of this modified procedure in Algorithm~\ref{alg:Initialize_Weighted}.

\begin{lemma} \label{lem: preprocessingWeighted}
Algorithm~\ref{alg:Initialize_Weighted} runs in $\Otil(m\beta^{-4}\epsilon^{-4})$ time and outputs a graph $H$ satisfying $\LL_H\approx_\epsilon \SC(G,K)$, with high probability. 
\end{lemma}
\begin{proof}
We first bound the running time of Algorithm~\ref{alg:Initialize_Weighted}.  By Lemma~\ref{lem: generateSingleWalk}, the time needed to generate a $\beta$-shorted walk is $\tilde{O}(\beta^{-4} \epsilon^{-2})$. Combining this with the fact that the algorithm generates $\rho m = \tilde{O}(m \epsilon^{-2})$ walks, it follows that the running time of the algorithms is dominated by $\tilde{O}(m \beta^{-4} \epsilon^{-4})$.

We next show the correctness of our procedure. First, note that procedure \textsc{GenerateSingleWalk} generates a valid $\beta$-shorted walk with its weight being approximated up to a $(1 + \epsilon)$ multiplicative error~(Lemma~\ref{lem: generateSingleWalk}). Assume for now that there is an oracle that fixes this approximate weight of a walk back to its original exact weight. Then the collection of generated walks from Algorithm~\ref{alg:Initialize_Weighted} forms the set $W$ of $\beta$-shorted walks, and let $\hat{H}$ be the corresponding output graph. By Theorem~\ref{thm:RandomWalkProperties},  with high probability, each of the walks that starts at a component containing a vertex in $K$ hits $K$. Conditioning on the latter, Theorem~\ref{thm:SparsifySchur} gives that with high probability, $\LL_{\hat{H}}\approx_\epsilon \SC(G,K)$. 

Finally, let $H$ be the graph where the edge weights are correct up to a $(1 + \epsilon)$ multiplicative error. In other words, the weight of each edge $e$ in $H$ differs from the corresponding weight $\ww_{\hat{H}}(e)$ in $\hat{H}$ by $\epsilon \ww_{\hat{H}}(e)$. Summing over all the edges we get that $\LL_H \approx_{\epsilon} \LL_{\hat{H}}$. Since $\LL_{\hat{H}} \approx_{\varepsilon} \SC(G,K)$ by the discussion above, we get that $\LL_H \approx_{O(\epsilon)} \SC(G,K)$. Scaling $\epsilon$ appropriately completes the correctness. 
\end{proof}

\begin{algorithm2e}[t]
\label{alg:Initialize_Weighted}
\caption{$\textsc{InitializeWeighted}(G, K', \beta)$}
\Input{Weighted graph $G=(V,E,\ww)$ with $\ww(e) = [1, n^{c}]$ for each $e \in E$ and $c>0$, set of vertices $K' \subseteq V$ such that $|K'| \le O(\beta m)$, and $\beta\in (0, 1)$ }
\Output{Approximate Schur Complement $H$ and union of $\beta$-shorted walks $W$}
Set $K \gets K'$, $H \gets (V,\emptyset)$ and $W \gets \emptyset$ \;
For each edge $e=(u,v)$ in $G$, let $K \gets K \cup \{u,v\}$ with probability $\beta$ \;
Let $\rho \gets O(\log n \epsilon^{-2})$ be the sampling overhead according to Theorem~\ref{thm:SparsifySchur} \;
\For{each edge $e=(u,v) \in E$ and each $i=1,\ldots,\rho$}
{
Using Algorithm \ref{alg:GenSingleWalkWeighted}, generate a random walk $w_1(e,i)$ from $u$ until $\Theta(\beta^{-1} \log n)$ different vertices have been hit, it reaches $K$, or it has hit every edge in its component \;
Using Algorithm \ref{alg:GenSingleWalkWeighted}, generate a random walk $w_2(e,i)$ from $v$ until $\Theta(\beta^{-1}\log n)$ different vertices have been hit, it reaches $K$, or it hast hit every edge in its component \;
\If{both walks reach $K$ at $t_1$ and $t_2$ respectively}
{
  Connect $w_1(e,i)$, $e$ and $w_2(e,i)$ to form a walk $w(e,i)$ between $t_1$ and $t_2$ \;
Let $s \gets s(w_1(e,i))+s(w_2(e_i))+ 1/\ww(e)$ \;
Add an edge $(t_1,t_2)$ with weight $1/(\rho s)$ to $H$ \;
Add $w(e,i)$ to $W$ \;
}
}
\Return $H$ and $W$
\end{algorithm2e}


We now have all the necessary tools to present our dynamic algorithm for maintaining the collection of walks $W$~(equivalently, the approximate Schur complement $H$), on weighted graphs.


\begin{proof}[Proof of Lemma~\ref{lem:WeightedDynamic}]
Similarly to the unweighted case, we give a two-level data-structure for dynamically maintaining Schur complements on weighted graphs. Specifically, we keep the terminal set $K$ of size $\Theta(m\beta)$. This entails maintaining
\begin{enumerate}
\itemsep0em
\item an approximate Schur complement $H$ of $G$ with respect to $K$~(Theorem~\ref{thm:SparsifySchur}),
\item a dynamic spectral sparsifier $\tilde{H}$ of $H$~(Lemma~\ref{lem:DynamicSpectralSparsifier}).
\end{enumerate}
We implement the procedure $\textsc{Initialize}$ by running Algorithm~\ref{alg:Initialize_Weighted}, which produces a graph $H$ and then compute a spectral sparsifier $\tilde{H}$ of $H$ using Lemma~\ref{lem:DynamicSpectralSparsifier}. Note that by construction of our data-structure, every update in $H$ will be handled by the black-box dynamic sparsifier $\tilde{H}$.

Similarly to the unweighted case, operations $\textsc{Insert}$ and $\textsc{Delete}$ are reduced to adding terminals to the set $K$ and we refer the reader to the previous section for details on this reduction. Thus, the bulk of our effort is devoted to implementing the procedure $\textsc{AddTerminal}$. Let $u$ be a non-terminal vertex that we want to append to $K$. We set $K \gets K \cup \{u\}$, and then shorten all the walks at the first location they meet $u$. This shortening of walks induces in turn edge insertions and deletions to $H$, which are then processed by $\tilde{H}$. To quickly locate the first appearances of $u$ in the random walks from $W$, we maintain a linked list $W_u$ for each $u \in V$. This linked list contains the first appearances of $w$ in the collections of random walks $W$. Note that constructing such lists can be performed at no additional cost during preprocessing phase, since Algorithm~\ref{alg:GenSingleWalkWeighted} directly gives the first appearances of vertices in every walk belonging to $W$. After locating the first appearances of $u$, we cut the walks in these locations, delete the corresponding affected walks (together with their weight from $H$), and insert the new shorter walks to $H$. Note that we can simply use arrays to represent each random walk in $W$.

We next analyze the performance of our data-structure. Let us start with the preprocessing time. First, by Lemma~\ref{lem: preprocessingWeighted} we get that the cost for constructing $H$ on a graph with $m$ edges is bounded by $\Otil(m \beta^{-4} \epsilon^{-4})$. Next, since $H$ has $\Otil(m \epsilon^{-2})$ edges, constructing $\tilde{H}$ takes $\Otil(m \epsilon^{-4})$ time. Thus, the amortized time of \textsc{Initialize} operation is bounded by $\Otil(m\beta^{-4} \epsilon^{-4})$. 

We now analyze the update operations. By the above discussion, note that it suffices to bound the time for adding a vertex to $K$, which in turn (asymptotically) bounds the update time for edge insertions and deletions. The main observation we make is that adding a vertex to $K$ only shortens the existing walks, and by the above discussion we can find such walks in time proportional to the amount of edges deleted from the walk. Since the walk needed to be generated in the \textsc{Initialize} operation, the deletion of these edges take equivalent time to generating them. Moreover, we note that (1) handling the updates in $\tilde{H}$ induced by $H$ introduces additional $O(\poly(\log n)\epsilon^{-2})$ overheads, and (2) adding or deleting $\rho$ edges until the next rebuild costs $\tilde{O}(\beta m \epsilon^{-2})$, since we process only up to $\beta m$ operations. These together imply that the amortized cost for adding a terminal can be charged against the preprocessing time, which is bounded by $\Otil(m\beta^{-4} \epsilon^{-4})$, up to poly-logarithmic factors. Thus it follows that the operations \textsc{AddTerminal}, \textsc{Insert} and \textsc{Delete} can be implemented in $\tilde{O}(1)$ amortized update time.
\end{proof}


\subsection{Dynamic All-Pair Effective Resistance on Weighted Graphs}

Following exactly the same arguments as in the proof of Theorem~\ref{thm:UnweightedER}, we can use the above data-structure to efficiently maintain effective resistances on weighted, undirected dynamic graphs.

\begin{theorem}\label{thm:WeightedER}
For any given error threshold $\epsilon > 0$,
there is a data-structure for maintaining a weighted, undirected multi-graph $G=(V,E,\ww)$ with up to $m$ edges that supports the following operations
in $\tilde{O}(m^{5/6}\epsilon^{-4})$ expected amortized time:
\begin{itemize}
\itemsep0em 
	\item \textsc{Insert}$(u,v, w)$: Insert the edge $(u,v)$ with resistance $1/w$ in $G$.
	\item \textsc{Delete}$(u,v)$: Delete the edge $(u,v)$ from $G$.
	\item \textsc{EffectiveResistance}$(s,t)$: Return a $(1 \pm \epsilon)$-approximation to the effective resistance between $s$ and $K$ in the current graph $G$. 
\end{itemize}
\end{theorem}

\begin{proof}
Let $\mathcal{D}(\tilde{H})$ denote the data structure that maintains a dynamic (sparse) Schur complement $\tilde{H}$ of $G$~(Lemma~\ref{lem:WeightedDynamic}). Since $\mathcal{D}(\tilde{H})$ supports only up to $\beta m$ operations, we rebuild $\mathcal{D}(\tilde{H})$ on the current graph $G$ after such many operations. Note that the operations \textsc{Insert} and \textsc{Delete} on $G$ are simply passed to $\mathcal{D}(\tilde{H})$. For processing the query operation $\textsc{EffectiveResistance}(s,t)$, we declare $s$ and $t$ terminals (using the operation \textsc{AddTerminal} of $\mathcal{D}(\tilde{H})$), which ensures that they are both now contained in $\tilde{H}$. Finally, we compute the (approximate) effective resistance between $s$ and $t$ in $\tilde{H}$ using Lemma~\ref{lemm:efficientEffectiveResistance}.

We now analyze the performance of our data-structure. Recall that the insertion or deletion of an edge in $G$ can be supported in $\tilde{O}(1)$ expected amortized time by $\mathcal{D}(\tilde{H})$. Since our data-structure is rebuilt every $\beta m$ operations, and rebuilding $\mathcal{D}(\tilde{H})$ can be implemented in $\tilde{O}(m\beta^{-4} \epsilon^{-4})$ time,  it follows that the amortized cost per edge insertion or deletion is 
\[
	\frac{\tilde{O}(m\beta^{-4} \epsilon^{-4})}{\beta m} = \tilde{O}(\beta^{-5} \epsilon^{-4}).
\]

The cost of any $(s,t)$ query is dominated by (1) the cost of declaring $s$ and $t$ terminals and (2) the cost of computing the $(s,t)$ effective resistance to $\epsilon$ accuracy on the graph $\tilde{H}$. Since (1) can be performed in $\tilde{O}(1)$ time, we only need to analyze (2). We do so by first giving a bound on the size of $K$. To this end, note that each of the $m$ edges in the current graph adds two vertices to $K$ with probability $\beta$ independently. By a Chernoff bound, the number of random augmentations added to $K$ is at most $2\beta m$ with high probability.
In addition, since $\mathcal{D}(\tilde{H})$ is rebuilt every $\beta m$ operations, the size of $K$ never exceeds $4\beta m$
with high probability. The latter also bounds the size of $\Htil$ by $\Otil(\beta m\epsilon^{-2})$
and gives that the query cost is $\tilde{O}(\beta m \epsilon^{-2})$.

Combining the above bounds on the update and query time, we obtain the following trade-off \[ \tilde{O}\left((\beta m + \beta^{-5})\epsilon^{-4}\right),\]
which is minimized when $\beta = m^{-1/6}$, thus giving an expected amortized update and query time of \[ \tilde{O}(m^{5/6}\epsilon^{-4}). \qedhere \]
\end{proof}

\section{Dynamic Laplacian Solver in Sub-linear Time}
\label{sec:DynamicSolver}

In this section we extend our dynamic approximate Schur complement algorithm to obtain a dynamic Laplacian solver for unweighted, bounded degree graphs. Specifically, as described in Theorem~\ref{thm:Solver}, our goal is to design a data-structure that maintains a solution to the Laplacian system $\LL \xx = \bb$ under updates to both the underlying graph and the demand vector vector $\bb$ while being able to query a few entries of the solution vector. For the sake of exposition, in what follows we assume that the underlying graph is always connected. 

Consider the dynamic Schur complement data-structure provided by Lemma~\ref{lem:Dynamic}. If the demand vector $\bb$ has up to $O(\beta m)$ non-zero entries, for some parameter $\beta \in (0,1)$, we can simply incorporate the vertices corresponding to these entries in the terminal set $K$ using operation $\textsc{AddTerminal}$ of the dynamic Schur complement data-structure~(Lemma~\ref{lem:Dynamic}). Upon receipt of a query index, we add the corresponding vertex to the Schur complement and (approximately) solve a Laplacian system on the maintained Schur complement. The obtained solution vector can then be lifted back to the Laplacian matrix using the following lemma, which we introduced in the preliminaries.  

\begin{lemma}[Restatement of Lemma~\ref{fac:solve_by_sc_and_proj}]
Let $\xx_K$ be a solution vector such that $\SC(G,K)\xx_K=\proj{G}{K}\bb$. Then there exists an extension $\xx$ of $\xx_K$ such that $\LL\xx=\bb$.
\end{lemma}

As we argued in Section~\ref{sec:Overview_DSC}, this approach leads to a dynamic Laplacian solver with $O(m^{3/4})$ amortized update time per operation. Moreover, note that the algorithm applies to any undirected, unweighted graph. However, the prime difficulty for constructing a dynamic solver is in handling the case where $\bb$ has a large number of non-zero entries, i.e., $\vecnorm{\bb}_0 = \Omega(n)$, thus preventing us from obtaining a sub-linear algorithm using the reduction above. We alleviate this by projection this demand vector onto the current set of terminals and showing that such a projection can be maintained dynamically while introducing controllable error in the approximation guarantee. At a high level, our solver can be viewed as an one layer version of sparsified block-Cholesky algorithms~\cite{KyngLPSS16}.

We next discuss specific implementation details. Recall that $\proj{G}{K}$ is the matrix projection of non-terminal vertices $F$ onto $K$. By Lemma~\ref{fac:solve_by_sc_and_proj}, it is sufficient to maintain a solution $\xx_{T} = \SC(G,T)^{\dagger} \proj{G}{T} \bb$ dynamically. Since Lemma~\ref{lem:Dynamic} already allows us to maintain $\SC{G,K}$, we need to devise a routine that maintains the projection $\proj{G}{T} \bb$ of $\bb$ under vertex additions to the terminal set.

To this end, we describe an algorithm that maintains such a projection which in turn allows us to again achieve sub-linear running times. The algorithm itself can be viewed as a numerically minded generalization of the approach for the small-support case. Let $S$ denote the current set of terminals that the algorithm maintains~($S$ and $K$ will always be equal, and we differentiate between them only for the sake of presentation). We initialize $S$ with $O(\beta m)$ vertices from the corresponding entries in $\bb$ that have the largest value. Our key structural observation is that if the entries of $\bb$ are small,
adding vertices to $S$ does not change the projection significantly. To measure the error incurred by declaring some vertex a terminal, we exploit the fact that the projection $\proj{G}{S} \bb$ itself is tightly connected to specific random walks in the underlying graph. We then argue that it is possible to reuse earlier projections, even when new terminals are added to $S$, while paying an error corresponding to the lengths of these random walks and the magnitude of entries in $\bb$. Finally, we analyze how to control the accumulation of these errors over a sequence of terminal additions, and also describe an initialization procedure that involves solving a Laplacian system for computing the starting (approximate) projection vector. These together lead to the main lemma of this section, whose implementation details and analysis are deferred to Subsection~\ref{subsec:DynamicProjection}.



\begin{restatable}{lemma}{DynamicProjection}
	\label{lem:DynamicProjection}
Given an error parameter $\epsilon > 0$, an unweighted unweighted bounded-degree $G=(V,E)$ with $n$ vertices, a vector $\bb\in \mathbb{R}^n$ in the image of $\LL$, a subset of terminal vertices $S'$ and a parameter $\beta \in (0,1)$ such that $|S'|=O(\beta m)$, there is a data-structure
	\textsc{DynamicProj}$(G,S',\beta)$ for maintaining a vector $\bbtil$ with $\vecnorm{\bbtil-\proj{G}{S}\bb}_{\LL^\dag} \le \epsilon \vecnorm{\bb}_{\LL^\dag}$ for some $S$ with $S'\subseteq S$, $|S|=O(\beta m)$, while supporting at most $\beta^3 m^{1/2} \epsilon (\poly \log n)^{-1}$ operations in the following running times: 
	\begin{itemize}
		\item \textsc{Initialize}$(G, S', \beta)$: Initialize the data-structure in $\tilde{O}(m)$ time.

		\item \textsc{Insert$(u,v)$}: Insert the edge $(u,v)$ to $G$ in $O(1)$ time while keeping $G$ bounded-degree.
		\item \textsc{Delete$(u,v)$}: Delete the edge $(u,v)$ from $G$ in $O(1)$ time.
		\item \textsc{Change$(u,\bb'(u), v, \bb'(v))$}: Change $\bb(u)$ to $\bb'(u)$ and changes $\bb(v)$ to $\bb'(v)$ while keeping $\bb$ in the range of $\LL$ in $O(1)$ time.
		\item \textsc{AddTerminal$(u)$}: Add $u$ to $S$ in $O(1)$ time.
		\item \textsc{Query$()$}: Output the maintained $\bbtil$ in $O(\beta m)$ time.
	\end{itemize}
\end{restatable}

The following lemma, whose proof will be shortly provided, allows us to combine the approximation guarantees of the data-structures (1) dynamic Schur complement and (2) dynamic Projection.

\begin{lemma}
	\label{lem:SolverApprox}
	Let $0<\epsilon\le \frac{1}{2}$. Let $k$ be a positive number such that $\vecnorm{\bb}_{\LL^\dag}\le k$. Suppose $\tilde{\LL}\approx_{\epsilon} \LL$, $\vecnorm{\bbtil-\bb}_{\LL^\dag}\le \epsilon k$ and $\vecnorm{\xxtil-\tilde{\LL}^\dag\tilde{\bb}}_{\LLtil}\le \epsilon \vecnorm{\tilde{\LL}^\dag\bbtil}_{\LLtil}$. Then $\vecnorm{\tilde{\xx}-\LL^\dag \bb}_{\LL}\le 10\epsilon k$.
\end{lemma}


We now have all the necessary tools to present the data-structure for solving Laplacian systems in bounded-degree graphs, which essentially entails combining Lemma~\ref{lem:Dynamic} and Lemma~\ref{lem:DynamicProjection}.

\begin{proof}[Proof of Theorem \ref{thm:Solver}] 
Let $\mathcal{D}(\tilde{H})$ and $\mathcal{D}(\bbtil)$ denote the data-structure that maintains a dynamic (sparse) Schur complement $\tilde{H}$ of $G$~(Lemma~\ref{lem:Dynamic}) and an approximate dynamic Projection $\bbtil$ of $\proj{G}{S}\bb$~(Lemma~\ref{lem:DynamicProjection}), respectively. Set $\epsilon \gets (\epsilon / 10)$ for both data-structures. Our dynamic solver simultaneously maintains $\mathcal{D}(\tilde{H})$ and $\mathcal{D}(\bbtil)$. Since $\mathcal{D}(\bbtil)$ supports only up to $\beta^3 m^{1/2} \epsilon (\poly \log n)^{-1}$, we rebuild both data-structures after such many operations. 

We now describe the implementation of the operations. First, we find the first $\beta m$ entries with maximum value in $\bb$. We then take the corresponding vertices and initialize $S'$ and $K'$ to be these $\beta m$ vertices. The implementation of these data-structures involves including the endpoints of each edge with probability $\beta$ to $S$ and $K$, respectively. We couple these algorithms such that $S = K$, and this property will be maintained throughout the algorithm. The operations \textsc{Insert} and \text{Delete} on $G$ are simply passed to $\mathcal{D}(\tilde{H})$ and $\mathcal{D}(\bbtil)$. The operation \textsc{Change$(u,\bb'(u), v, \bb'(v))$} is passed to $\mathcal{D}(\bbtil)$. Upon receipt of a query $\xx(u)$, for some vertex $u \in V$, i.e., operation $\textsc{Solve}(u)$, we declare $u$ a terminal (using the operation $\textsc{AddTerminal}(u)$ of both $\mathcal{D}(\tilde{H})$ and $\mathcal{D}(\bbtil)$). We then proceed by extracting an approximate Schur complement $\tilde{H}$ of $G$  from $\mathcal{D}(\tilde{H})$ and an approximate projection vector $\bbtil$ from $\mathcal{D}(\bbtil)$. Finally, using a black-box Laplacian solver~\cite{KoutisMP11}, we compute a solution vector $\xxtil_K$ to the system $\LL_{\tilde{H}} \xxtil_K = \bbtil$ and output $\xxtil_K(u)$ (this is possible since $u$ was added to $K$).

We next show the correctness of the operation $\textsc{Solve}(u)$. The Laplacian solver guarantees that the vector $\tilde{\xx}_K$ satisfies
\begin{equation}
\label{eq: laplacGuarantee}
\vecnorm{\xxtil_K-\LL_{\tilde{H}}^\dag\bbtil}_{\LL_{\tilde{H}}}\le (\epsilon/10) \vecnorm{\LL_{\tilde{H}}^\dag\bbtil}_{\LL_{\tilde{H}}}.
\end{equation}

Data-structure $\mathcal{D}(\bbtil)$ guarantees that
\begin{equation}
\label{eq: projGaurantee}
	\vecnorm{\bbtil - \proj{G}{K} \bb}_{\SC(G,T)^\dag} \leq (\epsilon/10) \vecnorm{\bb}_{\LL^\dag}.
\end{equation}

Note that $\vecnorm{\proj{G}{K} \bb}_{\SC(G,K)^\dag}\le \vecnorm{\bb}_{\LL^\dag}$. 
Bringing together Equations~(\ref{eq: laplacGuarantee}) and~(\ref{eq: projGaurantee}) and applying Lemma~\ref{lem:SolverApprox} with $k=\vecnorm{\bb}_{\LL^\dag}$, $\LL := \SC(G,K)$, $\bb := \proj{G}{K} \bb$, $\LL_{\tilde{H}}$ and $\bbtil$ yield
\[	\vecnorm{\xxtil_K - \SC(G,K)^\dag \proj{G}{K} \bb}_{\SC(G,K)}\le \epsilon k.
\]
Using Lemma~\ref{fac:solve_by_sc_and_proj} we can lift the vector $\xxtil_K$ to a solution $\xxtil$ such that
\[
	\vecnorm{\xxtil-\LL^{\dag}\bb}_{\LL}\le \epsilon k= \epsilon \vecnorm{\bb}_{\LL^\dag} = \epsilon \vecnorm{\LL^\dag\bb}_{\LL}.
\] 

Finally, we bound the running time of our dynamic solver. Changes in the demand vector $\bb$ can be performed in $O(1)$ times, thus having negligible affect in our running times.  The insertion or deletion of an edge in $G$ can be supported in $\tilde{O}(1)$ expected amortized time by both $\mathcal{D}(\tilde{H})$ and $\mathcal{D}(\bbtil)$. Since we build our data-structures every $\beta^3 m^{1/2} \epsilon (\poly \log n)^{-1}$ operations, and the total rebuild cost is dominated by $\tilde{O}(m \beta^{-2} \epsilon^{-4})$, it follows that the amortized cost per edge insertion or deletion is 
\[
	\frac{\tilde{O}(m \beta^{-2} \epsilon^{-4})}{\beta^3 m^{1/2} \epsilon (\poly \log n)^{-1}} = \tilde{O}(m^{1/2} \beta^{-5} \epsilon^{-5}).
\]

The cost of any query is dominated by (1) the cost of declaring the queried vertex $u$ a terminal and (2) the cost of extracting $\tilde{H}$ and $\bbtil$. Since (1) can be performed in $\tilde{O}(1)$ amortized time, we only need to analyze (2). Size of the terminal set $S=K$, which can be easily shown to be $O(\beta m)$ with high probability, immediately implies that the running time for (2) is dominated by $\tilde{O}(\beta m \epsilon^{-2}) = \tilde{O}(\beta m \epsilon^{-5})$, which in turn bounds the query cost.

Combining the above bounds on the query and update time, we obtain the following trade-off
\[
	\tilde{O}\left( (m^{1/2} \beta^{-5} + \beta m )\epsilon^{-5} \right)
\]
which is minimized when $\beta = m^{-1/12}$, thus giving an expected amortized update and query time of
\[
	\tilde{O}(m^{11/12} \epsilon^{-5}).
\]

We can replace $m$ by $n$ in the above running time guarantee since by our assumption, $G$ has bounded-degree throughout the algorithm.
\end{proof}

We next prove Lemma~\ref{lem:SolverApprox}.

\begin{proof}[Proof of Lemma~\ref{lem:SolverApprox}]
We will use triangle inequality to decompose the error as: 
\begin{align}
\vecnorm{\xxtil-\LL^\dag\bb}_{\LL}
=&\vecnorm{\xxtil-\LLtil^\dag\bbtil+\LLtil^\dag\bbtil-\LLtil^\dag\bb+\LLtil^\dag\bb-\LL^\dag\bb}_{\LL}
\nonumber \\
\le&\vecnorm{\xxtil-\LLtil^\dag\bbtil}_{\LL}+\vecnorm{\LLtil^\dag\bbtil-\LLtil^\dag\bb}_{\LL}+\vecnorm{\LLtil^\dag\bb-\LL^\dag\bb}_{\LL},
\label{eq:ErrorSplit}
\end{align}
and bound each of them separately.

\begin{enumerate}
\item The first term can be bounded by first
invoking the similarity of $\LL$ and $\LLtil$ to
change the norm to $\LLtil$,
and applying the guarantees of the solve involving $\LLtil$:
\[
\vecnorm{\xxtil-\LLtil^\dag\bbtil}_{\LL}
\leq
\sqrt{(1+2\epsilon)} \vecnorm{\xxtil-\LLtil^\dag\bbtil}_{\LLtil}
\leq
2
\vecnorm{\xxtil-\LLtil^\dag\bbtil}_{\LLtil}
\leq
2 \epsilon \vecnorm{\bbtil}_{\LLtil^\dag}.
\]
This norm can in turn be transferred back to $\LL$,
and the discrepancy between $\bb$ and $\bbtil$ absorbed
using triangle inequality:
\[
\leq
3 \epsilon \vecnorm{\bbtil}_{\LL^\dag}
\leq
3 \epsilon \left(\vecnorm{\bb}_{\LL^\dag}+\vecnorm{\bbtil-\bb}_{\LL^\dag}\right)
\le 3 \epsilon (1+\epsilon) k\\
\le 5 \epsilon k.
\]
\item The second term follows from combining the norms
in $\LLtil$ and $\LL$ using the approximations between
these matrices:
\[
\vecnorm{\LLtil^\dag\bbtil-\LLtil^\dag\bb}_{\LL}
=
\vecnorm{\bbtil-\bb}_{\LLtil^\dag\LL\LLtil^\dag}
\leq
2 \vecnorm{\bbtil-\bb}_{\LLtil^\dag\LLtil\LLtil^\dag}
=
2 \vecnorm{\bbtil-\bb}_{\LLtil^\dag},
\]
and once again converting the norm back from $\LLtil$ to $\LL$:
\[
\leq
4 \vecnorm{\bbtil-\bb}_{\LL^\dag}
\leq
4 \epsilon  k.
\]
\item The third term can first be written in terms
of the norm of $\bb$ against a matrix involving the
difference between $\LL$ and $\LLtil$:
\[
\vecnorm{\LLtil^\dag\bb-\LL^\dag\bb}_{\LL}
=
\vecnorm{\left(\LLtil^\dag-\LL^\dag\right)\bb}_{\LL}
=
\vecnorm{\LL^{\dag /2} \bb}_{
	\left( \LL^{1/2} \left(\LLtil^\dag-\LL^\dag\right) \LL^{1/ 2} \right)^2}
\]
where because $\LL^{1/2} \left(\LLtil^\dag-\LL^\dag\right) \LL^{1/ 2}$ is a symmetric matrix, we have by
the definition of operator norm:
\begin{equation}
\leq
\vecnorm{ \LL^{1/2} \left(\LLtil^\dag-\LL^\dag\right) \LL^{1/ 2}}_{2}^2
\vecnorm{\LL^{\dag/2}\bb}_{2}
=
\norm{ \LL^{1/2} \left(\LLtil^\dag-\LL^\dag\right) \LL^{1/ 2}}_{2}^2
\vecnorm{\bb}_{\LL^{\dag}}.
\label{eq:EigBound}
\end{equation}
Composing both sides of $\LLtil\approx_\epsilon \LL$
by $\LL^{1/2}$ gives $\LL^{1/2}\LLtil\LL^{1/2}\approx_\epsilon \LL^{1/2}\LL\LL^{1/2}$, or upon rearranging:
\[
-\epsilon \II \preceq
\LL^{1/2}(\LLtil^\dag-\LL^\dag)\LL^{1/2}
\preceq \epsilon \II,
\]
or $\norm{\LL^{1/2}(\LLtil^\dag-\LL^\dag)\LL^{1/2}}_2^2
\leq \epsilon$.
Substituting this bound into Equation~\ref{eq:EigBound}
above then gives the result.
%
\end{enumerate}
Summing up these three cases as
in Equation~\ref{eq:ErrorSplit}
then gives the overall result
\[
\vecnorm{\xxtil-\LL^\dag\bb}_{\LL}
\le 10\epsilon k
.
\]
\end{proof}

\subsection{Dynamic Projection}
\label{subsec:DynamicProjection}


We next discuss the main ideas behind the dynamic algorithm that maintains an approximate projection in Lemma~\ref{lem:DynamicProjection} and then formally describe the implementation of this data-structure together with its running time guarantees. To this end, suppose we are given an approximate projection $\bbtil$ of $\proj{G}{S} \bb$ satisfying the following inequality
\begin{equation}
\label{eq:proj_approx_guarantee}
\vecnorm{\bbtil-\proj{G}{S}\bb}_{\LL^\dag}
\leq
\epsilon \vecnorm{\bb}_{\LL^\dag}
\end{equation}
The crucial idea is to exploit the fact that the right hand side of the above inequality $\vecnorm{\bb}_{\LL^\dag}$ corresponds to the square root of the energy needed by the electrical flow to route the demand vector $\bb$ (see Lemma~2.1 in~\cite{MillerP13}). Since we assume that our dynamic graph $G$ has bounded-degree, this energy is lower-bounded by

\[
	\vecnorm{\bb}_{L^{\dagger}} \geq \sqrt{\sum_{u \in V} \left(\frac{|\bb(u)|}{\dd(u)} \right)} = \Omega \left(\sqrt{\sum_{u \in V} |\bb(u)|}\right).
\]

Let $S'$ be the set of $\beta m$ vertices such that their corresponding coordinates in $\bb$ have the largest values. Without loss of generality, scale all the entries in the vector $\bb$ such that 
\begin{equation}
\label{eq:scaling}
\abs{\bb(u)} \ge 1, \quad \forall u \in S' \quad \text{ and } \quad
\abs{\bb(u)} \le 1, \quad \forall u\in V \setminus S'
\end{equation}

By definition of $S'$, after 
up to $(\beta m)/2$ operations in our data-structure, we know that at least half of the vertices in $S'$ will keep their corresponding $\bb$ values.
Thus the allowable error from right hand side of
Equation~(\ref{eq:proj_approx_guarantee}), $\vecnorm{\bb}_{\LL^{\dag}}$, is lower bounded by $\Omega(\sqrt{\beta m})$. Our goal is to control the error between the maintained approximate projection $\bbtil$ and the true projection $\proj{G}{S} \bb$. Our algorithm has two main components. First, it shows how to use a Laplacian solver that computes an approximate projection $\bbtil$ of $\proj{G}{S} \bb$ satisfying Equation~(\ref{eq:proj_approx_guarantee}) in nearly-linear time. Second, it gives a way to control the error of the projection $\proj{G}{S} \bb$ under terminal additions to $S$ with respect to the $\vecnorm{\cdot}_{\LL^{\dag}}$ norm.  

We now state the initialization lemma, whose is proof is deferred to Subsection~\ref{subsec:Errors}.

\begin{lemma}
\label{lem:proj_init} 
Given an unweighted graph $G=(V,E)$ with $n$ vertices and $m$ edges,  a demand vector $\bb\in \mathbb{R}^n$, set of vertices $S\subseteq V$ and an error parameter $\epsilon > 0$, there is an $\tilde{O}(m)$ time algorithm that computes a vector $\bbtil$ such that \[ \vecnorm{\bbtil-\proj{G}{S} \bb}_{\LL^\dag}\le \epsilon \vecnorm{\bb}_{\LL^\dag}. \]
\end{lemma}


To elaborate on the second component of the algorithm, consider the error induced on $\proj{G}{S} \bb$ when we add a vertex $u$ to some terminal set $S$
\[
\vecnorm{\proj{G}{S}\bb-\proj{G}{\newS}\bb}_{\LL^\dag}, \quad \text{ where } \newS = S \cup \{u\}. 
\]

In other words, the above expression gives the error when we simply keep the same vector $\bb$ under a terminal addition to the set $S$. We will show that over a certain number of such additions we can bound the compounded error by $O(\sqrt{\beta m})$. Since the latter is a lower bound on $\vecnorm{\bb}_{\LL^\dag}$, it follows that the maintained projection still provide good approximation guarantee. The following lemma, whose proof is deferred to Subsection~\ref{subsec:Stability}, bounds the error after one terminal addition.

\begin{restatable}{lemma}{DoNothing}
	\label{lem:DoNothing} 
Consider an unweighted undirected bounded-degree graph $G=(V,E)$, a demand vector $\bb \in \mathbb{R}^{n}$ and a parameter $\beta \in (0,1)$. Let $S \subseteq V$ with $|S| = O(\beta m)$ and assume that $\abs{\bb(u)} \geq 1$ for all $u \in S$, and $\abs{\bb(u)}$ for all $u \in V \setminus S$. For each edge in $G$, include its endpoints to $S$ independently, with probability at least $\beta$. Then, for any vertex $u \in V \setminus S$, with high probability
	\[
	\vecnorm{\proj{G}{S}\bb-\proj{G}{\newS}\bb}_{\LL^\dag}=\tilde{O}(\beta^{-5/2}), \quad \text{ where } \newS = S \cup \{u\}.
	\]
\end{restatable}

We now have all the necessary tools to give a dynamic data-structure that maintains an approximate projection, i.e., prove Lemma~\ref{lem:DynamicProjection}.
\begin{proof}[Proof of Lemma~\ref{lem:DynamicProjection}]
Given the input demand vector $\bb$, let $S'$ be the set of $\beta m$ vertices such that their corresponding coordinates in $\bb$ have the largest values.  Without loss of generality, scale $\bb$ according to Equation~(\ref{eq:scaling}). For each edge in $G$ include its endpoints to $S'$ independently, with probability at least $\beta$. 

\sloppy We next describe the implementation of the operations. For implementing procedure  $\textsc{Initialize}(G,S',\beta)$, we invoke Lemma~\ref{lem:proj_init} with $\epsilon/2$. Let $\bbtil$ be the output approximate projection satisfying Equation~(\ref{eq:proj_approx_guarantee}) with error parameter $\epsilon/2$ and set $S = S'$. As we will shortly see, operations $\textsc{Insert}$ and $\textsc{Delete}$ will be reduced to adding terminals to the set $S$. Thus we first discuss the implementation of the operation $\textsc{AddTerminal}$. To this end, let $u$ be a non-terminal vertex that we want to append to $S$. We set $S = S \cup \{ u \}$ and simply add an entry $\bbtil(u) = 0$ to $\bbtil$ while keeping the rest of the entries unaffected. To insert or delete an edge from the current graph, we simply run $\textsc{AddTerminal}$ procedure for the edge endpoints. 

Consider the operation $\textsc{Change}(u, \bb(u)', v, \bb(v)')$. We first invoke $\textsc{AddTerminal}$ on both $u$ and $v$ and then add $\bb(u)' - \bb(u)$ to $\bbtil(u)$ and $\bb(v)' - \bb(v)$ to $\bbtil(v)$. Finally, to implement $\textsc{Query}$ we simply return the approximate projection $\bbtil$.

We next analyze the correctness of our data-structure which solely depends on the correctness of \textsc{AddTerminal} and \textsc{Change} operations.  We will show that after $k$ many such operations, our maintained approximate projection $\bbtil$ satisfies
\begin{equation}
\label{eq: compundedError}
\vecnorm{\bbtil-\proj{G}{S}\bb}_{\LL^\dag} \leq
\tilde{O}(k \beta^{-5/2}) + (\epsilon/2) \vecnorm{\bb}_{\LL^\dag},
\end{equation}
where $S$ denotes the set of terminals after $k$ operations. Note that when $(k = 0)$, the above inequality holds by Lemma~\ref{lem:proj_init} that implements the initialization. Let us analyze the error when a single terminal is added to $S$, i.e., $(k=1)$. Then Lemma~\ref{lem:DoNothing} implies that
\[
	\vecnorm{\proj{G}{S}\bb-\proj{G}{\newS}\bb}_{\LL^\dag}=\tilde{O}(\beta^{-5/2}), \quad \text{ where } \newS = S \cup \{u\}.
\]

Combining these two guarantees and applying triangle inequality, we get that the error after one terminal addition is
\begin{align*}
	\vecnorm{\bbtil-\proj{G}{\newS}\bb}_{\LL^\dag} & = \vecnorm{\bbtil-\proj{G}{\newS}\bb + \proj{G}{S}\bb - \proj{G}{S}\bb}_{\LL^\dag} \\
	& \leq \vecnorm{\bbtil -\proj{G}{S}\bb}_{\LL^\dag} + \vecnorm{\proj{G}{S}\bb-\proj{G}{\newS}\bb}_{\LL^\dag} \\
	& \leq \tilde{O}(\beta^{-5/2}) + (\epsilon/2) \vecnorm{\bb}_{\LL^\dag}.
\end{align*}

Next, to analyze the changes in the values of $\bb$, let the updated $\bb'$ be the updated vector $\bb$. Let $\bbtil'$ be the updated $\bbtil$ and let $\proj{G}{S} \bb '$ be the updated $\proj{G}{S} \bb$. Using the fact that $u$ and $v$ are added to $S$, we get that
\[
		(\bbtil - \bbtil') = (\bb - \bb') = (\proj{G}{S} \bb - \proj{G}{S} \bb'),
\]
which in turn implies that
\[
		(\bbtil - \proj{G}{S} \bb) = (\bbtil' - \proj{G}{S} \bb'),
\]
and thus the error vector does not change.

We showed that after each operation, either the correct vector moves by at most $\tilde{O}(\beta^{-5/2})$ with respect to its $\vecnorm{\cdot}_{\LL^{\dag}}$ norm, or $(\bbtil - \proj{G}{S} \bb)$ does not change. Thus repeating the above argument $k$ times yields Equation~(\ref{eq: compundedError}). Setting $k = c_{\textrm{EN}} \cdot m^{1/2} \beta^{3} \epsilon (\poly \log n)^{-1})$ such that $\vecnorm{\bb}_{\LL^\dag} = c_{\textrm{EN}} \cdot \sqrt{\beta m}$, we get that
\[
	\vecnorm{\bbtil-\proj{G}{S}\bb}_{\LL^\dag} \leq (\epsilon/2) c_{\textrm{EN}} \sqrt{\beta m} + (\epsilon/2) \vecnorm{\bb}_{\LL^\dag} \leq \epsilon \vecnorm{\bb}_{\LL^\dag}.
\]

For the running time, Lemma~\ref{lem:proj_init} implies that the initialization cost is bounded by $\tilde{O}(m)$. Since the size of the maintained vector $\bbtil$ is bounded by $|S|$, it follows that the query cost is $O(\beta m)$. All the remaining operations can be implemented in $O(1)$ time.
\end{proof}

\subsection{Initialization of Approximate Projection Vector}
\label{subsec:Errors}

In this subsection we show how to compute an initial approximate projection vector of $\proj{G}{S} \bb$, i.e., we prove Lemma~\ref{lem:proj_init}.




\begin{proof}[Proof of Lemma~\ref{lem:proj_init}]

Define $F = V \setminus S$ and let $G'$ be an $n'$-vertex graph obtained from $G$ by contracting all vertices in $S$ within $G$ into a single vertex $s$ and keeping parallel edges. Let $\LL'$ denote the corresponding Laplacian matrix of $G'$ and consider the induced vertex mapping $\gamma : V \rightarrow V(G')$ with $\gamma(u) = u$ for $u \in F$ and $\gamma(u)  = s$ for $u \in S$. Let $\bb' \in \mathbb{R}^{n'}$ be the corresponding demand vector in $G'$  such that for $u \in V$, $\bb'(\gamma(u)) = \bb(u)$ if $\gamma(u) = u$ and $\bb'(\gamma(u)) = \sum_{v \in S} \bb(v)$ otherwise. For the given error parameter $\epsilon > 0$, we can invoke a black-box Laplacian solver to compute an approximate solution vector $\tilde{\vv}'$ to $\vv' = \LL'^{\dag}\bb'$ such that
\begin{equation}
\label{eq:voltage_error}
	\vecnorm{\tilde{\vv}' - \vv'}_{\LL'} \leq \epsilon \vecnorm{\vv'}_{\LL'}.
\end{equation}

Now, to lift back the vector $\tilde{\vv}'$ to $G$ we define new vectors $\tilde{\vv}$ and $\vv$ such that for all $u \in V$
\[
	\tilde{\vv} (u) : = \tilde{\vv}'(\gamma(u)) \text{ and } \vv(u) : = \vv'(\gamma(u)).
\]

Observe that for any edge $e=(u,v)$ in $G$, we have that
\[ (\tilde{\vv} (u)- \tilde{\vv} (v)) = (\tilde{\vv}' (u)- \tilde{\vv} (v)) \text{ and } (\vv (u)- \vv (v)) = (\vv' (u)- \vv' (v)).
\]

The above relations imply that the approximation guarantee from Equation~(\ref{eq:voltage_error}) can be written as follows
\begin{equation}
\label{eq: liftedVoltageError}
	\vecnorm{\tilde{\vv} - \vv}_{\LL} \leq \epsilon \vecnorm{\vv}_{\LL}.
\end{equation}

It is well known that if we interpret $G$ as a resistor network, $\vv$ represents the voltage vector on the vertices induced by the electrical flow that routes a certain demand in the network~(see e.g.,~\cite{doyle84}). Thus, by linearity of electrical flows and our construction, we can view $\vv$ as being the sum of the voltage vectors corresponding to the electrical flows that route $\bb(u)$ amount of flow to $S$, where the sum is over all $u \in F$. By Lemma~\ref{lem:min_energy_to_S}, for each $u \in F$, the demand corresponding to the electrical flow that send $\bb(u)$ units of flow to $S$ is given by 
\[
  \bb(u) (\boldone_u - \proj{G}{S} \boldone_u).
\]

Summing over all $u \in F$ we get the demand vector corresponding to $\vv$
\begin{align*}
	\sum_{u \in F} \bb(u) (\boldone_u - \proj{G}{S} \boldone_u) & = \left(\bb|_F - \proj{G}{S} \bb|_F \right) = \left( \bb|_F - \proj{G}{S} (\bb - \bb|_S) \right) \\
	&  = \left( \bb|_F - \proj{G}{S} \bb - \bb|_S\right),
\end{align*}
where $\bb|_U$ is the restriction of $\bb$ on the subset $U$ with $\bb|_U (u) = \bb(u)$ if $u \in U$, and $\bb|_U(u) = 0$ otherwise. Since we determined the demand vector corresponding to $\vv$, we get that
\begin{equation}
\label{eq: relationApprox}
	\LL \vv = \left( \bb|_F - \proj{G}{S} \bb - \bb|_S\right).
\end{equation}

Define the approximate project vector $\tilde{\bb}$ that our algorithm outputs using the following relation
\begin{equation}
\label{eq: relationExact}
	\bbtil := \left( \bb|_F - \LL \tilde{\vv} - \bb|_S \right),
\end{equation}

where $\tilde{\vv}$ is the extended voltage vector we defined above. To complete the proof of the lemma, it remains to bound the difference between $\bbtil$ and $\proj{G}{S} \bb$ with respect to the $\LL^\dag$ norm. To this end, using Equations~(\ref{eq: relationApprox}) and~(\ref{eq: relationExact}) we have
\begin{align*}
	\vecnorm{\bbtil - \proj{G}{S} \bb}_{\LL^{\dag}} & = \vecnorm{\bb|_F - \LL \tilde{\vv} - \bb|_S - \left( \bb|_F - \LL \vv - \bb|_S\right)}_{\LL^{\dag}} \\
	& = \vecnorm{\LL \tilde{\vv} - \LL \vv}_{\LL^{\dag}} = \vecnorm{ \tilde{\vv} - \vv}_{\LL}.
\end{align*} 
Using the approximate guarantee in Equation~(\ref{eq: liftedVoltageError}) we have that
\[
  \vecnorm{\tilde{\vv} - \vv}_{\LL} \leq \epsilon \vecnorm{\vv}_{\LL} = \epsilon \vecnorm{\vv'}_{\LL'} = \epsilon \vecnorm{\bb'}_{\LL'^{\dag}} \leq \epsilon  \vecnorm{\bb}_{\LL^\dag},
\]
where the last inequality follows from the fact that the minimum energy needed to route $\bb$ becomes smaller when contracting vertices.
\end{proof}

\subsection{Stability of Projected Vectors}
\label{subsec:Stability}

In this subsection we prove our core structural observation, namely that the the projection vectors remain stable under the addition of a new terminal vertex, as stated in Lemma~\ref{lem:DoNothing}.

We start by considering the projection vector $\proj{G}{S} \boldone_u$, where $u \in F = V \setminus S$. Recall that for $s \in S$, Lemma~\ref{fac:StopVertexDistribution} gives that $[\proj{G}{S} \boldone_u](s)$ is the probability that the random walk that starts at $u$ hits the set $S$ at the vertex $s$. Equivalently, we can view the probability of this walk as routing a fraction of $\boldone_u$ from $u$ to $s$. Now, consider the operation of adding a non-terminal $u \in F$ to $S$, i.e., $\newS = S \cup \{u\}$. We observe that the fraction of $\boldone_u$ that we routed to some vertex $v$ in $S$ might have used the vertex $u \in F$. This indicates that this this fraction should have stopped at $u$, instead of going to other vertices in $S$, which in turn implies that the old projection vector $\proj{G}{S} \boldone_u$ is not valid anymore. We will later show that this change is tightly related to the load that random walks from other vertices in $F$ put on the new terminal vertex $u$. In the following we focus on showing a provable bound on this load, which in turn will allow us to control the error for the maintained projection vector.

Concretely, for each vertex $u \in F$, we want to bound the load incurred by the random walks of the other vertices $v \in F \setminus u$ to the set $S$. For the purposes of our proof, it will be useful to introduce some random variable. For $v \in F$, let $Z_v(S)$ be the set of vertices visited in a random walk starting at $v$ and ending at some vertex in $S$. For $t \geq 0$, let $X_v(t)$ be the set of vertices visited in a random walk starting at $v \in F$ after $K$ steps. For a demand vector $\bb$ and any two vertices $u,v \in F$, the contribution of $v$ to the load of $u$, denoted by $Y_v(u)$, is defined as follows
\[
	Y_v(u) = \bb(v) \cdot \boldone_{(u \in Z_v(S))}.
\] 

The \emph{load} of a vertex $u \in F$, denoted by $N_u$, is obtained by summing the contributions over all vertices in $F$, i.e.,
\[
	N_u = \sum_{v \in F} Y_v(u).
\]

The following lemma gives a bound on the expected load of every non-terminal vertex.

\begin{lemma} 
\label{lem:VertexLoad}
For a parameter $\beta \in (0,1)$ and every vertex $u \in F$ we have that $\expec{}{N_u} = \tilde{O}(\beta^{-2})$. 
\end{lemma}

For proving the above lemma it will be useful to rewrite the load quantity. To this end, recall that in the proof of Theorem~\ref{thm:RandomWalkProperties} we have shown that any random walk that start at a vertex $v$ of length $\ell = \tilde{O}(\beta^{-2})$ hits a vertex in the terminal set $K$ with probability at least $1 - 1/n^{c}$, for some large constant $c$. Note that by construction of $S$ in Lemma~\ref{lem:DoNothing}, the exact same argument applies to the set $S$. Thus, instead of terminating the random walks once they hit $S$, we can run all the walks from the vertices in $F$ up to $\ell$ steps. The latter together with the assumption $\bb(v) \leq 1$ for all $v \in F$~(provided by Lemma~\ref{lem:DoNothing}) give that
\begin{align*}
\label{eq: expectedloadU}
	\expec{}{N_u} & = \sum_{v \in F} \bb(v) \cdot \prob{v}{v \in Z_v(S)} \\
	    & \leq \sum_{v \in F} \left( \prob{v}{\text{walk $w$ from $v$ uses $u$ in its first $\ell$ steps}} + \prob{v}{|w| > \ell} \right) \\
	    & \leq \sum_{v \in F} \left(\sum_{0 \leq t \leq \ell}  \prob{v}{u \in X_v(t)} + 1/n^{c}\right) \\
	    & \leq \sum_{0 \leq t \leq \ell} \left( \sum_{v \in V} \deg(v)\cdot \prob{v}{u \in X_v(t)} \right) + o(1). \numberthis
\end{align*}

It turns out that that the term contained in the brackets of Equation~(\ref{eq: expectedloadU}) equals $\deg(u)$. Formally, we have the following lemma.

\begin{lemma}
\label{lem:LULZ}
Let $G$ be an undirected unweighted graph. For any vertex $u \in V$ and any length $t \geq 0$, we have \normalfont
\[
\sum_{v \in V} \deg\left(v\right) \cdot \prob{}{u \in X_v(t)}
= \deg\left( u \right).
\]
\end{lemma}

To prove this, we use the reversibility of random walks, along with the
fact that the total probability over all edges of a walk starting at
$e$ is $1$ at any time. Below we verify this fact in a more principled manner.

\begin{proof}[Proof of Lemma~\ref{lem:LULZ}]
The proof is by induction on the length of the walks $K$. When $t = 0$, we have
\[
\prob {}{u \in X_v(0)}
=
\begin{cases}
1 & \text{if $u = v$},\\
0 & \text{otherwise},
\end{cases}
\]
which gives a total of $\deg(u)$.

For the inductive case, assume the result is true for $t - 1$.
The probability of a walk reaching $u$ after $K$ steps can then be written in terms of its location at time $t - 1$, the neighbor $x$ of $u$, as well as the probability of reaching there:
\[
\prob{}{u \in X_v(t)}
=
\sum_{x: (u, x) \in E}
\frac{1}{\deg\left( x \right)}
\prob{} {x \in X_v(t-1)}
.
\]
Substituting this into the summation to get
\[
\sum_{v \in V} \deg\left(v\right) \cdot \prob{}{u \in X_v(t)}
=
\sum_{v \in V} \deg\left(v\right)
\sum_{x: (u, x) \in E}
\frac{1}{\deg\left( x \right)}
\prob{} {x \in X_v(t-1)},
\]
which upon rearranging of the two summations gives:
\[
\sum_{x: (u,x) \in E}
\frac{1}{\deg\left( x \right)}
\left( 
\sum_{v \in V} \deg\left(v\right) \cdot
\prob{} {x \in X_v(t-1)}\right).
\]
By the inductive hypothesis, the term contained in the bracket
is precisely $\deg(x)$, which cancels with the division,
and leaves us with $\deg(u)$.
Thus the inductive hypothesis holds for $K$ as well.
\end{proof}

Plugging Lemma~\ref{lem:LULZ} in Equation~(\ref{eq: expectedloadU}), along with the fact that by assumption $G$ has bounded degree we get that
\[
	\expec{}{N_u} \leq \deg(u) \cdot \ell = \tilde{O}(\beta^{-2}),
\]
thus proving Lemma~\ref{lem:VertexLoad}.

We now have all the tools to prove Lemma \ref{lem:DoNothing}.

\begin{proof}[Proof of Lemma \ref{lem:DoNothing}]
Recall that $\newS = S \cup \{u\}$, where $u$ is vertex in $F = V \setminus S$. We want to obtain a bound  on the difference $(\proj{G}{S} \bb - \proj{G}{\newS} \bb)$ with respect to the $\LL^{\dag}$ norm. We distinguish the following types of entries of the difference vector: (1) newly  added terminal $u$, (2) the old terminals $S$ and (3) the remaining non-terminal vertices $F \setminus \{u\}$. Note that $\proj{G}{S} \bb$ and $\proj{G}{\newS} \bb$ are not $n$-dimensional vectors, so we assume that all missing entries are appended with zeros. This also allows us to compute the $\LL^{\dag}$ norm.

In what follows, we will repeatedly make of the following relation by Lemma~\ref{fac:StopVertexDistribution} for vertices $u \in F$ and $v \in S$
\[
	\proj{G}{S}\boldone_u (v) = \sum_{\substack{u_0=u,\ldots,u_{k-1}\in F, \\ u_k=v}} \frac{\prod_{i=0}^{k-1} \ww(u_i,u_{i+1})}{\prod_{i=1}^{k-1} \dd(u_i)}.
\]

For the type (1) entry, i.e., newly added terminal $u$, using the definition of the load $N_u$, we get:
\begin{align*}
\label{eq: type1}
	[ \proj{G}{S} \bb - \proj{G}{\newS} \bb ] (u) & = -\sum_{\substack{u_0=u,\ldots,u_{k-1}\in F \setminus \{u\}, \\ u_k=u}} \bb(u_0) \cdot \frac{\prod_{i=0}^{k-1} \ww(u_i,u_{i+1})}{\prod_{i=1}^{k-1} \dd(u_i)} \\
	& = - \sum_{u_0 \in F} \expec{}{Y_{u_0}(u)} = - \expec{}{N_u}. \numberthis
\end{align*}

Note that for type (3) entries, i.e., the remaining non-terminals $v \in F \setminus \{u\}$, we have that
\begin{align*}
\label{eq: type3}
	[ \proj{G}{S}\bb - \proj{G}{\newS} \bb ] (v) = 0. \numberthis
\end{align*}

Finally, for type (2) entries, i.e., old terminals $v \in S$, similarly to the type (1) entries we get
\begin{align*} \label{eq: type2}
[ \proj{G}{S}\bb  - \proj{G}{\newS} \bb ] (v) & = \sum_{\substack{u_0=u,\ldots,u_{k-1}\in F, \\ u_k=v}} \bb(u_0) \cdot \frac{\prod_{i=0}^{k-1} \ww(u_i,u_{i+1})}{\prod_{i=1}^{k-1} \dd(u_i)} \\
& - \sum_{\substack{u_0=u,\ldots,u_{k-1}\in F \setminus \{u\}, \\ u_k=v}} \bb(u_0) \cdot \frac{\prod_{i=0}^{k-1} \ww(u_i,u_{i+1})}{\prod_{i=1}^{k-1} \dd(u_i)} \\
& = \sum_{\substack{u_0=u,\ldots,u_{k} = u, \\ u_{k+1},\ldots,u_{r-1} \in F, u_r = v}} \bb(u_0) \cdot \frac{\prod_{i=0}^{r-1} \ww(u_i,u_{i+1})}{\prod_{i=1}^{r-1} \dd(u_i)} \\
& = \sum_{\substack{u_0=u,\ldots,u_{k-1}\in F \setminus \{u\}, \\ u_k=v}} \bb(u_0) \cdot \frac{\prod_{i=0}^{k-1} \ww(u_i,u_{i+1})}{\prod_{i=1}^{k-1} \dd(u_i)} \\ 
& \quad \quad \sum_{\substack{u_0=u,\ldots,u_{k-1}\in F, \\ u_k=v}} \frac{\prod_{i=0}^{k-1} \ww(u_i,u_{i+1})}{\prod_{i=1}^{k-1} \dd(u_i)} \\
& = \expec{}{N_u} \cdot [\proj{G}{S} \boldone_u](v). \numberthis
\end{align*}

Bringing together Equations~(\ref{eq: type1}),~(\ref{eq: type3}) and~(\ref{eq: type2}) we get that
\[
	[ \proj{G}{S}\bb  - \proj{G}{\newS} \bb ] = -(\expec{}{N_u} (\boldone_u - \proj{G}{S} \boldone_u)).
\]

The right-hand side of the equation can be interpreted as routing $\expec{}{N_u}$ unit of flows from $u$ to $S$. Thus, to measure the error, we simply need to upper-bound the square root of the energy need to route $\expec{}{N_u}$ amount of flow from $u$ to $S$~(Lemma~\ref{lem:min_energy_to_S}),i.e.,
\[
	\vecnorm{\expec{}{N_u} (\boldone_u - \proj{G}{S} \boldone_u)}_{\LL^\dag}.
\] 

By the simplifying assumption that $G$ is connected and the fact that each endpoint of an edge in $E$ is added to $S$ independently, with probability at least $\beta$, it is easy to show that with high probability, there exists a path $p(v,S)$ from $u$ to $S$ that uses at most $O(\beta^{-1} \log n)$ edges. Hence, if we route $\expec{}{N_u}$ units of flow from $u$ to $S$ along the path $p(v,S)$, the energy of such a flow is upper-bounded by
\[
	(\expec{}{N_u})^2 \cdot \tilde{O}(\beta^{-1}) = \tilde{O}((\expec{}{N_u})^2 \beta^{-1}).
\]

Using the latter we get that 
\begin{align*}
\vecnorm{\proj{G}{S}\bb  - \proj{G}{\newS} \bb}_{\LL^\dag} & = \vecnorm{\expec{}{N_u} (\boldone_u - \proj{G}{S} \boldone_u)}_{\LL^\dag} \\ & \leq \tilde{O}\left(\sqrt{(\expec{}{N_u})^2 \beta^{-1}}\right) \\ & = \tilde{O}(\expec{}{N_u} \beta^{-1/2}) \\
& = \tilde{O}(\beta^{-5/2}),
\end{align*}
where the last inequality uses the fact that $\expec{}{N_u} = \tilde{O}(\beta^{-2})$ by Lemma~\ref{lem:VertexLoad}. This completes the proof the lemma.
\end{proof}

\section{Sampling Weights of a Random Walk}
\label{sec:approx_sample}

In this section, we show that given a random walk $w$ of length $\ell$ in a weighted $G$ with polynomially bounded weights, we can efficiently sample an approximation to $s(w) = \sum_{i=1}^{\ell} (1/\ww(w_{i-1},w_i))$. Concretely, we prove the following lemma from Section~\ref{sec:DynamicSCWeighted}.

\begin{lemma}[Restatement of Lemma~\ref{lem:approx_sample}]
Let $G=(V,E,\ww)$ be a undirected, weighted graph with $\ww(e) = [1,n^c]$ for each $e \in E$, where $c$ is a positive constant. For any finite random walk $w$ of length $\ell$ with $\ell \leq n^{d}$, where $d$ is a positive constant, let $s(w)$ be the sum of the inverse of its edge weights, i.e.,
\[
    s(w) = \sum_{i=1}^{\ell} \frac{1}{\ww(w_{i-1},w_i)}.
\] 
Moreover, for any $u,v \in V$, let
\[
\pmf{\ell}{u}{v}
\] 
be the probability mass function of $s(w)$ conditioning on \emph{(1)} $w$ being a random walk that starts at $u$ and ends at $v$, and \emph{(2)} length of the walk $\ell(w)$ is $\ell$ in $G$.  Then, for any pair $u,v \in V$, there exists an algorithm that that samples from $\pmf{\ell}{u}{v}$ and outputs a sampled $s(w)$ up to a $(1 + \epsilon)$ multiplicative error in $\tilde{O}(n^{3} \epsilon^{-2})$ time.
\end{lemma}

To prove the above lemma, we employ a doubling technique. Specifically, for any pair of vertices $u,v \in V$, and a random walk $w$ of length $\ell$ that starts at $u$ and ends at $v$, it is easy to see that
\[
	\pmf{\ell}{u}{v} = \sum_{y \in V} \left(\pmf{\ell/2}{u}{y} * \pmf{\ell/2}{y}{v} \right),
\]
where $*$ denotes the convolution between two probability mass functions. However, one challenge here is that we cannot afford dealing with exact representations of probability mass functions as this would be computationally expensive. Instead, we introduce an \emph{approximate} representation of such functions, and then give an algorithm that allows computing the convolution between such approximate representations. Before proceeding further, note that we can scale down the edge weights so that $\ww(e) \leq 1$, and thus $1/\ww(e) \geq 1$ for every $e \in E$. In addition, we remark that $\ww$ does not need to be integral. 


Let us introduce a compact way to represent any given probability mass function approximately $f$. The main idea is to `move' each number in the support of $f$ by $(1 + \epsilon)$, which in turn results in a $(1+ \epsilon)$ approximation of the sampled value for $f$. Formally, let $f$ be a probability mass function such that $f(x)=0$, for each $x\not\in \{0,\ldots,n^{c}\}$, where $c$ is a positive constant. For $j \geq 1$, let $I_k^{j}$ be the interval $[(1+\epsilon)^{k}, (1+\epsilon)^{k+j})$ for $k \in \{0, \ldots, L \}$ where $L=O\left((c+d)\epsilon^{-1}\log n\right)$. Note that the upper bound $L$ is chosen in such a way that $\cup_{k} I_k^{1}$ covers the range of $\pmf{\ell}{u}{\ell}$ for every possible triplet $(u,v,\ell)$. For $j \geq 1$ and $\epsilon > 0$, we say that $g$ is an $(\epsilon,j)$-approximation of a probability mass function $f$ iff there exists a matrix $\HH$ satisfying the following properties:


\begin{enumerate}[noitemsep, label=(\alph*), leftmargin=*, itemsep=0.4ex, before={\everymath{\displaystyle}}]%
  \item $\sum_{k=0}^{L} \HH_{x,k}=f(x),~\forall x \in \{0,\ldots,n^{c}\}$, \label{cond:1}
  \item $\sum_{x=0}^{n^{c}} \HH_{x,k}=g(k),~\forall k \in \{0,\ldots, L\}$,\label{cond:2}
  \item $\HH_{x,k}=0,~\forall x \not \in I_k^j$.\label{cond:3}
\end{enumerate}

Note that an $(\epsilon, j)$-approximation of $f$ is also an $\left(\epsilon, (j+1)\right)$-approximation of $f$. Moreover, observe that the intervals $\{I_k^1\}_{k \in \{0\} \cup L}$ are disjoint for different $k$ but $I_k^{j}$ overlaps with $I_{k'}^j$ whenever $j \geq 2$ and $|k-k'|< j$.

\begin{algorithm2e}[t]
\caption{\textsc{Convolute}$(g^{(1)}, g^{(2)}, \epsilon, j)$}
\label{alg:multiply}
\Input{Two $(\epsilon, j)$-approximations $g^{(1)}$ and $g^{(2)}$  of two probability mass functions $f^{(1)}$ and $f^{(2)}$}
\Output{An $\left(\epsilon, (j+1)\right)$-approximation $g := g^{(1)} * g^{(2)}$ of $f := f^{(1)} * f^{(2)}$}
Set $g \gets \boldzero$ \;
\For{$(k_1, k_2) \in \{0,\ldots,L\}^2$}
{
 Find $k_3$ such that $(1+\epsilon)^{k_1}+(1+\epsilon)^{k_2} \in I_{k_3}^1$ \; 
 Set $g(k_3) \gets g({k_3}) + g^{(1)}({k_1}) \cdot g^{(2)}({k_2})$ \;
}
\Return $g$
\end{algorithm2e}

Next we show how to compute the convolution of two probability mass functions under their approximate representations. Let and $g^{(1)}$ and $g^{(2)}$ be ($\epsilon, j$)-approximations of probability mass functions $f^{(1)}$ and $f^{(2)}$, respectively. Now consider two intervals $I_{k_1}^j$ and $I_{k_2}^j$. Without loss of generality, assume that $k_1 \le k_2$. If $x \in I_{k_1}^j$ and $y \in I_{k_2}^j$, then 
\[
x+y\in I' := [\textrm{le}, \textrm{ri}), \text{ where}
\]
\[ \textrm{le} :=\left((1+\epsilon)^{k_1}+(1+\epsilon)^{k_2}\right), \quad \textrm{ri} := \left((1+\epsilon)^{k_1+j}+(1+\epsilon)^{k_2+j}\right).
\]

Furthermore, let $I_{k_3}^{1}$ be an interval such that $\textrm{le} \in I_{k_3}^{1}$. The latter implies that $(1+\epsilon)^{k_3}\le \textrm{le} < (1+\epsilon)^{k_3+1}$. Since $\textrm{ri} = \textrm{le} \cdot (1+\epsilon)^j$, it follows that $\textrm{ri} < (1+\epsilon)^{k_3 + j + 1}$. Bringing together the above bounds we get that $(1+\epsilon)^{k_3} \leq \text{le} < (1+\epsilon)^{k_3 + j + 1}$, i.e.,  $I' \subseteq I_{k_3}^{j+1}$. Since $k_3$ depends on $k_1,k_2,$ and $j$ we sometimes write $k_3(k_1,k_2,j)$ instead of $k_3$. 

Since the above approach gives us a way to combine two different intervals, it is now straightforward to compute the convolution between two probability mass functions. This task is performed in the standard way and we review its implementation details in Algorithm~\ref{alg:multiply} for the sake of completeness.

\begin{lemma}
\label{lem:multiply}
Let $j \geq 1$ and $\epsilon > 0$ by two parameters. Given any two $(\epsilon,j)$-approximations $g^{(1)}$ and $g^{(2)}$ of probability mass functions $f^{(1)}$ and $f^{(2)},$ \textsc{Convolute}$(g^{(1)},g^{(2)}, \epsilon,j)$~\emph{(}Algorithm~\ref{alg:multiply}\emph{)} computes in $\tilde{O}(\epsilon^{2})$ time an $\left(\epsilon, (j+1) \right)$-approximation $g := g^{(1)} * g^{(2)}$ of the convolution $f : = f^{(1)} * f^{(2)}$. 
\end{lemma}
\begin{proof}
We first show the correctness. Since $g^{(1)}$ and $g^{(2)}$ are $(\epsilon,j)$-approximations to $f^{(1)}$ and $f^{(2)}$ by assumption of the lemma, we know that there exists matrices $\HH^{(1)}$ and $\HH^{(2)}$ satisfying properties \ref{cond:1}, \ref{cond:2}  and \ref{cond:3}. To show that the output $g$ is correct we need to construct a matrix $\HH$ that satisfies each of these properties. By construction of the algorithm, the new matrix $\HH$ is defined as follows:
\[
	\HH_{z,k_3} := \sum_{\substack{x \in I_{k_1}^{j}, x \in I_{k_2}^{j} \\ x+y = z, k_3 = k_3(k_1,k_2,j)}} \HH^{(1)}_{x,k_1} \cdot \HH^{(2)}_{y,k_2},\quad z \in \{0,\ldots,n^{c}\},~ k_{3} \in \{0,\ldots, L\}.
\]
We start by showing property \ref{cond:1} for $\HH$. Concretely, for any $z \in \{0,\ldots,n^{c}\}$ we get that
\begin{align*}
\sum_{k_3=0}^L \HH_{z,k_3} & = \sum_{k_3 = 0 }^L \sum_{\substack{x \in I_{k_1}^{j}, x \in I_{k_2}^{j} \\ x+y = z, k_3 = k_3(k_1,k_2,j)}} \HH^{(1)}_{x,k_1} \cdot \HH^{(2)}_{y,k_2}\\
& = \sum_{x \in I_{k_1}^j, y \in I_{k_2}^j, x+y=z} \HH^{(1)}_{x,k_1} \cdot \HH^{(2)}_{y,k_2}\\
& = \sum_{x+y=z} \left(\sum_{x\in I_{k_1}^j}\HH^{(1)}_{x,k_1}\right)\left(\sum_{y\in I_{k_2}^j}\HH^{(2)}_{y,k_2}\right)\\
& =\sum_{x+y=z} f^{(1)}(x) \cdot f^{(2)}(y)\\
& = \left(f^{(1)} * f^{(2)}\right)(z) = f(z).
\end{align*}
Next, $\HH$ satisfies property \ref{cond:2} since for any $k_3 \in \{0,\ldots,L\}$ we get that
\begin{align*}
\sum_{z = 0}^{n^{c} }\HH_{z,k_3} & =  \sum_{z = 0}^{n^{c}} \sum_{\substack{x \in I_{k_1}^{j}, x \in I_{k_2}^{j} \\ x+y = z, k_3 = k_3(k_1,k_2,j)}} \HH^{(1)}_{x,k_1} \cdot \HH^{(2)}_{y,k_2} \\
& = \sum_{x \in I_{k_1}^j, y \in d_{k_2}^j, k_3=k_3(k_1,k_2,j)} \HH^{(1)}_{x,k_1} \cdot \HH^{(2)}_{y,k_2}\\
& =  \sum_{k_3=k_3(k_1,k_2,j)} \left(\sum_{x\in I_{k_1}^j}\HH^{(1)}_{x,k_1}\right)\left(\sum_{y\in I_{k_2}^j}\HH^{(2)}_{y,k_2}\right)\\
& = \sum_{k_3=k_3(k_1,k_2,j)} g^{(1)}_{k_1} \cdot g^{(2)}_{k_2}\\
& = \left(g^{(1)} * g^{(2)}\right) (k_3) = g (k_3).
\end{align*} 

where the penultimate equality follows by Algorithm~\ref{alg:multiply}.

Finally, for every $x \not \in I_k^{j}$, we have that $\HH_{x,k} = 0$, i.e., property~\ref{cond:3} holds for $\HH$. The latter holds since $x \in I_{k_1}^j$ and $y \in I_{k_2}^j$ gives that $x+y \in I_{k_3(k_1,k_2,j)}^{j+1}$. Thus, by definition of approximate probability mass function, it follows that $g = g^{(1)} * g^{(2)}$ is an $\left(\epsilon, (j+1) \right)$-approximation of $f = f^{(1)} * f^{(2)}$. 

For the running time first recall that $L=O\left((c+d) \epsilon^{-1} \log n \right) = \tilde{O}(\epsilon^{-1})$. Since the cost for implementing \textsc{Convolute} is bounded by $\tilde{O}(L^{2})$, it follows that we can implement this procedure in $\tilde{O}(\epsilon^{-2})$ time.
\end{proof}

The last ingredient we need is to show that given a family of probability mass functions, and their corresponding approximations, choosing one of these functions according to some probability distribution yields a random approximation in the natural way. Specifically, for an index set $Q$, let $\{f^{(q)}\}_{q \in Q}$ be a set of probability mass functions. Let $\hat{q}$ be be a random variable~(independent from $\{f^{(q)}\}_{q \in Q}$) such that for every $q \in Q$, $\displaystyle \mbox{Pr}[\hat{q}=q] = \pp(q)$, and $\sum_{q \in Q} \pp(q) = 1$. Furthermore, define \[ f := f^{(\hat{q})} = \sum_{q \in Q} \pp(q)f^{(q)} \]. 

\begin{lemma}
\label{lem:addition}
Suppose $g^{(q)}$ is an $(\epsilon, j)$-approximation of the probability mass function $f^{q}$, for all $q\in Q$. Let $f$ be the probability mass function as defined above. Then \[ g := \sum_{q \in Q} \pp(q) g^{(q)} \] is an $(\epsilon, j)$-approximation of $f$.
\end{lemma}
\begin{proof} By definition of an $(\epsilon,j)$-approximation, we know that there exist matrices $\{\HH^{(q)}\}_{q\in Q}$ for $\{g^{(q)}\}_{ q \in Q}$ satisfying properties~\ref{cond:1},~\ref{cond:2} and ~\ref{cond:3}. We need to show that for $g$ as defined in the lemma, there exist a suitable matrix $\HH$ that satisifes each of these properties. To this end, define $\HH$ as follows 
\[
	\HH_{x,k} := \sum_{q\in Q} \pp(q) \HH^{(q)}_{x,k}, \quad x \in \{0,\ldots, n^{c}\},~k \in \{0,\ldots,L\}.
\]
We start by showing property~\ref{cond:1}. Concretely, for any $z \in \{0,\ldots,n^{c}\}$ we get that

\begin{align*}
\sum_{k=0}^L \HH_{x,k} = \sum_{k=0}^L \sum_{q \in Q} \pp(q) \HH^{(q)}_{x,k} =\sum_{q \in Q} \pp(q) \sum_{k=0}^L \HH^{(q)}_{x,k} =\sum_{q \in Q} \pp(q) f^{(q)}(x) = f(x).
\end{align*}

Next, $\HH$ satisfies property~\ref{cond:2} since for any $k \in \{0,\ldots, L\}$ we get that
 
\begin{align*}
\sum_{x=0}^{n^{c}} \HH_{x,k} = \sum_{x=0}^{n^{c}} \sum_{q \in Q} \pp(q) \HH^{(q)}_{x,k} = \sum_{q \in Q} \pp(q) \sum_{x=0}^{n^{c}} \HH^{(q)}_{x,k} = \sum_{q \in Q} \pp(q) g^{(q)}(k) = & g(k).
\end{align*}

Finally, for every $x\not\in I_k^j$, we have that $\HH_{x,k}=0$, i.e., property~\ref{cond:3} is satisfied for $\HH$. The latter holds since for all $x \not \in I_{k}^{j}$ we have that $\HH^{(q)}_{x,k}=0$ and thus $\HH_{x,k}=\sum_{q \in Q} \pp(q) \HH^{(q)}_{x,k} = 0$. As a result we conclude that $g$ is an $(\epsilon, j)$-approximation of $f$ with matrix $\HH$ satisfying all the required properties.
\end{proof}

We now describe how to compute a probability distribution that will in turn allow us to sample approximately from $\pmf{\ell}{u}{v}$. At a high level we accomplish this task by employing the ``doubling technique'' together with the approximate representations of the probability mass functions and their convolution. As an input, the algorithm receives a weighted graph $G$ with polynomially bounded weights, a length parameter $\ell \geq 1$, an error parameter $\epsilon > 0$ and two vertices $u,v \in V$. The procedure computes and outputs a vector $\left(j_{u,v},\pmfg{\ell}{u}{v}, p_{\ell}^{u,v}\right)$, where $j_{u,v} \geq 1$ is a precision parameter, $\pmfg{\ell}{u}{v}$ is an $(\epsilon,j_{u,v})$-approximation of $\pmf{\ell}{u}{v}$, and $p_{\ell}^{u,v} = \prob{u}{w_{\ell} = v}$ is the probability that the random walk $w$ that originates at $u$ hits $v$ after $\ell$ steps. 

If $(\ell=1)$, then there are two possibilities depending on whether $(u,v) \in E$ or not. If the former holds, then the algorithms simply returns $(1,\boldzero,0)$ as it is not possible to reach $v$ after performing one step of the random walk from $v$. Otherwise, we simply return $(1,\pmfg{1}{u}{v}, p_{\ell}^{u,v})$, where $\pmfg{1}{u}{v}(\frac{1}{\ww(u,v)}) = 1$ and $p_{\ell}^{u,v} = \frac{\ww(u,v)}{\dd(v)}$. 

However, if $(\ell > 1)$, then it first halves $\ell$ into two parts $\ell' = \lfloor \ell/2 \rfloor$ and $\ell'' = \lceil \ell/2 \rceil$. Next, for each $y \in V$ it recursively calls itself with input parameters $(G,u,y,\ell',\epsilon)$ and $(G,y,v,\ell'',\epsilon)$. The outputs from these two calls are then combined using the convolution manipulations described above to produce the final output. Exact details for implementing this procedure are summarized in Algorithm~\ref{alg:ComputeDistribution}. The following lemma proves the correctness and the running time of the algorithm.


\begin{algorithm2e}[t]
\caption{\textsc{ComputeDistrib}$(G,u,v,\ell,\epsilon)$}
\label{alg:ComputeDistribution}
\Input{Weighted graph $G=(V,E,\ww)$, with $\ww(e) = [1, n^{c}]$ for each $e \in E$ and $c > 0$, two vertices $u,v \in V$, a length parameter $\ell \in [1,n^{d}]$ and an error parameter $\epsilon > 0$}
\Output{A vector $(j_{u,v}, \pmfg{\ell}{u}{v}, p^{u,v}_{\ell})$, where $j_{u,v} \geq 1$ is a precision parameter, $\pmfg{\ell}{u}{v}$ is an $(\epsilon,j_{u,v})$-approximation of $\pmf{\ell}{u}{v}$, and $p^{u,v}_{\ell}$ is the probability that the random walk $w$ that starts at $u$ hits $v$ after $\ell$ steps}

\If{$(\ell=1)$}
{
 If $(u,v) \not\in E$, return $(1,\boldzero, 0)$ \;
 If $(u,v) \in E$, return $(1,\pmfg{\ell}{u}{v}, p^{u,v}_{\ell})$, where $g_1^{u,v}(\frac{1}{\ww(u,v)}) \gets 1$ and $p_{\ell}^{u,v} \gets \frac{\ww(u,v)}{\dd(v)}$ \;
}
\If{$(\ell \geq 2)$}
{
Set $\ell' \gets \lfloor \ell/2 \rfloor$ and $\ell'' \gets \lceil \ell/2 \rceil$ \;
\For{every $y \in V$}
{
 Invoke \textsc{ComputeDistrib}$(G,u,y,\ell',\epsilon)$ and \textsc{ComputeDistrib}$(G,y,v,\ell'',\epsilon)$ \;
 Let $(j_{u,y},\pmfg{\ell'}{u}{y},p_{\ell'}^{u,y})$ and $(j_{y,v},\pmfg{\ell''}{y}{v},p_{\ell''}^{y,v})$ be the corresponding outputs \;
}
 
 Set $\pmfg{\ell}{u}{v} \gets \sum_{y \in V} p_{\ell'}^{u,y} p_{\ell''}^{y,v} \cdot  \left( \pmfg{\ell'}{u}{y} * \pmfg{\ell''}{y}{v} \right)$ \;
 Return $\left((\max_{y \in V}\max(j_{u,y},j_{y,v}))+1, \pmfg{\ell}{u}{v}, \sum_{y\in V} p_{\ell'}^{u,y} \cdot p_{\ell''}^{y,v}\right)$ \;
}
\end{algorithm2e}

\begin{lemma}
\label{lem:correctness_alg_computedistribution}
Given a weighted graph $G=(V,E,\ww)$ with $\ww(e) \in [1,n^{c}]$ for each $e \in E$ and $c>0$, two vertices $u,v \in V$, a length parameter $\ell \in [1,n^{d}]$ and an error parameter $\epsilon > 0$, \textsc{ComputeDistrib}$(G,u,v,\ell,\epsilon)$~(Algorithm~\ref{alg:ComputeDistribution}) correctly computes a vector $(j_{u,v}, \pmfg{\ell}{u}{v}, p_{\ell}^{u,v})$ in $\tilde{O}(n^{3} \epsilon^{-2})$ time, where $\pmfg{\ell}{u}{v}$ is an $(\epsilon,j_{u,v})$-approximation to $\pmf{\ell}{u}{v}$ and $p_{\ell}^{u,v}$ is the probability that the random walk $w$ the starts at $u$ hits $v$ after $\ell$ steps. Moreover, the output $j_{u,v}$ cannot exceed $O(\log n)$.
\end{lemma}
\begin{proof}
We first prove that the third coordinate of the output vector equals $\prob{u}{w_{\ell} = v}$. We proceed by induction on the length of the walk $\ell$. If $(\ell=1)$, it is easy to check that the condition holds by construction of the algorithm. Next assume $(\ell \ge 2)$ and note that $(\ell'<\ell)$ and $(\ell'' < \ell)$. Applying induction hypothesis on each recursion call, we know $p^{u,y}_{\ell'}$ is $\prob{u}{w_{\ell}=y}$ and $p^{y,v}_{\ell''}$ is $\prob{y}{w_{\ell''} = u}$. The latter along with the fact that $(\ell'+\ell''=\ell)$ imply

\begin{align*}
\sum_{y\in V} \left(p^{u,y}_{\ell'} \cdot p^{y,v}_{\ell''}\right) =\sum_{y\in V} \left(\prob{u}{w_{\ell}=y} \cdot \prob{y}{w_{\ell''} = u}\right) = \prob{u}{w_{\ell}=v}.
\end{align*}

We next prove that the second coordinate $\pmfg{\ell}{u}{v}$ is an $(\epsilon, j)$-approximation of $\pmf{\ell}{u}{v}$. First, since $j_{u,v}=(\max_{y}\max(j_{v,y},j_{y,u}))+1$, Lemma~\ref{lem:multiply} implies that $\left(\pmfg{\ell'}{u}{y} * \pmfg{\ell''}{y}{v} \right)$ is an $(\epsilon, j_{u,v})$-approximation of $f^{u,y,v}_{s(w),\ell',\ell'}$ where we define $f^{u,y,v}_{s(w),\ell',\ell'}$ to be the probability mass function of $s(w)$, conditioning on $w\sim w_{v,u}$, $\ell(w)=\ell'+\ell''$ and $w_{\ell'}=y$. Second, consider the triplets $\{(u,y,v)\}_{y \in V}$, and let $g_y = \pmfg{\ell'}{u}{y} * \pmfg{\ell''}{y}{v}$ and $p_y = p_{\ell'}^{u,y}\cdot p_{\ell'}^{y,v}$. Then by Lemma~\ref{lem:addition} we get that $\pmfg{\ell}{u}{v} = \sum_{y \in V} p_y \cdot g_y$ is the desired $(\epsilon, j_{u,v})$-approximation.

Finally, we prove that $j_{u,v}= O(\log n)$. We will inductively show that the first coordinate $j_{u,v}$ of the output vector from \textsc{ComputeDistrib}$(G,u,v,\ell,\epsilon)$ is at most $k+1$, for $\ell \le 2^k$. For the base case $k=0$, which implies that $\ell =1$ and and the claim trivially holds. Now assume that $k \ge 1$. Since $\ell \le 2^k$, by construction we get that $\ell' \leq 2^{k-1}$ and $\ell'' \le 2^{k-1}$. By induction hypothesis, the first coordinates returned by all of the recursion calls are no more than $(k-1)+1=k$. Thus, the returned $j_{u,v}$ at most $k+1 = O(\log n)$.

For the running time, note that in all recursion calls of the procedure \textsc{ComputeDistrib} there are at most $n^2$ possible pairs $(u, v)$ and  $O(\log n)$ possible values of $\ell$. In each of these calls, we invoke the procedure $\textsc{Convolute}$ exactly $n$ times, where each invocation costs $\Otil(\epsilon^{-2})$ by Lemma~\ref{lem:multiply}. Thus the total running time is bounded by $\Otil(n^3\epsilon^{-2})$. 

\end{proof}

We now have all the tools to prove Lemma~\ref{lem:approx_sample}.
\begin{proof}[Proof of Lemma~\ref{lem:approx_sample}]

Our algorithm for sampling $s(w)$ is implemented as follows. First, it invokes the procedure $\textsc{computeDistrib}(G,u,v,\ell, \epsilon)$ and obtains the resulting vector $(j_{u,v}, \pmfg{\ell}{u}{v}, p^{u,v}_\ell)$. Then it samples from the distribution by choosing the interval $I^{j_{u,v}}_k = [\textrm{le}, \textrm{ri}]$ with probability $\pmfg{\ell}{u}{v}(k)$, where $\textrm{le} := (1+ O(\frac{\epsilon}{\log n}))^{k}$ and $\textrm{ri} := (1+ O(\frac{\epsilon}{\log n}))^{k+j_{u,v}}$. Finally the algorithm outputs $\text{ri}$. This procedure is summarized in Algorithm~\ref{alg:approx_sample}.

\begin{algorithm2e}[t]
\caption{\textsc{Sample}$(G,u,v,\ell,\epsilon)$}
\label{alg:approx_sample}
\Input{Weighted graph $G=(V,E,\ww)$, with $\ww(e) = [1,n^{c}]$ for each $e \in E$ and some $c > 0$, two vertices $u,v \in V$, a length parameter $\ell \in [1,n^{d}]$ and an error parameter $\epsilon > 0$} 
\Output{A sampled $s(w)$ up to a $(1+\epsilon)$ relative error, where $w$ is a random walk of length $\ell$ that starts at $u$ and ends at $v$}

Set $(j_{u,v}, \pmfg{\ell}{u}{v}, p^{u,v}_{\ell}) \gets $ \textsc{ComputeDistrib}$(G,u,v,\ell,O(\frac{\epsilon}{\log n}))$ \;
Let $k_0$ be the index of the interval $I^{j_{u,v}}_{k_0}$ that is sampled according to distribution $\pmfg{\ell}{u}{v}$ \;
Return $\left(1+ O(\frac{\epsilon}{\log n})\right)^{k_0 + j_{u,v}}$\;
\end{algorithm2e}

We next argue about the correctness. Note that by property~\ref{cond:2} in the definition of approximation $\pmfg{\ell}{u}{v}$ of $\pmf{\ell}{u}{v}$, this sampling process can be viewed as sampling the pair $(x,i)$ from the distribution $\HH_{x,i}$, without knowing $x$. Furthermore, by property~\ref{cond:1}, each $x$ is sampled with the correct probability $\pmf{\ell}{u}{v}(x)$. Since we can restrict to $\epsilon \le 1/2$ it follows by Lemma~\ref{lem:correctness_alg_computedistribution} that $\textrm{ri}/\textrm{le}=(1+O(\frac{\epsilon}{\log n}))^{j_{u,v}} \le (1+\epsilon)$. Thus by property~\ref{cond:3} we get that $\textrm{ri}$ is within $[x,(1+\epsilon)x]$ for the (unknown) sampled $x$. 

The running time of our sampling procedure is asymptotically dominated by the running time of $\textsc{ComputeDistrib}$, which is in turn bounded by $O(n^{3} \epsilon^{-2})$, as desired. 
\end{proof}

\section{Schur Complement Sparsifier from Sum of Random Walks}
\label{sec:SchurComplement}
In this section we prove Theorem~\ref{thm:SparsifySchur},
which states that sampling random walks generates sparsifiers of Schur complements:
\begin{lemma}[Restatement of Theorem~\ref{thm:SparsifySchur}]
Let $G=(V,E,w)$ be an undirected, weighted multi-graph with a subset of vertices $K$. Furthermore, let $\epsilon \in (0,1)$, and let $\rho$ be some parameter related to the concentration of sampling given by
\[
\rho = O\left( \log{n}  \epsilon^{-2} \right).
\]
Let $H$ be an initially empty graph, and for every edge $e=(u,v)$ of 
repeat $\rho$ times the following procedure:
\begin{enumerate}
\itemsep0em
\item Simulate a random walk starting from $u$ until
it \emph{first} hits $K$ at vertex $t_1$,
\item Simulate a random walk starting from $v$ until
it \emph{first} hits $K$ at vertex $t_2$,
\item Combine these two walks (including $e$) to get a walk $u = (t_1=u_0,\ldots,u_\ell=t_2)$, where $\ell$ is the length of the combined walk.
\item Add the edge $(t_1, t_2)$ to $H$ with weight
\[
	1/\left( \rho \sum_{i=0}^{\ell-1} \left(1/\ww(u_i,u_{i+1}) \right )\right)
\]
\end{enumerate}
The resulting graph $H$ satisfies $\normalfont \LL_H \approx_{\epsilon} \SC(G,K)$ with high probability.
\end{lemma}

Note that this rescaling by $1/\left( \rho \sum_{i=0}^{\ell-1} \left(1/\ww(u_i,u_{i+1}) \right )\right)$ is quite natural:
Consider the degenerate case where $K=V$. This routine generates $\rho$ copies of each edge weight, which then need to be rescaled by $1 / \rho$ to ensure approximation to the original graph. 

Similar to other randomized graph sparsification
algorithms~\cite{SpielmanS11,KoutisLP16,AbrahamDKKP16,DurfeePPR17:arxiv,JindalKPS17},
our sampling scheme directly interacts with Chernoff bounds. Our random matrices are `groups' of edges related to random walks
starting from the edge $e$. We will utilize Theorem 1.1 due to~\cite{Tropp12}, which we paraphrase in our notion of approximations.

\begin{theorem} 
Let $\normalfont \XX_{1}, \XX_{2} \ldots \XX_{k}$ be a set of random matrices satisfying the following properties:
\begin{enumerate}
\itemsep0em
\item Their expected sum is a projection operator onto some subspace, i.e., 
$ \normalfont \sum_{i} \expec{}{\XX_i} = \PPi. $
\item For each $\XX_{i}$, its entire support satisfies:
$ 0 \preceq \XX_{i} \preceq \frac{\epsilon^2}{O\left( \log{n} \right)} \II.$
\end{enumerate}
Then, with high probability, we have
\[
\sum_{i} \XX_{i} \approx_{\epsilon} \PPi.
\]
\end{theorem}

Re-normalizations of these bounds similar to the work of~\cite{SpielmanS11}
give the following graph theoretic interpretation of the theorem above.
\begin{corollary}
\label{cor:Sparsify}
Let $E_1 \ldots E_k$ be distributions over random edges satisfying the following properties:
\begin{enumerate}
\itemsep0em
\item Their expectation sums to the graph $G$, i.e., $ \sum_{i} \expec{}{E_i} = G. $
\item For each $E_{i}$, any edge in its support has
low leverage score in $G$, i.e., 
$
\normalfont \ww({e}) \er^{G} \left( e \right)
\leq
\frac{\epsilon^2}{O\left( \log{n} \right)}.$
\end{enumerate}
Then, with high probability, we have
\[
\sum_{i} \LL_{E_{i}} \approx_{\epsilon} \LL_{G}. 
\]
\end{corollary}

To fit the sampling scheme outlined in Theorem~\ref{thm:SparsifySchur}
into the requirements of Corollary~\ref{cor:Sparsify},
we need (1) a specific interpretation of Schur complements in terms of walks, and (2) a bound on the effective resistances between two vertices at a given distance.

Given a walk $w = u_0,\ldots,u_{\ell}$ of length $\ell$ in $G$ with a subset a vertices $K$, we say that $w$ is a \emph{terminal-free} walk iff $u_0,u_{\ell} \in K$ and $u_1,\ldots,u_{\ell-1} \in V \setminus K$.

\begin{fact}[\cite{DurfeePPR17:arxiv}, Lemma 5.4]\label{fact:WalkDecomposition}
For any undirected, unweighted graph $G$
and any subset of vertices $K \subseteq V$,
the Schur complement $\SC(G,K)$ is given as an union over all multi-edges corresponding to terminal-free walks $u_{0}, \ldots ,u_{\ell}$ with weight
\[
 \frac{\prod_{i = 0}^{\ell - 1}\ww(u_{i},u_{i+1})}{\prod_{i = 1}^{\ell - 1}\dd\left( u_{i} \right)}.
\]
\end{fact}
The fact below follows by repeatedly applying the triangle inequality of the effective resistances between two vertices.
\begin{fact}
\label{fact:ERBound}
In an weighted undirected graph $G$, the effective resistance
between two vertices that are connected by a path $p=(p_0,\ldots,p_\ell)$ is at most $\sum_{i=0}^{\ell-1}1/\ww(p_i, p_{i+1})$.
\end{fact}

Combining the above results gives the guarantees of our sparsification routine.

\begin{proof}[Proof of Theorem~\ref{thm:SparsifySchur}]
For every edge $e \in E$, let $W_e$ be the random graph corresponding the the terminal-free random walk that started at edge $e$. Define $H = \rho \cdot \sum_{e} W_e$ to be the output graph by our sparsification routine, where $\rho= O(\log n \epsilon^{-2})$ is the sampling overhead. To prove that $\LL_H \approx_{\epsilon} \SC(G,K)$ with high probability, we need to show that (1) $\expec{}{H} = \SC(G, K)$ and (2) for any edge $f$ in $W_e$, its leverage score $\ww(f) \er^{W_e}( f )$  is at most $\epsilon^{2}/ \log n$ (by Corollary~\ref{cor:Sparsify}). Note that (2) immediately follows from the effective resistance bound of Fact~\ref{fact:ERBound} and the choice of $\rho =O(\log n / \epsilon^2)$. We next show (1).

To this end, we start by describing the decomposition of $\SC(G, K)$ into
random multi-edges, which correspond to random terminal-free walks in Fact~\ref{fact:WalkDecomposition}. The main idea is to sub-divide each walk 
$u_0 \ldots u_{\ell}$ of length $\ell$ in $G$ into $\ell$ walks of the same length, each starting at one of the $\ell$ edges on the walk, and each having weight
\[
 \frac{\prod_{i = 0}^{\ell - 1}\ww(u_{i},u_{i+1})}{\prod_{i = 1}^{\ell - 1}\dd\left( u_{i} \right)}
\]
By construction of our sparsification routine, note that every random graph $W_e$ is a distribution over walks $u_0 \ldots u_{\ell}$, each picked with probability
\[
\frac{1}{\ww(e)} \frac{\prod_{i = 0}^{\ell - 1}\ww(u_{i},u_{i+1})}{\prod_{i = 1}^{\ell - 1}\dd\left( u_{i} \right)}.
\]
Thus, to retain expectation, when such a walk is picked, our routine correctly adds it to $H$ with weight $1/(\rho \sum_{i=0}^{\ell-1}1/\ww(u_i, u_{i+1})).$ 

Formally, we get the following chain of equalities
\begin{align*}
 & \expec{}{H} = \rho \cdot \sum_{e} \expec {}{W_e}  \\
             & = \rho \cdot \sum_{e} \sum_{ w = u_0, u_1 \ldots u_{\ell\left( w \right)} : w \ni e} \frac{1}{\rho \left(\sum_{i=0}^{\ell-1}1/\ww(u_i,u_{i+1})\right)} \cdot \frac{1}{\ww(e)} \cdot \frac{\prod_{i = 0}^{\ell - 1}\ww(u_{i},u_{i+1})}{\prod_{i = 1}^{\ell - 1}\dd\left( u_{i} \right)}  \\
             & = \sum_{w = u_0, u_1 \ldots u_{\ell\left( w \right)}} \sum_{e : e \in w} \frac{1}{\left(\sum_{i=0}^{\ell-1}1/\ww(u_i,u_{i+1})\right)} \cdot \frac{1}{\ww(e)} \cdot \frac{\prod_{i = 0}^{\ell - 1}\ww(u_{i},u_{i+1})}{\prod_{i = 1}^{\ell - 1}\dd\left( u_{i} \right)}  \\
             & = \sum_{w = u_0, u_1 \ldots u_{\ell\left( w \right)}} \frac{\prod_{i = 0}^{\ell - 1}\ww(u_{i},u_{i+1})}{\prod_{i = 1}^{\ell - 1}\dd\left( u_{i} \right)}  \\ 
             &= \SC(G,K).
\end{align*}
\end{proof}

\section{Conclusion}
In this chapter, we study algorithms for dynamically maintaining all-pairs effective resistances in undirected weighted graphs and Laplacian solvers in undirected, unweighted, bounded degree graphs. In particular, we obtain an algorithm with $O(m^{3/4} \epsilon^{-4})$ update and query time for $(1+\epsilon)$-approximating effective resistances in unweighted graphs, and an algorithm with $O(n^{11/12} \epsilon^{-5})$ update and query time for solving Laplacian systems approximately while allowing implicit access to few entries of the solution vector. Our key component is the dynamic maintenance of spectral vertex sparsifiers (also known as approximate Schur complements) with respect to a set of terminals of our choice. 

A natural attempt to improve the running times of our effective resistance data-structure is to employ a hierarchy of dynamic spectral vertex sparsifiers. However, this is not an easy task as there are many dependencies which one needs to deal with when employing a hierarchical approach. We believe that a careful analysis combined with a way to control the propagation of updates among levels might indeed lead to further improvements. 

Our dynamic Schur complement data-structure works only against an oblivious adversary. While this is a standard assumption in dynamic algorithms, especially when designing the first non-trivial algorithm for a particular problem, it is highly desirable to remove this assumption as this might lead to other algorithmic applications. A good starting step would be to design a randomized algorithm that works against an adaptive adversary.

Perhaps one of the most important problems is to remove our bounded-degree assumption for dynamic Laplacian solvers with demand vectors that have large non-zero support. Our current algorithm exploits this assumption in several places, the most critical one being the bound on the load on any vertex induced by the random walks that our algorithm maintains. This suggests that new approaches might be required to be able to remove this assumption.

\chapter[Dynamic Low-Stretch Trees via Dynamic Low-Diameter Decompositions][Dynamic Low-Stretch Trees]{Dynamic Low-Stretch Trees via Dynamic Low-Diameter Decompositions}\label{cha:STOC2019_LSST}

Spanning trees of low average stretch on the non-tree edges, as introduced by Alon et al.~\cite{AlonKPW95}, are a natural graph-theoretic object. In recent years, they have found significant applications in solvers for symmetric diagonally dominant (SDD) linear systems. In this work, we provide the first dynamic algorithm for maintaining such trees under edge insertions and deletions to the input graph. Our algorithm has update time $ n^{1/2 + o(1)} $ and the average stretch of the maintained tree is $ n^{o(1)} $, which matches the stretch in the seminal result of Alon et al.

Similar to Alon et al., our dynamic low-stretch tree algorithm employs a dynamic hierarchy of low-diameter decompositions (LDDs).
As a major building block we use a dynamic LDD that we obtain by adapting the random-shift clustering of Miller et al.~\cite{MillerPX13} to the dynamic setting.
%
The major technical challenge in our approach is to control the propagation of updates within our hierarchy of LDDs: each update to one level of the hierarchy could potentially induce several insertions \emph{and} deletions to the next level of the hierarchy.
We achieve this goal by a sophisticated amortization approach.
In particular, we give a bound on the number of changes made to the LDD per update to the input graph that is significantly better than the trivial bound implied by the update time.

We believe that the dynamic random-shift clustering might be useful for independent applications.
One of these applications is the dynamic spanner problem.
By combining the random-shift clustering with the recent spanner construction of Elkin and Neiman~\cite{ElkinN17}.
We obtain a fully dynamic algorithm for maintaining a spanner of stretch $ 2k - 1 $ and size $ O (n^{1 + 1/k} \log{n}) $ with amortized update time $ O (k \log^2 n) $ for any integer $ 2 \leq k \leq \log n $.
Compared to the state-of-the art in this regime Baswana et al. \cite{BaswanaKS12}, we improve upon the size of the spanner and the update time by a factor of $ k $.

\section{Introduction}

Graph compression is an important paradigm in modern algorithm design.
Given a graph $ G $ with $ n $ nodes, can we find a substantially smaller (read: sparser) subgraph $ H $ such that $ H $ preserves central properties of~$ G $?
Very often, this compression is ``lossy'' in the sense that the properties of interest are only preserved approximately.
A ubiquitous example of graph compression schemes are \emph{spanners}:
every graph $ G $ admits a spanner $ H $ with $ O (n^{1 + 1/k}) $ edges that has \emph{stretch}~$ 2 k - 1 $ (for any integer $ k \geq 2 $), meaning that for every edge $ e = (u, v) $ of~$ G $ not present in $ H $ there is a path from $ u $ to $ v $ in~$ H $ of length at most $ 2 k - 1 $.
Thus, when $ k = \log{n} $, very succinct compression with $ O (n) $ edges can be achieved at the price of stretch $ O (\log n) $.

The most succinct form of subgraph compression is achieved when $ H $ is a tree.
Spanning trees, for example, are a well-known tool for preserving the connectivity of a graph.
It is thus natural to ask whether, similar to spanners, one could also have spanning trees with low stretch for each edge.
This unfortunately is known to be false: in a ring of $ n $ nodes every tree will result in a stretch of $ n - 1 $ for the single edge not contained in the tree.
However, it turns out that a quite similar goal can be achieved by relaxing the concept of stretch:
every graph $ G $ admits a spanning tree $ T $ of \emph{average stretch} $ O (\log{n} \log \log n) $~\cite{AbrahamN12}, where the average stretch is the sum of the stretches of all edges divided by the total number of edges.
Such subgraphs are called \emph{low (average) stretch trees} and have found numerous applications in recent years, most notably in the design of fast solvers for symmetric diagonally dominant (SDD) linear systems~\cite{SpielmanT14,KoutisMP14,BlellochGKMPT14,KoutisMP11,KelnerOSZ13,CohenKMPPRX14}.
We believe that their fundamental graph-theoretic motivation and their powerful applications make low-stretch trees a very natural object to study as well in a dynamic setting, similar to spanners~\cite{AusielloFI06,Elkin11,BaswanaKS12,BodwinK16} and minimum spanning trees~\cite{Frederickson85,EppsteinGIN97,HenzingerK01,HolmLT01,Wulff-Nilsen17,NanongkaiSW17}.
Indeed, the design of a dynamic algorithm for maintaining a low-stretch tree was posed as an open problem by Baswana et al.~\cite{BaswanaKS12}, but despite extensive research on dynamic algorithms in recent years, no such algorithm has yet been found.

In this chapter, we give the first non-trivial algorithm for this problem in the \emph{dynamic} setting.
Specifically, we maintain a low-stretch tree $ T $ of a dynamic graph~$ G $ undergoing updates in the form of edge insertions and deletions in the sense that after each update to $ G $ we compute the set of necessary changes to $ T $.
The goal in this problem is to keep the time spent after each update small while still keeping the average stretch of $ T $ tolerable.
Our main result is a fully dynamic algorithm for maintaining a spanning tree of expected average stretch~$ n^{o(1)} $ with expected amortized update time~$ n^{1/2 + o(1)} $.
At a high level, we obtain this result by combining the classic low-stretch tree construction of Alon et al.~\cite{AlonKPW95} with a dynamic algorithm for maintaining low diameter decompositions (LDD) based on random-shift clustering~\cite{MillerPX13}.
Our LDD algorithm might be of independent interest, and we provide another application by using it to obtain a dynamic version of the recent spanner construction of Elkin and Neiman~\cite{ElkinN17}.
The resulting dynamic spanner algorithm improves upon one of the state-of-the-art algorithms by Baswana et al.~\cite{BaswanaKS12}.

Our overall approach towards the low-stretch tree algorithm -- \sloppy to use low-diameter decompositions based on random-shift clustering in the construction of Alon et al.~\cite{AlonKPW95} -- has been used before in parallel and distributed algorithms~\cite{BlellochGKMPT14,GhaffariKKLP15,HaeuplerL18}.
However, to make this approach work in the dynamic setting we need to circumvent some non-trivial challenges.
In particular, we cannot employ the following paradigm that often is very helpful in designing dynamic algorithms: design an algorithm that can only handle edge deletions and then extend it to the fully dynamic setting using a general reduction.
While we do follow this paradigm for our dynamic LDD algorithm, there are two obstacles that prevent us from doing so for the dynamic low-stretch tree:
First, many fully-dynamic-to-decremental reductions exploit some form of ``decomposability'', which does not hold for low-stretch trees, i.e., low-stretch trees of subgraphs of the input graph cannot be simply be combined to a single low-stretch tree of the full graph.
Second, in our dynamic low-diameter decomposition edges might start and stop being inter-cluster edges, even if the input graph is only undergoing deletions.
In the hierarchy of Alon et al.\ this leads to both insertions and deletions at the next level of the hierarchy.
As opposed to other dynamic problems~\cite{HenzingerKN16,AbrahamDKKP16}, one algorithm cannot simply enforce some type of ``monotonicity'' by not passing on insertions to the next level of the hierarchy (to stay within a deletions-only setting) as there might be too many such edges to ignore them. 
Thus, it seems that we really have to deal with the fully dynamic setting in the first place.
We show that this can be done by a sophisticated amortization approach that explicitly analyzes the number of updates passed on to the next level.

\paragraph*{Related Work.}

Low average stretch trees have been introduced by Alon et al.~\cite{AlonKPW95} who obtained an average stretch of $ 2^{O (\sqrt{\log n \log \log n})} $ and also gave a lower bound of $ \Omega (\log n) $ on the average stretch.
The first construction with polylogarithmic average stretch was given by Elkin et al.~\cite{ElkinEST08}.
Further improvements~\cite{AbrahamBN08,KoutisMP11} culminated in the state-of-the-art construction of Abraham and Neiman~\cite{AbrahamN12} with average stretch $ O(\log n \log \log n) $.
All these trees with polylogarithmic average stretch can be computed in time $ \tilde O (m) $. To the best of our knowledge, all known algorithms in parallel and distributed models of computation~\cite{BlellochGKMPT14,GhaffariKKLP15,HaeuplerL18} are based on the scheme of Alon et al.\ and thus do not provide polylogarithmic stretch guarantees.

The main application of low-stretch trees has been in solving symmetric, diagonally dominant (SDD) systems of linear equations.
It has been observed that iterative methods for solving these systems can be made faster by preconditioning with a low-stretch tree~\cite{Vaidya91,BomanH03,SpielmanW09}.
Consequently, they have been an important ingredient in the breakthrough result of Spielman and Teng~\cite{SpielmanT14} for solving SDD systems in nearly linear time.
In this solver, low-stretch trees are utilized for constructing ultra-sparsifiers, which in turn are used as preconditioners.
Beyond this initial breakthrough, low-stretch trees have also been used in subsequent, faster solvers~\cite{KoutisMP14,BlellochGKMPT14,KoutisMP11,KelnerOSZ13,CohenKMPPRX14}.
Another prominent application of low-stretch trees~(concretely, the variant of random spanning trees with low expected stretch) is the remarkable cut-based graph decomposition of R\"acke~\Cite{Racke08,AndersenF09}, which embeds any general undirected graph into convex combination of spanning trees, while paying only a $\tilde{O}(\log n)$ congestion for the embedding. This decomposition tool, initially aimed at giving the best competitive ratio for oblivious routing, has found several applications ranging from approximation algorithms for cut-based problems~(e.g., minimum bisection~\cite{Racke08}) to graph compression~(e.g., vertex sparsifiers~\cite{Moitra09}). 
Other classic problems in the realm of approximation algorithms that utilize the properties of low-stretch trees include the $ k $-server problem~\cite{AlonKPW95} and the minimum communication cost spanning tree problem~\cite{Hu74,PelegR98}.

In terms of dynamic algorithms, we are not aware of any prior work for maintaining low-stretch trees.
The closest related works are arguably dynamic algorithms for maintaining distance oracles and spanners, as they also aim preserving pairwise distances, and dynamic algorithms for maintaining minimum spanning trees, as they also are spanning trees with an additional property.


A distance oracle is a data structure that can answer queries for the (approximate) distance between a pair of nodes.
The fully dynamic distance oracle of Abraham, Chechik, and Talwar~\cite{AbrahamCT14} for unweighted, undirected graphs has expected amortized update time $ \tilde O (\sqrt{m} n^{1/k}) $, query time $ O (k^2 \rho^2) $, and stretch $ 2^{O (k \rho)} $, where the parameter $ k \geq 2 $ is integer and $ \rho = 1 + \lceil \tfrac{\log n^{1 - 1/k}}{\log(m / n^{1 - 1/k})} \rceil $.
To the best of our knowledge, the recent decremental distance oracle of Chechik~\cite{Chechik18} can be used to extend this result to weighted graphs and to improve the stretch and the query time, while leaving the update time essentially unchanged.

For dynamic spanner algorithms, the main goal is to maintain, for any given integer $ k \geq 2 $, a spanner of stretch $ 2k - 1 $ with $ \tilde O (n^{1 + 1/k}) $ edges.
Spanners of stretch $ 2k - 1 $ and size $ O (n^{1 + 1/k}) $ exist for every undirected graph~\cite{Awerbuch85}, and this trade-off is presumably tight under Erd\H{o}s's girth conjecture.
The dynamic spanner problem has been introduced by Ausiello et al.~\cite{AusielloFI06}.
They showed how to maintain a $3$- or $5$-spanner with amortized update time proportional to the maximum degree of the graph.
Using techniques from the streaming literature, Elkin~\cite{Elkin11} provided an algorithm for maintaining a $ (2k - 1) $-spanner with $ \tilde O (m n^{-1/k}) $ expected update time.
Faster update times were achieved by Baswana et al.~\cite{BaswanaKS12}: their algorithms maintain $ (2k - 1) $-spanners either with expected amortized update time $ O(1)^k $ or with expected amortized update time $ O (k^2 \log^2 n) $.
Later, Bodwin and Krinninger~\cite{BodwinK16} initiated the study of dynamic spanners with worst-case update times, and recently, Bernstein, Forster, and Henzinger~\cite{BernsteinFH19} presented a deamortization approach to maintain $ (2k - 1) $-spanners with high-probability worst-case update time $ O(1)^k \log^3 n $.
All of these algorithms exhibit the stretch/space trade-off mentioned above in unweighted graphs, up to polylogarithmic factors in the size of the spanner.

The first non-trivial algorithm for dynamically maintaining a minimum spanning tree was developed by Frederickson~\cite{Frederickson85} and had a worst-case update time of $ O (\sqrt{m}) $.
Using a general sparsification technique, this bound was improved to $ O (\sqrt{n}) $ by Eppstein et al.~\cite{EppsteinGIN97}.
In terms of amortized bounds, Holm et al.~\cite{HolmLT01} were the first to improve this bound and obtained polylogarithmic amortized update time.
A recent breakthrough of Nanongkai, Saranurak, and Wulff-Nilsen~\cite{Wulff-Nilsen17,NanongkaiS17,NanongkaiSW17}, who finally achieved a \emph{worst-case} update time of $ n^{o(1)} $.

\paragraph*{Our Results.}

Our main result is a dynamic algorithm for maintaining a low average stretch tree of an unweighted, undirected graph.

\begin{theorem}\label{thm:fully dynamic low stretch tree}
Given any unweighted, undirected graph undergoing edge insertions and deletions, there is a fully dynamic algorithm for maintaining a spanning forest of expected average stretch $ n^{o(1)} $ that has expected amortized update time $ m^{1/2 + o(1)} $.
These guarantees hold against an oblivious adversary.
\end{theorem}

This is the first non-trivial algorithm for this fundamental problem.
Our stretch matches the seminal construction of Alon et al.~\cite{AlonKPW95}, which is still the state of the art in parallel and distributed settings~\cite{BlellochGKMPT14,GhaffariKKLP15,HaeuplerL18}.

Similar to the approach of~\cite{KoutisLP16} in the \emph{static} setting, we can apply Theorem~\ref{thm:fully dynamic low stretch tree} to a cut sparsifier of the input graph, which has only $ \tilde O (n) $ edges, to improve the running time for dense graphs.
Such a cut sparsifier can be maintained with the dynamic algorithm of Abraham et al.~\cite{AbrahamDKKP16} that has polylogarithmic update time.

\begin{corollary}\label{cor:fully dynamic low stretch tree}
Given any unweighted, undirected graph undergoing edge insertions and deletions, there is a fully dynamic algorithm for maintaining a spanning forest of expected average stretch $ n^{o(1)} $ that has expected amortized update time $ n^{1/2 + o(1)} $.
These guarantees hold against an oblivious adversary.
\end{corollary}

Obtaining this improvement is non-trivial because cut sparsifiers are weighted graphs, even when the input graph is unweighted, and the algorithm of Theorem~\ref{thm:fully dynamic low stretch tree} only accepts unweighted graphs.
To deal with this issue, we deviate from the approach of~\cite{KoutisLP16} by interpreting the edge weights of the sparsifier as edge multiplicities in an unweighted graph.
A fine-grained analysis of the amount of change to edge the multiplicities per update to the input graph then allows us to get the desired benefits of combining both algorithms.


We additionally show that $ \sqrt{n} $ is not an inherent barrier to the update time, at least if very large stretch is tolerated.
A modification of our algorithm gives average stretch $ O(t) $ and update time $ \tfrac{n^{1 + o(1)}}{t} $ for $ t \geq \sqrt{n} $.

One of the main building blocks of our dynamic low-stretch tree algorithm is a dynamic algorithm for maintaining a low-diameter decomposition (LDD). Roughly speaking, for $\beta \in (0,1)$ and $\diam > 0$,  a $(\beta, \diam)$-decomposition of a graph is a partitioning of its nodes into node-disjoint clusters such that (1) any pair of nodes belonging to the same cluster are at distance at most $\diam$, and (2) the number of edges whose endpoints belong to different clusters is bounded by $\beta m$. The following theorem gives a dynamic variant of such decompositions.
 
\begin{theorem}\label{thm:fully dynamic LDD}
Given any unweighted, undirected multigraph undergoing edge insertions and deletions, there is a fully dynamic algorithm for maintaining a $ (\fract, O (\tfrac{\log{n}}{\fract})) $-decomposition (with clusters of strong diameter $ O (\tfrac{\log{n}}{\fract}) $ and at most $ \beta m $ inter-cluster edges in expectation) that has expected amortized update time $ O (\log^2{n} / \fract^2) $.
A spanning tree of diameter $ O (\tfrac{\log{n}}{\fract}) $ for each cluster can be maintained in the same time bound.
The expected amortized number of edges to become inter-cluster edges after each update is $ O (\log^2{n} / \fract) $.
These guarantees hold against an oblivious adversary.
\end{theorem}
Our algorithm is based on the random-shift clustering of Miller at al.~\cite{MillerPX13}, with many tweaks to make it work in a dynamic setting.
In our analysis of the algorithm, we bound the amortized number of changes to the clustering per update by $ \tilde O (1 / \fract) $, which is significantly smaller than the naive bound of $ \tilde O (1 / \fract^2) $ implied by the update time.
This is particularly important for hierarchical approaches, such as in our dynamic low-stretch tree algorithm, because a small bound on the number of amortized changes helps in controlling the number of induced updates to be processed within the hierarchy.
Independently, Saranurak and Wang~\cite{SaranurakW19} obtained a fully dynamic LLD algorithm with nearly the same guarantees (up to polylogarithmic factors).\footnote{The low-diameter decomposition of Saranurak and Wang can be maintained against an adaptive online adversary. However, the low-diameter spanning trees of their clustering can only be maintained against an oblivious adversary. Therefore, plugging in their dynamic LDD algorithm into our dynamic low-stretch tree construction does not yield any improvement over our guarantees.}
We believe that our solution is arguably simpler than their expander pruning approach.

The dynamic random-shift clustering underlying our dynamic LDD is of independent interest.
A direct consequence demonstrating the usefulness of our dynamic random-shift clustering algorithm is the following new result for the dynamic spanner problem.
\begin{theorem}\label{thm:fully dynamic spanner}
Given any unweighted, undirected graph undergoing edge insertions and deletions, there is a fully dynamic algorithm for maintaining a spanner of stretch $ 2k - 1 $ and expected size $ O (n^{1 + 1/k} \log{n}) $ that has expected amortized update time $ O (k \log^2{n}) $.
These guarantees hold against an oblivious adversary.
\end{theorem}
Recall that the fully dynamic algorithm of Baswana et al.~\cite{BaswanaKS12} maintains a spanner of stretch $ 2k - 1 $ and expected size $ O (k n^{1 + 1/k} \log n) $ with expected amortized update time $ O (k^2 \log^2{n}) $.
Our new algorithm thus improves both the size and the update time by a factor of~$ k $.
This is particularly relevant because the stretch/size trade-off of $ 2k - 1 $ vs.\ $ O (n^{1 + 1/k}) $ is tight under the girth conjecture.
We thus exceed the conjectured optimal size by a factor of only $ \log{n} $ compared to the prior $ k \log{n} $, where $ k $ might be as large as $ \log n $.
When we restrict ourselves to the decremental setting, we do achieve size $ O (n^{1 + 1/k}) $ with expected amortized update time $ O (k \log{n}) $.
Again, this saves a factor of~$ k $ compared to Baswana et al.~\cite{BaswanaKS12}.
To obtain Theorem~\ref{thm:fully dynamic spanner}, we employ our dynamic random-shift clustering algorithm in the spanner construction of Elkin and Neiman~\cite{ElkinN17} and combine it with the dynamic spanner framework of Baswana et al.~\cite{BaswanaKS12}.

\paragraph*{Structure of this Chapter.}

The remainder of this chapter is structured as follows.
We first settle the notation and terminology in Section~\ref{sec:prelim_LSST}.
We then give a high-level overview of our results and techniques in Section~\ref{sec:overview_LSST}.
Finally, we provide all necessary details for our dynamic low-stretch tree (Section~\ref{sec:low_stretch_tree}), our dynamic low-diameter decomposition (Section~\ref{sec:LDD}), and our dynamic spanner algorithm (Section~\ref{sec:spanner}).

\section{Preliminaries}\label{sec:prelim_LSST}

\paragraph*{Graphs.}

Let $G=(V,E,\ww_G)$ be an undirected weighted graph, where $n = |V|$, $m = |E|$ and $\ww_G: E \rightarrow \mathbb{R}_+$. If $\ww_G(e) = 1$ for all $e \in E$, then we say $G$ is an undirected unweighted graph.
If $ E $ is a multiset, i.e., every element of $E$ may have integer multiplicity greater than $ 1 $, then we call $ G $ a multigraph.
For a subset $C \subseteq V$ let $G[C]$ denote the subgraph of $G$ induced by  $C$. Throughout the chapter we call $C \subset V$ a \emph{cluster}. For any positive integer $k$, a \emph{clustering} of $G$ is a partition of $V$ into disjoint subsets $C_1, C_2, \ldots, C_k$. We say that an edge is an \emph{intra-cluster} edges if both its endpoints belong to the same cluster $C_i$ for some $i$; otherwise, we say that an edge is an \emph{inter-cluster} edge.

For any $u,v \in V$ let $\dist_G(u,v)$ denote the length of a shortest path between $u$ and $v$ induced by the edge weights $\ww_G$ of the graph $G$. When $G$ is clear from the context, we will omit the subscript. The \emph{strong diameter} of a cluster $C \subset V$ is the maximum length of the shortest path between two nodes in $G[C]$, i.e., $\max \set{\dist_{G[C]}(u,v)}{u,v \in C}$. In the following we define a low-diameter clustering of $G$. 

\begin{definition} Let $k$ be any positive integer, $\beta \in (0,1)$ and $\diam > 0$. Given an undirected, unweighted graph $G=(V,E)$, a $(\fract,\diam)$-decomposition of $G$ is a partition of $V$ into disjoint subsets $C_1, C_2, \ldots,C_k$ such that:
\begin{enumerate}
\item The strong diameter of each $C_i$ is at most $\diam$.
\item The number of edges with endpoints belonging to different subsets is at most $\fract m$.
\end{enumerate}
\end{definition}
In the $(\fract,\diam)$-decompositions of the randomized dynamic algorithms in this chapter, the bound in Condition~2 is in expectation.

Let $H = (V, F) $ be a subgraph of $G = (V, E, \ww_G)$. For any pair of nodes $u,v \in V$, we let $\dist_H(u,v)$ denote the length of a shortest path between $u$ and $v$ in $H$. We define the \emph{stretch} of an edge $(u,v) \in E$ with respect to $H$ to be 
\begin{equation*}
	\str_H(u,v) \coloneqq  \frac{\dist_H(u,v)}{\ww_G(u,v)} \, .
\end{equation*}
The stretch of $ H $ is defined as the maximum stretch of any of edge $(u,v) \in E$.
The \emph{average stretch} over all edges of $G$ with respect to $H$ is given by
\[
	\avestr_H(G) \coloneqq \frac{1}{|E|} \sum_{(u,v) \in E} \str_H(u,v).
\]

\paragraph*{Exponential Distribution.}

For a parameter $\lambda$, the probability density function of the \emph{exponential distribution} $\expdis(\lambda)$ is given by
\[
	f(x,\lambda) \coloneqq  \begin{cases} 
   \lambda e^{-\lambda x} & \text{if } x \geq 0 \\
   0       & \text{otherwise}.
  \end{cases}
\]
The mean of the exponential distribution is $1/\lambda$.

\paragraph*{Dynamic Algorithms.}
Consider a graph with $ n $ nodes undergoing updates in the form of edge insertions and edge deletions.
An \emph{incremental} algorithm is a dynamic algorithm that can only handle insertions, a \emph{decremental} algorithm can only handle deletions, and a \emph{fully dynamic} algorithm can handle both.
We follow the convention that a fully dynamic algorithm starts from an empty graph with $ n $ nodes.
The (maximum) running time spent by a dynamic algorithm for processing each update (before the next update arrives) is called \emph{update time}.
We say that a dynamic algorithm has \emph{(expected) amortized} update time $ u (n) $ if its total running time spent for processing a sequence of $ q $ updates is bounded by $ q \cdot u (n) $ (in expectation).
In this chapter, we assume that the updates to the graph are performed by an \emph{oblivious adversary} who fixes the sequences of updates in advance, i.e., the adversary is not allowed to adapt its sequence of updates as the algorithm proceeds.
This is a standard assumption in dynamic graph algorithms\footnote{For example, all known randomized dynamic spanner algorithms~\cite{Elkin11,BaswanaKS12,BodwinK16,BernsteinFH19} work under this assumption.} and in particular, it implies that for randomized dynamic algorithms the sequence of updates is independent from the random choices of the algorithm.

\section{Technical Overview}\label{sec:overview_LSST}

In the following, we provide some intuition for our approach and highlight the main ideas of this chapter.

\paragraph*{Low Average Stretch Tree.}
A first idea is to employ the dynamic low-diameter decomposition of Theorem~\ref{thm:fully dynamic LDD}.
This algorithm can maintain a \sloppy $ (\fract, O(\tfrac{\log{n}}{\fract})) $-decomposition, i.e., a partitioning of the graph into clusters such that there are at most $ \fract m $ inter-cluster edges and the (strong) diameter of each cluster is at most $ O (\tfrac{\log{n}}{\fract}) $.
In particular, each cluster has a designated center and the algorithm maintains a spanning tree of each cluster in which every node is at distance at most $ O (\tfrac{\log{n}}{\fract}) $ from the center.
Now consider the following simple dynamic algorithm:
\begin{enumerate}
\item Maintain a $ (\fract, O(\tfrac{\log{n}}{\fract})) $-decomposition of the input graph~$ G $.
\item Contract the clusters in the decomposition to single nodes and maintain a multigraph $ G' $ containing one node for each cluster and all inter-cluster edges.
\item Compute a low-stretch tree $ T' $ of $ G' $ after each update to~$ G $ using a static algorithm providing polylogarithmic average stretch.
\item Maintain $ T $ as the ``expansion'' of $ T' $ in which every node in $ T' $ is replaced by the spanning tree of diameter $ O (\tfrac{\log{n}}{\fract}) $ of the cluster representing the node.
\end{enumerate}
As the clusters are non-overlapping it is immediate that $ T $ is indeed a tree.
To analyze the average stretch of~$ T $, we distinguish between inter-cluster edges (with endpoints in different clusters) and intra-cluster edges (with endpoints in the same cluster).
Each intra-cluster edge has stretch at most $ O (\tfrac{\log{n}}{\fract}) $ as the spanning tree of the cluster containing both endpoints of such an edge is a subtree of $ T $.
Each inter-cluster edge has polylogarithmic average stretch in $ T' $ with respect to $ G' $.
By expanding the clusters, the length of each path in $ T' $ increases by a factor of at most $ O (\tfrac{\log{n}}{\fract}) $.
Thus, inter-cluster edges have an average stretch of $ O (\tfrac{\log{n}}{\fract} \polylog{n}) $ in $ T $.
As there are at most $ m $ intra-cluster edges and at most $ \fract m $ inter-cluster edges, the total stretch over all edges is at most $ O (m \cdot \tfrac{\log{n}}{\fract} + \fract m \cdot \tfrac{\log{n}}{\fract} \polylog{n}) = \tilde O (m \cdot \tfrac{1}{\fract}) $, which gives an average stretch of $ \tilde O (\tfrac{1}{\fract}) $.

To bound the update time, first observe that the number of inter-cluster edges is at most $ \fract m $.
Thus, $ G' $ has at most $ \fract m $ edges and therefore the static algorithm for computing $ T' $ takes time $ \tilde O (\fract m) $ per update.
Together with the update time of the dynamic LDD, we obtain an update time of $ \tilde O (\tfrac{1}{\fract^2} + \fract m) $.
By setting $ \fract = m^{1/3} $, we would already obtain an algorithm for maintaining a tree of average stretch $ \tilde O (m^{1/3}) $ with update time $ \tilde O (m^{2/3}) $.

We can improve the stretch and still keep the update time sublinear by a hierarchical approach in which the scheme of clustering and contracting is repeated $ k $ times.
Observe that the $i$-th contracted graph will contain at most $ \fract^i m $ many edges and, in the final tree $ T $, the stretch of each edge disappearing with the $(i + 1)$-th contraction is $ O (\tfrac{\log{n}}{\fract})^{i + 1} $, which can be obtained by expanding the contracted low-diameter clusters. 
After $ k $~contractions, there are at most $ \fract^k m $ edges remaining and they have polylogarithmic average stretch in~$ T' $ with respect to~$ G' $, which, again by expanding clusters, implies an average stretch of at most $ O (\tfrac{\log{n}}{\fract})^k \cdot \polylog{n} $ in~$ T $ with respect to~$ G $.
This leads to a total stretch of $ O (\sum_{0 \leq i \leq k - 1} \fract^i m \cdot O(\tfrac{\log{n}}{\fract})^{i + 1} + \fract^k m \cdot O (\tfrac{\log{n}}{\fract})^k \polylog{n}) = \tilde O (m \cdot \tfrac{O (\log n)^k}{\fract}) $, which gives an average stretch of $ \tilde O (\tfrac{O (\log n)^k}{\fract}) $.
To bound the update time, observe that updates propagate within the hierarchy as each change to inter-cluster edges of one layer will appear as an update in the next layer.
Each operation in the dynamic LDD algorithm will perform at most one change to the clustering, i.e., the number of changes propagated to the next layer of the hierarchy is at most $ \tilde O(\tfrac{1}{\fract^2}) $ per update to the current layer.
This will result in an update time of $ \tilde O ((\tfrac{\polylog{n}}{\fract})^{2(i - 1)} \cdot \frac{1}{\fract^2}) $ in the $i$-th contracted graph per update to the input graph.
The update time for maintaining the tree~$ T $ will then be $ \tilde O (\tfrac{1}{\fract^{2 k}} + \fract^k m) $, which is $ m^{2/3} $ at best, i.e., no better than the simpler approach above.
A tighter analysis can improve this update time significantly:
The second part of Theorem~\ref{thm:fully dynamic LDD} bounds the amortized number of edges to become inter-cluster edges by $ \tilde O (\tfrac{1}{\fract}) $.
This results in an update time of $ \tilde O ((\tfrac{\polylog{n}}{\fract})^{k + 1} + \fract^k m) $.
By setting $ k = \sqrt{\log{n}} $ and $ \fract = \tfrac{1}{m^{1 / (2k + 1)}} $ we can roughly balance these two terms in the update time and thus arrive at an update time of $ m^{1/2 + o(1)} $ while the average stretch is $ n^{o(1)} $.
The crux of our approach is thus an ``early stopping'' of the Alon et al.\ LDD hierarchy such that it does not ``exhaust'' the graph.
We crucially exploit that, for an unweighted input graph, the size of the contracted graph decreases geometrically, which allows us to partially compensate for the blow-up of propagated updates in the hierarchy.

We can use the following sparsification approach to further reduce the update time to $ n^{1/2 + o(1)} $:
The main idea is to maintain a cut sparsifier with $ \tilde O (n) $ edges and then run the algorithm on the cut sparsifier instead of the input graph to reduce the update time from $ m^{1/2 + o(1)} $ to $ n^{1/2 + o(1)} $.
The dynamic algorithm of Abraham et al.~\cite{AbrahamDKKP16} can maintain such a cut sparsifier with polylogarithmic update time.
Using a different cut sparsifier construction, Koutis, Levin, and Peng~\cite{KoutisLP16} showed in the static setting that a low-stretch tree of their cut sparsifier is also a low-stretch tree of the input graph (where the average stretch only increases multiplicatively by the approximation guarantee of the cut sparsifier).
However, we cannot use exactly the same approach because the cut sparsifier of Abraham et al.\ has edge weights, even though the input graph is unweighted.
We show that the main argument in~\cite{KoutisLP16} still goes through if we interpret the edge weights of the sparsifier as edge multiplicities in an unweighted graph.
We then show that the algorithm of Theorem~\ref{thm:fully dynamic low stretch tree} can also handle such graphs for updates that increment or decrement the multiplicity of some edge by $ 1 $.
A fine-grained analysis of the total multiplicity of edges of the sparsifier and its expected amount of change per update to the input graph then gives the desired result.

In Section~\ref{sec:low_stretch_tree}, where we present the details of our approach, we consider two slight generalizations:
First, we implicitly handle the case that the input graph could become disconnected by maintaining a low-stretch \emph{forest}.
Second, we give a parameterized analysis that also allows for a trade-off between stretch and update time.

\paragraph*{Low Diameter Decomposition.}

To obtain a suitable algorithm for dynamically maintaining a low-diameter decomposition, we follow the widespread paradigm of first designing a decremental -- i.e., deletions-only -- algorithm and then extending it to a fully dynamic one.
We can show that, for any sequence of at most $ m $ edge deletions (where $ m $ is the initial number of edges in the graph), a $ (\fract, O(\tfrac{\log{n}}{\fract})) $-decomposition can be maintained with expected total update time $ \tilde O (m / \fract) $.
Here, we build upon the work of Miller et al.~\cite{MillerPX13} who showed that \emph{exponential random-shift clustering} produces clusters of radius $ O (\log{n} / \fract) $ such that each edge has a probability of at most $ \fract $ to go between clusters.
This clustering is obtained by first having each node sample a random \emph{shift value} from the exponential distribution and then determining the cluster center of each node as the node to which it minimizes the difference between distance and (other node's) shift value.

In the parallel algorithm of~\cite{MillerPX13}, the clustering is obtained by essentially computing one single-source shortest path tree of maximum depth $ O (\log{n} / \fract) $.
To make this computation efficient\footnote{For their parallel algorithm, efficiency in particular means low depth of the computation tree.}, the shift values are rounded to integer values and the fractional values are only considered for tie-breaking.
We observe that one can maintain this bounded-depth shortest path tree with a simple modification of the well-known Even-Shiloach algorithm that spends time $ O ( \degree (v)) $ every time a node $ v $ increases its level (distance from the source) in the tree.
By rounding to integer edge weights, similar to~\cite{MillerPX13}, we can make sure that the number of level increases to consider is at most $ O (\log{n} / \fract) $ for each node.
Note however that this standard argument charging each node only when it increases its level is not enough for our purpose: the assignment of nodes to clusters follows the fractional values for tie-breaking, which might result in some node $ v $ changing its cluster -- and in this way also spend time $ O (\degree (v)) $ -- without increasing its level~(note that here the difficulty is not on maintaining the cluster that $v$ belongs to, but rather on bounding the number of cluster changes for $v$).
As has been observed in~\cite{MillerPX13}, the fractional values of the shift values effectively induce a random permutation on the nodes.
Using a similar argument as in the analysis of the dynamic spanner algorithm of Baswana et al.~\cite{BaswanaKS12}, we can thus show that in expectation each node changes its cluster at most $ O (\log{n}) $ times while staying at a particular level.
This results in a total update time of $ \tilde O (m / \fract) $.
Trivially, this also bounds the total number of times that edges become inter-cluster edges during the whole decremental algorithm by $ \tilde O (m / \fract) $.
Using a more sophisticated analysis we can obtain the stronger bound of $ \tilde O (m) $ on the latter quantity:
Intuitively, each endpoint of an edge changes its cluster at most $ \tilde O (\tfrac{1}{\fract}) $ times and after each cluster change the edge is an inter-cluster edge with probability at most $ \beta $, yielding a total of $ \tilde O ( m \cdot \tfrac{1}{\fract} \cdot \beta ) $ times that edges become inter-cluster edges.
The rigorous argument is however more complicated because we cannot guarantee that the event of being an inter-cluster edge might not be independent of the event of the endpoint changing its cluster.

To obtain a fully dynamic algorithm, we observe that any LDD can tolerate a certain number of insertions to the graph.
A $ (\fract, O(\tfrac{\log{n}}{\fract})) $-decomposition allows at most $ \fract m $ inter-cluster edges and thus, if we insert $ O (\fract m) $ edges to the graph without changing the decomposition, we still have an $ (O(\fract), O(\tfrac{\log{n}}{\fract})) $-decomposition.
We can exploit this observation by simply running a decremental algorithm, that is restarted from scratch after each phase of $ \Theta (\fract m) $ updates to the graph.
We then deal with edge deletions by delegating them to the decremental algorithm and we deal with edge insertions in a lazy way by doing nothing.
This results in a total time of $ \tilde O (m / \fract) $ that is amortized over $ \Theta (\fract m) $ updates to the graph, i.e., amortized update time $ \tilde O (1/\beta^2) $.
Similarly, the amortized number of edges to become inter-cluster edges after an update is $ \tilde O (1/\beta) $.

In our detailed description and analysis in Section~\ref{sec:LDD}, we first review the construction of Miller et al., and then present our decremental and fully dynamic algorithms.

\paragraph*{Dynamic Spanner via Exponential Random Shift Clustering} At a high level, the key idea behind our improved result on dynamic spanners is that a slight extension of the techniques we developed already leads to a deletions-only algorithm. Concretely, we show that it is possible to combine our decremental random-shift clustering with the recent spanner construction of Elkin and Neiman~\cite{ElkinN17} to design such an algorithm. Observe that this is sufficient for our purposes due to the decomposability property of spanners, which allows to extend decremental algorithms to fully dynamic ones while paying only a logarithmic factor in the size of the spanner and the update time of the data-structure (see e.g.,~\cite{BaswanaKS12}).

Inspired by the low diameter clustering algorithm of Miller et al.~\cite{MillerPX13}, Elkin and Neiman devised the following simple routine for constructing a spanner: (1) each node samples a random \emph{shift value} (which depends on some stretch parameter) from the exponential distribution and then it defines its cluster center to be the node which minimizes the difference between the distance of these two nodes and the other node's shift value, also known as the \emph{shifted distance}; (2) for each node all the neighbours that lie on a shortest path between the node and the set of nodes whose shifted distance is within $ 1 $ of the minimum one are added to the spanner. In comparison to the low-diameter clustering, where each node needs to determine the cluster it belongs to, keeping track of the spanner edges for each node might seem more challenging at first. Fortunately, we observe that determining these edges in the static setting still reduces to computing one single-source shortest path tree of bounded depth. Moreover, similar to the random-shift clustering for low-diameter decompositions, we exploit the structural properties of this tree to maintain the spanner edges under deletions using the well-known Even-Shiloach algorithm together with the rounding tricks that were tightly linked to defining a random permutation on the nodes. Details on the implementation of this algorithm are provided in Section~\ref{sec:spanner}.

\section{Dynamic Low Average Stretch Forest}\label{sec:low_stretch_tree}

Our dynamic algorithms for maintaining a low average stretch forest will use a hierarchy of low-diameter decompositions.
We first analyze very generally the update time for maintaining such a decomposition and explain how to obtain a spanning forest from this hierarchy in a natural way, similar to the construction of Alon et al.~\cite{AlonKPW95}.
We then analyze two different approaches for maintaining the tree, which will give us two complementary points in the design space of dynamic low-stretch tree algorithms.
Finally, we explain how to exploit input graph sparsification to improve the update time of our first algorithm.

\subsection{Generic Dynamic LDD Hierarchy}

Consider some integer parameter $ k \geq 1 $ and parameters $ \fract_0, \ldots, \fract_{k-1} \in (0, 1) $.
For each $ 0 \leq i \leq k-1 $, let $ \mathcal{D}_i $ be the fully dynamic algorithm for maintaining a $ (\fract_i, O (\tfrac{\log{n}}{\fract_i})) $-decomposition as given by Theorem~\ref{thm:fully dynamic LDD}.
Our \emph{LDD-hierarchy} consists of $ k+1 $ multigraphs $ G_0 = (V, E_0), \ldots, G_k = (V, E_k) $ where $ G_0 $ is the input graph $ G $ and, for each $ 0 \leq i \leq k-1 $, the graph $ G_{i+1} $ is obtained from contracting $ G_i $ according to a $ (\beta_i, O (\tfrac{\log{n}}{\fract_i}))$-decomposition of $ G_i $ as follows:
For every node $ v \in V $, let $ c_i (v) $ denote the center of the cluster to which $ v $ is assigned in the $ (\beta_i, O (\tfrac{\log{n}}{\fract_i}))$-decomposition of $ G_i $.
Now define $ E_{i+1} $ as the multiset of edges containing for every edge $ (u, v) \in E_i $ such that $ c_i (u) \neq c_i (v) $ one edge $ (c_i (u), c_i (v)) $, i.e., $ E_{i+1} = \{ (c_i (u), c_i (v)) : (u, v) \in E_i \text{ and } c_i (u) \neq c_i (v) \} $, where the multiplicity of each edge is equal to the number of edges between the corresponding clusters in~$ G_i $.
Remember that all graphs~$ G_i $ have the same set of nodes, but nodes that do not serve as cluster centers in $ G_{i-1} $ will be isolated in $ G_i $.
It might seem counter-intuitive at first that these isolated nodes are not removed from the graph, but observe that in our dynamic algorithm nodes might start or stop being cluster centers over time.
By keeping all nodes in all subgraphs, we avoid having to explicitly deal with insertions or deletions of nodes.\footnote{Note that it is easy to explicitly maintain the sets of isolated and non-isolated nodes by observing the degrees.}

Note that the $ (\fract_i, O (\tfrac{\log{n}}{\fract_i})) $-decomposition of $ G_i $ guarantees that $ | E_{i+1} | \leq \beta_i \cdot | E_i | $ in expectation, which implies the following bound.

\begin{observation}\label{obs:number of nodes and edges}
For every $ 0 \leq i \leq k $, $ |E_i| \leq m \cdot \prod_{0 \leq j \leq i - 1} \beta_j $ in expectation.\footnote{Note that for $ i = 0 $ the product $ \prod_{0 \leq j \leq i-1} \fract_j $ is empty and thus equal to $ 1 $.}
\end{observation}

We now analyze the update time for maintaining this LDD-hierarchy under insertions and deletions to the input graph~$ G $.
Note that for each level~$ i \leq k - 1 $ of the hierarchy, changes made to the graph $ G_i $ might result in the dynamic algorithm~$ \mathcal{D}_i $ making changes to the $ (\fract_i, O (\tfrac{\log{n}}{\fract_i})) $-decomposition of $ G_i $.
In particular, edges of $ G_i $ could start or stop being inter-cluster edges in the decomposition, which in turn leads to edges being added to or removed from~$ G_{i + 1} $.
Thus, a single update to the input graph~$ G $ might result in a blow-up of induced updates to be processed by the algorithms $ \mathcal{D}_1, \ldots, \mathcal{D}_{k -1} $.
To limit this blow-up, we use an additional property of our LDD-decomposition given in Theorem~\ref{thm:fully dynamic LDD}, namely the non-trivial bound on the number of edges to become inter-cluster edges after each update.

\begin{lemma}\label{lem:update time hierarchy}
The LDD-hierarchy can be maintained with an expected amortized update time of
\begin{equation*}
\tilde O \left( \sum_{0 \leq j \leq k-1} \frac{ O (\log n)^{2 (k-1)} }{\fract_j \prod_{0 \leq j' \leq j} \fract_{j'}} \right) \, .
\end{equation*}
\end{lemma}

\begin{proof}
For every $ 0 \leq i \leq k - 1 $ and every $ q \geq 1 $ define the following random variables:
\begin{itemize}
\item $ X_i (q) $: The total time spent by algorithm $ \mathcal{D}_i $ for processing any sequence of $ q $~updates to~$ G_i $.
\item $ Y_i (q) $: The total number of changes performed to $ G_{i+1} $ by $ \mathcal{D}_i $ while processing any sequence of $ q $~updates to~$ G_i $.
\item $ Z_i (q) $: The total time spent by algorithms $ \mathcal{D}_i, \ldots, \mathcal{D}_{k-1} $ for processing any sequence of $ q $~updates to~$ G_i $.
\end{itemize}
Note that the expected values of $ X_i (q) $ and $ Y_i (q) $ are bounded by Theorem~\ref{thm:fully dynamic LDD}~(the latter holds since only changes involving inter-cluster edges are propagated as updates to the next level).
We will show by induction on $ i $ that $ \mathbb{E} [Z_i (q)] = \tilde O (q \cdot \sum_{i \leq j \leq k-1} \tfrac{ O (\log n)^{2(k-i-1)} }{\fract_j \prod_{i \leq j' \leq j} \fract_{j'}}) $, which with $ i = 0 $ implies the claim we want to prove.

Before showing the proof, observe that our LDD-hierarchy uses multiple instances of the dynamic low-diameter decomposition. We can order these instances in a hierarchical manner such that changes in the instance $i$ only affect instances $i+1$ and above (this is possible because all changes propagate one way through the hierarchy). Since the random bits among levels are independent, we can think of the random bits in the previous level being fixed in advance, and hence the updates to the instance $i$ are fixed as well. The latter implies that each instance $i$ in the LDD-hierarchy is running in the oblivious adversary setting, as required by Theorem~\ref{thm:fully dynamic LDD}.

We next prove the claimed bound on $\mathbb{E} [Z_i (q)]$. In the base case $ i = k - 1 $, we know by Theorem~\ref{thm:fully dynamic LDD} that algorithm $ \mathcal{D}_{k-1} $ maintaining the $ (\fract_{k - 1}, O (\tfrac{\log{n}}{\fract_{k - 1}})) $-decomposition of $ G_{k-1} $ spends expected amortized time $ \tilde O (\frac{1}{\fract_{k-1}^2}) $ per update to $ G_{k-1} $, i.e., $ \mathbb{E} [Z_{k - 1} (q)] = \mathbb{E} [X_{k - 1} (q)] = \tilde O (q \cdot \frac{1}{\fract_{k-1}^2}) $ for any $ q \geq 1 $.
For the inductive step, consider some $ 0 \leq i < k - 1 $ and any $ q \geq 1 $.
Any sequence of $ q $ updates to $ G_i $ induces at most $ Y_i (q) $ updates to $ G_{i + 1} $.
Each of those updates has to be processed by the algorithms $ \mathcal{D}_{i+1}, \ldots, \mathcal{D}_{k-1} $.
We thus have $ Z_i (q) = X_i (q) + Z_{i+1} (Y_i (q)) $.

To bound $ \mathbb{E} [Z_i (q)] $, recall first the expectations of the involved random variables.
As by Theorem~\ref{thm:fully dynamic LDD} the algorithm~$ \mathcal{D}_i $ maintaining the $ (\fract_i, O (\tfrac{\log{n}}{\fract_i})) $-decomposition of $ G_i $ has expected amortized update time $ \tilde O (\tfrac{1}{\fract_i^2}) $, it spends an expected total time of $ \mathbb{E} [X_i (q)] = \tilde O (q \cdot \tfrac{1}{\fract_i^2}) $ for any sequence of $ q $ updates to $ G_i $.
Furthermore, over the whole sequence of $ q $~updates, the expected number of edges to ever become inter-cluster edges in the $ (\fract_i, O (\tfrac{\log{n}}{\fract_i})) $-decomposition of $ G_i $ is $ O (q \cdot \tfrac{ \log^2 n }{\fract_i}) $.
This induces at most $ O (q \cdot \tfrac{ \log^2 n }{\fract_i}) $ updates (insertions or deletions) to the graph $ G_{i+1} $, i.e., $ \mathbb{E} [Y_i (q)] = O (q \cdot \tfrac{ \log^2 n }{\fract_i}) $.
By the induction hypothesis, the expected amortized update time spent by $ \mathcal{D}_{i+1}, \ldots, \mathcal{D}_{k-1} $ for any sequence of $ q' $ updates to $ G_{i+1} $ is $ \mathbb{E} [Z_{i+1} (q')] = \tilde O (q' \cdot \sum_{i + 1 \leq j \leq k-1} \tfrac{ O (\log n)^{2(k-i-2)} }{\fract_j \prod_{i + 1 \leq j' \leq j} \fract_{j'}}) $.

Now by linearity of expectation we get
\begin{equation*}
\mathbb{E} [Z_i (q)] = \mathbb{E} \left[ X_i (q) + Z_{i+1} (Y_i (q)) \right] = \mathbb{E} \left[ X_i (q) \right] + \mathbb{E} \left[ Z_{i+1} (Y_i (q)) \right]
\end{equation*}
and by the law of total expectation we can bound $ \mathbb{E} \left[ Z_{i+1} (Y_i (q)) \right] $ as follows:
\begin{align*}
\mathbb{E} \left[ Z_{i+1} (Y_i (q)) \right] &= \sum_y \mathbb{E} \left[ Z_{i+1} (Y_i (q)) \mid Y_i (q) = y \right] \cdot \Pr [Y_i (q) = y] \\
 &= \sum_y \mathbb{E} \left[ Z_{i+1} (y) \right] \cdot \Pr [Y_i (q) = y] \\
 &= \sum_y \tilde O \left( y \cdot \sum_{i + 1 \leq j \leq k-1} \frac{ O (\log n)^{2(k-i-2)} }{\fract_j \prod_{i + 1 \leq j' \leq j} \fract_{j'}} \right) \cdot \Pr [Y_i (q) = y] \\
 &= \tilde O \left( \sum_{i + 1 \leq j \leq k-1} \frac{ O (\log n)^{2(k-i-2)} }{\fract_j \prod_{i + 1 \leq j' \leq j} \fract_{j'}} \right) \cdot \sum_y y \cdot \Pr [Y_i (q) = y] \\
 &= \tilde O \left( \sum_{i + 1 \leq j \leq k-1} \frac{ O (\log n)^{2(k-i-2)} }{\fract_j \prod_{i + 1 \leq j' \leq j} \fract_{j'}} \right) \cdot \mathbb{E} [Y_i (q)] \\
 &= \tilde O \left( \sum_{i + 1 \leq j \leq k-1} \frac{ O (\log n)^{2(k-i-2)} }{\fract_j \prod_{i + 1 \leq j' \leq j} \fract_{j'}} \right) \cdot O \left( q \cdot \frac{ \log^2 n }{\fract_i} \right) \\
 &= \tilde O \left( q \cdot \sum_{i+1 \leq j \leq k-1} \frac{ O (\log n)^{2(k-i-1)} }{\fract_j \prod_{i \leq j' \leq j} \fract_{j'}} \right)
\end{align*}
We thus get
\begin{align*}
\mathbb{E} [Z_i (q)] & = \tilde O (q \cdot \tfrac{1}{\fract_i^2}) + \tilde O \left( q \cdot \sum_{i+1 \leq j \leq k-1} \frac{ O (\log n)^{2(k-i-1)} }{\fract_j \prod_{i \leq j' \leq j} \fract_{j'}} \right) \\ 
            & = \tilde O \left( q \cdot \sum_{i \leq j \leq k-1} \frac{ O (\log n)^{2(k-i-1)} }{\fract_j \prod_{i \leq j' \leq j} \fract_{j'}} \right)
\end{align*}
as desired.
\end{proof}

Given any spanning forest $ T' $ of $ G_k $, there is a natural way of defining a spanning forest~$ T $ of~$ G $ from the LDD-hierarchy.
To this end, we first formally define the contraction of a node $ v $ of $ G $ to a cluster center $ v' $ of $ G_i $ (for $ 0 \leq i \leq k) $ as follows:
Every node $ v $ of $ G $ is contracted to itself in $ G_0 $, and, for every $ 1 \leq i \leq k $, a node $ v $ of $ G $ is contracted to $ v' $ in $ G_i $ if $ v $ is contracted to $ u' $ in $ G_{i-1} $ and $ c_{i-1} (u') = v' $.
Similarly, for every $ 0 \leq i \leq k $, an edge $ e = (u, v) $ of $ G $ is contracted to an edge $ e' = (u', v') $ of $ G_i $ if $ u $ is contracted to $ u' $ and $ v $ is contracted to $ v' $.
Now define $ T $ inductively as follows:
We let $ T_0 $ be the forest consisting of the spanning trees of diamteter $ O (\tfrac{\log{n}}{\fract_0}) $ of the clusters in the $ (\fract_0, O (\tfrac{\log{n}}{\fract_0}))$-decomposition of $ G_0 $.
For every $ 1 \leq i \leq k $, we obtain $ T_i $ from $ T_{i-1} $ and a $ (\fract_i, O(\tfrac{\log{n}}{\fract_i})) $-decomposition of $ G_i $ as follows: for every edge $ e' $ in a shortest path tree in one of the clusters, we include in $ T_i $ \emph{exactly one} edge $e$ of $G$ among the edges that are contracted to $ e' $ in $ G_i $.
Finally, $ T $ is obtained from $ T_k $ as follows: for every edge~$ e' $ in the spanning forest~$ T' $ of~$ G_k $, we include in~$ T $ the edge~$ e $ of~$ G $ contracted to~$ e' $ in~$ G_k $.
As the clusters in each decomposition are non-overlapping, we are guaranteed that $ T $ is indeed a forest.
Note that, apart from the time needed to maintain $ T' $, we can maintain $ T $ in the same asymptotic update time as the LDD-hierarchy (up to logarithmic factors).

We now partially analyze the stretch of $ T $ with respect to $ G $.

\begin{lemma}\label{lem:path length for nodes in same cluster}
For every $ 1 \leq i \leq k $, and for every pair of nodes $ u $ and~$ v $ that are contracted to the same cluster center in $ G_i $, there is a path from $ u $ to $ v $ in $ T $ of length at most $ \tfrac{O(\log{n})^i}{\prod_{0 \leq j \leq i-1} \fract_j} $.
\end{lemma}

\begin{proof}
The proof is by induction on~$ i $.
The induction base $ i = 1 $ is straightforward:
For $ u $ and $ v $ to be contracted to the same cluster center in $ G_1 $, they must be contained in the same cluster~$ C $ of the $ (\fract_0, O (\tfrac{\log{n}}{\fract_0})) $-decomposition of $ G_0 $ maintained by $ \mathcal{D}_0 $.
Remember that $ C $ has strong diameter at most $ O (\tfrac{\log{n}}{\fract_0}) $.
Thus, in the shortest path tree of $ C $ there is a path of length at most $ O (\tfrac{\log{n}}{\fract_0}) $ from $ u $ to $ v $ using edges of $ G_0 = G $.
By the definition of $ T $, this path is also present in $ T $.

For the inductive step, let $ 2 \leq i \leq k $ and let $ u' $ and $ v' $ denote the cluster centers to which $ u $ and $ v $ are contracted in $ G_{i - 1} $, respectively.
For $ u $ and $ v $ to be contracted to the same cluster center in $ G_i $, $ u' $ and $ v ' $ must be contained in the same cluster~$ C $ of the $ (\fract_{i - 1}, O (\tfrac{\log{n}}{\fract_{i - 1}})) $-decomposition of $ G_{i - 1} $ maintained by $ \mathcal{D}_{i - 1} $.
As $ C $ has strong diameter at most $ O (\tfrac{\log{n}}{\fract_{i - 1}}) $, there is a path $ \pi $ from $ u' $ to $ v' $ of length at most $ O (\tfrac{\log{n}}{\fract_{i - 1}}) $ in the shortest path tree of $ C $.
Let $ x_1, \ldots, x_t $ denote the nodes on $ \pi $, where $ x_1 = u' $ and $ x_t = v' $.
By the definition of our tree $T$ with respect to $ G $, there must exist edges $ (a_1, b_1), \ldots, (a_t, b_t) $ of~$ G $ such that
\begin{itemize}
\item $ (a_\ell, b_\ell) $ is contained in $ T $ for all $ 1 \leq \ell \leq t $,
\item $ u $ and $ a_1 $ are contracted to the same cluster center in $ G_{i-1} $,
\item $ b_t $ and $ v $ are contracted to the same cluster center in $ G_{i-1} $, and
\item $ b_\ell $ and $ a_{\ell+1} $ are contracted to the same cluster center in $ G_{i-1} $ for all $ 1 \leq \ell \leq t - 1 $.
\end{itemize}
By the induction hypothesis we know that for every $ 1 \leq \ell \leq t-1 $ there is a path of length at most $ \tfrac{O (\log{n})^{i - 1}}{\prod_{0 \leq j \leq i-2} \fract_j} $ from $ b_\ell $ to $ a_{\ell+1} $ in $ T $.
Paths of the same maximum length also exist from $ u $ to $ a_1 $ and from $ b_t $ to $ v $.
It follows that there is a path from $ u $ to $ v $ in $ T $ of length at most
\begin{align*}
(t + 1) \cdot \frac{O (\log{n})^{i - 1}}{\prod_{0 \leq j \leq i-2} \fract_j} + t & \leq 3 t \cdot \frac{O (\log{n})^{i - 1}}{\prod_{0 \leq j \leq i-2} \fract_j} \\
 & = O \left( \frac{\log{n}}{\fract_i} \right) \cdot \frac{O (\log{n})^{i - 1}}{\prod_{0 \leq j \leq i-2} \fract_j}
 = \frac{O (\log{n})^i}{\prod_{0 \leq j \leq i-1} \fract_j}
\end{align*}
as desired.
\end{proof}

To analyze the stretch of $ T $, we will use the following terminology:
we let the \emph{level} of an edge $ e $ of $ G $ be the largest $ i $ such that edge $ e $ is contracted to some edge~$ e' $ in $ G_i $.
Remember that $ E_i $ is a multiset of edges containing as many edges $ (u', v') $ as there are edges $ (u, v) \in E $ with $ u $ and $ v $ being contracted to different cluster centers $ u' $ and $ v' $ in $ G_i $, respectively.
Thus, the expected number of edges at level $ i $ is at most $ | E_i | $.
Note that for an edge $ e = (u, v) $ to be at level~$ i $, $ u $ and $ v $ must be contracted to the same cluster center in $ G_{i + 1} $.
Therefore, by \Cref{lem:path length for nodes in same cluster}, the stretch of edges at level~$ i $ in $ T $ with respect to $ G $ is at most $ \frac{O(\log{n})^{i+1}}{\prod_{0 \leq j \leq i} \fract_j} $.
The expected contribution to the total stretch of $ T $ by edges at level $ i \leq k - 1 $ is thus at most
\begin{equation}\label{eq:stretch of level i edges}
| E_i | \cdot \frac{O(\log{n})^{i+1}}{\prod_{0 \leq j \leq i} \fract_j} \leq \frac{m}{\beta_i} \cdot O (\log{n})^{i+1} \, .
\end{equation}

\subsection{Dynamic Low-Stretch Tree Algorithms}

To now obtain a fully dynamic algorithm for maintaining a low-stretch forest, it remains to plug in a concrete algorithm for maintaining $ T' $ together with suitable choices of the parameters.
We analyze two choices for dynamically maintaining $ T' $.
The first is the ``lazy'' approach of recomputing a low-stretch forest from scratch after each update to the input graph.
The second is a fully dynamic spanning forest algorithm with only trivial stretch guarantees.

\begin{theorem}[Restatement of Theorem~\ref{thm:fully dynamic low stretch tree}]
Given any unweighted, undirected graph undergoing edge insertions and deletions, there is a fully dynamic algorithm for maintaining a spanning forest of expected average stretch $ n^{o(1)} $ that has expected amortized update time $ m^{1/2 + o(1)} $.
These guarantees hold against an oblivious adversary.
\end{theorem}

\begin{proof}

We set $ k = \lceil \sqrt{\log{n}} \rceil $ and $ \fract_i = \fract = \tfrac{1}{m^{1 / (2k + 1)}} $ for all $ 0 \leq i \leq k-1 $ and maintain an LDD-hierarchy with these parameters.
Additionally, we maintain the graph $ G' $ induced by all non-isolated nodes of $ G_k $, which can easily be done by maintaining the degrees of nodes in $ G_k $.
After each update to $ G $, we compute a low-average stretch forest of $ T' $ of $ G' $.
Note that this recomputation is performed \emph{after} having updated all graphs in the hierarchy; we use the state-of-the-art static algorithm for computing a spanning forest of the multigraph $ G' $ with total stretch $ \tilde O (| E_k |) $ in time $ \tilde O (| E_k |) $.

By Equation~\eqref{eq:stretch of level i edges}, the contribution to the total stretch of $ T $ by edges at level $ i \leq k - 1 $ is at most $ m \cdot \tfrac{O (\log{n})^{i+1}}{\beta_i} $.
To bound the contribution of edges at level~$ k $, consider some edge $ e = (u, v) $ at level~$ k $ and let $ u' $ and $ v' $ denote the cluster centers to which $ u' $ and $ v' $ are contracted in $ G_k $, respectively.
Let $ \pi $ denote the path from $ u' $ to $ v' $ in $ T' $.
Using similar arguments as in the proof of Lemma~\ref{lem:path length for nodes in same cluster}, the contracted nodes and edges of $ \pi $ can be expanded to a path from $ u $ to~$ v $ in~$ T $ of length at most $ \tfrac{O (\log{n})^k}{\prod_{0 \leq i \leq k - 1} \fract_i} \cdot | \pi | $.
Thus, the contribution of edges at level~$ k $ is at most $ \tilde O (| E_k |) \cdot \tfrac{O (\log{n})^k}{\prod_{0 \leq i \leq k - 1} \fract_i} = \tilde O (m \cdot O (\log{n})^k) $ and the total stretch of~$ T $ with respect to~$ G $ is
\begin{align*}
\sum_{0 \leq i \leq k - 1} m \cdot \frac{O (\log{n})^{i+1}}{\beta} &  + \tilde O (m \cdot O (\log{n})^k) \\
 & = \tilde O \left( m \cdot \left( \frac{1}{\fract} \cdot \sum_{0 \leq i \leq k - 1} O (\log{n})^{i+1} + O (\log{n})^k \right) \right) \\
 &= \tilde O \left( m \cdot \frac{O (\log{n})^k}{\fract} \right) \\
 &= \tilde O \left( m m^{1 / (2k + 1)} \cdot O (\log{n})^k \right) \\
 &= m^{1 + o(1)} \, ,
\end{align*}
which gives an average stretch of $ m^{o(1)} = n^{o(1)} $.

By Observation~\ref{obs:number of nodes and edges}, $ G_k $ has at most $ m \beta^k $ edges in expectation and thus $ G' $ has at most $ m \beta^k $ nodes and edges in expectation.
Using the bound of Lemma~\ref{lem:update time hierarchy} for the update time of the LDD-hierarchy and the bound of $ \tilde O (m \beta^k ) $ for recomputing the low-stretch tree $ T' $ on $ G' $ from scratch, the expected amortized update time for maintaining $ T $ is
\begin{align*}
\tilde O \left( \sum_{0 \leq j \leq k-1} \frac{O (\log n)^{2 k}}{\fract_j \prod_{0 \leq j' \leq j} \fract_{j'}} + | E_k | \right) &= \tilde O \left( \sum_{0 \leq j \leq k-1} \frac{O (\log n)^{2 k}}{\fract^{j + 2}} + m \beta^k \right) \\
 &= \tilde O \left( \frac{O (\log n)^{2 k}}{\fract^{k + 1}} + m \beta^k \right) \\
 &= \tilde O (m^{(k + 1) / (2k + 1)} \cdot O (\log n)^{2 k}) \\
 &= \tilde O (m^{1/2 + 1 / (4k + 2)} \cdot O (\log n)^{2 k})  \\
 &= m^{1/2 + o(1)} \, . \qedhere
\end{align*}
\end{proof}

\begin{theorem}\label{thm:LST trade-off}
Given any unweighted, undirected graph undergoing edge insertions and deletions, there is a fully dynamic algorithm for maintaining a spanning forest of expected average stretch $ O (t + n^{1/3 + o(1)}) $ that has expected amortized update time $ \tfrac{n^{1 + o(1)}}{t} $ for every $ 1 \leq t \leq n $.
These guarantees hold against an oblivious adversary.
\end{theorem}

\begin{proof}
We set $ k = \lceil \log{\log{n}} \rceil $, $ \fract_0 = \sqrt{t/n} $ and $ \fract_i = \sqrt{\fract_{i-1}} $ for all $ 1 \leq i \leq k-1 $ and maintain an LDD-hierarchy with these parameters.
The spanning forest $ T' $ is obtained by fully dynamically maintaining a spanning forest of $ G_k $ using any algorithm with polylogarithmic update time. 

By Equation~\eqref{eq:stretch of level i edges}, the contribution to the total stretch of $ T $ by edges at level $ i \leq k - 1 $ is at most $ \frac{m}{\beta_i} \cdot O (\log{n})^{i+1} $.
For every edge $ e = (u, v) $ at level~$ k $ with $ u $ contracted to $ u' $ and $ v $ contracted to $ v' $ in $ G_k $, there is a path from $ u' $ to $ v' $ in $ T' $ that by undoing the contractions can be expanded to a path from $ u $ to $ v $ in $ T $, which trivially has length at most $ n - 1 $.
Thus, the contribution by each edge at level $ k $ is at most $ n - 1 $.
As for every $ 0 \leq i \leq k $ there are at most $ | E_i | = m \cdot \prod_{0 \leq j \leq i - 1} \beta_j $ edges at level $ i $ in expectation, we can bound the expected total stretch of $ T $ with respect to $ G $ as follows:
\begin{align*}
\sum_{0 \leq i \leq k-1}  | E_i | \cdot \frac{O(\log{n})^{i+1}}{\prod_{0 \leq j \leq i} \fract_j} & + | E_k | \cdot n  \\ & =
\sum_{0 \leq i \leq k-1} \frac{m \cdot O (\log{n})^{i+1}}{\fract_i} + m \cdot \prod_{0 \leq i \leq k-1} \fract_i \cdot n \\
 &= m \cdot \left( \sum_{0 \leq i \leq k-1} \frac{O (\log{n})^{i+1}}{\fract_i} + \prod_{0 \leq i \leq k-1} \fract_i \cdot n \right)
\end{align*}
This gives an average stretch of $ \sum_{0 \leq i \leq k-1} \tfrac{O (\log{n})^{i+1}}{\fract_i} + \prod_{0 \leq i \leq k-1} \fract_i \cdot n $.
We now simplify these two terms.
Exploiting that $ \fract_i \geq \fract_0 $ for all $ 1 \leq i \leq k-1 $, we get
\begin{align*}
\sum_{0 \leq i \leq k-1} \frac{O (\log{n})^{i+1}}{\fract_i} & \leq \sum_{0 \leq i \leq k-1} \frac{O (\log{n})^{i+1}}{\fract_0} = \frac{O (\log{n})^k}{\fract_0} \\ & = \frac{O (\log n)^k}{\sqrt{t/n}} = \sqrt{\frac{n^{1+o(1)}}{t}} \, .
\end{align*}
Furthermore, the geometric progression of the $ \fract_i $'s gives
\begin{align*}
\prod_{0 \leq i \leq k-1} \fract_i \cdot n & = \prod_{0 \leq i \leq k-1} \fract_0^{1/2^i} \cdot n = \fract_0^{\sum_{0 \leq i \leq k-1} 1/2^i} \cdot n = \fract_0^{2 - 1/2^{k-1}} \cdot n \\ 
& = \frac{t^{1 - 1/2^k}}{n^{1 - 1/2^k}} \cdot n \leq t \cdot n^{1/2^k} = O (t) \, .
\end{align*}
The average stretch of the forest maintained by our algorithm is thus at most $ O ( t + \sqrt{\tfrac{n^{1+o(1)}}{t}} ) $, which, after balancing the two terms, can be rewritten as $ O (t + n^{1/3 + o(1)}) $.

It remains to bound the update time of the algorithm.
By Lemma~\ref{lem:update time hierarchy}, the hierarchy can be maintained with an amortized update time of
\begin{align*}
\tilde O \left( \sum_{0 \leq j \leq k-1} \frac{O (\log n)^{2 k}}{\fract_j \cdot \prod_{0 \leq j' \leq j} \fract_{j'}} \right) & = \tilde O \left( \sum_{0 \leq j \leq k-1} \frac{O (\log n)^{2 k}}{\fract_0^{1 / 2^j} \cdot \fract_0^{2 - 1 / 2^j}} \right) \\
&  = \tilde O \left( \sum_{0 \leq j \leq k-1} \frac{O (\log n)^{2 k}}{\fract_0^2} \right)  \\ 
& = \frac{n \cdot O (\log n)^{2 k}}{t} = \frac{n^{1 + o(1)}}{t} \, .
\end{align*}
Since the amortized number of changes to $ G_k $ per update to $ G $ is trivially bounded by $ \tfrac{n^{1 + o(1)}}{t} $ as well and since $ T' $ can be maintained with polylogarithmic amortized time per update to $ G_k $, we can maintain $ T $ with amortized update time $ \tfrac{n^{1 + o(1)}}{t} $.
\end{proof}

Note that the algorithm of Theorem~\ref{thm:fully dynamic low stretch tree} is superior to the algorithm of Theorem~\ref{thm:LST trade-off} as long as $ t \leq \sqrt{n} $.
If $ t \geq \sqrt{n} $, then the algorithm of Theorem~\ref{thm:LST trade-off} provides stretch $ O (t) $ and update time $ \tfrac{n^{1 + o(1)}}{t} $.

\subsection{Input Graph Sparsification}\label{apx:sparsifier}

In the following, we explain how input graph sparsification can be performed to the algorithm of Theorem~\ref{thm:fully dynamic low stretch tree} by running the algorithm on a cut sparsifier, similar to the approach of Koutis et al.~\cite{KoutisLP16} in the \emph{static} setting.

\begin{corollary} [Restatement of Corollary~\ref{cor:fully dynamic low stretch tree}] \label{corApp:fully dynamic low stretch tree}
Given any unweighted, undirected graph undergoing edge insertions and deletions, there is a fully dynamic algorithm for maintaining a spanning forest of expected average stretch $ n^{o(1)} $ that has expected amortized update time $ n^{1/2 + o(1)} $.
These guarantees hold against an oblivious adversary.
\end{corollary}

To make the analysis rigorous, we introduce some additional notation for multigraphs.

\paragraph{Succinct Representation of Multigraphs.}
A multigraph $ G = (V, E) $ consists of a set of nodes~$ V $ and a multiset of edges~$ E $.
We denote by $ \bar{E} = \{ (u, v) \in \binom{V}{2} \mid (u, v) \in E \} $ the support of the multiset~$ E $.
This allows a multigraph $ G = (V, E) $ to be succinctly represented as its \emph{skeleton} $ \bar{G} = (V, \bar{E}, \mu_G) $ where $ \mu_G $ is a multiplicity function $ \mu_G : \bar{E} \rightarrow \mathbb{Z}^+ $ that assigns to each edge $ e $ its (positive integer) multiplicity $ \mu_G (e) $.
We denote by $ m \coloneqq | E | $ the number of multi-edges (considering multiplicities), and by $ \bar{m} \coloneqq | \bar{E} | $ the size of the support of $ E $ (disregarding multiplicities).
For simplicity, we assume that $ m $ is polynomial in~$ n $.
The total stretch of a spanning forest~$ T $ is defined with respect to~$ E $, i.e.,
\begin{equation}
\str_T (G) = \sum_{e = (u, v) \in E(G)} \dist_T (u, v) = \sum_{e = (u, v) \in \bar{E}(G)} \mu_G (e) \cdot \dist_T (u, v) \, . \label{eq:definition of total stretch}
\end{equation}

Our dynamic algorithm will exploit that, given the skeleton of a multigraph, a low-stretch forest of can be computed without (significant) dependence on the multiplicities.
\begin{lemma}\label{lem:multiplicity oblivious running time}
Given the skeleton $ \bar{G} $ of a multigraph $ G $, a spanning forest of $ G $ of total stretch $ m^{1 + o(1)} $ can be computed in time $ \tilde O (\bar{m}) $.
\end{lemma}
Such a guarantee can be achieved with a \emph{static} version of our algorithm, i.e., by combining the scheme of Alon et al.~\cite{AlonKPW95} with the LDD of Miller et al.~\cite{MillerPX13}.
Although we are not aware of any statement of such a ``multiplicity-oblivious'' running time in the literature, it seems plausible that the state-of-the art algorithms (achieving a total stretch of $ \tilde O (m) $) also have this property.
Note however that a stretch of $ m^{1 + o(1)} $ is anyway good enough for our purpose.

\paragraph{Refined Analysis of Dynamic Low-Stretch Tree Algorithm.}

We now restate the guarantees of our fully dynamic low-stretch forest algorithm when the input is a multigraph undergoing insertions and deletions of multi-edges (i.e., each update increases or decreases the multiplicity of some edge by~$ 1 $).
Our fully dynamic LDD algorithm maintains a clustering such that every edge is an inter-cluster edge with probability $ \beta $.
This implies that at most a $ \beta $-fraction of the edges are inter-cluster edges in expectation -- regardless of whether we consider multiplicities.
More precisely, contracting the clusters to single nodes yields a multigraph $ G' = (V', E') $ with $ | E' | \leq \beta | E | $ and $ | \bar{E}' | \leq \beta | \bar{E} | $.
Now, in particular the LDD hierarchy in the proof of Theorem~\ref{thm:fully dynamic low stretch tree} results in a multigraph $ G' = (V', E') $ with $ | E' | \leq \beta^k m $ and $ | \bar{E}' | \leq \beta^k \bar{m} $ (after $k$ levels).
For such a graph, if its skeleton is given explicitly, one can compute a spanning forest of total stretch $ O (|E'|^{1 + o(1)}) $ in time $ \tilde O (|\bar{E}'|) $ by Lemma~\ref{lem:multiplicity oblivious running time}.
Note that our dynamic algorithm can explicitly maintain the skeleton of $ G' $ with neglegible overheads in the update time.
It follows that our algorithm maintains a spanning forest of total stretch $ O (m^{1 + o(1)}) $ and has an update time of $ \tilde O (\bar{m}^{1/2 + o(1)}) $.

\paragraph{Cut Sparsifiers.}
For the definition of cut sparsifiers, we consider cuts of the form $ (U, V \setminus U) $ induced by a subset of nodes $ U \subset V $.
The capacity of such a cut $ (U, V \setminus U) $ in a graph $ G $ is defined as the total multiplicity of edges crossing the cut, i.e.,
\begin{equation*}
\capacity_G (U, V \setminus U) = \sum_{\substack{e = (u, v) \in \bar{E} \\ u \in U, v \in V \setminus U}} \mu_G (e)
\end{equation*}
A $ (1 \pm \epsilon) $-\emph{cut sparsifier}~\cite{BenczurK15} (with $ 0 \leq \epsilon \leq 1/2 $) of a multigraph $ G = (V, E) $ is a ``subgraph'' $ H = (V, F) $ with $ \bar{F} \subseteq \bar{E} $ such that for every $ U \subset V $ we have
\begin{equation*}
(1 - \epsilon) \capacity_G (U, V \setminus U) \leq \capacity_H (U, V \setminus U) \leq (1 + \epsilon) \capacity_G (U, V \setminus U) \, ,
\end{equation*}
i.e., $ H $ approximately preserves all cuts of $ G $.
Now let $ H $ be a $ (1 \pm \epsilon) $-\emph{cut sparsifier} of a multigraph $ G = (V, E) $ and let $ T = (V, E(T))$ be a (simple) spanning forest of $ H $.
For every edge $ e $ of the forest~$ T $, the nodes are naturally partitioned into two connected subsets upon removal of~$ e $.
Let these two subsets be denoted by $ V_e $ and $ V \setminus V_e $.
Emek~\cite{Emek11} and Koutis et al.~\cite{KoutisLP16}, observed that by rearranging the sum in~\eqref{eq:definition of total stretch}, one obtains the following cut-based characterization of the stretch:
\begin{equation*}
\str_T (G) = \sum_{e \in E(T)} \capacity_G (V_e, V \setminus V_e) \, .
\end{equation*}
Observe that the cut $ (V_e, V \setminus V_e) $ is approximately preserved in $ H $ and thus $ \capacity_G (V_e, V \setminus V_e) \leq \tfrac{1}{1 - \epsilon} \capacity_H (V_e, V \setminus V_e) \leq (1 + 2 \epsilon) \capacity_H (V_e, V \setminus V_e) $.
The stretch of $ G $ with respect to~$ T $ can now be bounded by
\begin{align*}
\str_T (G) &= \sum_{e \in E(T)} \capacity_G (V_e, V \setminus V_e) \\
&\leq (1 + 2 \epsilon) \sum_{e \in E(T)} \capacity_H (V_e, V \setminus V_e) \\
&= (1 + 2 \epsilon) \str_T (H) \, .
\end{align*}
Thus, computing the low-stretch forest on the sparsifier $ H $ instead of the original graph $ G $ only increases the total stretch by a constant factor if the number of multi-edges in $ H $ is proportional to the number of edges in $ G $.

\paragraph{Dynamic Cut Sparsifiers.}
The fully dynamic algorithm of Abraham et al.~\cite{AbrahamDKKP16} maintains, with high probability, a $ (1 \pm \epsilon) $-cut sparsifier~$ H = (V, F) $ of a simple graph~$ G = (V, E) $ such that $ | \bar{F} | = \tilde O (n / \epsilon^2) $ with update time $ \poly (\log n, \epsilon) $.
For each node $ v $, the degree in $ H $ exceeds the degree in $ G $ by at most a factor of $ (1 \pm \epsilon) $ because the cut $ (\{ v \}, V \setminus \{ v \}) $ is approximately preserved in $ H $.
We can thus bound the number of multi-edges in $ H $ (i.e., the sum of all edge multiplicities) by $ | F | = O ((1 + \epsilon) |E|) $.
The algorithm maintains a hierarchy of the edges with $ O (\log n) $ layers, where edges at level~$ i $ have multiplicity $ 4^i $ and each edge is at level~$ i $ with probability at most $ 1/4^i $.
After an update to the input graph, the dynamic algorithm adds or removes at most $ \poly(\log n, \epsilon) $ edges in each level.
Thus, we can bound the amount of change to $ H $ per update to $ G $ as follows: for every update to $ G $, the expected sum of the changes to the edge multiplicities of~$ H $ is at most $ \poly(\log n, \epsilon) $.

\paragraph{Putting Everything Together~(Proof of Corollary~\ref{corApp:fully dynamic low stretch tree}).}
We now first use the fully dynamic algorithm of Abraham et al. to maintain a cut sparsifier $ H = (V, F) $ of the input graph $ G = (V, E) $ (with $ \epsilon = 1/2 $) and second run our fully dynamic low-stretch tree algorithm on top of $ H $.
Here, $ G $ is a simple graph with $ m = | E | $ edges and $ H $ is a multigraph with $ |F| = O (m) $ and $ |\bar{F}| = \tilde O (n) $.
The spanning forest $ T $ maintained in this way gives expected total stretch at most $ |F|^{1 + o(1)} $ with respect to $ H $.
As argued above, this implies an expected total stretch of at most $ O ((1 + 2 \epsilon) |F|^{1 + o(1)}) = O (m^{1 + o(1)}) $ with respect to $ G $, i.e., an average stretch of $ m^{o(1)} = n^{o(1)} $.
Each update to the input graph results in $ \polylog{n} $ changes to the sparsifier in expectation, which are then processed as ``induced'' updates by our dynamic low-stretch tree algorithm.
Thus, we overall arrive at an expected amortized update time of $ \tilde O (|\bar{F}|^{1/2 + o(1)}) = O (n^{1/2 + o(1)}) $.

\section{Dynamic Low-Diameter Decomposition}\label{sec:LDD}

In this section we develop our dynamic algorithm for maintaining a low-diameter decomposition following three steps.
First, we review the static algorithm for constructing a low-diameter decomposition using the clustering due to Miller et al.~\cite{MillerPX13}.
Second, we design a decremental algorithm by extending the Even-Shiloach algorithm~\cite{EvenS81} in a suitable way.
Third, we lift our decremental algorithm to a fully dynamic one by using a ``lazy'' approach for handling insertions.

\subsection{Static Low-Diameter Decomposition}~\label{sec:staticLDD}

In the following, we review the static algorithm for constructing a low-diameter decomposition clustering due to Miller et al.~\cite{MillerPX13}. Let $G=(V,E)$ be an unweighted, undirected multigraph $G$, and let $\beta \in (0,1)$ be some parameter. Our goal is to assign each node $u$ to exactly one node $c(u)$ from $V$. Let $C(u) \subset V$ denote the set of nodes assigned to node $u$, i.e., $C(u) \coloneqq \set{v \in V}{ c(v) = u}$. For each node $u$, we initially set $C(u)= \varnothing $ and pick independently a \emph{shift} value $\delta_u$ from $\expdis(\beta)$. Next, we assign each node $u$ to a node $v$, i.e., set $c(u) = v$ and add $u$ to $C(v)$, if $v$ is the node that minimizes the \emph{shifted distance} $m_v(u)\coloneqq \dist(u,v) - \delta_v$. Finally, we output all clusters that are non-empty. The above procedure is summarized in Algorithm~\ref{alg:partition}.

\begin{algorithm2e}[htb!]
\caption{Partitioning Using Exponentially Shifted Shortest Paths}
\label{alg:partition}
\Input{Multigraph $G=(V,E)$, parameter $\beta \in (0,1)$}
\Output{Decomposition of $G$}
\BlankLine

For each $u \in V$, set $C(u) \gets \varnothing$ and pick $\delta_u$ independently from $\expdis(\beta)$ \;
Assign each $u \in V$ to $c(u) \gets \arg \min_{v \in V} \{\dist(u,v) - \delta_v\}$\;
For each $v \in V$, set $C(u) \gets \set{v \in V}{c(v) = u}$\;
Return the clustering $\set{C(u)}{C(u) \neq \varnothing}$
\end{algorithm2e}

The following theorem gives bounds on the strong diameter and the number of inter-cluster edges output by the above partitioning.

\begin{theorem}[\cite{MillerPX13}, Theorem~1.2] \label{thm: Miller_LDD} Given an undirected, unweighted multigraph graph $G=(V,E)$ and a parameter $\beta \in (0,1)$, Algorithm~\ref{alg:partition} produces a $(\beta, 2 d \cdot(\log n/\beta))$-decomposition such that the guarantee on the number of inter-cluster edges holds in expectation, while the diameter bound holds with probability at least $1-1/n^d$, for any $d \geq 1$.
\end{theorem}

Here, the the diameter bound holds when the maximum shift value of any node is at most $ d \log n/\beta $, which happens with probability $1-1/n^d$.
We remark that in the work of Miller et al., the above guarantees are stated only for undirected, unweighted \emph{simple} graphs. However, by~Lemma 4.4 in~\cite{MillerPX13}, we get that each edge $e \in E$~(regardless of whether $E$ allows parallel edges) is an inter-cluster edges with probability at most $\beta$. By linearity of expectation, it follows that the (expected) number of inter-cluster edges in the resulting decomposition is at most $\beta m$, thus showing that the algorithm naturally extends to multigraphs.

For technical reasons, it is not sufficient in the analysis of our decremental LDD algorithm to apply Theorem~\ref{thm: Miller_LDD} in a black-box manner.
We thus review the crucial properties of the clustering algorithm, which we will exploit for bounding the number of changes made to inter-cluster edges in the decremental algorithm. Following~\cite{MillerPX13}, for each edge $e=(u,v) \in E$, let $w$ be the \emph{mid-point} of $e$, i.e., the imaginary node in the ``middle'' of edge $ e $ that is at distance $ \tfrac{1}{2} $ to both $ u $ and $ v $. Lemma~4.3 in \cite{MillerPX13} states that if $u$ and $v$ belong to two different clusters, i.e., $c(u) \neq c(v)$, then the shifted-distances $m_{c(u)}(w)$ and $m_{c(v)}(w)$ are within $1$ of the minimum shifted distance to $w$.

\begin{lemma}[\cite{MillerPX13}]\label{lem:inter cluster edge implies mid-point condition}
Let $e=(u,v)$ be an edge with mid-point $w$ such that $c(u) \neq c(v)$ in Algorithm~\ref{alg:partition}. Then $m_{c(u)}(w)$ and $m_{c(v)}(w)$ are within $1$ of the minimum shifted distance to $w$. 
\end{lemma}

Lemma 4.4 of~\cite{MillerPX13} shows that the probability that the smallest and the second smallest shifted distances to $w$ are within $c$ of each other is at most $c \cdot \beta$.

\begin{lemma}[\cite{MillerPX13}]\label{lem:probability of mid-point condition}
Let $ e = (u,v) $ be an edge with mid-point $ w $.
Then \[ \Pr [| m_{c(u)}(w) - m_{c(v)}(w) | \leq c] \leq c \cdot \beta. \]
\end{lemma}

Setting $c=1$, this gives the desired bound of $ \beta $ for the probability of an edge being an inter-cluster edge in Theorem~\ref{thm: Miller_LDD}.

\paragraph*{Implementation.} 
Na\"ively, we could implement Algorithm~\ref{alg:partition} by computing $c(u)$ for each node $u \in V$ in $\tilde{O}(m)$, thus leading to a $\tilde{O}(mn)$ time algorithm. In the following, using standard techniques, we show that this running time can be reduced to $\tilde{O}(m)$.

To this end, let $\delta_{\max} \coloneqq \max_{u \in V} \{ \delta_u \}$. We begin with the following augmentation of the input graph $G$: add a new source $s$ to $G$ and edges $(s,u)$ of weight $(\delta_{\max} - \delta_u) \geq 0$, for every $u \in V$. Let $\hat{G}=(V \cup \{s\},\hat{E},\hat{\ww})$ denote the resulting graph. We claim that the sub-trees below the source $s$ in the shortest path tree of $\hat{G}$ rooted at $s$ give us the clustering output by Algorithm~\ref{alg:partition} for the graph $G$. To see this, suppose that we instead added edges of weight $-\delta_u$ to $s$, for every $u \in V$. Then it is easy to check that for every $u \in V$, the distance between $s$ and $u$ is exactly $ \min_{v \in V} (\dist(u,v) - \delta_v) = \min_{v \in V} m_v(u)$. Thus the node $v$ attaining the minimum is exactly the root of the sub-tree below the source $s$ that contains $v$. Now, adding $\delta_{\max}$ to all edges incident to the source increases all distances to $s$ by $\delta_{\max}$, and thus does not affect the shortest path tree.

Now, note that we could use Dijkstra's algorithm to construct the shortest path tree of $\hat{G}$, and modify it appropriately to output the clustering. However, for reasons that will become clear in the next section, we need to modify Dijkstra's algorithm in a specific way. This modification can be viewed as mimicking a BFS computation on a graph with special \emph{integral} edge lengths. 

We start by observing that due to the random shift values, the weight of the edges incident to the source $s$ in $\hat{G}$ are not integers. Since we only want to deal with \emph{integral} weights, we round down all the $\delta_u$ values to $\floor{\delta_u}$ and modify the weights of these edges using the new rounded values. Let $G'=(V \cup \{s\}, \hat{E}, \ww')$ denote the modified graph. Note that due to the rounding, we need to introduce some tie-breaking scheme in $G'$, such that  every clustering of $G'$ matches exactly the same clustering in $\hat{G}$, and vice versa. Naturally, the fractional parts of the rounded values, i.e., $\delta_u - \floor{\delta_u}$, define an ordering on the nodes (if they are sorted in ascending order), and this ordering can be in turn used to break ties whenever two rounded distances are equal in $G'$. In their PRAM implementation, Miller et al.~\cite{MillerPX13} observed that this ordering can emulated by a random permutation. This is due to the fact that the shifts are generated independently, and that the exponential distribution is memoryless. 

The main motivation for using random permutations in previous works was to avoid errors that might arise from the machine precision. In our work, breaking ties according to a random permutation on the nodes is one of their algorithmic ingredients that allows us to obtain an efficient dynamic variant of the clustering. Below, we give specific implementation details about how our clustering interacts with random priorities in the static setting. 

Given the graph $G'$ and a distinguished source node $s \in V'$,  Dijkstra's classical algorithm maintains an upper-bound on the shortest-path distance between each node $u \in V$ and $s$, denoted by $\ell(u)$. Initially, it sets $\ell(u) = \infty$, for each $u \in V$ and $\ell(s) = 0$. It also marks every node unvisited. Moreover, for each node $u \in V$, the algorithm also maintains a pointer $p(u)$ (initially set to $\nil$), which denotes the parent of $u$ in the current tree rooted at $s$. Using these pointers, we can maintain the cluster pointer $c(u)$, for each $u \in V$. This follows from the observation that in order to compute the cluster of $u$, it suffices to know the cluster of its parent. Formally we have the following rule.

\begin{observation} \label{obs: clusterUpdate}
Let $p(\cdot)$ be the parent pointers. Then for each $u \in V$, we can determine the cluster pointer $c(u)$ using the following rule:

\[
	c(u) =  \begin{cases} 
    u & \text{\emph{if} } p(u)=s \\
    c(p(u))      & \text{\emph{otherwise}}.
  \end{cases}
\]
\end{observation}

Now, at each iteration, Dijkstra's algorithm selects an unvisited node $u$ with the smallest $\ell(u)$, marks it as visited, and \emph{relaxes} all its edges. In the standard relaxation, for each edge $(u,v) \in E'$ the algorithm sets $\ell(v) \gets \min\{\ell(v), \ell(u) + w'(u,v)\}$ and updates $p(v)$ accordingly. Here, we present a relaxation according to the following tie-breaking scheme. Let $\pi$ be a random permutation on $V$. For $u, v \in V$, we write $\pi(u) < \pi(v)$ if $u$ appears before $v$ in the permutation $\pi$.
Now, when relaxing an edge $(u,v) \in E'$, we set $u$ to be the parent of $v$, i.e., $p(v) = u$, and $\ell(v) = \ell(u) + w'(u,v)$, if the following holds
\begin{align} \label{eq: breakingTies}
\begin{split}
\ell(v) & > \ell(u) + w'(u,v), \text{ or} \\
\ell(v) & = \ell(u) + w'(u,v) \text{ and } \pi(c(v)) > \pi(c(u)).
\end{split}
\end{align}

After each edge relaxation, we also update the cluster pointers using  Observation~\ref{obs: clusterUpdate}. We continue the algorithm until every node is visited. As usual, we maintain the unvisited nodes in a heap $Q$, keyed by the their estimates $\ell(v)$. This procedure is summarized in Algorithm~\ref{alg:ModDijkstra}.

\begin{algorithm2e}[htb!]
\caption{Modified Dijkstra}
\label{alg:ModDijkstra}
\Input{Graph $G'=(V \cup \{s\},E',\ww')$}
\Output{Decomposition of $G$}
\BlankLine

Generate random permutation $\pi$ on $V$\;
\ForEach{$u \in V$}{
Set $\ell(u) \gets \infty$\;
Set $p(u) \gets \nil$\;
Set $c(u) \gets \nil$\;
}

Set $\ell(s) \gets 0$\;
\BlankLine

Add every $u \in V \cup \{s\}$ into heap $Q$ with key $\ell(u)$\;
\While{heap $ Q $ is not empty}{
	    Take node $ u $ with minimum key $ \lev (u) $ from heap $ Q $ and remove it from $ Q $ \;
		\ForEach{neighbor $ v $ of $ u $}{
		\Relax{$u$, $v$, $w'$, $\fraction$} \;
		\eIf{$ p(v) = s $}{
			Set $ c (v) \gets v  $\;
		}{
			Set $ c (v) \gets c( p(v)) $\;
		}
	}
}

\Procedure{\Relax{$u$, $v$, $w'$, $\fraction$}}{
	\uIf{$\ell(v) > \ell(u) + w'(u,v)$}{
		Set $\ell(v) \gets \ell(u) + w'(u,v)$\;
		Set $p(v) \gets u$\;
		}
	\uElseIf{$\ell(v) = \ell(u) + w'(u,v)$ \emph{and} $\pi(c(v)) > \pi(c(u))$}{
     Set $p(v) \gets u$ \;}
}
\end{algorithm2e}

Correctness of Algorithm~\ref{alg:ModDijkstra} follows by our above discussion. Moreover, the running time of the algorithm is asymptotically bounded by the running time of Dijkstra's classical algorithm and the time to generate a random permutation. It is well known that the former runs in $\tilde{O}(m)$ time and the latter can be generated in $O(n)$ time (see e.g., Knuth Shuffle~\cite{BermanK76}), thus giving us a total $\tilde{O}(m)$ time.

\subsection{Decremental Low-Diameter Decomposition}\label{sec:ES tree}\label{sec:decremental LDD}

We now show how to maintain a lower-diameter decomposition under deletion of edges. Recall that in the previous section we observed that computing a lower-diameter decomposition of a undirected, unweighted graph can be reduced to the single-source shortest path problem in some modified graph. In the same vein, we observe that maintaining a low-diameter decomposition under edge deletions amounts to maintaining a bounded-depth single-source shortest path tree of some modified graph under edge deletions. 

Even and Shiloach~\cite{EvenS81} devised a data-structure for maintaining a bounded-depth SSSP-tree under edge deletions, which we refer to as \emph{ES-tree}. The ES-tree initially worked only for undirected, unweighted graphs. However, later works~\cite{HenzingerK95,King99} observed that it can be extended even to directed, weighted graphs with positive integer edges weights. The mere usage of the ES-tree as a sub-routine will not suffice for our purposes, due to the constraints that our clustering imposes. In the following we show how to augment and modify an ES-tree that maintains a valid clustering, without degrading its running time guarantee. 

Let $G=(V,E)$ be an undirected, unweighted graph for which we want to maintain a decremental $(\beta, \log n/ \beta)$ decomposition, for any parameter $\beta \in (0,1)$. Further, let $G'=(V\cup \{s\},E',\ww')$ be the undirected graph with integral edge weights, as defined in Section~\ref{sec:staticLDD}.
Let $\pi$ be a random permutation on $V$. By discussion in Section~\ref{sec:staticLDD}, in order to maintain a low-diameter decomposition of $G$ it suffices to maintain a clustering of $G'$ with $\pi$ used for tie-breaking.

We describe an ES-tree that efficiently maintains a clustering of $G'$ for a given root node $s$ and a given distance parameter $\Delta$. Here we set $\Delta = O(\log n / \beta)$, as by Theorem~\ref{thm: Miller_LDD}, the maximum distance that we run our algorithm to is bounded by $O(\log n / \beta)$. Our data-structure handles arbitrary edge deletions, and maintains the following information. First, for each node $u \in V \cup \{s\}$, we maintain a label $\lev(u)$, referred to as the \emph{level} of $u$. This level of $u$ represents the shortest path between the root $s$ and $u$, i.e., $\lev(u) = \dist(s,u)$. Next, for each node $u \in V$, we maintain pointers $p(u)$ and $c(u)$, which represent the parent of $u$ in the tree and the node that $u$ is assigned to, respectively. Finally, we also maintain the set of \emph{potential} parents $P(u)$, for each $u \in V$, which is the set of all neighbors of $u$ that are in the same level with the parent of $u$, and share the same clustering with $u$, i.e., a neighbor $v$ of $u$ belongs to $P(u)$ if $v$ minimizes $(\ell(v) + \ww'(u,v), \pi(c(v)))$ lexicographically, and $c(v) = c(u)$. Edge deletions in $G'$ can possibly affect the above information for several nodes. Our algorithm adjusts these information on the nodes so as to make them valid for the modified graph.

\paragraph*{Algorithm Description and Implementation.}
We give an overview and describe the implementation of Algorithm~\ref{alg:ES_tree}. The data-structures $\ell(\cdot)$, $p(\cdot)$ and $c(\cdot)$ are initialized using Algorithm~\ref{alg:ModDijkstra} in Section~\ref{sec:staticLDD}. Note that for each $u \in V$, $P(u)$ can be computed by simply considering all neighbors of~$u$ in turn, and adding a neighbor $v$ to $P(u)$ if $v$ is a potential parent. The algorithm also maintains a heap $Q$ whose intended use is to store nodes whose levels or clustering might need to be updated. (see procedure \Initialize{}). 

In our decremental algorithm, each node tries to maintain its level $\ell(u)$, which corresponds to its current distance to the root $s$, together with its cluster pointer $c(u)$ in the current graph. Concretely, we maintain the following invariant for each node $u \in V$:
\begin{align}
\ell(u)&=\min\set{\ell(v)+\ww'(u, v)}{v \text{ is a neighbor of } u} \label{eq:ES tree level condition}
\end{align}
where ties among neighbors are broken according to~(\ref{eq: breakingTies}). This invariant allows to compute the cluster pointer $c(u)$ using Observation~\ref{obs: clusterUpdate}. Deleting an edge incident to $u$ might lead to a change in the values of $\ell(u)$ and $c(u)$. If this occurs, all neighbors of $u$ are notified by $u$ about this change, since their levels and cluster points might also change. It is well-known that the standard ES-tree can efficiently deal with changes involving the levels $\ell(\cdot)$. However, in our setting, it might be the case that an edge deletion forces a node $u$ to change its cluster while the level $\ell(u)$ still remains the same under this deletion. This is the point where our algorithm differs from the standard ES-tree, and we next show that (1) such changes can be handled efficiently, and (2) the number of cluster changes per node, within the same level, is small in expectation.

Let us consider the deletion of an edge $(u,v)$ (see procedure $\Delete()$); assume without loss of generality that $\ell(v) \leq \ell(u)$. Now note that an edge deletion might lead to a cluster change only if $v \in P(u)$. If this is the case, the algorithm first removes $v$ from the set $P(u)$. If $P(u)$ is still non-empty, the clustering remains unaffected. However, if $P(u)$ is empty, the clustering of $u$ will change, and the algorithm inserts $u$ into the heap $Q$ with key $\ell(u)$. Observe that a change in clustering of $u$ might potentially lead to cluster changes for children of $u$, given that $u$ was their only potential parent. In this way, we observe that deleting $(u,v)$ might force changes in the clustering for many descendants of $v$. The algorithm handles such changes using procedure $\UpdateLevels()$, which we describe below.

Procedure $\UpdateLevels()$ considers the nodes in $Q$ in the order of their current level. At each iteration, it takes the node $y$ with the smallest level $\ell(y)$ from $Q$. The node $y$ computes the set of potential parents $P(y)$, by examining each neighbor of $y$ in turn, and then adding to $P(y)$ all neighbors $z$ that minimize $(\ell(z) + \ww'(y,z), \pi(c(z)))$ lexicographically. Next, $y$ sets $p(y)$ as one of nodes in $P(y)$, and updates its level by setting $\ell(y) = \ell(p(y)) + \ww'(y,p(y))$. Having computed its parent pointer, $y$ updates the cluster pointer using Observation~\ref{obs: clusterUpdate}. Specifically, if the parent of $y$ is the source node $v$, then $y$ form a new cluster itself, i.e., $c(y) = y$. Otherwise, $y$ shares the same cluster with its parent and sets $c(y) = c(p(y))$. Finally, the algorithm determines whether the change in the clustering of $y$ affected its neighbors. Concretely, for each neighbor $x$ of $y$, it checks whether $y \in P(x)$. If this is not the case, then there is no change in the clustering of $x$. Otherwise, $y$ is removed from $P(x)$, and if $P(x)$ becomes empty after this removal, the algorithm inserts $x$ into the heap $Q$ with key $\ell(x)$, given that $Q$ does not already contain $x$. 

\paragraph*{Running Time Analysis.}

We first concern ourselves with the number of cluster change per node in our decremental algorithm. For any node $v \in V$, we say that the clustering \emph{changes} for~$v$ due to an edge deletion if this deletion either increases the level $\ell(v)$ or forces a change in the cluster pointer $c(v)$. It is well-known that the ES-tree can handle a level increase for any node $v$ in time $O(\deg(v))$. As we will see next, we can also handle a cluster change for a node in the same level in $O(\deg(v) \log n )$ time. However, we need to ensure that the number of such cluster changes for any node and any fixed level is small, for our algorithm to be efficient. Below we argue that one can have a fairly good bound on the expected number of such changes, and this is due to the special tie-breaking scheme we use when assigning nodes to clusters.

Fix any node $v \in V$, and consider $v$ during the sequence of edge deletions. Note that since only deletions are allowed, the level $\ell(v)$ is non-decreasing. This induces a natural partitioning of the sequence of edge deletions into subsequences such that the $\ell(v)$ remains unaffected during each subsequence. Specifically, for every node $v \in V$ and every $0 \leq i \leq \Delta$, let $S(i)$ the be subsequence of edge deletions during which $\ell(v) = i$, where $\Delta \leq O(\log n / \beta)$. The following bound on the expected number of cluster changes of $v$ during $S(i)$ follows an argument by Baswana et al.~\cite{BaswanaKS12}.

\begin{lemma} \label{lem: nrClusterChanges}
For every node $ v \in V$ and every $0 \leq i \leq \Delta$, during the entire subsequence $S(i)$, the cluster $c(v)$ of $v$ changes at most $O(\log n)$ times, in expectation.
\end{lemma}
\begin{proof}
Let $N_{i-1}(v)$ be the neighbors of $v$ at level $(i-1)$, grouped according to the the clusters they belong to. This grouping naturally induces a family $\mathcal{P}$ of all potential parents sets $P$ of $v$ at level $(i-1)$, just before the beginning of subsequence $S(i)$.
Let $C$ be the set of the corresponding clusters centers, i.e., for each $P \in \mathcal{P}$ add $c(P)$ to $C$, and note that $v$ can only join those centers during $S(i)$. Since we are considering only edge deletions, observe that when $v$ leaves a cluster centred at some node $c \in C$, it cannot join later the same cluster $c$ during $S(i)$.  

We next bound the number of cluster changes. For each $c \in C$, there must exist an edge in the subsequence $S(i)$ whose deletion increases $\dist(v,c)$, and thus $c$ is no longer a valid cluster center for $v$ at level $i$. The latter is also equivalent to some $P$ with $c(P) = c$ becoming empty after this edge deletion. Let $\langle c_1,\ldots,c_t \rangle$ be the sequence of nodes of $C$ \emph{ordered} according to the time when $v$ has no edge to a node in $P_j$, $1 \leq j \leq t$. We want to compute the probability that $v$ ever joins the cluster centred at $c_j$ during $S(i)$. Note that this event is a consequence of $v$ changing its current cluster center $c_{j'}$ due to all parents in $P(j')$ increasing their level. According to our tie-breaking scheme in $(\ref{eq: breakingTies})$, for this to happen, $c_j$ must be the first among all potential cluster centers $\{c_j,\ldots,c_t\}$ in the random permutation~$\pi$. 
Since $\pi$ is a uniform random permutation, the probability that $c_j$ appears first is $1/(t-j+1)$. By linearity of expectation, the expected number of centers from $C$ whose clusters $v$ joins during $S(i)$ is $\sum_{j=1}^{t} \frac{1}{t-j+1} = O(\log t) = O(\log n)$. This also bounds the number of cluster changes of $v$ during $S(i)$. 
\end{proof}

We next bound the total update time of our decremental algorithm, and also give a bound on the total number of inter-cluster edges during its execution. 

\begin{theorem}\label{thm:decremental decomposition}
There is a decremental algorithm for maintaining a $ (\fract, O (\tfrac{\log{n}}{\fract})) $-decomposition with at most $ O (\fract n) $ clusters (in expectation) containing non-isolated nodes under a sequence of edge deletions in expected total update time $ O (m \log^3 n / \fract) $ such that, over all deletions, each edge becomes an inter-cluster edge at most $ O (\log^2 n) $ times in expectation.
\end{theorem}
\begin{proof}
In a preprocessing step, we first repeat the sampling of the shift values until the maximum shift value is $ \log n/\beta $.
This event happens with probability $1-1/n$ (compare Theorem~\ref{thm: Miller_LDD}) and thus, by the waiting time bound, we need to repeat the sampling only a constant number of times.
Therefore, this preprocessing takes time $ O (n) $, which is subsumed in our claimed bound on the total update time.

We first note that procedure $\Initialize()$ can be implemented in $O (m \log n)$ time. This is because (1) the data-structures $\ell(\cdot), p(\cdot)$ and $c(\cdot)$ are initialized using Algorithm~\ref{alg:ModDijkstra} whose running time is bounded by $O (m \log n)$, (2) for each $u \in V$, the set $P(u)$ can by computed in $O(\deg (u))$ time, which in turn gives that all such sets can be determined in $\sum_{u \in V} O(\deg(u)) = O(m)$ time.

We next analyze the total time over the sequence of all edge deletions. Consider procedure $\Delete(u,v)$ for deletion of an edge $(u,v)$. If edge $(u,v)$ does not lead to a change in the clustering of one of its endpoints, then it can be processed in $O(1)$ time. Otherwise, the end-point whose clustering has changed is inserted into heap $Q$, which can be implemented in $O(\log n)$ time. Now, observe that the computation time spent by procedure $\Delete(u,v)$ is bounded by the number of nodes processed by heap $Q$ after the deletion of edge $(u,v)$, during procedure $\UpdateLevels()$. By construction, the processed nodes are precisely those whose clustering has changed due to the deletion of $(u,v)$, and after the processing, their new clustering its computed. A node $y$ extracted from $Q$ is processed in $O(\deg(y) \log n)$ time, as we will shortly argue. Therefore, we conclude that over the entire sequence of edge deletions, a node $y$ will perform $O(\deg(y) \log n)$ amount of work, each time its clustering changes. By Lemma~\ref{lem: nrClusterChanges}, as long as the level of $y$ is not increased, the clustering of $y$ will change $O(\log n)$ times, in expectation. Since there are at most $\Delta = O(\log n / \beta)$ levels, the expected number of cluster changes for $y$ is bounded by $O((\log^{2} n) / \beta)$. As our analysis applies to any node $y \in V$, we conclude that the expected total update time of our decremental algorithm is
\begin{equation} \label{eqn: runningTime}
	\sum_{y \in V} O \left((\deg(y) \log^3 n) / \beta \right) = O\left((m \log^{3} n / \beta)\right) \, .
\end{equation}

To show our claim that each node $y$ extracted from $Q$ is processed in time $O(\deg(y) \log n)$, we need two observations. First, recall that $P(y)$ can be computed in $O(\deg(y))$ time, and thus the data-structures $\ell(\cdot), p(\cdot)$ and $c(\cdot)$ can be then updated in $O(1)$ time. Second, in the worst-case, $y$ affects the clustering of all its neighbors and inserts them into $Q$. This step can be implemented in $O(\deg(y) \log n)$ time.

We finally show that each edge becomes an inter-cluster edge at most $ O (\log^2 n) $ times in expectation.
Fix some arbitrary edge $ e = (x, y) $ and consider the graph $ G $ after an arbitrary number of the adversary's deletions.
We first formulate a necessary condition for $ e $ being an inter-cluster edge and give a bound on the probability of the corresponding event.
Let $ w $ denote the mid-point of $ e $, i.e., the imaginary node in the ``middle'' of edge $ e $ that is at distance $ \tfrac{1}{2} $ to both $ u $ and~$ v $.
Let $ m_{c(x)}(w) $ and $ m_{c(y)}(w) $ denote the shifted distance from $ w $ to $ c (x) $ and $ c (y) $ in $ G $, respectively.
We would like to argue that both $ m_{c(x)}(w) $ and $ m_{c(y)}(w) $ are close to the minimum shifted distance of the mid-point $ w $.
However, we cannot readily apply Lemma~\ref{lem:inter cluster edge implies mid-point condition} as our algorithm does not run on~$ G $; instead it runs on~$ G' $, in which the edge weights are rounded to integers.
However, we can apply Lemma~\ref{lem:inter cluster edge implies mid-point condition} on $ G' $ and get that $ \floor{m_{c(x)}(w)} $ and $ \floor{m_{c(y)}(w)} $ are within $ 1 $ of the minimum rounded shifted distance of the mid-point $ w $.
Thus, $ | \floor{m_{c(x)}(w)} - \floor{m_{c(y)}(w)} | \leq 1 $, which implies that $ | m_{c(x)}(w) - m_{c(y)}(w) | \leq 2 $.
This means that $ | m_{c(x)}(w) - m_{c(y)}(w) | \leq 2 $ is a necessary condition for $ e = (x, y) $ to be an inter-cluster edge.
As the adversary is oblivious to the random choices of our algorithm, we know by Lemma~\ref{lem:probability of mid-point condition} that $ \Pr [| m_{c(x)}(w) - m_{c(y)}(w) | \leq 2 ] \leq 2 \beta $ in each of the graphs created by the adversary's sequence of deletions.

Observe that for each of the endpoints ($ x $ and~$ y $) of $ e $ the level in our decremental algorithm is non-decreasing.
Let $ 0 \leq i \leq \Delta $, and let $ S (i) $, say of length~$ t $, be the (possibly empty) subsequence of edge deletions during which $ \ell (x) = i $.
We show below that the expected number of times that $ e $ becomes an inter-cluster edge during deletions in $ S (i) $ is $ O (\beta \log{n}) $.
It then follows that the total number times $ e $ becomes an inter-cluster edges is $ O (\log^2{n}) $ by linearity of expectation: sum up the number of times $ e $ becomes an inter-cluster edge in each subsequence $ S (i) $ for $ 0 \leq i \leq \Delta $ where $ \Delta \leq O (\log{n} / \beta) $, and repeat the argument for the other endpoint $ y $ of $ e $ as well.

For every $ 1 \leq j \leq t $ define the following events:
\begin{itemize}
\item $ A_j $ is the event that $ e $ becomes an inter-cluster edge after the $j$-th deletion in $ S (i) $, and was not an inter-cluster edge directly before this deletion.
\item $ B_j $ is the event that at least one of the endpoints of $ e $, $ x $ or $ y $, changes its cluster after the $j$-th deletion in $ S (i) $.
\item $ C_j $ is the event that $ e $ is an inter-cluster edge after the $j$-th deletion in $ S (i) $.
\item $ D_j $ is the event that $ | m_{c(x)}(w) - m_{c(y)}(w) | \leq 2 $ after the $j$-th deletion in $ S (i) $, where $ w $ is the mid-point of $ e $.
\end{itemize}

Note that $ e $ can only become an inter-cluster edge if at least one of its endpoints changes its cluster.
Thus, the event~$ A_j $ implies the event $ B_j \wedge C_j $ and therefore $ \Pr [A_j] \leq \Pr [B_j \wedge C_j] $.
Furthermore, by Lemma~\ref{lem:inter cluster edge implies mid-point condition}, the event $ C_j $ implies the event $ D_j $.
We thus have $ \Pr [B_j \wedge C_j] \leq \Pr [B_j \wedge D_j] $.
Observe that the event $ D_j $ only depends on the random choice of the shift values $ \delta $ and that, in the fixed subsequence of deletions $ S (i) $, the event $ B_j $ only depends on the random choice of the permutation~$ \pi $.
Thus, $ B_j $ and $ D_j $ are independent and therefore $ \Pr [B_j \wedge D_j] = \Pr [B_j] \cdot \Pr [D_j] $.
Finally, note that the expected number of indices $ j $ such that the event $ B_j $ happens is at most the expected number of cluster changes for both endpoints of $ e $, as bounded by Lemma~\ref{lem: nrClusterChanges}, and thus $ \sum_{1 \leq i \leq t} \Pr [B_j] = O (\log n) $ for the random permutation~$ \pi $.
It follows that the expected number of times edge $ e $ becomes an inter-cluster edge (i.e., the expected number of indices $ j $ such that event $ A_j $ happens) is
\begin{align*}
\sum_{1 \leq i \leq t} \Pr [A_j] & \leq \sum_{1 \leq i \leq t} \Pr [B_j \wedge C_j] \leq \sum_{1 \leq i \leq t} \Pr [B_j \wedge D_j] = \sum_{1 \leq i \leq t} \Pr [B_j] \cdot \Pr [D_j] \\ 
& \leq \sum_{1 \leq i \leq t} \Pr [B_j] \cdot 2 \beta = 2\beta \cdot \sum_{1 \leq i \leq t} \Pr [B_j] = O (\beta \log{n}) \, ,
\end{align*}
where the penultimate inequality follows from Lemma~\ref{lem:probability of mid-point condition}.
\end{proof}

Note that in this proof, to bound the number total number of inter-cluster edges, we exploited that our two sources of randomness, the random shifts~$ \delta $ and the random permutation~$ \pi $ have different purposes: $ \delta $ influences whether an edge $ e $ is an inter-cluster edge and $ \pi $ influences the number of cluster changes of the endpoints of $ e $.
We have deliberately set up the algorithm in such a way that the independence of the corresponding events can be exploited in the proof.
This is the reason why we explicitly introduced a new random permutation for tie-breaking instead of using the random shifts for this purpose as well.
\begin{remark} \label{remark: LDDrunningTime}
Note that Equation~(\ref{eqn: runningTime}) implies that the total expected update time of Theorem~\ref{thm:decremental decomposition} is $O(m \log^{3} n / \beta)$. For the sake of exposition, we have implemented the ES-tree using a heap, which introduces a $O(\log n)$ factor in the running time. \cite{King99} (Section~2.1.1) gives a faster implementation of the ES-tree that eliminates this extra $O(\log n)$ factor. Thus, using her technique, we can also bring down our running time to $O(m \log^{2} n / \beta)$. This improvement will be particularly useful when applying our dynamic low-diameter decomposition to the construction of dynamic spanners in Section~\ref{sec:spanner}.
\end{remark}

\begin{algorithm2e}[htb!]
\caption{Modified ES-tree}
\label{alg:ES_tree}

\tcp{\textrm{The modified ES-tree is formulated for weighted undirected graphs.}}
\BlankLine

\tcp{\textrm{Internal data structures:
\begin{itemize}
\item $ \pi $: random permutation on $V$
\item $ \delta_v $: random shift of $ v $
\item $ P (v) $: the set of potential parents in the tree
\item $ p (v) $: for every node $ v $ a pointer to its parent in the tree
\item $ c (v) $: for every node $ v $ a pointer to the cluster center
\item $ Q $: global heap whose intended use is to store nodes whose levels might need to be updated
\end{itemize}
}}
\vspace{-3ex}

\BlankLine

\Procedure{\Initialize{}}{
	Initialize using Algorithm~\ref{alg:ModDijkstra}\;
	Set $ \lev (v) $, $ P(v) $, $ p(v) $, $ c(v) $ for every node $ v $ accordingly\;
}

\Procedure{\Delete{$u$, $v$}}{
	\If{$ v \in P (u) $}{
		Remove $ v $ from $ P (u) $\;
		\If{$ P (u) = \emptyset$}{
			Insert $ u $ into heap $ Q $ with key $ \lev (u) $\;
			\UpdateLevels{}\;
		}
	}
}

\Procedure{\UpdateLevels{}}{
	\While{heap $ Q $ is not empty}{\label{line:while loop}
	    Take node $ y $ with minimum key $ \lev (y) $ from heap $ Q $ and remove it from $ Q $ \label{line: take y from Q}\;
		Compute $ P (y) $ as the set of neighbors $ z $ of $ y $ minimizing $ (\lev (z) + \ww (y, z), \pi(c(z))) $ lexicographically\;
		Set $ p (y) $ as one of the nodes in $ P (y) $\;
		Set $ \lev (y) \gets \lev (p(y)) + \ww' (y, p(y) $\;
		\eIf{$ p(y) = s $}{
			Set $ c (y) \gets y  $\;
		}{
			Set $ c (y) = c( p(y)) $\;
		}

		\ForEach{neighbor $ x $ of $ y $}{\label{line: update neighbors' heaps}
			\If{$ y \in P (x) $}{
				Remove $ y $ from $ P (x) $\;
				\If{$ P (x) = \emptyset$}{
					Insert $ x $ into heap $ Q $ with key $ \lev (x) $ if $ Q $ does not already contain $ x $\;
				}
			}
		}
	}
}
\end{algorithm2e}

\subsection{Fully Dynamic Low-Diameter Decomposition}\label{sec:fully dynamic LDD}

We finally show how to extend the decremental algorithm of Theorem~\ref{thm:decremental decomposition} to a fully dynamic algorithm, allowing also insertions of edges.
\begin{theorem}[Restatement of Theorem~\ref{thm:fully dynamic LDD}]
Given any unweighted, undirected multigraph undergoing edge insertions and deletions, there is a fully dynamic algorithm for maintaining a $ (\fract, O (\tfrac{\log{n}}{\fract})) $-decomposition (with clusters of strong diameter $ O (\tfrac{\log{n}}{\fract}) $ and at most $ \beta m $ inter-cluster edges in expectation) that has expected amortized update time $ O (\log^2{n} / \fract^2) $.
A spanning tree of diameter $ O (\tfrac{\log{n}}{\fract}) $ for each cluster can be maintained in the same time bound.
The expected amortized number of edges to become inter-cluster edges after each update is $ O (\log^2{n} / \fract) $.
These guarantees hold against an oblivious adversary.
\end{theorem}

\begin{proof}
The fully dynamic algorithm proceeds in phases, starting from an empty graph.
For every $ i > 1 $, let $ m_i $ denote the number of edges in the graph at the beginning of phase $ i $.
After $ \fract m_i / 3 $ updates in the graph we end phase $ i $ and start phase $ i+1 $.
At the beginning of each phase we re-initialize the decremental algorithm of Theorem~\ref{thm:decremental decomposition} for maintaining a $ (\fract / 3, 3 \cdot O (\tfrac{\log{n}}{\fract})) $-decomposition.\footnote{Note that for the first constant number of updates this basically amounts to recomputation from scratch at each update.}
Whenever an edge is deleted from the graph, we pass the edge deletion on to the decremental algorithm.
Whenever an edge is inserted to the graph, we do nothing, i.e., we deal with insertions of edges in a completely \emph{lazy} manner.

We first analyze the ratio of inter-cluster edges at any time during phase $ i $.
First observe that the number of inter-cluster edges is at most $ 2 \fract m_i / 3 $ in expectation, where at most $ \fract m_i / 3 $ edges in expectation are contributed by the $ (\fract / 3, 3 \cdot O (\tfrac{\log{n}}{\fract})) $-decomposition of the decremental algorithm and at most $ \fract m_i / 3 $ edges are contributed from inserted edges.
Second, the number of edges in the graph is at least $ m_i - \fract m_i / 3 $, as $ m_i $ is the initial number of edges and at most $ \fract m_i / 3 $ edges have been deleted.
Thus, the ratio of inter-cluster edges is at most
\begin{equation*}
\frac{2 \fract m_i / 3}{m_i - \fract m_i/3} = \frac{2 \fract}{3 - \fract} \leq \frac{2 \fract}{2 + \fract - \fract} = \fract \, .
\end{equation*}
Our fully dynamic algorithm therefore correctly maintains a $ (\fract, O (\tfrac{\log{n}}{\fract})) $-decomposition.

We now analyze the amortized update time of the algorithm.
Start with an empty graph and consider a sequence of $ q $ updates.
Let $ k $ denote the number of the phase after the $ q $-th update.
Then $ q $ can be written as $ q = \sum_{1 \leq i < k} \fract m_i / 3 + t $, where $ t $ is the number of updates in phase $ k $.
For every phase $ i $ that has been started, we spend time $ O (m_i \log^2 n / \fract) $ by Theorem~\ref{thm:decremental decomposition} and Remark~\ref{remark: LDDrunningTime}. We know that $ t \leq \fract m_k / 3 $ and in particular we also have $ m_k \leq \sum_{1 \leq i \leq k-1} \fract m_i / 3 $ as every edge that is contained in the graph at the beginning of phase $ k $ has been inserted in one of the previous phases.
We can thus bound the amortized spent by the algorithm for $ q $ updates by
\begin{multline*}
\frac{\sum_{1 \leq i \leq k-1} O (m_i \log^2 n / \fract) + O (m_k \log^2 n / \fract)}{\sum_{1 \leq i \leq k-1} \fract m_i / 3} \\
 \leq \frac{\sum_{1 \leq i \leq k-1} O (m_i \log^2 n / \fract) + O (\sum_{1 \leq i \leq k-1} m_i \log^2 n)}{\sum_{1 \leq i \leq k-1} \fract m_i / 3} = O \left( \frac{\log^2 n}{\fract^2} \right) \, .
\end{multline*}

Finally, we analyze the amortized number of edges to become inter-cluster edges per update.
For every phase $ i $ that has been started, we have a total number of $ O (m_i \log^2 n) $ edges that become inter-cluster edges in the decremental algorithm by Theorem~\ref{thm:decremental decomposition}.
Additionally, at most $ \fract m_i / 3 = O (m_i) $ inserted edges could also become inter-cluster edges.
We can thus bound the amortized number of edges to become inter-cluster per update by
\begin{multline*}
\frac{\sum_{1 \leq i \leq k-1} O (m_i \log^2 n) + O (m_k \log^2 n)}{\sum_{1 \leq i \leq k-1} \fract m_i / 3} \\
 \leq \frac{\sum_{1 \leq i \leq k-1} O (m_i \log^2 n) + O (\sum_{1 \leq i \leq k-1} \fract m_i \log^2 n)}{\sum_{1 \leq i \leq k-1} \fract m_i / 3} = O \left( \frac{\log^2 n}{\fract} \right) \, .
\end{multline*}
\end{proof}

\section{Dynamic Spanner Algorithm}\label{sec:spanner}

\subsection{Static Spanner Construction}
In the following we review and adapt the static algorithm for constructing sparse low-stretch spanners due to Elkin and Neiman~\cite{ElkinN17}. Let $G=(V,E)$ be an unweighted, undirected graph on $n$ nodes, and let $k \geq 1$ be an integer. For every $u \in V$, we denote by $N(u)$ the set of all nodes incident to $u$. Recall that $\expdis(\beta)$ denotes the exponential distribution with parameter $\beta$. In what follows, we set $\beta = \log (cn)/k$, where $c>3$ denotes the success probability. A $2 k - 1$-\emph{spanner} of $G$ is a a subgraph $H=(V,E')$ such that for every $u,v \in V$, $\dist_H(u,v) \leq 2 k - 1 \cdot \dist_G(u,v)$. We refer to $2 k - 1$ and $|E'|$ as the \emph{stretch} and \emph{size} of $H$, respectively. 

We next review some useful notation. Let $\delta_u$ be the shift value of node $u \in V$. For each $x, u \in V$, recall that $m_u(x) = \dist_G(x,u) - \delta_u$ is the shifted distance of $x$ with respect to $u$, and let $p_u(x)$ denote the neighbor of $x$ that lies on a shortest path from $x$ to $u$. Also, for every node $x \in V$, let $m(x) = \min_{u \in V} \{ m_u(x) \}$ be the minimum shifted distance. Using our clustering notation from Section~\ref{sec:LDD}, it follows that $c(x) = \arg \min_{u \in V} \{ m_u(x) \}$, and thus $m(x) = m_{c(x)}(x)$.

We now present an algorithm that constructs spanners using exponential random-shift clustering. Specifically, we initially set $H=(V,\emptyset)$, and for each node $u \in V$, we independently pick a shift value $\delta_u$ from $\expdis(\beta)$. Then, for every $x \in V$, we add to the spanner $H$ the following set of edges
\begin{equation}
C(x) = \set{(x, p_u(x))}{m_u(x) \leq m(x) + 1} \, . \label{eqn: SpannerEqn}
\end{equation}

The following theorem give bounds on the stretch and the size of the spanner output by the above algorithm.

\begin{theorem}[\cite{ElkinN17}] For any unweighted, undirected simple graph $G=(V,E)$ on $n$ nodes, any integer $k \geq 1$, $c \geq 3$, there is a randomized algorithm that with probability at least $ 1 - \tfrac{2}{c} $ computes a spanner $H$ with stretch $2k-1$ and size at most $ (cn)^{1+1/k} $.
\end{theorem}

Our analysis will rely on the following useful properties of the above algorithm.

\begin{claim}[\cite{ElkinN17}]
The expected size of $ H $ is at most $ (c n)^{1/k} \cdot n $.
\end{claim}

\begin{claim}[\cite{ElkinN17}]\label{claim:bound on random shifts}
With probability at least $1-1/c$, it holds that $\delta_u < k$ for all $u \in V$.
\end{claim}

\begin{claim}[\cite{ElkinN17}]\label{claim:bound on depth of tree} Assume $\delta_u < k$ for all $u \in V$. Then for any $x \in V$, if $u$ is the node minimizing $m_u(x)$, i.e., $u = c(x)$, then $\dist_G(u,x) < k$.
\end{claim}

As argued by Elkin and Neiman, Claim~\ref{claim:bound on depth of tree} implies that the stretch of the spanner is at most $ 2k - 1 $.
Thus, the reason reason why the stretch guarantee is probabilistic is Claim~\ref{claim:bound on random shifts}.

\paragraph*{Implementation.} In the description of the spanner construction, it is not clear how to compute in nearly-linear time the set of edges $C(x)$ in Equation~(\ref{eqn: SpannerEqn}), for every node $x \in V$. To address this, we give an equivalent definition of $C(x)$, which better decouples the properties that the edges belonging to this set satisfy. Specifically, we define the set of edges
\begin{equation}
	C'(x) = \set{(x,y)}{y \in N(x) \text{ and } m_{c(y)}(x) \leq m(x) + 1} \, ,\label{eqn: equivalentSpannerEqn}
\end{equation}
and then show that $C(x) = C'(x)$. 

To this end, we will show that (a) $C(x) \subseteq C'(x)$ and (b) $C'(x) \subseteq C(x)$. Let $(x,y) \in C(x)$, where $y = p_u(x)$. By definition of $p_u(x)$, we have that $y \in N(x)$. We next show that $m_{c(y)}(x) \leq m(x) + 1$, which in turn proves (a). Indeed, 
\[
	m_{c(y)}(x) = m_{c(y)}(y) + 1 = m(y) + 1 \leq m_u(y) + 1 = m_u(x) \leq m(x)+1,
\]
where the last inequality follows from Equation~(\ref{eqn: SpannerEqn}). For showing the other containment, i.e., proving (b), let $(x,y) \in C'(x)$. Then we need to prove that there exists some $u \in V$ such that $y=p_u(x)$ and $m_u(x) \leq m(x) +1$. This follows by simply setting $u=c(y)$ and using Equation~(\ref{eqn: equivalentSpannerEqn}).

Now, similarly to the static low-diameter decomposition in Section~\ref{sec:staticLDD}, we augment the input graph~$G$ by adding a new source $s$ to $G$ and edges $(s,x)$ of weight $(\delta_{max} - \delta_x) \geq 0$, for every $x \in V$, where $\delta_{\max} = \max_{x \in V} \{\delta_x \}$. Recall that in the resulting graph $\hat{G}=(V \cup \{s\},\hat{E},\hat{\ww})$, for every $x \in V$, the node $u$ attaining the minimum $m(x)$ is exactly the root of the sub-tree below the source $s$ that contains $u$. Thus, we could use Dijkstra's algorithm to construct the shortest path tree of $\hat{G}$, and augment it appropriately to output the edge sets $C'(x)$, which in turn give us the spanner $H$.

However, in the dynamic setting, it is crucial for our algorithm to deal only with integral edge weights. To address this, we round down all the $\delta_u$ values to $\floor{\delta_u}$ and modify the weights of the edges incident to the source $s$ in $\hat{G}$. Let $G'=(V \cup \{s\}, E',\ww')$ be the resulting graph, and let $\floor{m_u(x)}$ denote the rounded shifted distances. Whenever two rounded distances are the same, we break ties using the permutation $ \pi $ on the nodes induced by the fractional values of the random shift values. Thus, the edge set $C'(x)$ is given by
\begin{equation*} 
	C'(x) = \set{(x,y)}{y \in N(x),~\floor{m_{c(y)(x)}} \leq \floor{m(x)} + 1 \text{ and } \pi(c(y)) < \pi(c(x))}.
\end{equation*}

Finally, we observe that the definition of the above set can be further simplified by using the facts that $m_{c(y)}(x) = m_{c(y)}(y) + 1$ and $\floor{m_{c(y)}(x)} \geq \floor{m(x)}$, that is
\begin{equation} \label{eqn: roundedSpannerEqn}
\begin{aligned}
   C'(x) & = \{(x,y) \mid y \in N(x),~\floor{m(y)} = \floor{m(x)} - 1 \text{ or } \\ 
   & [\floor{m(y)} = \floor{m(x)} \text{ and } \pi(c(y)) < \pi(c(x))] \} \, . 	
\end{aligned}
\end{equation}

Interpreting the above set in terms of the shortest-path tree output by Dijkstra's algorithm, we get that for any $x \in V$, we add the edge $(x,y)$ to the spanner $H$, if $y$ is a neighbor one level above the level or $x$, or if $x$ and $y$ are at the same level, and the cluster $y$ belongs to appears before in the permutation when compared to the cluster $x$ belongs to.
By Claim~\ref{claim:bound on depth of tree} the shortest-path tree has depth at most $ 2 k $ with high probability.

Now observe that the randomized properties of this spanner construction only depend on the integer parts of the random shift values and the permutation~$ \pi $ on the nodes induced by the order statistics of the fractional parts of the random shift values.
Similar to the argument of Miller et al.~\cite{MillerPX13} for low-diameter decompositions, it can be argued that due to memorylessness of the exponential distribution, one might as well use a uniformly sampled random permutation~$ \pi $ instead to obtain a spanner with the same probabilistic properties.

\subsection{Dynamic Spanner Algorithm}

Spanners have a useful property called \emph{decomposability}: Assume we are given a graph $ G = (V, E) $ with a partition into two subgraphs $ G_1 = (V, E_1) $ and $ G_2 = (V, E_2) $.
If $ H_1 = (V, F_1) $ is a spanner of $ G_1 $ and $ H_2 = (V, F_2) $ is a spanner of $ G_2 $, both of stretch $ t $, then $ H = (V, F_1 \cup F_2) $ is a spanner of $ G $.
This property allows for a reduction that turns decremental algorithms into fully dynamic ones at the expense of logarithmic overhead in size and update time, as it has been observed by Baswana et al.~\cite{BaswanaKS12}.
\begin{lemma}[Implicit in~\cite{BaswanaKS12}]\label{lem:spanner decremental to fully dynamic}
If there is a decremental algorithm for maintaining a spanner of stretch $ t $ and expected size $ s (n) $ with total update time $ m \cdot u (m, n) $, then there is a fully dynamic algorithm for maintaining a spanner of stretch $ t $ and expected size $ s (n) \cdot O (\log{n}) $ with amortized update time $ u (m, n) \cdot O (\log{n}) $.
\end{lemma}

In the remainder of this section, we explain how the techniques we developed in Section~\ref{sec:LDD} allow for a decremental implementation of the spanner construction explained above.
\begin{theorem}
Given any unweighted, undirected graph undergoing edge deletions, there is a decremental algorithm for maintaining a spanner of stretch $ 2k - 1 $ and expected size $ O (n^{1 + 1/k}) $ that has expected total update time $ O (k m \log{n}) $.
These guarantees hold against an oblivious adversary.
\end{theorem}
Using the reduction of Lemma~\ref{lem:spanner decremental to fully dynamic}, these guarantees carry over to the fully dynamic setting.
\begin{theorem}[Restatement of Theorem~\ref{thm:fully dynamic spanner}]
Given any unweighted, undirected graph undergoing edge insertions and deletions, there is a fully dynamic algorithm for maintaining a spanner of stretch $ 2k - 1 $ and expected size $ O (n^{1 + 1/k} \log{n}) $ that has expected amortized update time $ O (k \log^2{n}) $.
These guarantees hold against an oblivious adversary.
\end{theorem}

The decremental algorithm is obtained as follows:
In a preprocessing step, the algorithm samples the random shift values for the nodes from the exponential distribution and additionally a uniformly random permutation $ \pi $ on the nodes.
The sampling of the random shift values is repeated until $ \delta_u < k $ for all $u \in V$.
By Claim~\ref{claim:bound on random shifts} this condition holds with probability at least $ 1 - 1/c $.
Thus, by the waiting time bound, we need to repeat the sampling at most a constant number of times for the condition to hold.
As each round of sampling takes time $ O (n) $, this preprocessing step requires an additional $ O (n) $ in the total update time.

We can then readily use Algorithm~\ref{alg:ES_tree} from Section~\ref{sec:decremental LDD} to maintain a shortest path tree up to depth $ 2 k $ from $ s $ in the graph $ G' $, as defined above.
For maintaining the spanner dynamically, we need to extend the algorithm to maintain the set $ C' (x) $ for every node $ x $.
Using the arguments introduced in Section~\ref{sec:decremental LDD}, this can be done in a straightforward way:
Every time a node $ x $ changes its level in the tree or changes its cluster $ c (x) $, it (1) recomputes the set $ C' (x) $ in time $ O (\deg (x)) $ and stores it in a hash set and (2) informs each neighbor about the change and updates the set $ C' (y) $ of each neighbor $ y $ by setting the entry corresponding to the edge $ (x, y) $ accordingly.
Both (1) and (2) require (expected) time $ O (\deg (x)) $.
As the maximum level in the tree is $ O (k) $ and at each node changes its clustering at a fixed level at most $ O (\log n) $ times in expectation, the expected total update time of our algorithm is $ O (k m \log{n}) $ as desired.

\section{Conclusion}
In this chapter, we showed a fully dynamic algorithm that maintains a $n^{o(1)}$-stretch spanning tree in an unweighted, undirected graph with $n^{1/2 + o(1)}$ amortized time per edge insertion or deletion. The core building block behind the algorithm is a dynamic algorithm that maintains a low-diameter clustering a graph. We also showed that this technique can be applied to the dynamic spanner problem, for which we improved upon the best-known update time and the size of the spanner. Our work leaves several open problems. One important problem is whether the running time can be brought down to $n^{o(1)}$. We believe that this is closely connected to how we deal with insertions in our dynamic clustering algorithm. In fact, any subroutine that outperforms our lazy insertion technique might lead to further improvements in the update time. Another interesting problem is extending our techniques to weighted, undirected graphs. A natural attempt is to extend the hierarchy due to Alon et al.~\cite{AlonKPW95} on weighted graphs to a dynamic setting. We remark that a black-box extension seems not to be feasible, so new ideas might be required to achieve this. Finally, the question of whether there are dynamic algorithms that maintain low-stretch trees with poly-logarithmic stretch and sub-linear update time remains a major open problem. 


\chapter[Incremental Exact Min-Cut in Poly-logarithmic Amortized Update Time][Incremental Min-Cut in Poly-logarithmic Time]{Incremental Exact Min-Cut in Poly-logarithmic Amortized Update Time}\label{cha:TALG2018_IMC}

We present a deterministic incremental algorithm for \textit{exactly} maintaining the size of a minimum cut with $O(\log^3 n \log \log^2 n)$ amortized time per edge insertion and $O(1)$ query time. This result partially answers an open question posed by Thorup~\cite{Thorup07}. It also stays in sharp contrast to a polynomial conditional lower-bound for the fully-dynamic weighted minimum cut problem. Our algorithm is obtained by combining a sparsification technique of Kawarabayashi and Thorup~\cite{KawarabayashiT19} or its recent improvement by Henzinger, Rao and Wang~\cite{HenzingerRW17}, and an exact incremental algorithm of Henzinger~\cite{Henzinger97}.

We also study space-efficient incremental algorithms for the minimum cut problem. Concretely, we show that there exists an ${O}(n\log n/\epsilon^2)$ space Monte-Carlo algorithm that can process a stream of edge insertions starting from an empty graph, and with high probability, the algorithm maintains a $(1+\epsilon)$-approximation to the minimum cut. The algorithm has $O((\alpha(n) \log^{3} n) / \epsilon^{2})$ amortized update-time and constant query-time, where $\alpha(n)$ stands for the inverse of Ackermann function.

\section{Introduction}

Computing a minimum cut of a graph is a fundamental algorithmic graph problem. While most of the focus has been on designing static efficient algorithms for finding a minimum cut, the dynamic maintenance of a minimum cut has also attracted increasing attention over the last two decades. The motivation for studying the dynamic setting is apparent, as real-life networks such as social or road network undergo constant and rapid changes.

Given an initial graph $G$, the goal of a dynamic graph algorithm is to build a data-structure that maintains $G$ and supports update and query operations. Depending on the types of update operations we allow, dynamic algorithms are classified into three main categories: (i) \emph{fully dynamic}, if update operations consist of both edge insertions and deletions, (ii) \emph{incremental}, if update operations consist of edge insertions only and (iii) \emph{decremental}, if update operations consist of edge deletions only. In this chapter, we study incremental algorithms for maintaining the size of a minimum cut of an unweighted, undirected graph (denoted by $\lambda(G) = \lambda$) supporting the following operations:
\begin{itemize}
\item \textsc{Insert}$(u,v)$: Insert the edge $(u,v)$ to $G$.
\item \textsc{QuerySize}: Return the exact (approximate) size of a minimum cut of the current $G$.
\end{itemize}
For any $\alpha \geq 1$, we say that an algorithm is an $\alpha$-approximation of $\lambda$ if \textsc{QuerySize} returns a positive number $k$ such that $\lambda \leq k \leq \alpha \cdot \lambda$. Our problem is characterized by two time measures; \emph{query time}, which denotes the time needed to answer each query and \emph{total update time}, which denotes the time needed to process \emph{all} edge insertions. We say that an algorithm has an $O(t(n))$ amortized update time if it takes $O(m(t(n)))$ total update time for $m$ edge insertions starting from an empty graph. 
\paragraph*{Related Work.} For over a decade, the best known static and deterministic algorithm for computing a minimum cut was due to Gabow~\cite{Gabow95} which runs in $O(m + \lambda^{2} n \log n)$ time.~Kawarabayashi and Thorup~\cite{KawarabayashiT19} devised an $O(m \log^{12} n)$ time algorithm which applies only to unweighted, undirected simple graphs. Recently, Henzinger et al.~\cite{HenzingerRW17} improved the running time to $O(m \log^{2} n \log \log^{2} n)$. Randomized Monte Carlo algorithms in the context of static minimum cut were initiated by Karger~\cite{Karger99}. The best known randomized algorithm is due to Karger~\cite{Karger00} and runs in $O(m \log^{3} n)$ time.  

Karger~\cite{KargerSODA94} was the first to study the dynamic maintenance of a minimum cut in its full generality. He devised a fully dynamic, albeit randomized, algorithm for maintaining a  $\sqrt{1+2/\epsilon}$-approximation of the minimum cut in $\tilde{O}(n^{1/2 + \epsilon})$ expected amortized time per edge operation. In the incremental setting, he showed that the update time for the same approximation ratio can be further improved to $\tilde{O}(n^{\epsilon})$. Thorup and Karger~\cite{ThorupSWAT00} improved upon the above guarantees by achieving an approximation factor of $\sqrt{2+o(1)}$ and an $\tilde{O}(1)$ expected amortized time per edge operation.

Henzinger~\cite{Henzinger97} obtained the following guarantees for the incremental minimum cut; for any $\epsilon \in (0,1]$, (i) an $O(1/\epsilon^{2})$ amortized update-time for a $(2+\epsilon)$-approximation, (ii) an $O(\log^{3} n / \epsilon^{2})$ expected amortized update-time for a $(1+\epsilon)$-approximation and (iii) an $O(\lambda \log n)$ amortized update-time for the exact minimum cut. 

For minimum cut up to some poly-logarithmic size, Thorup~\cite{Thorup07} gave a fully dynamic  Monte-Carlo algorithm for maintaining exact minimum cut in $\tilde{O}(\sqrt{n})$ time per edge operation. He also showed how to obtain an $1+o(1)$-approximation of an arbitrary sized minimum cut with the same time bounds. In comparison to previous results, it is worth pointing out that his work achieves \textit{worst-case} update times.

Lacki and Sankowski~\cite{LackiS11} studied the dynamic maintenance of the exact size of the minimum cut in planar graphs with arbitrary edge weights. They obtained a fully dynamic algorithm with $\tilde{O}(n^{5/6})$ worst-case query and update time.

There has been a growing interest in proving conditional lower bounds for dynamic problems in the last few years~\cite{abboud,HenzingerKNS15}. A recent result of Nanongkai and Saranurak~\cite{NanongkaiS16} shows the following conditional lower-bound for the \emph{exact weighted} minimum cut assuming the Online Matrix-Vector Multiplication conjecture: for any $\epsilon > 0$, there are no fully-dynamic algorithms with polynomial-time preprocessing that can simultaneously achieve $O(n^{1-\epsilon})$ update-time and $O(n^{2-\epsilon})$ query-time. 
\paragraph*{Our Results and Technical Overview.} We present two new incremental algorithms concerning the maintenance of the size of a minimum cut. Both algorithms apply to undirected, unweighted simple graphs. 

Our first and main result, presented in Section \ref{sec: exactMinCut}, shows that there is a deterministic incremental algorithm for \textit{exactly} maintaining the size of a minimum cut with $O(\log^3 n \log \log^2 n)$ amortized time per operation and $O(1)$ query time. This result allows us to  partially answer in the affirmative a question regarding efficient dynamic algorithms for exact minimum cut posed by Thorup~\cite{Thorup07}. Additionally, it also stays in sharp contrast to the polynomial conditional lower-bound for the fully-dynamic weighted minimum cut problem of Nanongkai and Saranurak~\cite{NanongkaiS16}. 

We obtain our result by heavily relying on a recent sparsification technique developed in the context of static minimum cut algorithms. Specifically, for a given simple graph $G$, Kawarabayashi and Thorup~\cite{KawarabayashiT19} (and subsequently Henzinger et al.~\cite{HenzingerRW17}) designed an $\tilde{O}(m)$ procedure that contracts vertex sets of $G$ and produces a multigraph $H$ with considerably fewer vertices and edges while preserving some family of cuts of size up to $(3/2)\lambda(G)$. Motivated by the properties of $H$, the crucial observation is that it is ``safe'' to work entirely with graph $H$ as long as the sequence of newly inserted edges do not increase the size of the minimum cut in $H$ by more than $(3/2) \lambda(G)$. If the latter occurs, we recompute a new multigraph $H$ for the current graph $G$. Since $\lambda(G) \leq n$, the number of such re-computations is $O(\log n)$. For maintaining the minimum-cut of $H$, we appeal to the exact incremental algorithm due to Henzinger~\cite{Henzinger97}. Our main technical contribution is to skilfully combine these two algorithms and formally argue that such combination leads to our desirable guarantees.

Motivated by the recent work on \textit{space-efficient} dynamic algorithms~\cite{BhattacharyaHNT15}, we also study the efficient maintenance of the size of a minimum cut using only $\tilde{O}(n)$ space. Concretely, we present a ${O}(n\log n / \epsilon^2)$ space Monte-Carlo algorithm that can process a stream of edge insertions starting from an empty graph, and with high probability, it maintains an $(1+\epsilon)$-approximation to the minimum cut in ${O}((\alpha(n) \log^3 n) /\epsilon^2)$ amortized update-time and constant query-time. 

Note that while the streaming model also allows only $\tilde{O}(n)$ space, it is less constrained than the space efficient dynamic model since streaming algorithms do not need to maintain an explicit sparsifier at every moment, but just have enough information to construct one at the end of the stream. There have been several streaming algorithms~\cite{AhnG09,KelnerL13,KyngPPS17} for maintaining a cut sparsifier, and thus $(1+\epsilon)$-approximating the minimum cut. The best bounds are due to Kyng et al.~\cite{KyngPPS17} who compute a stronger spectral sparsifier with $O(n \log n / \epsilon^2)$ size and $O(\log^2 n)$ amortized update-time. In comparison to our result, while our update-time is slightly worse, we can achieve constant query-time, whereas their algorithms requires $\Omega(n)$ time to answer a query. 

\section{Preliminaries}

Let $G = (V,E)$ be an undirected, unweighted multi-graph with no self-loops. Two vertices $x$ and $y$ are $k$-\textit{edge connected} if there exist $k$ edge-disjoint paths connecting $x$ and $y$. A graph $G$ is $k$-\textit{edge connected} if every pair of vertices is $k$-edge connected. The \textit{local edge connectivity} $\lambda(x,y,G)$ of vertices $x$ and $y$ is the largest $k$ such that $x$ and $y$ are $k$-edge connected in $G$. The \textit{edge connectivity} $\lambda(G)$ of $G$ is the largest $k$ such that $G$ is $k$-edge connected.  

For a subset $S \subseteq V$ in $G$, the \textit{edge cut} $E_G(S, V \setminus S)$ is a set of edges that have one endpoint in $S$ and the other in $ V \setminus S$. We may omit the subscript when clear from the context. Let $\lambda(S,G) = |E_G(S, V \setminus S)|$ be the size of the edge cut. If $S$ is a singleton, we refer to such cut as a \textit{trivial} cut. Two vertices $x$ and $y$ are \textit{separated} by $E(S, V \setminus S)$ if they belong to different connected components of the graph induced by $E \setminus E(S,V \setminus S)$. A \textit{minimum edge cut} of $x$ and $y$ is a cut of minimum size among all cuts separating $x$ and $y$. A \textit{global minimum cut} $\lambda(G)$ for $G$ (or simply $\lambda$ when $G$ is clear from the context) is the minimum edge cut over all pairs of vertices. By Menger's Theorem \cite{menger}, (a) the size of the minimum edge cut separating $x$ and $y$ is $\lambda(x,y,G)$, and (b) the size of the global minimum cut is equal to $\lambda(G)$.

Let $n$, $m_0$ and $m_1$ be the number of vertices, initial edges and inserted edges, respectively. The total number of edges $m$ is the sum of the initial and inserted edges. Moreover, let $\lambda$ and $\delta$ denote the size of the global minimum cut and the minimum degree in the final graph, respectively. Note that the minimum degree is always an upper bound on the edge connectivity, i.e., $\lambda \leq \delta$ and $m = m_0 + m_1 = \Omega{(\delta n)}$. 

A subset $U \subseteq V$ is \textit{contracted} if all vertices in $U$ are identified with some element of $U$ and all edges between them are discarded. For $G=(V,E)$ and a collection of vertex sets, let $H=(V(H),E(H))$ denote the graph obtained by contracting such vertex sets. Such contractions are associated with a mapping $h : V \rightarrow V(H)$. For an edge subset $N \subseteq E$, let $N_h= \{(h(a),h(b)) : (a,b) \in N\} \subseteq E(H)$ be its corresponding edge subset induced by $h$. 

Throughout, we will use the term with high probability (in short, w.h.p.) to denote the event that holds with probability at least $1-1/n^{c}$, for some positive constant $c$.

\section{Sparse certificates} \label{sparseCertificates}
In this section we review a useful sparsification tool, introduced by Nagamochi and Ibaraki~\cite{NagamochiI92}. We first give the following definition from Benczur and Karger~\cite{BenczurK15}, which also appeared implicitly in~\cite{NagamochiI92}.
\begin{definition} A \emph{sparse $k$-connectivity certificate}, or simply a \emph{$k$-certificate}, for an unweighted graph $G$ with $n$ vertices is a subgraph $G'$ of $G$ such that 
\begin{enumerate}
\itemsep0em
\item  $G'$ consists of at most $k(n-1)$ edges, and 
\item  $G'$ contains  all edges crossing cuts of size at most $k$.
\end{enumerate} \label{sparsedef}
\end{definition}
Given an undirected graph $G = (V,E)$, a \textit{(maximal) spanning forest decomposition (msfd)} $\mathcal{F}$ of order $k$ is a decomposition of $G$ into $k$ edge-disjoint spanning forests $F_i$, $1\leq i \leq k$, such that $F_i$ is a (maximal) spanning forest of $G \setminus (F_1 \cup F_2 \ldots \cup F_{i-1})$. Note that $G_k = (V, \bigcup_{i \leq k} F_i)$ is a $k$-certificate. An msfd fulfills the following property.

\begin{lemma}[\cite{NagamochiI}] \label{lemm: Nagamochi}
Let $\mathcal{F}=(F_1,\ldots,F_m)$ be an \emph{msfd} of order $m$ of a graph $G=(V,E)$, and let $k$ be an integer with $1 \leq k \leq m$. Then for any nonempty and proper subset $S \subset V$ ,
\[
	\lambda(S,G_k) \begin{cases}
    \geq k,& \text{if } \lambda(S,G) \geq k\\
    = \lambda(S,G)  & \text{if } \lambda(S,G) \leq k-1.
\end{cases}
\]
\end{lemma}

We next present a proof of the above lemma, which closely follows the work of Nagamochi and Ibaraki~\cite{NagamochiI}. We start by presenting the following helpful result.

\begin{lemma} \label{lemm: sparseHelpful}
Let $\mathcal{F} = (F_1,\ldots,F_m)$ be an \emph{msfd} of order $m$ of a graph $G=(V,E)$. Then for any edge $(u,v) \in F_j$
and any $i \le j$, it holds that $\lambda(u,v,\bigcup_{l \leq i}F_l) \geq i$.
\end{lemma}
\begin{proof}
Fix some edge $e=(u,v) \in F_j$. We first argue that for each $i=1,\ldots,j-1$, the forest $(V,F_i)$ contains some $(u,v)$-path. Indeed, by the maximality of the forest $(V,F_i)$, the graph $(V,F_i \cup \{e\})$ must have some cycle $C$ that contains $e$. Thus, $P=C \setminus e$ is the $(u,v)$-path in the forest $(V,F_i)$. It follows that $(V,\bigcup_{l \leq i}F_l)$ has $i$ edge-disjoint paths. Next, observe that $G_j=(V,\bigcup_{l \leq j}F_l)$ has $j$ edge-disjoint paths, namely the $j-1$ edge disjoint paths
in $G_{j-1}$ (which does not contain the edge $(u,v)$) and
the 1-edge path consisting of  the edge $(u,v)$. Hence, $\lambda(u,v,\bigcup_{l \leq i}F_l) \geq i$,
\end{proof}

\begin{proof}[Proof of Lemma \ref{lemm: Nagamochi}] Assume that $\lambda(S,G) \leq k-1$. Then by definition of $G_k$, we know that $G_k$ preserves any cut $S$ of size up to $k$. Thus $\lambda(S,G_k) = \lambda(S,G)$. 

For the other case, $\lambda(S,G) \geq k$ and assume that $\lambda(S,G_k) < \lambda(S,G)$ (otherwise the lemma follows). Then there is an edge $e=(u,v) \in E_G(S,V \setminus S) \setminus E_{G_k}(S, V \setminus S)$. Since $e \not \in \bigcup_{i \leq k} F_i$, this means that $e$ belongs to some forest $F_j$ with $j > k$. By Lemma \ref{lemm: sparseHelpful}, we have that $\lambda(u,v,G_k) \geq k$. Since $(S,V \setminus S)$ separates $u$ and $v$ in $G_k$, it follows that $\lambda(S,G_k) = |E_{G_k}(S, V \setminus S)| \geq \lambda (u,v,G_k) \geq k$. 
\end{proof}

Note that by Lemma~\ref{lemm: Nagamochi} we have that $\lambda(G_k) \leq \lambda(G)$ since $G_k$ is a subgraph of $G$. This implies that $\lambda(G_k) \geq \min(k,\lambda(G))$. 

Nagamochi and Ibaraki~\cite{NagamochiI92} presented an $O(m+n)$ time algorithm (which we call a \emph{decomposition algorithm} (DA)) to construct a special msfd, which we refer to as DA-msfd. 

\section{Incremental Exact Minimum Cut} \label{sec: exactMinCut}
In this section we present a deterministic incremental algorithm that exactly maintains $\lambda(G)$. The algorithm has $O(\log^{3} n \log \log^2 n)$ update time, $O(1)$ query time and it applies to any undirected, unweighted simple graph $G = (V,E)$. The result is obtained by carefully combining a recent static min-cut algorithm by Kawarabayashi and Thorup~\cite{KawarabayashiT19} or its recent improvement due to Henzinger et al.~\cite{HenzingerRW17}, and the incremental min-cut algorithm of Henzinger~\cite{Henzinger97}. We start by describing the maintenance of non-trivial cuts, that is, cuts with at least two vertices on both sides.    
\paragraph*{Maintaining non-trivial cuts.} Kawarabayashi and Thorup~\cite{KawarabayashiT19} devised a near-linear time algorithm that contracts vertex sets of a simple input graph $G$ and produces a sparse multi-graph $H$ preserving all non-trivial minimum cuts of $G$. We refer to such a graph $H$ as a $\textsc{KT-Sparsifier}$. Recently, Henzinger et al.~\cite{HenzingerRW17} improved the running time for constructing $H$ and provided better bounds on the size of $H$. We next define a slightly generalized version of a $\textsc{KT-Sparsifier}$, and then state the bounds achieved by these two algorithms. 

\begin{definition}[\textsc{KT-Sparsifier}] Let $G=(V,E)$ be an undirected, unweighted simple graph with $n$ vertices, $m$ edges and min-cut $\lambda$.  \sloppy A multi-graph $H=(V(H),E(H))$ is a \textsc{KT-Sparsifier} of $G$ if the following holds:
\begin{itemize}
\item $H$ has $n_H = \tilde{O}(n/\lambda)$ vertices and $m_H = \tilde{O}(m/\lambda)$ edges.
\item $H$ preserves all non-trivial cuts of size up to $(3/2) \lambda$ in $G$.
\item $H$ is obtained by contracting vertex sets in $G$.
\end{itemize}
\end{definition}

\begin{theorem}[\cite{KawarabayashiT19}] Given an undirected, unweighted simple graph $G=(V,E)$, there is an $O(m \log^{12} n)$ time algorithm to construct a \textsc{KT-Sparsifier} $H$ of $G$ such that $H$ has $O(n \log^{4} n / \lambda)$ vertices and $O(m \log^{4} n / \lambda)$ edges.
\label{SparsificationThm}
\end{theorem}
In what follows, whenever we invoke the algorithm \sloppy that constructs a \textsc{KT-Sparsifier}, we mean to invoke the algorithm from the theorem below. 

\begin{theorem}[\cite{HenzingerRW17}] Given an undirected, unweighted simple graph $G=(V,E)$, there is an $O(m \log^{2} n \log \log^{2} n)$ time algorithm to construct a \textsc{KT-Sparsifier} $H$ of $G$ such that $H$ has $O(n \log n / \lambda)$ vertices and $O(m \log n / \lambda)$ edges.
\label{SparsificationThm1}
\end{theorem}

As far as non-trivial cuts are concerned, Theorem~\ref{SparsificationThm1} implies that it is safe to work on $H$ instead of $G$ as long as the sequence of newly inserted edges satisfies $\lambda_H \leq (3/2) \lambda$. To incrementally maintain the correct $\lambda_H$, we apply Henzinger's algorithm~\cite{Henzinger97} on top of $H$. The basic idea to verify the correctness of the solution is to compute and store all min-cuts of $H$. Clearly, a solution is correct as long as an edge insertion does not increase the size of all min-cuts. If all min-cuts have increased, a new solution is computed using information about the previous solution. The steps above can be performed efficiently by making use of the cactus tree representation, which we will define shortly. The crucial observation is that whenever $\lambda_H$ increases (and assuming that we can efficiently check this), instead of recomputing the cactus tree from scratch, we update intermediate structures that remained from the previous cactus tree. We next show a precise implementation of these steps. 

The minimum edge cuts are stored using the \textit{cactus tree} representation introduced by Dinitz, Karzanov and Lomonosov~\cite{dinitz} (see also~\cite{fleiner2009quick} for a concise proof). A cactus tree of a graph $G=(V,E)$ is a weighted graph $G_c = (V_c, E_c)$ defined as follows: There is a mapping $\phi: V \rightarrow V_c$ such that:
\begin{enumerate}
\itemsep0em
\item Every node in $V$ maps to exactly one node in $V_c$ and every node in $V_c$ corresponds to a (possibly empty) subset of $V$.
\item $\phi(x) = \phi(y)$ iff $x$ and $y$ are $(\lambda(G)+1)$-edge connected.
\item Every min-cut in $G_c$ corresponds to a min-cut in $G$, and every min-cut in $G$ corresponds to \text{at least} one min-cut in $G_c$.
\item If $\lambda$ is odd, every edge of $E_c$ has weight $\lambda$ and $G_c$ is a tree. If $\lambda$ is even, no two simple cycles of $G_c$ intersect in more than one node. Furthermore, edges that belong to a cycle have weight $\lambda / 2$ while those not belonging to a cycle have weight $\lambda$.
\end{enumerate}

Dinitz and Westbrook~\cite{DinitzW98} showed that given a cactus tree, we can use the data structures from~\cite{GalilI93,Poutre00} to efficiently maintain the cactus tree for fixed minimum cut size $\lambda$ under edge insertions. This implies that this data-structure can be used to efficiently test whether min-cut has increased its value during edge insertions. The result is summarized in the theorem below. 

\begin{theorem}[\cite{DinitzW98}] Given a cactus tree, there is an algorithm that maintains the cactus tree for fixed minimum cut size $\lambda$ under $u$ edge insertions, reporting when the minimum cut size increase to $\lambda + 1$ in $O(u+n)$ total time.
\label{DinitzThm}
\end{theorem}
We now turn our attention to the efficient construction and update of the cactus tree representation of a given multigraph $G$. To construct the cactus tree we use an algorithm due to Gabow~\cite{Gabow91}, which proceeds as follows. It first computes a subgraph of $G$, called a \textit{complete $\lambda$-intersection} or $I(G,\lambda)$, with at most $\lambda n$ edges, and then uses $I(G,\lambda)$ to compute the cactus tree. In the theorem below we state the running time for the cactus tree construction dependent on the time for computing $I(G,\lambda)$.

\begin{theorem}[\cite{Gabow91}] Let $G=(V,E)$ be an undirected, unweighted multigraph, and assume there is an algorithm that computes $I(G,\lambda)$ in $O(T(m,n))$ time. Given $I(G,\lambda)$, the cactus tree representation of $G$ can be constructed in $O(m)$ time. Hence, the total time for constructing the cactus tree of $G$ is bounded by $O(T(m,n) + m)$.
\label{Gabow91Thm}
\end{theorem}

Gabow~\cite{Gabow95} devised an algorithm to compute $I(G,\lambda)$ in $O(m + \lambda^{2} n \log n)$ time. Moreover, his algorithm is incremental in the sense that whenever $I(G,\lambda)$ is given as an input, the new $I(G,\lambda+1)$ can be computed more efficiently, rather than just recomputing it from scratch. The precise statement and bounds are given in the following theorem.

\begin{theorem}[\cite{Gabow95}] Given an undirected, unweighted multigraph $G=(V,E)$, there is an algorithm that computes $I(G,\lambda)$ in $O(m + \lambda^{2} n \log n)$ time. Moreover, given $I(G,\lambda)$ and a sequence of edge insertions that increase the minimum cut by 1, the new $I(G,\lambda+1)$ can be computed in $O(m'\log n)$ time, where $m'$ is the number of edges in the current graph. 
\label{Gabow95Thm}
\end{theorem}

Note that by combining Theorems~\ref{Gabow95Thm} and~\ref{Gabow91Thm} we get that the cactus tree for the initial graph can be computed in $O(m_0 + \lambda^{2} n \log n)$ time, and the new cactus tree for some current graph whose minimum cut has increased can be computed in $O(m'\log n)$ time.

\paragraph*{Maintaining trivial cuts.} We remark that the multigraph $H$ from Theorem \ref{SparsificationThm1} preserves only non-trivial cuts of $G$. If $\lambda = \delta$, then we also need a way to keep track of a trivial minimum cut. We achieve this by maintaining a minimum heap $\mathcal{H}_G$ on the vertices, where each vertex is stored with its degree. When an edge insertion is performed, the values of the edge endpoints are updated accordingly in the heap. It is well known that constructing $\mathcal{H}_G$ takes $O(n)$ time. The supported operations \textsc{Min($\mathcal{H}_G$)} and \textsc{UpdateEndpoints($\mathcal{H}_G$,$e$)} can be implemented in $O(1)$ and $O(\log n)$ time, respectively (see \cite{Cormen}). This leads to Algorithm \ref{algo: ExactMinCut}.

\def\IF{\textbf{if}~}
\def\THEN{~\textbf{then}}
\def\ENDIF{\textbf{endif}~}
\def\WHILE{\textbf{while}~}
\def\ENDWHILE{\textbf{endwhile}~}
\def\ELSE{\textbf{else}~}
\def\GOTO{\textbf{Goto}~}
\def\SPACE{\quad\quad}

\begin{algorithm2e}
\label{algo: ExactMinCut}
\caption{Incremental Exact Minimum Cut}

Compute the size $\lambda_0$ of the min-cut of $G$ and set $\lambda^* \gets \lambda_0$ \;
\nonl Build a heap $\mathcal{H}_G$ on the vertices, where each vertex stores its degree as a key \;
\nonl Compute a \textsc{KT-sparsifier} $H$ of $G$ and a mapping $h : V \rightarrow V_H$ \;
\nonl Compute the size $\lambda_H$ of the min-cut of $H$, a DA-msfd $F_1, \ldots, F_m$ of order $m$ of $H$, \;
\nonl $I(H,\lambda_H)$, and a cactus-tree of $\bigcup_{i \leq \lambda_H+1} F_i$ \;
Set $N_h \gets \emptyset$ \; 
\nonl \tcp{\textrm{ Use the data-structure from Theorem~\ref{DinitzThm} to maintain the cactus tree}}
\nonl \While{there is at least one minimum cut of size \textsc{$\lambda_H$}}
 {
   \nonl \textbf{Receive the next operation} \;
   \nonl \uIf{it is a query} { 
   \nonl \Return $\min\{\lambda_H, \normalfont \textsc{Min}(\mathcal{H}_G)\}$ }
    \nonl \uElseIf{it is the insertion of an edge $(u,v)$} {
      \nonl Update the cactus tree according to the insertion of the new edge $(h(u),h(v))$ \;
      \nonl Add the edge $(h(u),h(v))$ to $N_h$ and update the degrees of $u$ and $v$ in $\mathcal{H}_G$ \;  	 
   	 }      
  }
\nonl Set $\lambda_H \gets \lambda_H + 1$ \; 
  
\uIf{$\min\{\lambda_H, \normalfont \textsc{Min}(\mathcal{H}_G) > (3/2) \lambda^{*}$} {
\nonl  \tcp{\textrm{Full Rebuild Step}}
\nonl Compute $\lambda(G)$ and set $\lambda^{*} \gets {\lambda(G)}$ \;
\nonl Compute a \textsc{KT-sparsifier} $H$ of the current graph $G$ \;
\nonl Update $\lambda_H$ to be the min-cut of $H$ \; 
\nonl Compute a DA-msfd $F_1, \ldots, F_m$ of order $m$ of $H$ \;
\nonl  and then $I(H, \lambda_H)$ and a cactus tree of $\bigcup_{i \leq \lambda_H+1} F_i$ \; 
}
\nonl \uElseIf{$\lambda_H  \leq (3/2) \lambda^{*}$} 
{
\nonl \tcp{\textrm{Partial Rebuild Step}}
\nonl Compute a DA-msfd $F_1, \ldots, F_m$ of order $m$ of $\bigcup_{i \leq (3/2) \lambda^{*} + 1} F_i \cup N_h$ and \;
\nonl call the resulting forests $F_1,\ldots,F_m$ \;
\nonl \tcp{\textrm{Update the cactus tree using Theorems~\ref{Gabow91Thm} and~\ref{Gabow95Thm}}}
\nonl Let $H' \gets (V(H),E')$ be a graph with $E' \gets I(H,\lambda_H - 1) \cup \bigcup_{i \leq \lambda_H + 1} F_i$ \;
\nonl Compute $I(H', \lambda_H)$, a cactus tree of $H'$ and set $H \gets H'$ \;
}
\nonl \Else{

\nonl \tcp{\textrm{Special Step}}

\nonl \While{\normalfont \textsc{Min($\mathcal{H}_G$)} $\leq (3/2) \lambda^*$}
{
\nonl \uIf{the next operation is a query}{ \nonl \Return \textsc{Min($\mathcal{H}_G$)} \; } 
\nonl \Else {\nonl Update the degrees of the edge endpoints in $\mathcal{H}_G$ \; }
\nonl \GOTO Step 3 \;
}
}
\nonl \GOTO Step 2
\end{algorithm2e}

\paragraph*{Correctness.}
Let $G$ be the current graph throughout the execution of the algorithm and let $H$ be the corresponding multigraph maintained by the algorithm. Recall that $H$ preserves some family of cuts from $G$. We say that $H$ is \textit{useful} if and only if there exists a minimum cut from $G$ that is contained in the union of (a)  all trivial cuts of $G$ and (b) all cuts in $H$. Note that we consider $H$ to be useful even in the \emph{Special Step} (i.e., when $\lambda_H > (3/2) \lambda^*$), where $H$ is not updated anymore since we are certain that the smallest trivial cut is smaller than any cut in $H$. 

To prove the correctness of the algorithm we will show that (1) it correctly maintains
a trivial min-cut at any time, (2) as long as $\lambda_H \leq (3/2) \lambda^*$, the algorithm correctly maintains all cuts of size up to $(3/2)\lambda^* + 1$ of $H$, and (3) $H$ is useful as long as  $\min\{\textsc{Min}(\mathcal{H}_G),\lambda_H\} \leq (3/2) \lambda^{*}$ (Note that when this condition fails we rebuild $H$).

\begin{lemma} \label{correctness3} The algorithm correctly maintains a trivial min-cut in $G$.
\end{lemma}
\begin{proof}
This follows directly from the min-heap property of $\mathcal{H}_G$.
\end{proof}

To simplify the notation, in the following we will refer to Step 1 as a \emph{Full Rebuild Step} (namely the initial \emph{Full Rebuild Step}). Let $G=(V,E)$ be the current graph, and let $H$ be the multigraph obtained by invoking \textsc{KT-sparsifier} on $G$, at the time of a \emph{Full Rebuild Step}. Now, as long as $\lambda_H \leq (3/2)\lambda^{*}$, suppose that the graph $G$ and its corresponding multigraph $H$ have undergone a sequence of edge insertions that triggered $k$ executions of \emph{Partial Rebuild Steps} (including Step 2), for some $k\geq 0$. Note that no \emph{Full Rebuild Step} is executed as long as $\lambda_H \leq (3/2) \lambda^*$.

Let $H^{(k)}=(V(H),E(H^{(k)}))$ be the multigraph $H$ \emph{after} the $k$-th partial rebuild and let $H^{(0)} = H$. Let $N_h^{(k)} \subseteq E(H^{(k)})$ be the set of inserted edges in $H$ that the algorithm maintains during the execution of the \textbf{while} loop in Step 2, after the $(k-1)$-st and before the $k$-th partial rebuild. Define $\tilde{H}^{(k)} = (V(H), \bigcup_{i \leq (3/2)\lambda^* + 1} F_i^{(k)}) \cup N_h^{(k)}$) to be the sparsified graph that the algorithm maintains, where $F_1^{(k)},\ldots,F_m^{(k)}$ is a DA-msfd for the graph $\tilde{H}^{(k-1)}$, and let $\tilde{H}^{(0)} = H$ be the multigraph right after the last full rebuild. We next show that $\tilde{H}^{(k)}$ preserves all cuts of size up to $(3/2) \lambda^* + 1$ of $H^{(k)}$.

\begin{lemma} \label{correctness1} For $k \geq 0$, let $H^{(k)}$ and $\tilde{H}^{(k)}$ be the multigraphs defined above. Then for any nonempty and proper subset $S \subset V(H)$,
\[
	\lambda(S,\tilde{H}^{(k)}) \begin{cases}
    \geq (3/2)\lambda^{*} + 1,& \text{if } \lambda(S,H^{(k)}) \geq (3/2)\lambda^*+1\\
    = \lambda(S,H^{(k)})  & \text{if } \lambda(S,H^{(k)}) \leq (3/2)\lambda^*.
\end{cases}
\]

\end{lemma}
\begin{proof} 
We proceed by induction on the number $k$ of partial rebuilds. We give the inductive step; the base case $(k=0)$ follows from the fact that $\tilde{H}^{(0)} = H = H^{(0)}$. 

Fix any cut $(S,V(H) \setminus S)$ in $H^{(k)}$, and note that $H^{(k)} = (V(H),E(H^{(k-1)}) \cup N^{(k)}_h)$. Define $A \coloneqq E_{H^{(k)}}(S, V(H) \setminus S) \cap N^{(k)}_h$ and $B \coloneqq E_{H^{(k)}}(S, V(H) \setminus S) \cap E(H^{(k-1)})$ such that $E_{H^{(k)}}(S, V(H) \setminus S) = A \uplus B$. Letting $F' = \bigcup_{i \leq (3/2)\lambda^*+1}F_i^{(k)}$, we similarly define edge sets $\tilde{A}$ and $\tilde{B}$ partitioning the edges $E_{\tilde{H}^{(k)}}(S,V(H)\setminus S)$ that cross the cut $(S,V(H)\setminus S)$ in $\tilde{H}^{(k)}$. Note that $A = \tilde{A}$ since edges of $N^{(k)}_h$ are always included in $\tilde{H}^{(k)}$ and $\lambda(S,H^{(k)}) = \abs{A} + \abs{B}$, $\lambda(S,\tilde{H}^{(k)})=\abs{\tilde{A}} + \abs{\tilde{B}}$. We distinguish two cases.

First, assume $\lambda(S,H^{(k)}) \leq (3/2)\lambda^{*}$. Then, since $H^{(k-1)} \subseteq H^{(k)}$ and by construction of $H^{(k)}$, $\lambda(S,H^{(k-1)}) = \abs{B}$, we get that $\lambda(S,H^{(k-1)}) \leq (3/2) \lambda^{*}$. By induction hypothesis, it follows that $\lambda(S,\tilde{H}^{(k-1)}) = \lambda(S,H^{(k-1)}) \leq (3/2)\lambda^{*}$. The latter along with Lemma~\ref{lemm: Nagamochi} implies that $\abs{\tilde{B}} = \lambda(S,\tilde{H}^{(k-1)})$, and thus $\lambda(S,\tilde{H}^{(k)}) = \abs{\tilde{A}} + \abs{\tilde{B}}  =  \abs{A} + \abs{B} = \lambda(S,H^{(k)})$. 

Second, assume $\lambda(S,H^{(k)}) \geq (3/2)\lambda^{*} + 1$.  Then either $\lambda(S,H^{(k-1)}) \leq (3/2)\lambda^*$ or $\lambda(S,H^{(k-1)}) \geq (3/2)\lambda^* + 1$. In the first case, by induction hypothesis it follows that $\lambda(S,\tilde{H}^{(k-1)}) = \lambda(S,H^{(k-1)}) \leq (3/2) \lambda^*$. This along with Lemma~\ref{lemm: Nagamochi} implies that $\abs{\tilde{B}} = \lambda(S,\tilde{H}^{(k-1)})$, and thus $\lambda(S,\tilde{H}^{(k)}) = \abs{\tilde{A}} + \abs{\tilde{B}}  =  \abs{A} + \abs{B} = \lambda(S,H^{(k)}) \geq (3/2)\lambda^* + 1$. In the second case, by induction hypothesis it follows that $\lambda(S,\tilde{H}^{(k-1)}) \geq (3/2)\lambda^* + 1$. The latter along with Lemma~\ref{lemm: Nagamochi} imply that $\abs{\tilde{B}} \geq (3/2)\lambda^* + 1$, and thus $\lambda(S,\tilde{H}^{(k)}) = \abs{\tilde{A}} + \abs{\tilde{B}} \geq (3/2)\lambda^* + 1$, which completes the proof.   
\end{proof}

We now show that the multigraphs $H^{(k)}$ and $\tilde{H}^{(k)}$ share the same set of minimum cuts.

\begin{lemma} \label{min-cut-equi}
Assume that $\lambda(H^{(k)}) \leq (3/2) \lambda^*$. Then a cut is a min-cut in $H^{(k)}$ iff it is a min cut in  $\tilde{H}^{(k)}$.
\end{lemma}
\begin{proof}
We first show that every non-min cut in $H^{(k)}$ is a non-min cut in $\tilde{H}^{(k)}$. By contrapositive, we get that a min-cut in $\tilde{H}^{(k)}$ is a min-cut in $H^{(k)}$.

To this end, let $(S, V(H) \setminus S)$ be a cut with $\lambda(S,H^{(k)}) \geq \lambda(H^{(k)}) + 1$ in $H^{(k)}$. Note that by assumption $\lambda(H^{(k)}) \leq (3/2)\lambda^*$. By Lemma~\ref{correctness1} we distinguish two cases. (1) If $\lambda(S,H^{(k)}) \leq (3/2)\lambda^*$, then $\lambda(S,\tilde{H}^{(k)}) = \lambda(S,H^{(k)}) \geq \lambda(H^{(k)}) + 1$. (2) If $\lambda(S,H^{(k-1)}) \geq (3/2)\lambda^* + 1$, then $\lambda(S,\tilde{H}^{(k)}) \geq (3/2)\lambda^* + 1 \geq \lambda(H^{(k)}) + 1$. The above cases along with $\lambda(H^{k}) \geq \lambda(\tilde{H}^{(k)})$ give that $\lambda(S,\tilde{H}^{(k)}) \geq \lambda(\tilde{H}^{(k)}) + 1$, which in turn implies that $(S,V(H) \setminus S)$ cannot be a min-cut in $\tilde{H}^{(k)}$.

For the other direction, consider a min-cut $(D,V(H) \setminus D)$ of size $\lambda(D,\tilde{H}^{(k)})$ in $\tilde{H}^{(k)}$. Considering the cut in $H^{(k)}$ we know that $\lambda(D,H^{(k)}) \geq \lambda(H^{(k)})$. Then, similarly as above, Lemma~\ref{correctness1} implies that $\lambda(D,\tilde{H}^{(k)}) \geq \lambda(H^{(k)})$. Since $(D,V(H) \setminus D)$ was chosen arbitrarily, we get that $\lambda(\tilde{H}^{(k)}) \geq \lambda(H^{(k)})$ must hold. The latter along with $\lambda(\tilde{H}^{(k)}) \leq \lambda(H^{(k)})$ imply that $\lambda(\tilde{H}^{(k)}) = \lambda(H^{(k)})$.

Now, let $(S,V(H) \setminus S)$ be a min-cut in $H^{(k)}$. Since $\tilde{H}^{(k)}$ is a subgraph of $H^{(k)}$ we know that $\lambda(S,\tilde{H}^{(k)}) \leq \lambda(S,H^{(k)})$. The latter along with $\lambda(\tilde{H}^{(k)}) = \lambda(H^{(k)})$ imply that 
\[ \lambda(S,\tilde{H}^{(k)}) \leq \lambda(S,H^{(k)}) = \lambda(H^{(k)}) = \lambda(\tilde{H}^{(k)}), \] 
or, $\lambda(S,\tilde{H}^{(k)}) \leq \lambda(\tilde{H}^{(k)})$. It follows that the inequality must hold with equality since $\lambda(\tilde{H}^{(k)})$ is the value of min-cut in $\tilde{H}^{(k)}$. Thus, $(S,V(H) \setminus S)$ is also a min-cut in $\tilde{H}^{(k)}$.
\end{proof}

\begin{lemma} \label{correctness2} For some current graph G, let $H$ be the current multigraph maintained by the algorithm and assume that $\lambda_H \leq (3/2) \lambda^{*}$, where $\lambda^{*}$ denotes the min-cut of $G$ at the last \emph{Full Rebuild Step}. Then the value $\lambda_H$ maintained by the algorithm satisfies $\lambda_H = \lambda(H)$.
\end{lemma}
\begin{proof} Let $\lambda(H^{(k)})$ be the value of $\lambda_H$ \emph{after} the $k$-th execution of partial rebuild step, for $k \geq 0$. Since, $\lambda(H^{(k)}) = \lambda(H)$, it suffices to show that $\lambda(H^{(k)})$ is correct. We proceed by induction on the number $k$ of partial rebuilds since the last full rebuild.

We first consider the base case $k=0$, i.e., the time right after the last full rebuild. At the beginning of a full rebuild, the algorithm computes a \textsc{KT-sparsifier} $H$ of $G$ that preserves all non-trivial min-cuts of $G$. The value of $\lambda_H$ is updated to $\lambda(H)$, a DA-msfd $F_1,\ldots,F_m$ is computed for $H$, and a cactus tree is constructed for $F' = \bigcup_{i \leq \lambda_H+1} F_i$. Lemma~\ref{lemm: Nagamochi} shows that a cut is a min-cut in $H$ iff it is a min-cut in $F'$. The latter implies that since the cactus tree preserves the min-cuts of $F'$, it also preserves those of $H$. The fact that the cactus tree algorithm correctly tells us when to increment $\lambda_H$ in Step 2, we conclude that the value of $\lambda_H$ after a full rebuild is set correctly.

We next give the inductive step. By induction hypothesis assume that $\lambda(H^{(k-1)})$ is correct. By Lemma~\ref{min-cut-equi} we get that a cut is a min-cut in $H^{(k-1)}$ iff it is a min-cut in $\tilde{H}^{(k-1)}$. Now, let $F^{(k)}_1,\ldots,F^{(k)}_m$ be the DA-msfd computed on $\tilde{H}^{(k-1)}$ during the $k$-th partial rebuild, and define $\tilde{F}^{(k)} = \bigcup_{i \leq \lambda(H^{(k-1)})+ 1} F_i^{(k)} $. Lemma~\ref{lemm: Nagamochi} shows that a cut is min-cut in $\tilde{H}^{(k-1)}$ iff it is a min-cut in $\tilde{F}^{(k)}$. The two equivalences above give that every min-cut in $H^{(k-1)}$ is a min-cut $\tilde{F}^{(k)}$, and thus the graph $H'^{(k)}$ (as defined in Algorithm~\ref{algo: ExactMinCut}) correctly preserves all min-cuts of $H^{(k-1)}$. Given the correctness of $\lambda(H^{(k-1)})$, the properties of the cactus trees, and the fact that the incremental cactus tree algorithm correctly tells us when to increment $\lambda(H^{(k-1)})$ in Step 2, we conclude that $\lambda(H^{(k)})$ is the correct min-cut value for the graph $H^{(k)} = (V(H), E(H^{(k-1)}) \cup N^{(k)}_h)$ after the $k$-th partial rebuild.
\end{proof}

Note that when $\lambda_H > (3/2) \la^*$, the above lemma is not guaranteed to hold as the algorithm does not execute a \emph{Partial Rebuild Step} in this case. However, we will show below that this is not necessary for the correctness of the algorithm. The fact that we do not need to execute a \emph{Partial Rebuild Step} in this setting is crucial for achieving our time bound.

\begin{lemma} \label{correctness4} If $\min\{\textsc{Min}(\mathcal{H}_G),\lambda_H\} \leq 3/2 \lambda^{*}$, then $H$ is useful. 
\end{lemma}
\begin{proof}
Let $(S',V \setminus S')$ be any non-trivial cut in $G$ that is not in $H$. Such a cut must have cardinality strictly greater than $(3/2) \lambda^{*}$ since otherwise it would be contained in $H$. We show that $(S',V \setminus S')$ cannot be a minimum cut as long as $\min\{\textsc{Min}(\mathcal{H}_G),\lambda_H\} \leq (3/2) \lambda^{*}$ holds. We distinguish two cases.
\begin{enumerate}
\itemsep0em
\item If $\lambda_H \leq (3/2) \lambda^*$, then by Lemma \ref{correctness2} the algorithm maintains $\lambda_H$ correctly. Since $H$ is obtained from $G$ by contracting vertex sets, there is a cut $(S,V_H,S)$ in $H$, and thus in $G$, of value $\lambda_H$. It follows that $(S',V \setminus S')$ cannot be a minimum cut of $G$ since $|E(S', V \setminus S')| > (3/2) \lambda^* \geq \lambda_H = \lambda(H) \geq \lambda(G)$, where the last inequality follows from the fact that $H$ is a contraction of $G$.

\item If $\textsc{Min}(\mathcal{H}_G) \leq (3/2) \lambda^{*}$, then by Lemma \ref{correctness3} there is a cut of size $\textsc{Min}(\mathcal{H}_G) = \delta$ in $G$. Similarly, $(S', V \setminus S')$ cannot be a minimum cut of $G$ since $|E(S', V \setminus S')| > (3/2) \lambda^{*} \geq \delta \geq \lambda(G)$.
\end{enumerate}
Appealing to the above cases, we conclude $H$ is useful since a min-cut of $G$ is either contained in $H$ or it is a trivial cut of $G$. 
\end{proof}

\begin{lemma} Let $G$ be some current graph. Then the algorithm correctly maintains $\lambda(G)$.
\end{lemma}
\begin{proof} 
Let $G$ be some current graph and $H$ be the current multigraph maintained by the algorithm.
We will argue that $\lambda(G) = \min\{\textsc{Min}(\mathcal{H}_G),\lambda_H\}$. 

If $\min\{\textsc{Min}(\mathcal{H}_G),\lambda_H\} \leq (3/2) \lambda^{*}$, then by Lemma \ref{correctness4}, $H$ is useful i.e., there exists a minimum cut of $G$ that is contained in the union of all trivial cuts of $G$ and all cuts in $H$. Lemma \ref{correctness3} guarantees that the algorithm correctly maintains $\textsc{Min}(\mathcal{H}_G)$, i.e., the trivial minimum cut of $G$. If $\lambda_H \leq (3/2) \lambda^*$, then Lemma \ref{correctness2} ensures that $\lambda_H = \lambda(H)$, and thus $\min\{\textsc{Min}(\mathcal{H}_G), \lambda_H\} = \lambda(G)$.  If, however, $\lambda_H > (3/2) \la^*$ but $\min\{\textsc{Min}(\mathcal{H}_G),\lambda_H\} \leq (3/2) \lambda^{*}$, then $\lambda_H  > \min\{\textsc{Min}(\mathcal{H}_G),\lambda_H\}$ which implies that $\min\{\textsc{Min}(\mathcal{H}_G), \lambda_H\} = \textsc{Min} (\mathcal{H}_G) = \lambda(G)$. As we argued above, the algorithm correctly maintains $\textsc{Min}(\mathcal{H}_G)$ at any time. Thus it follows that the algorithm correctly maintains $\lambda(G)$ in this case as well.

The only case that remains to consider is $\min\{\textsc{Min}(\mathcal{H}_G),\lambda_H\} > (3/2) \lambda^{*}$. But whenever this happens the algorithm performs a full rebuild step. After this full rebuild $\la(G) = \min\{\textsc{Min}(\mathcal{H}_G),\lambda_H\}$ trivially holds.
\end{proof}
 
\paragraph*{Running Time Analysis.}

\begin{theorem} Let $G$ be a simple graph with $n$ nodes and $m_0$ edges. Then the total time for inserting $m_1$ edges and maintaining a minimum edge cut of $G$ is \[ O((m_0 + m_1) \log^{3}n \log \log^{2}n).\] If we start with an empty graph, the amortized time per edge insertion is $O(\log^{3}n \log \log^{2}n)$. The size of a minimum cut can be answered in constant time.
\end{theorem}
\begin{proof}
We first analyse Step 1. Note that building the heap $\mathcal{H}_G$ and computing $\lambda_0$ take $O(n)$ and $O(m_0 \log^{2}n \log \log^{2}n)$ time, respectively. Recall that $m_0 \geq \lambda_0 n$. The total running time for constructing $H$, $I(H,\lambda_H)$ and the cactus tree is dominated by $O((m_0 + \lambda^{2}_0 \cdot ( n / \lambda_0))\log^{2}n \log \log^{2}n) = O(m_0 \log^{2}n \log \log^{2}n)$. Thus, the total time for Step 1 is $(m_0 \log^{2}n \log \log^{2}n)$.

Let $\lambda_H^0, \ldots, \lambda_H^f$ be the values that $\lambda_H$ assumes in Step 2  during the execution of the algorithm in increasing order. We define \text{Phase} $i$ to be all steps executed after Step $1$ while $\lambda_H = \lambda_H^{i}$, excluding \emph{Full Rebuild Steps} and \emph{Special Steps}. Additionally, let $\lambda^{*}_0, \ldots, \lambda^{*}_{O(\log n)}$ be the values that $\lambda^{*}$ assumes during the algorithm. 
We define \textit{Superphase} $j$ to consist of the $j$-th \emph{Full Rebuild Step} along with all steps executed until the next \emph{Full Rebuild Step}, i.e.,  while $\min\{\textsc{Min}(\mathcal{H}_G), \lambda_H\} \leq (3/2) \lambda^{*}_j$, where $\lambda^{*}_j$ is the value of $\lambda(G)$ at the $j$-th \emph{Full Rebuild Step}. Note that a superphase consists of a sequence of phases and potentially a final \emph{Special Step}. Moreover, the algorithm executes a phase if $\lambda_H \leq (3/2) \lambda^{*}$. 

We say that $\lambda_H^i$ \textit{belongs} to superphase $j$, if the $i$-th phase is executed during superphase $j$ and $\lambda_H^i\leq (3/2) \lambda_j^{*}$. We remark that the number of vertices in $H$ changes only at the beginning of a superphase, and remains unchanged during its lifespan.

Let $n_j$ denote the number of vertices in some superphase $j$. We bound this quantity as follows:
\begin{proposition} \label{fact}
Let $j$ be a superphase during the execution of the algorithm. Then, we have
\[
	n_j = O((n \log n) / \lambda_H^i), \text{ for all } \lambda_H^i \text{ belonging to superphase } j.
\]
\end{proposition}
\begin{proof}
From Step 3 and Theorem~\ref{SparsificationThm1} we know that $n_j = O((n \log n) / \lambda^{*}_j)$. Moreover, observe that $\lambda_j^{*} \leq \lambda_H^i$ and a phase is executed whenever $\lambda_H^i \leq (3/2) \lambda_j^{*}$. Thus, for all $\lambda_H^i$'s belonging to superphase $j$, we get the following relation
\begin{equation}
\label{MinCutRelation}
	\lambda^{*}_j \leq \lambda_H^i \leq (3/2) \lambda^{*}_j,
\end{equation}
which in turn implies that $n_j = O((n \log n) / \lambda^{*}_j) = O((n \log n)/ \lambda_H^i)$.
\end{proof}

For the remaining steps, we divide the running time analysis into two parts, one part corresponding to phases, and the other to superphases. 

\paragraph*{Part $1$.}For some superphase $j$, the $i$-th phase consists of the $i$-th execution of a \emph{Partial Rebuild Step} followed by the execution of Step 2. Let $u_i$ be the number of edge insertions in Phase $i$. By Theorem~\ref{DinitzThm} and the fact that heap-insertions are performed in $O(\log n)$ time, it follows that
the total time for Step 2 during the $i$-th phase is $O(n_j+u_i \log n) = O((n + u_i) \log n)$. Since $n_j = {O}((n \log n)/\lambda^{*}_j)$, we observe that $\bigcup_{i \leq (3/2)\lambda_j^* +1}F_i \cup N_h$ has size $O(u_{i-1} + \lambda_j^* n_j) = O((u_{i-1} + n)\log n)$. Thus, the total time for computing DA-msfd in a \emph{Partial Rebuild Step} is $O((u_{i-1} + n) \log n)$. Using Proposition~\ref{fact} note that $H'$ has $O(\lambda_H^{i} n_j) = O(n \log n)$ edges and thus it takes $O(n \log^2 n)$ time to compute $I(H',\lambda_{H}^i)$ and the new cactus tree.

The total time spent in Phase $i$ is $O((u_{i-1} + u_{i} + n) \log^{2} n)$. Let $\lambda$ and $\lambda_H$ denote the size of the minimum cut in the final graph and its corresponding multigraph, respectively. Note that $\sum_{i=1}^{\lambda} u_i \leq m_1$, $\lambda n \leq m_0 + m_1$ and recall Eqn. (\ref{MinCutRelation}).  This gives that the total work over all phases is
\begin{align*}
    \sum_{i = 1}^{\lambda_H} O((u_{i-1} & + u_{i} + n) \log^{2} n) \\
    & = \sum_{i = 1}^{\lambda} O((u_{i-1} + u_{i} + n) \log^{2} n) = O((m_0 + m_1)\log^{2} n) . 
\end{align*}

\paragraph*{Part $2$.}The $j$-th superphase consists of the $j$-th execution of a \emph{Full Rebuild Step} along with a possible execution of a \emph{Special Step}, depending on whether the condition is met. In a \emph{Full Rebuild Step}, computing $\lambda(G)$ takes $O((m_0+m_1)\log^2 n \log \log^{2} n)$ time. The total running time for constructing $H$, $I(H,\lambda^{*}_j)$ and the cactus tree is dominated by $O((m_0 + m_1 + (\lambda^{*}_j)^{2} \cdot (n / \lambda^{*}_j)) \log^2 n \log \log^{2} n) = O((m_0 + m_1)\log^2 n \log \log^{2} n)$. The running time of a \emph{Special Step} is $O(m_1 \log n)$.

Throughout its execution, the algorithm begins a new superphase whenever $\lambda(G) =\min$ $\{\textsc{Min}(\mathcal{H}_G), \lambda_H\} > (3/2)\lambda^{*}$. This implies that $\lambda(G)$ must be at least $(3/2)\lambda^{*}$, where $\lambda^{*}$ is the value of $\lambda(G)$ at the last \emph{Full Rebuild Step}. Thus, a new superphase begins whenever $\lambda(G)$ has increased by a factor of $3/2$, i.e., only $O(\log n)$ times over all insertions. 
This gives that the total time over all superphases is $O((m_0 + m_1)\log^3 n \log \log^{2} n)$. $\quad \square$
\end{proof}

\section{Incremental \texorpdfstring{$(1+\epsilon)$}{1+eps} Minimum Cut with \texorpdfstring{$\tilde{O}(\MakeLowercase{n})$}{O(n poly log n)} space} \label{sec: ApproxMinCut}
In this section we present two $\tilde{O}(n)$ space incremental Monte-Carlo algorithms that w.h.p.  maintain the size of a min-cut up to a $(1+\epsilon)$-factor. Both algorithms have $\tilde{O}(1)$ update-time and $\tilde{O}(1)$, resp.~$O(1)$ query-time. The first algorithm uses $O(n \log^{2}n / \epsilon^2)$ space, while the second one improves the space complexity to $O(n \log n / \epsilon^2)$. 

\subsection{An \texorpdfstring{$O(n \log^2 n / \epsilon^2)$}{O (n log2 n)} space algorithm}
Our first algorithm follows an approach that was used
in several previous work~\cite{Henzinger97,ThorupSWAT00,Thorup07}. The basic idea is to maintain the min-cut up to some size $k$ using small space. We achieve this by maintaining a sparse $(k+1)$-certificate and incorporating it into the incremental exact min-cut algorithm due to Henzinger~\cite{Henzinger97}, as described in Section \ref{sec: exactMinCut}. Finally we apply the well-known randomized sparsification result due to Karger~\cite{Karger99} to obtain our result.

\paragraph*{Maintaining min-cut up to size $k$ using $O(kn)$ space.} We incrementally maintain an msfd for an unweighted graph $G$ using $k+1$ union-find data structures $\mathcal{F}_1, \ldots, \mathcal{F}_{k+1}$ (see~\cite{Cormen}). Each $\mathcal{F}_i$ maintains a spanning forest $F_i$ of $G$. Recall that $F_1,\ldots,F_{k+1}$ are edge-disjoint. 
When a new edge $e=(u,v)$ is inserted into $G$, we define $i$ to be the first index such that $\mathcal{F}_i.$\textsc{Find}$(u)$ $\neq$ $\mathcal{F}_i.$\textsc{Find}$(v)$. If we found such an $i$, we append the edge $e$ to the forest $F_i$ by setting $\mathcal{F}_i.$\textsc{Union}$(u,v)$ and return $i$. If such an $i$ cannot be found after $k+1$ steps, we simply discard edge $e$ and return NULL. We refer to such procedure as $(k+1)$-\textsc{Connectivity}$(e)$.

It is easy to see that the forests maintained by $(k+1)$-\textsc{Connectivity}$(e)$ for every newly inserted edge $e$ are indeed edge-disjoint. Combining this procedure with techniques from Henzinger~\cite{Henzinger97} leads to the following Algorithm \ref{algo: ExactMinCutUpToK}. 

\begin{algorithm2e}
\label{algo: ExactMinCutUpToK}
\caption{Incremental Exact Min-Cut up to size $k$}
Set $\lambda \gets 0$, initialize $k+1$ union-find data structures $\mathcal{F}_1, \ldots, \mathcal{F}_{k+1}$, \; 
\nonl $k+1$ empty forests $F_1,\ldots,F_{k+1}$, $I(G,\lambda)$, and an empty cactus tree \;
\nonl \While{there is at least one minimum cut of size $\lambda$}
{
\nonl \textbf{Receive the next operation} \;
\nonl \uIf{it is a query}{\nonl \Return $\lambda$}
\nonl \uElseIf{it is the insertion of an edge $e$} 
{
\nonl Set $i \gets (k+1)$-\textsc{Connectivity}$(e)$ \;
\nonl \If{$i \neq $ \normalfont NULL}{
\nonl Set $F_i \gets F_i \cup \{e\}$ \;
\nonl Update the cactus tree according to the insertion of the edge $e$ \;
}
}
}
Set $\lambda = \lambda + 1$ \;
\nonl  Let $G' = (V,E')$ be a graph with $E' \gets I(G,\lambda - 1) \cup \bigcup_{i \leq \lambda+1} F_i$ \;
\nonl Compute $I(G',\lambda)$ and a cactus tree of $G'$ \;
\nonl \GOTO Step 2\;

\end{algorithm2e}
The correctness of the above algorithm is immediate from Lemmas \ref{correctness1} and \ref{correctness2}. The running time and query bounds follow from Theorem 8 of~\cite{Henzinger97}. For the sake of completeness, we provide here a full proof.

\begin{corollary} \label{cor: ExactPolyLog}
For $k > 0$, there is an $O(kn)$ space algorithm that processes a stream of edge insertions starting from any empty graph $G$ and maintains an exact value of $\min\{\lambda(G),k\}$. Starting from an empty graph, the total time for inserting $m$ edges is $O(km\alpha(n) \log n )$ and queries can be answered in constant time, where $\alpha(n)$ stands for the inverse of Ackermann function.
\end{corollary}
\begin{proof}
We first analyse Step $1$. Initializing $k+1$ union-find data structures takes $O(kn)$ time. The running time for constructing $I(G,\lambda)$ and building an empty cactus tree is also dominated by $O(kn)$. Thus, the total time for Step $1$ is $O(kn)$.

Let $\lambda_0, \ldots, \lambda_f$, where $\lambda_f \leq k$,  be the values that $\lambda$ assumes in Step $2$ during the execution of the algorithm in increasing order. We define Phase $i$ to be all steps executed while $\lambda = \lambda_i$. For $i\geq 1$, we can view Phase $i$ as the $i$-th execution of Step $3$ followed by the execution of Step $2$. Let $u_i$ denote the number of edge insertion in Phase $i$. The total time for testing the $(k+1)$-connectivity of the endpoints of the newly inserted edges, and updating the cactus tree in Step $2$ is dominated by $O(n + k \alpha(n) u_i)$. Since the graph $G'$ in Step $3$ has always at most $O(kn)$ edges, the running time to compute $I(G',\lambda)$ and the cactus tree of $G'$ is $O(kn \log n)$. Combining the above bounds, the total time spent in Phase $i$ is $O(k(\alpha(n)u_i + n \log n))$. Thus, the total work over all phases is $O(km\alpha(n) \log n)$. 

The space complexity of the algorithm is only $O(kn)$, since we always maintain at most $k+1$ spanning forests during its execution.
\end{proof}

\paragraph*{Dealing with min-cuts of arbitrary size.} We observe that Corollary \ref{cor: ExactPolyLog} gives polylogarithmic amortized update time only for min-cuts up to some polylogarithmic size. For dealing with min-cuts of arbitrary size, we use the well-known sampling technique due to Karger~\cite{Karger99}. This allows us to get an $(1+\epsilon)$-approximation to the value of a min-cut with high probability.

\begin{lemma}[\cite{Karger99}] \label{lemm: Karger} Let $G$ be any graph with minimum cut $\lambda$ and let $p \geq 12(\log n)/(\epsilon^{2}\lambda)$. Let $G(p)$ be a subgraph of $G$ obtained by including each of edge of $G$ to $G(p)$ with probability $p$ independently. Then the probability that the value of any cut of $G(p)$ has value more than $(1+\epsilon)$ or less than $(1-\epsilon)$ times its expected value is $O(1/n^{4})$.  
\end{lemma} 

For some integer $i \geq 1$, let $G_i$ denote a subgraph of $G$ obtained by including each edge of $G$ to $G_i$ with probability $1/2^{i}$ independently. We now have all necessary tools to present our incremental algorithm.

\begin{algorithm2e}
\label{algo: SmallSpaceMinCut1} 
\caption{$(1+\epsilon)$-Min-Cut with $O(n \log^2 n / \epsilon^{2})$ space}
\For{$i=0,\ldots, \lfloor \log n \rfloor$}
{ \nonl let $G_i$ be an initially empty sampled subgraph \; }
\textbf{Receive the next operation} \;
\nonl  \uIf{it is a query}
{
\nonl Find the minimum $j$ such that $\lambda(G_j) \leq k$ and  \Return $2^{j}\lambda(G_j)/(1-\epsilon)$ \; 
}
\nonl \uElseIf{it is the insertion of an edge $e$}{
\nonl Include edge $e$ to each $G_i$ with probability $1/2^{i}$ \;
\nonl Maintain the exact min cut of each $G_i$ up to size $k \gets 48 \log n / \epsilon^2$  \; 
\nonl using Algorithm \ref{algo: ExactMinCutUpToK} \;
}
\GOTO Step 2.
\end{algorithm2e}

\begin{theorem} \label{thm: space1}
There is an $O(n \log^{2} n/\epsilon^{2})$ space randomized algorithm that processes a stream of edge insertions starting from an empty graph $G$ and maintains a $(1+\epsilon)$-approximation to a min-cut of $G$ with high probability. The amortized update time per operation is $O(\alpha(n)\log^{3} n / \epsilon^{2})$ and queries can be answered in $O(\log n)$ time. 
\end{theorem}
\begin{proof}
We first prove the correctness of the algorithm. For an integer $t \geq 0$, let $G^{(t)} = (V,E^{(t)})$ be the graph after the first $t$ edge insertions. Further, let $\lambda(G^{(t)})$ denote the min-cut of $G^{(t)}$, $p^{(t)}=12(\log n)/(\epsilon^{2}\lambda^{(t)})$ and $\lambda(S,G) = |E_G(S, V \setminus S)|$, for some cut $(S, V \setminus S)$. For any integer $i \leq \lfloor \log_2 1 / p^{(t)} \rfloor$, Lemma \ref{lemm: Karger} implies that for any cut $(S,V \setminus S)$, $((1-\epsilon)/2^{i}) \lambda(S,G^{(t)}) \leq \lambda(S,G_{i}^{(t)}) \leq ((1+\epsilon)/2^{i}) \lambda(S,G^{(t)})$, with probability $1-O(1/n^{4})$. Let $(S^*, V \setminus S^*)$ be a min-cut of $G_{i}^{(t)}$, i.e.,  $\lambda(S^*, G_{i}^{(t)}) = \lambda(G_{i}^{(t)})$. Setting $i= \lfloor \log_2 1/p^{(t)} \rfloor$, we get that:
\[
 \mathbb{E}[\lambda(G^{(t)}_i)] \leq \lambda(G^{(t)})/2^{i}  \leq 2p^{(t)} \lambda(G^{(t)}) \leq 24 \log n/\epsilon^{2}.
\]
The latter along with the implication of Lemma \ref{lemm: Karger} give that for any $\epsilon \in (0,1)$, the size of the minimum cut in $G^{(t)}_{i}$ is at most $(1+\epsilon) 24 \log n / \epsilon^{2} \leq 48 \log n / \epsilon^{2}$ with probability $1-O(1/n^{4})$. Thus, $j \leq \lfloor \log_2 1 / p^{(t)} \rfloor$ with probability $1-O(1/n^{4})$. Additionally, we observe that the algorithm returns a $(1+O(\epsilon)) $-approximation to a min-cut of $G^{(t)}$ w.h.p. since by Lemma \ref{lemm: Karger}, $2^{i} \lambda(G_i^{(t)})/(1-\epsilon) \leq (1+\epsilon)/(1-\epsilon)\lambda(G^{(t)}) = (1+O(\epsilon))\lambda(G^{(t)})$ w.h.p. Note that for any $t$, $\lfloor \log_2 1 / p^{(t)} \rfloor \leq \lfloor \log n \rfloor$, and thus it is sufficient to maintain only $O(\log n)$ sampled subgraphs.

Since our algorithm applies to unweighted simple graphs, we know that $t \leq O(n^{2})$. Now applying union bound over all $t \in \{1,\ldots O(n^{2})\}$ gives that the probability that the algorithm does not maintain a $1 + O(\epsilon)$-approximation is at most $O(1/n^2)$.

The total expected time for maintaining a sampled subgraph is $O(m\alpha(n) \log^{2} n / \epsilon^{2})$ and the required space is $O(n \log n / \epsilon^{2})$ (Corollary \ref{cor: ExactPolyLog}). Maintaining $O(\log n)$ such subgraphs gives an $O(\alpha(n)\log^{3} n / \epsilon^{2})$ amortized time per edge insertion and an $O(n \log^2 n / \epsilon^{2})$ space requirement. The $O(\log n)$ query time follows as in the worst case we scan at most $O(\log n)$ subgraphs, each answering a min-cut query in constant time. 
\end{proof}

\subsection{Improving the space to \texorpdfstring{$O(n \log n / \epsilon^2)$}{O (n log n)}}
We next show how to bring down the space requirement of the previous algorithm to $O(n \log n / \epsilon^{2})$ without degrading its running time. The main idea is to keep a single sampled subgraph instead of $O(\log n)$ of them. 

Let $G=(V,E)$ be an unweighted undirected graph and assume each edge is given some random weight $p_e$ chosen uniformly from $[0,1]$. Let $G^{w}$ be the resulting weighted graph. For any $p > 0$, we denote by $G(p)$ the unweighted subgraph of $G$ that consists of all edges that have weight at most $p$. We state the following lemma due to Karger~\cite{kargerPHD}:

\begin{lemma} \label{lemm: Karger2} Let $k = 48 \log n / \epsilon^{2}$.
Given a connected graph $G$, let $p$ be a value such that $p \geq k/ (4 \lambda(G))$. 
Then with high probability, $\lambda(G(p)) \leq k$ and $\lambda(G(p))/p$ is an $(1+\epsilon)$-approximation to a min-cut of $G$.
\end{lemma}
\begin{proof}
Since the weight of every edge is uniformly distributed, the probability that an edge has weight at most $p$ is exactly $p$. Thus, $G(p)$ is a graph that contains every edge of $G$ with probability $p$. The claim follows from Lemma~\ref{lemm: Karger}.
\end{proof}

For any graph $G$ and some appropriate weight $p \geq k/(4\lambda(G))$, the above lemma tells us that the min-cut of $G(p)$ is bounded by $k$ with high probability. 
Thus, instead of considering the graph $G$ along with its random edge weights, we build a collection of $k+1$ minimum edge-disjoint spanning forests (using those edge weights). We note that such a collection is an msfd of order $k+1$  for $G$ with $O(kn)$ edges and by Lemma \ref{correctness1}, it preserves all minimum cuts of $G$ up to size $k$.

Our algorithm uses the following two data structures: 

(1) {\textsc{NI-Sparsifier}$(k)$ data-structure}:  Given a graph $G$, where each edge $e$ is assigned some weight $p_e$ and some parameter $k$, we devise an insertion-only data-structure that maintains a collection of $k+1$  minimum edge-disjoint spanning forests $F_1,\ldots,F_{k+1}$ with respect to the edge weights. Let $F = \bigcup_{i\leq k+1} F_i$. Since we are in the incremental setting, it is known that the problem of maintaining a single minimum spanning forest can be solved in time $O(\log n)$ per insertion using the dynamic tree structure of Sleator and Tarjan~\cite{SleatorT83}. Specifically, we use this data-structure to 
determine for each pair of nodes $(u,v)$ the maximum weight of an edge in the cycle that the edge $(u,v)$ induces in the
minimum spanning forest $F_i$. Let \text{max-weight}$(F_i(u,v))$ denote such a maximum weight. The update operation works as follows:  when a new edge $e = (u,v)$ is inserted into $G$, we first use the dynamic tree data structure to test whether $u$ and $v$ belong to the same tree. If no, we link their two trees with the edge $(u,v)$ and return the pair (TRUE, NULL) to indicate that $e$ was added to $F_i$ and no 
edge was evicted from $F_i$. Otherwise, we 
check whether $p_e > \text{max-weight}(F_i(e))$. If the latter holds, we make no changes in the forest and return 
(FALSE, $e$). Otherwise, we replace one of the maximum edges, say $e'$, on the path between $u$ and $v$ in the tree by $e$ and return (TRUE, $e'$). The boolean value that is returned indicates whether $e$ belongs to $F_i$ or not, the second value that is returned gives an edge that does not (or no longer) belong to $F_i$. Note that each edge insertion requires $O(\log n)$ time. We refer to this insert operation as \textsc{Insert-MSF}$(F_i, e, p_e)$. 

Now, the algorithm that maintains the weighted minimum spanning forests implements the following operations:
\begin{itemize}
\item \textsc{Initialize-NI}$(k)$: Initializes the data structure for $k+1$ empty minimum spanning forests.
\item \textsc{Insert-NI}$(e,p_e)$: Set $i \gets 1$, $e' \gets e$, taken $\gets$ FALSE. \\
\hspace*{2.8cm} \textbf{while} ($(i \le k+1$) and $e' \neq \text{NULL}$) \textbf{do}  \\
 \hspace*{3.5cm} Set ($t'$, $e''$) $\gets$ $\textsc{Insert-MSF}(F_i, e', p_{e'})$.\\
 \hspace*{3.5cm} \textbf{if} ($e' = e$) \textbf{then} set taken $\gets$ $t'$ \textbf{endif} \\
 \hspace*{3.5cm} Set $e' \gets e''$ and $i \gets i + 1$.\\
\hspace*{2.8cm} \textbf{endwhile}\\ 
\hspace*{2.8cm} \textbf{if} ($e' \ne e$) \textbf{then} \textbf{return} (taken, $e'$) \\
\hspace*{2.8cm} \textbf{else} \textbf{return} (taken, NULL).

\end{itemize}

The boolean value that is returned indicates whether $e$ belongs to $F$ or not, the second value returns an edge that is removed from $F$, if any. 

Recall that $F = \bigcup_{i \le \kk+1} F_i$. We use the abbreviation $\textsc{NI-Sparsifier}(k)$ to refer to this data-structure. Throughout the algorithm we will associate
a weight with each edge in $F$ and use $F^w$ to refer to this weighted version of $F$.

\begin{lemma} \label{lemma: NI} For $k > 0$ and any graph $G$, \textsc{NI-Sparsifier$(k)$} maintains a weighted mfsd of order $\kk+1$ of $G$ under edge insertions. The algorithm uses $O(kn)$ space and the total time for inserting $m$ edges is $O(k m\log n)$.
\end{lemma}
\begin{proof}
We first show that \textsc{NI-Sparsifier$(k)$} maintains a forest decomposition such that (1) the forests are edge-disjoint and (2) each forest is maximal. We proceed by induction on the number $m$ of edge insertions. 

For $m=0$, the forest decomposition is empty. Thus the edge-disjointness and maximality of forests trivially hold.
For $m>0$, consider the $m$-th edge insertion, which inserts an edge $e$. Let $F'$, resp. $F$, denote the union of forests before, resp. after, the insertion of edge $e$. By the inductive assumption, $F'$ satisfies (1) and (2). If $F = F'$, i.e., the edge $e$ was not added to any of the forests when \textsc{Insert-NI}$(e,p_e)$ was called, then $F$ also satisfies (1) and (2). Otherwise $F \neq F'$ and note that by construction, $e$ is appended to exactly one forest. Let $F'_j$, resp. $F_j$, denote such maximal forest before, resp. after, the insertion of $e$. We distinguish two cases. If $e$ links two trees of $F'_j$, then $F_j$ is also a maximal forest and forests of $F$ are edge-disjoint. Thus $F$ satisfies (1) and (2). Otherwise, the addition of $e$ results in  the deletion of another edge $e' \in F'_j$. It follows that $F_j$ is maximal and the current forests are edge-disjoint. Applying a similar argument to the addition of edge $e'$ in the remaining forests, we conclude that $F$ satisfies (1) and $(2)$. 

We next argue about time and space complexity. The dynamic tree data structure can be implemented in $O(n)$ space, where each query regarding $\text{max-weight}(F_i(u,v))$ can be answered in $O(\log n)$ time. Since the algorithm maintains $k+1$ such forests, the space requirement is $O(kn)$. The total running time follows since insertion of an edge can result in at most $k+1$ executions of the \textsc{Insert-MSF}$(F_i,e,p_e)$ procedures, each running in $O(\log n)$ time. 
\end{proof}

(2)  {\textsc{Limited Exact Min-Cut}$(k)$ data-structure}:  We use Algorithm \ref{algo: ExactMinCutUpToK} to implement the following operations for
any  unweighted graph $G$ and parameter $k$,
\begin{itemize}
\item \textsc{Insert-Limited}$(e)$:  Executes the insertion of edge $e$ using Algorithm \ref{algo: ExactMinCutUpToK}.
\item \textsc{Query-Limited}$()$: Returns $\lambda$.
\item \textsc{Initialize-Limited}$(G,k)$: Builds a data structure for $G$ with parameter $k$ by executing Step 1 of Algorithm \ref{algo: ExactMinCutUpToK} and then 
\textsc{Insert-Limited}$(e)$ for each edge $e$ in $G$. 
\end{itemize}
We use the abbreviation \textsc{Lim}$(k)$ to refer to such data-structure. Combining the above data-structures leads to Algorithm \ref{algo: SmallSpaceMinCut}.

\begin{algorithm2e}
\label{algo: SmallSpaceMinCut}
\caption{$(1+\epsilon)$-Min-Cut with $O(n \log n / \epsilon^{2})$ space}
Set $k \gets 48 \log n / \epsilon^{2}$ \;

\nonl Set $p \gets 12 \log n / \epsilon^{2}$ \;
\nonl Let $H$ and $F^w$ be empty graphs \; 
 \textsc{Initialize-Limited}$(H,k)$ \;
\nonl \While {\normalfont $\textsc{Query-Limited()} < k$}
{
\nonl \textbf{Receive the next operation} \;
\nonl \uIf{it is a query}{ \nonl \Return $\textsc{Query-Limited()}/\min\{1,p\}$ \; }
\nonl \uElseIf{it is the insertion of an edge $e$}{
\nonl Sample a random weight from $[0,1]$ for the edge $e$ and denote it by $p_e$ \;
\nonl \If{$p_e \leq p$} { \nonl \textsc{Insert-Limited}$(e)$ \; }
\nonl Set (taken, $e'$) $\gets$ \textsc{Insert-NI}$(e, p_e)$ \;
\nonl \If{\normalfont taken}{
\nonl Insert $e$ into $F^w$ with weight $p_e$ \;
\nonl \If{$e' \ne$ \normalfont NULL}{ \nonl Remove $e'$ from $F^w$ \;}
}
}
}

Set $p \gets p/2$  \quad \quad \tcp{\textrm{Rebuild Step}}
\nonl  Let $H$ be the unweighted subgraph of $F^w$ consisting of all edges of weight at most $p$ \;
\nonl \GOTO Step 2

\end{algorithm2e}
\paragraph*{Correctness and Running Time Analysis.}  

Throughout the execution of Algorithm \ref{algo: SmallSpaceMinCut}, $F$ corresponds exactly to the msfd of order $\kk+1$ of $G$ maintained by \textsc{NI-Sparsifier}($k$).   
In the following, let $H$ be the graph that is given as input to \textsc{Lim}($k$). 
Thus, by Corollary \ref{cor: ExactPolyLog}, \textsc{Query-Limited}$()$ returns $\min\{k,\lambda(H)\}$, i.e., it returns
$\lambda(H)$  as long as $\lambda(H) \leq k$. 
We now formally prove the correctness. 
\begin{lemma}\label{lem:1}
Let $\epsilon \le 1$, $k = 48 \log n / \epsilon^2$ and assume that the algorithm is started on an empty graph. As long as $\lambda(G) < k$, we have $H=G$, $p = k/4$, and \textsc{Query-Limited}$()$ returns $\lambda(G)$.
The first rebuild step is triggered after the first insertion that increases  $\lambda(G)$ to $k$ and at that time, it holds that $\lambda(G) = \lambda(H) = k$.
\end{lemma}
\begin{proof}
The algorithm starts with an empty graph $G$, i.e., initially $\lambda(G)= 0$. Throughout the sequence of edge insertions $\lambda(G)$ never decreases. 
We show by induction on the number $m$ of edge insertions that $H=G$  and $p = k/4$ as long as
$\lambda(G) < k$. 

Note that $k/4 \ge 1$ by our choice of $\epsilon$.
For $m = 0$, the graphs $G$ and $H$ are both empty graphs and  $p$ is set to $k/4$. 
For $m > 0$, consider the $m$-th edge insertion, which inserts an edge $e$. Let $G$ and $H$ denote the corresponding graphs after the insertion of $e$. By the inductive assumption, $p = k/4$
and 
 $G \setminus \{e\} = H \setminus \{e \}$. As $p \ge 1$,  $e$ is added to $H$ and, thus, it follows that $G = H$. Hence, $\lambda(H) = \lambda(G)$. If $\lambda(G) < k$ but $\lambda(G \setminus \{e\}) < k$, no rebuild is performed and $p$ is not changed. If $\lambda(G) = k$, then the last insertion was exactly the
 insertion that increased $\lambda(G)$ from $k-1$ to $k$. As $H = G$ before the rebuild,  \textsc{Query-Limited}$()$ returns $k$, triggering the first execution of the rebuild step.
\end{proof}

We next analyze the case that $\lambda(G) \ge k$. In this case, both $H$ and $p$ are random variables, as they depend on the randomly chosen weights for the edges.
Let $F(p)$ be the  unweighted subgraph of $F^w$ that contains all edges of weight at most $p$.

\begin{lemma}\label{lem:H}
Let $N_h(p)$ be the graph consisting of all edges that were inserted after the last rebuild and have weight at most $p$ and let $F^{\text{old}}(p)$ be
$F(p)$ right after the last rebuild. Then it holds that $H = F^{\text{old}}(p) \cup N_h(p)$.
\end{lemma}
\begin{proof}
Up to the first rebuild, $N_h = G$ and $p \ge 1$. Thus $N_h(p)  = N_h = G$.
Lemma \ref{lem:1} shows that until the first rebuild $H=G$. As $F^{\text{old}}(p) = \emptyset$, it follows that $H = G = N_h(p) \cup F^{\text{old}}(p)$ up to the first rebuild.

Immediately after each rebuild step, $N_h = \emptyset$ and $H$ is set to be $F(p)$, thus the claim holds.
After each subsequent edge insertion that does not trigger a rebuild, the newly inserted edge is added to $N_h(p)$ and to $H$ iff its weight is at most $p$. Thus, 
both $N_h(p)$ and $H$ change in the same way, which implies that $H = F^{\text{old}}(p) \cup N_h(p)$.
\end{proof}

\begin{lemma}\label{DA-msfd}
At the time of a rebuild $F(p)$ is an msfd of order $\kk+1$ of $G(p)$.
\end{lemma}
\begin{proof}
\textsc{NI-sparsifier} maintains a maximal spanning forest decomposition based on  minimum-weight spanning forests  $F_1, \dots F_{\kk+1}$ of $G$ using the weights $p_e$.
Now consider the hierarchical decomposition $F_1(p), \dots, F_{\kk+1}(p)$ of $G(p)$ induced by taking only the edges of weight at most $p$ of each forest $F_i$.
Note that \textsc{NI-sparsifier} would return exactly the same hierarchy $F_1(p), \dots, F_{k+1}(p)$ if only the edges of $G(p)$ were inserted into \textsc{NI-sparsifier}.
Thus $F_1(p), \dots, F_{\kk+1}(p)$ is an msfd of order $\kk+1$ of $G(p)$.
\end{proof}

In order to show that $\lambda(H)/\min\{1,p\}$ is an $(1+\epsilon)$-approximation of $\lambda(G)$ with high probability, we need to show that if $\lambda(G) \geq k$ then
(a) the random variable $p$
is at least $k / (4 \lambda(G))$ w.h.p., which implies that $\lambda(G(p))$ is a $(1 + \epsilon)$-approximation of $\lambda(G)$ w.h.p. and (b) that $\lambda(H) = \lambda(G(p))$ (by Lemma \ref{lemm: Karger2}).

\begin{lemma} Let $\epsilon \le 1$. If $\lambda(G) \geq k$, then (1) $p \geq k / (4\lambda(G))$ with probability $1- O(\log n/n^4)$ and
(2) $\lambda(H) = \lambda(G(p))$. 
\end{lemma}
\begin{proof}
For any $i \geq 0$, after the $i$-th rebuild we have $p = p^{(i)} := 12 \log n / (2^{i}\epsilon^{2})$. Let $\ell = \lfloor \log (12 \log n / \epsilon^2)\rfloor$ denote the index of the last rebuild at which $p^{(i)} \geq 1$. For any $i \geq \ell + 1$, we will show by induction on $i$ that (1)  $p^{(i)} = 12 \log n / (2^{i}\epsilon^{2}) \geq 12 \log n / (\epsilon^{2}\lambda(G))$ with probability $1-O((i-1-\ell)/n^{4})$,
which is equivalent to showing that $\lambda(G) \geq 2^i$ and that (2) at any point between 
the $(i-1)$-st and the $i$-th rebuild, $\lambda(H) = \lambda(G(p^{(i-1)}))$.

Once we have shown this, we can argue that the number of rebuild steps is small, thus giving the claimed probability in the lemma. Indeed, note that $\lambda(G) \leq n$ since $G$ is unweighted. Additionally, from above we get that after the $i$-th rebuild, $\lambda(G) \geq 2^{i}$ with high probability. Combining these two bounds yields $i \leq O(\log n)$ w.h.p., i.e., the number of rebuild steps is at most $O(\log n)$.

We first analyse $i=\ell+1$. Note that $\ell+1$ is the index of the first rebuild at which $p^{(i)} < 1$. Assume that the insertion of some edge $e$ caused the first rebuild. Lemma~\ref{lem:1} showed that (1) at the first rebuild $\lambda(G) = k$
and (2) that up to the first rebuild $G(p) = G = H$. We observe that (1) and (2) remain true up to the $(\ell+1)$-st rebuild. In addition, $\lambda(G) = k \geq 24 \log n / \epsilon^2 \geq 2^{i}$, which implies that $p^{(i)} \geq 1/2$. This shows the base case.

For the induction step ($i > \ell+1$), we inductively assume that (1) at the $(i-1)$-st rebuild, $p^{(i-1)} \geq 12 \log n / (\epsilon^{2} \lambda(G^{\text{old}}))$ with probability $1- O((i-2-\ell)/n^{4})$, where $G^{\text{old}}$ is the graph $G$ right before the insertion that triggered the $i$-th rebuild
(i.e., at the last point in time when \textsc{Query-Limited}$()$ returned a value less than $k$),
and (2) that $\lambda(H) = \lambda(G(p^{(i-2)}))$ at any time between the $(i-2)$-nd and the $(i-1)$-st rebuild.
Let $e$ be the edge whose insertion caused the $i$-th rebuild.
Define $G^{\text{new}} = G^{\text{old}} \cup \{e\}$. By induction hypothesis, with probability $1-O((i-2-\ell)/n^4)$, $p^{(i-1)} \geq 12 \log n / (\epsilon^{2}\lambda(G^{\text{old}})) \geq 12 \log n / (\epsilon^{2}\lambda(G^{\text{new}}))$ as $\lambda(G^{\text{old}}) \leq \lambda(G^{\text{new}})$. Thus, by Lemma \ref{lemm: Karger2}, we get that $\lambda(G^{\text{new}}(p^{(i-1)}))/p^{(i-1)} \leq (1+\epsilon) \lambda(G^{\text{new}})$ with probability $1-O(1/n^4)$. Applying an union bound, we get that the two previous statements hold simultaneously with probability $1-O((i-1-\ell)/n^4)$.

 We show below that $\lambda(G^{\text{new}}(p^{(i-1)})) = \lambda(H^{\text{new}})$, where
$H^{\text{new}}$ is the graph stored in \textsc{Lim}($k$) right before the $i$-th rebuild.
Thus, $\lambda(H^{\text{new}}) = k$, which implies that  
\begin{align*}
	\lambda(G^{\text{new}}(p^{(i-1)})) = k =48 \log n / \epsilon^{2} & \leq  (1+\epsilon) \lambda(G^{\text{new}}) \cdot  p^{(i-1)} \\
	 & = (1+\epsilon) \lambda(G^{\text{new}}) \cdot 12 \log n /(2^{i-1}\epsilon^{2}),
\end{align*}
with probability $1-O((i-1-\ell)/n^4)$. This in turn implies that with probability $1-O((i-1-\ell)/n^4)$, $\lambda(G^{\text{new}}) \geq 2^{i+1}/(1+\epsilon) \geq 2^{i}$ by our choice of $\epsilon$. 

It remains to show that $\lambda(G^{\text{new}}(p^{(i-1)})) = \lambda(H^{\text{new}})$. Note that this
is a special case of (2), which claims that at any point between that $(i-1)$-st and the $i$-th rebuild
$\lambda(H) = \lambda(G(p^{(i-1)}))$, where $H$ and $G$ are the current graphs. Thus, to complete the proof of the lemma it suffices to show (2).

 As $H$ is a subgraph of $G(p^{(i-1)})$, we know that $\lambda(G(p^{(i-1)})) \ge \lambda(H)$.
Thus, we only need to show that $\lambda(G(p^{(i-1)})) \le \lambda(H)$.
Let $G^{i-1}$, resp.~$F^{i-1}$,  resp.~$H^{i-1}$, be the graph $G$, resp.~$F$, resp.~$H$, right after rebuild $i-1$ and let $N_h$ be the set of
 edges inserted since, i.e., $G = G^{(i-1)} \cup N_h$. 
 As we showed in Lemma~\ref{lem:H}, $H = F^{i-1}(p^{(i-1)}) \cup N_h(p^{(i-1)})$. Thus, $H^{i-1} = F^{i-1}(p^{(i-1)})$.
 Additionally, 
 by Lemma~\ref{DA-msfd}, $F^{i-1}(p^{(i-1)})$ is an msfd of order $\kk+1$ of
 $G^{i-1}(p^{(i-1)})$. Thus by Lemma \ref{lemm: Nagamochi}, for every cut $(A, V\setminus A)$ of value at most $\kk$ in $H^{i-1}$, $\lambda(A,H^{i-1})  = \lambda(F^{i-1}(p^{(i-1)}),A) =\lambda(A,G^{i-1}(p^{(i-1)}))$,
 where $\lambda(A,G) = |E_G(A,V \setminus A)|$. 
 Now assume towards contradiction that $\lambda(G(p^{(i-1)})) > \lambda(H)$ and consider a minimum cut $(A, V\setminus A)$ in $H$, i.e., $\lambda(H) = \lambda(A,H)$.
 We know that at any time $k \ge \lambda(H).$ Thus $k \ge \lambda(H) = \lambda(A,H)$, which implies $k \ge \lambda(A,H^{i-1})$. By Lemma \ref{lemm: Nagamochi} it follows that $\lambda(A,H^{i-1}) = \lambda(A,G^{i-1}(p^{(i-1)}))$.
 Note that $H = H^{i-1} \cup N_h(p^{(i-1)})$ and $G(p^{(i-1)}) = G^{i-1}(p^{(i-1)}) \cup N_h(p^{(i-1)})$. Let $x$ be the number of edges of  $N_h(p^{(i-1)})$ that cross the cut $(A, V\setminus A)$.
 Then $\lambda(H) = \lambda(H,A) = \lambda(A,H^{i-1}) + x = \lambda(A,G^{i-1}(p^{(i-1)})) + x = \lambda(A,G(p^{(i-1)}))$, which contradicts the assumption that
 $\lambda(G(p^{(i-1)})) > \lambda(H)$.
\end{proof}

Since our algorithm is incremental and applies only to unweighted graphs, we know that there can be at most $O(n^{2})$ edge insertions. The above lemma implies that for any current graph $G$, Algorithm \ref{algo: SmallSpaceMinCut} returns a $(1+\epsilon)$-approximation to a min-cut of $G$ with probability $1-O(\log n/n^{4})$. Applying an union bound over $O(n^{2})$ possible different graphs, gives that the probability that the algorithm does not maintain a $(1+\epsilon)$-approximation is at most $O(\log n/n^2) = O(1/n)$. Thus, at any time we return a $(1+\epsilon)$-approximation with probability $1-O(1/n)$.

\begin{theorem} \label{thm: space2}
There is an $O(n \log n/\epsilon^{2})$ space randomized algorithm that processes a stream of edge insertions starting from an empty graph $G$ and maintains a $(1+\epsilon)$-approximation to a min-cut of $G$ with high probability. The 
total time for insertiong $m$ edges is 
$O(m \alpha(n)\log^{3} n / \epsilon^{2})$ 
and queries can be answered in constant time.
\end{theorem}
\begin{proof}
The space requirement is $O(n \log n/ \epsilon^{2})$ since at any point of time, the algorithm keeps $H$, $F^{w}$,
\textsc{Lim}($k$), and \textsc{NI-Sparsifier} ($k)$, each of size at most $O(n \log n/ \epsilon^{2})$ (Corollary \ref{cor: ExactPolyLog} and Lemma \ref{lemma: NI}).

When Algorithm \ref{algo: SmallSpaceMinCut} executes a \emph{Rebuild Step}, only the \textsc{Lim}($k)$ data-structure is rebuilt, but not \textsc{NI-Sparsifier}($k$). During the whole algorithm $m$ \textsc{Insert-NI} operations are
performed. Thus, by Lemma \ref{lemma: NI},  the total time for all operations involving $\textsc{NI-Sparsifier}(k)$ is $O(m\log^2 n / \epsilon^{2})$.

It remains to analyze Steps $2$ and $3$. By Corollary~\ref{cor: ExactPolyLog}, \textsc{Initialize-Limited}$(H,k)$ takes at most $O(m \alpha(n)\log^{2} n / \epsilon^{2})$ total time (Step 2). 
The running time of Step $3$ is $O(m)$ as well. Since the number of \emph{Rebuild Steps} is at most $O(\log n)$,
it follows that the total time for all
\textsc{Initialize-Limited}$(H,k)$ calls in Steps $2$ and the total time of Step $3$ throughout the execution of the algorithm is $O(m \alpha(n)\log^{3} n / \epsilon^{2})$.

We are left with analyzing the remaining part of Step 2. Each query operation executes one \textsc{Query-Limited}() operation, which takes constant time.
Each insertion executes one \textsc{Insert-NI}($e,p_e$) operation, which takes amortized time $O(\log^2n/ \epsilon)$. We maintain the edges of $F^w$ in a balanced binary tree so that each insertion and deletion takes $O(\log n)$ time.
As there are $m$ edge insertions the  remaining part of Step 2 takes total time
$O(m \log^{2} n / \epsilon^{2})$. Combining the above bounds gives the theorem. \end{proof}

\section{Conclusion}

We obtained two new algorithms for the incremental (global) minimum cut problem in undirected, unweighted graphs. Our first algorithm maintains exactly the value of a minimum cut and has an $O(\log^{3} n \log \log^{2} n)$ amortized time per edge insertion and $O(1)$ query time. The main techniques behind this algorithm are (1) constructing a small sparsifier that preserves the non-trivial minimum cuts (2) incrementally maintaining the value of the minimum cut on the sparsifier and (3) employing periodical rebuilds whenever the maintained sparsifier is not valid for the current graph. While we believe the maintained sparsifier might prove useful to extend our algorithm to less restrictive settings, techniques in (2) and (3) crucially exploit the fact that the underlying data-structure supports edge insertions. An important problem is whether there is a fully-dynamic algorithm for exactly maintaining the value of the minimum cut in sub-linear query and update time. Perhaps a good starting point is trying to come up with a deletions-only algorithm. 

Our second result maintains a $(1+\epsilon)$-approximation to the value of a minimum cut in poly-logarithmic update time while using only $O(n \log n / \epsilon^{2})$ space. The main idea behind our construction is to first maintain all minimum cuts up to a given threshold using small space and then apply the randomized sparsification result due to Karger~\cite{Karger99}. It is an interesting direction to explore whether similar guarantees can be achieved in the fully-dynamic or decremental setting. In fact, even in the less general setting, that ignores the space requirement, it is not known whether there are decremental algorithms that maintain the value of the minimum cut up to a $(1+\epsilon)$ multiplicative factor in poly-logarithmic update and query time.

\chapter[Fast Incremental Algorithms via Local Sparsifiers]{Fast Incremental Algorithms via Local Sparsifiers}\label{cha:Man2019_LS}

We show $n^{o(1)}$-approximation incremental algorithms with $n^{o(1)}$
\emph{worst-case} update and query time on an undirected weighted
$n$-node graph for many problems including all-pairs shortest paths,
all-pairs max flow and min cut, multi-commodity concurrent flow, and
uniform sparsest cut. By increasing the time to $n^{\epsilon}$ for any fixed
$\epsilon>0$ the approximation factors can be improved to $\mbox{polylog}(n)$, and for all-pairs shortest paths to $O(1)$. For the all-pairs shortest paths
problem, no previous algorithm with both $o(n)$
worst-case update and query time was known. For the other problems, even 
algorithms with both $o(n)$ \emph{amortized}
update and query time were not known.

As key to our result, we introduce a new notion of a sparsifier,
called \emph{local sparsifier}, for any graph property $\mathcal{P}$
and present a new general technique that converts any efficient construction
algorithm for a local sparsifiers for $\mathcal{P}$ into an \emph{incremental}
algorithm for approximately maintaining $\mathcal{P}$. This
technique connects several open problems between the fields of graph
sparsifiers and dynamic graph algorithms, and leads to challenging
new research questions for graph sparsifiers.

\section{Introduction}

In a recent study of the usage of graphs in practice~\cite{DBLP:journals/pvldb/SahuMSLO17}
it was shown that real-world graphs are usually very large and more
than half of the graphs in the survey change frequently, i.e., are
dynamic. Due to the large size of these graphs, a dynamic algorithm
needs to have \emph{sublinear} time per operation to be useful for
these applications. Another interesting finding of the study is that
more than 2/3 of the graph computations are for ``non-basic'' graph
problems, i.e., for problems for which no linear-time static algorithm
is known such as all-pairs shortest paths and various forms of graph
partitioning. However, the current state-of-the art in dynamic graph
algorithms is far from solving these ``non-basic'' graph problems
in sublinear time, for many of them not even a dynamic algorithm better
than recomputation from scratch is known. The reason for this ``lack''
of efficient dynamic algorithms became clear only recently: it has
been shown that under certain, widely accepted assumptions maintaining
the \emph{exact} answer of many ``non-basic'' graph problems is
not possible in sublinear time~\cite{AbboudW14,HenzingerKNS15,AbboudD16,Dahlgaard16}.
Thus, to design dynamic algorithms for these problems, it is necessary
to study \emph{approximation} algorithms for them. %

In this chapter, we study several ``non-basic'' graph problems including
all-pairs shortest paths, all-pairs max flow (and min cuts), multi-commodity
concurrent flow, and uniform sparsest cut (defined in \Cref{tab:problems}).
Despite an extensive research on dynamic all-pairs shortest paths
\cite{Sankowski05,Thorup05,Bernstein09,RodittyZ12,AbrahamCT14,Bernstein16,AbrahamCK17,Chechik18},
no previous algorithms were known with $o(n)$ worst-case update and query
time on a general graph with $n$ nodes. For other problems where near-optimal time algorithms in the static setting are 
well-studied (for example, max flow and multi-commodity concurrent flow~\cite{Madry10,Sherman13,KelnerLOS14,Peng16,Sherman17}, uniform sparsest cut~\cite{SpielmanT13,AndersenCL07,KhandekarRV09,Sherman09,Madry10})
even algorithms
with both $o(n)$ \emph{amortized} update and query time were not known. 
	
\paragraph{Our results.}
We show \emph{incremental} approximation algorithms for the above
problems. Incremental algorithms are data structures
that maintain information about a graph property while the graph is
modified by a sequence of edge insertions. Our algorithms significantly break the $o(n)$ bound by showing $n^{o(1)}$-approximation algorithms with $n^{o(1)}$ worst-case update time for all above problems. By increasing the time to $n^{\eps}$
for any constant $\eps>0$, the approximation factors can be improved to$\mbox{polylog}(n)$
and $O(1)$ for all-pairs shortest paths. The precise statement is
as follows:
\begin{theorem}
	\label{thm:main_inc} For any two parameters $r, \ell \geq 1$, there are incremental approximation algorithms
	on weighted~(capacitated) undirected $n$-node graphs for the following problems (as defined in Table~\ref{tab:problems}) with their corresponding guarantees: 
	\begin{enumerate}
	\itemsep0em
		\item All-pairs max flow and min cuts: $O(\log n)^{4\ell}$-approximation,
		$\tilde{O}(n^{2/(\ell+1)})$ worst-case update and query time.
		\item All-pairs shortest paths: $(2r-1)^{\ell}$-approximation, $\tilde{O}(n^{2/(\ell+1)}n^{2/r})$
		worst-case update and query time.
		\item Multi-commodity concurrent flow: $O(\log n)^{8\ell}$-approximation,
		$\tilde{O}(n^{2/(\ell+1)})$ worst-case update time, and $\tilde{O}(k^2)$
		query time when there are $k$ commodity pairs in the query.
		\item Uniform Sparsest Cut: $O(\log n)^{8\ell}$-approximation, $\tilde{O}(n^{2/(\ell+1)})$
		worst-case update time $O(1)$ query time.
		
	\end{enumerate}
	All the above algorithms are randomized, except the all-pairs shortest paths algorithm, which is deterministic.
\end{theorem}

\renewcommand{\arraystretch}{1.3}
\begin{table}
\label{tab:problems}
	\small{
		
		\begin{tabular}{>{\centering}p{0.3\textwidth}|>{\raggedright}p{0.6\textwidth}}
			Dynamic problems & Query\tabularnewline
			\hline 
			\hline 
			All-pairs max flow & Given $(s,t)$, return the value of max flow from $s$ to $t$.\tabularnewline
			\hline 			
			All-pairs shortest paths & Given $(s,t)$, return the distance from $s$ to $t$.\tabularnewline
			\hline 
			Multi-commodity concurrent flow & Given $\{(s_{i},t_{i},\dd(i))\}_{i=1}^{k}$, return the value $\alpha$
			where, concurrently for all $i$, $s_{i}$ can send $\alpha \dd(i)$ unit of flow to $t_{i}$.\tabularnewline
			\hline 
			Uniform Sparsest Cut & Return $\Phi_{G}=\min_{S \subset V}\frac{\capacity_G(S, V \setminus S)}{|S| \cdot |V \setminus S|}$.
			\tabularnewline
		\end{tabular}
		
	}
	
	\caption{List of dynamic problems and their corresponding query operation. For a weighted graph $G=(V,E,\ww)$, we have that $\capacity_G(S,V \setminus S)=\sum_{(u,v)\in E,u\in S,v \protect \notin S}\ww(u,v)$.}
\end{table}

\paragraph{Previous cut/flow algorithms.}

Despite the fact that all-pairs max flow and min cuts, multi-commodity concurrent flow, and uniform sparsest cut, are central problems in combinatorial optimization and have been extensively studied in the static setting,
there are essentially no fast algorithms in the dynamic setting. 
Using previous techniques, it is possible to get dynamic algorithms with $\tilde{O}(1)$
worst-case update time and $\tilde{O}(n)$ query time under the assumption 
that the adversary is oblivious.\footnote{We maintain a dynamic cut-sparsifier (against oblivious adversary)
	of size $\tilde{O}(n)$ due to \cite{AbrahamDKKP16} with $\tilde{O}(1)$
	update time, and when given a query, we execute the fastest static approximation algorithms on the sparsifier in $\tilde{O}(n)$ time (using, for example,
	\cite{Peng16} for $(1+\eps)$-approximate max flow, \cite{Sherman17}
	for $(1+\eps)$-approximate multi-commodity concurrent flow, and \cite{Sherman09}
	for $O(\sqrt{\log n})$-approximate uniform sparsest cuts).} To the best of our knowledge, there is no previous algorithm with both $o(n)$
update and query time, even when we are content with only amortized guarantees. 

Most closely related work to our work is the dynamic
algorithm due to \cite{CheungKL13} for explicitly maintaining all the values
of all-pairs min-cuts in $\tilde{O}(m^{2})$ update time. For $s$-$t$ max flow where $s$ and $t$ are fixed, there is an incremental algorithm with $O(n)$ amortized update time \cite{GuptaK18}. If we restrict to bipartite graphs with a certain specific structure, there is a $(1+\eps)$-approximation fully dynamic algorithm \cite{AbrahamDKKP16} with polylogarithmic worst-case update time. From the lower bound perspective, Dahlgaard \cite{Dahlgaard16} shows
a conditional lower bound of $\Omega(n^{1-o(1)})$ amortized update
time for exact incremental $s$-$t$ max flow in \emph{capacitated
	undirected} graphs. This shows that approximation is necessary to
achieve sublinear running times.

\paragraph{Previous distance algorithms.}

The dynamic all pairs shortest paths problem has been extensively
studied. Most previous work requires amortized update time. In particular, they either
have $\Omega(n)$ update time or need to assume an oblivious adversary
\cite{DemetrescuI04,BaswanaHS07,Bernstein09,RodittyZ12,AbrahamCT14,Bernstein16,HenzingerKN16,Chechik18}.
An exception here is the work due to Alstrup et al.~\cite{AlstrupDFTW17} that shows a very fast amortized deterministic algorithm for approximating the
distance between two nodes, but this works \emph{only if the queried
	distance are short}. 

For worst-case update time, all previous algorithms \cite{Sankowski05,Thorup05,AbrahamCK17}
give \emph{exact }answers but require  $\Omega(n^{1.8})$ update
time. If we allow a large approximation factor, then the best algorithm to
our knowledge is an $O(\sqrt{\log n})$-approximation algorithm with
$O(n^{1+o(1)})$ worst-case update time. This also assumes an oblivious
adversary\footnote{They maintain a dynamic $O(\sqrt{\log n})$-spanner \cite{BernsteinFH19}
	(against oblivious adversary) of size $O(n^{1+o(1)})$ with $n^{o(1)}$
	worst-case update time. Then, given a query, we run a static shortest
	path algorithm. }. To summarize, our all-pairs shortest paths algorithm is the first
algorithm with $o(n)$ worst-case update. Moreover, it is deterministic.

Even for the more restricted dynamic $s$-$t$ shortest path problem, the story is
similar as there is no $o(n)$ worst-case update algorithm. All previous
amortized algorithms either take at least $\Omega(n^{3/4})$ on sparse
graphs \cite{EvenS81,BernsteinC16,Bernstein17,BernsteinC17} or assume
an oblivious adversary \cite{BernsteinR11,HenzingerKN14}. There is
in fact, a conditional lower bound of $\Omega(n^{2-o(1)})$ worst-case
update time for the incremental dynamic $s$-$t$ exact shortest paths
on weighted graphs~\cite{AbboudW14,HenzingerKNS15}. This again shows
that approximation is necessary to obtain our worst-case update time.

\paragraph{Our techniques}
As a key to our results, we introduce a new notion of graph sparsifier,
called \emph{local sparsifier}. It is a stronger version of a well-studied
notion called \emph{vertex sparsifier} \cite{Moitra09,leighton,charikar,juliasteiner,mm10,EnglertGKRTT14}. Here, we give informal definitions.
Let $G=(V,E)$ be a graph and, for each $u,v\in V$, let $\Property(u,v,G)$
denote a \emph{property }between $u$ and $v$ in $G$. For example,
$\Property(u,v,G)$ is the distance or the size of the $u$-$v$ min cut. Let
$K\subseteq V$ be a set of nodes called \emph{terminals}. A \emph{vertex sparsifier}
of $G$ with respect to $K$ is a graph $H=(V',E')$ such that 1) $|V'|\approx|K|$
and 2) $\Property(u,v,H)\approx\Property(u,v,G)$ for all $u,v\in K$. That is,
$H$ has size close to $K$ but still ``approximately preserves''
the property $\Property$ between all terminal nodes. 

A local sparsifier of $G$ is a graph which contains possible
vertex sparsifiers with respect to \emph{any} given set of terminals. More precisely, a local sparsifier of $G$ is
a graph $H$ such that, for any terminal set $K$, there is a subgraph
of $H$, denoted by $H[K]$\footnote{This may not be a subgraph induced by $K$.}, where $H[K]$ is vertex sparsifier of
$G$ with respect to $K$. See \Cref{sec: localSparsifiers} for the formal
definition.

Our main technical contribution is a \emph{meta-theorem} which turns
any efficient construction for local sparsifiers for any property
$\Property$ into fast incremental algorithms for $\Property$. Our reduction gives
worst-case update time bounds and it is deterministic. Given a randomized
sparsifier construction, the resulting incremental algorithm is also randomized. Details on this construction can be found in \Cref{sec:meta_inc}.

Given the meta-theorem, we then show that existing efficient constructions
of vertex sparsifiers, such as algorithms for computing R\"{a}cke
trees in \cite{RackeST14} and Thorup-Zwick emulators in \cite{ThorupZ05,RodittyTZ05},
can be adapted to build local sparsifiers. Details on these constructions can be found in \Cref{sec:distance_inc,sec:maxflow_inc}. By plugging these constructions into our
framework, we obtain our results on all-pairs max flow and all-pairs shortest paths \Cref{thm:main_inc}.
In fact, it is simple to extend our data-structure and show an incremental algorithm for 
maintaining a {\em tree flow sparsifier} (i.e. R\"{a}cke tree \cite{RackeST14}) itself.

\begin{theorem}
	[Informal]\label{thm:raecke_informal}For any $\ell\ge1$, there
	is an incremental randomized algorithm with $\tilde{O}(n^{2/(\ell+1)})$ worst-case time for maintaining a \emph{tree flow sparsifier} of an $n$-node graph $G$ 
	with quality $O(\log^{8\ell}n)$ and depth $O(\ell \log^2 n)$.
	
\end{theorem}
See \Cref{sec:raecke} for the formal definition of a tree flow sparsifier
and its quality. Basically, it is a tree which ``approximately
preserves'' all the cut/flow information of the graph. From \Cref{thm:raecke_informal},
the simple structure of low-depth tree allows us to further implement other algorithms on the tree. 
Then, we easily obtain incremental algorithms for uniform sparsest cut and multi-commodity concurrent flow as stated \Cref{thm:main_inc}.

\paragraph{Offline Fully Dynamic Algorithms.}

An \emph{offline} dynamic algorithm is an algorithm where the whole
sequences of updates (edge insertions and deletions) and queries is given as an input, and the algorithm
needs to output information of the updated graph at every step that
is queried. We say that an offline dynamic algorithm has (average)
update and query time of $t$, if given a sequence of length $L$, then the total
running time is $t\cdot L$.

Although the offline setting is a weaker than the standard dynamic
setting, it is interesting for two reasons. First, offline algorithms are used to obtain fast static algorithms (e.g. \cite{BringmannKN19,LPYZ18}).
Second, many conditional lower bounds (e.g. \cite{AbboudW14,AbboudD16,Dahlgaard16})
for the standard dynamic setting also hold for the offline dynamic setting.
Thus, giving an efficient algorithm for the offline dynamic setting
shows that no such conditional lower bound is possible.

Simplifying the technique for incremental algorithms we can show an ``offline''
version of the meta-theorem which converts any efficient construction
of vertex sparsifiers to an \emph{offline} \emph{fully dynamic} algorithm
(see \Cref{sec:meta_offline}). Note that this version is incomparable
with the previous one: an offline fully dynamic algorithm is incomparable
to an online incremental algorithm. As this meta theorem only need
vertex sparsifiers which are weaker than local sparsifiers, we immediately
obtain the following.

\begin{corollary}
	[Informal]\label{cor:main_offline}There are offline fully dynamic
	approximation algorithms for the same problems with the same parameters
	as in \Cref{thm:main_inc}\footnote{To make sense of this, we in fact must replace ``worst-case update
		time'' with ``average update time''. There is also no concept of
		adversary in the offline setting.}. 
\end{corollary}
In fact, there were previous several offline algorithms in the literature
which are based on vertex sparsifiers. This includes the offline algorithms
for minimum spanning trees \cite{Eppstein91}, effective resistance
\cite{LPYZ18}, and 2/3-edge connectivity \cite{PengSS17}. Our offline
meta-theorem puts all their work into one framework: by just identifying
the efficient construction of vertex sparsifier from each of these woks, their
results can be immediately reproduced.

\paragraph{New directions for sparsifiers.}

Apart from the new algorithms we devised in \Cref{thm:main_inc} and \Cref{cor:main_offline},
we believe that our meta-theorems are valuable by themselves. They
explicitly connect open problems of the two fields, namely
dynamic algorithms and graph sparsifiers: any new upper or lower bounds
is immediately transferred via them (see \Cref{sec:hardness} for particularly
interesting examples.) This connection also motivates the following
research directions for constructing sparsifiers:

(1) \emph{Trading size for quality}: If there exists a near-linear
time construction of a vertex sparsifier with respect to terminals $K$ which
has size as large as $O(|K|n^{o(1)})$ but preserve a graph property within a
factor of $(1+\eps)$ for any $\eps>0$, then the resulting offline
dynamic algorithms would have $n^{o(1)}$ update time and approximation
factor only $(1+\eps')$ for any $\eps'>0$. A similar implication holds for
local sparsifier and incremental algorithms. This will give a significant
improvement over our results that have large approximation factor. To the best of our knowledge, this question has not been explored in the vertex
sparsifier literature since the research has concentrated on obtaining
a vertex sparsifier whose size depends only on $|K|$. In fact, even a vertex sparsifier of size $\text{poly}(|K|n^{o(1)})$ with the $(1+\eps)$ factor would still give an interesting implication for dynamic algorithms (see e.g. \Cref{thm:lb max flow}).

(2) \emph{Local sparsifier for effective resistance}: A near-linear
time construction for vertex sparsifiers for effective resistance
is known, i.e. an approximate Schur complement. This gives a very fast offline algorithm for effective resistance, as
observed in \cite{LPYZ18}. However, in order to get an incremental algorithm, we would need a local sparsifier with an efficient construction. We are not aware whether such sparsifiers exist and we pose this as an important open question.

(3) \emph{Speeding up existing constructions}: In this chapter, we only used existing sparsifiers that admit fast construction oracles. However, for example, there exist sparsifiers with better approximation quality for which
no fast construction algorithm is known (e.g. \cite{juliasteiner,EnglertGKRTT14}). Thus, it is an interesting research question to develop faster algorithms for constructing them.

\section{Local Sparsifiers}
\label{sec: localSparsifiers}

Let $G=(V,E)$ be graph. For any $u,v \in V$ we define $\mathcal{P}(u,v,G)$ to be a \emph{property} between vertices $u$ and $v$ in $G$. Throughout $\mathcal{P}(u,v,G)$\footnote{Our idea extends also to other graph properties, but we decided to work with minimization problems in order to simplify the presentation.} will be a solution to a minimization problem involving $u$ and $v$ in $G$. We next review several notions that allows us to reduce the size of $G$ while (approximately) retaining pair-wise information for some properties of $G$.

\begin{definition}[Sparsifiers] \label{def: propertySparsifier}
Let $G = (V,E)$ be a graph, and let $\alpha \geq 1$. A graph $H=(V', E')$ with $V \subseteq V'$ is an $\alpha$-sparsifier of $G$ iff for every $u,v \in V$
\[
	\mathcal{P}(u,v,G) \leq \mathcal{P}(u,v,H) \leq \alpha \cdot \mathcal{P}(u,v,H).
\] 
\end{definition}

The above notion captures different forms of sparsification. When $V' = V$ and $E' \subseteq E$, then $H$ is referred to as \emph{edge} sparsifier of $G$. Another example is when $H$ contains additional vertices and edges which do not appear in $G$ but $H$ has a simpler structure than $G$.

The following sparsification notion is particularly useful if the goal is to reduce the vertex count of the input graph $G$.

\begin{definition}[Vertex Sparsifiers] \label{def: propertyVertexSparsifier}
Let $G=(V,E)$ be a graph, with a \emph{terminal} set $K \subseteq V$, and let $\alpha \geq 1$. A graph $H=(V',E')$ with $K \subseteq V'$ is an $\alpha$-\emph{vertex sparsifier} of $G$ with respect to $K$ iff for every $u,v \in K$
\[
	\mathcal{P}(u,v,G) \leq \mathcal{P}(u,v,H) \leq \alpha \cdot \mathcal{P}(u,v,G).
\]
\end{definition} 

Our work requires that sparsifiers satisfy two important properties, namely transitivity and decomposability. While transitivity is obvious, decomposability gives the following useful fact: if a graph is a combination of two graphs on disjoint edge sets, combining the respective sparsifiers of these graphs gives a sparsifier for the original graph. We next make these statement more precise.

Given a graph $G=(V,E)$, a parameter $\alpha \geq 1$, and an $\alpha$-sparsifier $H$ of $G$, we define $S$ to be a mapping that takes $G$ and $\alpha$ as inputs and produces $H$, i.e., $H := S(G,\alpha)$. We call such a mapping a {\em sparsifier mapping.}
This leads to the following definition.


\begin{definition}[Transitivity] \label{ref: propertyTransitive}
Assume a sparsifier mapping $S$ fulfills the following condition:
For any graph $G$ and parameters $\alpha_1 \ge 1$ and $\alpha_2 \ge 1$ it holds that
when $H_1 = S(G,\alpha_1)$ and $H_2 = S(H_1,\alpha_2)$ then $H_2$ is an $\alpha_1 \alpha_2$-sparsifier of $G$. Then we say that the mapping $S$ is \emph{closed under {transitivity}.}
\end{definition}

\begin{definition}[Decomposability] \label{def: propertyDecomposability}
Assume a sparsifier mapping $S$ fulfills the following condition:
For any two edge-disjoing graphs $G_1 = (V,E_1)$ and $G_2=(V,E_2)$ over a set $V$ of nodes
when $H_1=S(G_1, \alpha)$ and $H_2=S(G_2,\alpha_2)$ then $H = H_1 \cup H_2$ is an $\max\{\alpha_1, \alpha_2\}$-sparsifier of $G$. Then we say that $S$ is {closed under \emph{decomposition}.}
\end{definition}

We next introduce a new notion of sparsification that captures properties of both sparsifiers and vertex sparsifiers.

\begin{definition}[Local Sparsifiers] \label{def: propertylocalSparsifier}
Let $G=(V,E)$ be a graph and $\alpha_1 \geq 1$ be a parameter. A graph $H=(V',E')$ with $V \subseteq V'$ is a \emph{local sparsifier} of $G$ with \emph{quality} $\alpha \geq 1$ iff the following hold:
\begin{enumerate}
\itemsep0em 
\item The graph $H$ is an $\alpha$-sparsifier of $G$,
\item For every $K \subseteq V$, there exists a subgraph $H[K]$ of $H$ such that $H[K]$ is $\alpha$-vertex sparsifier of $G$ with respect to $K$. Additionally, if $K$ is a proper subset of $V$, then $H[K]$ must be a proper subgraph of $H$.
\end{enumerate}
\end{definition}

In other words, the above definition suggests that local sparsifiers are sparsifiers from which we can extract vertex sparsifiers for any set of terminals $K$. Note that there are $\Theta(2^{n})$ different terminal sets, thus Condition (2) of local sparsifiers is very strong. The transitivity and decomposability notions readily extend to local sparsifiers.

Since we will exploit local sparsifiers to speed up dynamic graph algorithms, it is natural to define some notion that involves running times for manipulating local sparsifiers. We address this in the following definition, where we introduce a data-structure version of local sparsifiers. To avoid overloading the notation, we will simply refer to this data-structure as local sparsifiers. 

\begin{definition} \label{def: DSLocalSparsifier}
Given a graph $G=(V,E)$ and a parameter $\alpha \geq 1$, a \emph{data-structure local sparsifier} or simply a \emph{local sparsifier} $H=(V',E')$ with quality $\alpha$ of $G$ is a data-structure supporting the following operations:
\begin{itemize}
\itemsep0em 
\item \textsc{Preprocess$(G,\alpha)$}: compute an $\alpha$-sparsifier $H$ of $G$,
\item \textsc{QuerySparsifier$(G,K)$}: compute the subgraph $H[K]$ of $H$ and return $H[K]$ as an $\alpha$-vertex sparsifier of $G$ with respect to $K$. 
\end{itemize}
\end{definition}

The above data-structure is characterized by two important measures: \emph{preprocessing time}, which denotes the time for executing the operation \textsc{Preprocess$(G,\alpha)$}, and \emph{query time}, which denotes the time for executing the operation \sloppy \textsc{QuerySparsifier$(G,K)$}. Note that the data-structures always produces a mapping $S$ and and we will exploit properties of this mapping, specifically transitivity and decomposability, in our dynamic algorithms.

Since our goal is to design incremental algorithms with sub-linear update and query time, we will focus on building a local sparsifier with $\tilde{O}(m \cdot f(n))$ preprocessing time, while supporting queries in time $\tilde{O}(|K| \cdot g(n))$, where $f(n), g(n)$ are both sub-linear functions in $n$. In other words, this means that after computing a sparsifier of the input graph in time roughly proportional to its size, for any given set of terminals, we can construct a vertex sparsifier with respect to the terminals in time which depends \emph{only} on the number of terminals, up to sub-linear factors. We make precise this requirement in the following definition.

\begin{definition}\label{def: efficientLocalSparsifier}
Let $G=(V,E)$ be a graph, and let $f(n), g(n) \geq 1$ be functions. We say that $(H,\alpha,f(n),g(n))$ is an \emph{efficient local sparsifier} with quality $\alpha \geq 1$ of $G$ iff  $H$ is a local sparsifier of quality $\alpha$, and the preprocessing and query time of $H$ are bounded by $O(m \cdot f(n))$ and $O(|K| \cdot g(n))$, respectively.
\end{definition} 

To simplify the presentation, we will abuse the notation and sometimes write $(H,\alpha)$ instead of $(H,\alpha,f(n),g(n))$ when the runtime overheads are not important in specific contexts.

\section{From Local Sparsifiers to Incremental Algorithms}
\label{sec:meta_inc}
In this section we show how to use efficient local sparsifiers to design online (approximate) incremental algorithms for problems with certain properties while achieving fast worst-case update and query time. Roughly speaking, the key idea behind our result is to form a set $K$ out of the endpoints of all inserted edges since the last rebuild, and to use the efficient local sparsifiers with this set $K$ to build a suitable vertex sparsifier at query time on which we answer the query using a static algorithm.

\begin{theorem} \label{thm: metaTheorem}
Let $G=(V,E)$ be a graph, and for any $u,v \in V$, let $\mathcal{P}(u,v,G)$ be a solution to a minimization problem between $u$ and $v$ in $G$. Let $f(n),g(n),h(n) \geq 1$ be functions, $\alpha, \ell \geq 1$ be parameters associated with the approximation factor, and let $\beta_0,\beta_1,\ldots, \beta_{\ell}$ with $\beta_0 = m$ be parameters associated with the running time. Assume the following properties are satisfied
\begin{enumerate}
\itemsep0em 
\label{metaThM: P1}\item $G$ admits an efficient local sparsifier $(H,\alpha,f(n),g(n))$,
\label{metaThM: P2} \item $H$ is transitive and decomposable,
\label{metaThM: P3} \item The property $\mathcal{P}(u,v,G)$ can be computed in $O(m h(n))$ time in a graph with $m$ edges and $n$ vertices.
\end{enumerate}
Then there is an incremental (approximate) dynamic algorithm that maintains for every pair of nodes $u$ and $v$, an estimate $\delta(u,v)$, such that
\begin{equation} \label{eq: approxMeta}
	\mathcal{P}(u,v,G) \leq \delta(u,v) \leq \alpha^{\ell} \cdot \mathcal{P}(u,v,G),
\end{equation}
with worst-case update and query time of
\begin{equation} \label{eq: runningTimeMeta}	
	\tilde{O}\left (\left( \sum_{j=1}^{\ell} \left(\frac{\beta_{j-1}}{\beta_j} \right)f(n) + \beta_{\ell} h(n)\right) g(n)  \right) \quad \text{where } \beta_0 = m.
\end{equation}
\end{theorem}

To gain some intuition, we first consider just a two-level scheme and then explain how this scheme naturally generalizes to more levels. Given an initial graph $G=(V,E)$ and an approximation parameter $\alpha \geq 1$, we build a data-structure that maintains
\begin{enumerate}[noitemsep]
\item an efficient local sparsifier $(H,\alpha,f(n),g(n))$ of $G$~(Theorem~\ref{thm: metaTheorem}~Part~1), and 
\item a set of edges $E_1$, which is initially set to empty.
\end{enumerate}
Our data-structure is initialized using the \textsc{Preprocess}$(G,\alpha)$ operation of $(H,\alpha,f(n),g(n))$, and it is rebuilt every $\beta_1$ insertions, for some $\beta_1 \geq 0$ to be fixed later. Unless otherwise started, we will refer to $G$ as the current graph. We next describe the \textsc{Insert$(e)$} and \textsc{Query$(s,t)$} operations. Upon insertion of a new edge $e$ in $G$, we simply append edge $e$ to $E_1$. For answering $(s,t)$ queries, we first create the terminal set 
\begin{equation} \label{eq: defK}
K= \cup_{e \in E_1} V(e) \cup \{s, t\} ,
\end{equation}
where $V(e)$ are the endpoints of $e$, and then invoke \textsc{QuerySparsifier$(G,K)$} to get a vertex sparsifier $H[K]$ of $G \setminus E_1$ with respect to the terminal set $K$. Finally, we set $H' = H[K] \cup E_1$, and run on $H'$ a static algorithms that computes property $\mathcal{P}(s,t, H)$ between $s$ and $t$ in $H$, denoted by $\delta_{H'}(s,t)$, and return this value as an estimate.

We next argue that the $\delta_{H'}(s,t)$ approximates property $\mathcal{P}(s,t,G)$ up to an $\alpha$ factor. Note that it is sufficient to show that $H$ is an $\alpha$-vertex sparsifier of $G$ with respect to $K$. To this end, by definition of local sparsifiers, $H[K]$ is an $\alpha$-vertex sparsifier of $G \setminus E_1$ with respect to $K$, which in turn implies that $H' = H[K] \cup E_1$, is an $\alpha$-vertex sparsifier of $(G \setminus E_1) \cup E_1 = G$ with respect to $K$. The latter follows by decomposability of efficient local sparsifiers~(Theorem~\ref{thm: metaTheorem}~Part~2) and since endpoints of $E_1$ are added as terminals to $K$~(Equation~(\ref{eq: defK})).

We next analyze the update time. Note that the initialization time of our data-structure cost $O(m f(n))$~(Theorem~\ref{thm: metaTheorem}~Part~1), and recall that our data-structure is rebuilt every $\beta_1$ operations. Thus, the amortized update time per insertion is $O(mf(n)\beta_1^{-1})$. For the query time, note that the size of the terminal set $K$ at any time is $O(\beta_1)$. By Theorem~\ref{thm: metaTheorem}~Part~1, we get that the the size of the sparsifier $H$ of $G$ is $O(\beta_1 g(n))$. Finally, the query time is bounded by $O(\beta_1 g(n) h(n))$ assuming that $P(u,v,H)$ can be computed in $O(|E(H)| h(|V|)$ time.

Combining the above bounds on the update and query time, we obtain the following trade-off
\[
	O \left(\left(\frac{m}{\beta_1}\right)f(n) + \beta_1g(n)h(n)\right)
\] 
which in turn bounds the amortized update time and worst-case query time. The update time can be turned into a worst-case guarantee by a standard global rebuilding technique~(see, for example,~\cite{GoranciHP17a}, Section~3.3.2). 

We next explain the generalization of our approach to a multi-level hierarchy.   

\paragraph*{Data Structure.} 

Consider some integer parameter $\ell \geq 1$ and parameters $\beta_0 \geq \ldots \geq \beta_{\ell}$, with $\beta_0 = m$. Our data structure maintains
\begin{enumerate}[noitemsep]
\item a hierarchy of edge sets $\{E_i\}_{1 \leq i \leq \ell}$, each associated with the parameters $\{\beta_{i}\}_{1 \leq i \leq \ell}$,
\item a hierarchy of efficient local sparsifiers $\{(H_i,\alpha^{i+1})\}_{0 \leq i \leq \ell-1}$ for $\{G_i\}_{0 \leq i \leq \ell-1}$, where $G_0=G$, and remaining $G_i$'s are graphs that will be specified later,
\end{enumerate}

We initialize our data-structure by constructing an efficient local sparsifier $(H_0,\alpha_0)$ for the initial graph $G_0 = G$~(Theorem~\ref{thm: metaTheorem}~Part~1), and setting $H_i \gets H_0$. We also set $E_i \gets \emptyset$ for $1 \leq i \leq \ell$. 

We note that $\{E_i\}_{1 \leq i \leq \ell}$ will change over the course of the algorithm, as we will shortly make precise. For $1 \leq i \leq \ell$, we will use $E_i^{(t)}$, when necessary, to denote the set $E_i$ after the edge insertion at time $t$.

The hierarchy $\{E_i\}_{1 \leq i \leq \ell}$ keeps track of the inserted edges among different levels in our update sequence. 
Maintaining these edges will be useful when deciding to periodically rebuild parts of our data-structure. These periodical rebuilds will allow us to strictly reduce the running time at the cost of paying a multiplicative increase which is proportional to the number of levels $k$ in the hierarchy.

\paragraph*{Handling Insertions.}  Consider the insertion of edge $e=(u,v)$ in $G$. We maintain a variable $i$ that represents the level in the hierarchy (initially set to $1$), and a boolean variable rebuild~(initially set to \textsc{false}) that determines whether a rebuild is triggered at some level of the hierarchy when processing the insertion of $e$. While $i \leq \ell$ and rebuild equals \textsc{false}, we proceed as follows. We add $e$ to $E_i$, and test whether the size of $E_i$ exceeds $\beta_i$. If the latter holds, we set rebuild $\gets$ \textsc{true}, and distinguish two cases depending on whether $i = 1$ or $i \geq 2$. 

If $i=1$, we recompute from scratch an efficient local sparasifier $(H,\alpha)$ of the current graph $G$, set $G_0 \gets G$. Moreover, we set $H_j \gets H_0$ for $i \leq j \leq \ell-1$, and let $E_j \gets \emptyset$ for $1 \leq j \leq \ell$.

If $i \geq 2$, our goal will be to recompute efficient local sparsifier $H_{i-1}$ at level $(i-1)$ in the hierarchy. To this end, we first define the graph 
\[
	R_{i-1} :=  H_{i-2}[V(E_{i-1})] \cup E_{i-1},
\] 
where $H_{i-2}$ is the efficient local sparsifier that we maintain at level $(i-2)$, and $V(E_{i-1})$ denotes the endpoints of the edges in $E_{i-1}$. In other words, $R_{i-1}$ is obtained by taking the union over the edges stored at level $(i-1)$ and the vertex sparsifier $H_{i-2}[V(E_{i-1})]$ with respect to $V(E_{i-1})$ associated to the graph at level $(i-2)$. We then construct an efficient local sparsifier $(R'_{i-1},\alpha)$ of $R_{i-1}$. The efficient local sparsifier $H_{i-1}$ is updated using the following rule
\[
	H_{i-1} \gets \left(H_{i-2} \setminus H_{i-2}[V(E_{i-1})] \right) \cup R'_{i-1}.
\]

Finally, we update the efficient local sparsifiers in the levels $(i,\ldots,\ell-1)$ by setting $H_{j} \gets H_{i-1}$, for $i \leq j \leq \ell-1$. We also let $E_j \gets \emptyset$, for $i \leq j \leq \ell$, and increment $i$ by $1$. This algorithm is depicted in Figure~\ref{alg: IncrementalInsert}.

\begin{algorithm2e}[t!]
\caption{\textsc{Insert}$(e=(u,v))$}
\label{alg: IncrementalInsert}
Set $i \gets 1$ \\
Set $\text{rebuild} \gets \textsc{false}$ \\
Set $E \gets E \cup \{(u,v)\}$ \\
\While{$i \leq \ell$ and $\emph{rebuild} = \textsc{false}$} 
{
   $E_i \gets E_i \cup \{(u,v)\}$ \\
   \If{$|E_i| > \beta_i$}
   {  	  
	  $\text{rebuild} \gets \textsc{true}$   \\  	 
   	  \If{$i=1$}   	
   	  { \label{line: fullRebuild}
   	  	 Set $G_0 \gets G$ \\
   	  	 Compute an efficient local sparsifier $(H_0,\alpha)$ of $G_0$~(Theorem~\ref{thm: metaTheorem} Part~1) \\
   	  }
   	  \Else
   	  {
\label{line: partialRebuildStart}   	  	 Let $R_{i-1} \gets H_{i-2}[V(E_{i-1})] \cup E_{i-1}$\\
   	  	 Compute efficient local sparsifier $(R_{i-1}', \alpha)$ of $R_{i-1}$~(Theorem~\ref{thm: metaTheorem} Part~1) \\
\label{line: sparsifierRebuild} 	  	 Set $H_{i-1} \gets \left(H_{i-2} \setminus H_{i-2}[V(E_{i-1})] \right) \cup R'_{i-1}$ \\
\label{line: partialRebuildEnd}    	  	 
   	  }   
\label{line: sparsifierUpdate}   	  Set $H_j \gets H_{i-1}$, for $i \leq j \leq \ell-1$ \\	  
   	  Set $E_j \gets \emptyset$ for $i \leq j \leq \ell$ \\
   }
  Set $i \gets i+1$
} 
\end{algorithm2e}

\begin{algorithm2e}[t!]
\caption{\textsc{Query}$(s,t)$}
\label{alg: IncrementalQuery}
Set $K \gets \cup_{e \in E_{\ell}} V(e) \cup \{s,t\}$\\
Set $H \gets H_{\ell-1}[K] \cup E_{\ell}$ \\
Let $\delta_H(s,t)$ be the result obtained by the algorithm from Theorem~\ref{thm: metaTheorem}~Part~3. \\
\Return $\delta_H(s,t)$.
\end{algorithm2e}

\paragraph*{Handling Queries.} To answer the query for the approximate property $\mathcal{P}(u,v,G)$ between any pair of vertices $s$ and $t$ in $G$ we proceed as follows. We first create a terminal set using the endpoints of the edges stored at the last level $E_\ell$ together with $s$ and $t$, i.e.,
\[ K = \cup_{e \in E_{\ell}} V(e) \cup \{s,t\}, \]
where $V(e)$ are the endpoints of $e$. We then proceed by querying the vertex sparsifier $H_{\ell-1}[K]$ with respect to $K$, and union this with the maintained edge set $E_\ell$, i.e., we define an auxiliary graph
\[
	H := H_{\ell-1}[K] \cup E_\ell.
\]
Finally, we run the algorithm from Theorem~\ref{thm: metaTheorem}~Part~3 on $H$ to calculate the property $\mathcal{P}(u,v,H)$ between $u$ and $v$ in $H$, which we denote by $\delta_H(s,t)$, and return this value as an estimate. 

\paragraph*{Correctness.} Let $G$ be the current graph throughout the execution of the algorithm. We will show that as long as $|E_1| \leq \beta_1$, the efficient local sparsifier we maintain at level $(k-1)$ is sufficient to give a good approximation to the graph property $\mathcal{P}$ between any two pair of vertices from $G$. Note that whenever $|E_1| > \beta_1$, the entire data-structure is built from scratch, and in this case, the local sparsifier $H_0$ is already a good estimate for $G$. 

To make the above statements precise, we need to introduce some useful notation. First, recall that for $1 \leq i \leq \ell$, we use $E_i^{(t)}$ to denote the set $E_i$ after the edge insertion at time $t$ in our algorithm. Let $E$ be the set of inserted edges so far in our graph, i.e., $G = G_0 \cup E$, where $G_0$ is the initial graph from the last rebuild of the entire data-structure~(Line~\ref{line: fullRebuild} in Algorithm~\ref{alg: IncrementalInsert}) or from the beginning of the algorithm. For each $e \in E$, we let $\tau_e$ be the index of the lowest level edge set in the current hierarchy $\{E_i\}_{1 \leq i \leq \ell}$ that contains $e$, i.e., 
\[
	\tau_e = \max\{j \in \{1,\ldots, \ell \} \mid e \in E_j\}.
\]
This naturally induces a partitioning of $E$ defined as follows
\[
	E = \cup_{1 \leq i \leq \ell} \tilde{E}_i , \quad \text{ where } \tilde{E}_i = 
\{e \in E \mid \tau_e = i\} \text{ for } 1 \leq i \leq \ell.
\]

Let $\{H_{i}\}_{0 \leq i \leq \ell-1}$ be the hierarchy of current efficient local sparsifier that our data structure maintains. Using the partitioning of $E$, we next show that each of these local sparsifiers maintains information for property $\mathcal{P}$ with respect to some edge sets in the partition, where the size of the edge set increases with the number of levels. In particular, this implies that the lowest-level local sparsifier $H_{\ell-1}$ will be a good estimate to property $\mathcal{P}$ in the current graph $G$. This approach is formally summarized in the following lemma.

\begin{lemma} \label{lem: correctnessLocalSparsifier}
The graph $H_{i}$ at level $i$ is an $\alpha^{i+1}$-efficient local sparsifier of $G_0 \cup \left(E \setminus \cup_{i+1 \leq j \leq \ell} \tilde{E}_j\right)$ for $0 \leq i \leq \ell-1$.
\end{lemma}
\begin{proof}
We proceed by induction on the level $i$ of the hierarchy. For the base case, i.e., $i = 0$, by construction $H_0$ is a $\alpha^{0+1}$-efficient local sparsifier of $G_0 \cup (E \setminus \cup_{1 \leq j \leq \ell} \tilde{E}_i) = G_0$, and thus the claim holds.

Let $H_i$ the efficient local sparsifier that our algorithm maintains at level $i > 0$. We want to show that $H_i$ is an $\alpha^{i+1}$-efficient local sparsifier of $G_0 \cup \left(E \setminus \cup_{i+1 \leq j \leq \ell} \tilde{E}_j \right)$. To this end, note that it suffices to prove that $H_i$ is an $\alpha$-efficient local sparsifier of $H_{i-1} \cup \tilde{E}_i$. We show this claim
\begin{itemize}[noitemsep]
\itemsep0em 
\item using the induction hypothesis on $H_{i-1}$, i.e., that $H_{i-1}$ is an $\alpha^{i}$-efficient local sparsifier of $G_0 \cup \left(E \setminus \cup_{i \leq j \leq \ell} \tilde{E}_j\right)$, and 
\item using the transitivity on $H_i$ and $H_{i-1} \cup \tilde{E}_i$~(Theorem~\ref{thm: metaTheorem}~Part~2).
\end{itemize}
We these two facts and the decomposability of efficient local sparsifiers~(Theorem~\ref{thm: metaTheorem}~Part~2) we get that that $H_{i}$ is an $\alpha^{i+1}$-efficient local sparsifier of \[G_0 \cup \left(E \setminus\cup_{i \leq j \leq \ell} \tilde{E}_j \right) \cup \tilde{E}_i = G_0 \cup \left(E \setminus \cup_{i+1 \leq j \leq \ell} \tilde{E}_j\right).\]

Thus it remains to show that $H_i$ is an $\alpha$-efficient local sparsifier of $H_{i-1} \cup \tilde{E}_i$. We distinguish two cases. (1) If $\tilde{E}_i = \emptyset$, then we know that there was a rebuild at a level smaller than $i$ in the hierarchy, which implies that $H_i =H_{i-1}$~(Line~\ref{line: sparsifierUpdate} of Algorithm~\ref{alg: IncrementalInsert}). Thus $H_i$ is trivially an $\alpha$-efficient local sparsifier of $H_{i-1} \cup \tilde{E}_i$. (2) If $\tilde{E}_i \neq \emptyset$, let $t_i$ be the last time that $H_i$ was rebuilt with respect to the set $E_i$, i.e., Lines~\ref{line: partialRebuildStart}-\ref{line: partialRebuildEnd} in Algorithm~\ref{alg: IncrementalInsert} were executed at time $t_i$. We claim that $\tilde{E}_i = E^{(t_i)}_i$. Note that this follows by definition of $\tilde{E}_i$ since edges belonging to this set do not appear in the levels larger than $i$. To prove the claimed approximation guarantee on $H_i$, we first observe that the graph $H_{i-1} \cup \tilde{E}_i$ can be partitioned into edge-disjoint graphs as follows
\[
	H_{i-1} \cup \tilde{E}_i = \left(H_{i-1} \setminus H_{i-1}[V(\tilde{E}_i)]\right) \cup \left(H_{i-1}[V(\tilde{E}_i)] \cup \tilde{E}_i\right),
\]
where $R_{i} = H_{i-1}[V(\tilde{E}_i)] \cup \tilde{E}_i$ by our construction. Let $(R'_{i}, \alpha)$ be the efficient local sparsifier of $R_i$ computed by the algorithm. By definition of efficient local sparsifiers we know that $V(R_i) \subseteq V(R'_{i})$ and $R_i'$ is an $\alpha$-sparsifier of $R_i$. Moreover, recall that the algorithm updates $H_i$ as follows
\[
	H_{i} = \left(H_{i-1} \setminus H_{i-1}[V(\tilde{E}_i)]\right) \cup  R'_{i}.
\]
Applying the decomposability property of Theorem~\ref{thm: metaTheorem}~Part~2 on $H_{i-1} \setminus H_{i-1}[V(\tilde{E}_i)]$ and $R'_{i}$ we get that $H_{i}$ is an $\alpha$-efficient local sparsifier of $H_{i-1} \cup \tilde{E}_i$, which completes the proof. 
\end{proof}

We finally show that the estimate $\delta_H(s,t)$ returned by the query algorithm in Figure~\ref{alg: IncrementalQuery} approximates the property $\mathcal{P}$ of the current graph $G$ up to an $\alpha^{\ell}$ factor, thus proving the claimed estimate in Theorem~\ref{thm: metaTheorem}.

By Lemma~\ref{lem: correctnessLocalSparsifier}, we get that $H_{\ell-1}$ is an $\alpha^{\ell}$-efficient local sparsifier of graph $G_0 \cup (E \setminus \tilde{E}_\ell)$. Since $\tilde{E}_\ell=E_\ell$~(because $\ell$ is largest level), we get that $H_{\ell-1}[K]$ is a $\alpha^{\ell}$-vertex sparsifier of $G_0 \cup (E \setminus E_\ell)$ with respect to $K$. Using decomposability of efficient local sparsifiers~(Theorem~\ref{thm: metaTheorem}~Part~2), the latter implies that $H = H_{\ell-1} \cup E_\ell$ is a $\alpha^{\ell}$-vertex sparsifier of $G_0 \cup (E \setminus E_\ell) \cup E_\ell = G_0 \cup E = G$.

\paragraph*{Running Time.}



We first study the update time of our data structure. To this end, it will be useful to bound the size of each efficient local sparsifier in the hierarchy $\{H_i\}_{0 \leq i \leq \ell-1}$ at any given point of time.

\begin{lemma} \label{lem: sparsifierSizes}
At any point of time, for each $0 \leq i \leq (\ell-1)$ and $K \subseteq V$, we have that \[ |H_i[K]|  \leq \tilde{O}\left ( |K|\cdot g(n) \right). \]
\end{lemma}
\begin{proof}
We actually prove something stronger, namely that at any point of time, for each  $K \subseteq V$ and $0 \leq i \leq (\ell-1)$ we have that $|H_i[K]|  \leq O\left((i+1) |K| \cdot g(n)\right)$. As we will shortly see, the number of levels $\ell$ in the hierarchy does not exceed $O(\log n)$. Since $i \leq (\ell-1)$, we immediately get the claimed bound of the lemma.

At any point of time during the execution of our data-structure, note that the worst-case bound on the size of $H_i[K]$ at level $i$ is attained when the $H_j$ for $0 \leq j \leq i$ are different i.e., each of the efficient local sparsifier has undergone a rebuild with respect to the current edge set $E_j$. Thus, throughout we assume that this is indeed the case, as otherwise the bounds can only get better.

We now prove the claim by induction on the level $i$ of the hierarchy. For the base case, i.e., $i=0$, we know that $H_0$ is an efficient local sparsifier of $G_0$, and by querying $H_0$ with respect to $K$ it follows that $|H_0[K]| \leq O(|K| \cdot g(n))$, and hence the claim holds.

By induction hypothesis we get that for each $K \subseteq V$, it holds that $|H_{i-1}[K]| \leq O(i |K| \cdot g(n))$. We now show the inductive step. Let $K \subseteq V$ be any subset of vertices. To this end, let $H_i$ be the efficient local sparsifier that has undergone a rebuild with respect to $E_i$ at level $i > 0$. Let $R_i := H_{i-1}[V(E_{i})] \cup E_{i}$ be the intermediate graph which is used to rebuild $H_i$,  
and let $(R_{i}',\alpha)$ be the efficient local sparsifier of $R_{i}$, as defined in Algorithm~\ref{alg: IncrementalInsert}. Define $K' := K \cap V(R_i)$ and note that by construction $|V(R_i)| \leq n$. Then by querying $R_{i}'$ with respect to $K'$  we get that  
\[ |R_i'(K')| \leq O( |K'| \cdot g(|V(R_i)|)) \leq O(|K| \cdot g(n)). \]

Finally, since $H_i$ is formed by taking the union of $R'_i$ with some part of $H_{i-1}$ we get that
\[
	|H_i| \leq |H_{i-1} \cup R'_i| \leq O((i+1) |K| \cdot g(n)),
\]
where the bound on $H_{i-1}$ follows by induction hypothesis.

\end{proof}

The lemma below bounds the amortized update time of our data-structure. 

\begin{lemma} \label{lem: insertRunningTime}
The amortized time of \textsc{Insert$(e=(u,v))$} operation is bounded by \[
	\tilde{O}\left(\left(\sum_{j=0}^{\ell-1} \frac{\beta_j}{\beta_{j+1}}\right) f(n) g(n) \right).
\]
\end{lemma}
\begin{proof}
For $0 \leq i \leq \ell-1$, let $Y(i)$ be the amortized update time aggregated up to (and including) level $i$ in the hierarchy. Furthermore, let $Z(i)$ be the amortized update time at level $i$ in the hierarchy~(and excluding all other levels). We will show by induction on the number of levels $i$ that $Y(i) = \tilde{O}\left(\left(\sum_{j=0}^{i} \frac{\beta_j}{\beta_{j+1}}\right) f(n) g(n) \right)$, which with $i=(\ell-1)$ implies the claimed bound of the lemma.

For the base case, i.e., $i = 0$, recall that the cost for constructing an efficient local sparsifier $H_0$ of the current graph $G_0$ is $\tilde{O}(\beta_0 f(n))$~(Theorem~\ref{thm: metaTheorem}~Part~1), where $\beta_0 = m$ is the current number of edges. Moreover, the cost for updating the efficient local sparsifiers in the levels below $\{H_j\}_{1 \leq j \leq k-1}$ is bounded by $\tilde{O}(\ell \beta_0 f(n)) = \tilde{O}(\beta_0 f(n))$. Thus, the overall cost of a rebuild at level $i=0$ is $\tilde{O}(\beta_0 f(n))$. Since $H_0$ is rebuilt every $\beta_1$ insertions, we get that the amortized cost per insertion is $Y(0) = Z(0) = \tilde{O}\left(\left(\frac{\beta_0}{\beta_1}\right) f(n) \right) = \tilde{O}\left(\left(\frac{\beta_0}{\beta_1}\right) f(n) g(n)\right)$, as desired.

We next show the inductive step. Consider the maintained efficient local sparsifier $H_i$ at level $i$ that undergoes a rebuild with respect to $E_i$, and let $H_{i-1}$ be the efficient local sparsifier one level above~(recall that a rebuild at level $i$ is triggered by level $(i+1)$, i.e., because $|E_{i+1}| > \beta_{i+1}$). We want to bound the size of the intermediate graph $R_i = H_{i-1}[V(E_i)] \cup E_i$, as defined in Algorithm~\ref{alg: IncrementalInsert}, which in turn determines the cost for rebuilding $H_i$. To this end, first observe that by construction $|E_i| \leq \beta_i$. Second, by Lemma~\ref{lem: sparsifierSizes} we get that 
\[ |H_{i-1}[V(E_i)]| \leq \tilde{O}(|V(E_i)|g(n)) \leq \tilde{O}(\beta_i g(n)). \]

Combining these two bounds we get that $|R_i| \leq \tilde{O}(\beta_i g(n))$. We now bound the cost for computing $R'_i$ and updating the efficient local sparsifier $H_i$. As Algorithm~\ref{alg: IncrementalInsert} computes an efficient local sparsifier $(R'_i,\alpha)$ of $R_i$, by Theorem~\ref{thm: metaTheorem}~Part~1 we get that the cost for computing $R'_i$ is
\[
	\tilde{O}(|R_i|f(n)) = \tilde{O}(\beta_i g(n)) \cdot \tilde{O}(f(n)) = \tilde{O}(\beta_i f(n) g(n)).
\]

Consider the update of the efficient local sparsifier $H_i$, and assume that before the update, $H_i = H_{i-1}$ holds. Then we can simply update $H_i$ by deleting the edges $H_{i-1}[V(E_i)]$ from $H_i$ and adding the new edges $R'_i$ to $H_i$. Since by the above discussion the size of both $H_{i-1}[V(E_i)]$ and $R'_i$ is bounded by $\tilde{O}(\beta_i f(n) g(n))$, we claim the cost for updating $H_i$ is also bounded by $\tilde{O}(\beta_i f(n) g(n))$. 

Now, if $H_i \neq H_{i-1}$ holds before updating $H_i$, this means that $H_i$ has undergone already a rebuild with respect to $E_i$. We then reverse all the operations of the data-structure during the last rebuild until $H_i = H_{i-1}$, and proceed as above for updating $H_i$. Since by construction $|E_i| \leq \beta_i$, observe that the reversing cost cannot exceed the cost of updating $H_i$, which we showed to be at most $\tilde{O}(\beta_i f(n) g(n))$. Similarly, for updating the efficient local sparsifiers $\{H_j\}_{i-1 \leq j \leq \ell-1}$, we first reverse their the data-structure operations until $H_{i-1} = H_{i+1} = \ldots = H_{\ell-1}$, and then proceed as above for updating $\{H_j\}_{i-1 \leq j \leq \ell-1}$\footnote{Note that this is faster than copying $H_{i-1}$ into the data structures for $\{H_j\}_{i-1 \leq j \leq \ell-1}$}. Since there are at most $\ell$ levels below to update during the rebuild at level $i$, the total cost for updating the hierarchy  $\{H_j\}_{i \leq j \leq \ell-1}$ is bounded by 
\[ \tilde{O}(\ell \beta_i f(n) g(n)) = \tilde{O}(\beta_i f(n) g(n)). \]

Summing the cost for computing $R_i'$ and the cost for updating the hierarchy $\{H_j\}_{i \leq j \leq \ell-1}$, we conclude that the total cost for rebuilding $H_i$ with respect to $E_i$ is bounded by $\tilde{O}(\beta_i f(n) g(n))$. Since the emulator $H_i$ is rebuilt every $\beta_{i+1}$ operations, we get that the amortized cost per operation is
\[
	Z(i) = \tilde{O}\left(\left(\frac{\beta_i}{\beta_{i+1}}\right) f(n) g(n) \right).
\]

To complete the inductive step, note that by induction hypothesis \[ Y(i-1)= \tilde{O}\left(\left(\sum_{j=0}^{i-1}\frac{\beta_j}{\beta_{j+1}}\right) f(n) g(n)\right). \]

Summing over this and the bound on $Z(i)$ we get

\begin{align*}
Y(i) & = Y(i-1) + Z(i) \\ 
& = \tilde{O}\left(\left(\sum_{j=0}^{i-1}\frac{\beta_j}{\beta_{j+1}}\right)f(n)g(n)\right) + \tilde{O}\left(\left(\frac{\beta_i}{\beta_{i+1}}\right) f(n)g(n) \right) \\ 
& = \tilde{O}\left(\left(\sum_{j=0}^{i} \frac{\beta_j}{\beta_{j+1}}\right) f(n)g(n) \right).
\end{align*}

\end{proof}
We next study the query time of our data-structure.

\begin{lemma} \label{lem: queryTime}
The time for a $\textsc{Query}(s,t)$ operation is bounded by $\tilde{O}\left( \beta_{\ell} g(n) h(n) \right).$
\end{lemma}
\begin{proof}
Let $K = \cup_{e \in E_{\ell}} V(e) \cup \{s,t\}$ be the set of terminals defined in Algorithm~\ref{alg: IncrementalQuery}. By construction, we know that $|E_{\ell}| \leq \beta_{\ell}$, which in turn implies that $|K| \leq O(\beta_{\ell})$. Let $H = H_{\ell-1}[K] \cup E_{\ell}$ be the graph estimator as defined in Algorithm~\ref{alg: IncrementalQuery}, where $H_{\ell-1}$ is the efficient local sparsifier at level $(\ell-1)$ in the hierarchy. By Lemma~\ref{lem: sparsifierSizes} and the bound on the size of $T$, we get that $|H_{\ell-1}[K]| \leq \tilde{O}(\beta_\ell g(n))$, which in turn implies that
\[
	|H| = |H_{k-1}[K] \cup \beta_{\ell}| \leq \tilde{O}(\beta_{\ell} g(n)).
\]  
Since the algorithm for testing property $\mathcal{P}(s,t,G)$ runs in $\tilde{O}(|H| h(n))$ time by Theorem~\ref{thm: metaTheorem}~Part~3, we get the our query time is bounded by $\tilde{O}(\beta_{\ell} g(n) h(n))$.
\end{proof}

Combining the bounds on the update and query time from Lemmas~\ref{lem: insertRunningTime} and~\ref{lem: queryTime}, we obtain the following trade-off

\[
	\tilde{O}\left(\left(\sum_{j=0}^{\ell-1} \left(\frac{\beta_j}{\beta_{j+1}}\right)f(n) + \beta_{\ell} h(n )\right) g(n)\right), \quad \text{where } \beta_0 = m,
\]
which in turn proves the claimed update and query time in Theorem~\ref{thm: metaTheorem}.

Finally we show for what choice of parameters $\{\beta_i\}_{0 \leq i \leq \ell}$ the above trade-off is minimized, if we ignore functions $f(n),g(n)$ and $h(n)$.  As we will see in the subsequent sections, this simplification will be justified in all the applications of Theorem~\ref{thm: metaTheorem}.

\begin{lemma} \label{lem: optimalTradeoff}
For $1 \leq \ell \leq \log n$, let $\{\beta_i\}_{0 \leq i \leq \ell}$ be a family of parameters with $\beta_0 = m$. If we set \[ \beta_i = (\beta_{i-1})^{\frac{\ell-(i-1)}{\ell+1-(i-1)}},~1 \leq i \leq \ell \] then 
\[
	\tilde{O}\left(\sum_{j=0}^{\ell-1}\frac{\beta_j}{\beta_{j+1}} + \beta_{\ell} \right) = \tilde{O}\left(m^{1/k+1}\right).
\]
\end{lemma}
\begin{proof}
We claim that for each $i \geq 1$, it holds that $\beta_i = m^{1-\frac{i}{\ell+1}}$, and prove this by induction on $i$. For the base case, i.e., $i=1$, by the choice of  $\beta_1$ we have $\beta_1 = (\beta_{0})^{\frac{\ell}{\ell+1}} = m^{1-\frac{1}{\ell+1}}$. 

For the inductive step, we have
\begin{equation} \label{eq: tradeOff}
\begin{split}
	\beta_i = (\beta_{i-1})^{\frac{\ell-(i-1)}{\ell+1-(i-1)}} & = \left(m^{1-\frac{(i-1)}{\ell+1}}\right)^{\frac{\ell-(i-1)}{\ell+1-(i-1)}} = m^{\frac{\ell+1 - (i-1)}{\ell+1} \cdot \frac{\ell-(i-1)}{\ell+1-(i-1)}} \\ 
	& = m^{1 - \frac{i}{\ell+1}},
\end{split}
\end{equation}

where the second equality follows by induction hypothesis on $\beta_{i-1}$.

Plugging the choice of $\beta_i$ in Equation~\ref{eq: tradeOff} yields 

\begin{align*}
	\tilde{O} \left(\sum_{j=0}^{\ell-1} \frac{m^{1-\frac{j}{\ell+1}}}{m^{1-\frac{(j+1)}{\ell+1}}} + m^{\frac{1}{k+1}}\right)
	 & = \tilde{O} \left( \sum_{j=0}^{\ell-1} m^{\frac{1}{\ell+1}} + m^{\frac{1}{\ell+1}}\right) 
	 = \tilde{O}\left(\ell m^{\frac{1}{\ell+1}}\right) \\
	 & = \tilde{O}\left( m^{\frac{1}{\ell+1}} \right).
\end{align*} 

\end{proof}

\section{Incremental All Pair Shortest Paths}
\label{sec:distance_inc}

In this section we show how to use our general Theorem~\ref{thm: metaTheorem} to design online incremental algorithms for the approximate All-Pair Shortest Path Problem with fast worst-case update and query time. Concretely, we will show that that assumptions (1) and (2) in Theorem~\ref{thm: metaTheorem} are satisfied with certain parameters for shortest paths. Note that (3) follows immediately by any $\tilde{O}(m)$ time single pair shortest path algorithm. This results in the following theorem.
\begin{theorem} \label{thm: IncrementalApproximateAPSP}  Let $G=(V,E)$ be an undirected, weighted graph. For every $r, \ell \geq 1$, there is a deterministic incremental approximate APSP algorithm that maintains for every pair of nodes $u$ and $v$, a distance estimate $\delta(u,v)$ such that \[ \dist_G(u,v) \leq \delta(u,v) \leq (2r-1)^{\ell} \dist_G(u,v), \] 
with worst-case update and query time of \[ \tilde{O}(n^{2/(\ell+1)} n^{2/r}).\]

\end{theorem}
We start by introducing the usual definitions of sparsifiers and vertex sparsifiers for distances. Having defined these, the definition of local sparsifiers becomes apparent from the general definition we introduced in Section~\ref{sec: localSparsifiers}. Let $G=(V,E)$ be an undirected, weighted graph with a \emph{terminal} set $K \subseteq V$. For $u,v \in V$, let $\dist_G(u,v)$ denote the length of a shortest path between $u$ and $v$ in $G$.

\begin{definition}[Sparsifiers for Distances] Let $G=(V,E)$ be an undirected, weighted graph, and let $\alpha \geq 1$ be a \emph{stretch} parameter. A graph $H=(V',E')$ with $V \subseteq V'$ is an $\alpha$-\emph{sparsifier} of $G$ iff for all $u,v \in V$,
\[
	\dist_G(u,v) \leq \dist_{H}(u,v) \leq \alpha \cdot \dist_G(u,v).
\]
\end{definition}

\begin{definition}[Vertex Sparsifiers for Distances] \label{def: vertexDistanceSparsifier} Let $G=(V,E)$ be an undirected, weighted graph with a \emph{terminal} set $K \subseteq V$, and let $\alpha \geq 1$. A graph $H=(V',E')$ with $K \subseteq V'$ is an $\alpha$-\emph{(vertex) distance sparsifier} of $G$ with respect to $K$ iff for all $u,v \in K$, 
\[
	\dist_G(u,v) \leq \dist_{H}(u,v) \leq \alpha \cdot \dist_G(u,v).
\]
\end{definition}

We next show that the distance property in graphs admits efficient local sparsifiers. We achieve this by showing a deterministic variant of the distance oracle due to Thorup and Zwick~\cite{ThorupZ05}. While we closely follow the ideas presented in the deterministic oracle due to Roddity, Thorup and Zwick~\cite{RodittyTZ05}, we note that they only give a bound on the \emph{total} size of the oracle, which is not sufficient for our purposes.

\begin{lemma}[Efficient Distance Local Sparsifiers] \label{lem: efficientDistanceLocalSparsifier}
Given an undirected, weighted graph $G=(V,E)$, and a parameter $r \geq 1$, there is a \emph{deterministic} algorithm for constructing an \emph{efficient} distance local sparsifier with $(2r-1)$ stretch, $\tilde{O}(mn^{1/r})$ preprocessing time, and $O(|K| n^{1/r})$ query time, where $K$ is any set of queried terminals.
\end{lemma}

We start by reviewing the randomized algorithm for APSP due to Thorup and Zwick~\cite{ThorupZ05}(which is depicted in Figure~\ref{alg: randomizedPreprocessing}), and then derandomize that algorithm and show how it can be used to solve the above problem.

\begin{enumerate}
\itemsep0em 
\item  Set $A=V$ and $A_r = \emptyset$, and for $1 \leq i \leq r-1$ obtain $A_i$ by picking each node from $A_{i-1}$ independently, with probability $n^{-1/r}$. 

\item For each $1 \leq i < r$, and for each vertex $v \in V$, find the vertex $p_i(v) \in A_i$ (also known as the $i$-th \emph{pivot}) that minimizes the distance to $v$, i.e., \[ p_i(v) := \arg \min_{u \in A_i} \dist_G(u,v), \] and its corresponding distance value \[\dist_G(A_i,v) := \min\{\delta(w,v) \mid w \in A_i\} = \dist_G(v,p_i(v)). \]

\item For each vertex $v \in V$, define the \emph{bunch} $B(v) = \cup_{i=0}^{r-1} B_i(v)$, where
\begin{align*}
B_i(v) & := \{w \in A_i \setminus A_{i+1} \mid \dist_G(w,v) < \dist_G(A_{i+1},v) \}.
\end{align*}
\end{enumerate}

\begin{algorithm2e}[t!]
\caption{\textsc{HierarchyConstruct}$(G, r)$}
\label{alg: randomizedPreprocessing}
$A_0 \gets V$ ; $A_r \gets \emptyset$ \\
\For{$i \gets 1$ to $r-1$} 
{
	\label{algStep: random} $A_i \gets \textsc{Sample} \left( A_{i-1}, |V|^{-1/r} \right)$ \\
} 

\For{every $v \in V$}
{
	\For{$i \gets 0$ to $r-1$}
	{
	   Let $\dist_G(A_i,v) \gets \min \{ \dist_G(w,v) \mid w \in A_i \}$ \\
	   Let $p_i(v) \in A_i$ be such that $\dist_G(p_i(v),v) = \dist_G(A_i,v)$ \\
    }    
    \BlankLine
    $\dist_G(A_r,v) \gets \infty$ \\
    \BlankLine
    Let $B(v) \gets \cup_{i=0}^{r-1} \{w \in A_i \setminus A_{i+1} \mid \dist_G(w,v) < \dist_G(A_{i+1},v)\}$  
    
}

\end{algorithm2e}

Thorup and Zwick~\cite{ThorupZ05} showed that using the hierarchy of sets $(A_i)_{0 \leq i \leq r}$ chosen as above, the expected size of a bunch $\expec{}{|B(v)|}$ is $O(r n^{1/r})$, for each vertex $v \in V$. We note that the only place where their construction uses randomization is when building the hierarchy of sets~(the \textbf{for} loop in Step~\ref{algStep: random} in Figure~\ref{alg: randomizedPreprocessing}). Therefore, to derandomize their algorithm it suffices to design a deterministic algorithm that efficiently computes a hierarchy of set $(A_i)_{0 \leq i \leq r}$ such that $|B(v)| \leq \tilde{O}(r n^{1/r})$, for each $v \in V$ (note that compared to the randomized construction, we are content with additional poly-log factors on the size of the bunches).

We present a deterministic algorithm for computing the hierarchy of sets that closely follows the ideas presented in the deterministic construction of Roditty, Thorup, and Zwick~\cite{RodittyTZ05}. The main two ingredients of the algorithm are the \emph{hitting set} problem, and the \emph{source detection} problem. For the sake of completeness, we next review their definitions and properties.

\begin{definition}[Hitting set] \label{def: hittingSet} Let $U$ be a set of elements, and let $\mathbb{S} = \{S_1,\ldots,S_p\}$ be a collection of subsets of $U$. We say that $T$ is a \emph{hitting set} of $U$ with respect to $\mathbb{S}$ if $T \subseteq U$, and $T$ has a non-empty intersection with every set of $\mathbb{S}$, i.e., $T \cap S_i \neq \emptyset$ for every $1 \leq i \leq p$. 
\end{definition}

It is known that computing a hitting set of minimum size is an NP-hard problem. In our setting however, it is sufficient to compute approximate hitting sets. Since our goal is to design a deterministic algorithm, one way to deterministically compute such sets is using a variant of the well-known greedy approximation algorithm: (1) Form the set $T$ by repeatedly adding to $T$ elements of $U$ that `hit' as many `unhit' sets as possible, until only $|U|/s$ sets are unhit, where $|S_i| \geq s$ for each $1 \leq i \leq p$ ; (2) add an element from each one of the unhit sets to $T$. The lemma below shows that this algorithm finds a reasonably sized hitting set in time linear in the size of $U$ the collection $\mathbb{S}$. 


\begin{lemma} \label{lem: hittingSetAlgo}
Let $U$ be a set of size $u$ and let $\mathbb{S} = \{S_1,\ldots,S_p\}$ be the collection of subset of U, each of size at least $s$, where $s \leq p$. Then the above deterministic greedy algorithm runs in $O(u + ps)$ time and finds a hitting set $T$ of $U$ with respect to $\mathbb{S}$, whose size is bounded by $|T| = (u/s)(1+ \ln p)$.
\end{lemma}

Note that the size of this hitting set is within $O(\log n)$ of the optimum size since in the worst case $T$ has size at least $u /s$. 

\begin{definition}[Source Detection] \label{def: sourceDetect} Let $G=(V,E)$ be an undirected, weighted graph, let $U \subseteq V$ be an arbitrary set of sources of size $u$, and let $q$ be a parameter with $1 \leq q \leq u$. For every $v \in V$, we let $U(v,q,G)$ be the set of the $q$ vertices of $U$ that are closest to $v$ in $G$.
\end{definition}

Roditty, Thorup, and Zwick~\cite{RodittyTZ05} showed that the set $U(v,q,G)$ can be computed using $q$ single-source shortest path computations. We review their result in the lemma below.
\begin{lemma}[\cite{RodittyTZ05}] \label{lem: sourceDetectAlgo} For every $v \in V$, the set $U(v,q,G)$ can be computed in time $O(q m \log n)$.
\end{lemma}

Our algorithm for constructing the hierarchy of sets $(A_i)_{0 \leq i \leq r}$, depicted in Figure~\ref{alg: DeterministicHieararhcy}, is as follows. Initially, we set $A_0 = V$ and $A_r = \emptyset$. To construct the set $A_{i+1}$, given the set $A_i$, for $0 \leq i \leq p-2$, we first find the set $A_i(v,q,G)$, where $q=\tilde{O}(n^{1/r})$, using the source detection algorithm from Lemma~\ref{lem: sourceDetectAlgo}. Then we observe that the collection of sets $\{A_i(v,q,G)\}_{v \in V}$ can be viewed as an instance of the minimum hitting set problem over the set (universe) $A_i$, i.e., we want to find a set $A_{i+1} \subseteq A_i$ of minimum size such that each set $A_i(v,q,G)$ in the collection contains at least one node of $A_{i+1}$. We construct $A_{i+1}$ by invoking the deterministic greedy algorithm from Lemma~\ref{lem: hittingSetAlgo}, which produces a hitting set whose size is within $O(\log n)$ of the optimum one. We next prove the constructed hierarchy produces bunches whose sizes are comparable to the randomized construction, and also show that our deterministic construction can be implemented efficiently.

\begin{algorithm2e}[t!]
\label{alg: DeterministicHieararhcy}
\caption{\textsc{DetHierarhcy}$(G,r)$}
\Input{Undirected, weighted graph $G=(V,E)$, parameter $r \geq 1$}
\Output{Hierarchy of sets $(A_i)_{0 \leq i \leq r}$}
\BlankLine
$q \gets \lceil n^{1/r} (1+ \ln n) \rceil $\\
$A_0 \gets V$ ; $A_r \gets \emptyset$ \\
\label{step: ForLoop} \For{$i \gets 0$ to $r-2$} 
{
	Compute $A_i(v,q,G)$ for each $v \in V$ using the source detection algorithm~(Lemma~\ref{lem: sourceDetectAlgo}) \\
	Let $\{A_i(v,q,G)\}_{v \in V}$ be the resulting collection of sets \\
	Compute a hitting set $A_{i+1} \subseteq A_{i}$ with respect to $\{A_i(v,q,G)\}_{v \in V}$~(Lemma~\ref{lem: hittingSetAlgo}) \\
}

\Return $(A_i)_{0 \leq i \leq r}$
\end{algorithm2e}

\begin{lemma} \label{lem: deterministicHierarhcy} Given an undirected, weighted graph $G=(V,E)$, and a parameter $r \geq 1$, Algorithm~\ref{alg: DeterministicHieararhcy} computes deterministically, in $O(rmn^{1/r} \log n)$ time, a hierarchy of sets $(A_i)_{0 \leq i \leq r}$ such that for each $v \in V$,
\[
	|B(v)| = O(rn^{1/r} \log n).
\]
\end{lemma}
\begin{proof}
We start by showing the bound on the size of the bunches. To this end, we first prove by induction on $i$ that $|A_{i}| \leq n^{1-i/r}$ for all $0 \leq i \leq r-1$. For the base case, i.e., $i = 0$, the claim is true by construction since $A_0 = V$. We assume that $|A_{i}| \leq n^{1-i/r}$ for the induction hypothesis, and show that $|A_{i+1}| \leq n^{1-(i+1)/r}$ for the induction step. Note that by construction each set in the collection $\{A_i(v,q,G)\}_{i \in V}$ has size $q = \lceil n^{1/r} (1+\ln n) \rceil \geq n^{1/r} (1+\ln n)$. Invoking the greedy algorithm from Lemma~\ref{lem: hittingSetAlgo}, we get a hitting set $A_{i+1} \subseteq A_i$ of size at most 
\[
  \left(\frac{|A_i|}{q}\right) (1+\ln n)  \leq \left(\frac{n^{1-i/r}}{n^{1/r} (1+\ln n)} \right) (1+\ln n) = n^{1-(i+1)/r}.
\]

We next show that for each $v \in V$ and for each $0 \leq i \leq r-1$, $|B_i(v)| \leq O(n^{1/p} \log n)$, which in turn implies the claimed bound on the size of vertex bunches. Note that it suffices to show that $B_i(v) \subseteq A_i(v,q,G)$ since then $|B_i(v)| \leq |A_i(v,q,G)| \leq n^{1/p} (1+\ln n) = O(n^{1/p} \log n)$. Recall that for $1 \leq i \leq r-1$
\[
	B_i(v) = \{w \in A_i \setminus A_{i+1} \mid \dist_G(w,v) < \dist_G(A_{i+1},v) \}
\]

Now, by construction of $A_{i+1}$ we have that $A_{i+1} \cap A_i(v,q,G) \neq \emptyset$, which implies that $B_i(v) \subseteq A_i(v,q,G)$ by the definition of $B_i(v)$.

We finally analyze the running time. For $0 \leq i \leq r-2$, consider the sequence of steps in the $i$-th iteration of the \textbf{for} loop in Figure~\ref{alg: DeterministicHieararhcy}. By Lemma~\ref{lem: sourceDetectAlgo}, the time to construct the collection of sets $\{A_i(v,q,G)\}_{v \in V}$ is $O(mn^{1/r} \log n)$. Furthermore, since the size of each set in this collection is at least $q=O(n^{1/r} \log n)$, Lemma~\ref{lem: hittingSetAlgo} guarantees that the greedy algorithm for computing a hitting set $A_{i+1}$ takes $O(n^{1+1/r} \log n)$ time. Combining the above bounds, we get that the total time for the $i$-th iteration is $O(mn^{1/r} \log n)$. Since there are at most $r$ iterations, we conclude that the running time of the algorithm is $O(rmn^{1/r} \log n)$.
\end{proof}

We now have all the necessary tools to prove Lemma~\ref{lem: efficientDistanceLocalSparsifier}.

\begin{proof}[Proof of Lemma~\ref{lem: efficientDistanceLocalSparsifier}]
We first show how to implement the two operations of the efficient local distance sparsifier $(H,2r-1)$, and then analyze their running time. 

In the preprocessing phase, depicted in Figure~\ref{alg: PreprocessEmeregency}, given the graph $G$ and the stretch parameter $(2r-1)$,  we first invoke \textsc{HierarchyConstruct$(G,r)$} in Figure~\ref{alg: randomizedPreprocessing}, where Steps 1-3 are replaced by the deterministic algorithm for computing the hierarchy of sets \textsc{DetHierarchy$(G,r)$}. Note that this modification ensures that our preprocessing algorithm is deterministic. Next, for each vertex $v \in V$, we store its bunch $B(v)$ in a balanced binary search tree, where each vertex $w \in B(v)$ has as key the value $\dist_G(w,v)$~(this step could be implemented differently, but as we will shortly see, it will be useful in the subsequent applications of our algorithm). 

We next describe how to implement the query operation, depicted in Figure~\ref{alg: QueryDistanceSparsifier}. Let $K$ be the set of queried terminals. The main idea to construct a vertex distance sparsifier $H[K]$ of $G$ with respect to $K$ is to exploit the bunches that we stored in the preprocessing step. More concretely, let $H[K]$ be an initially empty graph. For each vertex $v \in K$, and every vertex in its bunch $u \in B(v)$, we add to $H[K]$ the edge $(u,v)$ with weight $\dist_G(u,v)$. To show that the resulting graph $H[K]$ is indeed a vertex distance sparsifier with respect to $K$, we briefly review the query algorithm in the construction of Thorup and Zwick~\cite{ThorupZ05}, and show that this immediately applies to our graph setting.

\begin{algorithm2e}[t!]
\caption{\textsc{Preprocess}$(G, 2r-1)$}
\label{alg: PreprocessEmeregency}
Invoke \textsc{HierarchyConstruct$(G,r)$}, where instead of Steps 1-3 invoke \textsc{DetHierarchy$(G,r)$} \\
\For{each $v \in V$}
{ Store each $B(v)$, where $w \in B(v)$ holds $\dist_G(v,w)$.
}
\end{algorithm2e}

\begin{algorithm2e}[t!]
\caption{\textsc{QuerySparsifier}$(G,K)$}
\label{alg: QueryDistanceSparsifier}
Set $H[K] \gets \emptyset$\\
\For{each $v \in K$}
{
    \For{every $u \in B(v)$}
    {
    	Add $(v,u)$ to $E(H[K])$ with weight $\dist_G(v,u)$
    }
}
\Return $H[K]$
\end{algorithm2e}

Let $u,v \in K$ by any two terminals. The algorithm uses the variables $w$ and $i$, and starts by setting $w \gets u$, and $i \gets 0$. Then it repeatedly increments the value of $i$, swaps $u$ and $v$, and sets $w \gets p_i(u) \in B(u)$, until $w \in B(v)$. Finally, it returns a distance estimate $\delta(u,v) = \dist_G(w,u) + \dist_G(w,v)$. Observe that $w = p_i(u) \in B(u)$ for some $0 \leq i \leq r-1$ and $w \in B(v)$. By construction of our vertex sparsifier $H[K]$, note that the edges $(w,u)$ and $(w,v)$, and their corresponding weights, $\dist_G(w,u)$ and $\dist_G(w,v)$, are added to $H[K]$. Thus, there must exist a path between $u$ and $v$ in $H[K]$ whose stretch is at most the stretch of the distance estimate $\delta(u,v)$. Since in \cite{ThorupZ05} it was shown that for every $u, v \in K$,
\[ \dist_G(u,v) \leq \delta(u,v) \leq (2r-1) \dist_G(u,v), \]
we immediately get that
\[
	\dist_G(u,v) \leq \dist_{H[K]}(u,v) \leq (2r-1) \dist_G(u,v).
\]

We finally analyze the running time for both operations. First, note that by Lemma~\ref{lem: deterministicHierarhcy}, the deterministic algorithm for constructing the hierarchy of sets \textsc{DetHierarhcy$(G,r)$} runs in $O(rmn^{1/r}\log n)$ time. Moreover, Thorup and Zwick~\cite{ThorupZ05} showed that given a hierarchy of sets, the bunches for all vertices in $G$ can be computed in $O(rmn^{1/r} \log n)$ time. Combining these two bounds we get that the operation \textsc{Preprocess$(G,r)$} runs in $O(rmn^{1/r}\log n) = \tilde{O}(mn^{1/r})$ time. For the running time of \textsc{QuerySparsifier$(G,K)$}, recall that $H[K]$ consists of the union over all bunches of terminal vertices in $K$. Since the size of a each individual vertex bunch is bounded by $O(rn^{1/r} \log n)$~(Lemma~\ref{lem: deterministicHierarhcy}), we get that the size of $H[K]$ is bounded by $O(|K| r n^{1/r} \log n) = \tilde{O}(|K| n^{1/r})$. The latter also bounds the time to output $H[K]$.
\end{proof}

We next show that local sparsifiers for distances are closed under transitivity and decomposition. While transitivity follows directly from the definition, for the sake of completeness we include the proof for decomposability.

\begin{lemma}[Transitivity] \label{lem: transitivityEmulator}
If $H_1$ is an $\alpha_1$-local sparsifier of $G$, and $H_2$ is a $\alpha_2$-beta local sparsifier of $H_1$, then $H_2$ is an $\alpha_1 \alpha_2$-local sparsifier of $G$.
\end{lemma}

\begin{lemma}[Decomposability] \label{lem: decomposabilityEmulator}
Let $G=(V,E)$ be an undirected, weighted graph, let $E_1,E_2$ be a partition of the edge set $E$, and let $H_i$ be an $\alpha_i$-local sparsifier of $G_i=(V,E_i)$, for each $1 \leq i \leq 2$. Then $H = H_1 \cup H_2$ is a $\max\{\alpha_1,\alpha_2\}$-local sparsifier of $H$.
\end{lemma}
\begin{proof} 
Let $u,v \in V$ be an arbitrary pair of vertices, and let $P_G(u,v)$ be the shortest path distance of weight $\dist_G(u,v)$ between $u$ and $v$ in $G$. Moreover, for each $1 \leq i \leq 2$  let $P_{G_i}(u,v)$ be the edges of $P_G(u,v)$ that belong to $E_i$. Since $H_i$ is an $\alpha_i$-sparsifier of $G_i$, for each $1 \leq i \leq r$, we know that for each edge $e \in P_{G_i}(u,v)$, there exists a path $P_{H_i}(e)$ in $H_i$ such that 
\begin{equation} \label{eq: edgeStretch} \ww(e) \leq \ww(P_{H_i}(e)) \leq \alpha_i \cdot \ww(e). 
\end{equation}
Define 
\[ \ww(P_{H}(u,v)) := \sum_{1 \leq i \leq 2}\sum_{e \in P_{G_i}(u,v)} \ww(P_{H_i}(e)),
\]
to be a path between $u$ and $v$ in $H$. We next show that weight of this path dominates as well as stretches $\dist_G(u,v)$ within a $\max\{\alpha_1,\alpha_2\}$ factor. 

Indeed, repeatedly applying Equation~\ref{eq: edgeStretch} we have that 
\begin{align*}
	\ww(P_{H}(u,v)) & = \sum_{1 \leq i \leq 2}\sum_{e \in P_{G_i}(u,v)} \ww(P_{H_i}(e))  
	 \geq \sum_{1 \leq i \leq 2} \sum_{e \in P_{G_i}(u,v)}  \ww(e) \\ 
	 & = \ww(P_G(u,v)) = \dist_G(u,v),
\end{align*}
and,
\begin{align*}
	\ww(P_{H}(u,v)) & = \sum_{1 \leq i \leq 2}\sum_{e \in P_{G_i}(u,v)} \ww(P_{H_i}(e)) 
	\leq \max\{\alpha_1,\alpha_2\} \cdot  \sum_{1 \leq i \leq 2}  \sum_{e \in P_{G_i}(u,v)}  \ww(e) \\ 
	 & = \max\{\alpha_1,\alpha_2\} \cdot \ww(P_G(u,v)) = \max\{\alpha_1,\alpha_2\} \cdot  \dist_G(u,v). \qedhere
\end{align*}
\end{proof}

We now have all the necessary tools to prove Theorem~\ref{thm: IncrementalApproximateAPSP}.
\begin{proof}[Proof of Theorem~\ref{thm: IncrementalApproximateAPSP}]
Let $(H,2r-1,\tilde{O}(n^{1/r}),\tilde{O}(n^{1/r}))$ be an efficient distance local sparsifier of $G$~(Lemma~\ref{lem: efficientDistanceLocalSparsifier}), which is closed under transitivity and decomposition~(Lemmas~\ref{lem: transitivityEmulator} and~\ref{lem: decomposabilityEmulator}). Plugging the parameters $\alpha = (2r-1)$, $f(n) = \tilde{O}(n^{1/r})$, $g(n) = \tilde{O}(n^{1/r})$, $h(n)=1$ into Theorem~\ref{thm: metaTheorem} we get an incremental algorithm such that for any pair of vertices $u$ and $v$ it reports a query estimate $\delta(u,v)$ with
\[
	\dist_G(u,v) \leq \delta(u,v) \leq (2r-1)^{\ell} \dist_G(u,v),
\]
and handles update and query operations in worst-case time of
\[
	\tilde{O}\left( \left(\sum_{j=1}^{\ell} \frac{\beta_{j-1}}{\beta_j} + \beta_{\ell} \right) n^{2/r} \right), \quad \text{where } \beta_0 = m.
\]

Note that the choice of parameters $\{\beta\}_{0 \leq i \leq \ell}$ does not depend on the factor $n^{2/r}$. Therefore, by ignoring this and applying Lemma~\ref{lem: optimalTradeoff} we get that there exists a choice of parameters $\{\beta\}_{0 \leq i \leq \ell}$ such that
\[
	\tilde{O} \left( \left(\sum_{j=1}^{\ell} \frac{\beta_{j-1}}{\beta_j} + \beta_{\ell} \right) n^{2/r} \right) = \tilde{O}\left( m^{1/(\ell+1)} n^{2/r} \right) = \tilde{O}(n^{2/(\ell+1)}n^{2/r}).
\]
\end{proof}

\section{Incremental All Pair Max-Flow}
\label{sec:maxflow_inc}

In this section we show how to use our general Theorem~\ref{thm: metaTheorem} to design online incremental algorithms for the approximate All-Pair Max-Flow Problem with fast worst-case update and query time. Concretely, we will show that that assumptions (1) and (2) in Theorem~\ref{thm: metaTheorem} are satisfied with certain parameters for flows. Note that (3) follows immediately by employing the $\tilde{O}(m)$ time approximate $(s,t)$-maximum flow algorithm due to Peng~\cite{Peng16}. We have the following theorem.

\begin{theorem}\label{thm:mincutInc}
Let $G=(V,E)$ be an undirected, weighted graph. For every $\ell \geq 1$, there is an incremental (randomized) approximate \emph{All Pair Max Flow} algorithm that maintains for every pair of nodes $u$ and $v$, a maximum flow estimate $\delta(u,v)$ such that
\[
	\frac{1}{O(\log^{4\ell} n)} \maxflow_G(u,v) \leq \delta(u,v) \leq \maxflow_G(u,v),
\]
with wort-case update and query time of
\[
	\tilde{O}(n^{2/(\ell+1)}).
\] 
\end{theorem}

We start by introducing the usual definitions of sparsifiers and vertex sparsifiers for flows. Having defined these, the definition of local sparsifiers becomes apparent from the general definition we introduced in Section~\ref{sec: localSparsifiers}. Let $G=(V,E)$ be a undirected, weighted graph with a \emph{terminal} set $K \subseteq V$. Let $\dd$ be a demand function over $K$ in $G$ such that $\dd(x,x') = \dd(x',x)$ and $\dd(x,x) = 0$ for all $x,x' \in K$. We denote by $P_{x,x'}$ the set of all paths between $x$ and $x'$ in $G$, for all $x,x' \in K$. Further, for each edge $e \in E$, let $P_e$ be the set of all paths using edge $e$. A \emph{concurrent}~(\emph{multi-commodity}) flow $\ff$ of \emph{congestion} $\lambda$ is function over terminal paths in $G$ such that (1) $\sum_{p \in P_{x,x'}} \ff(p) \geq \dd(x,x')$, for all distinct terminal pairs $x,x' \in K$, and (2) $\sum_{p \in P_e} \ff(p) \leq \lambda c(e)$, for all $e \in E$. We let $\congestion_G(\dd)$ denote the congestion of the concurrent flow that attains the smallest congestion. 

\begin{definition}[Sparsifiers for Flow] Let $G=(V,E)$ be an undirected, weighted graph. A graph $H=(V',E')$ with $V \subseteq V'$ is a \emph{flow sparsifier} of $G$ with quality $\alpha \geq 1$ iff for every demand function $\dd$ among any pair of vertices in $V$
\[
	\congestion_{H}(\dd) \leq \congestion_G(\dd) \leq \alpha \cdot \congestion_{H}(\dd).
\]
\end{definition}

\begin{definition}[Vertex Sparsifiers for Flows] \label{def: flowSparsifiers}
Let $G=(V,E)$ be an undirected, weighted graph with a \emph{terminal} set $K \subset V$. A graph $H=(V,E')$ with $K \subseteq V'$ is a \emph{(vertex) flow sparsifier} of $G$ with \emph{quality} $\alpha \geq 1$ iff for every demand function $\dd$ among any pair of vertices in $K$
\[
	\congestion_{H}(\dd) \leq \congestion_G(\dd) \leq \alpha \cdot \congestion_{H}(\dd).
\]
\end{definition}

We next show that the flow property in graphs admits efficient local sparsifier with desirable guarantees. 

\begin{lemma}[Efficient Flow Local Sparsifiers]\label{lem: emergencyFlowSparsifier}
Given an undirected, weighted graph $G=(V,E)$, there is a randomized algorithm that constructs an \emph{efficient flow} local sparsifier with $O(\log^4 n)$ quality, $\tilde{O}(m)$ preprocessing time, and $\tilde{O}(|K|)$ query time,  where $K$ is any set of queried terminals.	
\end{lemma}

We prove the above lemma by using and slightly extending the fast cut-based decomposition tree due to R\"acke, Shah and T\"aubig~\cite{RackeST14}, and Peng~\cite{Peng16}. We remark that their result is stated only for unweighted graphs, but it easily extends to the weighted case. 


\begin{theorem}[\cite{RackeST14,Peng16}] \label{thm: fastFlowSparsifier}
Given an undirected, weighted graph $G=(V,E)$, there is an $\tilde{O}(m)$ time randomized algorithm $\textsc{FlowSparsify}(G)$ that with high probability computes a flow sparsifier $H=(V',E')$ with $V \subseteq V'$ satisfying the following properties
\begin{enumerate}
\itemsep0em 
\item $H$ is a bounded degree rooted tree
\item $H$ has quality $O(\log^{4} n)$
\item The leaf nodes of $H$ correspond to nodes in $G$,
\item The height of $H$ is at most $O(\log^2 n)$.
\end{enumerate} 
\end{theorem}
\begin{proof}
The original construction of R\"acke et al.~\cite{RackeST14} produces a rooted tree $H'$ which satisifes the above properties, except that $H'$ has unbounded degree and the height of the tree is $O(\log n)$. Since we will exploit the bounded degree assumption in the subsequent applications of our data-structure, here we present a standard reduction from $H'$ to a bounded degree $H$ at the cost of increasing the height of the tree by a logarithmic factor. 

Let $H'$ be the rooted tree we described above. Let $u \in H'$ be an internal node of degree larger than $2$ and let $C(u)$ be its children. We start by removing all edges incident to the children $C(u)$ from $H'$, and record all their corresponding edge weights. Next, we create a bounded degree rooted tree $\tilde{H}$ where the children $C(u)$ are the leaf nodes, i.e., $L(\tilde{H}) = C(u)$, and $u$ is the root of $\tilde{H}$. To complete the construction of $\tilde{H}$ we need to define its edge weights. To this end, for any subtree $R \subseteq \tilde{H}$ let $E(L(R))$ denote the set of edges incident to leaf nodes in $R$. We distinguish the following two cases. (1) If $e=(x,y) \in E(L(\tilde{H}))$ and $x \in L(\tilde{H}) = C(u)$, we set $\ww_{\tilde{H}}(x,y) = \ww_{H'}(x,u)$. (2) If $e=(x,y) \not \in E(L(\tilde{H}))$, then let $\tilde{H}_x$ and $\tilde{H}_y$ be the trees obtained after deleting the edge $e$ from $\tilde{H}$. Further, for any subtree $R \subseteq \tilde{H}$ define
\[ \ww(R) := \sum_{e\in E(L(R))} \ww_{\tilde{H}}(e).\]

Finally, for $e=(x,y) \not \in E(L(\tilde{H}))$ and $e \in \tilde{H}$ we set
\[
	\ww_{\tilde{H}}(x,y) = \min\{\ww(\tilde{H}_x), \ww(\tilde{H}_y)\}.
\]
Note that the weight sums $\ww(\tilde{H}_x)$ and $\ww(\tilde{H}_x)$ can be calculated since we first defined the weights for edges in $E(L(\tilde{H}))$. Also observe that $H'$ remains a tree because we simply removed children of $u$~(which could be viewed as a star) and replaced this by another bounded degree tree $\tilde{H}$. We repeat the above process for every internal node of $H'$ until $H'$ becomes a bounded degree rooted tree, and denote by $H$ the final resulting tree.

We claim that $H$ has depth at most $O(\log^2 {n})$. Recall that the initial height of $H'$ was $O(\log n)$, and every replacement of the star centered at a non-terminal with a bounded degree tree increases the height by an additive of $O(\log n)$. 
Summing up over $O(\log n)$ levels, we get the claimed bound.

Finally, it is easy to see that $H$ is flow sparsifier of quality $1$ for $H'$ with respect to all leaf nodes of $H'$, which in turn correspond to the nodes of graph $G$. Thus, $H$ is also a flow sparsifier for $G$ with quality $O(\log^{4} n)$.
\end{proof}

We now have all the necessary tools to prove Lemma~\ref{lem: emergencyFlowSparsifier}.

\begin{proof}[Proof of Lemma~\ref{lem: emergencyFlowSparsifier}]
We show how to implement the two operations of the efficient local flow sparsifier $(H,O(\log^{4} n))$, argue about its correctness, and then analyze the running time of each operation.

In the preprocessing phase, given a graph $G$, we simply invoke \textsc{FlowSparsify$(G)$} from Theorem~\ref{thm: fastFlowSparsifier} and let $H$ be the resulting sparsifier. For implementing the query operation, let $K$ denote the set of queried terminals. The main idea for constructing a (vertex) flow sparsifier $H[K]$ of $G$ with respect to $K$ is to exploit the fact that $H$ is a tree. Concretely, let $H[K]$ be an initially empty graph. For $v \in K$, let $P(v,r,H)$ be the path between $v$ and $r$ in $H$, where $r$ is the root of $H$~(since $v \in K \subseteq V$, recall that $v$ is a leaf node of $H$ by Property~(3) in Theorem~\ref{thm: fastFlowSparsifier}). For each $v \in K$, and every edge $e \in P(v,r,H)$, we add $e$ with weight $w_H(e)$ to $H[K]$. Finally, we return $H[K]$ as a (vertex) flow sparsifier of $G$ with respect to $K$.

We now argue about the correctness of $H[K]$. First, we show that $H[K]$ is a quality $1$ (vertex) flow sparsifier of $H$ with respect to $K$. To see this, note that since $H$ is a tree, every (multi-commodity) flow among any two leaf vertices $(u,v)$ is routed according to the unique shortest path between between $u$ and $v$ in $H$, denoted by $P(u,v,H)$. Since $H[K]$ is formed taking the union of the paths $P(v,r,H)$, for each $v \in K$, and $P(u,v,H) \subseteq \left(P(v,r,H) \cup P(u,r,H)\right)$, it follows that $P(u,v,H)$ is also contained in $H[K]$. Thus every flow we can route in $H$ among any two pairs in $K$, we can feasible route in $H[K]$. For the next direction, observe that by construction $H[K] \subseteq K$. Therefore, any flow among any two pairs in $K$ that can be feasibly routed in $H[K]$, can also be routed in $H$~(this follows since $H$ has more edges than $H[K]$, and thus the routing in $H$ has more flexibility). Combining the above we get that $H[K]$ is a quality $1$ (vertex) flow sparsifier of $H$. Since $H$ is flow sparsifier of $G$ with quality $O(\log^{4} n)$~(Property~(2) in Theorem~\ref{thm: fastFlowSparsifier}) and $K \subseteq V$, applying transitivity on $H[K]$ and $H$~(which we will shortly prove) we get that $H[K]$ is a quality $O(\log^{4} n)$ (vertex) flow sparsifier of $G$ with respect to $K$.

We finally analyze the running time for both operations. Recall that the operation \textsc{Preprocess$(G)$} is implemented by simply invoking $\textsc{FlowSparsifiy(G)}$. By Theorem~\ref{thm: fastFlowSparsifier}, we know that the latter can be implemented in $\tilde{O}(m)$, which in turn bounds the running time of our preprocessing step. For the running time of $\textsc{QuerySparsifier}(G,K)$, recall that $H[K]$ consists of the union over the paths $P(v,r,H)$, for each $v \in K$. Since the length of each such path is bounded by $O(\log^{2} n)$~(Property~$(4)$ in Theorem~\ref{thm: fastFlowSparsifier}), we get that the size of $H[K]$ is bounded by $O(|K| \log^2 n) = \tilde{O}(|K|)$. Note that after having access to any leaf vertex $v$, the path $P(v,r,H)$ can be retrieved from $H$ in time proportional to its length. This implies that the time to output $H[K]$ is also bounded by $\tilde{O}(|K|)$.
\end{proof}

We next show that local sparsifiers for flows are closed under transitivity and decomposition. While transitivity follows directly from the definition, for the sake of completeness we include the proof for decomposability.

\begin{lemma}[Transitivity]
If $H_1$ is an $\alpha_1$-local sparsifier of $G$, and $H_2$ is an $\alpha_2$-local sparsifier of $H_1$, then $H_2$ is an $\alpha_1 \alpha_2$-local sparsifier of $G$.
\label{lem: transitivityFlowSparsifier}
\end{lemma}

\begin{lemma}[Decomposability] \label{lem: decomposabilityFlowSparsifier}
Let $G=(V,E)$ be an undirected, weighted graph, let $E_1,E_2$ be a partition of the edge set $E$, and let $H_i$ be an $\alpha_i$-local sparsifier of $G_i=(V,E_i)$, for each $1 \leq i \leq 2$. Then $H = H_1 \cup H_2$ is an $\max\{\alpha_1,\alpha_2\}$-local sparsifier of $H$.
\end{lemma}
\begin{proof}


Consider a demand $\dd$ among any pair of vertices $u,v \in V$ that is routable in $G$. Let $\ff$ a (multi-commodity) flow that routes $\dd$, and let $D = \{(p_1, \ff(p_1)), (p_2, \ff(p_2)), \ldots, (p_{\ell}, \ff(p_{\ell}))\}$ be a flow-decomposition, where $p_i$ is a path, and $\ff(p_i)$ is the amount of flow set along this path. Note that a flow path decomposition also specifies a demand since for any $u,v \in V$, $\dd(u,v)= \sum_{ p \in D(u,v)} \ff(p)$, where $D(u,v)$ is all the paths in $D$ whose endpoints are exactly $u$ and $v$. Fix any path $p \in D$, and let $p^{(1)}$ and $p^{(2)}$ be the set of subpaths of $p$ that use only edges from $G_1$ and $G_2$, respectively~(note that $p^{(1)}$ and $p^{(2)}$ partition $p$). Note that the set of paths $p^{(1)}$ and $p^{(2)}$ induce demands in $G_1$ and $G_2$. Taking the union over all paths $p \in D$ will induce demands $\dd_1$ in $G_1$ and $\dd_2$ in $G_2$ with $\dd = \dd_1 + \dd_2$, and these demands are routed among flow paths that lie entirely within $G_1$ or $G_2$. By the definition of flow-sparsifier, these demands are also routable in $H_1$ and $H_2$, and hence the demand $\dd_1 + \dd_2 = \dd$ is routable in $H$.

For the other direction assume that a demand $\dd$ among any pair of vertices $u, v \in V$ is routable in $H$. Similarly to above, let $D$ be the corresponding path decomposition of the flow $\ff$ that routes $\dd$. Fix any path $p \in D$, and let $p^{(1)}$ and $p^{(2)}$ be the set of subpaths of $p$ that use only edges from $H_1$ and $H_2$. Note that $H_1$ and $H_2$ might have extra vertices that do not belong to $V$. However, the endpoints of every path $p'$ belonging to $p^{(1)} \cup p^{(2)}$ must be from $V$. This means that the these paths induced in $H_1$, and $H_2$ are among pairs of vertices in $V$. Thus, taking the union over all paths $p \in D$ will induce demands $\dd_1$ in $H_1$ and $\dd_2$ in $H_2$ with $\dd = \dd_1 + \dd_2$, which are routed among flows path that lie entirely in $H_1$ and $H_2$, respectively. By the definition of flow sparsifiers, these demands routed in $G_1$ and $G_2$ with congestion $\max\{\alpha_1,\alpha_2\}$, respectively. Thus we can also route their sum $\dd = \dd_1 + \dd_2$ with congestion $\max\{\alpha_1,\alpha_2\}$ in $G$.
\end{proof}

We now have all the necessary tools to prove Theorem~\ref{thm:mincutInc}.

\begin{proof}[Proof of Theorem~\ref{thm:mincutInc}]
Let $(H,O(\log^{4} n),\tilde{O}(1),\tilde{O}(1))$ be an efficient flow local sparsifier of $G$~(Lemma~\ref{lem: emergencyFlowSparsifier}), which is closed under transitivity and decomposition~(Lemmas~\ref{lem: transitivityFlowSparsifier} and~\ref{lem: decomposabilityFlowSparsifier}). Plugging the parameters $\alpha = O(\log^{4} n)$, $f(n) = \tilde{O}(1)$, and $g(n) = \tilde{O}(1)$ in Theorem~\ref{thm: metaTheorem} we get an incremental algorithm such that for any pair of vertices $u$ and $v$ it reports a query estimate $\delta(u,v)$ with
\[
	 \frac{1}{\tilde{O}(\log^{4\ell} n)} \maxflow_G(u,v) \leq \delta(u,v) \leq \maxflow_G(u,v),
\]
and handles update and query operations in worst-case time of
\[
	\tilde{O}\left( \left(\sum_{j=1}^{\ell} \frac{\beta_{j-1}}{\beta_j} + \beta_{\ell} \right) \right), \quad \text{where } \beta_0 = m.
\]

Note that the choice of parameters $\{\beta\}_{0 \leq i \leq \ell}$ does not depend on the factor $\poly(\log n))^2$. Therefore, by ignoring this and applying Lemma~\ref{lem: optimalTradeoff} we get that there exists a choice of parameters $\{\beta \}_{0 \leq i \leq \ell}$ such that
\[
	\tilde{O} \left( \left(\sum_{j=1}^{\ell} \frac{\beta_{j-1}}{\beta_j} + \beta_{\ell} \right) \right) = \tilde{O}\left( m^{1/(\ell+1)} \right) = \tilde{O}(n^{2/(\ell+1)}).
\]
\end{proof}



\section{Incremental Tree Flow Sparsifier (R\"acke Tree)}
\label{sec:raecke}

In this section we show that a slightly modified version of the algorithm used to prove Theorem~\ref{thm: metaTheorem} and a few extensions allow us to design a fast incremental algorithm for maintaing a (multi-commodity) flow sparsifier $H$ of a graph $G$ with poly-logarithmic quality. Most importantly $H$ will be a tree graph that satisfies certain interesting properties that we will exploit to maintain other dynamic problems. 

Our extensions build upon the following two main ideas. First, we want to argue that the efficient local sparsifier is a tree. Indeed, observe that the efficient local sparsifier $H$ produced by Lemma~\ref{lem: emergencyFlowSparsifier} produces a tree~(Property (1)), and moreover, by definition of local sparsifiers, the vertex sparsifier $H[K]$ that we query from $H$ with respect to any set of terminals $K$ must also be a tree. Throughout we will refer to $H$ as a \emph{tree flow sparsifier}. Now, recall that in Algorithm~\ref{alg: IncrementalInsert} we have an update rule for rebuilding tree flow sparsifiers. Our goal is to show that under this update rule, the updated sparsifiers still remain trees. We observe that this becomes clear once one formalizes the update process, as shown below.

Let $H$ be a tree flow sparsifier of $G=(V,E)$, let $E_0$ be some set of edges with $V(E_0) \subseteq V$, and let $H[V(E_0)]$ be a (vertex) flow sparsifier of $G$ obtained by querying $H$ with respect to $V(E_0)$. Moreover, let $H'[V(E_0)]$ be a tree flow sparsifier of $H[V(E_0)] \cup E_0$. Then we have that
\[
	H' := \left( H \setminus H[V(E_0)] \right) \cup  H'[V(E_0)]
\]
is indeed a tree flow sparsifier of $G \cup E_0$. 

The second idea we need is to ensure that at any point of time our incremental algorithm maintains a tree flow sparsifier. Note that this is not the case in Algorithm~\ref{alg: IncrementalInsert} since for answering queries~(see Algorithm~\ref{alg: IncrementalQuery}) it was sufficient to consider the sparsifier $H_{\ell-1}$ plus the edge set $E_\ell$. To overcome this, we simply maintain an additional tree flow sparsifier $H_\ell$ at level $\ell$ of the hierarchy, and after each edge insertion we rebuild $H_{\ell}$. Concretely, $H_{\ell}$ is updated by the above rule using the sparsifier $H_{\ell-1}$, the edge set $E_\ell$ and the (vertex) flow sparsifier $H_{\ell-1}[V(E_\ell)]$ that is obtained by querying $H_{\ell-1}$ with respect to $V(E_{\ell})$. This modification gives that $H_{\ell}$ is tree flow sparsifier of $G$ at any point of time at the cost of increasing the quality guarantee by a poly-logarithmic factor but not affecting our running time guarantee.

Combining the above ideas leads to the following theorem.

\begin{theorem}\label{thm: incrementalRackeTree}
Let $G=(V,E)$ be an undirected, weighted graph. For every $\ell \geq 1$, there is an incremental (randomized) algorithm that maintains a \emph{tree flow sparsifier} $H$ of $G$ with quality $O(\log^{8\ell} n)$ and depth $O(\ell \log^2 n)$. The worst-case update time is
$
	\tilde{O}(n^{2/(\ell+1)}).
$
\end{theorem}

\subsection{Applications of tree flow sparsifiers}
We next show how to apply \Cref{thm: incrementalRackeTree} for designing efficient incremental algorithm for cut/flow based problems.

\paragraph{Incremental Maximum Concurrent Flow.}

Recall from \Cref{sec:maxflow_inc} that $\congestion_G(\dd)$ is the congestion of the concurrent flow that attains the smallest congestion among all flows that route demand $\dd$ supported on the terminals $K$. 
Recall from \Cref{tab:problems} that given $k$ demand pairs $\{(s_{i},t_{i},\dd(i))\}_{i=1}^{k}$, the vector $\dd$ will have $O(k)$ non-zero entries.
In the \emph{Maximum Concurrent Flow Problem} the we want to find a flow that minimizes $\congestion_G(\dd)$.

The fastest approximation algorithm for solving the Maximum Concurrent Flow Problem is due to Sherman~\cite{Sherman17}.

\begin{theorem}[\cite{Sherman17}] Let $\varepsilon >0$. Given an undirected, weighted graph $G=(V,E)$ and a demand vector $\dd$ describing $k$ demand pairs, there is an $\tilde{O}(m k)$ algorithm that approximates $\congestion_G(\dd)$ within a $(1+ \varepsilon)$ factor.  
\end{theorem}

In the dynamic version of this problem, we want to construct a data-structure that supports the following operations
\begin{itemize}
\item \textsc{Insert$(u,v)$}: insert the edge $(u,v)$ in the graph, and
\item \textsc{Query$(\dd)$}: return the congestion $\congestion_G(\dd)$ for routing demand $\dd$ in the current graph $G$.
\end{itemize}

Now, given \Cref{thm: incrementalRackeTree}, we just maintain tree flow sparsifier $H$.
Then, given a query $\{(s_{i},t_{i},\dd_{i})\}_{i=1}^{k}$ describing $k$ demand pairs, we do the following. Let $K$ be the terminals including all $s_i$ and $t_i$. Then, we just run Sherman's algorithm on $H[K]$ (which is the union of root-to-leaf paths of all nodes in $K$). This leads to the following corollary.

\begin{corollary}
For every $\ell \geq 1$, there is an incremental (randomized) approximate \emph{Maximum Concurrent Flow} algorithm that maintains for every demand $\dd$ describing $k$ demand pairs an estimate $\delta(\dd)$ such that
\[
	\congestion_G(\dd) \leq \delta(\dd) \leq O(\log^{8\ell} n) \congestion_G(\dd),
\]
with wort-case update of $\tilde{O}(n^{2/(\ell+1)})$ and query time of $\tilde{O}(k^{2}))$. 
\end{corollary}

\paragraph{Uniform sparsest cut and cut oracles.}

Recall that the uniform sparsest cut $\Phi_{G}$ of a weighted graph $G$ is defined as
\[
\ensuremath{\Phi_{G}=\min_{\emptyset\neq S\subset V}\frac{\capacity_G(S, V \setminus S)}{|S| \cdot |V \setminus S|}}
\]
where $\capacity_G(S,V \setminus S)=\sum_{(u,v)\in E,u\in S,v\notin S}\ww(u,v)$.

In the \emph{dynamic uniform sparsest cut} problem, we want to approximate
$\Phi_{G}$ when given a query. In the \emph{dynamic cut oracle} problem,
we want to maintain a data structure such that, given a set of nodes
$S$, we can approximate $\capacity_G(S,V \setminus S)$ in time proportional to $|S|$.

By \Cref{thm: incrementalRackeTree}, the above problems reduce to solving them on a tree that undergoes changes. More importantly, this tree has only polylogarithmic depth. Employing standard techniques for maintaining information on a dynamic tree (e.g. ET tree \cite{HenzingerK99} link/cut tree \cite{SleatorT83}
or top tree \cite{AlstrupHLT05}) leads to the following corollaries.
\begin{corollary}
	For every $\ell \geq 1$, there is an incremental (randomized) $O(\log^{8\ell} n)$-approximate \emph{uniform sparsest cut} algorithm with wort-case update of $\tilde{O}(n^{2/(\ell+1)})$. Given a query, the algorithm returns a $O(\log^{8\ell} n)$-approximation to the uniform sparsest cut in $O(1)$ time. 
\end{corollary}
\begin{corollary}
	For every $\ell \geq 1$, there is an incremental (randomized) \emph{cut oracle} algorithm with wort-case update of $\tilde{O}(n^{2/(\ell+1)})$. Given a set $S$ of nodes, the algorithm returns an $O(\log^{8\ell} n)$-approximation to the size of the cut induced by $S$, i.e. $\capacity_G(S,V \setminus S)$, in time $\tilde{O}(|S|)$. 
\end{corollary}

\section{From Vertex Sparsifiers to Offline Dynamic Algorithms}
\label{sec:meta_offline}
In this section we show how to use efficient vertex sparsifier constructions to design \emph{offline} (approximate) dynamic algorithms for graph problems with certain properties while achieving fast amortized update and query time. To achieve this we use a framework that has been exploited for solving offline $3$-connectivity~\cite{PengSS17}. Our main contribution is to show that this generalizes to a much wider class of problems, leading to several interesting bounds which are not yet known in the \emph{online} dynamic graph literature.  

We start by defining the model. We are given an undirected graph $G=(V,E)$ and an \emph{offline} sequence of events or operations $x_1,\ldots,x_m$, where $x_i$ is ether an edge update~(insertion or deletion), or a query $q_i$ which asks about some graph property in $G$ at time $i$. The goal is to process this sequence of updates in $G$ while spending total time proportional to $O(m f(m))$, where $f(m)$ is ideally some sub-linear function in $m$.

We next show that an analogue to Theorem~\ref{thm: metaTheorem} can also be obtained in the offline graph setting. Our algorithm makes use of the notion of vertex sparsifiers as well as their useful properties including transitivity and decomposability~(see Section~\ref{sec: localSparsifiers}). In our construction we want graph properties that admit (1) fast algorithms for computing vertex sparsifiers and (2) guarantee that the size of such sparsifers is reasonably small. We formalize these requirements in the following definition. 

\begin{definition} \label{def: efficientVertexSparsifier}
Let $G=(V,E)$ be a graph, with a \emph{terminal} set $K \subseteq V$ and let $f(n), g(n) \geq 1$ be functions. We say that $(G',\alpha,f(n),g(n))$ is an $\alpha$-\emph{efficient vertex sparsifier} of $G$ with respect to $K$ iff  $G'$ is an $\alpha$-vertex sparsifier of $G$, the time to construct $G'$ is $O(m \cdot f(n))$, and the size of $G'$ is $O(|K| \cdot g(n))$.
\end{definition}

\begin{theorem} \label{thm: metaTheorem2}
Let $G=(V,E)$ be a graph, and for any $u,v \in V$, let $\mathcal{P}(u,v,G)$ be a solution to a minimization problem between $u$ and $v$ in $G$. Let $f(n),g(n),h(n) \geq 1$ be functions, $\alpha, \ell \geq 1$ be parameters associated with the approximation factor, and let $\beta_0,\beta_1,\ldots, \beta_{\ell}$ with $\beta_0 = m$ be parameters associated with the running time. Assume the following properties are satisfied
\begin{enumerate}
\itemsep0em 
\label{metaThM2: P1}\item $G$ admits an efficient vertex sparsifier $(G',\alpha,f(n),g(n))$,
\label{metaThM2: P2} \item $G'$ is transitive and decomposable,
\label{metaThM2: P3} \item The property $\mathcal{P}(u,v,G)$ can be computed in $O(m h(n))$ time in a graph with $m$ edges and $n$ vertices.
\end{enumerate}
Then there is an offline (approximate) dynamic algorithm that maintains for every pair of nodes $u$ and $v$, an estimate $\delta(u,v)$, such that
\begin{equation} \label{eq: approxMeta2}
	\mathcal{P}(u,v,G) \leq \delta(u,v) \leq \alpha^{\ell} \cdot \mathcal{P}(u,v,G).
\end{equation}
The total time for processing a sequence of $m$ operations is:
\begin{equation} \label{eq: runningTimeMeta2}	
	\tilde{O}\left (\beta_0\left( \sum_{j=1}^{\ell} \left(\frac{\beta_{j-1}}{\beta_j} \right)f(n) + \beta_{\ell} h(n)\right) g(n)  \right) \quad \text{where } \beta_0 = m.
\end{equation}
\end{theorem}

Before describing the underlying data-structure upon which the above theorem builds, we reduce the arbitrary sequence of operations into a more structured one, and also build a particular view for the problem. These will allow us to greatly simplify the presentation. 

Concretely, first we may assume that each edge is inserted and deleted exactly once during the sequence of operations. We achieve this by simply treating each edge instance as a new edge, i.e., we assume that each insertion of an edge $e=(u,v)$ inserts a new edge that is different from all previous instances of $(u,v)$. 

Second, since we are given the entire sequence of operations, for each edge $e$ we associate an interval $[i_e,d_e]$ which indicates the insertion and deletion time of $e$ in the operation sequence. Furthermore, we denote by $q_t$ the time when query $q$ was asked in the operation sequence.  Let $[1,m]$ denote the interval covering the entire event sequence. If we are interested in processing updates from a given interval $[r,s]$, we will define graphs that consists of two types of edges with respect to this interval: 
\begin{enumerate} \itemsep0em 
\item \emph{non-permanent edges}, which are edges affected by an event in this interval, i.e., $E^{p}_{[r,s]} = \{e \mid i_e \text{ or } d_e \in [r,s]\}, $
\item \emph{permanant edges}, which are edges present throughout the entire interval, i.e., $E^{np}_{[r,s]} = \{e \mid i_e < r \leq s < d_e\}.$
\end{enumerate} 
Additionally, it will be useful to define the queried vertex pairs within the interval $[r,s]$: $Q_{[r,s]} = \{q \mid q_t \in [r,s]\}$.


\paragraph*{Data Structure.} We now describe a generic tree data-structure $T$, which allows us to unify our framework and thus greatly simplify the presentation. This tree structure is obtained by hierarchically partitioning the operation sequence into smaller disjoint intervals. These intervals induce graphs that are suitable for applying vertex sparsifiers, which in turn allow us to process updates in a fast way, while paying some error in the accuracy of the query operations. 

Consider some integer parameter $\ell \geq 1$ and parameters $\beta_0, \beta_1, \ldots, \beta_{\ell}$ with $\beta_0 = m$. The tree $T$ has $\ell+1$ levels, where each level $i$ is associated with the parameter $\beta_i$, $i=0,\ldots,\ell$. Each node of the tree stores some interval from the event sequence. Formally, our decomposition tree $T$ satisfies the following properties:
\begin{enumerate}
\itemsep0em 
\item The root of the tree stores the interval $[1,m]$.
\item The intervals stored at nodes of same level are disjoint.
\item Each interval $[r,s]$ stored at a node in $T$ is associated with 
\begin{itemize}
\itemsep0em 
\item a graph $G_{[r,s]} = \left(V,E^{p}_{[r,s]}\right)$,
\item a graph of \emph{new permanent edges} $H_{[r,s]} = G_{[r,s]} \setminus G_{[q,t]}$, where $G_{q,t}$ is the parent of $G_{[r,s]}$ in $T$ (if any). 
\item a set of \emph{boundary vertices} $\partial_{[r,s]} = V(E^{np}_{[r,s]}) \cup V(Q_{[r,s]})$.
\end{itemize}
\item If $[r,s] \subseteq [q,t]$ then it holds that (a) $\partial_{[r,s]} \subseteq \partial_{[q,t]}$, and (b) $E^{p}_{[q,t]} \subseteq E^{p}_{[r,s]}$.
\item The length of the interval stored at a node at level $i$ is $\beta_i$.
\item A node at level $i$ has $\beta_{i}/\beta_{i+1}$ children.
\item The number of nodes at level $i$ is at most $O(\beta_0/\beta_i)$.
\end{enumerate}

The lemma below shows that a decomposition tree can be constructed in time proportional to the length of the operation sequence times the height of the tree.

\begin{lemma} \label{lemm: decomposTree}
Let $G=(V,E)$ be a dynamic graph where the sequence of operations is revealed upfront. Then there is an algorithm that computes the decomposition tree $T$ in $O(\ell m)$ time, where $m$ denotes the length of the operation sequence and $\ell$ is the height of the tree.
\end{lemma}
\begin{proof}
Let $T$ be a tree with a single node (corresponding to its root) that stores the interval $[1,m]$. We augment $T$ in the following natural way: (a) We partition the interval $[1,m]$ into $\beta_0/\beta_1 = m/\beta_1$ disjoint intervals, each of length $\beta_1$. (b) For each of these intervals we create a node in the tree $T$, and connect each node with the root of $T$, i.e., those nodes form the children of the root, and thus the nodes at level $1$ of $T$. (c) We recursively apply steps (a) and (b) to the newly generated nodes until we reach the $(\ell+1)$-st level of the tree. 

By the construction above, it easily follows that the generated tree $T$ satisfies properties (1), (2), (4), (5), (6) and (7). Thus, it remains to show how to compute the quantities in (3). This can be achieved by (a) computing the intervals $[i_e,d_e]$, for every edge $e$ in the sequence (note that this is possible because we assumed that every edge is inserted and deleted exactly once within the interval $[1,m]$), and (2) for each node in the tree, computing the sets $E^{np}_{[r,s]}$ and $E^{p}_{[r,s]}$. 

For the running time, observe that computing the intervals $[i_e,d_e]$ takes $O(m)$ time. Having computed these intervals, we can level-wise compute the permanent and non-permanent edges for each node in that particular level. By disjointedness of the intervals, the amount of work we perform per level is $O(m)$. Since there are most $O(\ell)$ levels, it follows that the running time for constructing the decomposition tree is $O(\ell m)$.
\end{proof}
 
\paragraph*{Computing vertex sparsifiers in the hierarhcy.} We next show how to efficiently compute a vertex sparsifier $G'_{[r,s]}$ for each node $G_{[r,s]}$ from the decomposition tree $T$. The main idea behind this algorithm is to leverage the sparsifier computed at the parent nodes as well as apply the efficient vertex sparsifiers from Theorem~\ref{thm: metaTheorem2}~Part~1. The procedure accomplishing this task for a single node of the tree $T$ is formally given in Algorithm~\ref{alg: vertexNodeSparsiy}. To compute the vertex sparsifier for every node, we simply apply it in a top-down fashion to the nodes of $T$. 

\begin{algorithm2e}[h]
\caption{\textsc{VertexSparsify}$(G_{[r,s]})$}
\label{alg: vertexNodeSparsiy}
\If{$G_{[r,s]}$ is the root node}
{
	$G''_{[r,s]} = G'_{[r,s]} \gets (V,\emptyset)$, i,e, the empty graph.
}
\Else
{	Let $G_{[q,t]}$ be the parent of $G_{[r,s]}$ in $T$ \\
	$G''_{[r,s]} \gets \left(G'_{[q,t]} \cup  H_{[r,s]} \right)$, where $G_{[q,t]}'$ is an efficient vertex sparsifier of $G_{[q,t]}$ with respect to $\partial_{[q,t]}$\\
	Let $G'_{[r,s]}$ be an $\alpha$-efficient vertex sparsifier of $G''_{[r,s]}$ with respect to $\partial_{[r,s]}$~(Theorem~\ref{thm: metaTheorem2}~Part~1)
}
\Return $G'_{[r,s]}$
\end{algorithm2e}

To argue about the usefulness of Algorithm~\ref{alg: vertexNodeSparsiy}, we need to bound the quality of sparsifiers produced at the nodes of $T$. The lemma below show that the quality grows multiplicatively with the number of levels in $T$.

\begin{lemma} \label{lem: qualitySparsifier}
Let $G_{[r,s]}$ be a node of $T$ at level $i \geq 0$ . Then $G'=\textsc{VertexSparsify}(G_{[r,s]})$ outputs an $\alpha^{i}$-efficient vertex sparsifier of $G_{[r,s]}$ with respect to $\partial_{[r,s]}$
\end{lemma}
\begin{proof}
We proceed by induction on $i$. For the base case, i.e., $i = 0$, $G_{[1,m]}$ is the root node. Since $E^{p}_{[1,m]} = \emptyset$ by definition of permanent edges, we get that $G'_{[1,m]} = G_{[1,m]}$, i.e., $G_{[1,m]}$ is a sparsifier of itself. 

Let $G_{[r,s]}$ be a node at level $i > 0$. Let $G_{[q,t]}$ be the parent of $G_{[r,s]}$ in $T$, and let $G'_{[q,t]}$ be its cut sparsifier at level $(i-1)$, as defined in Algorithm~\ref{alg: vertexNodeSparsiy}. By Property (4) of $T$ note that $E^{p}_{[q,t]} \subseteq E^{p}_{[r,s]}$ since $[r,s] \subseteq [q,t]$. Also recall that $H_{[r,s]} = G_{[r,s]} \setminus G_{[q,t]}$. By induction hypothesis, we know that $G'_{[q,t]}$ is an $\alpha^{i-1}$-efficient vertex sparsifier of $G_{[q,t]}$ with respect to $\partial_{[q,t]}$. This together with the decomposability property in Theorem~\ref{thm: metaTheorem2}~Part~2 imply that that $G''_{r,s} = G'_{[q,t]} \cup (G_{[r,s]} \setminus G_{[q,t]})$ is an $\alpha^{i-1}$-efficient vertex sparsifier of $G_{[q,t]} \cup (G_{[r,s]} \setminus G_{[q,t]}) = G_{[r,s]}$ with respect to $\partial_{[q,t]}$. Now, by Theorem~\ref{thm: metaTheorem2}~Part~1 we get that $G'_{[r,s]}$ is an $\alpha$-efficient vertex sparsifier of $G''_{[r,s]}$ with respect to $\partial_{[r,s]}$. Since $\partial_{[r,s]} \subseteq \partial_{[q,t]}$, and applying the transitivity property~(Theorem~\ref{thm: metaTheorem2}~Part~2) on $G'_{[r,s]}$ and $G''_{[r,s]}$, we get that $G'_{[r,s]}$ is an $\alpha^{i-1+1} = \alpha^i$-efficient vertex sparsifier of $G_{[r,s]}$.
\end{proof}

We now state a crucial property of the nodes in the decomposition tree $T$, which allows us to get a reasonable bound on the running time for computing vertex sprasifiers for the nodes in $T$.

\begin{lemma} \label{lem: boundNewPermanentedges}
Let $G_{[r,s]}$ be a node in the decomposition tree $T$, and let $G_{[q,t]}$ be its parent. Then we have that the number of new permanent edges of $G_{[r,s]}$ is bounded by the number of non-permanent edges of its parent, i.e., $|E\left(H_{[r,s]}\right)| \leq |E^{np}_{[q,t]}|$.
\end{lemma}
\begin{proof}
If an edges in in $H_{[r,s]}$, then it is not in $G^{p}_{[r,s]}$, thus it is a non-permanent edge in $G_{[q,t]}$.
\end{proof}

The lemma below gives a bound on the running time for computing vertex sparsifers in $T$.

\begin{lemma} \label{lem: runningTimeForSparsifiers}

The total running time for computing the vertex sparsifiers for each node in the decomposition tree $T$ of height $\ell$ is bounded by 
\[
	\tilde{O} \left(\beta_0 \cdot \left(\sum_{j=1}^{\ell} \frac{\beta_{j-1}}{\beta_j}\right)\right), \quad \text{where }\beta_0 = m.
\]
\end{lemma}
\begin{proof}
For $i \geq 1$, let $Y(i)$ be the total time for computing the vertex sparsifiers for all the nodes in $T$ up to (and including) level $i$. Furthermore, let $Z(i)$ be the total time for computing the vertex sparsifier of the nodes at level $i$ in $Y$~(and excluding other levels). We will show by induction on the number of levels $i$ that $T(i) = O\left(\beta_0 \cdot \left(\sum_{j=1}^{i} \frac{\beta_{j-1}}{\beta{j}}\right)f(n)g(n)\right)$, which with $i=k$ implies the claim we want to prove.

For the base case, i.e., $i=1$, consider any node $G_{[r,s]}$ at level $1$ of $T$. By construction of $T$, $G_{[r,s]}$ contains at most $O(\beta_0)$ permanent edges. Furthermore, note that the parent of $G_{[r,s]}$ is the root node $G_{[1,m]}$, for which $G'_{[1,m]} = (V,\emptyset)$. Thus, by Theorem~\ref{thm: metaTheorem2}~Part~1 we get that the time to compute an efficient vertex sparsifier per node is $O(\beta_0 \cdot f(n))$. By Property (7) of $T$, the number of nodes at level $1$ is $O(\beta_0/\beta_1)$, implying that the total running time is $Y(1) = Z(1) = O\left(\beta_0 \left( \frac{\beta_0}{\beta_1} \right) f(n)\right) = O\left(\beta_0 \left( \frac{\beta_0}{\beta_1} \right) f(n)g(n)\right)$, as desired. 

We next show the inductive step. Let $G_{[r,s]}$ be a node at level $i > 1$, and let $G_{[q,t]}$ be its parent. We want to bound the size of the intermediate graph $G''_{[r,s]} = (G'_{[q,t]} \cup H_{[r,s]})$, as defined in Algorithm~\ref{alg: vertexNodeSparsiy}, which in turn determines the running time for computing an efficient vertex sparsifier of $G_{[r,s]}$. To this end, first observe that Theorem~\ref{thm: metaTheorem2}~Part~1 implies that the size of sparsifier $G'_{[q,t]}$ of $G_{[q,t]}$ is bounded by 
\[ O(|\partial_{[q,t]}| \cdot g(n)) \leq |V(E^{np}_{[r,s]}) \cup V(Q_{[r,s]})| \cdot g(n) \leq O(\beta_{i-1} \cdot g(n)),\]
since the number of non-permanent edges and queries is proportional to the length of the interval being considered. Second, by Lemma~\ref{lem: boundNewPermanentedges}, we also have that $|E(H_{[r,s]})| \leq |E^{np}_{q,t}| \leq O(\beta_{i-1})$, thus giving that $|G''_{[r,s]}| \leq O(\beta_{i-1} \cdot g(n))$. As Algorithm~\ref{alg: vertexNodeSparsiy} runs \textsc{CutSparsify} on the graph $G''_{[r,s]}$, Theorem~\ref{thm: metaTheorem2}~Part~1 gives that the running time to compute an efficient vertex sparsifier for the node $G_{[r,s]}$ is $O(\beta_{i-1} \cdot f(n) g(n))$, and that its size is $O(\beta_{i-1} \cdot g(n))$. Combining this together with the fact that the number of nodes at level $i$ is at most $O(\beta_0/\beta_i)$ (Property (7) of $T$) imply that \[Z(i) = O\left(\beta_0 \cdot \frac{\beta_{i-1}}{\beta_i} f(n) g(n)\right). \] 

To complete the inductive step, note that by induction hypothesis, \[Y(i-1) = O\left(\beta_0 \cdot \left(\sum_{j=1}^{i-1} \frac{\beta_j-1}{\beta_j}\right)f(n)g(n)\right).\] Summing over this and the bound on $Z(i)$ we get
\begin{align*}
	Y(i) & = Y(i-1) + Z(i) \\
	&  = O\left(\beta_0 \cdot \left(\sum_{j=1}^{i-1} \frac{\beta_{j-1}}{\beta_j}\right) f(n)g(n) \right) + O\left(\beta_0 \cdot \left(\frac{\beta_{i-1}}{\beta_i}\right)f(n)g(n)\right) \\
	 & = O\left(\beta_0 \cdot \left(\sum_{j=1}^{i} \frac{\beta_{j-1}}{\beta{j}}f(n)g(n)\right)\right).
\end{align*}
\end{proof}

\paragraph*{Processing operations in the hierarchy.} So far we have shown how to reduce the sequence of operations into smaller intervals in a hierarchical manner, while (approximately) preserving the properties of the edges and queries involved in the offline sequence. In what follows, we observe that for processing these events, it is sufficient to process the nodes (and their corresponding intervals) stored at the last level $\ell$ of the tree decomposition $T$ (note that this is possible because intervals at level $\ell$ form a partitioning of the event sequence $[1,m]$, and all vertex pairs within intervals that will be involved in edge updates or queries are preserved using vertex sparsifiers).



The algorithm for processing the updates is quite simple: for every node $G_{[r,s]}$ at level $\ell$ of $T$, we process all operations in the interval consecutively: for each edge insertion or deletion we add or remove that suitable edges to $G'_{[r,s]}$, and for each query $(x,y)$ we run on the vertex sparsifier $G'_{[r,s]}$ the static algorithm from Theorem~\ref{thm: metaTheorem2}~Part~3 to calculate the property $\mathcal{P}(x,y,G'_{[r,s]})$ between $x$ and $y$ in $G'_{[r,s]}$. (note that this is possible since $\partial_{[r,s]} \supseteq \{x,y\}$ by construction of $T$). 

We next analyze the total time for processing the sequence of events in the last level of $T$.

\begin{lemma} \label{lem: procesingUpdates}
The total time for processing the whole sequence of operations at level $\ell$ of the decomposition tree $T$ is $\tilde{O}(\beta_0 \beta_\ell \cdot g(n) h(n))$, where $\beta_0 = m$.
\end{lemma}
\begin{proof}
As in the worst-case there can be at most $O(\beta_\ell)$ queries within the interval, and since the size of $G'_{[r,s]}$ is also bounded by $O(\beta_\ell g(n))$, by Theorem~\ref{thm: metaTheorem2}~Part~3 it follows that answering all the queries and processing the non-permanent edges within a single interval at level $\ell$ is bounded by $\tilde{O}(\beta_\ell^{2} g(n) h(n))$. Combining this with the fact that the number of nodes at level $\ell$ is $O(\beta_0/\beta_\ell)$ (Property (7) of $T$), we get that the total cost for processing the queries is $\tilde{O}(\beta_0 \beta_\ell \cdot g(n) h(n))$.
\end{proof}

Combining Lemma~\ref{lem: runningTimeForSparsifiers} and Lemma~\ref{lem: procesingUpdates} leads to an overall performance of
\[
	\tilde{O}\left (\beta_0\left( \sum_{j=1}^{\ell} \left(\frac{\beta_{j-1}}{\beta_j} \right)f(n) + \beta_{\ell} h(n)\right) g(n)  \right) \quad \text{where } \beta_0 = m,
\]
which proves the claimed total update time in Theorem~\ref{thm: metaTheorem2}.

We finally prove the correctness of our algorithm. Concretely, we show that the estimate we return when processing any query $(x,y)$ in the last level of the hierarchy approximates the property $\mathcal{P}$ of the graph $G$ up to an $\alpha^{\ell}$ factor, thus proving the claimed estimate in Theorem~\ref{thm: metaTheorem2}.

To this end, let $q_i$ be a query in the sequence of operations $[1,m]$. Since the intervals at level $\ell$ of $T$ form a partitioning of $[1,m]$, there must exist an interval $[r,s]$ that contains the query $q_i$. Let $(x,y)$ be the queried vertex pair of $q_i$. By Lemma~\ref{lem: qualitySparsifier}, we get that the graph $G'_{[r,s]}$ at level $\ell$ is an $\alpha^{\ell}$-vertex sparsifier of $G_{[r,s]}$ with respect to $\partial_{[r,s]}$. Since by construction $\partial_{[r,s]} \supseteq \{x,y\}$, we get that the $G'_{[r,s]}$ approximates the property $\mathcal{P}(x,y,G)$ of $G_{[r,s]}$ up to an $\alpha^{\ell}$ factor. Finally, recall that we run the algorithm from Theorem~\ref{thm: metaTheorem2}~Part~3 on $G'_{[r,s]}$, thus worsening the approximation in the worst-case by at most a constant factor, which yields the claimed bound.



\subsection{Applications to Offline Shortest Paths and Max Flow}
In this section we show how to use our general Theorem~\ref{thm: metaTheorem2} to design offline dynamic algorithms for the approximate All Pair Shortest Paths and All Pair Max Flow with reasonably small total update time. 

We first consider shortest paths. Recall that our goal is to show that assumptions (1), (2) and (3) from Theorem~\ref{thm: metaTheorem2} are satisfied with certain parameters for the shortest path measure. For (1) we make the following observation: given a graph $G$, a subset of terminals $K$, and a parameter $r \geq 1$, we can construct an \emph{efficient} (vertex) distance sparsifier $(H,(2r-1),\tilde{O}(n^{1/r}),\tilde{O}(n^{1/r}))$ by simply constructing an efficient local sparsifier for $G$ using Lemma~\ref{lem: efficientDistanceLocalSparsifier} and querying it with respect to $K$. Also note that assumption (2) is satisfied by the transitivity and decomposability of $H$, and finally recall that (3) follows by any $\tilde{O}(m)$ time single pair shortest path algorithm. These together imply the following result.

\begin{theorem}
\label{thm: offlineShortestPaths}
Let $G=(V,E)$ be an undirected, weighted graph. For every $r,\ell \geq 1$, there is an \emph{offline} fully dynamic approximate \emph{All Pair Shortest Path} algorithm that maintains for every pair of nodes $u$ and $v$, a distance estimate $\delta(u,v)$ such that
\[
	\dist_G(u,v) \leq \delta(u,v) \leq (2r-1)^{\ell} \dist_G(u,v).
\]
The total time for processing a sequence of $m$ operations is
\[
	\tilde{O}(m \cdot m^{1/(\ell+1)}n^{2/r}).
\]
\end{theorem}

We now proceed with max flow. Following essentially the same idea as with shortest paths, we need to show that assumptions (1), (2) and (3) from Theorem~\ref{thm: metaTheorem2} are satisfied with certain parameters for the max flow measure. For (1) we have the following: given a graph $G$, a subset of terminals $K$, we can construct an \emph{efficient} (vertex) flow sparsifier $(H,O(\log^{4} n),\tilde{O}(1),\tilde{O}(1))$ by simply constructing an efficient flow local sparsifier for $G$ using Lemma~\ref{lem: emergencyFlowSparsifier} and querying it with respect to $K$. Also note that assumption (2) is satisfied by the transitivity and decomposability of $H$, and finally recall that (3) follows by employing the $\tilde{O}(m)$ time (approximate) $(s,t)$-maximum flow algorithm due to Peng~\cite{Peng16}. These together imply the following theorem.

\begin{theorem}
\label{thm: offlineMaxFlow}
Let $G=(V,E)$ be an undirected, weighted graph. For every $\ell \geq 1$, there is an \emph{offline} fully dynamic approximate \emph{All Pairs Max Flow} algorithm that maintains for every pair of nodes $u$ and $v$, a flow estimate $\delta(u,v)$ such that
\[
	\frac{1}{\tilde{O}(\log^{4\ell} n)} \maxflow_G(u,v) \leq \delta(u,v) \leq \maxflow_G(u,v).
\]
The total time for processing a sequence of $m$ operations is
\[
	\tilde{O}(m \cdot m^{1/(\ell+1)}).
\]
\end{theorem}

\section{Implications on Hardness of Approximate Dynamic Problems}
\label{sec:hardness}

\subsection{Approximate max flow and cut sparsifiers}

Assuming the OMv conjecture, Dahlgaard \cite{Dahlgaard16} show that
any incremental exact max flow algorithm on undirected graphs must
have amortized update time at least $\Omega(n^{1-o(1)})$. However,
the hardness of approximation is not known\footnote{However, on directed graphs, the hardness of approximation is known.
	This is because even dynamic reachability is hard under several conjectures
	\cite{AbboudW14,HenzingerKNS15}.}: 
\begin{proposition}
	There is no polynomial lower bound for dynamic $\omega(\mbox{polylog}(n))$-approximate
	max flow in the offline setting (and also in the online incremental
	setting).
\end{proposition}
This follows directly from \Cref{thm:mincutInc,thm: offlineMaxFlow}. Thus the important open problem is whether we can prove a hardness
for dynamic $(1+\epsilon)$-approximate max flow algorithms on undirected
graphs for a constant $\epsilon>0$. 

On the other hand, it is not known whether, given
a set of $k$ terminals, there is a $(1+\epsilon)$-approximate cut
(vertex) sparsifier of size $\mbox{poly}(k,1/\epsilon)$ or even $\mbox{poly}(k,1/\epsilon,\log n)$. 
If a cut sparsifier
can only contain terminals as nodes, then the approximation ratio
must be at least $\Omega(\sqrt{\log k}/\log\log k)$~\cite{mm10}. If we need an
exact cut sparsifier, then the size must be at least $2^{\Omega(k)}$~\cite{krauthgamer2017refined}.

In what follows we draw a connection between these two open problems; if there is a very
efficient algorithm for the above cut sparsifier, then there cannot
be a $\Omega(n^{1-o(1)})$ lower bound in the offline setting for the 
dynamic approximate max flow. Moreover, if the cut-sparsifier has
size almost best possible, then there cannot be even a super-polylogarithmic
lower bound. Concretely, we show the following.
\begin{theorem}
	\label{thm:lb max flow}If there is an algorithm that, given a undirected
	graph $G=(V,E)$ with $m$ edges and a set $T\subset V$ of $k$ terminals,
	constructs an $(1+\epsilon)$-approximate cut vertex sparsifier of size $s=\poly(k,1/\epsilon,\log n)$
	in time $O(m\poly(\log n,1/\epsilon))$, there is an offline
	dynamic algorithm for maintaining $(1+\epsilon')$-approximate
	value of max flow with update time $u=O(n^{1-\gamma}\poly(1/\ensuremath{\epsilon}'))$
	for some constant $\gamma>0$. Moreover, if the size of the sparsifier
	$s=k\cdot\poly(1/\epsilon,\log n)$, then we obtain the update
	time of $u=O(\poly(\log n,1/\epsilon'))$. The dynamic algorithm
	is Monte Carlo randomized and it is correct with high probability.\end{theorem}
\begin{proof}
	Let us assume $\epsilon'$ is a constant for simplicity. The proof
	generalizes easily when $\epsilon'$ is not a constant. 
	
	First, we only need to consider offline dynamic algorithms where the underling graph has $m=\tilde{O}(n)$ edges at every time step
	and the length of the update sequences is $n$. This is because there
	is a dynamic algorithm by \cite{AbrahamDKKP16} that can maintain
	a cut sparsifier $H=(V,E')$ of a graph $G=(V,E)$ when the terminal
	set is $V$ with $\tilde{O}(1)$ worst-case update. So we can work
	on $H$ instead, and divide the update sequences into segements of length $n$. If we have an offline dynamic algorithm with update time
	$u$ on average on each period, then the average update time is $\tilde{O}(u)$
	over the whole sequence.
	
	We set $\epsilon=\epsilon'/10\log n$. Suppose that the sparsifier
	from the assumption has size only $s=k\cdot\mbox{poly}(1/\epsilon,\log n)=\tilde{O}(k)$.
	Then, we apply the same proof as in \Cref{thm: offlineMaxFlow},  except that the number
	of levels of the decomposition tree will be $\log n$ instead of $O(\sqrt{\log n})$. The quality of the cut-sparsifier
	at any level is at most $(1+\epsilon)^{\log n}=(1+\epsilon'/10\log n)^{\log n}\le(1+\epsilon')$.
	The total running time will be $\tilde{O}(m^{1+\frac{1}{\log n+1}})=\tilde{O}(n)$.
	The latter implies that update time on average is $O(\mbox{polylog}(n))$.
	
	Assume that $s=k^{c}\cdot\mbox{poly}(1/\epsilon,\log n)=\tilde{O}(k^{c})$
	for some constant $c>1$. Then, we can apply again the same proof from \Cref{thm: offlineMaxFlow}. By using only two levels of the decomposition tree, we can
	obtain an update time of $\tilde{O}(n^{1-\frac{1}{c+1}})$. Concretely, if we set $\beta_{0}=m$ and $\beta_{1}=m^{1/(c+1)}$ then the  time for computing
	the decomposition tree is $\frac{\beta_{0}}{\beta_{1}}\cdot\tilde{O}(\beta_{0})=\tilde{O}(n^{2-\frac{1}{c+1}})$.
	The total time for running approximate max flow on the cut-sparsifier
	in the second level at each step is $\beta_{0}\cdot\tilde{O}(\beta_{1}^{c})=\tilde{O}(n^{2-\frac{1}{c+1}})$. Thus it follows that the update time is $\tilde{O}(n^{1-\frac{1}{c+1}})$ on average.
\end{proof}

\subsection{Approximate distance oracles on general graphs}

There are previous hardness results for approximation algorithms for
dynamic shortest path problems (including single-pair, single-source
and all-pairs problems) \cite{HenzingerKNS15}. All such results show
a very high lower bound, e.g. $\Omega(n^{1-\epsilon})$ or $\Omega(n^{1/2-\epsilon})$
time on an $n$-node graph. However, they hold only when the approximation
factor is a small constant. It is open whether one can obtain weaker
polynomial lower bounds for larger approximation factors. We show
that it is impossible to show super-constant factor lower-bounds in several settings.
\begin{proposition}
	There is no polynomial
	lower bound for dynamic $\omega(1)$-approximate distance oracles
	in the offline setting (and also in the online incremental setting). 
	
	More formally, for any lower bound stating that $\omega(1)$-approximate
	offline dynamic distance oracle algorithm on $n$-node graphs requires
	at least $u(n)$ update time or $q(n)$ query time, then we have $u(n)=n^{o(1)}$
	and $q(n)=n^{o(1)}$. The same holds for online incremental algorithm
	with worst-case update time.
\end{proposition}
This follows directly from \Cref{thm: IncrementalApproximateAPSP,thm: offlineShortestPaths}.

\subsection{Approximate distance oracles on planar graphs}

Similar to the situations above, assuming the APSP conjecture, Abboud
and Dahlgaard \cite{AbboudD16} show that any offline fully dynamic
algorithm for exact distance oracles on planar graph requires either
update time or query time of $\Omega(n^{1/2-o(1)})$. 
We can still hope for a hardness result for $(1+\epsilon)$-approximate distance oracles,
but this remains an important open problem in the field of dynamic
algorithms.

Recall the definition of \emph{distance approximating minors} from Chapter~\ref{cha:ICALP2016_DM}, which are vertex distance sparsifiers that are required to be minors of the input graph. In the exact setting, Krauthgamer et al. \cite{KrauthgamerNZ14} showed that any distance preserving minor with respect to $k$ terminals, even when restricted to planar graphs, must have size $\Omega(k^{2})$ size. Cheung et al.~\cite{cheung2016} showed that for planar graphs there is a $(1+ \epsilon)$-distance approximating minor of size $\tilde{O}(k^{2} \epsilon^{-2})$. 
The natural question is whether there is a $(1+\epsilon)$-approximate
{\em minor} distance sparsifier for $k$ terminals that has size $k^{1.99}\cdot\mbox{poly}(1/\epsilon,\log n)$.

We again draw a connection between dynamic graph algorithms and vertex sparsifiers; if there is a very
efficient algorithm for such distance sparsifiers, then we cannot
extend the $\Omega(n^{1/2-o(1)})$ lower bound to the approximate
setting. Moreover, if the sparsifier has the (almost) best possible size,
then there cannot be even a super-polylogarithmic lower bound. More
precisely, we show the following.
\begin{theorem}
	\label{thm:lb distance}Let $G$ be an undirected graph $G=(V,E)$ with $m$ edges and a set $K\subset V$ of $k$ terminals. If there is an algorithm that constructs a $(1+\epsilon)$-distance approximating minor
	of size $s=k^{2/(1+3\gamma)}\cdot\poly(1/\epsilon,\log n)$,
	for some constant $0<\gamma\le1/3$, in time $O(m\poly(\log n,1/\epsilon))$,
	then there is an offline dynamic $(1+\epsilon')$-approximate distance
	oracle algorithm for with update and query time $u=O(n^{1/2-\gamma/2})$.
	In fact, if the size of the sparsifier is $s=k\cdot\poly(1/\epsilon',\log n)$,
	then we obtain an update and query time of $u=O(\poly(\log n))$.
\end{theorem}
The proof will be very similar to the one in \Cref{thm:lb max flow}
except that we need to be more careful about planarity. Thus we first
proving the following useful lemma.
\begin{lemma}
	Each vertex sparsifier $G'_{[r_{p},s_{p}]}$ corresponding to a node
	in our decomposition tree is planar. \end{lemma}
\begin{proof}
	First, consider a sequence of $H_{[r_{1},s_{1}]},H_{[r_{2},s_{2}]},\dots,H_{[r_{p},s_{p}]}$
	corresponding to a path in the decomposition tree, where $H_{[r_{1},s_{1}]}$
	is a child of the root\footnote{Note that the graph $H_{[r,s]}$ is not defined at the root.},
	and $H_{[r_{i},s_{i}]}$ is a parent of $H_{[r_{i+1},s_{i+1}]}$.
	Observe that $\cup_{1\le i\le p}H_{[r_{i},s_{i}]}=G_{[r_{p},s_{p}]}$
	which is planar. 
	
	From \Cref{alg: vertexNodeSparsiy}, we unfold the recursion and obtain
	that 
	\[
	G'_{[r_{p},s_{p}]}=\cutsp(\cutsp(\ldots)\cup H_{[r_{p-1},s_{p-1}]})\cup H_{[r_{p},s_{p}]}).
	\]
	Note that we omit the second parameter of $\cutsp$ only for readability.
	We assume by induction $G'_{[r_{p-1},s_{p-1}]}=\cutsp(\cutsp(...)\cup H_{[r_{p-1},s_{p-1}]})$
	is planar. We will prove that $G'_{[r_{p},s_{p}]}$ planar. To this
	end, observe that $G'_{[r_{p-1},s_{p-1}]}$ is a minor of $\cup_{1\le i\le p-1}H_{[r_{i},s_{i}]}$.
	Next, we need the following observation.
	\begin{claim}
		\label{claim:preserve planar}Let $G_{1}$ be a minor of $G_{2}$.
		Let $(u,v)$ be an edge such that $u,v\in V(G_{1})\cap V(G_{2})$,
		i.e., the endpoints are nodes of both $G_{1}$ and $G_{2}$. Then,
		$G_{1} \cup \{(u,v)\}$ is a minor of $G_{2} \cup \{(u,v)\}$. In particular, if $G_{2} \cup \{(u,v)\} (u,v)$
		is planar, then so is $G_{1}\cup \{(u,v)\}$.
	\end{claim}

We apply \Cref{claim:preserve planar} where $G_{2}=\cup_{1\le i\le p-1}H_{[r_{i},s_{i}]}$
and $G_{1}=G'_{[r_{p-1},s_{p-1}]}$. As the endpoints of $H_{[r_{i},s_{i}]}$ are in both $G_{1}$ and $G_{2}$ by construction and $G_{2}\cup H_{[r_{p},s_{p}]}=\cup_{1\le i\le p}H_{[r_{i},s_{i}]}$
is planar, then $G_{1}\cup H_{[r_{p},s_{p}]}$ is planar. Finally,
$G'_{[r_{p},s_{p}]}=\cutsp(G_{1}\cup H_{[r_{p},s_{p}]})$ is a minor
of $G_{1}\cup H_{[r_{p},s_{p}]}$, so $G'_{[r_{p},s_{p}]}$ is planar.
\end{proof}
Now, we prove \Cref{thm:lb distance}.
\begin{proof}
	[Proof of \Cref{thm:lb distance}]We first prove the case when $s=k\cdot\mbox{poly}(1/\epsilon,\log n)$.
	We again prove the theorem when $\epsilon'$ is a constant for simplicity.
	Set $\epsilon=\epsilon'/10\log n$. We build the corresponding decomposition
	tree with $\log n$ levels. The quality of the sparsifier at any level
	is at most $(1+\epsilon)^{\log n}=(1+\epsilon'/10\log n)^{\log n}\le(1+\epsilon')$.
	The total running time will be $\tilde{O}(m^{1+\frac{1}{\log n+1}})=\tilde{O}(n)$
	using the same argument as in Lemma~\ref{lem: insertRunningTime}. That is the update
	time on average is $O(\mbox{poly}(\log n))$.
	
	For the case when $s=k^{2/(1+3\gamma)}\cdot\mbox{poly}(1/\epsilon,\log n)$,
	the proof is the same except the parameters need to be carefully chosen.
	We set $\epsilon=\epsilon'\gamma/2$. We choose $\beta_{0}=m=O(n)$,
	$\beta_{1}=n^{(1+\gamma)/2}$, and $\beta_{i+1}=n^{(1+\gamma-2\gamma i)/2}$
	for $i\ge0$. We get that there will be at most $1/\gamma$ levels in the
	decomposition tree and thus the quality at each level is at most 
	\[
	(1+\epsilon)^{1/\gamma}\le e^{\epsilon/\gamma}=e^{\epsilon'/2}\le(1+\epsilon')
	\]
	because $(1+x)\le e^{x}$ for any $x$ and $e^{x/2}\le(1+x)$ for
	$0\le x\le1$. 
	
	For each $i$, the total time to build the sparsifiers in level $i+1$
	by running the algorithm sparsifier at level $i$ is $n/\beta_{i+1}\cdot\tilde{O}(\beta_{i}^{2/(1+3\gamma)})$.
	This is because there are $n/\beta_{i+1}$ many sparsifiers, and the algorithm is applied on a graph of size $\tilde{O}(\beta_{i}^{2/(1+3\gamma)})$.
	By direct calculation we have that  
	\[
	n/\beta_{i+1}\cdot\beta_{i}^{2/(1+3\gamma)}=n^{1-(1+\gamma-2i\gamma)/2+\frac{(1+\gamma-2(i-1)\gamma)}{(1+3\gamma)}}\le n^{1.5-\gamma/2}.
	\]
	To see this, note that $2/(1+3 \gamma) \geq 1$ and consider the following chain of inequalities:
	\begin{align*}
	 \frac{(1+\gamma-2(i-1)\gamma)}{(1+3\gamma)}&-(1+\gamma-2i\gamma)/2\\
	& \le  \frac{1+\gamma}{1+3\gamma}-(i-1)\gamma-\frac{1+\gamma}{2}+i\gamma  \\
	&\le  (1-\gamma)+\gamma-1/2-\gamma/2\\
	& =  1/2-\gamma/2.
	\end{align*}
	It follows that the total time over all levels is $\frac{1}{\gamma}\cdot O(n^{1.5-\gamma/2})$, which is turn implies an average update time of $O(n^{0.5-\gamma/2})$. This completes the proof. 
\end{proof}

\section{Conclusion}
In this chapter, we showed a fast incremental algorithm for approximating all-pairs shortest paths, all-pairs max flow, multi-commodity concurrent flow and uniform sparsest cut. Our algorithmic constructions require poly-logarithmic approximation while achieving sub-linear time for all these problems, except shortest path, for which our approximate ratio improves to constant. The key building block behind our meta algorithm is a new sparsification notion, referred to as a local sparsifier, that generalizes the well-known notion of vertex sparsification. We also systemically study the power of (classic) vertex sparsification in the design of efficient offline dynamic algorithms, where the sequence of updates and queries is given beforehand.

Our work motivates the study of several important research directions. First, an important open problem is whether one can construct efficient local sparsifiers for cuts with constant quality, even when restricted to planar graphs. Recall that our construction uses trees and that there is a lower bound of $\Omega({\log n})$ on the quality when approximating the cut structure of a graph by a tree~\cite{RackeS14}. 

Second, an interesting problem is to construct efficient local sparsifiers for effective resistances. At first, this problem seems promising as there are already near-linear time construction of vertex resistance sparsifiers~\cite{DurfeeKPRS17}, known as approximate Schur complements. However, this construction employs approximate Gaussian elimination and thus it is highly sequential. It is worth investigating whether there are other ways of constructing such sparsifiers that would extend to the local setting.

\chapter[Graph Minors for Preserving Terminal Distances Approximately – Lower and Upper Bounds][Distance Approximating Minors]{Graph Minors for Preserving Terminal Distances Approximately – Lower and Upper Bounds}\label{cha:ICALP2016_DM}

Given a graph where vertices are partitioned into $k$ terminals and non-terminals,
the goal is to compress the graph (i.e., reduce the number of non-terminals) using minor operations while preserving terminal distances approximately.
The distortion of a compressed graph is the maximum multiplicative blow-up of distances between all pairs of terminals.
We study the trade-off between the number of non-terminals and the distortion.
This problem generalizes the Steiner Point Removal (SPR) problem, in which all non-terminals must be removed.

We introduce a novel black-box reduction to convert any lower bound on distortion for the SPR problem
into a super-linear lower bound on the number of non-terminals, with the same distortion, for our problem.
This allows us to show that there exist graphs such that every minor with distortion less than $2$, $5/2$ and $3$
must have $\Omega(k^2)$, $\Omega(k^{5/4})$, and $\Omega(k^{6/5})$ non-terminals, respectively, plus more trade-offs in between.
The black-box reduction has an interesting consequence: if the tight lower bound on distortion for the SPR problem is super-constant,
then allowing any $\calO(k)$ non-terminals will \emph{not} help improving the lower bound to a constant.

We also build on the existing results on spanners, distance oracles and connected 0-extensions
to show a number of upper bounds for general graphs, planar graphs, graphs that exclude a fixed minor and bounded treewidth graphs.
Among others, we show that any graph admits a minor with $\calO(\log k)$ distortion and $\calO(k^{2})$ non-terminals,
and any planar graph admits a minor with $1+\epsilon$ distortion and $\tilde{O}(k^{2} \epsilon^{-2} \log^2 k)$ non-terminals.

\section{Introduction}

\emph{Graph compression} generally describes a transformation of a \emph{large} graph $G$ into a \emph{smaller} graph $H$ that preserves,
either exactly or approximately, certain features (e.g., distance, cut, flow) of $G$.
Its algorithmic value is apparent, since the compressed graph can be computed in a preprocessing step of an algorithm,
so as to reduce subsequent running time and memory.
Some notable examples are graph spanners, distance oracles and cut/flow sparsifiers.

In this chapter, we study compression using minor operations, which has attracted increasing attention in recent years.
Minor operations include vertex/edge deletions and edge contractions.
It is naturally motivated since it preserves certain structural properties of the original graph, e.g., any minor of a planar graph remains planar, while
reducing the size of the graph.
We are interested in \emph{vertex sparsification}, where $G$ has a designated subset $K$ of $k$ vertices called the \emph{terminals},
and the goal is to reduce the number of non-terminals in $H$ while preserving some feature among the terminals.
Recent work in this field studied preserving cuts and flows.
Our focus here is on preserving terminal distances approximately in a multiplicative sense, i.e.,
we want that for any pairs of terminals $u,v \in K$, $\dist_G(u,v) \leq \dist_{H}(u,v) \leq \alpha \cdot \dist_G(u,v)$, for a small \emph{distortion} $\alpha$.
This problem, called {\em Approximate Terminal Distance Preservation (ATDP) problem}, has natural applications in
multicast routing~\cite{ChuRZ2000} and network traffic optimization~\cite{ScharfWZ2015}.
It was also suggested in~\cite{KrauthgamerNZ14} that to solve the \emph{subset travelling salesman problem},
one can compute a compressed minor with a small distortion as a preprocessing step for algorithms
that solve the travelling salesman problem for planar graphs.

ATDP was initiated by Gupta~\cite{Gupta01}, who introduced the related {\em Steiner Point Removal (SPR) problem}:
Given a tree $G$ with both terminals and non-terminals,
output a weighted tree $G'$ {\em with terminals only} which minimizes the distortion.
Gupta gave an algorithm that achieves a distortion of $8$.
Chan et al.~\cite{ChanXKR06} observed that Gupta's algorithm returned always  a minor of $G$.
For general graphs, Kamma et al.~\cite{KammaKN15} gave an algorithm to construct a minor
with distortion $\calO(\log^5 k)$. This bound has been recently improved to $O(\log^{2} k)$ by Cheung~\cite{Cheung18} and finally to $O(\log k)$ by Filtser~\cite{Filtser18}.
Krauthgamer et al.~\cite{KrauthgamerNZ14} studied ATDP and
showed that every graph has a minor with $\calO(k^4)$ non-terminals and distortion $1$.
It is then natural to ask, for different classes of graphs,
what is the trade-off between the distortion and the number of non-terminals.
In this chapter, for different classes of graphs, and with respect to different allowed distortions, we provide lower and upper bounds on the number of non-terminals needed.



\paragraph*{Further Related Work.}

Basu and Gupta~\cite{BasuG2008} showed that for outer-planar graphs,
SPR can be solved with distortion $\calO(1)$.
When randomization is allowed, Englert et al.~\cite{EnglertGKRTT14} showed that for graphs that exclude a fixed minor,
one can construct a randomized minor  for SPR with $\calO(1)$ expected distortion.
It remains open whether similar guarantees can be obtained in the deterministic setting.
%
Krauthgamer et al.~\cite{KrauthgamerNZ14} showed that solving ATDP with distortion $1$ for planar graphs
needs $\Omega(k^2)$ non-terminals.

In the past few years, there has been a considerable amount of work on vertex sparsifiers that preserve cuts~\cite{Moitra09, leighton, charikar, mm10, EnglertGKRTT14, juliasteiner, andoni, RackeST14}. In this setting, the goal is to compress the graph only on the termials while approximately preserving all possible terminal minimum cuts. This problem is closely connected to distance sparsification, and there exist techniques to construct vertex sparsifiers for cuts using distance sparsifiers, and vice versa~\cite{Racke08, EnglertGKRTT14}. 

A related graph compression is spanners, where the objective is to reduce the number of edges by edge deletions only.
We will use a spanner algorithm (e.g.,~\cite{AlthoferDDJS93}) to derive our upper bound results for general graphs.
Although spanner operation enjoys much less freedom than minor operation,
proving a lower bound result for it is notably difficult.
Assuming the Erd\"{o}s girth conjecture~\cite{Erdos1963}, there are lower bounds that match the best known upper bounds,
but the conjecture seems far from being settled \cite{Wenger1991}.
Woodruff~\cite{Woodruff2006} showed a lower bound result bypassing the conjecture, but only for \emph{additive} spanners.



\paragraph*{Our Results.}

For various classes of graphs, we show lower and upper bounds on the number of non-terminals needed
in the minor for low distortion.
The table below summarizes our results. 
\begin{table}
\begin{center}
\begin{tabular}{l|c|c}
Graph &  Upper Bound & Lower Bound   \\ \hline
      &  (distortion, size) & (distortion, size)   \\ \hline  \hline
General &   $(2q-1,O(k^{2+2/q}))$ & $(2-\varepsilon, \Omega(k^2))$   \\
General & $-$ & $(2.5-\varepsilon, \Omega(k^{5/4}))$  \\
 &   & $(3-\varepsilon, \Omega(k^{6/5}))$ \\
 &  & \small{(see Theorem \ref{thm:lower-bound-distortion-25} for more guarantees)}    \\
B.-Treewidth $p$ &  $(2q-1,O(p^{1+2/q}k))$ &  $(1, \Omega(pk))$~\cite{KrauthgamerNZ14}    \\ 
Exc.-Fix.-Minor &  $(O(1),\tilde{O}(k^{2})$  & $-$    \\ 
Planar &  $(1+\varepsilon,\tilde{O}((k/\varepsilon)^{2})$ &  $(1+o(1), \Omega(k^{2}))$~\cite{KrauthgamerNZ14}    \\ \hline
General & $(O(\log k),0)$~\cite{Filtser18} & $-$ \\
Outerplanar & $(O(1),0)$~\cite{BasuG2008} & $-$ \\
Trees & $(8,0)$~\cite{Gupta01} & $(8-o(1),0)$~\cite{ChanXKR06} \\ \hline
General & $(O(\log k),0)$-rand~\cite{EnglertGKRTT14} & $-$ \\
Exc.-Fix.-Minor & $(O(1),0)$-rand~\cite{EnglertGKRTT14} & $(2-o(1),0)$-rand
\\
\end{tabular}
\caption{The results which are \emph{not} followed by a reference are shown in this chapter.
The guarantees with the extension ``-rand'' refer to \emph{randomized} distance approximating minors;
``size'' refers to the number of non-terminals in the minor.}
\end{center}
\end{table}
For our lower bound results, we use a novel black-box reduction to convert any lower bound on distortion for the SPR problem
into a super-linear lower bound on the number of non-terminals for ATDP with the same distortion.
Precisely, we show that given any graph $G^*$ such that solving its SPR problem leads to a minimum distortion of $\alpha$,
we use $G^*$ to construct a new graph $G$ such that every minor of $G$ with distortion less than $\alpha$ must have
at least $\Omega(k^{1+\delta(\Gs)})$ non-terminals, for some constant $\delta(\Gs) > 0$.
The lower bound results in the above table are obtained by using for $G^*$ a complete ternary tree of height $2$,
which was shown that solving its SPR problem leads to minimum distortion $3$~\cite{Gupta01}.
More trade-offs are shown by using for $G^*$ a complete ternary tree of larger heights.

The black-box reduction has an interesting consequence.
For the SPR problem on general graphs, there is a huge gap between the best known lower and upper bounds, which are
$8$~\cite{ChanXKR06} and $\calO(\log k)$~\cite{Filtser18}; it is unclear what the asymptotically tight bound would be.
Our black-box reduction allows us to prove the following result concerning the tight bound:
for general graphs, if the tight bound on distortion for the SPR problem is super-constant,
then for any constant $c>0$, even if $ck$ non-terminals are allowed in the minor, the lower bound will remain super-constant.
See Theorem \ref{thm:spr-vs-lspr} for a formal statement of this result.

We also build on the existing results on spanners, distance oracles and connected 0-extensions to show a number of upper bound results for general graphs, planar graphs and graphs that exclude a fixed minor. Our techniques, combined with an algorithm in Krauthgamer et al.~\cite{KrauthgamerNZ14}, yield an upper bound result for graphs with bounded treewidth. In particular, our upper bound on planar graphs implies that allowing quadratic number of non-terminals, we can construct a deterministic minor with arbitrarily small distortion. 


\section{Preliminaries}

Let $G=(V,E,\ww)$ denote an undirected graph with terminal set $K \subset V$ of cardinality $k$,
where $\ww: E \ra \rr^+$ is the weight~(length) function over edges $E$. 
A graph $H$ is a \emph{minor} of $G$ if $H$ can be obtained from $G$ by performing a sequence of vertex/edge deletions and edge contractions,
but no terminal can be deleted, and no two terminals can be contracted together.
In other words, all terminals in $G$ must be \emph{preserved} in $H$.

Besides the above standard description of minor operations,
there is another equivalent way to construct a minor $H$ from $G$~\cite{KammaKN15},
which will be more convenient for presenting some of our results.
A partial partition of $V(G)$ is a collection of pairwise disjoint subsets of $V(G)$ (but their union can be a proper subset of $V(G)$).
Let $S_1,\cdots,S_m$ be a partial partition of $V(G)$ such that (1) each induced graph $G[S_i]$ is connected,
(2) each terminal belongs to exactly one of these partial partitions, and (3) no two terminals belong to the same partial partition.
Contract the vertices in each $S_i$ into one single ``super-node'' in $H$.
For any vertex $u\in V(G)$, let $S(u)$ denote the partial partition that contains $u$;
for any super-node $u\in V(H)$, let $S(u)$ denote the partial partition that is contracted into $u$.
In $H$, super-nodes $u_1,u_2$ are adjacent \emph{only if} there exists an edge in $G$ with one of its endpoints in $S(u_1)$ and the other in $S(u_2)$.
We denote the super-node that contains terminal $u$ by $u$ as well.

\begin{definition}
The graph $H=(V',E',\ww')$ is an $\alpha$-distance approximating minor (abbr.~$\alpha$\emph{-DAM}) of $G=(V,E,\ww)$
if $H$ is a minor of $G$ and for any $u,v \in K$,  $\dist_G(u,v) \leq \dist_H(u,v) \leq \alpha \cdot \dist_G(u,v)$.
$H$ is an $(\alpha,y)$\emph{-DAM} of $G$ if $H$ is an $\alpha$\emph{-DAM} of $G$ with at most $y$ non-terminals.
\end{definition}

We note that the SPR problem is equivalent to finding an $(\alpha,0)$-DAM.
One can also define a randomized version of distance approximating minor:

\newcommand{\calD}{\eta}

\begin{definition}
Let $\calD$ be a probability distribution over minors of $G=(V,E,\ww)$.
We call $\calD$  an $\alpha$-randomized distance approximating minor (abbr.~$\alpha$\emph{-rDAM}) of $G$
if for any $u,v \in K$,
 \[ \mathbb{E}_{H\sim\calD}\left[\dist_H(u,v)\right] \leq \alpha \cdot \dist_G(u,v), \]
and for every minor $H$ in the support of $\calD$, $\dist_H(u,v) \geq \dist_G(u,v)$.
Furthermore, we call $\calD$ an $(\alpha,y)$\emph{-rDAM} if $\calD$ is an $\alpha$\emph{-rDAM} of $G$,
and every minor in the support of $\calD$ has at most $y$ non-terminals.
\end{definition}

%


\section{Deterministic Lower Bounds}\label{sect:lower-bound}

For all the lower bound results, we use a tool in combinatorial design called \emph{Steiner system}
(or alternatively, \emph{balanced incomplete block design}). Let $[k]$ denote the set $\{1,2,\cdots,k\}$.

\begin{definition}
Given a ground set $K = [k]$, an $(s,2)$-Steiner system (abbr.~$(s,2)$-SS) of $K$ is a collection of $s$-subsets of $K$,
denoted by $\calT = \left\{K_1,\cdots,K_r\right\}$,
where $r = \binom{k}{2}\left/\binom{s}{2}\right.$,
such that every $2$-subset of $K$ is contained in \emph{exactly} one of the $s$-subsets.
\end{definition}

\begin{lemma}[\cite{Wilson75}]\label{lem:SSS-exist}
For any integer $s\geq 2$, there exists an integer $M_s$ such that for every $q\in\nn$,
the set $[M_s + qs(s-1)]$ admits an $(s,2)$-\emph{SS}.
\end{lemma}

Our general strategy is to use the following black-box reduction, which proceeds by taking a \emph{small} connected graph $\Gs$ as input,
and it outputs a \emph{large} graph $G$ which contains many disjoint embeddings of $\Gs$.
Here is how it exactly proceeds:
\begin{itemize}
\item Let $\Gs$ be a graph with $s\geq 2$ terminals and $q\geq 1$ non-terminals.
Let $k$ be an integer, as given in Lemma \ref{lem:SSS-exist}, such that the terminal set $K = [k]$ admits an $(s,2)$-SS $\calT$.
\item We construct $\calT' \subseteq \calT$ that satisfies \emph{certain} property depending on the specific problem.
For each $s$-set in $\calT'$, we add $q$ non-terminals to the $s$-set, which altogether form a \emph{group}.
The union of vertices in all groups is the vertex set of our graph $G$.
We note that each terminal may appear in many groups, but each non-terminal appears in one group only.
\item \emph{Within} each of the groups, we embed $\Gs$ in the natural way.
\end{itemize}

The following two lemmas describe some basic properties of all minors of $G$ output by the black-box above. Before presenting their proofs, we need to introduce some helpful notation. Let $G$ be an output graph from the black-box. In any minor $H$ of $G$, we say a super-node is of Type-A if $S(u)$ contains only non-terminals in $G$;
any other super-node $u$, for which $S(u)$ contains \emph{exactly} one terminal, is of Type-B. Here are two simple facts:
\begin{enumerate}
\item[(a)] If $u$ is of Type-A, since $G[S(u)]$ is connected, the non-terminals in $S(u)$ must belong to the same group.
\item[(b)] If $u$ is of Type-B, let $t$ be the terminal in $S(u)$. If $S(u)$ contains a vertex from some group $R$, then $t\in R$.
\end{enumerate}

%
%
%

\begin{lemma}\label{lem:unique-group-for-edge}
Let $H$ be a minor of $G$. Then for each edge $(u_1,u_2)$ in $H$, there exists exactly one group $R$ in $G$
such that $S(u_1)\cap R$ and $S(u_2)\cap R$ are both non-empty.
\end{lemma}
\begin{proof}
Existence of $R$ is easy to prove by a simple induction on the minor operation sequence that generates $H$ from $G$.
To show the uniqueness, we proceed to a case analysis.
In the first case, either $u_1$ or $u_2$ is of Type-A. Then the uniqueness is trivial by fact (a).

In the second case, both $u_1,u_2$ are of Type-B. For $i=1,2$, let $v_i$ be the terminal in $S(u_i)$.
Suppose there are two groups $R_a,R_b$ that intersect both $S(u_1)$ and $S(u_2)$.
Then by fact (b), $v_1,v_2$ are in both $R_a$ and $R_b$, a contradiction.
\end{proof}
The above lemma permits us to legitimately define the notion $R$-edge:
an edge $(u_1,u_2)$ in $H$ is an $R$-edge if $R$ is the unique group that intersects both $S(u_1)$ and $S(u_2)$.

\begin{lemma}\label{lem:interchange-at-terminal}
Suppose that in a minor $H$ of $G$,
$(u_1,u_2)$ is a $R_1$-edge and $(u_2,u_3)$ is $R_2$-edge, where $R_1\neq R_2$.
Then $R_1$ and $R_2$ intersect, and $S(u_2)$ contains the terminal in $R_1\cap R_2$.
\end{lemma}
\begin{proof}
Since $S(u_2)$ contains vertices from both $R_1$ and $R_2$, $u_2$ must be of Type-B, i.e., $S(u_2)$ contains exactly one terminal $v$.
By fact (b), $v$ is in both $R_1$ and $R_2$.
\end{proof}

We will show that for any minor $H$ with low distortion, at least one of the non-terminals in each group must be retained,
and thus $H$ must have at least $|\calT '|$ non-terminals.
We now present our main theorems on lower bounds and then prove them.

\begin{theorem} \label{thm:lower-bound-star}
For infinitely many $k\in \nn$, there exists a bipartite graph with $k$ terminals
which does not have a $(2-\epsilon,k^2/7)$-\emph{DAM}, for all $\epsilon > 0$.
\end{theorem}

\begin{theorem}\label{thm:lower-bound-distortion-25}
There exists a constant $c_1 > 0$, such that for infinitely many $k\in \nn$,
there exists a quasi-bipartite graph with $k$ terminals
which does not have an $(\alpha-\epsilon,c_1 k^\gamma)$-\emph{DAM}, for all $\epsilon>0$,
where $\alpha,\gamma$ are given in the table below.

\begin{center}
\begin{tabular}{c|c|c|c|c|c|c|c}
$\alpha$ & $2.5$ & $3$ & $10/3$ & $11/3$ & $4$ & $4.2$ & $4.4$ \\
\hline
$\gamma$ & $5/4$ & $6/5$ & $10/9$ & $11/10$ & $12/11$ & $21/20$ & $22/21$\\
\end{tabular}
\end{center}
\end{theorem}



\subsection{Distortion $2$ Lower Bound}

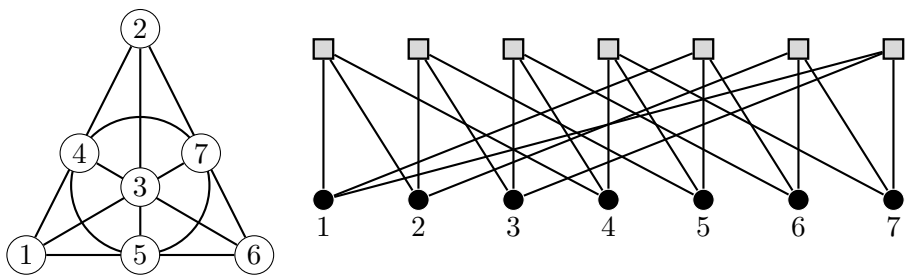
\begin{figure}[t]
\centering
\begin{minipage}{.3\textwidth}
  \centering
\begin{tikzpicture}
\tikzstyle{vertex}=[circle, fill=white, draw = black, minimum size = 6pt, inner sep=2pt]
\tikzstyle{vertex1}=[fill = white, draw = white]

\tikzstyle{edge}=[-,thick ]
\tikzstyle{elipse}=[-,thick ]
	
  \node[vertex] (n1) at (0,0) {$1$} ;
  \node[vertex] (n2) at (3,0)  {$6$} ;
  \node[vertex] (n3) at (1.5,3)  {$2$} ;
  \draw[edge] (n1) -- (n2) ;
  \draw[edge] (n2) -- (n3) ;
  \draw[edge] (n3) -- (n1) ;
  \draw[elipse] (1.5,0.92) ellipse (0.91 and 0.91) ;
  \node[vertex] (n4) at (1.5,0) {$5$};
  \node[vertex] (n5) at (0.7,1.35) {$4$};
  \node[vertex] (n6) at (2.3,1.35) {$7$};

  \draw[edge] (n3) -- (n4) ;
  \draw[edge] (n2) -- (n5) ;
  \draw[edge] (n1) -- (n6) ;
  \node[vertex] (n7) at (1.5,0.9)  {$3$} ;

\end{tikzpicture}
\end{minipage}%
\begin{minipage}{.65\textwidth}
  \centering
\begin{tikzpicture}
\tikzstyle{vertex}=[circle,draw = white, fill=black, minimum size = 8pt, inner sep=2pt]
\tikzstyle{vertex1}=[fill = white, draw = white]

\tikzstyle{edge}=[-,thick]
\tikzstyle{edge1}=[-,thick, gray]
\tikzstyle{vertex2}=[thick, draw = black, fill = gray!30, line width = 0.3mm] 
 
  \node[vertex] (n1) at (0,0) {} ;
  \node[vertex] (n2) at (1.25,0) {} ;
  \node[vertex] (n3) at (2.5,0) {} ;
  \node[vertex] (n4) at (3.75,0) {} ;
  \node[vertex] (n5) at (5,0) {};
  \node[vertex] (n6) at (6.25,0) {};
  \node[vertex] (n7) at (7.5,0) {} ;
  
  \node[vertex2] (n124) at (0,2) {} ;
  \node[vertex2] (n235) at (1.25,2) {} ;
  \node[vertex2] (n346) at (2.5,2) {} ;
  \node[vertex2] (n457) at (3.75,2) {} ;
  \node[vertex2] (n561) at (5,2) {} ;
  \node[vertex2] (n672) at (6.25,2) {} ;
  \node[vertex2] (n713) at (7.5,2) {} ;
  
  \draw[edge] (n124) -- (n1) node[below = 2.5pt] {$1$};
  \draw[edge] (n124) -- (n2) node[below = 2.5pt] {$2$};
  \draw[edge] (n124) -- (n4) node[below = 2.5pt] {$4$};
  
  \draw[edge] (n235) -- (n2) ;
  \draw[edge] (n235) -- (n3) node[below = 2.5pt] {$3$};
  \draw[edge] (n235) -- (n5) node[below = 2.5pt] {$5$};
  
  \draw[edge] (n346) -- (n3) ;
  \draw[edge] (n346) -- (n4) ;
  \draw[edge] (n346) -- (n6) node[below = 2.5pt] {$6$};
  
  \draw[edge] (n457) -- (n4) ;
  \draw[edge] (n457) -- (n5) ;
  \draw[edge] (n457) -- (n7) node[below = 2.5pt] {$7$};
  
  \draw[edge] (n561) -- (n5) ;
  \draw[edge] (n561) -- (n6) ;
  \draw[edge] (n561) -- (n1) ;
  
  \draw[edge] (n672) -- (n6) ;
  \draw[edge] (n672) -- (n7) ;
  \draw[edge] (n672) -- (n2) ;
  
  \draw[edge] (n713) -- (n7) ;
  \draw[edge] (n713) -- (n1) ;
  \draw[edge] (n713) -- (n3) ;
\end{tikzpicture}
\end{minipage}
\caption{On the left side: a Fano plane corresponding to a $(3,2)$-SS with $k=7$. On the right side: the bipartite graph of the Fano plane constructed using our black-box reduction. Numbered vertices are \emph{terminals} while square-shaped vertices are \emph{non-terminals}.}
\end{figure}

We next prove Theorem \ref{thm:lower-bound-star}. Let us start by reviewing the lower bound for SPR problem on stars due to Gupta~\cite{Gupta01}.
 
\begin{lemma}\label{lm: gupta1}
Let $\Gs = (K \cup \{v\}, E)$ be an unweighted star with $k \geq 3$ terminals, in which $v$ is the center of the star.
Then, every edge-weighted graph only on the terminals $K$ with fewer than $\binom{k}{2}$ edges has distortion at least $2$.
\end{lemma}

We construct $G$ using the black-box reduction above. Let $k\in\nn$ be such that the terminals $K=[k]$ admits a $(3,2)$-SS, denoted by $\calT$ (see the figure above).
Here, we set $\calT' = \calT$ and $\Gs$ to be the star with $3$ terminals, as described in Lemma \ref{lm: gupta1}.

By the definition of Steiner system, the shortest path between every pair of terminal $u,v$ in $G$ is unique,
which is the $2$-hop path within the group that contains both terminals, i.e., $\dist_G(u,v) = 2$ for all $u,v \in K$.
Every other simple path between $u,v$ must pass through an extra terminal, so the length of such simple path is at least $4$.



Let $H$ be a minor of $G$. Suppose that the number of non-terminals in $H$ is less than $r$, then there exists a group $R$ in which
its non-terminal is not retained (which means that it is either deleted, or contracted into a terminal in that group).
By Lemma \ref{lm: gupta1}, there exists a pair of terminals in that group
such that every simple path within $R$ (which means a path comprising of $R$-edges only) between the two terminals has length at least $4$.
And every other simple path must pass through an extra terminal (just as in $G$), so again it has length at least $4$.
Thus, the distortion of the two terminals is at least $2$.

Therefore, every $(2-\epsilon)$-DAM of $G$ must have $r > k^{2}/7$ non-terminals.

\subsection{Higher Distortion Lower Bounds}
We now prove Theorem \ref{thm:lower-bound-distortion-25}. Concretely, we will give the proof for the case $\alpha = 2.5$ here, and discuss how to generalize this proof for other distortions.
We will first define the notions of \emph{detouring graph} and \emph{detouring cycle},
and then use them to construct the graph $G$ that allows us to show the lower bound.

\paragraph*{Detouring Graph and Detouring Cycle.}
For any $s\geq 3$, let $k\in\nn$ be such that the terminal set $K = [k]$ admits an $(s,2)$-SS.
Let $\calT = \{K_1,\cdots,K_r\}$ be such an $(s,2)$-SS.
A \emph{detouring graph} has the vertex set $\calT$.
By the definition of Steiner system, $\left|K_i\cap K_j\right|$ is either zero or one.
In the detouring graph, $K_i$ is adjacent to $K_j$ if and only if $\left|K_i\cap K_j\right| = 1$.
Thus, in the detouring graph, it is legitimate to give each edge $(K_i,K_j)$ a \emph{terminal label}, which is the terminal in $K_i\cap K_j$.
A \emph{detouring cycle} is a cycle in the detouring graph such that no two neighboring edges of the cycle have the same terminal label.

\begin{fact} Suppose that two edges in the detouring graph have a common vertex,
and their terminal labels are different, denoted by $u,v$.
Then the common vertex must be an $s$-set in $\calT$ containing both $u,v$.
By the definition of Steiner system, the $s$-set is uniquely determined.
\end{fact}


\begin{claim}\label{claim:no-dcycle}
In the detouring graph, number of detouring cycles of size $\ell\geq 3$ is at most $k^\ell$.
\end{claim}

\begin{proof}
Let $(u_1,\cdots,u_\ell)$ be an $\ell$-tuple, where each entry is a terminal, that represents the terminal labels of a detouring cycle.
By the Fact above, the $\ell$-tuple determines uniquely all the vertices in the detouring cycle.
By trivial counting, the number of possible $\ell$-tuples is at most $k^\ell$,
and hence also the number of detouring cycles of size $\ell$.
\end{proof}


Our key lemma is shows that for any $L\geq 3$, we can retain $\Omega_s(k^{L/(L-1)})$ vertices in the detouring graph,
such that the induced graph on these vertices has \emph{no} detouring cycle of size $L$ or less.

\begin{lemma}\label{lem:large-detouring-graph}
For any integer $L\geq 3$, given a detouring graph with vertex set $\calT = \{K_1,\cdots,K_r\}$,
there exists a subset $\calT '\subset \calT$ of cardinality $\Omega_{s}(k^{L/(L-1)})$
such that the induced graph on $\calT '$ has no detouring cycle of size $L$ or less.
\end{lemma}

\begin{proof}
We choose the subset $\calT '$ by the following randomized algorithm:
\begin{enumerate}
\item Each vertex is picked into $\calT '$ with probability $\delta k^{-(L-2)/(L-1)}$, where $\delta = \delta(s) < 1$ is a positive constant
which we will derive explicitly later.
\item While (there is a detouring cycle of size $L$ or less in the induced graph of $\calT '$)\\
\hspace*{0.3in}Remove a vertex in the detouring cycle from $\calT '$
\end{enumerate}

After Step 1, $\expect{|\calT '|} = r \cdot \delta k^{-(L-2)/(L-1)} \ge \frac{\delta}{2s(s-1)} k^{L/(L-1)}$.
Using Claim \ref{claim:no-dcycle}, the expected number of detouring cycles of size $L$ or less is at most
$$\sum_{\ell=3}^L k^\ell \cdot (\delta k^{-(L-2)/(L-1)})^\ell \leq 2 \delta^3 k^{L/(L-1)}.$$
Thus, the expected number of vertices removed in Step 2 is at most $2 \delta^3 k^{L/(L-1)}$.
Now, choose $\delta = 1/\sqrt{8s(s-1)}$. By the end of the algorithm,
$$\expect{|\calT '|} \geq \frac{\delta}{2s(s-1)} k^{L/(L-1)} - 2 \delta^3 k^{L/(L-1)} = \Omega(k^{L/(L-1)}).\vspace*{-0.2in}$$
\end{proof}

\paragraph*{Construction of $G$ and the proof.}
Recall the black-box reduction.
Let $k$ be an integer such that $K=[k]$ admits a $(9,2)$-SS $\calT$.
By Lemma \ref{lem:large-detouring-graph}, we choose $\calT '$ to be a subset of $\calT$ with $|\calT'| = \Omega(k^{5/4})$,
such that the induced graph on $\calT '$ has no detouring cycle of size $5$ or less.
We choose $\Gs$ to be a complete ternary tree of height $2$, in which the $9$ leaves are the terminals.
For each $K_i\in \calT '$, we add four non-terminals to $K_i$, altogether forming a \emph{group}.

The following lemma is a direct consequence that the induced graph on $\calT'$ has no detouring cycle of size $5$ or less.

\begin{lemma}\label{lem:types-of-simple-paths}
For any two terminals $u,v$ in the same group, let $R$ denote the group.
Then, in any minor $H$ of $G$, every simple path from $u$ to $v$ either comprises of $R$-edges only,
or it comprises of edges from at least $5$ groups other than $R$.
\end{lemma}

\begin{proof}[Proof of Theorem~\ref{thm:lower-bound-distortion-25}]
Let $H$ be a $(2.5-\epsilon)$-DAM of $G$, for some $\epsilon>0$.
Suppose that there exists a group such that all its non-terminals are not retained in $H$.
By \cite{Gupta01}, there exists a pair of terminals $u,v$ in that group such that
every simple path between $u$ and $v$, which comprises of edges of that group only, has length at least $3 \cdot \dist_G(u,v)$.

By Lemma \ref{lem:types-of-simple-paths} and Lemma \ref{lem:interchange-at-terminal},
any other simple path $P$ between $u$ and $v$ passes through at least $4$ other terminals,
say they are $u_a,u_b,u_c,u_d$ in the order of the direction from $u$ to $v$.
We denote this path by $P := u\ra u_a\ra u_b\ra u_c\ra u_d\ra v$, by ignoring the non-terminals along the path.
Between every pair of consecutive terminals in $P$, the length is at least $2$.
Thus, the length of $P$ is at least $10$.
Since $\dist_G(u,v)\leq 4$, the length of $P$ is at least $2.5 \cdot \dist_G(u,v)$.

Thus, the length of \emph{every} simple path from $u$ to $v$ in $H$ is at least $2.5\cdot \dist_G(u,v)$, a contradiction.
Therefore, at least one non-terminal in each group is retained in $H$. As there are $\Omega(k^{5/4})$ groups, we are done.
\end{proof}

For the other results in Theorem \ref{thm:lower-bound-distortion-25},
we follow the above proof almost exactly, with the following modifications.
Set $s=3^h$ for some $h\geq 2$, and set $\Gs$ to be a complete ternary tree with height $h$, in which the leaves are the terminals.
Let $\alpha_h$ be a lower bound on the distortion for the SPR problem on $\Gs$.
Apply Lemma \ref{lem:large-detouring-graph} with some integer $h < L \leq \lceil\alpha_h h\rceil$.\footnote{Any choice of $L$
larger than $\lceil\alpha_h h\rceil$ will not improve the result.}
Following the above proof, attaining a distortion of $\min\left\{\frac{L}{h},\alpha_h\right\}-\epsilon$ needs $\Omega(k^{L/(L-1)})$ non-terminals.

The last puzzle we need is the values of $\alpha_h$.
Chan et al.~\cite{ChanXKR06} considered unweighted complete binary tree with height $h$, and showed that as $h$ tends to infinity, the minimum distortion of SPR problem tends to $8$. However, it is not clear from their proof how the minimum distortion depends on $h$,
which is needed for Theorem \ref{thm:lower-bound-distortion-25}.
In what follows, we use their ideas on unweighted complete ternary trees to derive such a dependence.

\newcommand{\drl}{\text{\textsf{DRL}}}
\newcommand{\calS}{\mathcal{S}}

Let $T_h$ denote a unweighted complete ternary tree of height $h$, where the leaves are the terminals.
Let $\calS_h$ denote the collection of all minors of $T_h$.
For each of its node $u$, let $T(u)$ denote the sub-tree rooted at $u$, and let $t(u)$ denote the terminal which $u$ contracts into.
Denote the root by $r$, and its three children by $x,y,z$.
Without loss of generality, we assume that $r$ is contracted into a terminal $t_r$ in $T(x)$, i.e., $t(r) = t_r$.
Then, let\footnote{Formally speaking,
there can be infinitely many minors (with weights) of $T_h$ with distortion at most $\alpha$,
so we should use $\inf$ instead of $\min$ in the definition.
Yet, for each fixed minor without weight, the standard restriction~\cite[Definition 1.3]{KammaKN15} is the optimal weight assignment.
Since there are only finitely many minors of $T_h$ (without weights), we can replace $\inf$ by $\min$.}
$$\drl(h,\alpha) := \min_{H\in \calS_h,\text{ distortion }\leq \alpha}~~\max_{\text{terminal }t\in T(y)\cup T(z)}~~\dist_H(t_r,t).$$
If there is not such a minor $H$, then $\drl(h,\alpha) = +\infty$ by default.
Note that when $\alpha$ increases, $\drl(h,\alpha)$ decreases.


Let $H\in\calS_h$ be a minor of $T_h$ with distortion $\leq \alpha$.
Let $w$ denote a deepest node in $T(y)\cup T(z)\cup \{r\}$ such that $t(w) = t_r$.
Let $\ell$ be the distance between $r$ and $w$ in $T_h$.
Let $w_1,w_2$ be two children of $w$ which are not in $T(x)$.

Then, by the definition of $\drl$, there exist two terminals $t_1\in T(w_1)$ and $t_2\in T(w_2)$ such that for $i=1,2$,
$\dist_H(t_i,t(w_i)) \geq \drl(h-\ell-1,\alpha)$.
Also, for $i=1,2$, $\dist_H(t(w_i),t_r) \geq \dist_{T_h}(t(w_i),t_r) = 2h$.
Hence,
\begin{align*}
	\dist_H(t_1,t_2) & =  \dist_H(t_1,t(w_1)) + \dist_H(t(w_1),t_r) + \dist_H(t_r,t(w_2)) \\ 
	& \quad + \dist_H(t(w_2),t_2)  \\
	& \geq  2\left[\drl(h-\ell-1,\alpha) + 2h\right] .
\end{align*}


Recall that $\dist_{T_h}(t_1,t_2) = 2(h-\ell)$.
Hence, the distortion w.r.t.~$t_1,t_2$ is at least
$$\frac{\drl(h-\ell-1,\alpha) + 2h}{h-\ell}.$$
This quantity cannot be larger than $\alpha$.

We are ready to give a recurrence relation that bounds $\drl(h,\alpha)$ from below:
\begin{equation}\label{eq:drl-recur}
\drl(h,\alpha) \geq \min_{\ell\in [0,h-1]:~\frac{\drl(h-\ell-1,\alpha) + 2h}{h-\ell} \leq \alpha} \drl(h-\ell-1,\alpha) + 2h,
\end{equation}
while the initial conditions are: $\forall \alpha\geq 1,~\drl(0,\alpha) = 0$, and
$$\drl(1,\alpha) =
\begin{cases}
+\infty, & \text{if }\alpha < 2;\\
2, & \text{if }\alpha \geq 2.
\end{cases}$$

Let $\alpha_h$ denote the minimum distortion of $T_h$.
By letting $\ell$ run over all possible distances between $r$ and $w$, we obtain the following lower bound on $\alpha_h$:
\begin{equation}\label{eq:alpha-recur}
\alpha_h \geq \min_\alpha~~\max\left\{\alpha , \left(\min_{\ell\in [0,h-1]} \frac{\drl(h-\ell-1,\alpha) + 2h}{h-\ell}\right)\right\}.
\end{equation}

We compute the lower bounds in \eqref{eq:drl-recur} and \eqref{eq:alpha-recur} using math software.
In the table below, we give the lower bounds on $\alpha_h$ for $h\in [3,10]$ and $h=1000$.

\smallskip

\begin{center}
\begin{tabular}{c|c|c|c|c|c|c|c}
$h$ & 2 & 3,4 & 5 & 6,7 & 8 & 9,10 & 1000\\
\hline
$\alpha_h$ & 3 & 4 & $22/5$ & $14/3$ & 5 & $26/5 $ & $257/35$\\
\end{tabular}
\end{center}


%

\subsection{Generalizing the Lower Bound and its Implication}

Indeed, we can set $\Gs$ as \emph{any} graph.
In our above proofs we used a tree for $\Gs$  because the only known lower bounds on distortion for the SPR problem are for trees.
If one can find a graph $\Gs$ (either by a mathematical proof, or by computer searches)
such that its distortion for the SPR problem is at least $\alpha$,
applying the black-box reduction with this $\Gs$, and  reusing the above proof show that
there exists a graph $G$ with $k$ terminals such that attaining a distortion of $\alpha-\epsilon$
needs $\Omega(k^{1+\delta(\Gs)})$ non-terminals, for some $\delta(\Gs) > 0$.

\begin{theorem}\label{thm:lower-bound-superlinear}
Let $\Gs$ be a graph with $s$ terminals, and the distance between any two terminals is between $1$ and $\beta$.
Suppose the distortion for the \emph{SPR} problem on $\Gs$ is at least $\alpha$.
Then, for any positive integer $\max\{2,\lceil\beta\rceil\} \leq L\leq \left\lceil \alpha\beta \right\rceil$,
there exists a constant $c_4 := c_4(s) > 0$, such that for infinitely many $k\in \nn$,
there exists a graph with $k$ terminals which does not have a
$\left(\min\left\{L/\beta,\alpha\right\}-\epsilon,c_4 k^{L/(L-1)}\right)$-\emph{DAM}, for all $\epsilon>0$.
\end{theorem}

The above theorem has an interesting consequence. For the SPR problem on general graphs, the best known lower bound is $8$,
while the best known upper bound is $\calO(\log k)$~\cite{Filtser18}.
There is a huge gap between the two bounds, and it is not clear where the tight bound locates in between.
Suppose that the tight lower bound on SPR is super-constant.
Then for any positive constant $\alpha$, there exists a graph $\Gs_\alpha$ with $s(\alpha)$ terminals and some non-terminals, such that the distortion is larger than $\alpha$.
By Theorem \ref{thm:lower-bound-superlinear}, $\Gs_\alpha$ can be used to construct a family of graphs with $k$ terminals,
such that to attain distortion $\alpha$, the number of non-terminals needed is super-linear in $k$.
Recall that in SPR, no non-terminal can be retained. In other words, Theorem \ref{thm:lower-bound-superlinear} implies that:
\emph{if retaining no non-terminal will lead to a super-constant lower bound on distortion,
then having the power of retaining any linear number of non-terminals will not improve the lower bound to a constant}.

\newcommand{\lspr}{\text{\textsf{LSPR}}}

Formally, we define the following generalization of SPR problem.
Let $\lspr_y$ denote the problem that for an input graph with $k$ terminals,
find a DAM with at most $yk$ non-terminals so as to minimize the distortion;
the SPR problem is equivalent to $\lspr_0$.
\begin{theorem}\label{thm:spr-vs-lspr}
For general graphs, \emph{SPR} has super-constant lower bound on distortion
if and only if for any constant $y\geq 0$, \emph{$\lspr_y$} has super-constant lower bound on distortion.
\end{theorem}

\section{Minor Construction for General Graphs}\label{sect:UB-general}
In this section we give  minor constructions that present numerous trade-offs between the distortion and size of DAMs.
Our results are obtained by combining the work of Coppersmith and Elkin~\cite{CoppersmithE06} on sourcewise distance preservers with the well-known notion of spanners.

Given an undirected graph $G=(V,E, \ww)$, we let $\pi_{u,v}$ denote the shortest path  between $u$ and $v$ in $G$.
Without loss of generality, we assume that for any pair of vertices $(u,v)$, the shortest path connecting $u$ and $v$ is \textit{unique}.
This can be achieved by slightly perturbing the original edge lengths of $G$ such that no paths have exactly the same length (see \cite{CoppersmithE06}). The perturbation implies a \emph{consistent} tie-breaking scheme: whenever $\pi$ is chosen as the shortest path, every subpath of $\pi$ is also chosen as the shortest path.

Let $N_G(u)$ denote the vertices incident to $u$ in a graph $G$. We say that two paths $\pi$ and $\pi'$ branch at a vertex $u \in V(\pi) \cap V(\pi')$ iff $|N_{\pi \cup \pi'}(u)| > 2$. We call such a vertex $u$ a \textit{branching} vertex.
Let $\mathcal{P}$ denote the set of shortest paths corresponding to every pair of vertices in $G$. 
We review the following result proved in \cite[Lemma 7.5]{CoppersmithE06}. 
\begin{lemma} \label{lemma: elkin} 
Any pair of shortest paths $\pi, \pi' \in \mathcal{P}$ has at most \emph{two} branching vertices.
\end{lemma}

To simplify our exposition, we introduce the notion of \emph{terminal path covers}.

\begin{definition}[Terminal Path Cover] Given $G = (V,E, \ww)$ with terminals $K$, a set of shortest paths $\mathcal{P}' \subset \mathcal{P}$ is an $(\alpha, f(k))$-terminal path cover (abbr.~$(\alpha, f(k))$-TPc) of $G$ with respect to $K$ if
\begin{enumerate} 
\item $\mathcal{P}'$ covers the terminals, i.e. $K \subseteq V(H)$, where $H = \bigcup_{\pi \in \mathcal{P}'} E(\pi)$, 
\item  $|\mathcal{P}'| \leq f(k)$ and for all $u,v \in K$, $\dist_G(u,v) \leq \dist_H (u,v) \leq \alpha \cdot \dist_G(u,v)$.
\end{enumerate}
\end{definition}
We remark that the endpoints of the shortest paths in $\mathcal{P}'$ are not necessarily terminals. The above definition naturally leads to the following algorithm, which is a slight generalization of the upper-bound technique employed by Krauthgamer et al.~\cite{KrauthgamerNZ14}.

\begin{algorithm2e}
\label{algo: minor_DP}
\caption{\textsc{MinorSparsifier}$(G,K,\mathcal{P}')$}
\Input{Graph $G=(V,E,\ww)$, terminals $K$, $(\alpha, f(k))$-TPc $\mathcal{P'}$ of $G$}
\Output{Distance Approximating Minor $H$ of $G$}
 Set $H \gets \emptyset$ \;
 Add all shortest paths from the path cover $\mathcal{P}'$ to $H$ \;
 
 \While{there exists a degree two non-terminal $v$ incident to edges $(v,u)$ and $(v,w)$} { 
 Contract the edge $(u,v)$ \;
 Set the length of the edge $(u,w)$ to $\dist_G(u,w)$ \;
}
 \Return $H$

\end{algorithm2e}

The following lemma gives an upper bound on the size of the DAM output by Algorithm \ref{algo: minor_DP}. It is an easy generalization of a lemma in \cite[Lemma 2.2]{KrauthgamerNZ14} and we review it here for the sake completeness.

\begin{lemma} \label{lemma: branching}
For a given graph $G = (V,E, \ww)$ with terminals $K \subset V$ and an $(\alpha,f(k))$-\emph{TPc} $\mathcal{P}'$ of $G$, \textsc{MinorSparsifier($G$,$K$,$\mathcal{P}'$)} outputs an $(\alpha,f(k)^2)$-\emph{DAM} of $G$. 
\end{lemma}
\begin{proof}
First, it is clear that the union over paths of $\mathcal{P}' \subset \mathcal{P}$ is a minor of $G$ (this can be alternatively viewed as deleting non-terminals and edges that do not participate in any of the shortest paths in $\mathcal{P}'$). Further, the algorithm performs only edge contractions. Thus, the produced graph $H$ is a minor of $G$. 

Since contracting edges incident to non-terminals of degree two does not affect any distance in $H$, the distortion guarantee follows directly from that of the cover $\mathcal{P}'$. Thus, it only remains to show the bound on the size of $H$.

To this end, consider any two paths $\pi, \pi'$ from $\mathcal{P}'$. From Lemma \ref{lemma: elkin}, we know that $\pi$ and $\pi'$ branch in at most two vertices. Let $u_1$ and $u_2$ denote such vertices. Due to the tie-breaking scheme in $G$, we know that the shortest path $\pi_{u_1,u_2}$ is unique, and thus it must be shared by both $\pi$ and $\pi'$. The latter implies that every vertex in the subpath must have degree degree exactly $2$. Therefore, the only non-terminals in $\pi \cup \pi'$ are vertices $u_1$ and $u_2$, since non-terminals of degree two are removed from the edge contractions performed in the algorithm. 

There are $O (f(k)^2)$ pairs of shortest paths from $\mathcal{P}'$, each having at most $2$ non-terminals. Hence, the number of non-terminals in $H$ is $O (f(k)^2)$.
\end{proof}

A trivial \textit{exact} terminal path cover for any $k$-terminal graph is to take the union of all terminal shortest paths,
which we refer to as the $(1,\calO(k^{2}))$-TPc $\mathcal{P}'$ of $G$.
Krauthgamer et al.~\cite{KrauthgamerNZ14} used this $(1,\calO(k^{2}))$-TPc to construct an $(1,O(k^{4}))$-DAM.
Here, we study the question of whether increasing the distortion slightly allows us to obtain a cover of size $o(k^2)$.
We answer this question positively, by reducing it to the well-known spanner problem.

Let $q \geq 1$ be an integer and let $G = (V,E,\ww)$ be an undirected graph. A $q$-spanner of $G$ is a subgraph $S = (V,E_S, \ww)$ such that
for all $u,v$ in $V$, $\dist_G(u,v) \leq \dist_S(u,v) \leq q \cdot \dist_G(u,v)$. We refer to $q$ and $|E_S|$ as the \textit{stretch} and \textit{size} of spanner $S$, respectively. It is well-known that a simply greedy algorithm achieves the following guarantees.
\begin{lemma}[\cite{AlthoferDDJS93}] \label{lemma: spanner}
Let $q \geq 1$ be an integer. Any graph $G = (V,E,\ww)$ admits a $(2q-1)$-spanner $S$ of size $O(|V|^{1+1/q})$.
\end{lemma}

We use the above lemma as follows. Given a graph $G=(V,E,\ww)$ with terminals $K$, we compute the complete graph $Q_K = (K,\binom{K}{2},d_G|K)$, where $d_G|K$ denotes the distance metric of $G$ restricted to the point set $K$ (In other words, for any pair of terminals $u,v \in K$, the weight of the edge connecting them in $Q_K$ is given by $\ww_{Q_K}(u,v) = d_G(u,v)$). Recall that all shortest paths in $G$ are unique.

Using Lemma \ref{lemma: spanner}, we construct a $(2q-1)$-spanner $S$ of size $O (k^{1+1/q})$ for $Q_K$. Observe that each edge of $S$ corresponds to a unique (terminal) shortest path in $G$ since $S$ is a subgraph of $Q_K$. Thus, the set of shortest paths corresponding to edges of $S$ form a $(2q-1, O (k^{1+1/q}))$-TPc $\mathcal{P}'$ of $G$. Using $\mathcal{P}'$ with Lemma \ref{lemma: branching} gives the following theorem.

\begin{theorem}
Let $q \geq 1$ an integer. Any graph $G=(V,E,\ww)$ with $K \subset V$ admits a $(2q-1,O (k^{2+2/q}))$-\emph{DAM}.
\end{theorem}
We mention two trade-offs from the above theorem. When $q=2$, we get an $(3,O (k^3))$-DAM. When $q=\log k$, we get an $(O(\log k),O(k^2))$-DAM. The above method allows us to have improved guarantees for bounded treewidth graphs. In particular, we prove that any graph $G$ with treewidth at \sloppy most $p$ admits an $(O(\log p),O(p k))$-DAM.

\begin{theorem}\label{thm:treewidth}
Let $q \geq 1$ be an integer. Any graph $G=(V,E,\ww)$ with treewidth at most $p$, terminals $K \subset V$ and $k \geq p$
admits a $(2q-1,O(p^{1+2/q} k))$-\emph{DAM}.
\end{theorem}
\begin{proof}
We crucially exploit the fact that such graphs admit small separators:
given a graph $G$ of bounded treewidth $p$ and any nonnegative vertex weight function $w(\cdot)$, there exists a set $S \subset V(G)$ of at most $p+1$ vertices
whose removal separates the graph into two connected components, $G_1$ and $G_2$, each with $w(V(G_i)) \leq 2/3 w(V(G))$ (see \cite{bodlaender95}).

Krauthgamer et al.~\cite{KrauthgamerNZ14} use the above fact to construct an $(1, \mathcal{O}(p^{3}k))$-DAM for graphs of treewidth at most $p$.
We show that with two modifications, their algorithm can be extended to constructing distance approximating minors. The first modification is Step 2 of the algorithm \textsc{ReduceGraphTW} in \cite{KrauthgamerNZ14}.
For any integer $q\geq 1$, we replace their call to \textsc{ReduceGraphNaive}$(H,K \cup B)$\footnote{We remark that they use $R$ to denote the set of terminals.}
by our procedure \textsc{MinorSparsifier}$(H,K \cup B, \mathcal{P}')$, where $ \mathcal{P}'$ is a $(2q-1, \mathcal{O}(p^{1+1/q}))$-TPc of $G$.

The second modification is a generalization of Lemma 4.2 in \cite{KrauthgamerNZ14}. The main idea is to use the small separator set $S$ to decompose the graph into smaller almost-disjoint graphs $G_1$ and $G_2$, compute their DAMs recursively, and then combine them using the separator $S$ into a DAM of $G$. This implies that the separator $S$ must belong to each $G_i$, i.e. all non-terminal vertices of $S$ must be counted as additional terminals in each $G_i$. Below we give a formal definition of this decomposition/composition process.

Let $G_1 = (V_1, E_1, \ww_1)$ and $G_2 = (V_2,E_2, \ww_2)$ be graphs on disjoint sets of non-terminals, having terminal sets $K_1 = \{s_1,s_2,\ldots,s_{a_1}\}$ and $K_2 = \{t_1,t_2,\ldots,t_{a_2}\}$, respectively. Further, let $\phi(s_i)=t_i$, for all $i=1,\ldots,c$ be an one-to-one correspondence between some subset of $K_1$ and $K_2$ (this correspondence is among the separator vertices). The \textit{$\phi$-merge} (or $2$-sum) of $G_1$ and $G_2$ is the graph $G = (V,E,\ww)$ with terminal set $K = K_1 \cup \{t_{c+1},\ldots,t_{a_2}\}$ formed by identifying the terminals $s_i$ and $t_i$, for all $i=1,\ldots,c$, where $\ww(e) = \min\{\ww_1(e),\ww_2(e)\}$ (assuming infinite length when $\ww_i(e)$ is undefined). We denote this operation by $ G := G_1 \oplus_{\phi} G_2$.

Below we state the main lemma whose proof goes along the lines of \cite[Lemma 4.2]{KrauthgamerNZ14}.

\begin{lemma} Let $G = G_1 \oplus_{\phi} G_2$. For $j=\{1,2\}$, let $H_j$ be an $(\alpha_j,f(a_j))$-\emph{DAM} for $G_j$. Then the graph $H = H_1 \oplus_{\phi} H_2$ is an $(\max\{\alpha_1,\alpha_2\}, f(a_1) + f(a_2))$-\emph{DAM} of G.  
\end{lemma}

In~\cite{KrauthgamerNZ14} it is shown that the size of the minor returned by the algorithm \textsc{ReduceGraphTW} is bounded by the number of leaves the in the recursion tree of the algorithm. Further, they prove that there are at most $\calO(k/p)$ such leaves. Plugging our bounds from the modification of Step 2 along with the above lemma yields our claimed result.
\end{proof}

\section{Minor Construction for Fixed Minor-Free Graphs}
In this section we give improved guarantees for distance approximating minors for special families of graphs. Specifically, we show that graphs that exclude a fixed minor admit an $(O(1),\tilde{O}(k^{2}))$-DAM. This family of graphs includes, among others, planar graphs. 

The reduction to spanner in Section \ref{sect:UB-general} does not consider the structure of $Q_K$, which is inherited from the input graph.
We exploit this structure by employing the randomized Steiner Point Removal Problem, which is equivalent to finding an $(\alpha, 0)$-rDAM. Let us start by reviewing the following result of Englert et al.~\cite{EnglertGKRTT14},
which shows that for graphs that exclude a fixed minor, there exists a randomized minor with constant distortion.
\begin{theorem}[\cite{EnglertGKRTT14}, Theorem 14]
\label{thm2}
Let $\alpha = \calO(1)$. Given a graph that excludes a fixed minor $G = (V,E, \ww)$ with $K \subset V$,
there is a probability distribution $\calD$ over minors $H = (K,E', \ww')$ of $G$,
such that for all $u,v$ in $K$, $\mathbb{E}_{H\sim\calD} [\dist_H(u,v)] \leq \alpha \cdot \dist_G(u,v)$ and for every minor $H$ in the support of $\calD$, $\dist_H(u,v) \geq \dist_G(u,v)$.
\end{theorem}

Given a graph $G$ that excludes a fixed minor, any minor $H$ of $G$ only on the terminals also excludes the same fixed minor. Thus $H$ has $\calO(k)$ edges~\cite{Thomason1984}. This leads to the corollary below.

\begin{corollary}
\label{cor1} Let $\alpha = \calO(1)$. Given a graph that excludes a fixed minor $G = (V,E, \ww)$ with $K \subset V$ and $Q_K$ previously defined,
there exists a probability distribution $\calD$ over subgraphs $H=(K,E', \ww')$ of $Q_K$, each having at most $O(k)$ edges, such that for all
$u,v$ in $K$, $\mathbb{E}_{H\sim\calD}[\dist_H(u,v)] \leq \alpha \cdot \dist_{Q_K}(u,v)$.
\end{corollary}
\begin{proof}
Let $\calD$ be the distribution over minors of $G$ from Theorem \ref{thm2}, then every minor in its support is clearly a subgraph of $Q_K$ with $\calO(k)$ edges. Since during the construction of these minors we may assume that for all $(u,v)$ in E'$,~\ww'(u,v) = \dist_G(u,v)$, the corollary follows.
\end{proof}

\begin{lemma} \label{lemma: planar 1}
Given a graph that excludes a fixed minor $G = (V,E, \ww)$ with $K \subset V$, and $Q_K$ as previously defined, there exists an $\calO(1)$-spanner $S$ of size $O (k \log k)$ for $Q_K$.
\end{lemma}
\begin{proof}
Let $\calD$ be the probability distribution over subgraphs $H$ from Corollary \ref{cor1}. Set $S = \emptyset$.
First, we sample independently $q = 3 \log k$ subgraphs $H_1, \ldots, H_q$ from $\calD$.
We then add the edges from all these subgraphs to the graph $S$, i.e., $E_S = \bigcup_{i=1}^{q} E_{H_i}$.
Fix an edge $(t,t')$ from $Q_K$ and a subgraph $H_i$. By Corollary \ref{cor1} and the Markov inequality,
$\mathbb{P} [\dist_{H_i}(u,v) \geq 2 \alpha \cdot \dist_{Q_K}(u,v)] \leq  2^{-1}$, and hence
\begin{align*}
	\mathbb{P}[\dist_{S}(u,v) \geq 2 \alpha \cdot & d_{Q_K}(u,v)] \\
	&  = \prod_{i=1}^{q} \mathbb{P}[\dist_{H_i}(u,v) \geq 2 \alpha \cdot d_{Q_K}(t,t')]  \leq 2^{-q} = k^{-3}.
\end{align*}
Applying union bound overall all edges from $Q_K$ yields
$$
	\mathbb{P}[\exists (u,v) \in E(Q_K) \text{ with } \dist_{S}(u,v) \geq 2\alpha \cdot \dist_{Q_K}(u,v)] \leq k^{2} \cdot k^{-3} = k^{-1}.
$$
Hence, for all edges $(u,v)$ from $Q_K$, with probability at least $1 - 1/k$, we preserve the shortest path distance between $u$ and $v$
up to a factor of $2 \alpha = \calO(1)$ in $S$. Since $S$ is a subgraph of $Q_K$, this implies that there exists a $\calO(1)$-spanner $S$ of size $\calO(k \log k)$ for $Q_K$.
\end{proof}
Similar to the last section, the set of shortest paths corresponding to edges of $S$ is an $(\calO(1), \calO(k \log k))$-TPc  $\mathcal{P}'$ of $G$.
Using  $\mathcal{P}'$ with Lemma \ref{lemma: branching} gives the following theorem.
\begin{theorem}
Any graph that excludes a fixed minor $G = (V,E, \ww)$ with $K \subset V$ admits an $(O (1),\tilde{O} (k^{2}))$-\emph{DAM}.
\end{theorem}

\section{Minor Construction for Planar Graphs}\label{sect:UB-planar}
In this section, we show that for planar graphs one can improve the constant guarantee bound on the
distortion to $3$ and $1+\epsilon$, respectively,
without affecting the size of the minor. Our work builds on existing techniques used in the context of approximate distance oracles,
thereby bypassing our previous spanner reduction. Both results use essentially the same ideas
and rely heavily on the fact that planar graphs admit separators with special properties.

We say that a graph $G=(V,E,\ww)$ admits a $\lambda$-separator if there exists a set $R \subseteq V$
whose removal partitions $G$ into connected components, each of size at most $\lambda n$, where $1/2 \leq \lambda <1$.
Lipton and Tarjan~\cite{lipton79} showed that every planar graph has a $2/3$-separator $R$ of size $O (\sqrt{n})$.
Later on, Gupta et al.~\cite{guptaKR04} and Thorup~\cite{ThorupJACM04} independently observed that one can modify their construction
to obtain a $2/3$-separator $R$, with the additional property that $R$ consists of vertices belonging to shortest paths from $G$
(note that this $R$ is not guaranteed to be small). We briefly review the construction of such \textit{shortest path separators}. 

Let $G=(V,E,\ww)$ be a triangulated planar graph (the triangulation is guaranteed by adding infinity edge lengths among the missing edges). Further, let us fix an arbitrary shortest path tree $A$ rooted at some vertex $r$. Then, it can be inferred from the work of Lipton and Tarjan~\cite{lipton79} that there always exists a non-tree edge $e=(u,v)$ of $A$ such that the fundamental cycle $\mathcal{C}$ in $A \cup \{e\}$, formed by adding the non-tree edge $e$ to $A$, gives a $2/3$-separator for $G$. Because $A$ is a tree, the separator will consist of two paths from the $\text{lca}(u,v)$ to $u$ and $v$. We denote such paths by $P_1$ and $P_2$, respectively. Both paths are shortest paths as they belong to $A$. We will show how to use such separators to obtain terminal path covers. Before proceeding, we give the following preprocessing step.

\paragraph*{Preprocessing Step.} Given a planar graph $G = (V,E,\ww)$ with $K \subset V$, the algorithm \textsc{MinorSparsifier}($G$, $K$, $\mathcal{P}'$) with $\mathcal{P}'$ being the $(1,O(k^{2}))$-TPc of $G$, produces an $(1,O(k^{4}))$-DAM $G'$ for $G$. To simplify our notation, we will use $G$ instead of $G'$ in the following, i.e., we assume that $G$ has at most $O(k^{4})$ vertices.


\subsection{Distortion-$3$ Guarantee}

When solving a graph problem it is often the case that the solution is much easier on simpler graph instances, e.g., trees.
Driven by this, it is desirable to reduce the problem from arbitrary graphs to one or several tree instances, possibly allowing a small loss in the quality of the solution. Along the lines of such an approach, Gupta et al.~\cite{guptaKR04} gave the following definition in the context of shortest path distances.
\begin{definition}[Forest Cover] Given a graph $G=(V,E,\ww)$, a forest cover (with stretch $\alpha$) of $G$ is a family $\mathcal{F}$ of subforests $\{F_1,F_2,\ldots,F_k\}$ of $G$ such that for every $u,v \in V$, there is a forest $F_i \in \mathcal{F}$ such that $\dist_{G}(u,v) \leq \dist_{F_i}(u,v) \leq \alpha \cdot \dist_G(u,v)$.
\end{definition}

If we restrict our attention to planar graphs, Gupta et al.~\cite{guptaKR04} used shortest path separators (as described above) to give a divide-and-conquer algorithm for constructing forest covers with small guarantees on the stretch and size. Here, we slightly modify and adopt their construction for our specific application. 
Before proceeding to the algorithm, we give the following useful definition.
\begin{definition}
Let $t$ be a terminal and let $\pi$ be a shortest path in $G$. Then $t_{\min}^{\pi}$ denotes the vertex of $\pi$ that minimizes $\dist_G(t,p)$, for all $p \in V(\pi)$, breaking ties lexicographically. 
\end{definition}

\begin{algorithm2e}
\label{algo: forestcover}
\caption{\textsc{ForestCover}$(G,K)$}
\Input{Planar graph $G=(V,E,\ww)$, terminals $K$}
\Output{Forest cover $\mathcal{F}$ of $G$}
\If{$|V(G)| \leq 1$} {	\Return $V(G)$ \; }
Compute a $2/3$-separator $\mathcal{C}$ consisting of shortest paths $\pi_1$ and $\pi_2$ for $G$ \;
\For{$i=1,2$}{ 
Contract $\pi_i$ to a single vertex $p_i$ and compute a shortest path tree $L_i$ from $p_i$ \;
Expand back the contracted edges in $L_i$ to get the tree $L_i'$ \;
\For{every terminal $t \in K$}{
  Add $t_{\min}^{\pi_i}$ as a terminal in the tree $L_i'$ \;
}
}
Let $(G_1,K_1)$ and $(G_2,K_2)$ be the resulting connected graphs from $G \setminus \mathcal{C}$, \; 
\nonl where $K_1$ and $K_2$ are disjoint subsets of the terminals $K$ induced by $\mathcal{C}$ \;
\tcp{Note that all distances involving terminals from $\mathcal{C}$ are taken care of}
\Return $\bigcup_{i=1}^{2} L_i' \cup \bigcup_{i=1}^{2} \textsc{ForestCover}(G_i,K_i)$ \;
\end{algorithm2e}

\begin{algorithm2e}
\label{algo: minor_DP1}
\caption{\textsc{PlanarTPc-1} $(G,K)$}
\Input{Planar graph $G=(V,E,\ww)$, terminals $K$}
\Output{Terminal path cover $\mathcal{P}'$ of $G$}
Set $\mathcal{P}' \gets \emptyset$ \;
Set $\mathcal{F} \gets \text{\textsc{ForestCover}}(G,K)$ \; 
\For{each forest $F_i \in \mathcal{F}$} {
Let $R_i$ be the terminal set of $F_i$ and let $\mathcal{P}'_i$  be the (trivial) $(1,O(k^{2}))$-TPc of $F_i$ \;
Compute $F_i'$ $gets$ \textsc{MinorSparsifier}($F_i$, $R_i$, $\mathcal{P}_i'$) \;
Add the shortest paths corresponding to the edges of $F_i'$ to $\mathcal{P}'$ \;
}
\Return $\mathcal{P}'$ \;
\end{algorithm2e}

Gupta et al.~\cite{guptaKR04} showed the following guarantees for Algorithm \ref{algo: forestcover}.

\begin{theorem}[\cite{guptaKR04}, Theorem 5.1] \label{thm: guptacover}
Given a planar graph $G=(V,E,\ww)$ with $K \subset V$, \textsc{ForsetCover}$(G,K)$ produces a stretch-$3$ forest cover with $O (\log |V|)$ forests. 
\end{theorem} 

We note that the original construction does not consider terminal vertices, but this does not worsen neither the stretch nor the size of the cover. The only difference here is that we need to add at most $k$ new terminals to each forest compared to the original number of terminals in the input graph. This modification affects our bounds on the size of a minor only by a constant factor. 

Below we show that using the above theorem one can obtain terminal path covers for planar graphs. 

\begin{lemma} Given a planar graph $G = (V,E, \ww)$ with $K \subset V$, \textsc{PlanarTPc-1}$(G,K)$ produces an $(3,O (k \log k))$-\emph{TPc} $\mathcal{P}'$ for $G$. 
\end{lemma}
\begin{proof}
We first review the following simple fact, whose proof can be found in \cite{KrauthgamerNZ14}.
\begin{fact} \label{fact: treecover}
Given a forest $F=(V,E,\ww)$ with terminals $K \subset V$ and $\mathcal{P}'$ being the (trivial) $(1,O(k^{2}))$-\emph{TPc} of $F$, the procedure \textsc{MinorSparsifier}$(F,K,\mathcal{P}')$ outputs an $(1,k)$-\emph{DAM}.  
\end{fact}

Let us proceed with the analysis. Observe that from the Preprocessing Step our input graph $G$ has at most $O(k^{4})$ vertices. Thus, applying Theorem \ref{thm: guptacover} on $G$ gives a stretch-$3$ forest cover $\mathcal{F}$ of size $O(\log k)$. In addition, recall that all shortest paths are unique in $G$.

Next, let $F_i$ by any forest from $\mathcal{F}$. By construction, we note that each tree belonging to $F_i$ has the nice property of being a concatenation of a given shortest path with another shortest path tree. We will exploit this in order to show that every edge of the minor $F_i'$ for $F_i$ corresponds to the (unique) shortest path between its endpoints in $G$.


To this end, let $e' = (u,v)$ be an edge of $F_i'$ that does not exist in $F_i$. Since $F_i'$ is a minor of $F_i$, we can map back $e'$ to the path $\pi_{u,v}$ connecting $u$ and $v$ in $F_i$. Because of the additional terminals $u_{\min}^{P_i}$ added to $F_i$, we claim that $\pi_{u,v}$ is entirely contained either in some shortest path tree $L_j$ or some shortest path separator $P_j$. Using the fact that subpaths of shortest paths are shortest paths, we conclude that the length of the path $\pi_{u,v}$ (or equivalently, the length of edge $e'$) corresponds to the unique shortest path connecting $u$ and $v$ in $G$. The same argument is repeatedly applied to every such edge of $F_i'$.

By construction we know that $F_i$ has at most $2k$ terminals. Using Fact \ref{fact: treecover} we get that $F_i'$ contains at most $4k$ edges. Since there are $O (\log k)$ forests, we conclude that the terminal path cover $\mathcal{P}'$ consists of $O (k \log k)$ shortest paths. The stretch guarantee follows directly from that of cover $\mathcal{F}$, since $F_i'$ exactly preserves all distances between terminals in $F_i$.
\end{proof}

\begin{theorem}
Any planar graph $G=(V,E,\ww)$ with $K \subset V$ admits a $(3, \tilde{O}(k^{2}))$-\emph{DAM}.
\end{theorem}


\subsection{Distortion-$(1+\epsilon)$ Guarantee}

Next we present our best trade-off between distortion and size of minors for planar graphs. Our idea is to construct terminal path covers using the construction of Thorup~\cite{ThorupJACM04} in the context of approximate distance oracles in planar graphs. Here, we modify a simplified version due to Kawarabayashi et al.~\cite{sommer11}. 

The construction relies on two important ideas. Similarly to the distortion-$3$ result, the first idea is to recursively use shortest path separators to decompose the graph. The second consists of approximating shortest paths that cross a shortest path separator. Below we present some necessary modification to make use of such a construction for our purposes.

Let $\pi$ be a shortest path in $G$. For a terminal $t \in K$, we let the pair $(p,t)$, where $p \in V(\pi)$, denote the \textit{portal} of $t$ with respect to the path $\pi$. An $\epsilon$-cover $C(t,\pi)$ of $t$ with respect to $\pi$ is a set of portals with the following property:
\begin{itemize}
\item for all $p \in V(\pi)$, thee exsits $q \in C(t,\pi)$ such that \[ \dist_G(t,q) + \dist_G(q,p) \leq (1+\epsilon) \dist_G(t,p). \]
\end{itemize}
Let $(t,t')$ by any terminal pair in $G$. Let $\pi_{t,t'}$ be the (unique) shortest path that crosses the path $\pi$ at vertex $w$. Then using the $\epsilon$-covers $C(t,\pi)$ and $C(t',\pi)$, there exist portals $(t,p)$ and $(p',t')$ such that the new distance between $t$ and $t'$ is
\begin{align} \label{eqn: stretch}
\begin{split}
	\dist_G(t,p) & + \dist_G(p,p')  + \dist_G(p',t') \\
	& \leq \dist_G(t,p) + \dist_G(p,w) + \dist_G(w,p') + \dist_G(p',t') \\
	& \leq (1+\epsilon) \dist_G(t,t').
\end{split}
\end{align}
The new distance clearly dominates the old one. The next result due to Thorup~\cite{ThorupJACM04} shows that maintaining a small number of portals per terminal suffices to approximately preserve terminal shortest paths.

\begin{lemma} \label{lemm: portal} 
Let $\epsilon > 0$. For a given terminal $t \in K$ and a shortest path $\pi$, there exists an $\epsilon$-cover $C(t,\pi)$ of size $O(1/\epsilon)$.
\end{lemma}

The above lemma leads to the following recursive procedure.

\begin{algorithm2e}
\label{algo: minor_DP2}
\caption{\textsc{PlanarTPc-2} $(G,K)$}
\Input{Planar graph $G=(V,E,\ww)$, terminals $K$}
\Output{Terminal path cover $\mathcal{P}'$ of $G$}
\If{$|V(G)| \leq 1$ or $K = \emptyset$}
{	\Return $\emptyset$ \; }

Set $\mathcal{B} \gets \emptyset$ \; 

Compute a $2/3$-separator $\mathcal{C}$ consisting of shortest paths $\pi_1$ and $\pi_2$ \;
Add $\pi_1$ and $\pi_2$ to $\mathcal{B}$ \;
\For{every terminal $t \in K$}{
Compute $\epsilon$-covers $C(t,\pi_1)$ and $C(t,\pi_2)$ \;
\For{every portal $(t,p) \in C(t,\pi_1) \cup C(t,\pi_2)$}{
 Add the shortest path $\pi_{t,p}$ to $\mathcal{B}$ \;
}
}
Let $(G_1,K_1)$ and $(G_2,K_2)$ be the resulting connected graphs from $G \setminus \mathcal{C}$, \;
\nonl where $K_1$ and $K_2$ are disjoint subsets of the terminals $K$ induced by $\mathcal{C}$ \;
\tcp{Note that all distances involving terminals from $\mathcal{C}$ are taken care of}
\Return $\mathcal{B} \cup \bigcup_{i=1}^{2} \textsc{PlanarTPc-2}(G_i,K_i)$ \;
\end{algorithm2e}

\begin{lemma}\label{lem:tpc-1-plus-ep}
Given a planar graph $G = (V,E, \ww)$ with $K \subset V$, \textsc{PlanarTPc-2}$(G,K)$ outputs an $(1+\epsilon,O (k \epsilon^{-1} \log k ))$-\emph{TPc} $\mathcal{P}'$ for $G$. 
\end{lemma}

\begin{proof}
From the Preprocessing Step we know that $G$ has at most $O(k^{4})$ vertices. Further,  recall that removing the vertices that belong to the shortest path separators from $G$ results into two graphs $G_1$ and $G_2$, whose size is at most $2/3 \cdot |G|$. Thus, there are at most $O(\log k)$ levels of recursion for the above procedure.  

Let $\mathcal{P}'$ be the terminal path cover output by \text{PlanarTPc-2}$(G,K)$. We first bound the number of separator shortest paths added in Step 3. Note that at any level of the recursion there at most $k$ terminals and, thus the number of recursive calls per level is at most $k$. Since we added two paths per recursive call, we get that there are at most $O (k \log k)$ paths overall.

We now continue with the counting or portals. Let $t \in K$ be any terminal and consider any recursive call applied on the current graph $(G',K')$. If $t \not\in K'$, then we simply ignore $t$. Otherwise, $t$ either belongs to one of the separator shortest paths in $G'$ or one of the partitions induced by the separators. In the first case, we know that $t$ is retained because we added $\pi_1$ and $\pi_2$ to $\mathcal{P}'$ and these are already counted. In the second case, using Lemma \ref{lemm: portal}, we add $O(1/\epsilon)$ shortest paths connecting portals from $C(t,\pi_1)$ and $C(t,\pi_2)$. Therefore, in any recursive call, we maintain at most $O(1/\epsilon)$ shortest paths per terminal. Since every terminal can participate in at most $O(\log k)$ recursive calls, we get that the total number of portal-shortest paths is at most $O (k \log k / \epsilon)$. Combining both bounds, it follows that the size of $\mathcal{P}'$ is at most $O(k \log k / \epsilon)$. 

It remains to show the stretch guarantee of $\mathcal{P}'$. Let $R$ be the recursion tree of the algorithm, where every node corresponds to a recursive call. For any pair $t,t' \in L$, let $a \in V(R)$ associated with $(G_a, K_a)$ be the leafmost node such that $t,t' \in K_a$. Then, it follows that among all ancestors of $a$ in the tree $R$, there must exist a separator path $\pi_i$, $i=1,2$ that crosses $\pi_{t,t'}$ and attains the minimum length. The stretch guarantee follows directly from (\ref{eqn: stretch}).
\end{proof}

\begin{theorem}\label{thm:planar-oneplus}
Any planar graph $G=(V,E,\ww)$ with $K \subset V$ admits an $(1+\epsilon,O(k^{2} \epsilon^{-2} \log^{2} k))$-\emph{DAM}.
\end{theorem} 

\section{Conclusion}
In this chapter, we introduced the notion of distance approximating minors, which are vertex sparsifiers that are minors of the input graph and approximately preserve shortest path distances among a designated subset of vertices, referred to as terminals. This notion is a natural generalization of the Steiner Point Removal problem~\cite{Gupta01}, where the sparsifier must contain only terminals and the Distance Preserving Minor problem~\cite{KrauthgamerNZ14}, where we want to exactly preserve pair-wise terminal distances while allowing additional non-terminal vertices in the sparsifier. We study distance approximating minors from both upper and lower bound perspective. For example, we show that for $k$-terminal general graphs and distortion $3-\epsilon$, one needs to retain at least $\Omega(k^{6/5})$ non-terminal vertices. For planar graphs, we show an algorithm that computes a $(1+\epsilon)$-distance approximating minors with $\tilde{O}(k^2 \epsilon^{-2})$ non-terminals. Our lower-bound and algorithmic constructions bring together techniques from distance oracles, branching events in shortest path computations and Steiner systems from combinatorics. 

There remain gaps between some of the best upper and lower bounds, e.g., for distortion $3-\epsilon$, the lower bound is $\Omega(k^{6/5})$, while for distortion $3$, our upper bound is $\tilde{O}(k^3)$. Therefore, understanding the trade-off between distortion and the size of the sparsifiers is an interesting open problem. In the same vein, it is interesting to explore whether the size of the sparisifer in the planar setting can be improved to $\tilde{O}(k^{2-o(1)})$, while keeping the same approximation guarantee. As we demonstrate in Chapter~\ref{cha:Man2019_LS}, this question is particularly relevant due to its connection to the offline dynamic APSP problem in planar graphs.

Another important problem in this area is to design fast algorithms for constructing distance preserving minors. While most of the vertex sparsification studies in the literature have focused on understanding the trade-off between distortion and size, we believe that the running time for constructing such sparsifiers is an important aspect that better serves the general purpose of using sparsification to speed up algorithmic constructions.

\chapter[Reachability Preserving Minors and Sparsifiers for Cuts and Distances][Sparsification for Reachability, Cuts and Distances]{Reachability Preserving Minors and Sparsifiers for Cuts and Distances}\label{cha:ESA2017_RM}

Graph Sparsification aims at compressing large graphs into smaller ones while preserving important characteristics of the input graph. In this chapter we study Vertex Sparsifiers, i.e., sparsifiers whose goal is to reduce the number of vertices. We focus on the following notions:

(1) Given a digraph $G=(V,E)$ and terminal vertices $K \subset V$ with $|K| = k$, a (vertex) reachability sparsifier of $G$ is a digraph $H=(V',E')$, $K \subset V'$ that preserves all reachability information among terminal pairs. In this chapter we introduce the notion of reachability-preserving minors (RPMs) , i.e., we require $H$ to be a minor of $G$. We show any directed graph $G$ admits a RPM $H$ of size $O(k^3)$, and if $G$ is planar, then the size of $H$ improves to $O(k^{2} \log k)$. We complement our upper-bound by showing that there exists an infinite family of grids such that any RPM must have $\Omega(k^{2})$ vertices.

(2) Given a weighted undirected graph $G=(V,E)$ and terminal vertices $K$ with $|K|=k$, an exact (vertex) cut sparsifier of $G$ is a graph $H$ with $K \subset V'$ that preserves the value of minimum-cuts separating any bipartition of $K$. We show that planar graphs with all the $k$ terminals lying on the same face admit exact cut sparsifiers of size $O(k^{2})$ that are also planar. Our result extends to flow and distance sparsifiers. It improves the previous best-known bound of $O(k^22^{2k})$ for cut and flow sparsifiers by an exponential factor, and matches an $\Omega(k^2)$ lower-bound for this class of graphs.

\section{Introduction}

Very large graphs or networks are ubiquitous nowadays, from social networks to information networks. One natural and effective way of processing and analyzing such graphs is to compress or sparsify the graph into a smaller one that well preserves certain properties of the original graph. Such a sparsification can be obtained by reducing the number of \emph{edges}. Typical examples include cut sparsifiers~\cite{BenczurK96}, spectral sparsifiers~\cite{SpielmanT11}, spanners~\cite{ThorupZ05} and transitive reductions~\cite{AhoGU72}, which are subgraphs defined on the same vertex set of the original graph $G$ while having much smaller number of edges and still well preserving the cut structure, spectral properties, pairwise distances and transitive closure of $G$, respectively. 

Another way of performing sparsification is by reducing the number of \emph{vertices}, which is most appealing when only the properties among a subset of vertices (which are called \emph{terminals}) are of interest~(see e.g., \cite{Moitra09,andoni,KrauthgamerNZ14}). We call such small graphs \emph{vertex sparsifiers} of the original graph. In this chapter, we will particularly focus on vertex reachability sparsifiers for \emph{directed} graphs and cut (and other related) sparsifiers for \emph{undirected} graphs. 

Vertex reachability sparsifiers in directed graphs is an important and fundamental notion in Graph Sparsification, which has been implicitly studied in the dynamic graph algorithms community~\cite{Subramanian93,DiksS07}, and explicitly in~\cite{katriel2005reachability}. Specifically, given a digraph $G=(V,E)$, $K \subset V$, a digraph $H=(V',E')$, $K \subset V'$ is a (\emph{vertex}) \emph{reachability sparsifier} of $G$ if for any $x ,x' \in K$, there is a directed path from $x$ to $x'$ in $H$ iff there is a directed path from $x$ to $x'$ in $G$. If $|K|=k$, we call the digraph $G$ a \emph{$k$-terminal digraph}. Note that any $k$-terminal digraph $G$ always admits a trivial reachability vertex sparsifier $H$, which corresponds to the transitive closure restricted to the terminals. 

In this chapter, we initiate the study of \emph{reachability-preserving minors}, i.e., vertex reachability sparsifiers with $H$ required to be a minor of $G$. 
The restriction on $H$ being a minor of $G$ is desirable as it makes sure that $H$ is structurally similar to $G$, e.g., any minor of a planar graph remains planar. We ask the question whether general graphs admit reachability-preserving minors whose size can be bounded independently of the input graph $G$, and study it from both the lower- and upper-bound perspective.

For the notion of cut (and other related) sparsifiers, we are given a capacitated undirected graph $G=(V,E,\cc)$, and a set of terminals $K$ and our goal is to find a (capacitated undirected) graph $H=(V',E',\cc')$ with as few vertices as possible and $K \subseteq V'$ such that the quantities like, cut value, multi-commodity flow and distance among terminal vertices in $H$ are the same as or close to the corresponding quantities in $G$. If $|K|=k$, we call the graph $G$ a \emph{$k$-terminal graph}. 

We say $H$ is a \emph{quality-$q$} (\emph{vertex}) \emph{cut sparsifier} of $G$, if for every bipartition $(U,K \setminus U)$ of the terminal set $K$, the value of the minimum cut separating $U$ from $K \setminus U$ in $G$ is within a factor of $q$ of the value of minimum cut separating $U$ from $K \setminus U$ in $H$. If $H$ is a quality-$1$ cut sparsifier, then it will be also called a \emph{mimicking network}~\cite{HagerupKNR98}. Similarly, we define flow and distance sparsifiers that (approximately) preserve multicommodity flows and distances among terminal pairs, respectively~(see Section~\ref{sec: UpperFlow} for formal definitions). These type of sparsifiers have proven useful in approximation algorithms~\cite{Moitra09} and also find applications in network routing~\cite{Chuzhoy12a}.

\paragraph*{Our Results.} Our first and main contribution is the study \sloppy of reachability-preserving minors. Although reachability is a weaker requirement in comparison to shortest path distances, directed graphs are usually much more cumbersome to deal with from the perspective of graph sparsification. Surprisingly, we show that general digraphs admit reachability-preserving minors with $O(k^{3})$ vertices, which is in contrast to the bound of $O(k^4)$ on the size of distance-preserving minors in undirected graphs by Krauthgamer et al.~\cite{KrauthgamerNZ14}. 

\begin{theorem} \label{thmi: general_reachability} Given a $k$-terminal digraph $G$, there exist a reachability-preserving minor $H$ of $G$ with size $O(k^{3})$.
\end{theorem}

It might be interesting to compare the above result with the construction of \emph{reachability preserver} by Abbound and Bodwin~\cite{AB18reachability}, where the reachability preserver for a pair-set $P$ in a graph $G$ is defined to be a \emph{subgraph} of $G$ that preserves the reachability of all pairs in $P$. The size (i.e., the number of edges) of such preservers is shown to be at least $\Omega(n^{2/(d+1)}|P|^{(d-1)/d})$, for any integer $d\geq 2$, which is in sharp contrast to our upper bound $O(|P|^{3/2})$ on the size of reachability-preserving minors by taking $P$ to be the pair-set of all terminals. 

Furthermore, by exploiting a tight integration of our techniques with the compact distance oracles for planar graphs by Thorup~\cite{ThorupJACM04}, we can show the following theorem regarding the size of reachability-preserving minors for planar digraphs. 

\begin{theorem} \label{thm: ubPlanar}
	Given a $k$-terminal planar digraph $G$, there exists a reachability-preserving minor $H$ of $G$ with size $O(k^{2} \log k)$.
\end{theorem}

We complement the above result by showing that there exist instances where the above upper-bound is tight up to a $O(\log k)$ factor. 


\begin{theorem} \label{thm: lbPlanar}
	For infinitely many $k \in \mathbb{N}$ there exists a $k$-terminal acyclic directed grid $G$ such that any reachability-preserving minor of $G$ must use $\Omega(k^{2})$ non-terminals.
\end{theorem}

Our second contribution is new algorithms for constructing quality-$1$ (exact) cut, flow and distance sparsifiers for $k$-terminal planar graphs, where all the terminals are assumed to lie on the same face. We call such $k$-terminal planar graphs \emph{Okamura-Seymour} (OS) instances. They are of particular interest in the algorithm design and optimization community, due to the classical Okamura-Seymour theorem that characterizes the existence of feasible concurrent flows in such graphs~(see e.g., \cite{OkamuraS81,chekuri09,chekuri2010flow,LeeMM13}).

We show that the size of quality-$1$ sparsifiers can be as small as $O(k^2)$ for such instances, for which only exponential (in $k$) size of cut and flow sparisifiers were known before~\cite{KrauthgamerR13,andoni}. Formally, we have the following theorem.
\begin{theorem} \label{thm: mainThm}
	For any \emph{OS} instance $G$, i.e., a $k$-terminal planar graph in which all terminals lie on the same face, there exist quality-$1$ cut, flow and distance sparsifers of size $O(k^2)$. Furthermore, the resulting sparsifiers are also planar.
\end{theorem} 

We remark that all the above sparsifiers can be constructed in polynomial time (in $n$ and $k$), but we will not optimize the running time here. As we mentioned above, previously the only known upper bound on the size of quality-$1$ cut and flow sparsifiers for OS instance was $O(k^22^{2k})$, given by~\cite{KrauthgamerR13,andoni}. Our upper bound for cut sparsifier also matches the lower bound of $\Omega(k^2)$ for OS instance given by~\cite{KrauthgamerR13}. More specifically, in \cite{KrauthgamerR13}, an OS instance (that is a grid in which all terminals lie on the boundary) is constructed, and used to show that any mimicking network for this instance needs $\Omega(k^2)$ edges, which is thus a lower bound for planar graphs (see the table below for an overview). Note that that even though our distance sparsifier is not necessarily a minor of the original graph $G$, it still shares the nice property of being planar as $G$. Furthermore, Krauthgamer and Zondiner~\cite{KrauthgamerZ12} proved that there exists a $k$-terminal planar graph $G$ (not necessarily an OS instance), such that any quality-$1$ distance sparsifier of $G$ that is planar requires at least $\Omega(k^2)$ vertices. 

\begin{table}[h]
\renewcommand{\arraystretch}{1.1}
\begin{center}
\begin{tabular}{c|c|c|c}
Graph & Type of sparsifier &  Upper Bound &  Lower Bound  \\ \hline \hline
Planar & Cut (minor) & $O(k2^{2k})$~\cite{KrauthgamerR13} & $\Omega(k^{2})$~\cite{KrauthgamerR13} \\
Planar ($\gamma$) & Cut (minor) & $O(\gamma^5 2^{2 \gamma} k^4)$~\cite{krauthgamer2017refined} & $\Omega(2^{k})$~\cite{karpov2017exponential} \\
Planar OS & Cut (planar) & $\mathbf{O(k^2})$ & $\Omega(k^{2})$~\cite{KrauthgamerR13} \\ \hline 
Planar OS & Distance (minor) & $O(k^{4})$~\cite{KrauthgamerNZ14} & $\Omega(k^2)$~\cite{KrauthgamerNZ14}   \\
Planar OS &  Distnace (planar) & $\mathbf{O(k^{2})}$ & $\Omega(k^2)$~\cite{KrauthgamerZ12} 
\end{tabular}
\caption{An overview on the best-known results for mimicking networks and distance sparsifiers. The results which are \emph{not} followed by a reference are shown in this chapter.}
\end{center}
\end{table}
We further provide a lower bound on the size of any \emph{data structure} (not necessarily a graph) that approximately preserves pairwise terminal distances of \emph{general} $k$-terminal graphs, which gives a trade-off between the distance stretch and the space complexity. 
 
\begin{theorem}~\label{thm:incompressibility}
For any $\varepsilon > 0$ and $t \geq 2$, there exists a (sparse) $k$-terminal $n$-vertex graph such that $k=o(n)$, and any data structure that approximates pairwise terminal distances within a multiplicative factor of $t - \varepsilon$ or an additive error $2t-3$ must use $\Omega(k^{1+1/(t-1)})$ bits. 
\end{theorem}

\paragraph*{Remark.} Recently and independently of our work, Krauthgamer and Rika~\cite{krauthgamer2017refined} constructed quality-$1$ cut sparsifiers of size $O(\gamma^{5}2^{2\gamma}k^4)$ for planar graphs whose terminals are incident to at most $\gamma=\gamma(G)$ faces. In comparison with our upper-bound which only considers the case $\gamma = 1$, the size of our sparsifiers from Theorem~\ref{thm: mainThm} is better by a $\Omega(k^{2})$ factor. Moreover, their work focuses on constructing sparsifiers that are minors of the originial input graph, while our construction only guarantee that the resulting sparsifiers are planar graphs. Subsequent to our work, Karpov et al.~\cite{karpov2017exponential} proved that there exists edge-weighted $k$-terminal planar graphs that require $\Omega(2^k)$ edges in any exact cut sparsifier, which implies that it is necessary to have some additional assumption (e.g., $\gamma=O(1)$) to obtain a cut sparsifier of $k^{O(1)}$ size. 

\paragraph*{Our Techniques.} Our results for reachability-preserving minors are obtained by exploiting a technique of counting ``branching'' events between shortest paths in the directed setting (this technique was introduced by Coppersmith and Elkin~\cite{CoppersmithE06}, and has also been recently leveraged by Bodwin~\cite{Bodwin17} and Abboud and Bodwin~\cite{AB18reachability}). Using this and a consistent tie-breaking scheme for shortest paths, we can efficiently construct a RPM for general digraphs of size $O(k^4)$ and by using a more refined analysis of branching events (see \cite{AB18reachability}), we can further reduce the size to be $O(k^3)$. We then combine our construction with a decomposition for planar digraphs (see~\cite{ThorupJACM04}), to show that it suffices to maintain the reachability information among $O(k\log k)$ terminal pairs, instead of the naive $O(k^2)$ pairs, and then construct a RPM for planar digraphs with $O(k^2\log k)$ vertices. 

The lower-bound follows by constructing a special class of $k$-terminal directed grids and showing that any RPM for such grids must use $\Omega(k^2)$ vertices. Similar ideas for proving the lower bound on the size of distance-preserving minors for undirected graphs have been used by Krauthgamer et al.~\cite{KrauthgamerNZ14}.

We construct our quality-$1$ cut and distance sparsifiers by repeatedly performing \emph{Wye-Delta transformations}, which are local operations that preserve cut values and distances and have proven very powerful in analyzing electrical networks and in the theory of circular planar graphs~(see e.g., \cite{curtis98,feo1993}). Khan and Raghavendra~\cite{KhanR14} used Wye-Delta transformations to construct quality-$1$ cut sparsifiers of size $O(k)$ for trees and outerplanar graphs, while our case (i.e., the planar OS instances) is more general and complicated and previously it was not clear at all how to apply such transformations to a more broad class of graphs. Our approach is as follows. Given a $k$-terminal planar graph with terminals lying on the same face, we first embed it into some large grid with terminals lying on the boundary of the grid. Next, we show how to embed this grid into a ``more suitable'' graph, which we will refer to as ``half-grid''. Finally, using the Wye-Delta operations, we reduce the ``half-grid'' into another graph whose number of vertices can be bounded by $O(k^{2})$. Since we  argue that the above graph reductions preserve exactly all terminal minimum cuts, our result follows. Gitler~\cite{Gitler91} proposed a similar approach for studying the reducibility of multi-terminal graphs with the goal to classify all Wye-Delta reducible graphs, which is very different from our motivation of constructing small vertex sparsifiers with good quality. 

The distance sparsifiers can be constructed similarly by slightly modifying the Wye-Delta operation. Our flow sparsifiers follow from the construction of cut sparsifiers and the flow/cut gaps for OS instances (which has been initially observed by Andoni et al.~\cite{andoni}). Our lower bound on the space complexity of any compression function approximately preserving terminal pairwise distance is derived by combining extremal combinatorics construction of Steiner Triple System that was used to prove lower bounds on the size of distance approximating minors (see~\cite{cheung2016}) and the incompressibility technique from~\cite{matousek96}.

\medskip
\noindent \textbf{Related Work. } There has been a long line of work on investigating the tradeoff between the quality of the vertex sparsifier and its size~(see e.g., \cite{EnglertGKRTT14,KrauthgamerR13,andoni}). (Throughout, cut, flow and distance sparsifiers will refer to their vertex versions.) Quality-$1$ \emph{cut sparsifiers} (or equivalently, mimicking networks) were first introduced by Hagerup et al.~\cite{HagerupKNR98}, who proved that for any graph $G$, there always exists a mimicking network of size $O(2^{2^k})$. 
Krauthgamer and Rika~\cite{KrauthgamerR13} showed how to build a mimicking network of size $O(k^22^{2k})$ for any planar graph $G$ that is minor of the input graph. They also proved a lower bound of $\Omega(k^2)$ on the number of edges of the mimicking network of planar graphs, and a lower bound of $2^{\Omega(k)}$ on the number of vertices of the mimicking network for general graphs. 

Quality-$1$ vertex flow sparsifiers have been studied in~\cite{andoni, GoranciR16}, albeit only for restricted families of graphs like quasi-bipartite, series-parallel, etc. It is not known if any general undirected graph $G$ admits a constant quality flow sparsifier with size independent of $|V(G)|$ and the edge capacities. For the quality-$1$ distance sparsifiers, Krauthgamer et al.~\cite{KrauthgamerNZ14} introduced the notion of \emph{distance-preserving minors}, and showed an upper-bound of size $O(k^4)$ for general undirected graphs. They also gave a lower bound of $\Omega(k^2)$ on the size of such a minor for planar graphs. Abboud et al.~\cite{AbboudGMW17} show how to compress a planar graph metric using only $\tilde{O}(\min\{k^2,\sqrt{k \cdot n}\})$ bits. Recently,~ Chang et al.~\cite{ChangGMW18} extended their compressing scheme to a graph sparsifer which matches their bound.

Over the last two decades, there has been a considerable amount of work on understanding the tradeoff between the sparsifier's quality $q$ and its size for $q>1$, i.e., when the sparsifiers only \emph{approximately} preserve the corresponding properties~\cite{juliasteiner,andoni,Moitra09,leighton,charikar,EnglertGKRTT14,mm10,Gupta01,
chekuri2006embedding,ChanXKR06,KammaKN15,cheung2016,Cheung18,Filtser18,
GajjarR17,BernsteinDDKMS18}. 

\section{Preliminaries} \label{sec: preli_RM}
Let $G=(V,E)$ be a directed graph with terminal set $K \subset V$, $|K|=k$, which we will refer to as a \emph{$k$-terminal digraph}. We say $G$ is a \emph{$k$-terminal} DAG if $G$ has no directed cycles. The \emph{in-degree} of a vertex $v$, denoted by $\deg_{G}^{-}(v)$, is the number of edges directed towards $v$ in $G$. A digraph $H=(V',E')$, $K \subset V'$ is a (\emph{vertex}) \emph{reachability sparsifier} of $G$ if for any $x ,x' \in K$, there is a directed path from $x$ to $x'$ in $H$ iff there is a directed path from $x$ to $x'$ in $G$. If $H$ is obtained by performing minor operations in $G$, then we say that $H$ is a \emph{reachability-preserving minor} of $G$. We define the \emph{size} of $H$ to be the number of non-terminals in $H$, i.e. $|V' \setminus K|$.

Let $G=(V,E,\cc)$ be an undirected graph with terminal set $K \subset V$ of
cardinality $k$, where $\cc: E \rightarrow \mathbb{R}_{\geq 0}$ assigns a non-negative
capacity to each edge. We will refer to such a graph as a \emph{k-terminal graph}. 
Let $U \subset V$ and $S \subset K$. We say that a cut $(U, V \setminus U)$ is
$S$-separating if it separates the terminal subset $S$ from its complement $K
\setminus S$, i.e., $U \cap K$ is either $S$ or $K \setminus S$. We will refer to such cut as a \emph{terminal cut}. The cutset
$\delta(U)$ of a cut $(U, V \setminus U)$ represents the edges that have one
endpoint in $U$ and the other one in $V \setminus U$. The cost
$\capacity_G(\delta(U))$ of a cut $(U, V \setminus U)$ is the sum over all
capacities of the edges belonging to the cutset. We let $\text{mincut}_{G}(S, K
\setminus S)$ denote the minimum cost of any $S$-separating cut of
$G$. A graph $H = (V', E', \cc')$, $K \subset V'$ is a \emph{quality-$q$} (\emph{vertex}) \emph{cut sparsifier} of $G$ with $q \geq 1$ if for any $S \subset K,
~ \mincut_G(S, K \setminus S) \leq \mincut_H(S, K \setminus S) \leq q \cdot
\mincut_G(S, K \setminus S).$

\section{Reachability-Preserving Minors for General Digraphs} \label{sec: minorsGeneralDigraphs}
In this section, we provide two constructions for reachability-preserving minors for general digraphs. The resulting minor from the first construction has size $O(k^4)$, which is larger than the size $O(k^3)$ of the minor from the second construction. However, our first construction can be implemented in polynomial time (in $n$), while the second one requires exponential running time.
\subsection{A Warm-up: An Upper Bound of $O(k^4)$}
In this section we show that any $k$-terminal digraph admits a reachability-preserving minor of size $O(k^{4})$. We accomplish this by first restricting our attention to DAGs, and then showing how to generalize the result to any digraph. 


We start by introducing the following definition. Given a digraph $G$ with a terminal set $K$ of size $k$ and a pair-set $P\subseteq K\times K$, we say that $H$ is a reachability-preserving minor with respect to $P$, if $H$ is a minor of $G$ that preserves the reachability information only among the pairs in $P$. Note that in the definition of vertex reachability sparsifiers, the \emph{trivial} pair-set $P$ contains $k(k-1)$ terminal-pairs, i.e., for any pair $x,x' \in K$, both $(x,x')$ and $(x',x)$ belong to $P$. Whenever we omit $P$, we mean to preserve the reachability information among all possible terminal pairs.

We next review a useful scheme for breaking ties between shortest paths connecting some vertex pair from $P$. This tie-breaking is usually achieved by slightly perturbing the edge lengths of the original graph such that no two paths have the same length (note that in our case, edge lengths are initially one). The perturbation gives a \emph{consistent} scheme in the sense that whenever $\pi$ is chosen as a shortest path, every sub-path of $\pi$ is also chosen as a shortest path. Below we formalize these ideas using two definitions and a lemma from~\cite{Bodwin17}.

\begin{definition}[Tie-breaking Scheme] Given a $k$-terminal $G$, a \emph{shortest path tie breaking scheme} is a function $\pi$ that maps every pair of vertices $(s,t)$ to some shortest path between $s$ and $t$ in $G$. For any pair-set $P$, we let $\pi(P)$ denote the union over all shortest paths between pairs in $P$ with respect to the scheme $\pi$.
\end{definition}

\begin{definition}[Consistency] A tie-breaking scheme is consistent if, for all vertices $y,x,x',y' \in V$, if $x,x' \in \pi(y,y')$ with $d(y,x) < d(y,x')$, then $\pi(x,x')$ is a sub-path of $\pi(y,y')$.
\end{definition}


\begin{lemma}[\cite{Bodwin17}] \label{lemm: consistency} For any $k$-terminal G, there is a consistent tie-breaking scheme in $G$.
\end{lemma}

We remark that for any $k$-terminal graph with $n$ vertices, the consistent tie-breaking scheme can be constructed in polynomial (in $n$) time~\cite{CoppersmithE06}. 

Let $G$ be a $k$-terminal DAG. Given a tie-breaking scheme $\pi$, the first step to construct a reachability-preserving minor is to start with an empty graph $H$ and then for every pair $p \in P$, repeatedly add the shortest-path $\pi(p)$ to $H$. We can alternatively think of this as deleting vertices and edges that do not participate in any shortest path among terminal-pairs in $P$ with respect to the scheme $\pi$. Clearly, the DAG $H=(V', E')$, $E' := \pi(P)$, is a minor of $G$ and preserves all reachability information among pairs in $P$. We next review the notion of a branching event, which will be useful to bound the size of $H$.  

\begin{definition}[Branching Event] A \emph{branching event} is a set of two distinct directed edges $\{e_1=(u_1,v), e_2=(u_2,v)\}$ that enter the same node $v$.
\end{definition}  

\begin{lemma} \label{lemm: branching} The \emph{DAG} $H$ has at most $|P|(|P|-1|)/2$ branching events.
\end{lemma}
\begin{proof}
	First, note that by construction of $H$, we can associate each edge $e \in E'$ with some pair $p \in P$ such that $e \in \pi(p)$. To prove the lemma, it suffices to show that for any two terminal-pairs $p_1, p_2 \in P$, there is at most one branching event in the graph induced by $\pi(p1) \cup \pi(p_2)$. Suppose towards contradiction that there exist two terminal pairs $p_1, p_2$ that have two branching events in $\pi(p1) \cup \pi(p2)$. More specifically, we assume there exist two branching events
	\[
	b := \{e_1 = (u_1,v), e_2=(u_2,v)\} \text{ and } b' := \{e_1 = (u_1',v'), e_2=(u_2',v')\},
	\]
	where $e_i$ and $e_i'$ lie on the dipath $\pi(p_i)$, for $i = 1,2$.
	
	Assume without loss of generality that the vertex $v$ appears before $v'$ in the dipath $\pi(p_1)$. We then claim that $v$ must also appear before $v'$ in the dipath $\pi(p_2)$, since otherwise we would have a directed cycle between $v$ and $v'$, thus contradicting the fact that $H$ is acyclic. Since the tie-breaking scheme $\pi$ is consistent (Lemma \ref{lemm: consistency}), it follows that the dipaths $\pi(p_1)$ and $\pi(p_2)$ must share the subpath $\pi(v,v')$. Thus, $\pi(p_1)$ and $\pi(p_2)$ use the same edge that enters the node $v'$, i.e., $e_1' = e_2'$. However, by definition of a branching event, the edges that enter a node must be distinct, contradicting the fact that $b'$ is a branching event. This implies that there cannot be two branching events for the terminal pairs $p_1$ and $p_2$, thus proving the lemma.  
\end{proof}
We now have all the tools to present our algorithm for constructing reachability-preserving minors for DAGs. 

\begin{algorithm2e}[h]
\label{algo: minor}
\caption{\textsc{MinorSparsifyDag} $(G,P)$}
\Input{$k$-terminal DAG $G$, pair-set $P$}
\Output{Reachability preserving minor $H$ of $G$ with respect to $P$}
Set $H \gets \emptyset$ \;
Compute a consistent tie-breaking scheme $\pi$ for shortest paths in $G$ \;
\label{alg:minor_2} For each $p \in P$,  add the shortest path $\pi(p)$ to $H$ \;
\While{there is an edge $(u,v)$ directed towards a non-terminal $v$ with $\deg_{H}^{-}(v) = 1$}{
\label{alg:minor_5} Contract the edge $(u,v)$ \;
}
 \Return $H$ \;
\end{algorithm2e}


\begin{lemma} \label{lemma: DAG}
Given a $k$-terminal \emph{DAG} $G$ with a pair-set $P$, \emph{Algorithm~\ref{algo: minor}} outputs a reachability-preserving minor $H$ of size $O(|P|^2)$ for $G$ with respect to $P$.
\end{lemma}
\begin{proof}
	We first argue that $H$ is a reachability-preserving minor with respect to the terminals. Indeed, after Line \ref{alg:minor_2} of the algorithm, graph $H$ can viewed as deleting vertices and edges from $G$ that do not lie on any of the shortest path among terminal pairs in $P$, chosen according to the scheme $\pi$. Thus, at this point $H$ is clearly a minor of $G$ that preserves the reachability information among the pairs in $P$. The edge contractions we perform in the remaining part of the algorithm guarantee that the resulting $H$ remains a reachability-preserving minor of $G$ with respect to $P$. 
	
	To bound the size of $H$, note that every non-terminal $v \in V' \setminus K$ has in-degree at least $2$, and thus it corresponds to at least one branching event. Lemma \ref{lemm: branching} shows that the number of branching events is at most $O(|P|^2)$. Observing that edge contractions in Line \ref{alg:minor_5} do not affect this number, 
	we get that the size of $H$ is $O(|P|^2)$.
\end{proof}
%


We next show how the construction of reachability-preserving minors can be reduced from general digraphs to DAGs, and prove the following theorem.

\begin{theorem} \label{thmi: general1}
Given a $k$-terminal digraph G with a pair-set $P$, there exists a polynomial time algorithm that outputs a reachability-preserving minor $H$ of size $O(|P|^2)$ with respect to $P$.
\end{theorem}
Taking $P$ to be the trivial pair-set, we get a reachability-preserving minor of size $O(k^{4})$. 

\begin{proof}[Proof of Theorem~\ref{thmi: general1}]
Recall that a digraph is \emph{strongly connected} if there is a directed path between all pair of vertices. We proceed by first finding a decomposition of the graph into strongly connected components (SCCs)~\cite{Tarjan72}. We observe that each SCC that contains terminals can be contracted into a smaller component only on the terminals. Then contracting each SCC into a single vertex to obtain a DAG and invoking Algorithm~\ref{algo: minor} on the resulting DAG gives some intermediate reachability-preserving minor. Finally, we show that this minor can be expanded back to produce a reachability-preserving minor for the original digraph. These steps are formally given in the procedure Algorithm~\ref{algo: minorSparsify}.

\begin{algorithm2e}[t]
\label{algo: minorSparsify}
	\caption{\textsc{MinorSparsify} $(G,P)$}
\Input{$k$-terminal digraph $G$, pair-set $P$}
\Output{Reachability preserving minor $H$ of $G$ with respect to $P$}
Compute a strongly connected component decomposition $\mathcal{D}$ of $G$ \;
Let $f$ be some initially empty labelling that records the SCC of every vertex \;
\For{each \normalfont{SCC} $C \in \mathcal{D}$} 
{
		\eIf{$C$ contains some terminal $x \in K$}
		{For all $v \in C$, set $f(v) \gets x$ \; }
		{Choose some arbitrary $u \in C$, and set $f(v) \gets u$, for all $v \in C$. \;}
}
\tcp{Preprocessing Step}
Let $\mathcal{D}_{K}$ denote the set of SCCs containing terminals in $G$ \;
\For{all \normalfont{SSC} $C \in \mathcal{D}_{K}$} 
{ \While{$C$ contains some non-terminal $v$}
       {
		\label{alg:ms_13} Choose some directed edge $(v,u)$ leaving $v$ inside $C$, and contract $v$ into $u$ \; 
	   }
}
Let $\hat{G}=(\hat{V}, \hat{E})$ and $\hat{\mathcal{D}}$ denote the resulting graph and the SCC decomposition \;
\tcp{Main Procedure}
Contract each SSC in $\hat{\mathcal{D}}$ into a single vertex, producing the DAG $G'=(V',E')$ \;
Let $K' \gets \emptyset$ and $P' \gets \emptyset$ be the terminal set and pair-set of $G'$, respectively \;
For all $k \in K$, add $f(k)$ to $K'$ and remove duplicates, if any \;
For all $(s,t) \in P$, add $(f(s),f(t))$ to $P'$ if $f(s) \neq f(t)$ \;
Set $H' =$\textsc{MinorSparsifyDag}($G',P'$) \;
\label{alg:ms_23} Let $H$ be the graph obtained by expanding back all contracted SCCs in $\hat{\mathcal{D}}_{K}$ in $H'$ \;
\Return $H$
\end{algorithm2e}

The main intuition behind the correctness of the above reduction lies on two important observations. First, vertices belonging to the same strongly connected components can always reach each other. Second, vertices belonging to different strongly connected components can reach each other if the corresponding vertices in the contracted graph can do so. 
We have the following useful observation.

\begin{fact} \label{lemma: edgecontraction}
	For any strongly connected digraph $G=(V,E)$, contracting any edge $e \in E$ results in another strongly connected digraph $G'=(V',E')$.
\end{fact}

Now we show that the graph $H$ output by \textsc{MinorSparsify} is a reachability-preserving minor of $G$. It is easy to verify that the produced graph $H$ is indeed a minor of $G$. To show the correctness, we will prove that $H$ preserves the reachability information among all pairs from $P$ in $G$. Before doing that, observe that the graph $\hat{G}$ obtained after the preprocessing step is a reachability preserving minor of $G$ with respect to $P$. Indeed, this can be inferred by a repeated application of Fact~\ref{lemma: edgecontraction} to each SSC containing terminal vertices.
	
	Now, let $(s,t) \in P$ be any terminal-pair in $G$. Assume that $t$ is reachable from $s$ in $G$. We distinguish two cases:
	\begin{enumerate}
		\item If $s$ and $t$ belong to the same SCC in $\mathcal{D}$, they do also belong to the corresponding SCC in $\hat{\mathcal{D}}$. In Line \ref{alg:ms_13}, $s$ and $t$ are contracted into a single terminal. However, since the contracted SSC contains terminals, it is expanded back to its original form in $\hat{\mathcal{D}}$ in Line \ref{alg:ms_23}. Thus, it follows that $t$ is reachable from $s$ in the output graph $H$. 
		\item If $s$ and $t$ do not belong to the same SCC in $\mathcal{D}$, they must also not belong to the same SCC in $\hat{\mathcal{D}}$. Let $f(s)$ and $f(t)$ denote the terminals in the DAG $G'$ obtained by contracting their corresponding components in $\hat{\mathcal{D}}$ (Line \ref{alg:ms_13}). Since $t$ is reachable from $s$ in $\hat{G}$, note that $f(t)$ must also be reachable from $f(s)$ in $G'$. By Lemma \ref{lemma: DAG}, it follows that $f(t)$ is reachable from $f(s)$ in the reachability-preserving minor $H'$ of $G'$. Expanding back the SCCs that contain terminals in $H'$ (Line \ref{alg:ms_23}), we can construct the directed path $s \rightsquigarrow f(s) \rightsquigarrow f(t) \rightsquigarrow t$ in $H$, which shows that $t$ is also reachable from $s$ in the output graph $H$.
	\end{enumerate}
	When $t$ is not reachable from $s$ in $G$, we can similarly show that $t$ is also not reachable from $s$ in $H$, thus concluding the correctness proof.
	
	We now bound the size of $H$. Since the DAG $G'$ has $|P'| \leq |P|$ pairs, it follows by Lemma \ref{lemma: DAG} that $H'$ has size at most $O(|P|^2)$. After expanding back the SCCs in Line 19, we get that each SSC in $H$ contains at most $k_i$ terminals, where $k = \sum_{i} k_i$. Note that this does not contribute to the size of $H$. Therefore, we get that the size of the output graph $H$ is at most $O(|P|^2)$.
\end{proof}

\subsection{An Improved Bound of $O(k^3)$}
Using the recent work due to Abboud and Bodwin~\cite{AB18reachability}, we next show how to get a polynomial improvement on the number of branching events from Lemma~\ref{lemm: branching}. This in turn gives a polynomial improvement on the size of reachability-preserving minor from Theorem~\ref{thmi: general1}.

Specifically, given a $k$-terminal DAG $G$ with a pair-set $P$, let $H=(V,E')$ be the subgraph of $G$ with minimum number of edges that preserves all reachability information among the pairs in $P$. We call such an $H$ the \emph{sparsest reachability preserver} of $G$.  The following lemma is implicit in~\cite{AB18reachability}, and we include it here for the sake of completeness.

\begin{lemma} The \emph{DAG} $H=(V,E')$ has at most $k \cdot |P|$ branching events.
\end{lemma}
\begin{proof}
For each pair $(s,t) \in P$, we associate a directed path $s \rightsquigarrow t$ in $H$, and let $\tilde{\pi}(s,t)$ denote such a path. Note that since $H$ is acyclic, every $\tilde{\pi}(s,t)$ is acyclic as well. Moreover, using the fact that $H$ is the sparsest reachability preserver, it follows that for every edge $e \in E'$, there must be some pair $(s,t) \in P$ such that deleting $e$ from $H$ implies that $s$ cannot reach $t$, i.e., $s \not \rightsquigarrow t$ in $H \setminus \{e\}$. This naturally leads to a relationship between edges and pairs. Specifically, we say that every edge $e \in E'$ is \emph{owned} by one such pair $(s,t) \in P$. 

Next, for each $(s,t) \in P$, we let $B^{H}_{(s,t)}$ denote the set of all branching events $\{e_1,e_2\}$ in $H$ such that either $e_1$ or $e_2$ (but not both) is owned by $(s,t)$. We claim that $\bigcup \set{B^{H}_{(s,t)}}{(s,t) \in P}$ contains all branching events in $H$. Indeed, suppose towards contradiction that $\{e_1,e_2\}$ is a branching event in $H$ but not in $\bigcup \set{B^{H}_{(s,t)}}{(s,t) \in P}$. Then by definition of $B^{H}_{(s,t)}$ there must be some pair $(s,t) \in P$ such that $e_1$ and $e_2$ are both owned by $(s,t)$. The latter implies that we can construct two directed paths from $s$ to $t$, where one path uses $e_1$ and the other uses $e_2$. Delete edge $e_1$ w.l.o.g. Then we still have another directed path from $s$ to $t$, thus contradicting the assumption that $e_1$ is owned by $(s,t)$. 

Now, to prove the lemma it suffices to show that $\abs{B^{H}_{(s,t)}} \leq k$, for every $(s,t) \in P$. Suppose towards contradiction that there exists a pair $(s,t) \in P$ such that $\abs{B^{H}_{(s,t)}} \geq k+1$.  Then by the pigeonhole principle, there exist two branching events 
\[
	\{(x_1,b_1),(x_2,b_1)\},~\{(y_1,b_2),(y_2,b_2)\} \in B^{H}_{(s,t)}
\]
entering the nodes $b_1$ and $b_2$, such that $(s,t)$ owns $(x_1,b_1)$ and $(y_1,b_2)$, and the other edges are owned by pairs that share a common left terminal, i.e., 
\[
	(x_2,b_1) \text{ is owned by } (u,v_1) \text{ and } (y_2,b_2) \text{ is owned by } (u,v_2)
\]
for some $u \in K$ and $(u,v_1), (u,v_2) \in P$. Note that by the definition of $B^H_{(s,t)}$, $y_1$ and $y_2$ are different vertices. We further assume w.l.o.g. that node $b_1$ appears before $b_2$ in $\tilde{\pi}(s,t)$.
Now, since the pair $(u,v_2)$ owns the edge $(y_2,b_2)$, every path $u \rightsquigarrow v_2$ must use the edge $(y_2,b_2)$, which further implies that every path $u \rightsquigarrow b_2$ must use the edge $(y_2,b_2)$. We can form a path $u \rightsquigarrow b_2$ by first taking the path $\tilde{\pi}(u,v_1)[u \rightsquigarrow b_1]$\footnote{Let $x,y,x',y' \in V$, $\tilde{\pi}(x,y)$ be a directed path from $x$ to $y$, and suppose $x',y' \in \tilde{\pi}(x,y)$ with $x'$ appearing before $y'$. Then $\tilde{\pi}(x,y)[x' \rightsquigarrow y']$ denotes the directed subpath from $x'$ to $y'$ in $\tilde{\pi}(x,y)$.} and then extend it by concatenating it with the path $\tilde{\pi}(s,t)[b_1 \rightsquigarrow b_2]$. This implies one of the following cases: (1) $(y_2,b_2) \in \tilde{\pi}(s,t)[b_1 \rightsquigarrow b_2]$ or (2) $(y_2,b_2) \in \tilde{\pi}(u,v_1)[u,b_1]$. We show that (2) cannot happen, thus only (1) holds. To this end, suppose towards contradiction that $(y_2,b_2) \in \tilde{\pi}(u,v_1)[u,b_1]$. Then we can find a directed path $b_2 \rightsquigarrow b_1$. But since $b_1$ appears before $b_2$, we get the cycle $b_2 \rightsquigarrow b_1 \rightsquigarrow b_2$, which contradicts the fact that $H$ is acyclic. 

Finally, case (1) implies that $(y_2,b_2) \in \tilde{\pi}(s,t)$. Therefore, the path $\tilde{\pi}(s,t)$ contains both $(y_1,b_2)$ and $(y_2,b_2)$. On the other hand, since $\tilde{\pi}(s,t)$ is acyclic, there cannot be two vertices entering $b_2$, which is a contradiction.  
\end{proof}

The above lemma leads to the following algorithm.

\begin{algorithm2e}[H]
\label{algo: minor2_Reach}
\caption{\textsc{MinorSparsifyDag2} $(G, P)$}
\Input{$k$-terminal DAG $G$, pair-set $P$}
\Output{Reachability preserving minor $H$ of $G$ with respect to $P$}
Set $H=(V,E')$ be the sparsest reachability preserver with respect to $P$ \;
Remove isolated non-terminal vertices from $H$, if any \;
For each $p \in P$,  add the shortest path $\pi(p)$ to $H$ \;
\While{there is an edge $(u,v)$ directed towards a non-terminal $v$ with $\deg_{H}^{-}(v) = 1$}
{ Contract the edge $(u,v)$ \;
}
\Return $H$
\end{algorithm2e}

We remark that the above construction is built upon the sparest reachability preserver $H$, which we can find in exponential time (say, by a brute-force approach). By using similar arguments as in the proof of Lemma~\ref{lemma: DAG} and Theorem~\ref{thmi: general1}, we have the following guarantees.
\begin{lemma} Given a $k$-terminal \emph{DAG} $G$ with a pair-set $P$, \emph{Algorithm \ref{algo: minor2_Reach}} outputs a reachability-preserving minor $H$ of size $O(k \cdot \abs{P})$ for $G$ with respect to $P$.
\end{lemma}

\begin{theorem} \label{thmi: general}
Given a $k$-terminal digraph G with a pair-set $P$, there exists an algorithm that outputs a reachability-preserving minor $H$ of size $O(k \cdot \abs{P})$ with respect to $P$.
\end{theorem}
Taking $P$ to be the trivial pair-set we get a reachability-preserving minor of size $O(k^{3})$, which proves Theorem~\ref{thmi: general_reachability}. We note that in contrast to Theorem~\ref{thmi: general1}, the above theorem guarantees only an exponential-time algorithm in the worst-case. As discussed above, this comes from the assumption that we have access to the sparsest reachability preserver. It is conceivable that a similar approach that appears in~\cite{AB18reachability} could be employed to achieve a better running-time. However, the focus of our work is on optimizing the size of reachability-preserving minors.

\section{Reachability-Preserving Minors for Planar Digraphs}
In this section we show that any $k$-terminal planar digraph $G$ admits a reachability-preserving minor of size $O(k^{2} \log{k})$ and thus prove Theorem~\ref{thm: ubPlanar}. This matches the lower-bound of Theorem~\ref{thm: lbPlanar} up to an $O(\log {k})$ factor. The main idea is as follows. Given a $k$-terminal planar digraph $G$ with the trivial pair-set $P$, $|P|=k(k-1)$, our goal will be to slightly increase the number of terminals while considerably reducing the size of the pair-set $P$, under the condition that no reachability information is lost among the terminal-pairs in $P$.

\vspace{-0.3cm}
\paragraph*{Preprocessing Step.} Given a $k$-terminal digraph $G$, we apply Theorem~\ref{thmi: general_reachability} to get a reachability-preserving minor $G'$. To simplify the notation, we will use $G$ instead of $G'$, i.e., throughout we assume that $G$ has at most $O(k^{3})$ vertices. 
\vspace{-0.3cm}
\paragraph{Decomposition into Path-Separable Digraphs and the Algorithm.} We say that a graph $G=(V,E)$ admits an \emph{$\alpha$-separator} if there exists a set $S \subset V$ whose removal partitions $G$ into connected components, each of size at most $\alpha \cdot |V|$, where $1/2 \leq \alpha < 1$. If the vertices of $S$ consist of the union over $r$ paths of $G$, for some $r \geq 1$, we say that $G$ is $(\alpha, r)$-\emph{path separable}. We now review the following reduction due to Thorup~\cite{ThorupJACM04}.
\begin{theorem}[\cite{ThorupJACM04}] \label{thm: thorup}
Given a digraph $G$, we can construct a series of digraphs $G_0,\ldots,G_{b}$ for some $b\leq n$ such that the number of vertices and edges over all $G_i$'s is linear in the number of vertices and edges in $G$, and
\begin{enumerate}
	\setlength{\itemsep}{2pt}
	\setlength{\parskip}{2pt}
	\item Each vertex and edge of $G$ appears in at most two $G_i$'s.
	\item\label{item:thorup_2} For all $u,v \in V$, if there is a dipath $R$ from $u$ to $v$ in $G$, there is a $G_i$ that contains $R$.  
	\item\label{item:thorup_3} Each $G_i=(V_i,E_i)$ is $(1/2,6)$-path separable.
	\item\label{item:thorup_4} Each $G_i$ is a minor of $G$. In particular, if $G$ is planar, so is $G_i$.
\end{enumerate}
\end{theorem}

Now we review how directed reachability can be efficiently represented by separator dipaths. Let $G$ be a $k$-terminal directed graph $G$ that contains some directed path $Q$. Assume that the vertices of $Q$ are ordered in increasing order in the direction of $Q$. For each terminal $x \in K$, let $\texttt{to}_x[Q]$ be the first vertex in $Q$ that can be reached by $x$, and let $\texttt{from}_x[Q]$ be the last vertex in $Q$ that reaches $x$. Let $(s,t)$ be a terminal pair and let $R$ be the directed path from $s$ to $t$ in $G$. We say that $R$ intersects $Q$ iff $s$ can reach $\texttt{to}_s[Q]$  and $t$ can be reached from $\texttt{from}_t[Q]$ in $Q$, and $\texttt{to}_s[Q]$ precedes $\texttt{from}_t[Q]$ in $Q$. 

We now are going to combine the above tools to give our labelling algorithm aimed at reducing the size of the trivial pair-set $P$. By Theorem~\ref{thm: thorup}, we restrict our attention only to the digraphs $G_i$. Let $K_i := V(G_i) \cap K$ be the set of terminals restricted to the graph $G_i$.
\begin{algorithm2e}[t]
\label{alg:reducepairset}
\caption{\textsc{ReducePairSet} $(G_i, K_i)$}
\Input{planar digraph $G$, terminals $K_i$}
\Output{Pair-set $P_i$ with respect to $K_i$}
\If{$|V(G_i)| \leq 1$ or $K_i = \emptyset$}{ \Return $\emptyset$ \; }
Let $P_i' \gets \emptyset$ be a new pair-set \; 
Compute a $1/2$-separator $S$ of $G_i$ consisting of $6$ dipaths by Item \ref{item:thorup_4} of Theorem~\ref{thm: thorup} \;
\For{each dipath $Q \in S$}
{
		\tcp{Addition of terminal connections with $Q$}
		Let $Q'$ be the set of existing terminals of $Q$ \;
     	\For{each terminal $x \in K_i$}
     	{
          \label{alg:rps_8} Compute $\texttt{to}_x[Q]$ and $\texttt{from}_x[Q]$ \;  Declare $\texttt{to}_x[Q]$ and $\texttt{from}_x[Q]$ terminals and add them to $Q'$ \;
          \label{alg:rps_9} Add $(x, \texttt{to}_x[Q])$ and $(\texttt{from}_x[Q],x)$ to $P_i'$ \;
        }
        \tcp{Sparsification of $Q$ using $Q'$}
        Remove all vertices in $Q \setminus Q'$ \; 
		\label{alg:rps_12} Define directed pairs $(s,t)$, where $s$ and $t$ are consecutive terminals of $Q'$, \; 
		\nonl according to the ordering of $Q$ and add all these pairs to $P_i'$\;
}
     Let $(G_i^{(1)},K_i^{(1)})$ and $(G_i^{(2)},K_i^{(2)})$ be the resulting graphs from $G \setminus S$, \;
     \nonl where $K_i^{(1)}$ and $K_i^{(2)}$ are disjoint subsets of the terminals $K$ separated by $S$ \;
     \tcp{Note that reachability info. about terminals in $S$ are taken care of.}
 	  \Return $P_i' \cup \bigcup_{j=1}^{2}\textsc{ReducePairSet}(G_i^{(j)},K_i^{(j)})$ \;
\end{algorithm2e}

\begin{lemma} Let $G$ be a $k$-terminal planar digraph. Let $P' := \cup_{i=0}^{b} P_i'$ be the union over all pair-sets output by running \emph{Algorithm \ref{alg:reducepairset}} on each digraph $G_i$. Then the size of $|P'|$ is at most $O(k \log k)$. Moreover, if $H$ is a reachability-preserving minor of $G$ with respect to $P'$, then $H$ is a reachability-preserving minor of $G$ with respect to all terminal pairs. 
\end{lemma} 
\begin{proof}
	By preprocessing, $G$ has at most $O(k^{3})$ vertices. Throughout, it will be useful to think of the above algorithm as simultaneously running it on each digraph $G_i$. By Item \ref{item:thorup_2} of Theorem~\ref{thm: thorup}, each terminal appears in at most two $G_i$'s. Thus at each recursive level, there will be at most $O(k)$ active $G_i$'s. Also, note that the separator properties imply that there are $O(\log k)$ recursive calls overall.
	
	We next bound the size of the pair-set $P'$. Let $q$ denote the total number of newly added terminals in Line \ref{alg:rps_8} per recursive level. Since there are $O(k)$ terminals, each adding at most $O(1)$ new terminals, it follows that $q=O(k)$. First, we argue about the number of pairs added in Line \ref{alg:rps_9}. Since this is bounded by $O(q)$, it follows that there are $O(k \log k)$ pairs overall. Second, we bound the number of pairs added when sparsifying the separator paths, i.e., pair additions in Line \ref{alg:rps_12}. For all the separators in the same recursive level, we can write $q := \sum_{i}{|Q'_j|}$, where $Q_j'$ denotes the set newly added terminals for some separator dipath. By Line \ref{alg:rps_12}, it follows that we need only $(|Q'_j| - 1)$ pairs to represent each such dipath. Thus, per recursive call, the total number of newly added pairs is $O(q) = O(k)$. Summing these overall $O(\log k)$ levels, and combining this with the previous bound, gives the claimed bound on $|P'|$.
	
	Finally, we argue that $P'$ is a pair-set that can recover reachability information among terminals. Fix any terminal pair $(s,t)$ and let $R$ be a directed path from $s$ to $t$ in $G$. By Item \ref{item:thorup_2} of Theorem~\ref{thm: thorup}, there is some digraph $G_i$ that contains $R$. Then, $R$ must intersect with some separator dipath $Q$, at some level of the recursion of the above algorithm on $G_i$. The above argument gives that $P'$ contains all the necessary information to give a (possibly) another directed path from $s$ to $t$ in $G$.
\end{proof}
Applying Theorem~\ref{thmi: general} on the digraph $G$ with pair-set $P'$, as defined by the above lemma, we get Theorem~\ref{thm: ubPlanar}.

\subsection{Lower-bound for Planar DAGs} \label{app: minorLB}
In this section we prove that there exists an infinite family of $k$-terminal acyclic directed grids such that any reachability-preserving minor for such graphs needs $\Omega(k^2)$ non-terminals (i.e., prove Theorem~\ref{thm: lbPlanar}). We achieve this by adapting the ideas of Krauthgamer et al.~\cite{KrauthgamerNZ14}, from their lower-bound proof on distance-preserving minors for undirected graphs.

We start by defining of our lower-bound instance. Fix $k$ such that $r=k/4$ is an integer. Construct an initially undirected $(r+1) \times (r+1)$ grid, where all the $k$ terminals lie on the boundary, except at the corners, and declare all non-boundary vertices non-terminals. Remove the four corner vertices, and then all boundary edges connecting the terminals. Now, make the graph directed by first directing each horizontal edge from left to right, and then directing each vertical edge from top to bottom. Let $G$ denote the resulting $k$-terminal directed grid. It is easy to verify that $G$ is acyclic. 

\begin{theorem}
	For infinitely many $k \in \mathbb{N}$ there exists a $k$-terminal acyclic directed grid $G$ such that any reachability-preserving minor of $G$ must use $\Omega(k^{2})$ non-terminals.
\end{theorem}
\begin{proof}
	Let $G$ be the $k$-terminal grid defined as above. Note that there are $r$ terminals on each side of the grid. Let $H$ be any reachability-preserving minor of $G$. Recall that $H$ contains all terminal vertices from $G$. Furthermore, let $x_1,x_2,\ldots,x_{r}$ be the terminals on the left-side of the grid, ordered from top to bottom. Similarly, define $y_1,y_2,\ldots,y_{r}$ to be the terminals on the right-side. Note that by construction of $G$, for an index pair $(i,j)$ with $i < j$, there is no directed path from $x_j$ to $y_i$. Finally, define $P^{i}_H$ to be the directed path from $x_i$ to $y_i$ in $H$, for $i=1,\ldots,r$. 
	Throughout we will refer to such paths as \emph{horizontal}.
	
	\begin{claim} \label{lemm: horizontal} The horizontal directed paths $P^{1}_H,P^{2}_H,\ldots,P^{r}_H$ are vertex disjoint in $H$.
	\end{claim}
	\begin{proof}
		Suppose towards contradiction that there exist some $i$ and $j$ with $i < j$ such that $P^{i}_H$ and $P^{j}_H$ intersect at some vertex $z$ in $H$. This implies that there are directed paths from $x_i$ and $x_j$ to $z$, and from $z$ to $y_i$ and $y_j$. The latter implies that there is a directed path from $x_j$ to $y_i$ in $H$. However, by construction of $G$, we know that $x_j$ cannot reach $y_i$ for $i < j$, contradicting the fact that $H$ is a reachability-preserving minor of $G$.
	\end{proof}
	We can apply symmetric argument to the \emph{vertical} paths in $H$. More specifically, define $u_1,u_2,\ldots,u_{r}$ to be the terminal on the top-side of the grid, order from left to right. Similarly, define $v_1,v_2,\ldots,u_{r}$ to be the terminals on the bottom-side. Note that by construction of $G$, for an index pair $(i,j)$ with $i<j$, there is no directed path from $u_j$ to $v_i$. Finally, define $Q^{i}_H$ to be the directed path from $u_i$ to $v_i$ in $H$, for $i=1,\ldots,r$. Then we get the following symmetric claim.
	\begin{claim} \label{lemm: vertical} The vertical directed paths $Q^{1}_H,Q^{2}_H,\ldots,Q^{r}_H$ are vertex disjoint in $H$.
	\end{claim} 
	We next argue that all the horizontal and the vertical paths must intersect with each other. 
	\begin{claim} \label{lemm: intersect} Any pair of horizontal and vertical paths $P^{i}_H$ and $Q^{j}_H$ intersect in $H$.
	\end{claim}
	\begin{proof}
		Since $H$ is a minor of $G$, any dipath that connects two terminals in $H$ can be mapped back to a dipath connecting two terminals in $G$. Let $P_i$ and $Q_j$ be the corresponding dipaths in $G$ that are obtained by expanding back the dipaths $P^{i}_H$ and $Q^{j}_H$ in $H$. By construction of $G$, the horizontal and vertical dipaths between terminals are unique, implying that $P_i$ and $Q_j$ must intersect at some vertex of $G$. By performing the backtracked minor-operations on this vertex yields an intersection vertex between $P^{i}_H$ and $Q^{j}_H$ in $H$.
	\end{proof}
	The last claim we need shows that no pair of horizontal and the vertical paths intersects intersect at a terminal vertex.
	\begin{claim} \label{lemm: noterminal}
		No pair of horizontal and vertical paths $P^{i}_H$ and $Q^{j}_H$ intersects at a terminal vertex in $G$.
	\end{claim}
	\begin{proof}
		Consider the terminal pairs $(x_i,y_i)$ and $(u_j,v_j)$ corresponding to the paths $P^{i}_H$ and $Q^{j}_H$. Note that by construction of $G$, the set of terminals reachable from both $x_i$ and $u_j$ in $G$ is $\{y_i, y_{i+1}, \ldots, y_{r}\} \cup \{v_j, v_{j+1}, \ldots, v_{r}\}$. Since $H$ is a reachability-preserving minor of $G$, $x_i$ and $u_j$ must also be able to reach this terminal-set in $H$ and also $P^{i}_H$ and $Q^{j}_H$ cannot intersect \sloppy at any terminal in $\{y_1,\ldots,y_{i-1}\} \cup \{v_1,\ldots,v_{j-1}\}$. Now, suppose towards contradiction that $P^{i}_H$ and $Q^{j}_H$ intersect at some terminal $y_k$, for $k \in \{i+1,\ldots,r\}$. This implies that in the path $P^{i}_H$, there is a directed path from $y_k$ to $y_i$, for $k > i$, giving a contradiction by construction of $G$. Furthermore, observe that $P^{i}_H$ and $Q^{j}_H$ cannot intersect at $y_i$, as otherwise we would have a directed path from $y_i$ to $v_j$, which is a contradiction by construction of $G$. Applying a similar argument to the case when paths intersect at some terminal $v_\ell$, for $k \in \{j+1,\ldots,r\}$, gives the claim.
	\end{proof}
	We know have all the necessary tools to prove the theorem. Claim~\ref{lemm: intersect} shows that the paths $P^{i}_H$ and $Q^{j}_H$ intersect in $H$ and let $z_H^{i,j}$ denote one of the intersection vertices. Now, we must show that all these vertices are distinct. To this end, assume that $z_{H}^{i_1,j_1} = z_{H}^{i_2,j_2}$. Since these vertices belong to both $P^{i_1}_{H}$ and $P^{i_2}_{H}$, by Claim~\ref{lemm: horizontal} we get that $i_1 = i_2$. Similarly, by Claim~\ref{lemm: vertical} we get that $j_1 = j_2$. Thus, we have that all vertices $z_{H}^{i,j}$, for $i,j=1,2,\ldots, r$ are distinct. Since Claim~\ref{lemm: noterminal} implies that none of this intersection vertices is a terminal, we conclude that $H$ must contain at least $r^{2} = (k/4)^{2}$ non-terminals.
\end{proof}

\section{An Exact Cut Sparsifier of Size $O(k^2)$} \label{sec: upperCut}
In this section we show that given a $k$-terminal planar graph, where all terminals lie on the same face, one can construct a quality-$1$ cut sparsifier of size $O(k^{2})$. Note that it suffices to consider the case when all terminals lie on the \emph{outer} face. 
We first present some basic tools. 
\subsection{Basic Tools}
\paragraph*{Wye-Delta Transformations. }
In this section we investigate the applicability of some graph reduction techniques that aim at reducing the number of non-terminals in a $k$-terminal graph. 
We start by reviewing the so-called \emph{Wye-Delta} operations in graph reductions. These operations consist of five basic rules, which we describe below. (See Fig.~\ref{fig:wyedelta} for illustrations.)
\begin{enumerate}
	\setlength\itemsep{0.1em}
	\item \emph{Degree-one reduction:} Delete a degree-one non-terminal and its incident edge.
	\item \emph{Series reduction:} Delete a degree-two non-terminal $y$ and its incident edges $(x,y)$ and $(y,z)$, and add a new edge $(x,z)$ of capacity $\min\{\cc(x,y), \cc(y,z)\}$.
	\item \emph{Parallel reduction:} Replace all parallel edges by a single edge whose capacity is the sum over all capacities of parallel edges.
	\item \emph{Wye-Delta transformation:} Let $x$ be a degree-three non-terminal with neighbours $\delta(x) = \{u,v,w\}$. Assume w.l.o.g.\footnote{Suppose there exist a pair $(u,v) \in \delta(x)$ with $\cc(u,x) + \cc(v,x) < \cc(w,x)$, where $w \in \delta(v)\setminus \{u,v\}$. Then we can simply set $\cc(w,x) = \cc(u,x) + \cc(v,x)$, since any terminal minimum cut would cut the edges $(u,x)$ and $(v,x)$ instead of the edge $(w,x)$.} that for any pair $(u,v) \in \delta(x)$, $\cc(u,x) + \cc(v,x) \geq \cc(w,x)$, where $w \in \delta(v)\setminus \{u,v\}$. Then we can delete $x$ (along with all its incident edges) and add edges $(u,v),(v,w)$ and $(w,u)$ with capacities $(\cc(u,x)+\cc(v,x)-\cc(w,x))/2$, $(\cc(v,x)+\cc(w,x)-\cc(u,x))/2$ and $(\cc(u,x)+\cc(w,x)-\cc(v,x))/2$, respectively. 
	\item \emph{Delta-Wye transformation:} Delete the edges of a triangle connecting $x$, $y$ and $z$, introduce a new non-terminal vertex $w$ and add new edges $(w,x)$, $(w,y)$ and $(w,z)$ with edge capacities $\cc(x,y) + \cc(x,z),$ $\cc(x,y) + \cc(y,z)$ and $\cc(x,z) + \cc(y,z)$ respectively.
\end{enumerate}

\begin{figure}[t]
\centering
\scalebox{.8}{
\begin{tikzpicture}
\tikzstyle{vertex}=[circle,draw = white, fill=black, minimum size = 7pt, inner sep=2pt]
\tikzstyle{vertex1}=[fill = white, draw = white]
\tikzstyle{edge}=[-,thick ]
\tikzstyle{elipse}=[-,thick ]
\tikzstyle{vertex2}=[circle,draw = black, fill=white, minimum size = 7pt, inner sep=2pt]

  \node[vertex2] (x) at (-0.75,0.75) {$1$};

  \node[vertex1] (n1) at (0,0) {} ;
  \node[vertex1] (n2) at (0.5,0.5) {} ;
  \node[vertex1] (n3) at (0,1) {} ;
  
  \node[vertex] (n4) at (2.5,1.5) {};
    
  \node[vertex] (n5) at (2.5,0.5) {} ;
  \node[vertex1] (n6) at (2,0) {} ;
  \node[vertex1] (n7) at (3,0) {} ;

  \draw[edge] (n4) node[above = 2.5pt] {$y$} -- (n5)  ;

  \draw[edge] (n6) -- (n5) ;
  \draw[edge] (n5) -- (n7) ;

  \node[vertex1] (n11) at (4, 0.5) {};
  \node[vertex1] (n12) at (5, 0.5) {};
  \draw[edge,->,very thick,>=stealth] (n11) -- (n12) ;
  
  \node[vertex1] (n1a) at (6,0) {} ;
  \node[vertex] (n2a) at (6.5,0.5) {} ;
  \node[vertex1] (n3a) at (7,0) {} ;

  \node[vertex1] (n5a) at (8.5,0.5) {} ;
  \node[vertex1] (n6a) at (9,1) {} ;
  \node[vertex1] (n7a) at (9,0) {} ;

  \draw[edge] (n1a) -- (n2a) ;
  \draw[edge] (n2a) -- (n3a) ;
  


\end{tikzpicture}
}

\scalebox{.8}{

\begin{tikzpicture}
\tikzstyle{vertex}=[circle,draw = white, fill=black, minimum size = 7pt, inner sep=2pt]
\tikzstyle{vertex1}=[fill = white, draw = white]

\tikzstyle{edge}=[-,thick ]
\tikzstyle{elipse}=[-,thick ]
\tikzstyle{vertex2}=[circle,draw = black, fill=white, minimum size = 7pt, inner sep=2pt]

  \node[vertex2] (x) at (-0.75,0.5) {$2$};
  \node[vertex1] (n1) at (0,0) {} ;
  \node[vertex] (n2) at (0.5,0.5) {} ;
  \node[vertex1] (n3) at (0,1) {} ;
  
  \node[vertex] (n4) at (1.5,0.5) {};
    
  \node[vertex] (n5) at (2.5,0.5) {} ;
  \node[vertex1] (n6) at (3,1) {} ;
  \node[vertex1] (n7) at (3,0) {} ;
  
  \draw[edge] (n1) -- (n2) node[above = 2.5pt] {$x$} ;
  \draw[edge] (n2) -- (n3) ;
  
  \draw[edge] (n2) -- (n4) node[above = 2.5pt] {$y$} ;
  \draw[edge] (n4) -- (n5) node[above = 2.5pt] {$z$};

  \draw[edge] (n6) -- (n5) ;
  \draw[edge] (n5) -- (n7) ;

  \node[vertex1] (n11) at (4, 0.5) {};
  \node[vertex1] (n12) at (5, 0.5) {};
  \draw[edge,->,very thick,>=stealth] (n11) -- (n12) ;
  
  \node[vertex1] (n1a) at (6,0) {} ;
  \node[vertex] (n2a) at (6.5,0.5) {} ;
  \node[vertex1] (n3a) at (6,1) {} ;

  \node[vertex] (n5a) at (8.5,0.5) {} ;
  \node[vertex1] (n6a) at (9,1) {} ;
  \node[vertex1] (n7a) at (9,0) {} ;

  \draw[edge] (n1a) -- (n2a) ;
  \draw[edge] (n2a) -- (n3a) ;
  
  \draw[edge] (n2a) node[above = 2.5pt] {$x$} -- (n5a) node[above = 2.5pt] {$z$};

  \draw[edge] (n5a) -- (n6a) ;
  \draw[edge] (n5a) -- (n7a) ;
 
\end{tikzpicture}
}

\scalebox{.8}{

\begin{tikzpicture}
\tikzstyle{vertex}=[circle,draw = white, fill=black, minimum size = 7pt, inner sep=2pt]
\tikzstyle{vertex1}=[fill = white, draw = white]

\tikzstyle{edge}=[-,thick ]
\tikzstyle{elipse}=[-,thick ]
\tikzstyle{vertex2}=[circle,draw = black, fill=white, minimum size = 7pt, inner sep=2pt]

  \node[vertex2] (x) at (-0.75,0.5) {$3$};
  \node[vertex1] (n1) at (0,0) {} ;
  \node[vertex] (n2) at (0.5,0.5) {} ;
  \node[vertex1] (n3) at (0,1) {} ;
    
  \node[vertex] (n5) at (2.5,0.5) {} ;
  \node[vertex1] (n6) at (3,1) {} ;
  \node[vertex1] (n7) at (3,0) {} ;
  
  \draw[edge] (n1) -- (n2) ;
  \draw[edge] (n2) -- (n3) ;
  
  \path (n2) edge [bend left] node {} (n5) ;
  \path (n2) edge [bend right] node {} (n5) ;

  \draw[edge] (n6) -- (n5) ;
  \draw[edge] (n5) -- (n7) ;

  \node[vertex1] (n11) at (4, 0.5) {};
  \node[vertex1] (n12) at (5, 0.5) {};
  \draw[edge,->,very thick,>=stealth] (n11) -- (n12) ;
  
  \node[vertex1] (n1a) at (6,0) {} ;
  \node[vertex] (n2a) at (6.5,0.5) {} ;
  \node[vertex1] (n3a) at (6,1) {} ;

  \node[vertex] (n5a) at (8.5,0.5) {} ;
  \node[vertex1] (n6a) at (9,1) {} ;
  \node[vertex1] (n7a) at (9,0) {} ;

  \draw[edge] (n1a) -- (n2a) ;
  \draw[edge] (n2a) -- (n3a) ;
  
  \draw[edge] (n2a) -- (n5a);

  \draw[edge] (n5a) -- (n6a) ;
  \draw[edge] (n5a) -- (n7a) ;
 
\end{tikzpicture}
}

\scalebox{.8}{

\begin{tikzpicture}
\tikzstyle{vertex}=[circle,draw = white, fill=black, minimum size = 7pt, inner sep=2pt]
\tikzstyle{vertex1}=[fill = white, draw = white]

\tikzstyle{edge}=[-,thick ]
\tikzstyle{elipse}=[-,thick ]
\tikzstyle{vertex2}=[circle,draw = black, fill=white, minimum size = 7pt, inner sep=2pt]

  \node[vertex2] (x) at (-0.75,1.5) {$4$};

  \node[vertex1] (n1) at (0,0) {} ;
  \node[vertex] (n2) at (0.5,0.5) {} ;
  \node[vertex1] (n3) at (0,1) {} ;
  
  \node[vertex] (n4) at (1.5,1.25) {} ;
  
  \node[vertex] (n5) at (2.5,0.5) {} ;
  \node[vertex1] (n6) at (3,1) {} ;
  \node[vertex1] (n7) at (3,0) {} ;
  
  \node[vertex] (n8) at (1.5,2.5) {} ;
  \node[vertex1] (n9) at (1,3) {} ;
  \node[vertex1] (n10) at (2,3) {} ;
  
  \draw[edge] (n1) -- (n2) ;
  \draw[edge] (n2) -- (n3) ;
  \draw[edge] (n2) node[right = 2.5pt] {$u$} -- (n4) node[right = 2.5pt] {$x$};
  
  \draw[edge] (n4) -- (n5) ;
  \draw[edge] (n5) node[left = 2.5pt] {$v$} -- (n6) ;
  \draw[edge] (n5) -- (n7) ;

  \draw[edge] (n4) -- (n8) node[right = 2.5pt] {$w$} ;
  \draw[edge] (n8) -- (n9) ;
  \draw[edge] (n8) -- (n10) ;

  \node[vertex1] (n11) at (4, 1.75) {};
  \node[vertex1] (n12) at (5, 1.75) {};
  \draw[edge,->,very thick,>=stealth] (n11) -- (n12) ;
  
  \node[vertex1] (n1a) at (6,0) {} ;
  \node[vertex] (n2a) at (6.5,0.5) {} ;
  \node[vertex1] (n3a) at (6,1) {} ;
  
  
  \node[vertex] (n5a) at (8.5,0.5) {} ;
  \node[vertex1] (n6a) at (9,1) {} ;
  \node[vertex1] (n7a) at (9,0) {} ;
  
  \node[vertex] (n8a) at (7.5,2.5) {} ;
  \node[vertex1] (n9a) at (7,3) {} ;
  \node[vertex1] (n10a) at (8,3) {} ;
  
  \draw[edge] (n1a) -- (n2a) ;
  \draw[edge] (n2a) -- (n3a) ;
  \draw[edge] (n2a) node[below = 2.5pt] {$u$} -- (n5a) node[below = 2.5pt] {$v$};
  
  \draw[edge] (n5a) -- (n8a) ;
  \draw[edge] (n5a) -- (n6a) ;
  \draw[edge] (n5a) -- (n7a) ;

  \draw[edge] (n2a) -- (n8a) node[right = 2.5pt] {$w$} ;
  \draw[edge] (n8a) -- (n9a) ;
  \draw[edge] (n8a) -- (n10a) ;
 
\end{tikzpicture}
}

\scalebox{.8}{

\begin{tikzpicture}
\tikzstyle{vertex}=[circle,draw = white, fill=black, minimum size = 7pt, inner sep=2pt]
\tikzstyle{vertex1}=[fill = white, draw = white]

\tikzstyle{edge}=[-,thick ]
\tikzstyle{elipse}=[-,thick ]
\tikzstyle{vertex2}=[circle,draw = black, fill=white, minimum size = 7pt, inner sep=2pt]

  \node[vertex2] (x) at (-0.75,1.5) {$5$};

  \node[vertex1] (n1) at (0,0) {} ;
  \node[vertex] (n2) at (0.5,0.5) {} ;
  \node[vertex1] (n3) at (0,1) {} ;
  
  
  \node[vertex] (n5) at (2.5,0.5) {} ;
  \node[vertex1] (n6) at (3,1) {} ;
  \node[vertex1] (n7) at (3,0) {} ;
  
  \node[vertex] (n8) at (1.5,2.5) {} ;
  \node[vertex1] (n9) at (1,3) {} ;
  \node[vertex1] (n10) at (2,3) {} ;
  
  \draw[edge] (n1) -- (n2) ;
  \draw[edge] (n2) -- (n3) ;
  \draw[edge] (n2) node[below = 2.5pt] {$x$} -- (n5) node[below = 2.5pt] {$y$};
  
  \draw[edge] (n8) -- (n5) ;
  \draw[edge] (n5) -- (n6) ;
  \draw[edge] (n5) -- (n7) ;

  \draw[edge] (n2) -- (n8) node[right = 2.5pt] {$z$} ;
  \draw[edge] (n8) -- (n9) ;
  \draw[edge] (n8) -- (n10) ;

  \node[vertex1] (n11) at (4, 1.75) {};
  \node[vertex1] (n12) at (5, 1.75) {};
  \draw[edge,->,very thick,>=stealth] (n11) -- (n12) ;
  
  \node[vertex1] (n1a) at (6,0) {} ;
  \node[vertex] (n2a) at (6.5,0.5) {} ;
  \node[vertex1] (n3a) at (6,1) {} ;
  
  \node[vertex] (n4a) at (7.5,1.25) {} ;
  
  \node[vertex] (n5a) at (8.5,0.5) {} ;
  \node[vertex1] (n6a) at (9,1) {} ;
  \node[vertex1] (n7a) at (9,0) {} ;
  
  \node[vertex] (n8a) at (7.5,2.5) {} ;
  \node[vertex1] (n9a) at (7,3) {} ;
  \node[vertex1] (n10a) at (8,3) {} ;
  
  \draw[edge] (n1a) -- (n2a) ;
  \draw[edge] (n2a) -- (n3a) ;
  \draw[edge] (n2a) node[right = 2.5pt] {$x$} -- (n4a) ;
  
  \draw[edge] (n5a) node[left = 2.5pt] {$y$} -- (n4a) ;
  \draw[edge] (n5a) -- (n6a) ;
  \draw[edge] (n5a) -- (n7a) ;

  \draw[edge] (n8a) node[right = 2.5pt] {$z$} -- (n4a) node[right = 2.5pt] {$w$} ;
  \draw[edge] (n8a) -- (n9a) ;
  \draw[edge] (n8a) -- (n10a) ;
 
\end{tikzpicture}
}

\caption{Wye-Delta operations: 1. Degree-one reduction; 2. Series reduction; 3. Parallel reduction; 4. Wye-Delta transformation; 5. Delta-Wye transformation.}
~\label{fig:wyedelta}
\end{figure}
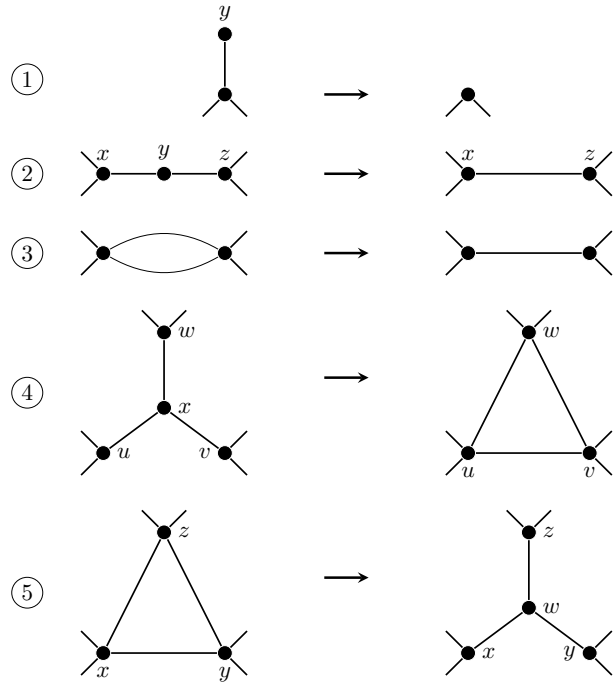

The following lemma (which follows from the above definitions) shows that the above rules preserve exactly all terminal minimum cuts.

\begin{lemma} \label{lemm: rules} Let $G$ be a $k$-terminal graph and $G'$ be a $k$-terminal graph obtained from $G$ by applying one of the rules $1-5$. Then $G'$ is a quality-$1$ cut sparsifier of $G$.
\end{lemma}

For our application, it will be useful to enrich the set of rules by introducing two new operations. These operations can be realized as series of the operations 1-5. (See Fig.~\ref{fig:edgedeletion} and \ref{fig:edgereplacement} for illustrations.)
\begin{enumerate}[resume]
	\item \emph{Edge deletion (with vertex $x$):} For a degree-three non-terminal with neighbours $u,v$, the edge $(u,v)$ can be deleted, if it exists. To achieve this, we use a Delta-Wye transformation followed by a series reduction.

	\item \emph{Edge replacement:} For a degree-four non-terminal vertex with neighbours $x,u,v,w$, if the edge $(x,u)$ exists, then it can be replaced by the edge $(v,w)$. To achieve this, we use a Delta-Wye transformation followed by a Wye-Delta transformation.
\end{enumerate}

\begin{figure}[H]
\begin{center}
\scalebox{.8}{
\begin{minipage}{.30\textwidth}
\centering

\begin{tikzpicture}
\tikzstyle{vertex}=[circle,draw = white, fill=black, minimum size = 7pt, inner sep=2pt]
\tikzstyle{vertex1}=[fill = white, draw = white]

\tikzstyle{edge}=[-,thick ]
\tikzstyle{elipse}=[-,thick ]
\tikzstyle{vertex2}=[circle,draw = black, fill=white, minimum size = 7pt, inner sep=2pt]

  \node[vertex2] (x) at (-0.75,1.5) {$6$};

  \node[vertex1] (n1) at (0,0) {} ;
  \node[vertex] (n2) at (0.5,0.5) {} ;
  \node[vertex1] (n3) at (0,1) {} ;
  
  \node[vertex] (n4) at (1.5,1.25) {} ;
  
  \node[vertex] (n5) at (2.5,0.5) {} ;
  \node[vertex1] (n6) at (3,1) {} ;
  \node[vertex1] (n7) at (3,0) {} ;
  
  \node[vertex] (n8) at (1.5,2.5) {} ;
  \node[vertex1] (n9) at (1,3) {} ;
  \node[vertex1] (n10) at (2,3) {} ;
  
  \draw[edge] (n1) -- (n2) node[below right] {$u$} ;
  \draw[edge] (n2) -- (n3) ;
  \draw[edge] (n2) -- (n4) node[right = 2.5pt] {$x$};
  
  \draw[edge] (n2) -- (n5); 
  
  \draw[edge] (n4) -- (n5) node[below left] {$v$} ;
  \draw[edge] (n5) -- (n6) ;
  \draw[edge] (n5) -- (n7) ;

  \draw[edge] (n4) -- (n8) ;
  \draw[edge] (n8) -- (n9) ;
  \draw[edge] (n8) -- (n10);
  
\end{tikzpicture}
\end{minipage} 
\hfill
\begin{minipage}{.25\textwidth}
\centering
\begin{tikzpicture}
\tikzstyle{vertex}=[circle,draw = white, fill=black, minimum size = 7pt, inner sep=2pt]
\tikzstyle{vertex1}=[fill = white, draw = white]

\tikzstyle{edge}=[-,thick ]
\tikzstyle{elipse}=[-,thick ]

  \node[vertex1] (n1) at (0,0) {} ;
  \node[vertex] (n2) at (0.5,0.5) {} ;
  \node[vertex1] (n3) at (0,1) {} ;
  
  \node[vertex] (n4) at (1.5,1.25) {} ;
  \node[vertex] (n11) at (1.5,0.625) {};
  
  \node[vertex] (n5) at (2.5,0.5) {} ;
  \node[vertex1] (n6) at (3,1) {} ;
  \node[vertex1] (n7) at (3,0) {} ;
  
  \node[vertex] (n8) at (1.5,2.5) {} ;
  \node[vertex1] (n9) at (1,3) {} ;
  \node[vertex1] (n10) at (2,3) {} ;
  
  \draw[edge] (n1) -- (n2) node[below right] {$u$} ;
  \draw[edge] (n2) -- (n3) ;
  \draw[edge] (n2) -- (n11) node[below = 2.5pt] {$w$} ;
 
  \draw[edge] (n4) node[right = 2.5pt] {$x$} -- (n11);
  
  \draw[edge] (n11) -- (n5) node[below left] {$v$} ;
  \draw[edge] (n5) -- (n6) ;
  \draw[edge] (n5) -- (n7) ;

  \draw[edge] (n4) -- (n8) ;
  \draw[edge] (n8) -- (n9) ;
  \draw[edge] (n8) -- (n10);
\end{tikzpicture}
\end{minipage}
\hfill
\begin{minipage}{.25\textwidth}
\centering
\begin{tikzpicture}
\tikzstyle{vertex}=[circle,draw = white, fill=black, minimum size = 7pt, inner sep=2pt]
\tikzstyle{vertex1}=[fill = white, draw = white]

\tikzstyle{edge}=[-,thick ]
\tikzstyle{elipse}=[-,thick ]

  \node[vertex1] (n1) at (0,0) {} ;
  \node[vertex] (n2) at (0.5,0.5) {} ;
  \node[vertex1] (n3) at (0,1) {} ;
  
  \node[vertex] (n4) at (1.5,1.25) {} ;
  
  \node[vertex] (n5) at (2.5,0.5) {} ;
  \node[vertex1] (n6) at (3,1) {} ;
  \node[vertex1] (n7) at (3,0) {} ;
  
  \node[vertex] (n8) at (1.5,2.5) {} ;
  \node[vertex1] (n9) at (1,3) {} ;
  \node[vertex1] (n10) at (2,3) {} ;
  
  \draw[edge] (n1) -- (n2) node[right = 2.5pt] {$u$};
  \draw[edge] (n2) -- (n3) ;
  \draw[edge] (n2) -- (n4) node[right = 2.5pt] {$w$}; 
  
  \draw[edge] (n4) -- (n5) node[left = 2.5pt] {$v$} ;
  \draw[edge] (n5) -- (n6) ;
  \draw[edge] (n5) -- (n7) ;

  \draw[edge] (n4) -- (n8) ;
  \draw[edge] (n8) -- (n9) ;
  \draw[edge] (n8) -- (n10);
  
\end{tikzpicture}
\end{minipage}
}
\end{center}
\caption{Edge deletion transformation. Edge capacities are omitted.}
\label{fig:edgedeletion}

\end{figure}
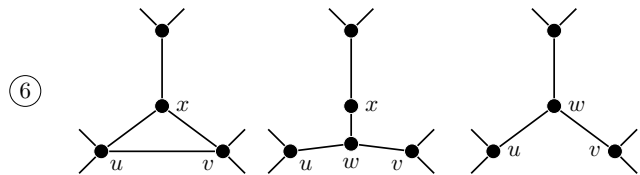

\begin{figure}[H]
\begin{center}
\scalebox{.8}{
\begin{minipage}{.30\textwidth}
\centering

\begin{tikzpicture}
\tikzstyle{vertex}=[circle,draw = white, fill=black, minimum size = 7pt, inner sep=2pt]
\tikzstyle{vertex1}=[fill = white, draw = white]

\tikzstyle{edge}=[-,thick ]
\tikzstyle{elipse}=[-,thick ]
\tikzstyle{vertex2}=[circle,draw = black, fill=white, minimum size = 7pt, inner sep=2pt]

  \node[vertex2] (x) at (-0.75,1.5) {$7$};

  \node[vertex1] (n1) at (0,0) {} ;
  \node[vertex] (n2) at (0.5,0.5) {} ;
  \node[vertex1] (n3) at (0,1) {} ;
  
  \node[vertex] (n4) at (1.5,1.25) {} ;
  
  \node[vertex] (n5) at (2.5,0.5) {} ;
  \node[vertex1] (n6) at (3,1) {} ;
  \node[vertex1] (n7) at (3,0) {} ;
  
  \node[vertex] (n8) at (0.5,2.5) {} ;
  \node[vertex1] (n9) at (0,3) {} ;
  \node[vertex1] (n10) at (1,3) {} ;
  
  \node[vertex] (n11) at (2.5,2.5) {} ;
  \node[vertex1] (n12) at (2,3) {} ;
  \node[vertex1] (n13) at (3,3) {} ;
  
  \draw[edge] (n1) -- (n2) node[below right] {$x$}  ;
  \draw[edge] (n2) -- (n3) ;
  \draw[edge] (n2) -- (n4) ;
  
  \draw[edge] (n2) -- (n5); 
  
  \draw[edge] (n4) -- (n5) node[below left] {$u$} ;
  \draw[edge] (n5) -- (n6) ;
  \draw[edge] (n5) -- (n7) ;

  \draw[edge] (n4) -- (n8) node[left = 2.5pt] {$w$};
  \draw[edge] (n8) -- (n9) ;
  \draw[edge] (n8) -- (n10);
  
  \draw[edge] (n4) -- (n11) node[right = 2.5pt] {$v$};
  \draw[edge] (n11) -- (n12) ;
  \draw[edge] (n11) -- (n13);
  
\end{tikzpicture}
\end{minipage} 
\hfill
\begin{minipage}{.25\textwidth}
\centering
\begin{tikzpicture}
\tikzstyle{vertex}=[circle,draw = white, fill=black, minimum size = 7pt, inner sep=2pt]
\tikzstyle{vertex1}=[fill = white, draw = white]

\tikzstyle{edge}=[-,thick ]
\tikzstyle{elipse}=[-,thick ]

  \node[vertex1] (n1) at (0,0) {} ;
  \node[vertex] (n2) at (0.5,0.5) {} ;
  \node[vertex1] (n3) at (0,1) {} ;
  
  \node[vertex] (n4) at (1.5,1.25) {} ;
  \node[vertex] (n11) at (1.5,0.625) {};
  
  \node[vertex] (n5) at (2.5,0.5) {} ;
  \node[vertex1] (n6) at (3,1) {} ;
  \node[vertex1] (n7) at (3,0) {} ;
  
  \node[vertex] (n8) at (0.5,2.5) {} ;
  \node[vertex1] (n9) at (0,3) {} ;
  \node[vertex1] (n10) at (1,3) {} ;
  
  \node[vertex] (n12) at (2.5,2.5) {} ;
  \node[vertex1] (n13) at (2,3) {} ;
  \node[vertex1] (n14) at (3,3) {} ;
  
  \draw[edge] (n1) -- (n2) node[below right] {$x$} ;
  \draw[edge] (n2) -- (n3) ;
  \draw[edge] (n2) -- (n11);
 
  \draw[edge] (n4) -- (n11);
  
  \draw[edge] (n11) -- (n5) node[below left] {$u$};
  \draw[edge] (n5) -- (n6) ;
  \draw[edge] (n5) -- (n7) ;

  \draw[edge] (n4) -- (n8) node[left = 2.5pt] {$w$};
  \draw[edge] (n8) -- (n9) ;
  \draw[edge] (n8) -- (n10);
  
  \draw[edge] (n4) -- (n12) node[right = 2.5pt] {$v$};
  \draw[edge] (n12) -- (n13) ;
  \draw[edge] (n12) -- (n14);
\end{tikzpicture}
\end{minipage}
\hfill
\begin{minipage}{.25\textwidth}
\centering
\begin{tikzpicture}
\tikzstyle{vertex}=[circle,draw = white, fill=black, minimum size = 7pt, inner sep=2pt]
\tikzstyle{vertex1}=[fill = white, draw = white]

\tikzstyle{edge}=[-,thick ]
\tikzstyle{elipse}=[-,thick ]

  \node[vertex1] (n1) at (0,0) {} ;
  \node[vertex] (n2) at (0.5,0.5) {} ;
  \node[vertex1] (n3) at (0,1) {} ;
  
  \node[vertex] (n4) at (1.5,1.25) {} ;
  
  \node[vertex] (n5) at (2.5,0.5) {} ;
  \node[vertex1] (n6) at (3,1) {} ;
  \node[vertex1] (n7) at (3,0) {} ;
  
  \node[vertex] (n8) at (0.5,2.5) {} ;
  \node[vertex1] (n9) at (0,3) {} ;
  \node[vertex1] (n10) at (1,3) {} ;
  
  \node[vertex] (n11) at (2.5,2.5) {} ;
  \node[vertex1] (n12) at (2,3) {} ;
  \node[vertex1] (n13) at (3,3) {} ;
  
  \draw[edge] (n1) -- (n2) node[below right] {$x$};
  \draw[edge] (n2) -- (n3) ;
  \draw[edge] (n2) -- (n4) ;
  
  \draw[edge] (n8) -- (n11); 
  
  \draw[edge] (n4) -- (n5) node[below left] {$u$};
  \draw[edge] (n5) -- (n6) ;
  \draw[edge] (n5) -- (n7) ;

  \draw[edge] (n4) -- (n8) node[left = 2.5pt] {$w$};
  \draw[edge] (n8) -- (n9) ;
  \draw[edge] (n8) -- (n10);
  
  \draw[edge] (n4) -- (n11) node[right = 2.5pt] {$v$};
  \draw[edge] (n11) -- (n12) ;
  \draw[edge] (n11) -- (n13);
  
\end{tikzpicture}
\end{minipage}
}
\end{center}
\caption{Edge replacement transformation. Edge capacities are omitted.}
\label{fig:edgereplacement}
\end{figure}
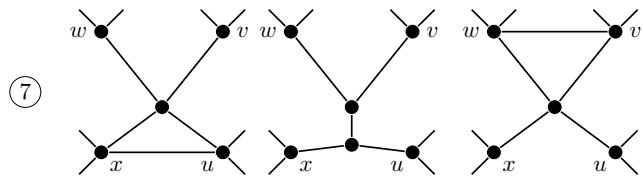

A $k$-terminal graph $G$ is \emph{Wye-Delta} reducible to another $k$-terminal graph $H$, if $G$ is reduced to $H$ by repeatedly applying one of the operations 1-7. 

\begin{lemma} \label{lemm: cutReducability} Let $G$ and $H$ be $k$-terminal graphs. Moreover, let $G$ be Wye-Delta reducible to $H$. Then $H$ is a quality-$1$ cut sparsifier of $G$.
\end{lemma}
\begin{proof}
	Observe that the rules 1-7 do not affect any terminal vertex and each rule preserves exactly all terminal minimum cuts by Lemma \ref{lemm: rules}. An induction on the number of rules needed to reduce $G$ to $H$ proves the claim.
\end{proof}
\paragraph*{Grid Graphs.} 
A \emph{grid} graph is a graph with $n \times n$ vertices $\{(u,v) : u,v = 1,\ldots, n\}$, where $(u,v)$ and $(u',v')$ are adjacent if $|u' -u| + |v'-v| = 1$. For $k < n$, a \emph{half-grid} graph with $k$ terminals is a graph $T^{n}_k = (V,E)$ with $K \subset V$ and $n(n+1)/2$ vertices $\{(i,j) : i \leq j \text{ and } i,j=1,\ldots,n  \}$, where $(i,j)$ and $(i',j')$ are connected by an edge if $|i'-i| + |j'-j| = 1$, and additional diagonal edges between $(i,i)$ and $(i+1,i+1)$ for $i = 1,\ldots, n-1$. Moreover, each terminal vertex in $T^{n}_k$ must be one of its diagonal vertices, i.e., every $x \in K$ is of the form $(m,m)$ for some $m \in \{1,\ldots,n\}$. Let $\hat{T}^{n}_{k}$ be the same graph as $T^{n}_k$ but excluding the diagonal edges. 

\paragraph*{Graph Embeddings.}\label{sec: graphEmbeddings} 
Throughout this chapter, we will be dealing with the embedding of a planar graph into a square \emph{grid} graph. One way of drawing graphs in the plane are \emph{orthogonal grid-embeddings}~\cite{Valiant81}. In such a setting, the vertices correspond to distinct points and edges consist of alternating sequences of vertical and horizontal segments. Equivalently, one can view this as drawing our input graph as a subgraph of some grid. Formally, a \emph{node-embedding} $\rho$ of $G_1=(V_1,E_1)$ into $G_2=(V_2,E_2)$ is an injective mapping that maps $V_1$ into $V_2$, and $E_1$ into paths in $G_2$, i.e., $(u,v)$ maps to a path from $\rho(u)$ to $\rho(v)$, such that every pair of paths that correspond to two different edges in $G_1$ is vertex-disjoint (except possibly at the endpoints). If $G_2$ is a planar graph, then $\rho(G_1)$ and $G_1$ are also planar. Thus, if $G_1$ and $G_2$ are planar we then refer to $\rho$ as an \emph{orthogonal embedding}. Moreover, given a planar graph $G_1$ drawn in the plane, the embedding $\rho$ is called \emph{region-preserving} if $\rho(G_1)$ and $G_1$ have the same planar topological embedding.


Let $G_1$ be a $k$-terminal graph. Since the embedding does not affect the vertices of $G_1$, the terminals of $G_1$ are also terminals in $\rho(G_1)$. Although the embedding does not consider capacity of the edges in $G_1$, we can still guarantee that such an embedding preserves all terminal minimum cuts, for which we make use of the following operation:
\begin{enumerate}
	\item \emph{Edge subdivision: } Let $(u,v)$ be an edge of capacity $\cc(u,v)$. Delete $(u,v)$, introduce a new vertex $w$ and add edges $(u,w)$ and $(w,v)$, each of capacity $\cc(u,v)$. 
\end{enumerate}

The following lemma shows that a node-embedding is a cut preserving mapping.
\begin{lemma} \label{lemm: cutembedding}
	Let $\rho$ be a node-embedding and let $G_1$ and $\rho(G_1)$ be $k$-terminal graphs defined as above. Then $\rho(G_1)$ preserves exactly all terminal minimum cuts of $G$.
\end{lemma}
\begin{proof}
	We can view each path obtained from the embedding as taking the edge corresponding to the path endpoints in $G_1$ and performing edge subdivisions finitely many times. We claim that such subdivisions preserve all terminal cuts. 
	
	Indeed, let us consider a single edge subdivision for $(u,v)$ (the general claim then follows by induction on the number of edge subdivisions). Fix $S \subset K$ and consider some $S$-separating minimum cut $(U,V \setminus U)$ in $G_1$ cutting $(u,v)$. Then, in the transformed graph $\rho(G_1)$, we can simply cut either the edge $(u,w)$ or $(w,v)$. Since by construction, the new edge has the same capacity as the subdivided edge, we get that $\capacity_{\rho(G_1)}(\delta(U)) = \capacity_{G_1}(\delta(U))$, and in particular $\mincut_{\rho(G_1)}(S,K \setminus S) \leq \mincut_{G_1}(S,K \setminus S)$. 
	
	Furthermore, since $G_1$ is obtained by contracting two edges of the same capacity of $\rho(G_1)$, for any $S$-separating minimum cut $(U, V \setminus U)$ in $\rho(G_1)$, we have $\capacity_{\rho(G_1)}(\delta(U)) \geq \capacity_{G_1}(\delta(U))$, and in particular $\mincut_{\rho(G_1)}(S,K \setminus S) \geq \mincut_{G_1}(S,K \setminus S)$. Combining the above gives the lemma. 
\end{proof}
%
%
\subsection{Our Construction}
In this section  we construct our exact cut sparsifier and prove that any planar $k$-terminal graph with all terminals lying on the same face admits a cut sparsifier of size $O(k^2)$ that is also planar.
\subsubsection{Embedding into Grids} \label{EmbeddingGrid}

It is well-known that one can obtain an orthogonal embedding of a planar graph with maximum-degree at most three into a grid (see Valiant~\cite{Valiant81}). However, our input planar graph can have arbitrarily large maximum degree. In order to be able to make use of such an embedding, we need to first reduce our input graph to a bounded-degree graph while preserving planarity and all terminal minimum cuts. We achieve this by making use of a \emph{vertex splitting} technique, which we describe below. 

Given a $k$-terminal planar graph $G'=(V',E',c')$ with $K \subset V'$ lying on the outer face, vertex splitting produces a $k$-terminal planar graph $G=(V,E,c)$ with $K \subset V$ such that the maximum degree of $G$ is at most three. Specifically, for each vertex $v$ of degree $d>3$ with neighboring vertices $u_1,\ldots,u_{d}$, we delete $v$ and introduce new vertices $v_1, \ldots, v_d$ along with edges $\{(v_i,v_{i+1}) : i=1,\ldots,d-1\}$, each of capacity $C+1$, where $C=\sum_{e \in E'} c'(e)$. Further, we replace the edges $\{(u_i,v) : i = 1,\ldots,d\}$ with $\{(u_i,v_i) : i = 1,\ldots,d\}$, each of corresponding capacity. If $v$ is a terminal vertex, we set one of the $v_i$'s to be a terminal vertex. It follows that the resulting graph $G$ is planar and terminals can be still embedded on the outer face. Note that while the degree of every vertex $v_i$ is at most $3$, the degree of any other vertex is not affected. 

\begin{claim} \label{claim: embedding}
Let $G'$ and $G$ be $k$-terminal graphs defined as above. Then $G$ preserves exactly all minimum terminal cuts of $G'$, i.e., $G$ is a quality-$1$ cut sparsifier of $G'$.
\end{claim} 
\begin{proof}
	It suffices to prove the case where $G$ is obtained from $G'$ by a single vertex splitting. Then the claim follows by induction on the number of vertex splittings required to transform $G'$ to $G$.
	
	Let $S \subset K$ and $(U, V \setminus U)$ be an $S$-separating cut in $G$ of size $\mincut_G(S,K \setminus S)$. Suppose towards contradiction that $\delta(U)$ contains an edge of the form $(v_j,v_{j+1})$, for some $j$, which in turn gives that $\capacity(\delta(U)) \geq C + 1$. Then we can move all the points $v_i$ to one of the sides of the cut $(U, V \setminus S)$ and obtain a new $S$-separating cut in $G$ of cost at most $C$, contradicting the fact that $(U, V \setminus U)$ is a minimum terminal cut. Hence, it follows that $\delta(U)$ uses either edges that are in both $G$ and $G'$ or edges of the form $(u_i,v_i)$, which by construction have the same capacity as the edges $(u_i,v)$ in $G'$. Thus, an $S$-separating minimum cut in $G$ corresponds to an $S$-separating minimum cut in $G'$ of the same cost. Since $S$ is chosen arbitrarily, the claim follows.
\end{proof}
%

Let $G=(V,E)$ be a $k$-terminal graph obtained by vertex splitting of all vertices of degree larger than $3$ of $G'=(V',E')$. Further, let $n' = |V'|$, $m' = |E'|$, $n = |V|$ and $m = |E|$. Then it is easy to show that $n \leq 2 m'$ and $m \leq m' + n \leq 3m'$. Since $G'$ is planar, we have that $n = O(n')$ and $m = O(n')$. Thus, by just a linear blow-up on the size of vertex and edge sets, we may assume w.l.o.g. that our input graph is a planar graph of degree at most three. 

Valiant~\cite{Valiant81} and Tamassia et al.~\cite{TamassiaT89} showed that a $k$-terminal planar graph $G$ with $n$ vertices and degree at most three admits an orthogonal region-preserving embedding into some square grid of size $O(n) \times O(n)$. By Lemma \ref{lemm: cutembedding}, we know that the resulting graph exactly preserves all terminal minimum cuts of $G$. We remark that since the embedding is region-preserving, the outer face of the input graph is embedded to the outer face of the grid. Therefore, all terminals in the embedded graph lie on the outer face of the grid. Performing appropriate edge subdivisions, we can make all the terminals lie on the boundary of some possibly larger grid. Further, we can add dummy non-terminals and zero edge capacities to transform our graph into a full-grid $H$. We observe that the latter does not affect any terminal min-cut. The above leads to the following:
\begin{lemma} \label{lemm: embeddingGrids}
Given a $k$-terminal planar graph $G$, where all terminals lie on the outer face, there exists a $k$-terminal grid graph $H$, where all terminals lie on the boundary such that $H$ preserves exactly all terminal minimum cuts of $G$. The resulting graph has $O(n^2)$  vertices and edges.
\end{lemma}

\subsubsection{Embedding Grids into Half-Grids}

Next, we show how to embed square grids into half-grid graphs (see Section \ref{sec: preli_RM}), which will facilitate the application of Wye-Delta transformations. The existence of such an embedding was claimed in the thesis of Gitler~\cite{Gitler91}, but no details on its construction were given. 

Let $G$ be a $k$-terminal square grid on $n \times n$ vertices where terminals lie on the boundary of the grid. We obtain the following:
\begin{figure}[t] 
\begin{center}
\includegraphics[scale=1]{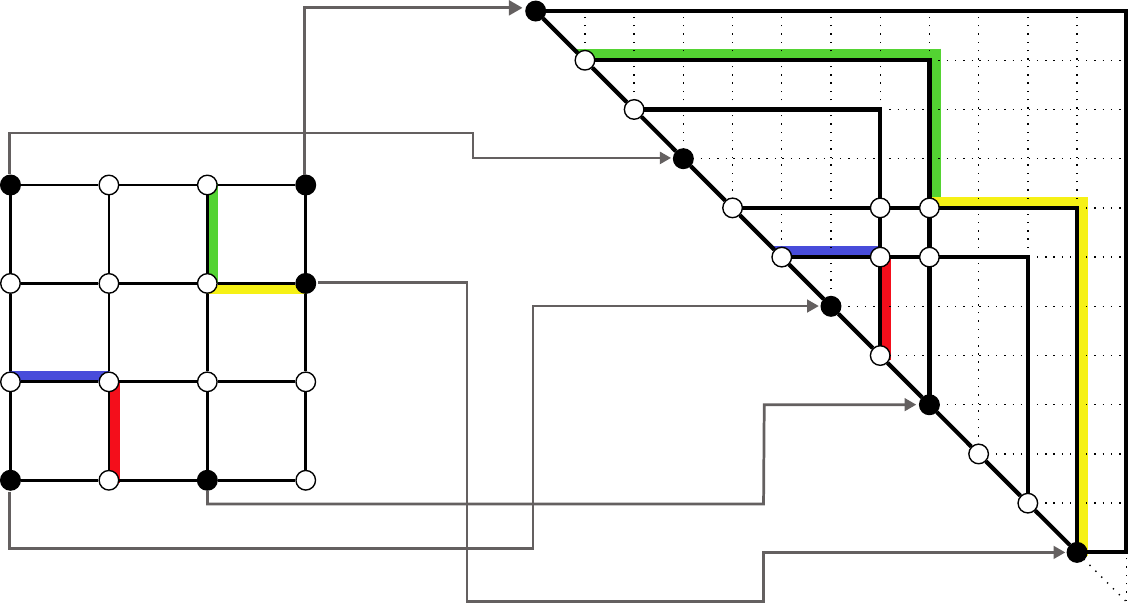}
\end{center}
\caption{Embedding grid into half-grid. Black vertices represent terminals while white vertices represent non-terminals. The counter-clockwise ordering starts at the top right terminal. Coloured edges and paths correspond to the mapping of the respective edges: blue for edges $((i,1),(i,2))$, red for edges $((n-1,j),(n,j))$, green for edges $((1,j),(2,j))$ and yellow for edges $((i,n-1), (i,n))$, where $i,j = 2,\ldots,n-1$. }
\label{fig: halfgridConst}
\end{figure}

\begin{lemma} \label{lemm: embeddingHalfGrid}
There exists a node embedding of the grid $G$ into $T^{\ell}_{k}$, where $\ell  = 4n-3$. 
\end{lemma}
\begin{proof}
Our construction works as follows (See Fig.~\ref{fig: HalfGridRed} for an example). We first fix an ordering on the vertices lying on the boundary of the grid in the order induced by the grid. Then we embed each vertex according to that order into the diagonal vertices of the half-grid, along with the edges that form the boundary of the grid. The sub-grid obtained by removing all boundary vertices is embedded appropriately into the upper-part of the half-grid. Finally, we show how to embed edges between the boundary and the sub-grid vertices and argue that such an embedding is indeed vertex-disjoint for any pair of paths.

\input{chapters/reachabilityminors/figure_halfgridReduction}

We start with the embedding of the vertices of $G$. Let us first consider the boundary vertices. The ordering imposed on these vertices can be viewed as starting with the upper-right vertex $(1,n)$ and visiting the rest of vertices in a counter-clockwise direction until reaching the vertex $(2,n)$. We map the vertices on the boundary as follows.
\begin{enumerate}
\setlength\itemsep{0.1em}
\item The vertex $(1,j)$ is mapped to the vertex $(n-j+1,n-j+1)$ for $j=2,\ldots,n$, 
\item The vertex $(i,1)$ is mapped to the vertex $(n+i-1,n+i-1)$ for $i=1,\ldots,n-1$,
\item The vertex $(n,j)$ is mapped to the vertex $(2n+j-2,2n+j-2)$ for $j=1,\ldots,n-1$,
\item The vertex $(i,n)$ is mapped to the vertex $(4n-i-2,4n-i-2)$  for $i=2,\ldots,n$.
\end{enumerate} 
Now we consider the vertices that belong to the induced sub-grid $S$ of $G$ of size $(n-2)^2$ when removing the boundary vertices of our input grid. We map the vertex $(i,j)$ to the vertex $(n+i-1, 2n+j-2)$ for $i,j=2,\ldots,n-1$. In other words, for every vertex of $S$ we make a vertical shift by $n-1$ units and an horizontal shift by $2n-2$ units. By construction, it is not hard to check that every vertex of $G$ is mapped to a different vertex of $T^{\ell}_{k}$ and all terminal vertices lie on the diagonal of $T^{\ell}_{k}$. 

We continue with the embedding of the edges of $G$. First, every edge between two boundary vertices in $G$ is embedded to the edge between the corresponding mapped diagonal vertices of $T^{\ell}_{k}$, except the edge between $(1,n)$ and $(2,n)$. For this edge, we define an edge embedding between the corresponding vertices $(1,1)$ and $(4n-4,4n-4)$ of $T^{\ell}_{k}$ by using the path:
\begin{align*} 
(1,1) \rightarrow (1,2) \rightarrow \ldots & \rightarrow (1,4n-3) \rightarrow (2,4n-3)  \\ 
&\rightarrow \ldots \rightarrow (4n-4,4n-3) \rightarrow (4n-4,4n-4).
\end{align*}

Next, every edge of the sub-grid $S$ is embedded in to the edge connecting the mapped endpoints of that edge in $T^{\ell}_{k}$. In other words, if $(i,j)$ and $(i',j')$ were connected by an edge $e$ in $S$, then $(n+i-1, 2n+j-2)$ and $(n+i'-1, 2n+j'-2)$ are connected by an edge $e'$ in $T^{\ell}_{k}$ and $e$ is mapped to $e'$. Finally, the only edges that remain are those connecting a boundary vertex of $G$ with a boundary vertex of $S$. We distinguish four cases depending on the edge position. 
\begin{enumerate}
\setlength\itemsep{0.1em}
\item The edge $((i,2), (i,1))$ is mapped to the horizontal path given by:
\begin{align*}
	(n+i-1,2n) & \rightarrow (n+i-1, 2n-1) \\ 
	& \rightarrow \ldots \rightarrow (n+i-1,n+i-1) \text{ for } i = 2,\ldots,n-1 
\end{align*}
\item The edge $((n-1,j), (n,j))$ is mapped to the vertical path given by:
\begin{align*}
	(2n-2,&~2n+j-2)  \rightarrow (2n-1, 2n+j-2) \\
	& \rightarrow \ldots \rightarrow (2n+j-2,2n+j-2) \text{ for } j = 2,\ldots,n-1.
\end{align*}
\item The edge $((2,j), (1,j))$ is mapped to the $L$-shaped path:
\begin{align*}
	(n+1,2n+&j-2)  \rightarrow (n, 2n+j-2) \rightarrow \ldots \rightarrow (n-j+1,2n+j-2) \\
	&  \rightarrow (n-j+1,2n+j-3)  \rightarrow \ldots \rightarrow (n-j+1,n-j+1) \\
	& \text{ for } j = 2,\ldots,n-1.
\end{align*}

\item The edge $((i,n-1), (i,n))$ is mapped to the $L$-shaped path:
\begin{align*}
	(n+i-1,3n&-3)  \rightarrow (n+i-1, 3n-2) \rightarrow \ldots \rightarrow (n+i-1,4n-i-2) \\
	&  \rightarrow (n+i,4n-i-2)  \rightarrow \ldots \rightarrow (4n-i-2,4n-i-2) \\
	& \text{ for } i = 2,\ldots,n-1.
\end{align*}
\end{enumerate} 
By construction, it follows that the paths in our edge embedding are vertex disjoint.
\end{proof}

\subsubsection{Reducing Half-Grids and Bringing the Piece Together}
We now review the construction of Gitler~\cite{Gitler91}, which shows how to reduce half-grids to much smaller half-grids (excluding diagonal edges) whose size depends only on $k$. For the sake of completeness, we provide a full proof here. Recall that $\hat{T}^{n}_{k}$ is the graph $T^{n}_k$ without the diagonal edges. 

\begin{lemma}[\cite{Gitler91}] \label{lemm: Gitler}
For any positive $k,n$ with $k<n$, $T^{n}_k$ is Wye-Delta reducible to $\hat{T}^{k}_{k}$.
\end{lemma}
\begin{proof}
	For sake of simplicity, we assume w.l.o.g that the four vertices $(1,1)$, $(2,2)$, $(n-1,n-1)$ and $(n,n)$ are terminals\footnote{If they are not terminals, we can simply define them as terminals, thus increasing the number of terminals to $k+4 = O(k)$.}. Furthermore, we say that two terminals $(i,i)$ and $(j,j)$ are \textit{adjacent} iff $i<j$ and there is no terminal $(\ell,\ell)$ such that $i < \ell < j$. 
	
	We next describe the reduction procedure. Also see Fig.~\ref{fig: HalfGridRed} for an example. The reduction procedure starts by removing the diagonal edges of $T^{n}_k$, thus producing the graph $\hat{T}^{n}_k$ . Specifically, the two edges $((1,1),(2,2))$ and $((n-1,n-1), (n,n))$ are removed using an edge deletion operation. For each remaining diagonal edge of the form $((i,i), (i+1,i+1))$, $i=2,\ldots,n-2$ we repeatedly apply an edge replacement operation until the edge is incident to a boundary vertex $(1,j)$ or $(j,n)$ of the grid, where an edge deletion operation with one of the neighbours of $(1,j)$ resp. $(j,n)$ as vertex $x$ is applied.
	
	Now, we know that all non-terminals of the form $(i,i)$ are degree-two vertices, thus a series reduction is applied on each of them. This produces new diagonal edges, which are effectively reduced by the above procedure. We keep removing the newly-created degree-two non-terminal vertices and the newly-created edges until no further removals are possible. At this point, the only degree-$2$ vertices are terminal vertices. 
	
	The resulting graph has a staircase structure, where for every pair of adjacent terminals $(i,i)$ and $(j,j)$, there is a non-terminal $(i,j)$ of degree three or four, namely, the intersection vertex, and a (possibly empty) sequence of degree-three non-terminals that lie on the boundary path from $(i,i)$ to $(j,j)$. For $k = i+1,\ldots,j-1$, let $(i,k)$ and $(k,j)$ be the degree-three non-terminals lying on the row and the column subpath, respectively. Additionally, for $k = i+1,\ldots,j-1$, let $C^{i}_{k}=\{(i',k) : i'=i,\ldots,1\}$, resp. $R^{j}_{k}=\{(k,j') : j'=j,\ldots,n\}$ be the vertices sharing the same column, resp. row with $(i,k)$, resp. $(k,j)$. We next show that the vertices belonging to $C^{i}_{k}$ and $R^{j}_{k}$ can be removed. 
	
	The removal process works as follows. For $k=i+1,\ldots,j-1$, we start by choosing a degree $3$ vertex $(i,k)$ and its corresponding column $C^{i}_{k}$. Then we apply a Wye-Delta transformation on $(i,k)$, thus creating two new diagonal edges. Similarly as above, we remove such edges by repeatedly applying an edge replacement operation until they have been pushed to the boundary of the grid, where an edge deletion operation is applied. In the resulting graph, the vertex $(i-1,k) \in C^{i}_{k}$ is now a degree-three non-terminal. We apply the same procedure to this vertex. Applying such a procedure to all remaining vertices of $C^{i}_{k}$, we eliminate a column of the grid. Symmetrically, the same process applies to the case when we want to remove the row $R^{j}_{k}$ corresponding to the vertex $(k,j)$.
	
	Applying the above removal process for every adjacent terminal pair and the corresponding degree-three non-terminals, we end up with the graph $\hat{T}^{k}_{k}$, where every diagonal vertex is a terminal. By definition, it follows that $\hat{T}^{k}_{k}$ has at most $O(k^2)$ vertices.
\end{proof}

Combining the above reductions leads to the following theorem:
\begin{theorem} \label{thm: CutSparsifier}
Let $G$ be a $k$-terminal planar graph where all terminals lie on the outer face. Then $G$ admits a quality-$1$ cut sparsifier of size $O(k^{2})$, which is also a planar graph.
\end{theorem}
\begin{proof}
Let $n$ denote the number of vertices in $G$. First, we apply Lemma \ref{lemm: embeddingGrids} on $G$ to obtain a grid graph $H$ with $O(n^{2})$ vertices, which preserves exactly all terminal minimum cuts of $G$. We then apply Lemma \ref{lemm: embeddingHalfGrid} on $H$ to obtain a node embedding $\rho$ into the half-grid $T^{\ell}_{k}$, where $\ell = 4n-3$. By Lemma \ref{lemm: cutembedding}, $\rho(H)$ preserves exactly all terminal minimum cuts of $H$. We can further extend $\rho(H)$ to the full half-grid $T^{\ell}_{k}$, if dummy non-terminals and zero edge capacities are added. Finally, we apply Lemma \ref{lemm: Gitler} on $T^{\ell}_{k}$ to obtain a Wye-Delta reduction to the reduced half-grid graph $\hat{T}^{k}_{k}$. It follows by Lemma \ref{lemm: cutReducability} that $\hat{T}^{k}_{k}$ is a quality-$1$ cut sparsifier of $T^{\ell}_{k}$, where the size guarantee is immediate from the definition of $\hat{T}^{k}_{k}$.
\end{proof}

\section{Extensions to Planar Flow and Distance Sparsifiers} \label{sec: UpperFlow}
In this section we show how to extend our result for cut sparsifiers to flow and distance sparsifiers. 
\subsection{An Upper Bound for Flow Sparsifiers}
We first review the notion of Flow Sparsifiers. Let $\dd$ be a demand function over terminal pairs in $G$ such that $\dd(x,x')=\dd(x',x)$ and $\dd(x,x)=0$ for all $x,x' \in K$. We denote by $P_{xx'}$ the set of all paths between vertices $x$ and $x'$, for all $x,x' \in K$. Further, let $P_{e}$ be the set of all paths using edge $e$, for all $e \in E$ . A \emph{concurrent} (\textit{multi-commodity}) flow $\ff$ of \emph{throughput} $\lambda$ is a function over terminal paths in $G$ such that (1) $\sum_{p \in P_{xx'}} \ff(p) \geq \lambda \dd(x,x')$, for all distinct terminal pairs $x,x' \in K$ and (2) $\sum_{p \in P_e} \ff(p) \leq \cc(e)$, for all $e \in E$. We let $\lambda_G(d)$ denote the \emph{throughput of the concurrent flow} in $G$ that attains the largest throughput and we call a flow achieving this throughput the \emph{maximum concurrent flow}. A graph $H = (V', E', \cc')$, $K \subset V'$ is a \emph{quality-$1$} (\emph{vertex}) \emph{flow sparsifier} of $G$ with $q \geq 1$ if for every demand function
$\dd$, $\lambda_G(\dd) \leq \lambda_H(\dd) \leq q \cdot \lambda_H(\dd).$

Next we show that given a $k$-terminal planar graph, where all terminals lie on the outer face, one can construct a quality-$1$ flow sparsifier of size $O(k^{2})$. Our result follows from combining the observation of Andoni et al.~\cite{andoni} for constructing flow-sparsifiers using flow/cut gaps and the flow/cut gap result of Okamura and Seymour~\cite{OkamuraS81}.

Given a $k$-terminal graph and a demand function $\dd$, recall that $\lambda_G(\dd)$ is the maximum fraction of $\dd$ that can be routed in $G$. We define the \emph{sparsity} of a cut $(U, V \setminus U)$ to be
\[
\Phi_G(U,\dd) := \frac{\capacity(\delta(U))}{\sum_{i,j: |\{i,j\} \cap U|=1}\dd_{ij}}
\]
and the \emph{sparsest cut} as $\Phi_G(\dd) := \min_{U \subset V} \Phi_G(U,\dd)$. Then the \emph{flow-cut} gap is given by
\[
\gamma(G) := \max \{\Phi_G(\dd) / \lambda_G(d) : \dd \in \mathbb{R}^{\binom{k}{2}}_{+}\}.
\]

We will make use of the following theorem:
\begin{theorem}[\cite{andoni}] \label{thm: FlowCutGap}
	Given a $k$-terminal graph $G$ with terminals $K$, let $G'$ be a quality-$\beta$ cut sparsifier for $G$ with $\beta \geq 1$. Then for every demand function $\dd \in \mathbb{R}^{\binom{k}{2}}_{+}$,
	\[
	\frac{1}{\gamma(G')} \leq \frac{\lambda_{G'}(\dd)}{\lambda_G(\dd)} \leq \beta \cdot \gamma(G).
	\]
	Therefore, the graph $G'$ with edge capacities scaled up by $\gamma(G')$ is a quality-$\beta \cdot \gamma(G) \cdot \gamma(G')$ flow sparsifier of size $|V(G')|$ for $G$.
\end{theorem}
This leads to the following corollary.
\begin{corollary} \label{cor: FlowSparsifiers}
	Let $G$ be a $k$-terminal planar graph where all terminals lie on the outer face. Then $G$ admits a quality-$1$ flow sparsifier of size $O(k^{2})$.
\end{corollary}
\begin{proof}
	Given a $k$-terminal planar graph where all terminals lie on the outer face, Theorem \ref{thm: CutSparsifier} shows how to construct a cut sparsifier $G'$ with quality $\beta = 1$ and size $O(k^2)$, which is also a planar graph with all the $k$ terminals lying on the outer face. Okamura and Seymour~\cite{OkamuraS81} showed that for every $k$-terminal planar graph $G$ with terminals lying on the outer face the flow-cut gap is $1$. This implies that $\gamma(G) = 1$ and $\gamma(G') = 1$. Invoking Theorem \ref{thm: FlowCutGap} we get that $G'$ is a quality-$1$ flow sparsifier of size $O(k^2)$ for $G$.
\end{proof}

\subsection{An Upper Bound for Distance Sparsifiers} \label{sec: UpperDist}
We first review the notion of Vertex Distance Sparsifiers. Let $G=(V,E,\ww)$ with $K \subset V$ be a $k$-terminal graph, where we replace the capacity function $\cc$ with a weight or length function $\ww : E \rightarrow \mathbb{R}_{\geq 0}$. For a terminal pair $(x,x') \in K$, let $\dist_G(x,x')$ denote the shortest path with respect to the edge lengths $\ww$ in $G$. A graph $H=(V',E',\ww')$ is a \emph{quality-$q$} (\emph{vertex}) \emph{distance sparsifier} of $G$ with $q \geq 1$ if for any $x,x' \in K$, $\dist_G(x,x') \leq \dist_H(x,x') \leq q \cdot \dist_G(x,x')$.

Next we argue that a symmetric approach applies to the construction of vertex sparsifiers that preserve distances. Concretely, we prove that given a $k$-terminal planar graph, where all terminals lie on the outer face, one can construct a quality-$1$ distance sparsifier of size $O(k^{2})$, which is also a planar graph. It is not hard to see that almost all arguments that we used about cut sparsifiers go through, except some adaptations regarding edge lengths in the Wye-Delta rules, edge subdivision operation and vertex splitting operation.

We start adapting the Wye-Delta operations. 

\begin{enumerate}
	\setlength\itemsep{0.1em}
	\item \emph{Degree-one reduction:} Delete a degree-one non-terminal and its incident edge.
	\item \emph{Series reduction:} Delete a degree-two non-terminal $y$ and its incident edges $(x,y)$ and $(y,z)$, and add a new edge $(x,z)$ of length $\ww(x,y) + \ww(y,z)$.
	\item \emph{Parallel reduction:} Replace all parallel edges by a single edge whose length is the minimum over all lengths of parallel edges.
	\item \emph{Wye-Delta transformation:} Let $x$ be a degree-three non-terminal with neighbours $\delta(x) = \{u,v,w\}$. 
	Delete $x$ (along with all its incident edges) and add edges $(u,v),(v,w)$ and $(w,u)$ with lengths $\ww(u,x) + \ww(v,x)$, $\ww(v,x) + \ww(w,x)$ and $\ww(w,x) + \ww(u,x)$, respectively. 
	\item \emph{Delta-Wye transformation:} Let $x$, $y$ and $z$ be the vertices of the triangle connecting them. Assume w.l.o.g.\footnote{Suppose there exists a triangle edge $(x,y)$ with $\ww(x,y) > \ww(x,z) + \ww(y,z)$, where $z$ is the other triangle vertex. Then we can simply set $\ww(x,y) = \ww(x,z) + \ww(y,z)$, since any shortest path between terminal pairs would use the edges $(x,z)$ and $(y,z)$ instead of the edge $(x,y)$.} that for any triangle edge $(x,y)$, $\ww(x,y) \leq \ww(x,z) + \ww(y,z)$, where $z$ is the other triangle vertex. Delete the edges of the triangle, introduce a new vertex $w$ and add new edges $(w,x)$, $(w,y)$ and $(w,z)$ with edge lengths $(\ww(x,y) + \ww(x,z) - \ww(y,z))/2,$ $(\ww(x,z) + \ww(y,z) - \ww(x,u))/2$ and $(\ww(x,y) + \ww(y,z) - \ww(x,z))/2$, respectively.
\end{enumerate}
The following lemma shows that the above rules preserve exactly all shortest path distances between terminal pairs. 

\begin{lemma} Let $G$ be a $k$-terminal graph and $G'$ be a $k$-terminal graph obtained from $G$ by applying one of the rules 1-5. Then $G'$ is a quality-$1$ distance sparsifier of $G$.
\end{lemma}
We remark that there is no need to re-define the Edge deletion and replacement operations, since they are just a combination of the above rules. An analogue of Lemma \ref{lemm: cutReducability} can also be shown for distances. We now modify the Edge subdivision operation, which is used when dealing with graph embeddings (see Section \ref{sec: graphEmbeddings}).
\begin{enumerate}
	\item \emph{Edge subdivision}: Let $(u,v)$ be an edge of length $\ww(u,v)$. Delete $(u,v)$, introduce a new vertex $w$ and add edges $(u,w)$ and $(w,v)$, each of length $\ww(u,v)/2$. 
\end{enumerate}

We now prove an analogue to Lemma \ref{lemm: cutembedding}.

\begin{lemma} \label{lemm: DistancePreservation}
	Let $\rho$ be a node embedding and let $G_1$ and $\rho(G_1)$ be $k$-terminal graphs as defined in Section~\ref{sec: graphEmbeddings}. Then $\rho(G_1)$ preserves exactly all shortest path distances between terminal pairs.
\end{lemma}
\begin{proof}
	We can view each path obtained from the embedding as taking the edge corresponding to that path endpoints in $G_1$ and performing edge subdivisions finitely many times. We claim that such subdivisions preserve all terminal shortest paths.
	
	Indeed, let us consider a single edge subdivison for $(u,v)$ (the general claim then follows by induction on the number of edge subdivions). Fix $x,x' \in K$ and consider some shortest path $p(x,x')$ in $G_1$ that uses $(u,v)$. We can construct in $\rho(G_1)$ a path $q(x,x')$ of the same length as follows: traverse the subpath $p(x,u)$, traverse the edges $(u,w)$ and $(w,v)$ and finally traverse the subpath $p(v,x')$. It follows that $\sum_{e \in p(x,x')} \ww(e) = \sum_{e \in q(x,x')} \ww(e)$, and thus $\dist_{\rho(G_1)}(s,t) \leq \dist_{G_1}(s,t)$.
	
	On the other hand, fix $x,x' \in K$ and consider some shortest path $p'(x,x')$ in $\rho(G_1)$ that uses the two subdivided edges $(u,w)$ and $(w,v)$ (note that it cannot use only one of them). We can construct in $G_1$ a path $q'(x,x')$ of the same length as follows: traverse the subpath $p'(x,u)$, traverse the edge $(u,v)$ and finally traverse the subpath $p'(v,x')$. It follows that $\sum_{e \in p'(x,x')} \ww(e) = \sum_{e \in q'(x,x')} \ww(e)$ and thus $\dist_{G_1}(s,t) \leq \dist_{\rho(G_1)}(s,t)$. Combining the above gives the lemma.
\end{proof}

We next consider vertex splitting for graphs whose maximum degree is larger than three. For each vertex $v$ of degree $d > 3$ with $u_1,\ldots,u_d$ adjacent to $v$, we delete $v$ and introduce new vertices $v_1, \ldots, v_d$ along with edges $\{(v_i,v_{i+1}) : i = 1,\ldots,d-1\}$, each of length $0$. Furthermore, we replace the edges $\{(u_i,v) : i=1,\ldots,d\}$ with $\{(u_i,v_i) : i = 1,\ldots, d\}$, each of corresponding length. If $v$ is a terminal vertex, we make one of the $v_i$'s be a terminal vertex. An analogue to Claim \ref{claim: embedding} gives that the resulting graph preserves all terminal shortest path distances.

We finally note that whenever we add dummy edges of capacity $0$ in the cut setting, we replace them by edges of length $D+1$ in the distance setting, where $D$ is the sum over all edge lengths in the graph we consider. Since any shortest path in the graph does not use the added edges, the terminal shortest path remain unaffected. The above discussion leads to the following theorem. 

\begin{theorem} \label{thm: DistanceSparsifier}
	Let $G$ be a $k$-terminal planar graph where all terminals lie on the outer face. Then $G$ admits a quality-$1$ distance sparsifier of size $O(k^{2})$, which is also a planar graph. 
\end{theorem}

\subsection{Incompressibility of Distances in $k$-Terminal Graphs} \label{sec: lowerBound_Reach}

In this section we prove the following incompressibility result (i.e., Theorem~\ref{thm:incompressibility}) concerning the trade-off between quality and size of any compression function when estimating terminal distances in $k$-terminal graphs: for every $\varepsilon > 0$ and $t \geq 2$, there exists a (sparse) $k$-terminal $n$-vertex graph such that $k=o(n)$, and that any compression algorithm that approximates pairwise terminal distances within a factor of $t - \varepsilon$ or an additive error $2t-3$ must use $\Omega(k^{1+1/(t-1)})$ bits. Our lower bound is inspired by the work of Matou{\v{s}}ek~\cite{matousek96}, which has also been utilized in the context of distance oracles~\cite{ThorupZ05}. Our arguments rely on the recent extremal combinatorics construction (see~\cite{cheung2016}) that was used to prove lower bounds on the size of distance approximating minors. 

\paragraph*{Discussion on our result.} Note that for any $k$-terminal graph $G$, if we do not have any restriction on the structure of the distance sparsifier, then $G$ always admits a trivial quality $1$ distance sparsifier $H$ which is the complete weighted graph on $k$ terminals with each edge weight being equal to the distance between the two endpoints in $G$. Furthermore, by the well-known result of Awerbuch~\cite{Awerbuch85}, such a graph $H$ in turn admits a multiplicative $(2t-1)$-\emph{spanner} $H'$ with $O(k^{1+1/t})$ edges, that is, all the distances in $H$ are preserved up to a multiplicative factor of $2t-1$ in $H'$, for any $t\geq 1$. This directly implies that the $k$-terminal graph $G$ has a quality $2t-1$ distance sparsifier with $k$ vertices and $O(k^{1+1/t})$ edges. On the other hand, though \emph{unconditional} lower bounds of type similar to our result have been known for the number of edges of spanners~\cite{lazebnik1995,Woodruff2006}, we are not aware of such lower bounds for the size of \emph{data structure} that preserves pairwise terminal distances for any $k$-terminal $n$-vertex graph when $k=o(n)$. In the extreme case when $k=n$ (i.e., all the vertices are terminals), the recent work by Abboud and Bodwin~\cite{abboudBodwin16} shows that any data structure that preserves the distances with an additive error $t$ needs $\Omega(n^{4/3-\varepsilon})$ bits, for any $\varepsilon>0, t=O(n^{\delta})$ and $\delta=\delta(\varepsilon)$ (see also the follow-up work~\cite{abboud2017}).

We start by reviewing a classical notion in combinatorial design.
\begin{definition} [Steiner Triple System] Given a ground set $K=[k]$, an $(3,2)$-Steiner system (abbr. $(3,2)$-\emph{SS}) of $K$ is a collection of $3$-subsets of $K$, denoted by $\mathcal{S} = \{S_1,\ldots,S_r\}$, where $r = \binom{k}{2}\left/3\right.$, such that every $2$-subset of $K$ is contained in \emph{exactly} one of the $3$-subsets.
\end{definition}

\begin{lemma}[\cite{Wilson75}] For infinity many $k$, the set $K=[k]$ admits an $(3,2)$-\emph{SS}.
\end{lemma}

Roughly speaking, our proof proceeds by forming a $k$-terminal bipartite graph, where terminals lie on one side and non-terminals on the other. The set of non-terminals will correspond to some subset of a Steiner Triple System $\mathcal{S}$, which will satisfy some \emph{certain} property. One can equivalently view such a graph as taking union over \emph{star} graphs. Before delving into details, we need to review a couple of other useful definitions and the construction from~\cite{cheung2016}. 

\paragraph*{\textbf{Detour Graph and Cycle.}} 
Let $k$ be an integer such that $K=[k]$ admits an $(3,2)$-SS. Let $\mathcal{S}$ be such an $(3,2)$-SS. We associate $\mathcal{S}=\{S_1,\ldots,S_r\}$ with a graph whose vertex set is $\mathcal{S}$. We refer to such graph as a \emph{detouring graph}.  By the definition of Steiner system, it follows that $|S_i \cap S_j|$ is either zero or one. Thus, two vertices $S_i$ and $S_j$ are adjacent in the detouring graph iff $|S_i \cap S_j|=1$. It is also useful to label each edge $(S_i, S_j)$ with the terminal in $S_i \cap S_j$. A \emph{detouring cycle} is a cycle in the detouring graph such that no two neighbouring edges in the cycles have the same terminal label. Observe that the detouring graph has other cycles which are not detouring cycles. 

Ideally, we would like to construct detouring graphs with long detouring cycles while keeping the size of the graph as large as possible. One trade-off is given in the following lemma.

\begin{lemma}[\cite{cheung2016}] \label{lemm: detouringGraph} For any integer $t \geq 3$, given a detouring graph with vertex set $\mathcal{S}$, there exists a subset $\mathcal{S}' \subset \mathcal{S}$ of cardinality $\Omega(k^{1+1/(t-1)})$ such that the induced graph on $\mathcal{S}'$ has no detouring cycles of size $t$ or less.
\end{lemma}

Now we are ready to prove our incompressibility result regarding approximately preserving terminal pairwise distances.
\paragraph*{\textbf{Proof of Theorem~\ref{thm:incompressibility}:}} 
Let $k$ be an integer such that $K=[k]$ admits an $(3,2)$-SS $\mathcal{S}$. Fix some integer $t \geq 3$, some positive constant $c$ and use Lemma \ref{lemm: detouringGraph} to construct a subset $\mathcal{S}'$ of $\mathcal{S}$ of size $\Omega(k^{1+1/(t-1)})$ such that the induced graph on $\mathcal{S}'$ has no detouring cycles of size $t$ or less. We may assume w.l.o.g. that $\ell = |\mathcal{S}'| = c \cdot k^{1+1/(t-1)}$ (this can be achieved by repeatedly removing elements from $\mathcal{S}'$, as the property concerning the detouring cycles is not destroyed). Fix some ordering among $3$-subsets of $\mathcal{S}'$ and among terminals in each $3$-subset.

We define the $k$-terminal graph $G$ as follows:
\begin{itemize}
\item For each $e_i \in \mathcal{S}'$ create a non-terminal vertex $v_i$. Let $V_{\mathcal{S}'}$ denote the set of such vertices. The vertex set of $G$ is $K \cup V_{\mathcal{S}'}$, where $K=[k]$ denotes the set of terminals.
\item For each $e_i \in \mathcal{S}'$, connect $v_i$ to the three terminals $\{x^{i}_{1},x^{i}_{2},x^{i}_{3}\}$ belonging to $e_i$, i.e., add edges $(v_i,x^{i}_j)$, $j=1,2,3$.
\end{itemize}
Note that $G$ is sparse since both the number of vertices and edges are $\Theta(\ell)$, and it also holds that $k=o(|V(G)|)$.

For any subset $R \subseteq \mathcal{S}'$, we define the subgraph $G_R=(V(G), E_R)$ of $G$ as follows. For each $e_i \in S'$, if $e_i \in R$, perform no changes. If $e_i \not \in R$, delete the edge $(v_i,x^{i}_1)$. Note that there are $2^{\ell}$ subgraphs $G_R$. We let $\mathcal{G}$ denote the family of all such subgraphs. 

We say a terminal pair $(x,x')$ \emph{respects $\mathcal{S'}$} if in the $(3,2)$-SS $\mathcal{S}$, the unique $3$-subset $e$ that contains $x$ and $x'$ belongs to $\mathcal{S'}$. Given $R \subseteq \mathcal{S}'$ and some terminal pair $(x,x')$, we say that $R$ \emph{covers} $(x,x')$ if both $x$ and $x'$ are connected to some non-terminal $v$ in $G_R$. 

\begin{claim}~\label{claim:cover} 
	For all $R \subseteq \mathcal{S}'$ and terminal pairs $(x,x')$ covered by $R$ we have that $\dist_{G_R}(x,x')=2$.
\end{claim}
\begin{proof}
By the definition of Steiner system and the construction of $G_R$, the shortest path between $x$ and $x'$ is simply a $2$-hop path, i.e., $\dist_{G_R}(x,x') = 2$. 
\end{proof}

\begin{claim}~\label{claim:non_cover}
For all $R \subseteq \mathcal{S}'$ and any terminal pair $(x,x')$ that respects $\mathcal{S}'$ and is \emph{not} covered by $R$, we have that $\dist_{G_R}(x,x')\geq 2t$.
\end{claim}
\begin{proof}
Since $(x,x')$ respects $\mathcal{S'}$, there exists $e_i=(x^i_1,x^i_2,x^i_3)\in \mathcal{S}'$ that contains both $x$ and $x'$. By construction of $G_R$ and the fact that $(x,x')$ is not covered by $R$, it follows that $e_i\in \mathcal{S}'\setminus R$, and one of $x,x'$ corresponds to $x^i_1$ and the other corresponds to $x^i_2$ or $x^i_3$. W.l.o.g., we assume $x=x^i_1$ and $x'=x^i_2$. Note that there is no edge connecting $x^{i}_1$ with the non-terminal $v_i$ that corresponds to $e_i$. Now by Lemma~\ref{lemm: detouringGraph}, the detouring graph induced on $\mathcal{S}'$ has no detouring cycles of size $t$ or less, which implies that any other simple path between $x^{i}_1$ and $x^{i}_2$ in $G$ must pass through at least $t-1$ other terminals. Let $w_1,\ldots,w_{t-1}$ be such terminals and let $P:=x^{i}_1 \rightarrow w_1,\ldots,w_{t-1} \rightarrow x^{i}_2$ denote the corresponding path, ignoring the non-terminals along the path. Between any consecutive terminal pairs in $P$, the shortest path is at least $2$. Thus, the length of $P$ is at least $2t$, i.e., $\dist_{G_R}(x^{i}_1,x^{i}_2) \geq 2t$.
\end{proof}

Fix any two subsets $R_1, R_2 \subseteq \mathcal{S}'$ with $R_1 \neq R_2$. It follows that there exists a $3$-subset $e_i= (x^i_1,x^i_2,x^i_3)\in \mathcal{S}'$ such that either $e \in R_1 \setminus R_2$ or $e \in R_2 \setminus R_1$. Assume w.l.o.g that $e \in R_2 \setminus R_1$. Note that $(x^i_1,x^i_2)$ respects $\mathcal{S}'$ and it is covered in $R_2$ but not in $R_1$. By~Claim~\ref{claim:cover} and~\ref{claim:non_cover}, it holds that $\dist_{G_{R_{2}}}(x^i_1,x^i_2) = 2$ and $\dist_{G_{R_1}}(x^i_1,x^i_2) \geq 2t$. In other words, there exists a set $\mathcal{G}$ of $2^{\ell}$ different subgraphs on the same set of nodes $V(G)$ satisfying the following property: for any $G_1,G_2 \in \mathcal{G}$, there exists a terminal pair $(x,x')$ such that the distances between $x$ and $x'$ in $G_1$ and $G_2$ differ by at least a $t$ factor as well as by at least $2t-2$. On the other hand, for any compression function that approximates terminal path distances within a factor of $t-\varepsilon$ or an additive error $2t-3$ and produces a bitstring with less than $\ell$ bits, there exist two different graphs $G_1,G_2 \in \mathcal{G}$ that map to the same bit string. Hence, any such compression function must use at least $\Omega(\ell) = \Omega(k^{1+1/(t-1)})$ bits if we want to preserve terminal distances within a $t-\varepsilon$ factor or an additive error $2t-3$. 



To complete our argument, we need to show the claim for quality $t=2$. The only significant modification we need is the usage of an $(3,2)$-SS in the construction of graph $G$ (instead of using a subset of it). The remaining details are similar to the above proof and we omit them here. 

\section{Conclusion}
In this chapter, we studied vertex sparsifiers for preserving reachability information, cuts, and distances. Our first contribution is studying the notion of reachability preserving minors, which are sparsifiers that preserve reachability information among a given set of terminals and are obtained by performing minor operations on given input graphs. We show that any $k$-terminal planar graph admits a reachability preserving minor of size $O(k^2 \log k)$, and then prove that this result is up to a logarithmic factor in grid graphs. For general graphs we obtain an upper bound of $O(k^{3})$. The algorithmic and lower bound constructions behind these results bring together techniques from reachability oracles and counting branching events in shortest path computations. Interesting open problems include closing the gap between the best-known upper and lower bounds in general graphs and improving the running time of our algorithms.

Our second contribution is studying vertex sparsifiers that preserve cuts and distances when restricted to planar graphs with terminals lying on the same faces, which are sometimes referred to as Okamura-Seymour (OS) graphs. For any $k$-terminal OS graph, we show that there exist quality-$1$ cut and distance sparsifiers that at the same time preserve planarity. The main idea behind these results is to adapt a local reduction technique, known as Why-Delta transformation, to the cut and distance measure. An important open problem is whether one can extend this technique to remove the assumption on the location of terminal vertices, or prove a non-trivial bound in the more general setting where terminals lie on a bounded number of faces, similar to Krauthgamer and Rika~\cite{krauthgamer2017refined}.

\printbibliography



\end{document}